\documentclass[aps,prd,twocolumn,showpacs,amsmath,amssymb,superscriptaddress,longbibliography,showkeys]{revtex4-1}
\usepackage{amsmath}
\usepackage{epsfig}
\usepackage{graphicx}
\usepackage{subfigure}
\usepackage{epstopdf}
\usepackage{color}
\usepackage{multirow}
\usepackage{setspace}
\usepackage{overpic}
\usepackage{amssymb}
\usepackage[bookmarksnumbered, pdfstartview=FitH,colorlinks,urlcolor=blue, citecolor=blue,linkcolor=blue] {hyperref}
\usepackage{lineno}
\usepackage{bm}
\usepackage{rotating}
\usepackage{siunitx}
\usepackage[utf8]{inputenc}
\usepackage{orcidlink}

\let\oldequation\equation
\let\oldendequation\endequation

\renewenvironment{equation}
  {\linenomathNonumbers\oldequation}
  {\oldendequation\endlinenomath}

\usepackage{makecell}

\begin{document}

\title{\bf \boldmath
Review of experimental studies of charmed meson decays at BESIII}

\author{Yijia Zeng\orcidlink{0009-0005-3279-0304}}
\affiliation{Institute of High Energy Physics, Beijing 100049, People's Republic of China}
\author{Xiang Pan\orcidlink{0000-0002-0423-8986}}
\email{panxiang@seu.edu.cn}
\affiliation{School of Physics, Southeast University, Nanjing 210094, People’s Republic of China}

\author{Hailong Ma\orcidlink{0000-0001-9771-2802}}
\email{mahl@ihep.ac.cn}
\affiliation{Institute of High Energy Physics, Beijing 100049, People's Republic of China}

\begin{abstract}
Experimental measurements of different decays of charmed mesons ($D^0$, $D^+$, and $D^+_s$ as well as their excitations $D^{*0}$, $D^{*+}$, and $D_s^{*+}$)
have been extensively performed at BESIII.
Precision measurements of absolute branching fractions of different decays, the decay constants of $D^+$ and $D^+_s$ mesons,
hadronic transition form factors of $D$ transitions to light hadrons ($K$, $\pi$, $\eta$, $\eta^\prime$, $K^*(892)$, $\rho$, $\omega$, $\phi$, $K_1(1270)$, $f_0(500)$, $f_0(980)$, $a_0(980)$),
$c\to s(d)$ Cabibbo-Kobayashi-Maskawa (CKM) matrix elements,
tests of lepton flavor universality with various (semi)leptonic $D$ decays,
precision measurements of strong phase difference between $D^0$ and $\bar D^0$ decays,
amplitude analyses of multibody hadronic $D_{(s)}$ decays,
searches for rare $D$ decays have been reported.
The reported results offer important information to test different theoretical calculations, to test the unitarity of the CKM matrix, and to search for new physics effects beyond the standard model~(SM). This paper reviews experimental studies of different decays of $D^0$, $D^+$, and $D^+_s$
as well as their excitations at BESIII as of April 15, 2026.
Based on existing results of (semi)leptonic $D$ decays from all experiments, we have presented the most precise averages for the CKM matrix elements $|V_{cs}|=0.9686\pm0.0009\pm0.0038$ and $|V_{cd}|=0.2259\pm0.0014\pm0.0013$, the decay constants of $D^+$ and $D^+_s$ $f_{D^+}=(213.1\pm2.0\pm1.5)$ MeV and $f_{D^+_s}=(253.2\pm1.2\pm1.6)$ MeV, as well as the hadronic transition form factors $f^{D\to K}_+(0)=0.7376\pm0.0007\pm0.0013$, $f^{D\to \pi}_+(0)=0.6337\pm0.0053\pm0.0037$, $f^{D\to \eta}_+(0)=0.351\pm0.009\pm0.005$, $f^{D\to \eta^\prime}_+(0)=0.263\pm0.025\pm0.006$, $f^{D_s\to \eta}_+(0)=0.4653\pm0.0058\pm0.0069$, $f^{D_s\to \eta^\prime}_+(0)=0.535\pm0.020\pm0.011$, and $f^{D_s\to K^0}_+(0)=0.627\pm0.036\pm0.009$, where the first and second uncertainties are statistical and systematic, respectively.
\end{abstract}


\maketitle

\tableofcontents

\oddsidemargin  -0.2cm
\evensidemargin -0.2cm

\section{Introduction}

The charm quark was first observed in November 1974 through the discovery of the $J/\psi$ particle~\cite{SLAC-SP-017:1974ind,E598:1974sol}. In the quark model, a charm quark can combine with antilight quarks ($\bar u$, $\bar d$, and $\bar s$) to form open-charm charmed mesons ($D^0$, $D^+$, and $D^+_s$).
The $D^0$ and $D^+$, together with their vector excitations $D^{*0}$ and $D^{*+}$, were subsequently observed in 1976~\cite{Goldhaber:1976xn,Peruzzi:1976sv}. Several years later, the $D^+_s$~\cite{DASP:1977eel,DASP:1978gcx} and its vector excitation $D_s^{*+}$~\cite{MARK-III:1987hge} were discovered.

Experimental studies of (semi)leptonic decays of these charmed mesons (denoted as $D$ for simplicity)
are important to explore weak and strong effects in the charm decays
for the following reasons~\cite{Silverman:1988gc,Becher:2005bg,Faustov:2019mqr}.
Theoretically, strong effects in (semi)leptonic $D$ decays are parameterized as  decay constants
or hadronic transition form factors,
while weak effects are parameterized as the $c\to q$ CKM matrix element $|V_{cq}|$~($q=s$ or $d$).
Experimental studies of these decays offer an ideal opportunity to determine decay constants
or hadronic transition form factors and $|V_{cq}|$,
which are important to test the lattice QCD~(LQCD) calculations of decay constants
or hadronic transition form factors and the unitarity of the CKM matrix.
In addition, precision measurements of the ratios of branching fractions of (semi)leptonic $D$ decays with different lepton families
are important to test lepton flavor universality in the charm sector.

Hadronic or semileptonic $D$ decays provide an invaluable window into the spectroscopy and dynamics of light hadrons.
Intensive amplitude analyses of multi-body hadronic $D$ decays help to disentangle the underlying substructures and extract detailed
information about intermediate resonant states, including vector, scalar, axial-vector, and tensor mesons,
which are important for advancing our understanding of low-energy QCD and properties of light hadrons~\cite{Ryd:2009uf}.
Especially, the quantum correlation of the $D^0\bar D^0$ meson pairs produced at the $\psi(3770)$ provides
a unique way to probe the amplitudes of the $D$ decays, the $D^0\bar D^0$  mixing parameters and potential
$CP$ violation in $D^0$ decays~\cite{Xing:1996pn}.
Furthermore, the determination of the strong-phase difference between the Cabibbo-favored (CF) and doubly
Cabibbo-suppressed (DCS) amplitudes in the decay of quantum-correlated $D^0\bar D^0$ meson pairs has several motivations: understanding the non-perturbative QCD effects in the charm sector; serving as essential inputs for extracting the angle $\gamma$ of the CKM unitarity triangle (UT); and relating the measured mixing parameters in hadronic decays $(x', y')$ to the mass and width difference parameters $(x, y)$~\cite{HFLAV:2016hnz,HFLAV:2024ctg}.

In the SM, rare decays of charmed mesons are heavily suppressed due to the Glashow-Iliopoulos-Maiani (GIM) mechanism and the absence of large penguin contributions, rendering their branching fractions extremely small~\cite{Fajfer:2001sa, Paul:2011ar,Cappiello:2012vg}. However, various new physics scenarios beyond the SM predict enhancements in these decay channels, potentially bringing them within reach of experimental sensitivity~\cite{Fajfer:2001sa, Paul:2011ar}.
The search for rare $D$ decays therefore serves as a powerful probe for physics beyond the SM,
thereby providing stringent constraints on the parameter space of new physics models.

This paper reviews the experimental studies of various decays of $D^0$, $D^+$, $D^+_s$, and their excited states at BESIII, as of April 15, 2026.
This paper is arranged as follows. Section~\ref{sec:experiment} describes the experiments that have studied charmed meson decays. Section~\ref{sec:leptonic} discusses experimental measurements of leptonic decays of $D^+$, $D^+_s$, and their excited states. Section~\ref{sec:semileptonic} presents measurements of semileptonic decays of $D^0$, $D^+$, and $D^+_s$. Section~\ref{sec:strongphase} provides measurements of strong phase differences between $D^0$ and $\bar D^0$.  Section~\ref{sec:hadronBFs} reports absolute branching fraction measurements of hadronic decays of $D^0$, $D^+$, and $D^+_s$. Section~\ref{sec:amplitude} covers amplitude analyses of multibody hadronic decays of $D^0$, $D^+$, and $D^+_s$. Section~\ref{sec:raredecays} discusses searches for rare decays of $D^0$, $D^+$, and $D^+_s$. 
 Section~\ref{sec:others} discusses other measurements of $D^{*0}$, $D^{*+}$, and $D^{*+}_s$.
Section~\ref{sec:summary} gives prospects for future studies of charm meson decays and a brief summary. Throughout this paper,
the $D$ or $D_{(s)}$ denotes $D^0$, $D^+$, and $D^+_s$; all decays include their charge conjugates; and uncertainties are given as (first: statistical, second: systematic, third: external) unless stated otherwise.
At BESIII, the charmed meson samples are relatively small compared to those from $B$ factories and large hadron collider experiments; therefore, no competitive results on $D^0$-$\bar D^0$ mixing and $CP$ violation have been reported and the related topics are not included in this article.

\section{Experiments}
\label{sec:experiment}

Experimentally, charmed mesons, $D^0$, $D^+$, and $D^+_s$ as well as their excited states,
can be produced in various environments, including $e^+e^-$ collisions, fixed-target interactions involving hadrons,
photons, or neutrinos with nuclear targets, and hadron collisions.

\subsection{$e^+e^-$ collision near threshold}

The $e^+e^-$ collision experiments operating around the $\psi(3770)$ resonance, corresponding to a center-of-mass energy of approximately 3.773 GeV,
where the $\psi(3770)$ decays predominantly into $D^0\bar D^0$ and $D^+D^-$ pairs, provide the cleanest environment for studying
decays of $D^0$ and $D^+$ mesons~\cite{BES:2008rkr,BESIII:2018iev}. Historically, the MARKI/II/III~(operated in the 1970s and 1980s)~\cite{Larsen:1976ju,Bernstein:1983wk}, BESII~(1997-2003)~\cite{BES:1994bjo,BES:2001vqx}, CLEO-c~(2003-2008)~\cite{Peterson:2002sk,CLEO:1991qyy,Artuso:2002ya}, and BESIII~(operated from 2009 to present)~\cite{BESIII:2009fln} experiments collected data samples at this resonance.
The integrated luminosities of data samples taken at MARKIII, BESII, CLEO-c and BESIII are 9.3~pb$^{-1}$, 33~pb$^{-1}$, 818~pb$^{-1}$, and 20.3~fb$^{-1}$~\cite{Ablikim:2013ntc,BESIII:2024lbn}, respectively.

Around the $\psi(3770)$, $D^0\bar D^0$ and $D^+D^-$ pairs are produced without accompanying hadrons. This feature enables the determination of absolute branching fractions for $D^0$ or $D^+$ meson decays using the double-tag method first developed by MARKIII~\cite{MARK-III:1990bbt}. In this method, one $\bar D$ meson is reconstructed in a set of “golden” hadronic decay modes. Events in which a $D$ decay is identified on the side recoiling against the selected single-tag $\bar D$ meson are called double-tag $D\bar D$ events. By combining the single-tag and double-tag samples, the absolute branching fractions of different $D$ decays can be measured.

To extract the yields of single-tag $\bar D$ mesons, two key variables are typically defined: the energy difference $\Delta E$ and the beam-constrained mass $M_{\rm BC}$.
Single-tag $\bar D^0$ mesons are reconstructed using six hadronic decay channels: $\bar D^0\to K^+\pi^-$, $K^+\pi^-\pi^0$, $K^+2\pi^-\pi^+$, $K^+\pi^-2\pi^0$, $K^+2\pi^-\pi^+\pi^0$, and $K^0_S\pi^+\pi^-$.
A total of about 21.4 million single-tag $\bar D^0$ mesons are obtained, with the relative fractions of 18.4\%, 36.7\%, 24.7\%, 8.8\%, 5.7\%, and 5.8\%, respectively~\cite{BESIII:2026uin,BESIII:2026ydr}.
Nine hadronic decay modes $D^-\to K^{+}2\pi^{-}$, $K^0_{S}\pi^{-}$, $K^{+}2\pi^{-}\pi^{0}$, $K^0_{S}\pi^{-}\pi^{0}$, $K^0_{S}\pi^{+}2\pi^{-}$, $K^{+}K^{-}\pi^{-}$, $K^0_SK^-$, $2\pi^-\pi^+$, and $K^+\pi^+3\pi^-$ are used to reconstruct single-tag $D^-$ mesons. Approximately 11.5 million single-tag $D^-$ mesons are identified, with fractions of 49.9\%, 6.0\%, 15.0\%, 12.6\%, 7.1\%, 4.3\%, 1.2\%, 1.9\%, and 2.0\%, respectively~\cite{BESIII:2013iro,BESIII:2024kvt}.
 Most measurements employ three “golden” $\bar D^0$ modes of $\bar D^0\to K^+\pi^-$, $K^+\pi^-\pi^0$, and $K^+2\pi^-\pi^+$ and six golden $D^-$ modes of $D^-\to K^{+}2\pi^{-}$, $K^0_{S}\pi^{-}$, $K^{+}2\pi^{-}\pi^{0}$, $K^0_{S}\pi^{-}\pi^{0}$, $K^0_{S}\pi^{+}2\pi^{-}$, and $K^{+}K^{-}\pi^{-}$.
It should be noted that measurements based on the 2.93~fb$^{-1}$ dataset taken in 2010 and 2011 used different photon selection criteria as well as different $\Delta E$ and $M_{\rm BC}$ windows, leading to differences of several percent to over ten percent in the yields for each single-tag mode.

For $D^+_s$ decays, the ideal center-of-mass energy for $e^+e^-$ collisions is near 4.03, 4.17, or 4.26~GeV,
where $D^+_s D^-_s$, $D^\pm_s D^{*\mp}_s$, or $D^{*+}_s D^{*-}_s$ pairs are produced dominantly,
according to the charm production cross sections between 3.97 and 4.26 GeV~\cite{BES:1999vjc,CLEO:2008ojp}.
BESI accumulated 22.3~pb$^{-1}$ of data at 4.03~GeV and 1.5~pb$^{-1}$ of data at 4.14~GeV~\cite{BES:1999vjc},
CLEO-c collected 600~pb$^{-1}$ of data at 4.17~GeV, and BESIII accumulated 482~pb$^{-1}$ of data at 4.009~GeV and
7.33~fb$^{-1}$ of data at 4.128-4.226~GeV~\cite{BESIII:2015qfd,BESIII:2022dxl},
and 10.64~fb$^{-1}$ of data at 4.237-4.699~GeV.
The technique for studying $D^+_s$ decays via $e^+e^-\to D^+_s D^-_s$ at 4.009~GeV and $e^+e^-\to D^{*+}_s D^{*-}_s$ at 4.237-4.699~GeV,
where $D^+_s D^-_s$ or $D^{*+}_s D^{*-}_s$ pairs are produced, is almost identical to that used for $D^{0(+)}$ decays at 3.773~GeV. Usually, single-tag $D_s^-$ mesons are reconstructed using sixteen hadronic $D_s^-$ decay modes:
$K^+K^-\pi^-$,
$K^+K^-\pi^-\pi^0$,
$\pi^+2\pi^-$,
$K_S^0K^-$,
$K_S^0K^-\pi^0$,
$K^-\pi^+\pi^-$,
$2K_S^0\pi^-$,
$K_S^0K^+2\pi^-$,
$K_S^0K^-\pi^+\pi^-$,
$\eta_{\gamma\gamma}\pi^-$,
$\eta_{\pi^+\pi^-\pi^0}\pi^-$,
$\eta^\prime_{\pi^+\pi^-\eta_{\gamma\gamma}}\pi^-$,
$\eta^\prime_{\gamma\rho(770)^0}\pi^-$,
$\eta_{\gamma\gamma}\rho(770)^-$,
$\eta_{\pi^+\pi^-\pi^0}\rho(770)^-$, and
$\eta_{\gamma\gamma}\pi^+2\pi^-$~\cite{BESIII:2023cym},
where the subscripts on the $\eta(\eta^{\prime})$ represent the decay modes used to reconstruct the $\eta(\eta^{\prime})$.
From these two data samples, however, only about 15k single-tag $D^-_s$ mesons and 0.124 million single-tag $D^{*-}_s$ mesons are obtained.
The most precise measurements of different $D^+_s$ decays at BESIII are based on the data samples at 4.128-4.226~GeV,
where those at 4.128, 4.157, 4.178, 4.189, 4.199, 4.209, 4.219, and 4.226 GeV
correspond to 0.40, 0.41, 3.19, 0.57, 0.53, 0.57, 0.57, and 1.09 fb$^{-1}$, respectively.
In general, background levels around 4.178~GeV are much higher than those at 4.009~GeV. Nevertheless, around 4.178~GeV, $D^+_s$ mesons are primarily produced via the $e^+e^- \to D^\pm_s D^{*\mp}_s$ process followed by the $D^{*\mp}_s \to \gamma D^\mp_s$ decay.
The cross section for $e^+e^- \to D^\pm_s D^{*\mp}_s$ around 4.178~GeV is about three times that of $e^+e^- \to D^+_s D^-_s$ at 4.009~GeV~\cite{CLEO:2008ojp}, and the branching fraction of $D^{*\mp}_s \to \gamma D^\mp_s$ is around 94\%~\cite{ParticleDataGroup:2024cfk}.
Mainly due to the higher effective $D^+_s$ production rate, larger $D^+_s$ samples can be obtained at 4.178~GeV than around 4.009~GeV for data samples of the same integrated luminosity.
Taking analyses at 4.128-4.226 GeV as an example, a total of 0.89 million single-tag $D^-_s$ mesons are obtained;
the fractions of 4.128, 4.157, 4.178, 4.189, 4.199, 4.209, 4.219, and 4.226 GeV are
4.3\%, 6.2\%, 49.1\%, 8.2\%, 7.7\%, 7.7\%, 6.6\%, and 10.3\%, respectively; while the fractions of
$K^+K^-\pi^-$,
$K^+K^-\pi^-\pi^0$,
$\pi^+2\pi^-$,
$K_S^0K^-$,
$K_S^0K^-\pi^0$,
$K^-\pi^+\pi^-$,
$2K_S^0\pi^-$,
$K_S^0K^+2\pi^-$,
$K_S^0K^-\pi^+\pi^-$,
$\eta_{\gamma\gamma}\pi^-$,
$\eta_{\pi^+\pi^-\pi^0}\pi^-$,
$\eta^\prime_{\pi^+\pi^-\eta_{\gamma\gamma}}\pi^-$,
$\eta^\prime_{\gamma\rho(770)^0}\pi^-$,
$\eta_{\gamma\gamma}\rho(770)^-$,
$\eta_{\pi^+\pi^-\pi^0}\rho(770)^-$, and
$\eta_{\gamma\gamma}\pi^+2\pi^-$ are
31.6\%, 9.7\%, 8.4\%, 7.1\%,
2.6\%, 3.8\%, 1.2\%, 3.4\%,
1.7\%, 4.4\%, 1.3\%, 2.2\%,
5.7\%, 9.1\%, 2.4\%, and 5.4\%, respectively~\cite{BESIII:2023cym}.
Experimental studies of various $D^+_s$ analyses were performed with different data samples,
tag combinations, and particle selection criteria.

Because the missing neutrino is undetectable at BESIII, the (semi)leptonic $D$ decays can only be selected in the presence of single-tag $\bar D$ candidates.
The information on the (semi)leptonic $D$ decays is inferred from a kinematic quantity defined as $M^2_{\mathrm{miss}}\equiv E^2_{\mathrm{miss}}-|\vec{p}_{\mathrm{miss}}|^2$ or $U_{\mathrm{miss}}\equiv E_{\mathrm{miss}}-|\vec{p}_{\mathrm{miss}}|$~\cite{BESIII:2023cym,BESIII:2024kvt,BESIII:2026ydr}. Here, $E_{\mathrm{miss}}\equiv E_{\mathrm{beam}}-E_{\ell^{+}}(-E_{P})$ and $\vec{p}_{\mathrm{miss}}\equiv \vec{p}_{D}-\vec{p}_{\ell^{+}}(-\vec{p}_{P})$ are the missing energy and momentum of the double-tag event in the $e^+e^-$ center-of-mass frame, in which $E_{P\,(\ell^+)}$ and $\vec{p}_{P\,(\ell^+)}$ are the energy and momentum of the $P$\,($\ell$) candidates, respectively.
For the correctly reconstructed signal events, the $M^2_{\mathrm{miss}}$ or $U_{\mathrm{miss}}$ distributions are expected to concentrate around zero.
The signal yields of the (semi)leptonic $D$ decays were determined by fitting these distributions.
The hadronic $D$ decays can be reconstructed via the single-tag method or the double-tag method.
For studies of hadronic $D^{0(+)}$ and $D^+_s$ decays with the single-tag method,
the signal yields are extracted by fitting individual beam-constrained mass $M_{\rm BC}$ and invariant mass $M_{\rm inv}$
on the tag side.
Meanwhile, yields of hadronic $D^{0(+)}$ and $D^+_s$ decays reconstructed with the double-tag method
are usually determined from
two-dimensional (2-D) fits to the $M_{\rm BC}$ or $M_{\rm inv}$ distributions on both the tag and signal sides or
one-dimensional fits to the $M_{\rm BC}$ or $M_{\rm inv}$ distributions of the signal sides.

\subsection{$e^+e^-$ collision at higher energy}

Charmed mesons, $D^0$, $D^+$, and $D^+_s$, can also be produced through quark fragmentation.
As a result, their decays can be studied using data samples collected by $e^+e^-$ experiments operating at center-of-mass energies around
10.5~GeV and 91~GeV, known as $B$ and $Z_0$ factories, respectively.
ARGUS~(1982-1987)~\cite{ARGUS:1988bds} operated at the DORISII storage ring at DESY,
CLEOI/II/III~(1979-2003)~\cite{Peterson:2002sk,CLEO:1991qyy,Artuso:2002ya} at the CESR storage ring,
BaBar(operated from 1999 to 2008)~\cite{BaBar:2001yhh} at PEP-II and Belle~(operated from 1999 to 2010)~\cite{Belle:2000cnh} at the asymmetric energy KEKB collider,
accumulated large data samples near 10.5~GeV.
Using these large $e^+e^- \to c\bar c$ samples, measurements of charmed meson decays have been carried out.
At $B$ factories, the analysis method for studying charmed meson decays requires unfolding the fragmentation process.
The total yield of $D$ mesons in the data sample is estimated by reconstructing the four-momentum of $D$ candidates recoiling against the rest of the event,
while the number of reconstructed events is obtained by identifying the signal $D$ decay and reconstructing the entire event.
Measurements of (semi)leptonic, hadronic, and rare $D$ decays as well as $D^0$-$\bar D^0$ mixing parameters,
and searches for $CP$ violation in charm decays have been reported.
A few studies were also reported by the ALEPH~\cite{Anelli:1989yw}, L3~\cite{L3:1989aa}, and OPAL~\cite{OPAL:1990yff} experiments at LEP,
operated from 1989 to 2000, based on data samples of hadronic $Z_0$ decay events at 91~GeV;
and the HRS~(High Resolution Spectrometer, 1980-1986)~\cite{Derrick:1981wx} experiment at PEP,
based on data at 29~GeV.

\subsection{Fixed-target experiments}

Early experimental studies of (semi)leptonic, hadronic, and rare $D$ decays were also from
fixed-target experiments, where charm mesons are produced via the interaction of incident particles with target nuclei.
From the secondary particles generated in such interactions, charmed mesons can be identified, enabling detailed studies of their
decay properties. Although the charm production cross sections in fixed-target experiments are larger than those at $e^+e^-$ colliders,
the non-charm background is also substantially higher.

During the 1980s and 1990s, there were a series of high-energy physics fixed-target experiments conducted at Fermilab in the US.
E653~\cite{TaggedPhotonSpectrometer:1987poq} employed a 600~GeV proton beam on a copper/tungsten target instrumented with hybrid emulsion,
achieving micron scale vertex resolution.
E691~\cite{TaggedPhotonSpectrometer:1987poq} used a tagged photon beam of 100–200~GeV on a beryllium target, and it became a true charm factory.
E687~\cite{E-687:1990ses} succeeded with a similar photon beam energy but upgraded to a silicon-microstrip target that improved vertexing.
E791~\cite{Amato:1992em} took a different approach with a 500~GeV $\pi^-$ beam on a segmented carbon/platinum target.
E771~\cite{E771:1995hgv} operated with an 800~GeV proton beam on a silicon target, aimed at both bottom and charm physics.
E789~\cite{E789:1999lnm} used an 800~GeV proton beam but employed a thin, segmented target coupled with a specialized trigger designed
to select low-multiplicity events.
SELEX (E781)~\cite{SELEX:1998lcb} operated with 600 GeV $\Sigma^-$ and $\pi^-$ beams, as well as a 540 GeV proton beam, directed onto copper and diamond targets.
FOCUS~(E831, 1996-1997)~\cite{FOCUS:2001ahw} used a photon beam with energies up to
approximately 300 GeV incident on a beryllium oxide (BeO) target.

There were also some high-energy physics fixed-target experiments operated at the CERN SPS in Europe.
WA75~(1982-1984) operated with a 360~GeV/$c$ $\pi^-$ beam directed onto a thick nuclear emulsion target.
WA82~(1987-1989)~\cite{Adamovich:1989ss} operated with a mixed hadron
beam~(340~GeV/$c$ $\pi^-$ and 370~GeV/$c$ protons) incident on a segmented silicon, copper, and tungsten target assembly.
WA92 or BEATRICE~(1991-1993)~\cite{Adinolfi:1992zw}, also known as the follow-up to WA82, operated with a 350~GeV/$c$ $\pi^-$ beam on copper and tungsten targets.
NA142~(1986-1989)~\cite{Barate:1984uj}, in contrast, used a 100~GeV/$c$ mean-energy photon beam on a beryllium target.
All four of these experiments exploited the $\Omega$ spectrometer or its upgrade.

Besides, there were four other fixed-target experiments operated at CERN.
EMC (European Muon Collaboration, 1978-1985)~\cite{EuropeanMuon:1980nje} operated at CERN with a high-energy muon beam on nuclear targets
(including hydrogen, deuterium, and heavy nuclei).
SPEC~(1980-1983)~\cite{Biino:1985ej} used the LEBC (Lexan Bubble Chamber) as a high
resolution hydrogen bubble chamber combined with the European Hybrid Spectrometer,
and operated with 400 GeV/$c$
proton beams and also 360 GeV/$c$ $\pi^-$ beams on a hydrogen target in the bubble chamber.
ACCMOR (NA32, 1982-1986)~\cite{ACCMOR:1992ypn}, operated at the CERN SPS, used a hadron beam (pions and protons)
on a copper target.
CHORUS~(1994-1997)~\cite{CHORUS:1997wxi} used the Wide Band Neutrino Beam from the SPS,
with an average energy of 27 GeV, directed onto a 770 kg nuclear emulsion target segmented into
four stacks.

Two other fixed-target experiments are LGW (Lead-Glass Wall, 1976-1977)~\cite{Augustin:1974xq,Barbaro-Galtieri:1977kfn}
at SLAC that used a photon beam on a beryllium target and
HERA-B~(2000-2003)~\cite{HERA-B:2002zbs} at DESY that used 920 GeV protons from the HERA proton beam halo
interacting with an internal wire target composed of carbon, titanium, and tungsten wires.

\subsection{Hadron collision experiments}

Precision measurements of hadronic $D_{(s)}$ decays and searches for rare $D_{(s)}$ decays came
also from hadron collision experiments, e.g., CDF and LHCb,
where charm mesons are mainly produced through hard scattering processes via the strong interaction, including gluon-gluon fusion, quark-antiquark annihilation, and other mechanisms.
These experiments are also powerful for precision measurements of $D^0$-$\bar D^0$ mixing parameters and searches for $CP$ violation in charm decays.
CDF~(1999-2010)~\cite{CDF:1996yrf} operated at the Tevatron proton-antiproton collider at Fermilab, with collisions at a center-of-mass energy of 1.96 TeV.
Its extremely large charm production cross-section combined with a flexible
trigger system capable of selecting fully hadronic final states allows for millions of charmed meson decays.

LHCb~(operated from 2010 to present)~\cite{LHCb:2008vvz,LHCb:2014set} is a dedicated heavy-flavor experiment at the Large Hadron Collider (LHC) at CERN, operating in proton-proton collisions at center-of-mass energies ranging from 7 to 13.6 TeV.
The production cross section for charm quark pairs at LHCb is about two orders of magnitude larger than that at the Tevatron.
To date, LHCb has accumulated the world's largest samples of charmed hadron decays.

\section{Leptonic decays}
\label{sec:leptonic}

Leptonic decays of charmed mesons offer an important test-bed to access the quark mixing-matrix elements and test lepton flavor universality. In the SM of particle physics, the fully radiative inclusive decay rate of $D_{(s)}\to \ell^+\nu_\ell$ ($\ell=e$, $\mu$ or $\tau$) can be written as~\cite{Silverman:1988gc}
\begin{equation}
\begin{split}
\Gamma_{D_{(s)}\to\ell^+\nu_\ell}&=\Gamma_{D_{(s)}\to\ell^+\nu_\ell}^{(0)}\left [1+\frac{\alpha}{\pi}C_p\right ]\\
&=\frac{G_F^2f^2_{D_{(s)}}m^3_{D_{(s)}}}{8\pi}|V_{cq}|^2\mu^2_{\ell}\left (1-\mu^2_{\ell} \right )^2\left [1+\frac{\alpha}{\pi}C_p\right ],
\label{eq1}
\end{split}
\end{equation}
where
$G_F$ is the Fermi coupling constant,
$f_{D_{(s)}}$ is the $D_{(s)}$ decay constant,
$|V_{cq}|$ is the magnitude of the $c\to q$
CKM matrix element,
$\mu_{\ell}$ is the ratio of the $\ell^+$ lepton mass to the $D_{(s)}$ meson mass ($m_{D_{(s)}}$),
and $[1+\frac{\alpha}{\pi}C_p]$ represents the radiative correction term~\cite{ParticleDataGroup:2024cfk}.

The ratio of the $D_{(s)}\to \tau^+\nu_\tau$ to $D_{(s)}\to \mu^+\nu_\mu$
decay branching fractions is given by
\begin{eqnarray}
R^{D_{(s)}}_{\tau/\mu}\ \equiv\
\frac{{\mathcal B}(D_{(s)}\to\tau^+\nu_{\tau})}{{\mathcal B}(D_{(s)}\to\mu^+\nu_{\mu})} &  = &
\left(\frac{m_{\tau}^2}{m_{\mu}^2}\right)
\frac{(m^2_{D_{(s)}}-m^2_{\tau})^2}{(m^2_{D_{(s)}}-m^2_{\mu})^2}\,,
\end{eqnarray}
and equals $9.74\pm0.01$ for $D_s^+$ decays and
$2.66\pm0.01$ for $D^+$ decays, based on the well-measured
values of $m_\mu$, $m_\tau$, and $m_{D_{(s)}}$~\cite{ParticleDataGroup:2024cfk}.
A significant deviation from this expectation would be
interpreted as lepton flavor universality violation in charged currents, which
signifies new physics.

\subsection{Leptonic $D^+$ decays}

Before BESIII, absolute measurements of the branching fractions of $D^+\to \mu^+\nu_\mu$ were reported by MARKIII~\cite{Adler:1987ty},
BESII~\cite{BES:1998iue,BES:2004ufx} and CLEO-c~\cite{CLEO:2004pwu,CLEO:2005jsh,CLEO:2008ffk}. Among them, the most precise measurement of $D^+\to \mu^+\nu_\mu$ was from CLEO-c
based on 150 signal events; no significant signals of the $D^+\to e^+\nu_e$ and $D^+\to \tau^+\nu_\tau$ decays were reported.

\begin{figure*}[htbp]
  \centering
  \includegraphics[width=0.3\textwidth]{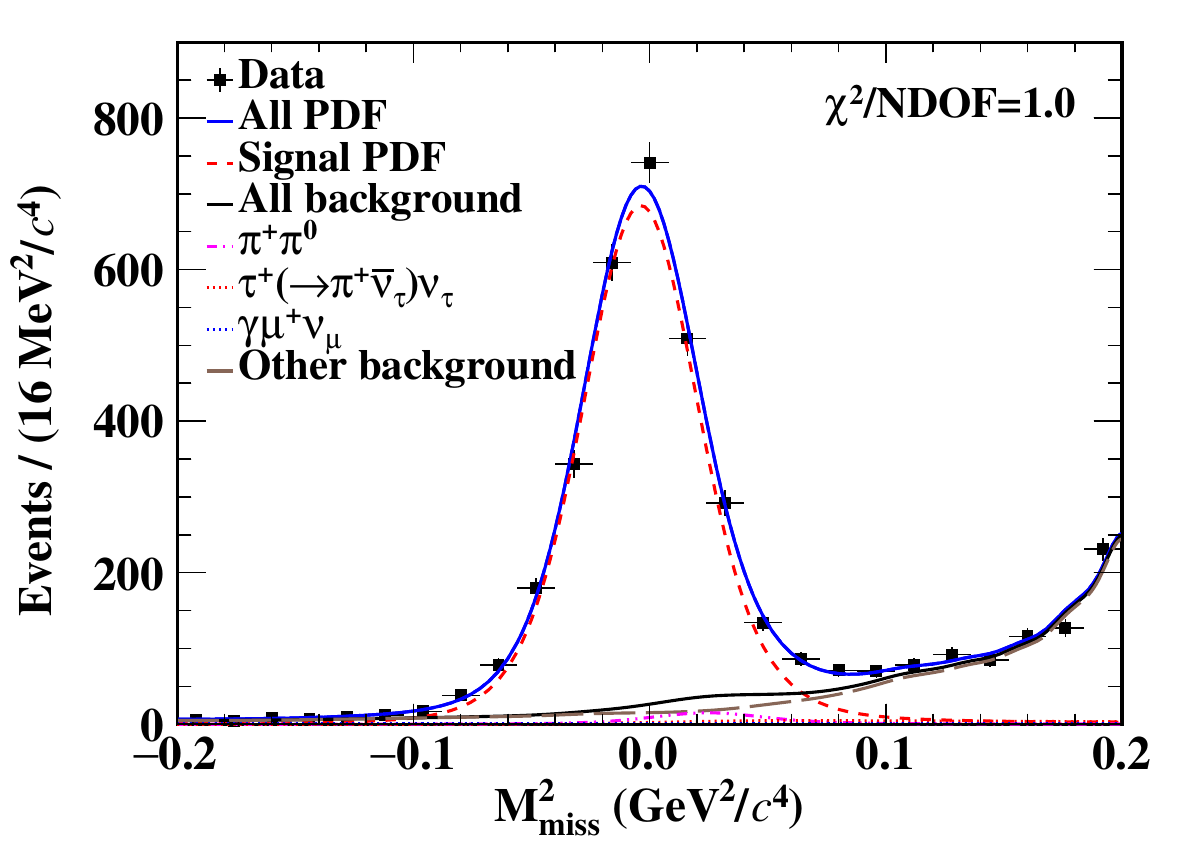}
  \includegraphics[width=0.3\textwidth]{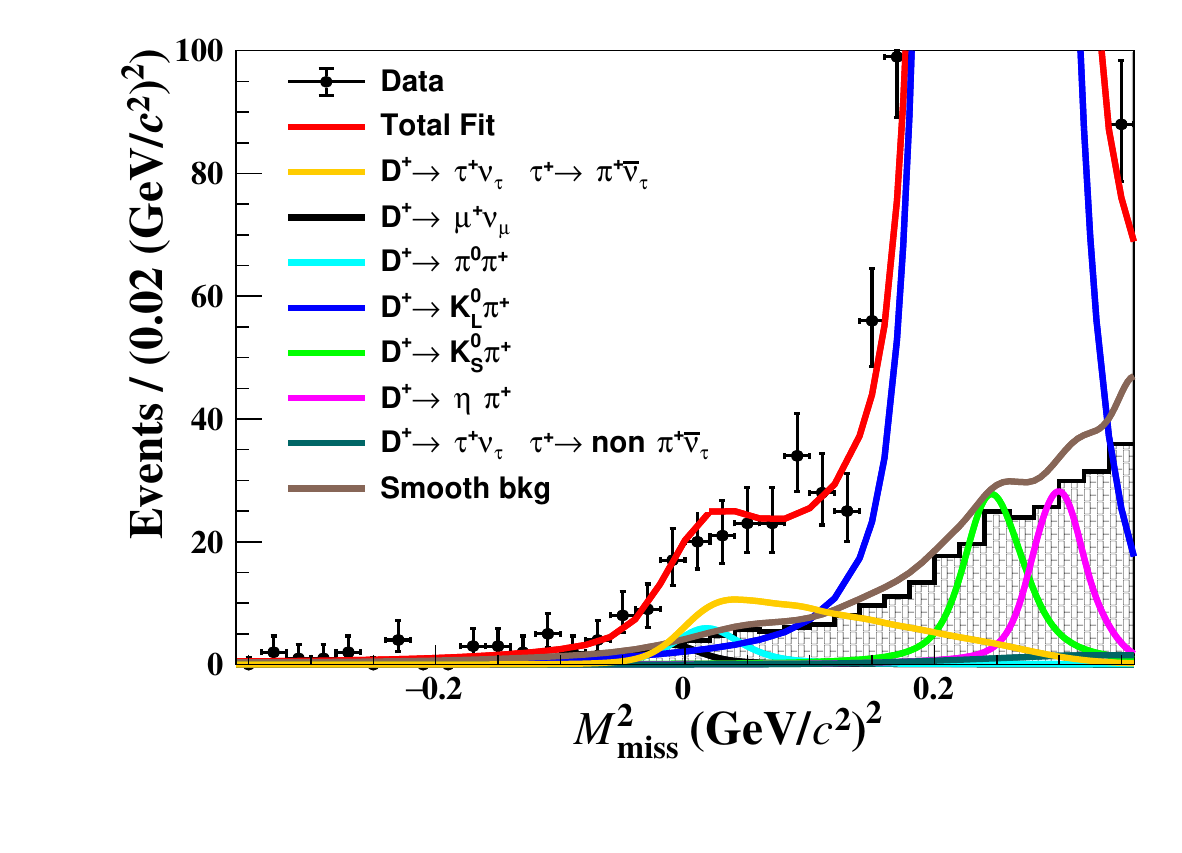}
  \includegraphics[width=0.3\textwidth]{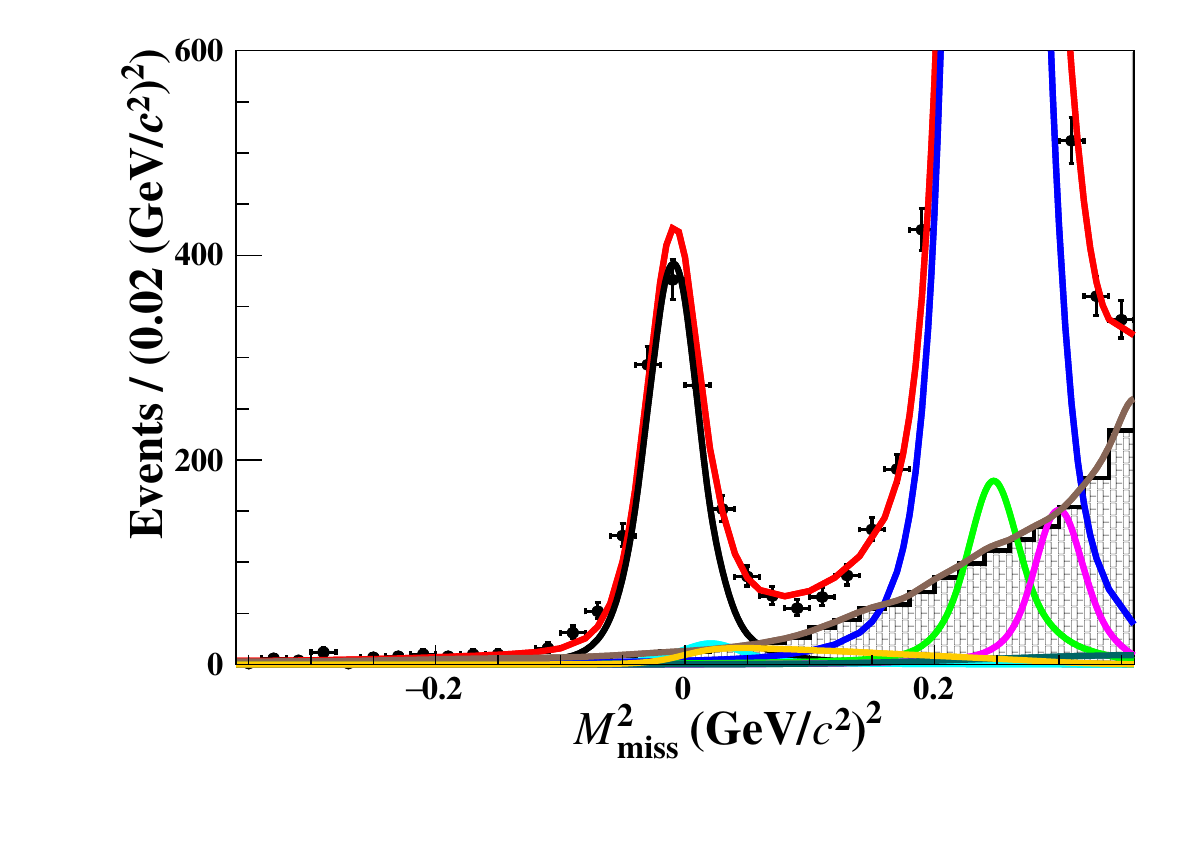}
  \caption{The $M_{\rm miss}^{2}$ distributions of the accepted candidates for (left) $D^+\to \mu^+\nu_\mu$~\cite{BESIII:2024kvt} as well as
  (middle) $\pi$-like and (right) $\mu$-like candidates for $D^+\to \tau^+(\to \pi^+\bar \nu_\tau)\nu_\tau$~\cite{BESIII:2024vlt}.
}
  \label{fig:Dp_lv}
\end{figure*}

In 2014, BESIII found $409\pm 21$ $D^+\to \mu^+\nu_\mu$ signal events from an analysis of 2.93 fb$^{-1}$ of data at 3.773~GeV~\cite{BESIII:2013iro}.
These correspond to a branching fraction for the leptonic decay $D^+\to \mu^+\nu_\mu$ of ${\cal B}(D^+\to\mu^+\nu_\mu)=(3.71\pm0.19\pm0.06)\times 10^{-4}$.
In 2025, an updated measurement of $D^+\to \mu^+\nu_\mu$ was reported with the full 20.3 fb$^{-1}$ of data at 3.773 GeV.
From this sample, $2833\pm 57$ $D^+\to \mu^+\nu_\mu$ signal events are obtained, corresponding to ${\cal B}(D^+\to\mu^+\nu_\mu)=(4.034\pm0.080\pm0.040)\times10^{-4}$~\cite{BESIII:2024kvt}, which supersedes the previous BESIII measurement.
In these measurements, signals with very low background are obtained owing to the unique muon counter at BESIII.
Figure~\ref{fig:Dp_lv}(left) shows the $M_{\rm miss}^{2}$ distributions of the accepted candidates for
$D^+\to \mu^+\nu_\mu$.
After taking into account necessary radiative corrections, BESIII determines $f_{D^+}|V_{cd}|=(48.02\pm0.48\pm0.24\pm0.12\pm0.15_{\rm EM})~\mathrm{MeV}$, where the fourth uncertainty arises from the radiative corrections. The precision is improved by a factor of 2.3 compared with the previous best measurement.
Using the value of the magnitude of the $c\to d$ CKM element $|V_{cd}|=0.22487\pm0.00068$ given by the global SM fit~\cite{ParticleDataGroup:2024cfk}, we obtain
$f_{D^+}=(213.5\pm2.1\pm1.1\pm0.8\pm0.7_{\rm EM})$~MeV. Alternatively, using the value of $f_{D^+}=(212.0\pm0.7)$~MeV
from a precise LQCD calculation~\cite{FlavourLatticeAveragingGroupFLAG:2021npn}, we extract $|V_{cd}|=0.2265\pm0.0023\pm0.0011\pm0.0009\pm0.0007_{\rm EM}$. The measured $f_{D^+}$ is consistent with the LQCD calculation~\cite{FlavourLatticeAveragingGroupFLAG:2021npn} within 1$\sigma$.
From the same data sample, a search for the $D^+\to e^+\nu_e$ decay is carried out~\cite{BESIII:2025ows};
no significant signal is observed, and the upper limit on its decay branching fraction is set to be
${\cal B}(D^+\to e^+\nu_e)<9.7\times 10^{-7}$.

In 2019, the first observation of $D^+\to \tau^+\nu_\tau$ reconstructed with $\tau^+\to \pi^+\bar\nu_\tau$
was reported by analyzing 2.93 fb$^{-1}$ of data at 3.773 GeV,
with a branching fraction of ${\cal B}(D^+\to\tau^+\nu_\tau) = (1.20\pm0.24\pm0.12)\times10^{-3}$~\cite{BESIII:2019vhn}.
In 2025, an updated measurement of $D^+\to \tau^+\nu_\tau$ was reported using 7.9 fb$^{-1}$ of data at 3.773 GeV.
283 signal events are obtained, and correspond to ${\cal B}(D^+\to\tau^+\nu_\tau) = (9.9\pm 1.1\pm 0.5)\times10^{-4}$~\cite{BESIII:2024vlt}.
Figure~\ref{fig:Dp_lv}(middle) and \ref{fig:Dp_lv}(right) show the $M_{\rm miss}^{2}$ distributions of the accepted candidates for
$D^+\to \tau^+(\to \pi^+\bar \nu_\tau)\nu_\tau$.
Using the value of the magnitude of the $c\to d$ CKM element $|V_{cd}|=0.22487\pm0.00068$ given by the global SM fit~\cite{ParticleDataGroup:2024cfk}, we obtain
$f_{D^+}=(204\pm11\pm5\pm1)$~MeV. Alternatively, using the value of $f_{D^+}=(212.0\pm0.7)$ MeV
from a precise LQCD calculation~\cite{FlavourLatticeAveragingGroupFLAG:2024oxs}, we extract $|V_{cd}|=0.216\pm0.012\pm0.006\pm0.001$.
The obtained ${\cal B}(D^+\to\ell^+\nu_\ell)$~($\ell=\mu$ or $\tau$) lead to the branching fraction ratio $R_{\tau/\mu} = \Gamma(D^+\to\tau^+\nu_{\tau})/\Gamma(D^+\to\mu^+\nu_{\mu})= 2.49\pm0.31$, which is well consistent with the SM prediction of $2.66\pm0.01$.

\begin{figure}[htbp]
  \centering
  \includegraphics[width=0.4\textwidth]{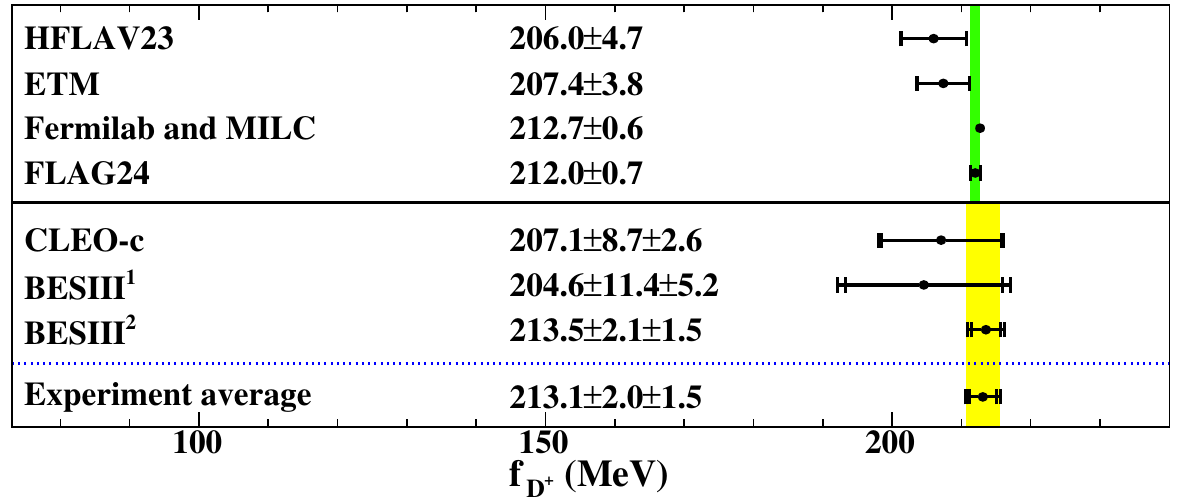}
  \put(-145,79){\tiny~\cite{HeavyFlavorAveragingGroupHFLAV:2024ctg}}
  \put(-145,71.33){\tiny~\cite{Carrasco:2014poa}}
  \put(-145,63.67){\tiny~\cite{Bazavov:2017lyh}}
  \put(-145,56){\tiny~\cite{FlavourLatticeAveragingGroupFLAG:2024oxs}}
  \put(-145,45){\tiny~\cite{CLEO:2008ffk}}
  \put(-145,37.33){\tiny~\cite{BESIII:2024vlt}}
  \put(-145,29.67){\tiny~\cite{BESIII:2024kvt}}
   \caption{Comparison of $f_{D^+}$ from experimental measurements of CLEO-c~\cite{CLEO:2008ffk} and BESIII~\cite{BESIII:2024kvt,BESIII:2024vlt}, LQCD calculations of ETM~\cite{Carrasco:2014poa} and Fermilab and MILC~\cite{Bazavov:2017lyh} as well as HFLAV23~\cite{HeavyFlavorAveragingGroupHFLAV:2024ctg} and FLAG24~\cite{FlavourLatticeAveragingGroupFLAG:2024oxs}.
The green band is the $\pm 1\sigma$ region of FLAG24 and the yellow band denotes the $\pm 1\sigma$ region of the result averaged over all measurements of $D^+\to \ell^+\nu_\ell$.
}
  \label{fig:fDp}
\end{figure}

\begin{figure}[htbp]
  \centering
  \includegraphics[width=0.4\textwidth]{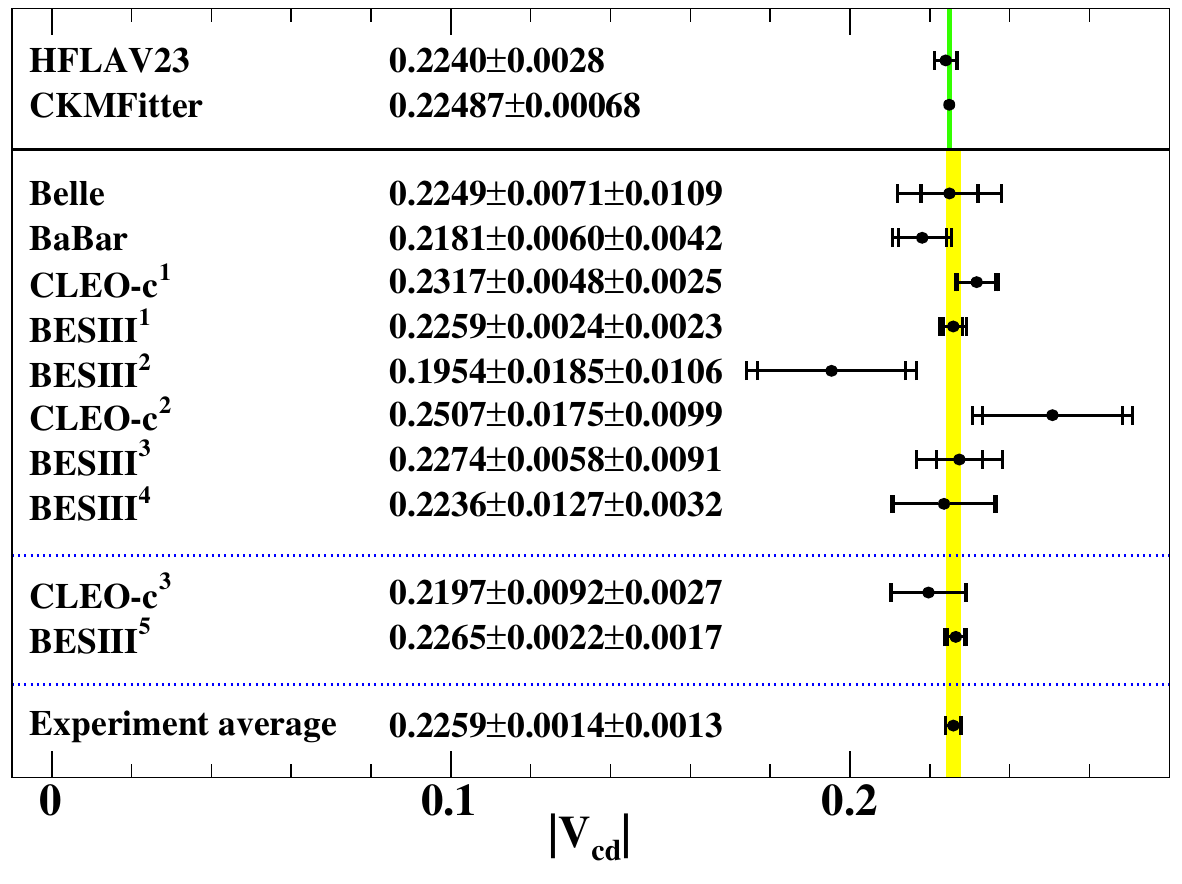}
  \put(-165,141){\tiny~\cite{HeavyFlavorAveragingGroupHFLAV:2024ctg}}
  \put(-165,133){\tiny~\cite{ParticleDataGroup:2024cfk}}
  \put(-165,117.5){\tiny~\cite{Belle:2006idb}}
  \put(-165,109.929){\tiny~\cite{BaBar:2014xzf}}
  \put(-165,102.357){\tiny~\cite{CLEO:2009svp}}
  \put(-165,94.786){\tiny~\cite{Ablikim:2015ixa,BESIII:2017ylw}}
  \put(-165,87.214){\tiny~\cite{BESIII:2025hjc}}
  \put(-165,79.643){\tiny~\cite{CLEO:2010pjh}}
  \put(-165,72.071){\tiny~\cite{BESIII:2024njj}}
  \put(-165,64.5){\tiny~\cite{BESIII:2025gov}}
  \put(-165,49){\tiny~\cite{CLEO:2008ffk}}
  \put(-165,41.429){\tiny~\cite{BESIII:2024kvt,BESIII:2024vlt}}
   \caption{Comparison of $|V_{cd}|$ from experimental measurements with leptonic $D$ decays
   (CLEO-c~\cite{CLEO:2008ffk} and BESIII~\cite{BESIII:2024kvt,BESIII:2024vlt})
   and semileptonic $D$ decays (Belle~\cite{Belle:2006idb}, BaBar~\cite{BaBar:2014xzf},
    CLEO-c~\cite{CLEO:2009svp,CLEO:2010pjh} and BESIII~\cite{Ablikim:2015ixa,BESIII:2017ylw,BESIII:2025hjc,BESIII:2024njj,BESIII:2025gov})
    as well as HFLAV23~\cite{HeavyFlavorAveragingGroupHFLAV:2024ctg} and CKMfitter~\cite{ParticleDataGroup:2024cfk}.
The green band is the $\pm 1\sigma$ region of CKMfitter and the yellow band denotes the $\pm 1\sigma$ region of the result averaged over all measurements with (semi)leptonic $D$ decays.
}
  \label{fig:Vcd}
\end{figure}
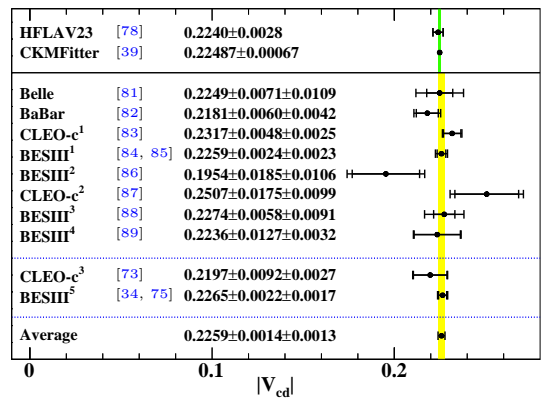

 A summary of experimental results and world averages for
${\cal{B}}(D^+\to \ell^+\nu_{\ell})$ and $f_{D^+}|V_{cd}|$ is given in Table~\ref{tab:DpLeptonic}. The BESIII averaged results, together with the other experimental measurements, take full account of correlated systematic uncertainties.
Figure~\ref{fig:fDp} shows a comparison of $f_{D^+}$ from different experiments and recent LQCD calculations,
and Fig.~\ref{fig:Vcd} presents the comparison of $|V_{cd}|$ measured with leptonic and semileptonic $D$ decays  from different experiments.

\begin{table*}[htbp]
\caption{Experimental results and world averages for ${\cal{B}}(D^+\to \ell^+\nu_{\ell})$ and $f_{D^+}|V_{cd}|$.
The SM-constrained result from CLEO-c is shown. The ``EM'' uncertainty of the BESIII result~\cite{BESIII:2024kvt} has been combined into the systematic uncertainty. The ``BESIII average'' values are obtained by combining results after considering the correlated effects. The ``Average" values  are obtained by weighting both statistical and systematic uncertainties, but excluding the third uncertainty of the $D^+$ lifetime. The first uncertainties of ``Average"  $f_{D^+}|V_{cd}|$ are the total experimental uncertainties combined from statistical and systematic effects, and the second uncertainties of ``Average"  $f_{D^+}|V_{cd}|$ are due to the input uncertainty of the quoted lifetime of $D^+$. 
\label{tab:DpLeptonic}}
\def\1#1#2{\multicolumn{#1}{#2}}
\begin{tabular}{l c c c c c}
\hline\hline
Experiment & $E_{\rm cm}$ (GeV) & Mode & $D^+$ decay & $\cal B$~($10^{-4}$)& $f_{D^+}|V_{cd}|$ (MeV)\\
\hline
CLEO-c~\cite{CLEO:2008ffk} &  3.774 & $D^+D^-$  & $\mu^+\nu_\mu$  & $3.82\pm0.32\pm 0.09$& $46.57\pm1.95\pm0.55\pm0.11$ \\
BESIII~\cite{BESIII:2024kvt} &  3.773 & $D^+D^-$  & $\mu^+\nu_\mu$  & $4.034\pm0.080\pm0.040$ & $48.02\pm0.48\pm0.28\pm0.12$  \\
BESIII~\cite{BESIII:2024vlt} &  3.773 & $D^+D^-$  & $\tau^+\nu_\tau$  & $9.9\pm 1.1\pm 0.5$  & $46.01\pm2.56\pm1.16\pm0.11$  \\ \hline
BESIII average                  &        &          & $\ell^+\nu_\ell$ & & $48.02\pm0.47\pm0.30\pm0.12$ \\ 
 Average                  &        &          & $\ell^+\nu_\ell$ & & $47.92\pm0.54\pm0.12$ \\ \hline
BESIII~\cite{BESIII:2025ows} &  3.773 & $D^+D^-$  & $e^+\nu_e$  & $<0.0097$ at 90\% C.L. & ...  \\
\hline
\hline
\end{tabular}
\end{table*}

\subsection{Leptonic $D^+_s$ decays}

Before 2001, early studies of leptonic $D^+_s$ decays were from the
WA75~\cite{WA75:1992xsf}, BES~\cite{BES:1994egf}, E653~\cite{FermilabE653:1996gzw}, L3~\cite{L3:1996vqf}, CLEO~\cite{CLEO:1997var}, BEATRICE~\cite{BEATRICE:2000wqz}
and OPAL~\cite{OPAL:2001wly} experiments.
Between 2001 and 2014, the absolute measurements of the branching fractions of $D^+_s\to \mu^+\nu_\mu$ were reported by
ALEPH~\cite{ALEPH:2002fge}, CLEO-c~\cite{CLEO:2007fgd,CLEO:2009lvj}, Belle~\cite{Belle:2007wtt,Belle:2013isi}, and BaBar~\cite{BaBar:2010ixw};
while the absolute measurements of the branching fractions of $D^+_s\to \tau^+\nu_\tau$ were reported by
ALEPH~\cite{ALEPH:2002fge}, OPAL~\cite{OPAL:2001wly}, L3~\cite{L3:1996vqf}, CLEO-c~\cite{CLEO:2007fgd,CLEO:2007qjo,CLEO:2009lvj,CLEO:2009vke,CLEO:2009jky}, Belle~\cite{Belle:2013isi} and BaBar~\cite{BaBar:2010ixw}. Among them, the Belle experiment provided
the most precise measurements with about 0.5k $D^+_s\to \mu^+\nu_\mu$ signal events and 2.2k $D^+_s\to \tau^+\nu_\tau$ signal events;
and no significant signal of $D^+_s\to e^+\nu_e$ was reported.

\begin{figure}[htbp]
  \centering
  \includegraphics[width=0.4\textwidth]{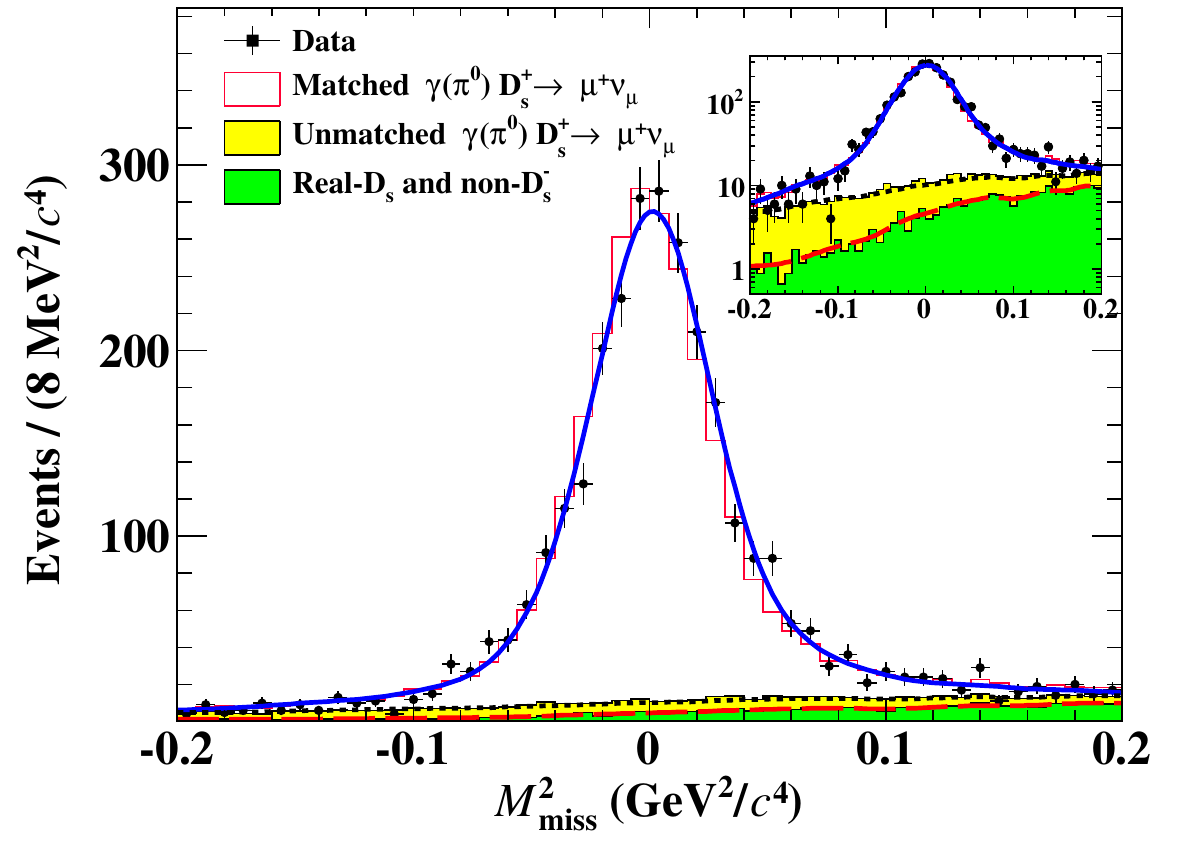}
  \caption{The $M_{\rm miss}^{2}$ distribution of the accepted candidates for $D^+_s\to \mu^+\nu_\mu$~\cite{BESIII:2023fhe}.
}
  \label{fig:Ds_munu}
\end{figure}

In 2018, BESIII reported a measurement of the branching fraction ${\cal B}(D^+_s\to\mu^+\nu_\mu)$
with 3.19~fb$^{-1}$ of data at 4.178~GeV.
From this sample, about 1.1k $D^+_s\to\mu^+\nu_\mu$ signal events, with low background owing to muon counter information, are obtained.
These correspond to ${\cal B}(D^+_s\to\mu^+\nu_\mu)=(5.49\pm0.16\pm0.15)\times 10^{-3}$~\cite{BESIII:2018hhz}.
In 2021, a measurement of ${\cal B}(D^+_s\to\mu^+\nu_\mu)=(5.35\pm0.13\pm0.16)\times10^{-3}$ was reported with
6.3~fb$^{-1}$ of data  at 4.178-4.226~GeV~\cite{BESIII:2021anh};
the statistical uncertainty is improved, but the systematic uncertainty is worsened without using
muon counter information.
In 2023, a further improved measurement of ${\cal B}(D^+_s\to\mu^+\nu_\mu)=(5.294\pm0.108\pm0.085)\times 10^{-3}$ with muon counter information was reported
with 7.33~fb$^{-1}$ of data  at 4.178-4.226~GeV~\cite{BESIII:2023cym},
with about 2.5k $D^+_s\to\mu^+\nu_\mu$ signal events.
Figure~\ref{fig:Ds_munu} shows the $M_{\rm miss}^{2}$ distributions of the accepted candidates for $D^+_s\to \mu^+\nu_\mu$.

\begin{figure}[htbp]
  \centering
  \includegraphics[width=0.225\textwidth]{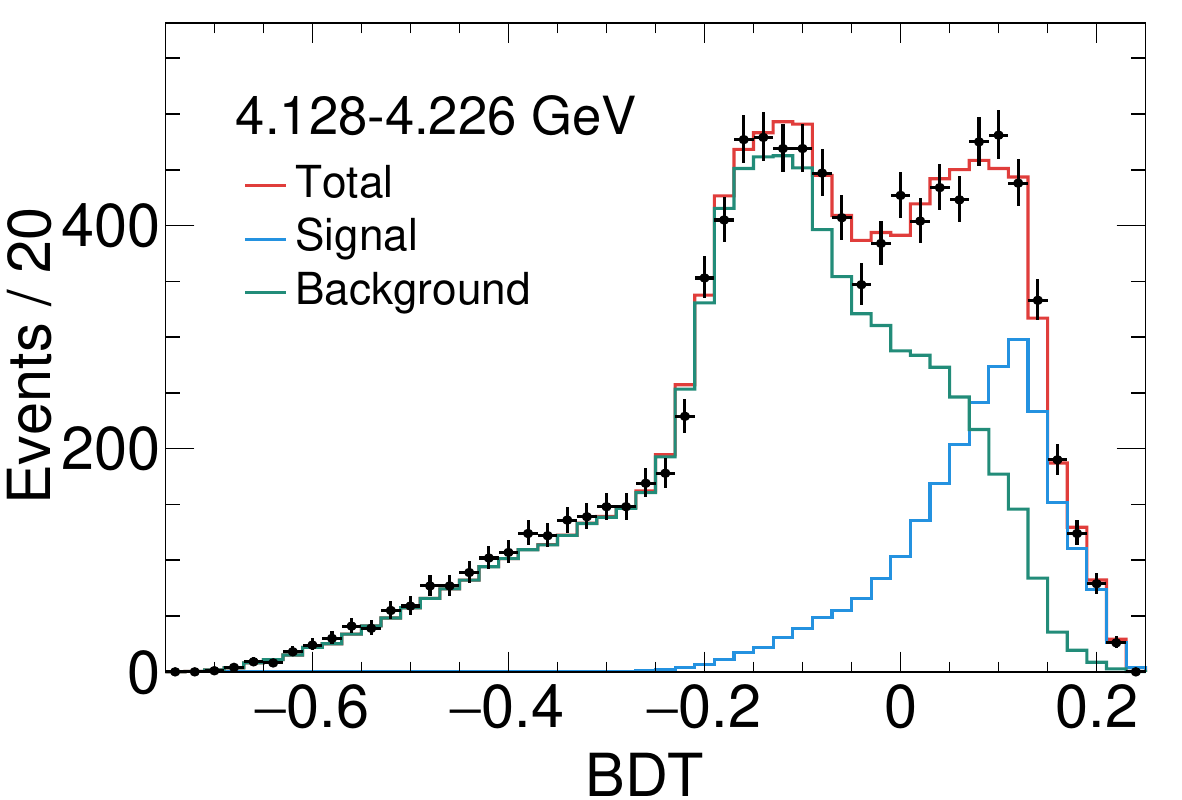}
  \includegraphics[width=0.225\textwidth,height=0.15\textwidth]{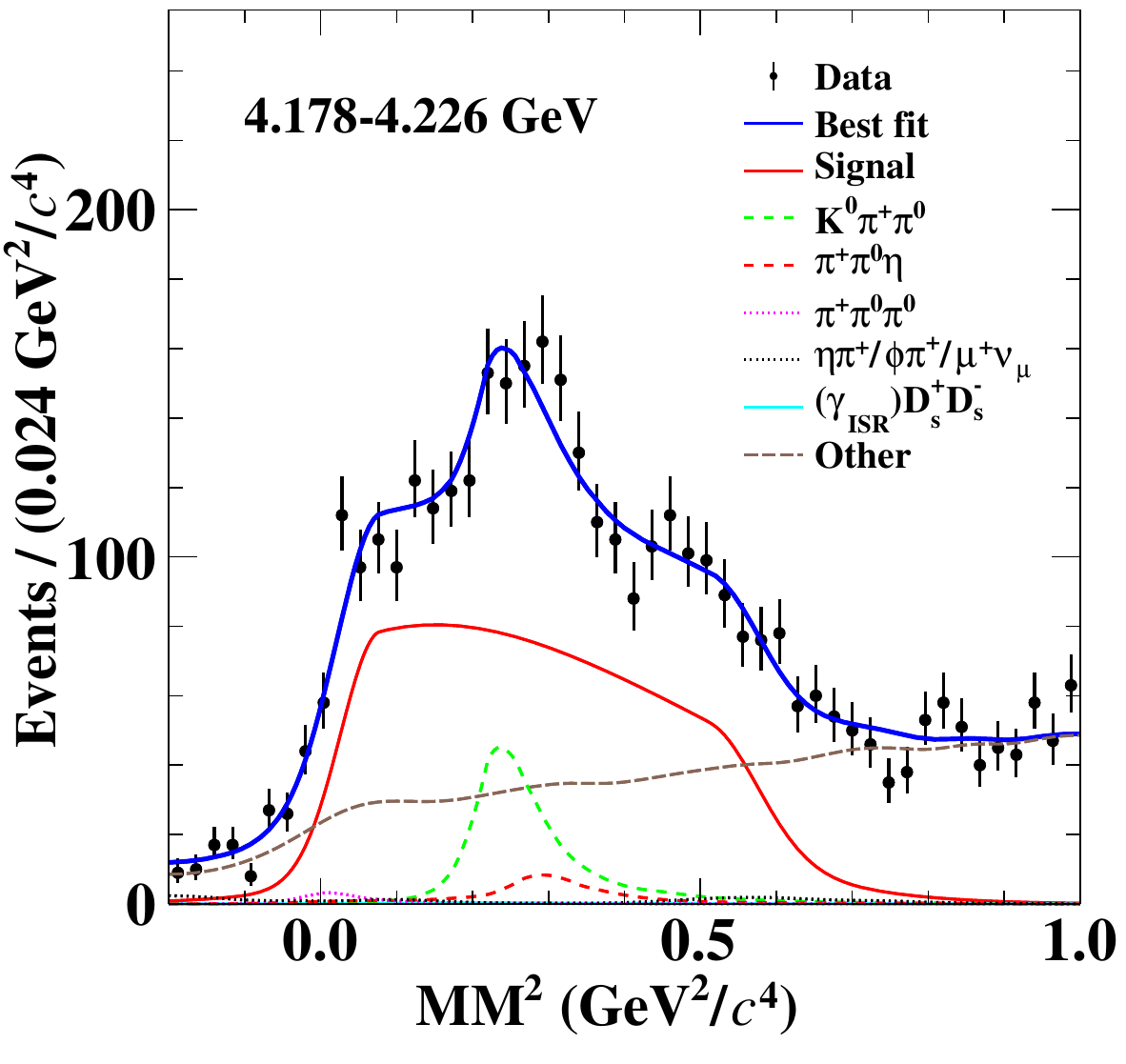}
  \includegraphics[width=0.225\textwidth]{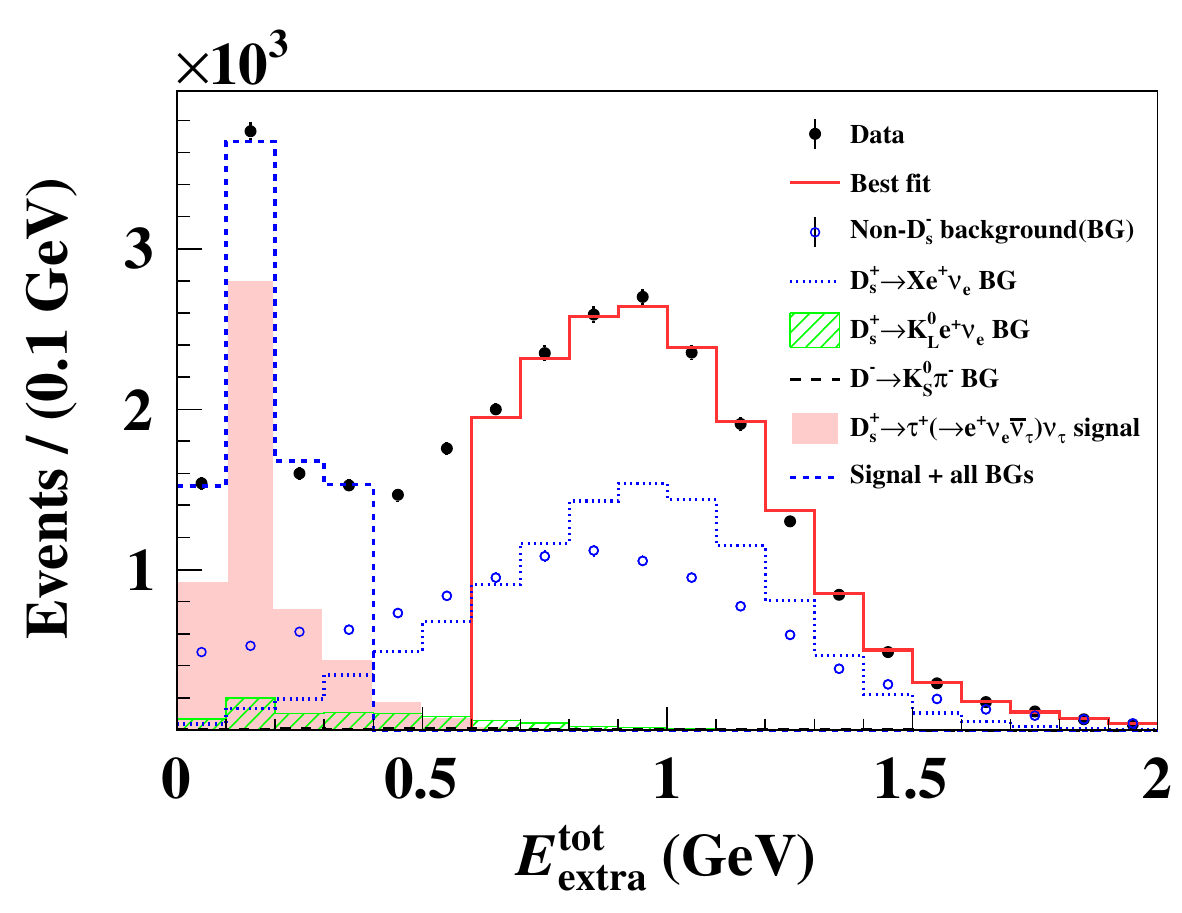}
  \includegraphics[width=0.225\textwidth]{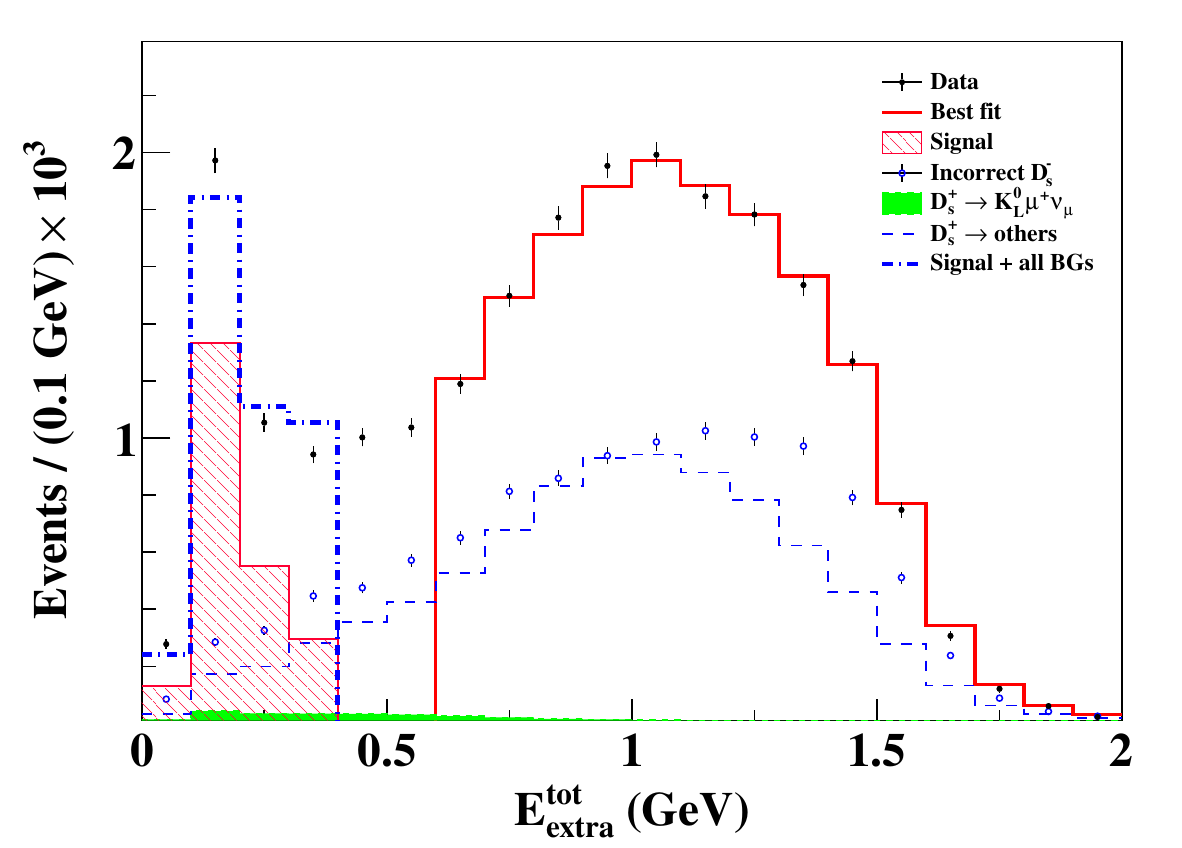}
  \put(-90,70){\tiny 4.128-4.226 GeV}
  \put(-205,70){\tiny 4.178-4.226 GeV}
  \caption{The BDT, MM$^2$ or $E_{\rm extra,\gamma}^{\rm tot}$ distributions of the accepted candidates for
  $D^+_s\to \tau^+(\to \pi^+\bar \nu_\tau)\nu_\tau$~\cite{BESIII:2023fhe},
  $D^+_s\to \tau^+(\to \rho(770)^+ \bar \nu_\tau)\nu_\tau$~\cite{BESIII:2021wwd},
  $D^+_s\to \tau^+(\to e^+\nu_e\bar \nu_\tau)\nu_\tau$~\cite{BESIII:2021bdp}, and
  $D^+_s\to \tau^+(\to \mu^+\nu_\mu\bar \nu_\tau)\nu_\tau$~\cite{BESIII:2023ukh},
  which are combined from all used tag modes and energy points.
}
  \label{fig:Ds_tauv}
\end{figure}

Since 2021, individual measurements of ${\cal B}(D^+_s\to\tau^+\nu_\tau)$ were made by using the
$\tau^+\to \pi^+\bar \nu_\tau$~\cite{BESIII:2023fhe}, $\tau^+\to \rho(770)^+\bar \nu_\tau$~\cite{BESIII:2021wwd},
$\tau^+\to e^+\nu_e\bar\nu_\tau$~\cite{BESIII:2021bdp}, and $\tau^+\to \mu^+\nu_\mu\bar\nu_\tau$~\cite{BESIII:2023ukh} decays,
based on 7.33 fb$^{-1}$ of data at 4.178-4.226 GeV, 6.3 fb$^{-1}$ of data at 4.178-4.226 GeV, 6.3 fb$^{-1}$ of data at 4.178-4.226 GeV,
and 7.33 fb$^{-1}$ of data at 4.128-4.226 GeV, respectively.
Here, it should be noted that the results reported in Ref.~\cite{BESIII:2023fhe} supersede those
in Ref.~\cite{BESIII:2021anh} based on 6.3 fb$^{-1}$ of data at 4.178-4.226 GeV, benefiting from a boosted decision tree method.
In these measurements,
about 2.4k, 1.7k, 4.9k, and 2.3k signal events of $D^+_s\to\tau^+\nu_\tau$ are reconstructed via
$\tau^+\to \pi^+\bar \nu_\tau$, $\tau^+\to \rho(770)^+\bar \nu_\tau$,
$\tau^+\to e^+\nu_e\bar\nu_\tau$, and $\tau^+\to \mu^+\nu_\mu\bar\nu_\tau$,
corresponding to
$\mathcal{B}(D_s^+ \to \tau^+\nu_\tau) = (5.44 \pm 0.17 \pm 0.13)\%$,
$\mathcal{B}(D_s^+ \to \tau^+\nu_\tau) = (5.29 \pm 0.25 \pm 0.20)\%$,
$\mathcal{B}(D_s^+ \to \tau^+\nu_\tau) = (5.27 \pm 0.10 \pm 0.12)\%$,
$\mathcal{B}(D_s^+ \to \tau^+\nu_\tau) = (5.37 \pm 0.17 \pm 0.15)\%$, respectively.
Figure~\ref{fig:Ds_tauv} shows the BDT, MM$^2$ or $E_{\rm extra,\gamma}^{\rm tot}$ distributions of the accepted candidates
for $D^+_s\to \tau^+\nu_\tau$ with $\tau^+$ reconstructed with four different decay modes.

Two independent  measurements of the branching fractions of $D^+_s\to \ell^+\nu_\ell$~($\ell=\mu$ or $\tau$) were also performed
by using 15k tagged $D_s^-$ mesons with $e^+e^-\to D^+_sD^-_s$  at 4.009~GeV~\cite{BESIII:2016cws} and
124k tagged $D_s^{*-}$ mesons with $e^+e^-\to D^{*+}_sD^{*-}_s$ at 4.237-4.699~GeV~\cite{BESIII:2024dvk}.
In both measurements, the $D_s^+ \to \mu^+\nu_\mu$ is selected with muon counter information.
The $\tau^+$ is reconstructed via $\tau^+\to \pi^+\bar \nu_\tau$ in the former article;
via $\tau^+\to \pi^+\bar \nu_\tau$, $\tau^+\to \rho(770)^+\bar \nu_\tau$,
$\tau^+\to e^+\nu_e\bar\nu_\tau$, and $\tau^+\to \mu^+\nu_\mu\bar\nu_\tau$ in the latter article.
Based on dozens of signal events, Ref.~\cite{BESIII:2016cws} reported
$\mathcal{B}(D_s^+ \to \mu^+\nu_\mu) = (5.17 \pm 0.75 \pm 0.21)\times 10^{-3}$ and
$\mathcal{B}(D_s^+ \to \tau^+\nu_\tau) = (3.28 \pm 1.83 \pm 0.37)\%$.
In Ref.~\cite{BESIII:2024dvk}, about 0.5k and 2.9k signal events of $D_s^+ \to \mu^+\nu_\mu$
and $D_s^+ \to \tau^+\nu_\tau$ give
$\mathcal{B}(D_s^+ \to \mu^+\nu_\mu) = (5.47 \pm 0.26 \pm 0.16)\times 10^{-3}$ and
$\mathcal{B}(D_s^+ \to \tau^+\nu_\tau) = (5.60 \pm 0.16 \pm 0.20)\%$.

After taking into account the correlation of systematic uncertainties,
the average branching fraction of $D_s^+\to\mu^+\nu_\mu$ is obtained by re-weighting the results reported in Refs.~\cite{BESIII:2016cws,BESIII:2023cym,BESIII:2024dvk};
while that of $D_s^+\to\tau^+\nu_\tau$ is obtained by re-weighting the results reported in Refs.~\cite{BESIII:2016cws,BESIII:2023fhe,BESIII:2021wwd,BESIII:2021bdp,BESIII:2023ukh,BESIII:2024dvk}.
The averaged branching  fractions are ${\cal B}(D_s^+\to\mu^+\nu_\mu)= (0.5310\pm0.0099\pm0.0053)\%$ and
${\cal B}(D_s^+\to\tau^+\nu_\tau) = (5.359\pm0.067\pm0.075)\%$.
The obtained ${\cal B}(D^+_s\to\ell^+\nu_\ell)$~($\ell=\mu$ or $\tau$) lead to the branching fraction ratio $R_{\tau/\mu} = \Gamma(D^+_s\to\tau^+\nu_{\tau})/\Gamma(D^+_s\to\mu^+\nu_{\mu})= 10.09\pm0.28$, which is well consistent with the SM prediction of 9.74$\pm$0.01.

Based on the averaged branching fractions and the value of the magnitude of the $c\to s$ CKM element $|V_{cs}|=0.97349\pm0.00016$ given by the global SM fit~\cite{ParticleDataGroup:2024cfk}, one obtains
$f_{D^+_s}=(249.4\pm2.3\pm1.2\pm0.5)_{\mu\nu}$~MeV and
$f_{D^+_s}=(253.9\pm1.6\pm1.8\pm0.6)_{\tau\nu}$~MeV.
Alternatively, using the value of $f_{D^+_s}=(249.9\pm0.5)$~MeV
from a precise LQCD calculation~\cite{FlavourLatticeAveragingGroupFLAG:2024oxs}, one extracts
$|V_{cs}|=(0.972\pm0.009\pm0.005\pm0.003)_{\mu\nu}$ and
$|V_{cs}|=(0.989\pm0.006\pm0.007\pm0.003)_{\tau\nu}$.
Further averaging these two results based on $D^+_s\to \mu^+\nu_\mu$ and $D^+_s\to \tau^+\nu_\tau$
gives
$f_{D^+_s}=(252.1\pm1.3\pm1.7\pm0.5)$~MeV
and
$|V_{cs}|=0.982\pm0.005\pm0.007\pm0.003$. The obtained $f_{D^+_s}$ is consistent with the LQCD calculation~\cite{FlavourLatticeAveragingGroupFLAG:2024oxs} within 1$\sigma$.

\begin{figure}[htbp]
  \centering
  \includegraphics[width=0.4\textwidth]{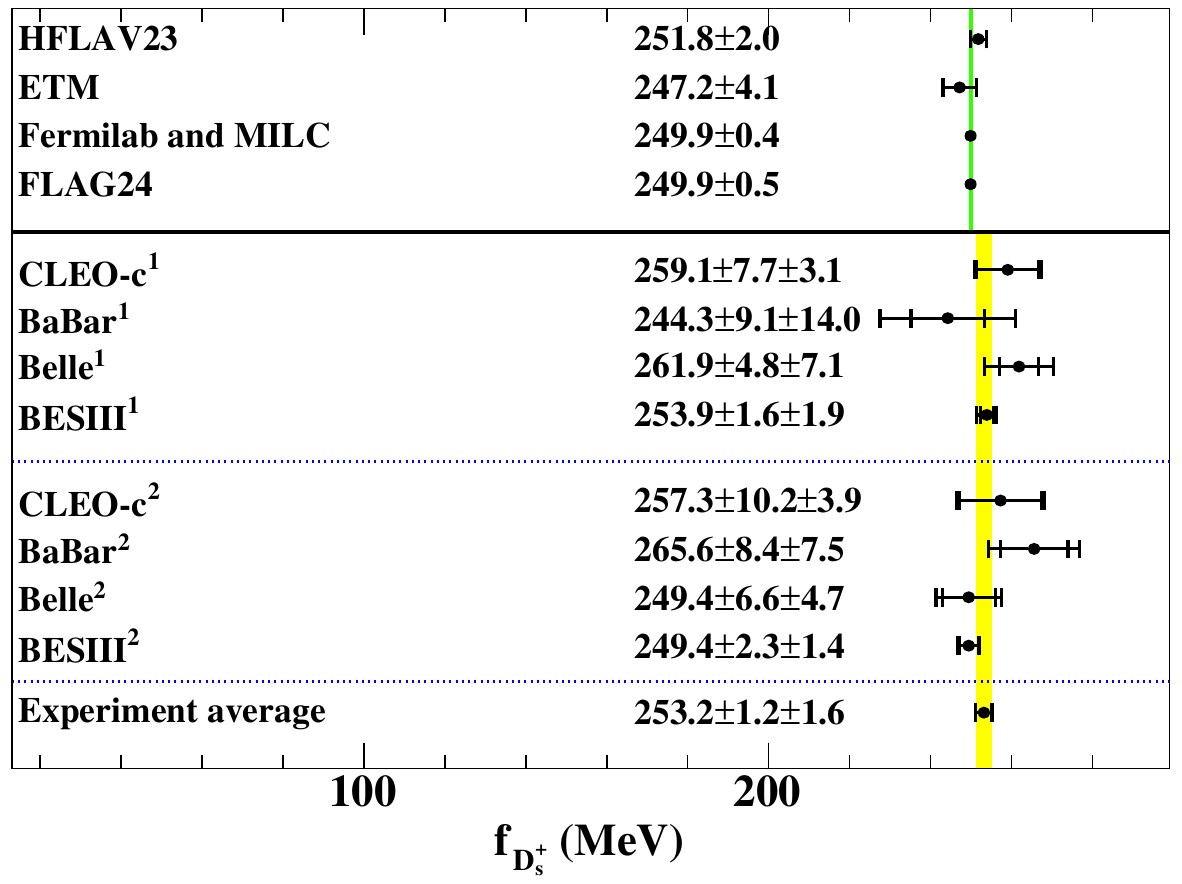}
 \put(-145,143){\tiny~\cite{HeavyFlavorAveragingGroupHFLAV:2024ctg}}
  \put(-145,135.333){\tiny~\cite{Carrasco:2014poa}}
  \put(-145,127.666){\tiny~\cite{Bazavov:2017lyh}}
  \put(-145,120){\tiny~\cite{FlavourLatticeAveragingGroupFLAG:2024oxs}}
  \put(-145,104.5){\tiny~\cite{CLEO:2009jky,CLEO:2009vke,CLEO:2009lvj}}
  \put(-145,96.1666){\tiny~\cite{BaBar:2010ixw}}
  \put(-145,87.833){\tiny~\cite{Belle:2013isi}}
  \put(-145,79.5){\tiny~\cite{BESIII:2023fhe,BESIII:2021wwd,BESIII:2021bdp,BESIII:2023ukh,BESIII:2016cws,BESIII:2024dvk}}
  \put(-145,65){\tiny~\cite{CLEO:2009lvj}}
  \put(-145,56.666){\tiny~\cite{BaBar:2010ixw}}
  \put(-145,48.333){\tiny~\cite{Belle:2013isi}}
  \put(-145,40){\tiny~\cite{BESIII:2023cym,BESIII:2016cws,BESIII:2024dvk}}
  \caption{Comparison of $f_{D^+_s}$ from experimental measurements of CLEO-c$^{1,2}$~\cite{CLEO:2009jky,CLEO:2009vke,CLEO:2009lvj}, BaBar$^{1,2}$~\cite{BaBar:2010ixw}, Belle$^{1,2}$~\cite{Belle:2013isi}, and BESIII$^{1,2}$~\cite{BESIII:2023cym,BESIII:2023fhe,BESIII:2021wwd,BESIII:2021bdp,BESIII:2023ukh,BESIII:2016cws,BESIII:2024dvk}, LQCD calculations of ETM~\cite{Carrasco:2014poa} and Fermilab and MILC~\cite{Bazavov:2017lyh} as well as HFLAV23~\cite{HeavyFlavorAveragingGroupHFLAV:2024ctg} and FLAG24~\cite{FlavourLatticeAveragingGroupFLAG:2024oxs}.
The combined $D^+_s\to \tau^+\nu_\tau$ result of CLEO-c is based on~\cite{CLEO:2009jky,CLEO:2009vke,CLEO:2009lvj};
the combined $D^+_s\to \tau^+\nu_\tau$ result of BESIII is based on~\cite{BESIII:2023fhe,BESIII:2021wwd,BESIII:2021bdp,BESIII:2023ukh,BESIII:2016cws,BESIII:2024dvk};
and the combined $D^+_s\to \mu^+\nu_\mu$ result of BESIII is based on~\cite{BESIII:2023cym,BESIII:2016cws,BESIII:2024dvk}.
The green band is the $\pm 1\sigma$ region of FLAG24 and the yellow band denotes the $\pm 1\sigma$ region of the result averaged over all measurements of $D_s^+\to\ell^+\nu_\ell$.
}
  \label{fig:fDs}
\end{figure}

\begin{figure}[htbp]
  \centering
  \includegraphics[width=0.4\textwidth]{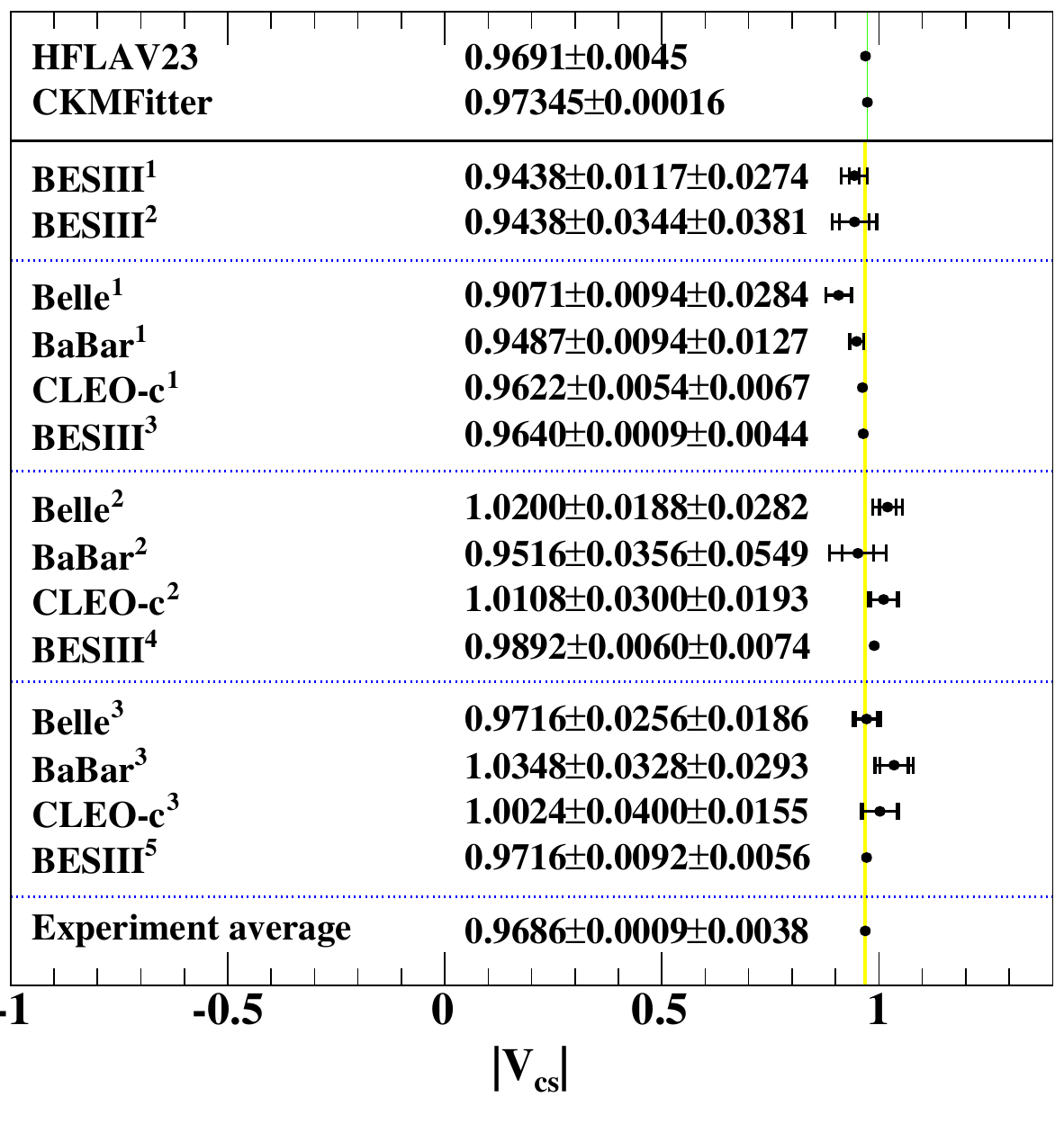}
\put(-160,205){\tiny~\cite{HeavyFlavorAveragingGroupHFLAV:2024ctg}}
  \put(-160,196.5){\tiny~\cite{ParticleDataGroup:2024cfk}}
  \put(-160,181.5){\tiny~\cite{BESIII:2023ajr,BESIII:2024mot,BESIII:2023gbn}}
  \put(-160,173){\tiny~\cite{BESIII:2023ajr,BESIII:2024mot,BESIII:2023gbn}}
  \put(-160,159.5){\tiny~\cite{Belle:2006idb}}
  \put(-160,150.5){\tiny~\cite{BaBar:2007zgf}}
  \put(-160,141.5){\tiny~\cite{CLEO:2009svp}}
  \put(-160,132.5){\tiny~\cite{BESIII:2026uin,BESIII:2026ydr,BESIII:2015jmz}}
  \put(-160,119.0){\tiny~\cite{Belle:2013isi}}
  \put(-160,110){\tiny~\cite{BaBar:2010ixw}}
  \put(-160,101){\tiny~\cite{CLEO:2009jky,CLEO:2009vke,CLEO:2009lvj}}
  \put(-160,92){\tiny~\cite{BESIII:2023fhe,BESIII:2021wwd,BESIII:2021bdp,BESIII:2023ukh,BESIII:2016cws,BESIII:2024dvk}}
  \put(-160,78){\tiny~\cite{Belle:2013isi}}
  \put(-160,69){\tiny~\cite{BaBar:2010ixw}}
  \put(-160,60){\tiny~\cite{CLEO:2009lvj}}
  \put(-160,51){\tiny~\cite{BESIII:2023cym,BESIII:2016cws,BESIII:2024dvk}}
     \caption{Comparison of $|V_{cs}|$ from experimental results measured with
semileptonic $D$ decays
(Belle$^1$~\cite{Belle:2006idb}, BaBar$^1$~\cite{BaBar:2007zgf}, CLEO-c$^1$~\cite{CLEO:2009svp}, and BESIII$^{1, 2, 3}$~\cite{BESIII:2026uin,BESIII:2026ydr,BESIII:2015jmz,BESIII:2023ajr,BESIII:2023gbn,BESIII:2024mot})
and leptonic $D$ decays
(CLEO-c$^2$~\cite{CLEO:2009jky,CLEO:2009vke,CLEO:2009lvj}, BaBar$^2$~\cite{BaBar:2010ixw}, Belle$^2$~\cite{Belle:2013isi}, and BESIII$^{4,5}$~\cite{BESIII:2023cym,BESIII:2023fhe,BESIII:2021wwd,BESIII:2021bdp,BESIII:2023ukh,BESIII:2016cws,BESIII:2024dvk},
in which the combined $D^+_s\to \tau^+\nu_\tau$ result of CLEO-c$^2$ is based on~\cite{CLEO:2009jky,CLEO:2009vke,CLEO:2009lvj};
the combined $D^+_s\to \tau^+\nu_\tau$ result of BESIII$^4$ is based on~\cite{BESIII:2023fhe,BESIII:2021wwd,BESIII:2021bdp,BESIII:2023ukh,BESIII:2016cws,BESIII:2024dvk};
and the combined $D^+_s\to \mu^+\nu_\mu$ result of BESIII$^5$ is based on~\cite{BESIII:2023cym,BESIII:2016cws,BESIII:2024dvk})
    as well as HFLAV23~\cite{HeavyFlavorAveragingGroupHFLAV:2024ctg} and CKMfitter~\cite{ParticleDataGroup:2024cfk}.
The green band is the $\pm 1\sigma$ region of CKMfitter and the yellow band denotes the $\pm 1\sigma$ region of the result averaged over all measurements with (semi)leptonic $D$ decays. 
}
  \label{fig:Vcs}
\end{figure}

Table~\ref{tab:DsLeptonic} presents comparisons of experimental results and world averages for
${\cal{B}}(D^+_s\to \ell^+\nu_{\ell})$ and $f_{D^+_s}|V_{cs}|$.
The comparison of $f_{D^+_s}$ from different experiments and recent LQCD calculations is given in Fig.~\ref{fig:fDs},
and the comparison of $|V_{cs}|$ measured with leptonic and semileptonic $D$ decays from different experiments is
shown in Fig.~\ref{fig:Vcs}.

\begin{table*}[htb]\centering
	\caption{Comparisons of the branching fractions  and the corresponding products of $f_{D_s^+}|V_{cs}|$ from various experiments. The ``BESIII average'' values are obtained by combining results after considering the correlated effects.
		The ``Average" values  are obtained by weighting both statistical and systematic uncertainties, but excluding the third uncertainty of the $D_s^+$ lifetime. The uncertainties of ``Average'' branching fractions  and the first uncertainties of ``Average"  $f_{D^+_s}|V_{cs}|$ are the total experimental uncertainties combined from statistical and systematic effects, and the second uncertainties of ``Average"  $f_{D^+_s}|V_{cs}|$ are due to the input uncertainty of the quoted lifetime of $D^+_s$. 
}
	\label{tab:DsLeptonic}
	\def\1#1#2{\multicolumn{#1}{#2}}
		\begin{tabular}{l c c c c c}
			\hline\hline
			Experiment & $E_{\rm cm}$ (GeV) & Mode & $D_s^+$ decay & $\cal B$~(\%)& $f_{D_s^+}|V_{\rm cs}|$ (MeV)\\
			\hline
				BESIII~\cite{BESIII:2023fhe} &  4.128-4.226 & $D^\pm_sD^{*\mp}_s$  & $\tau^+_\pi{\nu}_\tau$  & $5.44\pm0.17\pm0.13$ & $249.0\pm3.9\pm3.0\pm0.5 $ \\
			BESIII~\cite{BESIII:2021wwd}     & 4.178-4.226 &  $D^\pm_sD^{*\mp}_s$ & $\tau^+_\rho\nu_\tau$  & $5.30\pm0.25\pm0.20$ & $245.8\pm5.8\pm4.6\pm0.5$ \\
				BESIII~\cite{BESIII:2021bdp} & 4.178-4.226 & $D^\pm_sD^{*\mp}_s$ & $\tau^+_e\nu_\tau$  & $5.27\pm0.10\pm0.13$ & $245.1\pm2.3\pm3.0\pm0.5$ \\
				BESIII~\cite{BESIII:2023ukh} & 4.128-4.226 &  $D^\pm_sD^{*\mp}_s$ & $\tau^+_\mu{\nu}_\tau$  & $5.37\pm0.17\pm0.15$ & $247.4\pm3.9\pm3.5\pm0.5$ \\
			BESIII~\cite{BESIII:2016cws}            & 4.009 &  $D_s^+D_s^-$  & $\tau^+_\pi \nu_\tau$   & $3.28\pm1.83\pm0.37$ & $193.4\pm53.9\pm10.9\pm0.5$ \\
			BESIII~\cite{BESIII:2024dvk}&4.237-4.699& $D_{s}^{*\pm}D_{s}^{*\mp}$  & $\tau^+_e\nu_\tau,\tau^+_\mu\nu_\tau, \tau^+_\pi \nu_\tau,  \tau^+_\rho\nu_\tau$  & $5.60\pm0.16\pm0.20$ & $252.7\pm3.6\pm4.5\pm0.6$ \\
			\hline
		    BESIII average   & {$\cdot\cdot\cdot$ }& { $\cdot\cdot\cdot$}  &{$\tau^+\nu_\tau$ }  & {$5.359\pm0.067\pm0.075$}  &{ $247.2\pm1.5\pm1.7\pm0.5$ }\\
			\hline
			CLEO-c~\cite{CLEO:2009jky}   & 4.170 & $D^\pm_sD^{*\mp}_s$   & $\tau^+_e\nu_\tau$ & $5.30\pm0.47\pm0.22$ & $245.8\pm10.9\pm5.1\pm0.5$ \\
			CLEO-c~\cite{CLEO:2009vke}  & 4.170 &  $D^\pm_sD^{*\mp}_s$  & $\tau^+_\rho\nu_\tau$ & $5.52\pm0.57\pm0.21$ & $250.9\pm13.0\pm4.8\pm0.6$ \\
			CLEO-c~\cite{CLEO:2009lvj}  & 4.170 & $D^\pm_sD^{*\mp}_s$   & $\tau^+_\pi \nu_\tau$   & $6.42\pm0.81\pm0.18$ & $270.5\pm17.1\pm3.8\pm0.6$ \\
			BaBar~\cite{BaBar:2010ixw}  & 10.56 &  $DKX\gamma D^{-}_s$ & $\tau^+_e\nu_\tau ,\tau^+_\mu\nu_\tau$ & $4.96\pm0.37\pm0.57$ & $237.8\pm8.9\pm13.7\pm0.5$ \\
			Belle~\cite{Belle:2013isi}  & 10.56 &  $DKX\gamma D^{-}_s$   & $\tau^+_\pi \nu_\tau, \tau^+_e\nu_\tau,\tau^+_\mu\nu_\tau$ & $5.70\pm0.21^{+0.31}_{-0.30}$ & $254.9\pm4.7\pm7.0\pm0.6$\\
			\hline
			Average&&&&$5.38\pm0.09$&$247.7\pm2.1\pm0.5$\\
			\hline
			\hline
			BESIII~\cite{BESIII:2023cym}   & 4.128-4.226&  $D_s^{\pm}D_s^{*\mp}$ & $\mu^+\nu_\mu$ & $0.5294\pm0.0108\pm0.0085$ & $242.5\pm2.5\pm1.9\pm0.5$ \\
			BESIII~\cite{BESIII:2016cws} & 4.009 &  $D_s^+D_s^-$ & $\mu^+\nu_\mu$ & $0.517\pm0.075\pm0.021$ & $239.6\pm17.4\pm4.9\pm0.5$ \\
			BESIII~\cite{BESIII:2024dvk} & 4.237-4.699&$ D_{s}^{*\pm}D_{s}^{*\mp}$    & $\mu^+\nu_\mu$ & $0.547\pm0.026\pm0.016$ & $246.5\pm5.9\pm3.6\pm0.5$  \\
			\hline
			BESIII average  & {$\cdot\cdot\cdot$ }& { $\cdot\cdot\cdot$  }&{ $\mu^+\nu_\mu$  }& {$0.5310\pm0.0099\pm0.0053$ } & {$242.8\pm2.3\pm1.2\pm0.5$ }\\
			BESIII average  & {$\cdot\cdot\cdot$} & { $\cdot\cdot\cdot$  }&   { $\tau^+\nu_\tau,\mu^+\nu_\mu$}& {$\cdot\cdot\cdot$}& {$245.4\pm1.3\pm1.7\pm0.5
				$ }\\
		
			\hline
			CLEO-c~\cite{CLEO:2009lvj}   & 4.170 & $D^\pm_sD^{*\mp}_s$     & $\mu^+\nu_\mu$ & $0.565\pm0.045\pm0.017$ & $250.5\pm10.0\pm3.8\pm0.5$ \\
			BaBar~\cite{BaBar:2010ixw}   & 10.56 &  $DKX\gamma D^{-}_s$   & $\mu^+\nu_\mu$ & $0.602\pm0.038\pm0.034$ & $258.6\pm8.2\pm7.3\pm0.5$ \\
			Belle~\cite{Belle:2013isi}   & 10.56 &  $DKX\gamma D^{-}_s$     & $\mu^+\nu_\mu$ & $0.531\pm0.028\pm0.020$ & $242.8\pm6.4\pm4.6\pm0.5$ \\
			\hline	
			Average&&&&$0.539\pm0.009$&$244.6 \pm2.0\pm0.5$\\
\hline\hline
		\end{tabular}
\end{table*}

\subsection{Radiative leptonic decays of $D^+$ and $D^+_s$}

In 2017, BESIII reported the first search for the radiative leptonic decay $D^+\to\gamma e^+\nu_e$ by using 2.93 fb$^{-1}$ data at 3.773 GeV~\cite{BESIII:2017whk};
and reported the search for $D^+_s\to\gamma e^+\nu_e$ by analyzing 3.19 fb$^{-1}$ of data taken at 4.178~GeV~\cite{BESIII:2019pjk}.
The former result is superseded by a new search for $D^+\to\gamma e^+\nu_e$ with 20.3 fb$^{-1}$ of data at 3.773 GeV~\cite{BESIII:2025mnc} in 2025.
No significant signals are observed for both decays, and the upper limits on their partial decay branching fractions
with $E_\gamma>10$~MeV at the 90\% confidence level are set as
${\cal B}(D^+\to\gamma e^+\nu_e)<1.2\times 10^{-5}$
and
${\cal B}(D^+_s\to\gamma e^+\nu_e)<1.3\times 10^{-4}$. The measured result of $D^+\to\gamma e^+\nu_e$ excludes most existing theoretical predictions~\cite{Lu:2021ttf,Barik:2009zza,Korchemsky:1999qb,Yang:2014rna,Yang:2016wtm} and supports the theoretical predictions~\cite{Geng:2000if,Lu:2002mn,Desiderio:2020oej,Yang:2012jp}.
The measured result of $D^+_s\to\gamma e^+\nu_e$  is compatible with the theoretical predictions in Refs.~\cite{Geng:2000if,Lu:2002mn,Atwood:1994za,Korchemsky:1999qb,Yang:2012jp}.
  Table~\ref{tab:DToGamENu} summarizes the experimental results and theoretical predictions for ${\cal{B}}(D^+_{(s)}\to \gamma e^+\nu_e)$. 
\begin{table*}[htbp]
\caption{Experimental results and theoretical predictions for ${\cal{B}}(D^+_{(s)}\to \gamma e^+\nu_e)$.
\label{tab:DToGamENu}}
\def\1#1#2{\multicolumn{#1}{#2}}
\begin{tabular}{l c c c}
\hline\hline
Model &  ${\cal B}(D^+\to\gamma e^+\nu_e) (\times 10^{-5})$&${\cal B}(D^+_s\to\gamma e^+\nu_e) (\times 10^{-4})$\\
\hline
Light front quark model~\cite{Geng:2000if} & 0.69&$0.77$\\
Non-relativistic quark model~\cite{Lu:2002mn} & 0.46&$0.18$\\
Relativistic quark model~\cite{Barik:2009zza} & 3.34&---\\ 
Perturbative QCD~\cite{Korchemsky:1999qb} & $8.2\pm6.5$&$9\pm8$\\
Lattice QCD~\cite{Desiderio:2020oej} & $0.09\pm0.04$&---\\
QCD factorization ~\cite{Yang:2014rna} & 2.81&---\\
QCD factorization ~\cite{Yang:2016wtm} & 1.92&---\\
QCD factorization ~\cite{Lu:2021ttf} & $\left(1.88^{+0.36}_{-0.29},\, 2.31^{+0.65}_{-0.54}\right)$&---\\
Non-relativistic quark model~\cite{Atwood:1994za}&---&1.7\\
Vector meson dominance model~\cite{Yang:2012jp}&0.75&$1$\\
\hline
BESIII~\cite{BESIII:2025mnc,BESIII:2019pjk} &   $<1.2$ @90\% C. L.& $<1.3$ @90\% C. L.\\
\hline
\hline
\end{tabular}
\end{table*}

\subsection{Leptonic $D^{*+}_s$ and $D^{*+}$ decays}

In the SM, the decay width of $D_{(s)}^{*+}\to \ell^+\nu_{\ell}$ can be written as~\cite{Donald:2013sra}
\begin{linenomath*}
\small
\begin{multline}
  \Gamma(D_{(s)}^{*+}\to \ell^+\nu_{\ell})=\\
  \frac{G_F^2}{12\pi}\vert V_{cq}\vert^2f_{D_{(s)}^{*+}}^2m_{D_{(s)}^{*+}}^3\left (1-\frac{m_{\ell^+}^2}{m_{D_{(s)}^{*+}}^2} \right )^2
  \left (1+\frac{m_{\ell^+}^2}{2m_{D_{(s)}^{*+}}^2}\right ),
  \label{eq:Gamma}
\end{multline}
\end{linenomath*}
\\
where
$G_F$ is the Fermi coupling constant,
$\vert V_{cq}\vert$ is the $c\to q$~($q=s$ or $d$) CKM matrix element,
$m_{\ell^+}$ is the lepton mass,
and $m_{D_{(s)}^{*+}}$ is the $D_{(s)}^{*+}$ mass.
The branching fraction of $D^{*+}_s\to \ell^+\nu_{\ell}$ is predicted to be up to $10^{-5}$ according to Refs.~\cite{Donald:2013sra,Cheng:2022mvd,Yang:2021crs}.
The decay constant $f_{D_s^{*+}}$ has been calculated via
the nonrelativistic quark model, the relativistic quark model, the light-front quark model,
QCD sum rules, and LQCD with predicted values varying from 212 to 447 MeV, as summarized in Ref.~\cite{BESIII:2023zjq}.

\begin{figure}[htbp]
  \centering
  \includegraphics[width=0.4\textwidth]{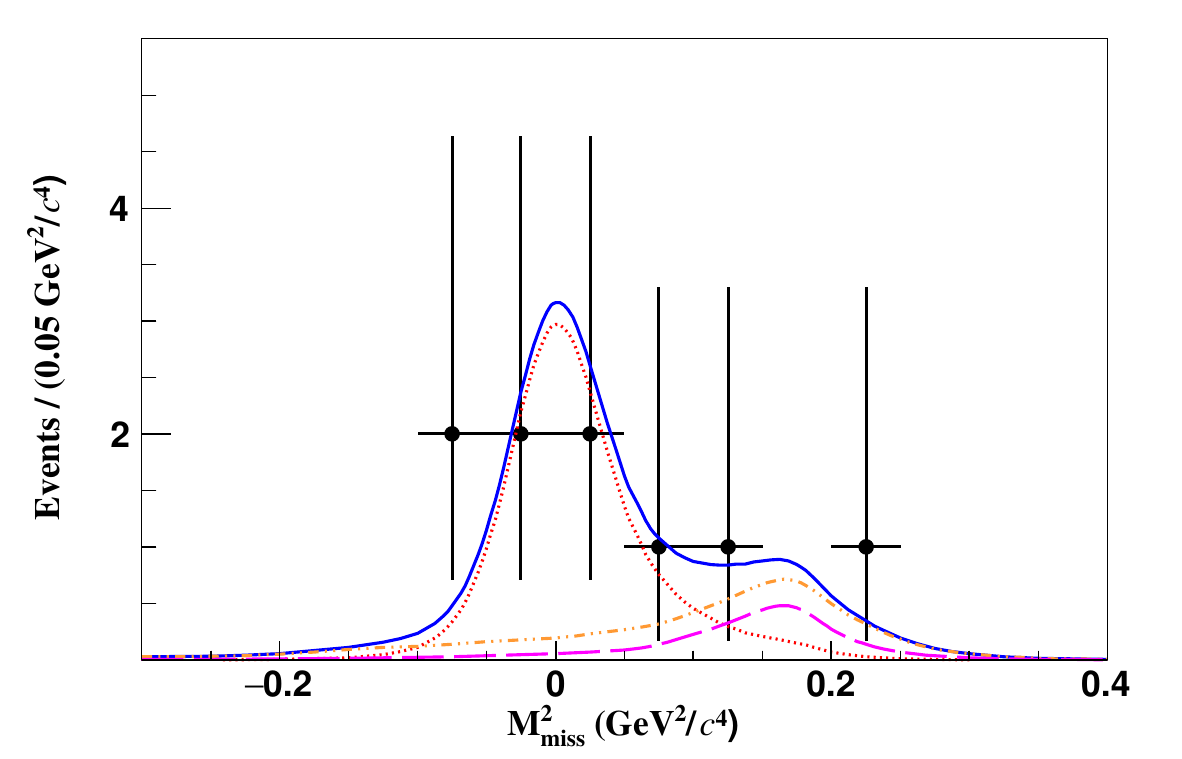}
  \caption{The $M_{\rm miss}^{2}$ distributions of the accepted candidates for $D^{*+}_s\to e^+\nu_e$~\cite{BESIII:2023zjq}.
}
  \label{fig:Dsstar_enu}
\end{figure}

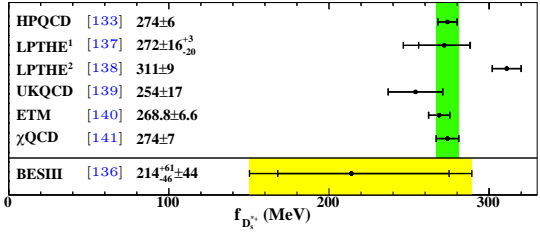
\begin{figure}[htbp]
  \centering
  \includegraphics[width=0.4\textwidth]{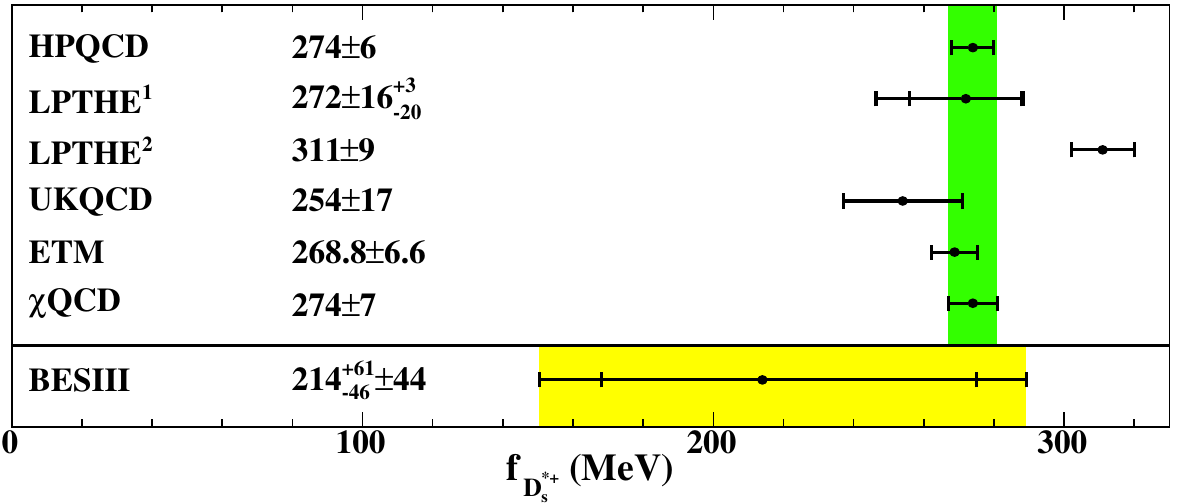}
 \put(-175,77.5){\tiny~\cite{Donald:2013sra}}
  \put(-175,68.6){\tiny~\cite{Becirevic:1998ua}}
  \put(-175,59.7){\tiny~\cite{Becirevic:2012ti}}
  \put(-175,50.8){\tiny~\cite{Bowler:2000xw}}
  \put(-175,41.9){\tiny~\cite{Lubicz:2017asp}}
  \put(-175,33){\tiny~\cite{Chen:2020qma}}
  \put(-175,20){\tiny~\cite{BESIII:2023zjq}}    
   \caption{Comparison of $f_{D^{*+}_s}$ reported by BESIII~\cite{BESIII:2023zjq}
   as well as LQCD calculations of
    HPQCD~\cite{Donald:2013sra},
   LPTHE~\cite{Becirevic:1998ua,Becirevic:2012ti},
   UKQCD~\cite{Bowler:2000xw},
   ETM~\cite{Lubicz:2017asp},
   and $\chi$QCD~\cite{Chen:2020qma}. The green band is the $\pm 1\sigma$ region of $\chi$QCD and the yellow band denotes the $\pm 1\sigma$ region of the BESIII result~\cite{BESIII:2023zjq}.}
  \label{fig:fDsstar}
\end{figure}

In 2024, a hint of $D^{*+}_s\to e^+\nu_e$ was obtained with about 0.38 million tagged $D_s^-$ mesons via $e^+e^-\to D_sD^*_s$
at 4.178-4.226~GeV. Figure~\ref{fig:Dsstar_enu} shows the
$M_{\rm miss}^{2}$ distribution of the accepted candidates for $D^{*+}_s\to e^+\nu_e$.
6.2 $D^{*+}_s\to e^+\nu_e$ signal events are obtained with a significance of $2.9\sigma$~\cite{BESIII:2023zjq},
corresponding to ${\cal B}(D^{*+}_s\to e^+\nu_e)=(2.1^{+1.2}_{-0.9}\pm0.2)\times10^{-5}$.
Taking  the total width of the $D^{*+}_s$ [$(0.070\pm0.028)$~keV] predicted with the radiative $D^{*+}_s$ decays
from the LQCD calculation~\cite{Donald:2013sra} and the value of $|V_{cs}|$ given by the global SM fit as input,
the $f_{D_s^{*+}}$ is determined to be $(214^{+61}_{-46}\pm44)$~MeV for the first time.
Figure~\ref{fig:fDsstar} shows a comparison of the $f_{D^{*+}_s}$ reported by BESIII and those predicted by theoretical calculations.

In addition, the first searches for $D^{*+}\to e^+\nu_e$ and $D^{*+}\to \mu^+\nu_\mu$ were also presented
with 0.33 million tagged $D^{*-}$ mesons via $e^+e^-\to D^{*+}D^{*-}$ from 6.3 fb$^{-1}$ of data at 4.178-4.226 GeV~\cite{BESIII:2024kxe}.
No significant signals were obtained; the upper limits on their decay branching fractions are set to be ${\cal B}(D^{*+}\to e^+\nu_e)< 1.1\times 10^{-5}$ and
${\cal B}(D^{*+}\to \mu^+\nu_\mu)<4.3\times 10^{-6}$ at the 90\% confidence level, respectively.

\section{Semileptonic decays}
\label{sec:semileptonic}

\subsection{Theoretical formula}

\subsubsection{Formula for $D\to P\ell^+\nu_\ell$ or $D\to S\ell^+\nu_\ell$}

The theoretical formula for the semileptonic decays $D\to S\ell^+\nu_\ell$ ($S$ denotes a scalar meson) is similar to those for $D\to P\ell^+\nu_\ell$ ($P$ denotes a pseudoscalar meson).
Therefore, we take the $D \to P \ell^+\nu_\ell$ decays for example in this article.
Their differential decay widths can be expressed as~\cite{Becher:2005bg,Faustov:2019mqr}
\begin{equation}
\label{ffk_function1}
\begin{array}{l}
   \displaystyle \frac{d\Gamma_{i}}{dq^{2}} =
   \frac{G_{F}^{2}|V_{cq}|^{2}}{24\pi^{3}}
   \frac{\left(q^{2}-m^{2}_{\ell}\right)^2|\vec p_{P}|}{q^{4}m^{2}_{D}}
   \displaystyle\left[\left(1+\frac{m^{2}_{\ell}}{2q^{2}}\right)m^{2}_{D}|\vec p_{P}|^{2}\right.\\\times|f^{D\to P}_{+}\left(q^{2}\right)|^{2}
   \displaystyle\left.+\frac{3m^{2}_{\ell}}{8q^{2}}\left(m^{2}_{D}-m^{2}_{P}\right)^{2}|f^{D\to P}_{0}\left(q^{2}\right)|^{2}\right],
\end{array}
\end{equation}
where $q$ is the four-momentum transfer to the $\ell^+\nu_\ell$ system, $|\vec p_{P}|$ is the modulus of the $P$ meson three-momentum in the $D$ rest frame, $G_F$ is the Fermi
constant, $f^{D\to P}_{+}(q^2)$ and $f^{D\to P}_{0}(q^2)$ are the vector and scalar hadronic transition form factors of $D\to P$, respectively.

In the early days, a single pole of the form
\begin{equation}
     f^{D\to P}_+(q^2)=\frac{f^{D\to P}_+(0)}{1-\frac{q^2}{M^2_{\rm pole}}}
\label{pole}
\end{equation}
was widely adopted to describe the vector form factor across different approaches,
including the constituent quark model, lattice gauge calculations, and QCD sum rules.
This was exemplified by the K\"orner Schuler~\cite{Korner:1987kd,Korner:1989tb,Korner:1989qb,Korner:1990ri} and
Bauer-Stech-Wirbel~\cite{Wirbel:1985ji,Bauer:1988fx} models.

The modified pole model is commonly written as
\begin{equation}
f^{D\to P}_+(q^2)=\frac{f^{D\to P}_+(0)}{(1-\frac{q^2}{M^2_{\rm pole}})(1-\alpha \frac{q^2}{M^2_{\rm pole}})},
\label{modifies-pole_model}
\end{equation}
with $\alpha$ a free parameter.
This parametrization, known as the Becirevic-Kaidalov form~\cite{Becirevic:1999kt}, was widely
adopted in subsequent LQCD computations and experimental analyses of semileptonic $D$ decays.

Currently, the series expansion~\cite{Becher:2005bg} is the most popular parameterization,
which has the form
\begin{equation}
     f^{D\to P}_{+}(q^2)=\frac{1}{P(q^2)\Phi(q^2)} \sum_{k=0}^{\infty} a_k[z(q^2)]^k,
 \label{eq:series}
\end{equation}
where $a_k$ are the real coefficients,
the functions $P(q^2)$, $\Phi(q^2)$, and $z(q^2)$ are defined following Ref.~\cite{CLEO:2009svp}.
In practice, one often takes $k_{\rm max}=1$ or $k_{\rm max}=2$ in Eq.~(\ref{eq:series}).

The scalar form factor $f^{D\to P}_{0}(q^2)$ is similar to $f^{D\to P}_{+}(q^2)$ but with a one-parameter series expansion, which is given by~\cite{Faustov:2019mqr}
\begin{equation}
f^{D\to P}_0(q^2) = \frac{1}{P(q^2)\Phi(q^2)}f^{D\to P}_0(0)P(0)\Phi(0).
\label{equation:ff0}
\end{equation}
Here, $f^{D\to P}_0(q^2)$ has the same normalization at $q^2=0$ as $f^{D\to P}_+(q^2)$, {\it i.e.}
\begin{equation}
f^{D\to P}_0(0) = f^{D\to P}_+(0),
\end{equation}
but with a different pole mass $m_{D_{s0}^{*}(2317)^+}$ for $c\to s \ell^+\nu_\ell$
and $m_{D_{0}^{*}(2300)}$  for $c\to d \ell^+\nu_\ell$ in $P(q^2)$.

\subsubsection{Formula for $D\to V\ell^+\nu_\ell$}

\begin{figure}[htp]
  \begin{center}
  \includegraphics[width=0.8\linewidth]{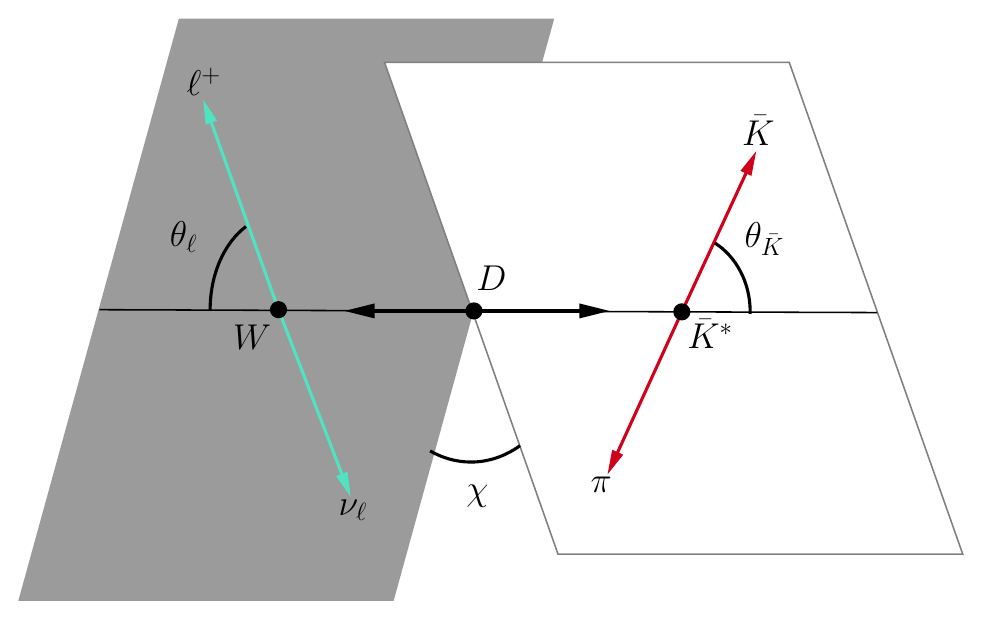}
  \caption{Definitions of the angular variables for $D\to \bar K\pi\ell^+\nu_\ell$.}
  \label{fig:phys_angle}
  \end{center}
\end{figure}

For $D\to V\ell^+\nu_\ell$ ($V$ denotes a vector meson), the four-body decays $D^{0(+)} \to \bar K \pi\ell^+ \nu_\ell$ are taken for example.
These decays can be usually characterized by five kinematic variables: the invariant-mass squared of $\bar K \pi$~($m^{2}$), the invariant-mass squared of $\ell^+ \nu_\ell$~($q^{2}$),
the angle between the three-momentum of the $\bar K$ in the $\bar K\pi$ rest frame and the direction of flight of the $\bar K\pi$ in the $D$ rest frame~($\theta_{\bar K}$), the angle between the three-momentum of $\ell^+$ in the $\ell^+ \nu_\ell$ rest frame and the direction of flight of the $\ell^+ \nu_\ell$ in the $D$ rest frame ($\theta_{\ell}$), and the angle between the two decay planes~($\chi$). As an example, Fig.~\ref{fig:phys_angle} shows the angular variables for $D\to \bar K\pi\ell^+\nu_\ell$.
The sign of $\chi$ is changed when analyzing $\bar D$ decays to ensure $CP$ conservation. Neglecting the positron mass, the differential decay width for the $D \to \bar K\pi e^+\nu_e$ decay is written as~\cite{Lee:1992ih}
\begin{equation}
\begin{aligned}
d^{5}\Gamma &~=~\frac{G^{2}_{F}||V_{cq}||^{2}}{(4\pi)^{6}m^{3}_{D}}X\beta{\cal I}(m^{2}, q^{2}, \theta_{\bar K}, \theta_{\ell}, \chi)\\&
\quad \times dm^{2} dq^{2} d\cos(\theta_{\bar K}) d\cos(\theta_{\ell})d\chi.
\end{aligned}
\end{equation}
In this expression, $G_F$ is the Fermi constant, $|V_{cq}|$~($q=s$ or $d$) is the $c \to q$ CKM matrix element, $X=p_{\bar K\pi}m_{D}$, and $\beta= 2p^{*}/m$, where $p^{*}$ denotes the momentum of $\bar K$ in the $\bar K\pi$ rest frame.
The dependence of $\cal I$ on $\theta_{\ell}$ and $\chi$ is given by~\cite{Lee:1992ih}
\begin{equation}
\begin{aligned}
  {\cal I}~=~& {\cal I}_{1}+{\cal I}_{2}\cos2\theta_{\ell}+{\cal I}_{3}\sin^{2}\theta_{\ell}\cos2\chi+{\cal I}_{4}
  \sin2\theta_{\ell}\cos\chi+\\
  &{\cal I}_{5}\sin\theta_{\ell}\cos\chi +{\cal I}_{6}\cos\theta_{\ell}+{\cal I}_{7}\sin\theta_{\ell}\sin\chi+\\&
    {\cal I}_{8}\sin2\theta_{\ell}\sin\chi
  +{\cal I}_{9}\sin^{2}\theta_{\ell}\sin2\chi,
\end{aligned}
\label{eq:decay_intensity}
\end{equation}
where the quantities ${\cal I}_{1, \ldots,9}$ depend on $m^{2}$, $q^{2}$, and $\theta_{\bar K}$,
and can be expressed in terms of three form factors ${\cal F}_{1,2,3}$~\cite{Lee:1992ih}.
Then one can expand  ${\cal F}_{1,2,3}$ into partial wave amplitudes including $\cal S$-wave
(${\cal F}_{10}$), $\cal P$-wave (${\cal F}_{i1}$) and $\cal D$-wave (${\cal F}_{i2}$), to show their explicit dependences on
$\theta_{\bar K}$.
The form factors are usually written as~\cite{BaBar:2010vmf,BESIII:2015hty}

\begin{equation}
\begin{split}
	\mathcal{F}_{1} &= \mathcal{F}_{10} + \mathcal{F}_{11} \cos\theta_{\bar K} + \mathcal{F}_{12} \frac{3\cos^2\theta_{\bar K}-1}{2}, \\
	\mathcal{F}_{2} &= \frac{1}{\sqrt{2}}\mathcal{F}_{21} + \sqrt{\frac{3}{2}}\mathcal{F}_{22} \cos\theta_{\bar K},\\
	\mathcal{F}_{3} &= \frac{1}{\sqrt{2}}\mathcal{F}_{31} + \sqrt{\frac{3}{2}}\mathcal{F}_{32} \cos\theta_{\bar K}.
\end{split}
\end{equation}

The ${\cal S}$-wave form-factor is described by~\cite{BaBar:2010vmf,BESIII:2015hty}
\begin{equation}
  \begin{split}
 \mathcal{F}_{10} &= p_{\bar K\pi} m_D \frac{1}{1-\frac{q^2}{m_A^2}} \times \mathcal{A}_S(m), 
 \end{split}
\label{eq:form_factor_S}
\end{equation}
where $\mathcal{A}_{\cal S}(m)$ corresponds to the mass-dependent ${\cal S}$-wave amplitude. The expression
$\mathcal{A}_{\cal S}(m) = r_{\cal S} P(m){\rm e}^{i\delta}_{\cal S}(m)$ from~\cite{BaBar:2010vmf,BESIII:2015hty} is adopted, in which
$P(m) = 1 + xr^{(1)}_{\cal S}$ with $x = \sqrt{(\frac{m}{m_{\bar K}+m_\pi})^2-1}$, where $r_{\cal S}$ and $r^{(1)}_{\cal S}$
are the relative intensity and the dimensionless
coefficient, respectively. The ${\cal S}$-wave phase takes the form of $\delta_{\cal S}(m) = \delta^{1/2}{\rm BG}$.
The mass dependence of $\delta^{1/2}{\rm BG}$ is described with $\cot(\delta^{1/2}{\rm BG})=1/(a_{\cal S,\rm BG}^{1/2}p^*)+b_{\cal S,\rm BG}^{1/2}p^*/2$,
where $a_{\cal S,\rm BG}^{1/2}$ and $b_{\cal S,\rm BG}^{1/2}$ are the scattering length and the effective range, respectively.

The $\cal P$-wave form factors $\mathcal{F}_{11}$, $\mathcal{F}_{21}$, and $\mathcal{F}_{31}$ are related to the helicity basis form factors $H_{0,\pm,t}(q^2)$ by
\begin{equation}
\begin{split}
	\mathcal{F}_{11} &= 2\sqrt{2}\alpha q H_0 \times \mathcal{A}(m), \\
	\mathcal{F}_{21} &= 2\alpha q (H_+ + H_-) \times \mathcal{A}(m), \\
	\mathcal{F}_{31} &= 2\alpha q (H_+ - H_-) \times \mathcal{A}(m).
\end{split}
  \label{eq:form_factor}
\end{equation}
Here $\alpha$ is a constant factor $\alpha=\sqrt {3\pi{\cal B}_{\bar K^*}/(p_0^*\Gamma_0)}$ with ${\cal B}_{\bar K^*}={\cal B}(\bar K^*(892)\to \bar K\pi)$, the $p^*_0$ is the momentum of the $\bar K$ at the
pole mass of the $\bar K^*(892)$ resonance, and the $\Gamma_0$ is the total width of the $\bar K^*$. ${\cal A}(m)$ denotes the amplitude of the ${\cal P}$-wave component, taken
as the Breit-Wigner (BW) form as in the following equation
\begin{equation}
{\cal A}(m)=\frac{m_0\Gamma_0F_J(m)}{m_0^2-m^2-im_0\Gamma(m)},
\end{equation}
where $m_0$ is the pole mass of the resonance. This parameterization is applicable to resonances of different angular momenta denoted by $J$. In the case of the ${\cal P}$-wave, $J=1$. The mass-dependent width $\Gamma(m)$ is given by $\Gamma(m)=\Gamma_0\frac{p^*}{p^*_0}\frac{m_0}{m}F_J^2(m)$, $F_J=(\frac{p^*}{p^*_0})^J\frac{B_J(p^*)}{B_J(p^*_0)}$. $B_J$ is the Blatt-Weisskopf damping factor given by $B_0(p)=1$, $B_1(p)=1/\sqrt{1+r_{\rm BW}^2p^2}$, and $B_2(p)=1/\sqrt{(r_{\rm BW}^2p^2-3)^2-9r_{\rm BW}^2p^2}$, where the barrier factor $r_{\rm BW}$ for light mesons is about 0.5 fm.
 The helicity form factors can be related to
the two axial-vector form factors, $A_1(q^2)$ and $A_2(q^2)$, as well as the vector form factor $V(q^2)$.
All three of them are taken as a one-pole form $A_{1,2}(q^2) = A_{1,2}(0)/(1-q^2/M^2_A)$
and $V (q^2) = V(0)/(1-q^2/M_V^2)$, with pole masses $M_V = M_{D_s^*(1^-)} = 2.1121$~GeV/$c^2$ and
$M_A = M_{D_{s1}(1^+)} = 2.4595$~GeV/$c^2$, which are the world average values. The form factor $A_1(q^2)$ is common to all three helicity
amplitudes. Therefore, it is natural to define two form factor ratios as $r_V = V(0)/A_1(0)$ and
$r_2 = A_2(0)/A_1(0)$ at the momentum transfer square $q^2 = 0$.

The $\mathcal{D}$-wave related form factors $\mathcal{F}_{12}$, $\mathcal{F}_{22}$ and $\mathcal{F}_{32}$ are necessary,
and they take the form of
\begin{linenomath*}
\begin{eqnarray}
\mathcal{F}_{12} &=& \frac{m_{D}p_{\bar K\pi}}{3} \left[-\frac{m^2_{D} \, p^2_{\bar K\pi}}{m_{D}+m_{\bar K\pi}} \, T_2(q^2) \right.\nonumber \\
                 &+&(m^2_{D}-m^2_{\bar K\pi}-q^2)(m_{D}+m_{\bar K\pi}) \nonumber \\
  & & \left.  T_1(q^2) \right] \, \mathcal{A}^{\prime}(m), \nonumber \\
  \mathcal{F}_{22} &=& \sqrt{\frac{2}{3}}\, m_{D} \, m_{\bar K\pi} \, q \, p_{\bar K\pi}(m_{D}+m_{\bar K\pi}) \nonumber \\
                 & & T_1(q^2) \, \mathcal{A}^{\prime}(m), \nonumber \\
\mathcal{F}_{32} &=& \sqrt{\frac{2}{3}} \, \frac{2m^2_{D} \, m_{\bar K\pi} \, q \, p^2_{\bar K\pi}}{m_{D}+m_{\bar K\pi}} \, T_V(q^2) \, \mathcal{A}^{\prime}(m) \, . \nonumber
\end{eqnarray}
They are assumed to follow the simple pole model and contain one vector and two axial-vector form factors which are denoted as $T_V(q^2)$ and $T_{1,2}(q^2)$.
The ratios of the form factor parameters are then defined as $r_V^{\cal D}=T_V(0)/T_1(0)$ and $r_2^{\cal D}=T_2(0)/T_1(0)$ at $q^2=0$, where $r_V^{\cal D}$ and $r_2^{\cal D}$ are expected to be 1~\cite{CLEO:2004jkt,BaBar:2010vmf}. The amplitude $\mathcal{A}^{\prime}(m)$ takes a relativistic BW form with a mass-dependent width,
$$ \mathcal{A}^{\prime}(m) = \frac{r_D \, m_{\bar K_2^*(1430)} \, \Gamma^0_{\bar K_2^*(1430)} \, F_2(m)}{m_{\bar K_2^*(1430)}^2 - m^2 - i \, m_{\bar K_2^*(1430)} \, \Gamma_{\bar K_2^*(1430)}(m)},$$
\end{linenomath*}
where $r_D$ denotes the magnitude of the $\mathcal{D}-$wave amplitude, $F_2(m)=\left(\frac{p^*}{p_0^*}\right)^2\frac{B_2(p^*)}{B_2(p_0^*)}$ with $B_2(p^*)=1/\sqrt{(r_{\rm BW}^2p^{*2}-3)^2+9r_{\rm BW}^2p^{*2}}$, and $\Gamma_{\bar K_2^*(1430)}(m)=\Gamma^0_{\bar K_2^*(1430)}\left(\frac{p^*}{p^*_0}\right)\frac{m_{\bar K_2^*(1430)}}{m_{\bar{K}\pi}} \, F^2_2(m)$ with the mass $m_{\bar K_2^*(1430)}$ and width $\Gamma^0_{\bar K_2^*(1430)}$ for $\bar K_2^*(1430)$ fixed to the world average values.

\subsubsection{Formula for  $D\to A\ell^+\nu_\ell$}

For $D\to A\ell^+\nu_\ell$ ($A$ denotes an axial-vector meson), the decays $D \to \bar K_1(1270)\ell^+ \nu_\ell$ are taken for example.
The covariant tensor amplitude is constructed as
\begin{equation}
\begin{aligned}
\mathcal{M} & =(V-A)^{\mu\eta}\cdot
[\sum_{\lambda_{W}}\epsilon^{*}(\lambda_{W})_{\mu}\epsilon(\lambda_{W})_{\rho}] \cdot\\
&
[\sum_{\lambda_{\bar K_1}}\epsilon^{*}(\lambda_{\bar K_1})_{\eta}\epsilon(\lambda_{\bar K_1})_{\sigma}]\cdot \mathcal{R}_{\bar{K}_1}\cdot J^{\sigma}\cdot\bar{u}_\nu\gamma^{\rho}(1-\gamma_5)v_l.
\end{aligned}
\end{equation}
Here, $(V-A)^{\mu\eta}\epsilon^{*}(\lambda_{\bar K_1})_{\eta}$ is the current for $D\to {\bar K}_1 W^*$ following the convention in Ref.~\cite{Bian:2021gwf}, written as
\begin{equation}
\begin{aligned}
V^{\mu\eta} \epsilon^{*}(\lambda_{\bar K_1})_{\eta}  =&
-(m_D-M_{\bar K_1}) V_1(q^2)\epsilon^{*\mu}(\lambda_{\bar K_1}) \\
&+V_2(q^2) \left(\frac{q\cdot\epsilon^{*}(\lambda_{\bar K_1})}{m_D-M_{\bar K_1}}\right)(p_D+p_{\bar K_1})^\mu,\\
A^{\mu\eta} \epsilon^{*}(\lambda_{\bar K_1})_{\eta} =&-\frac{2 i A(q^2)}{m_D-M_{\bar K_1}} \epsilon^{\mu \epsilon^*(\lambda_{\bar K_1}) p^{D}p^{\bar K_1}}. \\
\end{aligned}
\end{equation}
Both vector  and axial-vector form factors take a single pole form, written as  $V_{1,2}(q^2)= \frac{V_{1,2}(0)}{1-q^2/m_V^2}$ and $A(q^2)=\frac{A(0)}{1-q^2/m_A^2}$ with $q^\mu=p^\mu_{D}-p_{\bar K_1}^\mu$;
$\epsilon(\lambda_{W})$ and $\epsilon(\lambda_{\bar K_1})$ are the polarization vectors of the $W$-boson and the $\bar{K}_1$ meson, respectively; $\mathcal{R}_{\bar{K}_1}=\frac{1}{s-m_0^2+im_0\Gamma_0}$ is the BW function of $\bar{K}_1$ with mass $m_0$ and width $\Gamma_0$; $J^\sigma$ is the hadronic current of $\bar{K}_1\to K^-\pi^+\pi^{0(-)}$, which includes several intermediate states like $\rho(770)$ and $K^*(892)$; $W^*\to e^+\nu_e$ is described with $\bar{u}_\nu\gamma^{\rho}(1-\gamma_5)v_l$. The current statistics is not sensitive to the non-zero mass of the lepton, hence the approximation $m_{\ell}\to0$ is applied in the simplification. See more details about the amplitude formula in
the Appendix of Ref.~\cite{BESIII:2025hdt}. For the charge conjugate decay modes, the three-momenta of the final states from $\bar{D}$ decay are inverted  to incorporate them with the $D$ by assuming charge-parity conservation.
The isospin relationships are imposed to constrain the form factors in the semileptonic decays and complex coupling coefficients in the hadronic current $J^\sigma$ as shown in the Appendix of Ref.~\cite{BESIII:2025hdt}. Here, $V_{1}(0)$ is fixed to be 1 with the ratios $r_V=V_{2}(0)/V_{1}(0)$ and $r_A=A(0)/V_{1}(0)$ allowed to float, and the mass and width of $\bar K_1$ are free in the fit.

\subsection{Previous experimental studies}

Earlier measurements of exclusive semileptonic $D^{0(+)}$ decays into $\bar K$, $\pi$, $\bar K^*(892)$, or $\rho(770)$ were made relative
to the reference decays. For $D^{0(+)}\to \bar K\ell^+\nu_\ell$ or $D^{0}\to \pi^-\ell^+\nu_\ell$,
E691~\cite{TaggedPhotonSpectrometer:1988hnm},
CLEO~\cite{CLEO:1991ljl},
CLEOII~\cite{CLEO:1993isi}, and
BaBar~\cite{BaBar:2007zgf}
measured
the branching fraction of $D^0\to K^-e^+\nu_e$ relative to $D^0\to K^-\pi^+$;
CLEO~\cite{CLEO:1991ljl} and E687~\cite{E687:1993tpl,E687:1995rqi} measured
the branching fraction of $D^0\to K^-\mu^+\nu_\mu$ relative to $D^0\to K^-\pi^+$;
E653~\cite{FermilabE653:1991cdx,E653:1994wvf} measured
the branching fraction of $D^0\to K^-\mu^+\nu_\mu$ relative to $D^0\to X\mu^+\nu_\mu$;
while FOCUS~\cite{FOCUS:2004uby} reported the
branching fraction of $D^+\to \bar K^0\mu^+\nu_\mu$ relative to $D^+\to K^-2\pi^+$;
BaBar~\cite{BaBar:2014xzf} measured
the branching fraction of $D^0\to \pi^-e^+\nu_e$ relative to $D^0\to K^-\pi^+$.
CLEOII~\cite{CLEO:1995uht},
E687~\cite{E687:1996mzs}, and
CLEO-c~\cite{CLEO:2004arv} measured
the branching fraction of $D^0\to \pi^-e^+\nu_e$ relative to $D^0\to K^-e^+\nu_e$;
while FOCUS~\cite{FOCUS:2004qux} measured
the branching fraction of $D^0\to \pi^-\mu^+\nu_\mu$ relative to $D^0\to K^-\mu^+\nu_\mu$.

For $D^{0(+)}\to \bar K(892)^*\ell^+\nu_\ell$ or $D^{+}\to \rho(770)^0\ell^+\nu_\ell$,
CLEOII~\cite{CLEO:1993isi} reported the branching fraction of $D^0\to K^{*}(892)^-e^+\nu_e$ relative to $D^0\to K^0_S\pi^+\pi^-$;
while
FOCUS~\cite{FOCUS:2004zbs} reported the
branching fraction of $D^0\to K^{*}(892)^-\mu^+\nu_\mu$ relative to $D^0\to K^0_S\pi^+\pi^-$.
E691~\cite{FOCUS:2004uby},
FOCUS~\cite{TaggedPhotonSpectrometer:1988klg},
WA82~\cite{WA82:1991wsl}, and
CLEOII~\cite{CLEO:2002nbz} reported the
branching fraction of $D^+\to \bar K^*(892)^0e^+\nu_e$ relative to $D^+\to K^-2\pi^+$;
BaBar~\cite{BaBar:2010vmf} reported the
branching fraction of $D^+\to K^-\pi^+e^+\nu_e$ relative to $D^+\to K^-2\pi^+$;
while FOCUS~\cite{Focus:2002twy},
E653~\cite{FermilabE653:1992fap},
E687~\cite{E687:1993qiw},
and CLEOII~\cite{CLEO:2002nbz} reported the
branching fraction of $D^+\to \bar K^*(892)^0\mu^+\nu_\mu$ relative to $D^+\to K^-2\pi^+$.
Moreover,
FOCUS~\cite{FOCUS:2004uby} reported the branching fraction of $D^+\to \bar K^*(892)^0\mu^+\nu_\mu$  relative to $D^+\to \bar K^0\mu^+\nu_\mu$.
In addition,
E687~\cite{E687:1996tkt},
E791~\cite{E791:1996doj}, and
FOCUS~\cite{FOCUS:2005yyc} reported
the branching fraction of $D^+\to \rho(770)^0\mu^+\nu_\mu$ relative to $D^+\to \bar K^*(892)^0\mu^+\nu_\mu$;
while E791~\cite{E791:1996doj} reported
the branching fraction of $D^+\to \rho(770)^0e^+\nu_e$ relative to $D^+\to \bar K^*(892)^0e^+\nu_e$.

Previous direct measurements of  exclusive semileptonic decays of charmed mesons were from the MARKIII, BESII, and CLEO-c experiments.
MARKIII and BESII measured the branching fractions of
$D^0 \to K^- e^+ \nu_e$ and $D^0 \to \pi^- e^+ \nu_e$~\cite{MARK-III:1989dea,BES:2004rav}, as well as
$D^+ \to \bar K^0 e^+ \nu_e$~\cite{BES:2004obp} and $D^+ \to \bar K^*(892)^0 e^+ \nu_e$~\cite{MARK-III:1990bbt,BES:2006kzp}.
BaBar reported measurements of the branching fractions for
$D^0 \to K^- e^+ \nu_e$~\cite{BaBar:2007zgf} and $D^0 \to \pi^- e^+ \nu_e$~\cite{BaBar:2014xzf},
along with the hadronic transition form factors for $D \to K$ and $D \to \pi$.
Belle reported similar measurements for $D^0 \to K^- \ell^+ \nu_\ell$ and $D^0 \to \pi^- \ell^+ \nu_\ell$~\cite{Belle:2006idb},
including the hadronic transition form factors for $D \to K$ and $D \to \pi$.
Earlier CLEO-c studies reported branching fractions for
$D^0 \to h^- e^+ \nu_e$ ($h = K, \pi, K(892)^*, \rho(770)$)~\cite{CLEO:2005rxg} and
$D^+ \to h e^+ \nu_e$ ($h = \bar K^0, \pi^0, \rho(770)^0, \omega, \eta, \eta^\prime$)~\cite{CLEO:2005cuk},
based on a little portion of CLEO-c data at $3.774~\text{GeV}$, where several of these decay modes were observed for the first time.
Subsequently, improved measurements of the branching fractions for
$D \to \bar K e^+ \nu_e$, $D \to \pi e^+ \nu_e$, $D^+ \to \bar K^*(892)^0 e^+ \nu_e$, and $D \to \rho(770)e^+ \nu_e$,
along with the hadronic transition form factors for $D \to K$~\cite{CLEO:2007ntr,CLEO:2009svp},
$D \to \pi$~\cite{CLEO:2007ntr,CLEO:2009svp}, $D^+ \to \bar K^*$~\cite{CLEO:2010enr},
$D \to \rho$~\cite{CLEO:2011ab}, and $D^+ \to \eta$~\cite{CLEO:2010pjh},
were reported by CLEO-c using their full data at $3.774~\text{GeV}$.
In addition,
MARKIII~\cite{MARK-III:1990bbt} and BESII~\cite{BES:2006zzd} reported the branching fractions of
$D^0\to \bar K\pi e^+\nu_e$, and $D^+\to K^-\pi^+e^+\nu_e$;
and CLEO-c~\cite{CLEO:2007oer} reported the branching fraction of $D^0\to K^-\pi^+\pi^-e^+\nu_e$.

For  exclusive semileptonic $D^+_s$ decays,
CLEO~\cite{CLEO:1990lnz},
ARGUS~\cite{ARGUS:1990ptq},
E687~\cite{E687:1993uke},
CLEOII~\cite{CLEO:1994qdf}, and
FOCUS~\cite{Focus:2002twy}
reported the branching fraction of $D^+_s\to \phi e^+\nu_e$ relative to $D^+_s\to \phi\pi^+$;
CLEOII~\cite{CLEO:1995igy} reported
the branching fractions of $D^+_s\to \eta e^+\nu_e$ and $D^+_s\to \eta^\prime e^+\nu_e$ relative to $D^+_s\to \phi e^+\nu_e$.
Later, the branching fraction measurements of
$D^+_s \to \eta e^+\nu_e$,
$D^+_s \to \eta^\prime e^+\nu_e$,
$D^+_s \to \phi e^+\nu_e$,
$D^+_s \to K^*(892)^0 e^+\nu_e$,
$D^+_s \to K^0 e^+\nu_e$, and
$D^+_s \to f_0(980) e^+\nu_e$
were performed based on CLEO-c data at 4.170 GeV~\cite{CLEO:2009dyb, CLEO:2009ugx, Hietala:2015jqa},
and CLEO-c also reported a search for $D^+_s \to \omega e^+\nu_e$~\cite{CLEO:2011gyw}.
In addition, BaBar~\cite{BaBar:2008gpr} measured the branching fraction of $D^+_s \to \phi e^+\nu_e$ and
hadronic transition form factors of $D^+_s\to \phi$.

As for inclusive measurements of semileptonic decays of $D^0$, $D^+$, and $D^+_s$ mesons,
MARKIII~\cite{MARK-III:1984mab}, BESII~\cite{BES:2007dfh}, and CLEO-c~\cite{CLEO:2006ivk} reported measurements of the inclusive branching fractions for $D^0 \to X e^+ \nu_e$ and $D^+ \to X e^+ \nu_e$. Subsequently, CLEO-c~\cite{CLEO:2009uah} reported updated measurements of the inclusive decays $D^0 \to X e^+ \nu_e$, $D^+ \to X e^+ \nu_e$, and $D_s^+ \to X e^+ \nu_e$, where the first two superseded their previous results~\cite{CLEO:2006ivk} based on a larger data sample.
Measurements of the inclusive branching fractions for $D^0 \to X \mu^+ \nu_\mu$ were reported by ARGUS~\cite{ARGUS:1995rxw} and CHORUS~\cite{CHORUS:2005nog}; BESII~\cite{BES:2008pve} reported measurements for both $D^0 \to X \mu^+ \nu_\mu$ and $D^+ \to X \mu^+ \nu_\mu$.

\subsection{Results of $D\to P\ell^+\nu_\ell$ at BESIII}

\subsubsection{Cabibbo-favored decays}

The earlier studies of $D\to \bar K\ell^+\nu_\ell$ at BESIII were based
on 2.93~fb$^{-1}$ of data. See details of individual studies of $D^0\to K^-e^+\nu_e$~\cite{Ablikim:2015ixa},
$D^0\to K^-\mu^+\nu_\mu$~\cite{Ablikim:2018evp},
$D^+\to \bar K^0(\to K_S^0\to\pi^+\pi^-)e^+\nu_e$~\cite{BESIII:2017ylw},
$D^+\to \bar K^0(\to K^0_L)e^+\nu_e$~\cite{BESIII:2015jmz},
$D^+\to \bar K^0(\to K_S^0\to\pi\pi)\mu^+\nu_\mu$~\cite{BESIII:2016gbw},
$D^+\to \bar K^0(\to K^0_S\to 2\pi^0)e^+\nu_e$~\cite{BESIII:2016hko},
$D^{0(+)}\to \bar K e^+\nu_e$~\cite{BESIII:2021mfl}.
Among them, Refs.~\cite{BESIII:2016gbw,BESIII:2016hko,BESIII:2021mfl} only reported branching fraction measurements due to smaller samples or unique techniques,
while other works report both branching fractions and hadronic transition form factors.
In 2024, BESIII reported an updated analysis of $D^0\to K^-e^+\nu_e$, $D^0\to K^-\mu^+\nu_\mu$,  $D^+\to \bar K^0e^+\nu_e$, and $D^+\to \bar K^0\mu^+\nu_\mu$ based on 7.9~fb$^{-1}$ of
data at 3.773 GeV~\cite{BESIII:2024slx}, in which the branching fractions for each signal decay were measured separately and
the product of $f_{+}^{D\to K}(0)|V_{cs}|$ was extracted from a simultaneous fit to the partial decay rates of these four decays.
With the same analysis strategy, the most precise measurements of these four decays were reported
with 20.3~fb$^{-1}$ of data in early 2026~\cite{BESIII:2026uin,BESIII:2026ydr}.
From a sample of 20.2 million tagged $\bar D^0$ mesons and 10.6 million tagged $D^-$ mesons,
488k $D^0\to K^-e^+\nu_e$, 402k $D^0\to K^-\mu^+\nu_\mu$, 149k $D^+\to \bar K^0e^+\nu_e$, and 125k $D^+\to \bar K^0\mu^+\nu_\mu$ signal
events are observed.
The obtained branching fractions are
${\cal B}(D^0 \to K^-e^+\nu_e)=(3.548\pm 0.006\pm0.017)\%$,
${\cal B}(D^0 \to K^-\mu^+\nu_\mu)=(3.445\pm 0.007\pm0.017)\%$,
${\cal B}(D^+ \to\bar K^0e^+\nu_e)=(8.928\pm 0.025\pm0.050)\%$, and
${\cal B}(D^+ \to\bar K^0\mu^+\nu_\mu)=(8.770\pm 0.029\pm0.053)\%$.

The partial decay rates
$\frac{d\Gamma}{dq^2}$ and the forward-backward asymmetries $A_{\rm
FB}(q^2)$ of the semileptonic decays $D^{0(+)}\to \bar K\ell^+\nu_{\ell}$ are sensitive to the
scalar current contributions in the $c\to s\ell^+\nu_\ell$
transition. Therefore, experimental studies of these decays are
important to explore the parameter space of the scalar current, which
may reveal unknown scalar boson contributions beyond the
SM~\cite{Iguro:2022uzz,Blanke:2022pjy,Sakaki:2013bfa,Becirevic:2018afm}. The $q^2$-binned forward-backward asymmetry in the $i$-th $q^2$
interval
\begin{equation} A_{{\rm FB},i}=\int^{q^2_{{\rm
max}(i)}}_{q^2_{{\rm min}(i)}}{A_{{\rm FB}}(q^2)\frac{d\Gamma}{dq^2}
dq^2}\bigg{/}\int^{q^2_{{\rm max}(i)}}_{q^2_{{\rm
min}(i)}}{\frac{d\Gamma}{dq^2} dq^2} \end{equation}
is measured by 
$A_{{\rm FB},i} = \frac{N^{i}_{\rm prd,f}-N^{i}_{\rm prd,b}}{N^{i}_{\rm
prd,f}+N^{i}_{\rm prd,b}}$, where $N^{i}_{\rm prd,(f,b)}$ is the
number of produced forward~($\cos\theta_W>0$) and
backward~($\cos\theta_W<0$) events in the $i$-th $q^2$ interval
obtained using the similar approach to that of partial decay rates, where
$\theta_W$ is the angle between the flight direction of the lepton and the
opposite flight direction of $D$ in the $\ell^+\nu_\ell$ rest
frame.

In addition to the $(V-A)$ current $\mathcal{O}^\ell_{\rm SM}=(\bar{s}\gamma_{\mu}P_{L}c)(\bar{\nu}_\ell \gamma^\mu P_L\ell)$ allowed in the SM, further search for beyond
the SM scalar current $\mathcal{O}^\ell_{R(L)}=(\bar{s}P_{R(L)}c)(\bar{\nu}_\ell P_R\ell)$ is performed, with the corresponding contribution quantified by the complex Wilson coefficient $c^\ell_{R(L)}$.

The scalar current contribution is obtained via a simultaneous fit to
the partial decay rates $\frac{d\Gamma}{dq^2}$ and forward-backward
asymmetries $A_{\rm FB}(q^2)$, the explicit expressions of which are
written as~\cite{Fajfer:2015ixa,Faustov:2019mqr}
\begin{equation}
\begin{split}
\frac{d\Gamma}{dq^2} &= \mathcal{N}(q^2)\left(1-\frac{m_\ell^2}{q^2}\right)^2\\
&\times\left[\left(1+\frac{m_\ell^2}{2q^2}\right)|h_0(q^2)|^2 + \frac{3m_\ell^2}{2q^2}|h_t(q^2)|^2\right]
\end{split}
\end{equation}
and
\begin{equation}
\begin{split}
A_{\rm FB}(q^2)&=\frac{3\mathcal{N}(q^2)}{2}\frac{1}{d\Gamma/dq^2}\\
&\times\left(1-\frac{m_\ell^2}{q^2}\right)^2\frac{m_\ell^2}{q^2} {\rm Re}\left(h_0(q^2)h^*_t(q^2)\right),
\end{split}
\end{equation}
respectively. Here,
$\mathcal{N}(q^2)=\frac{G_F^2|V_{cs}|^2|\textbf{q}|q^2}{96\pi^3
m^2_D}$ is the factor including the Fermi coupling constant $G_F$, the
CKM matrix element $|V_{cs}|$, and the mass of the $D$ meson $m_{D}$; $q$ is
the four momentum of the $\ell^+\nu_\ell$ system in the $D$ rest frame
and $|\textbf{q}|$ is the magnitude of the three momentum of $q$;
$m_\ell$ is the mass of the lepton. With scalar currents included, the
hadronic helicity amplitudes $h_{0(t)}(q^2)$ are modified as
\begin{equation}
\begin{split}
\label{eq:helicity_amp}
h_0(q^2) &= \frac{2 m_D |\textbf{q}|}{\sqrt{q^2}}f_{+}(q^2),\\
h_t(q^2) &= \left(1+c_S^\ell\frac{q^2}{m_\ell(m_s-m_c)}\right)\frac{m_D^2-m_K^2}{\sqrt{q^2}}f_0(q^2),
\end{split}
\end{equation}
where $f_{+,0}(q^2)$ are the hadronic transition form factors; $c_S^\ell=c^{\ell}_R+c^{\ell}_L$ is the scalar combination of Wilson coefficients for the right and left-handed scalar currents~\cite{Fajfer:2015ixa}; $m_c=1.273$ GeV/$c^2$ and $m_s=93.5$ MeV/$c^2$~\cite{ParticleDataGroup:2024cfk} are the $\overline{\rm MS}$ masses at the energy scale $\mu=m_c$ and 2 GeV/$c^2$, respectively; $m_K$ is the mass of the kaon.

\begin{figure*}[htbp]
	\centering
	\includegraphics[width=0.8\linewidth]{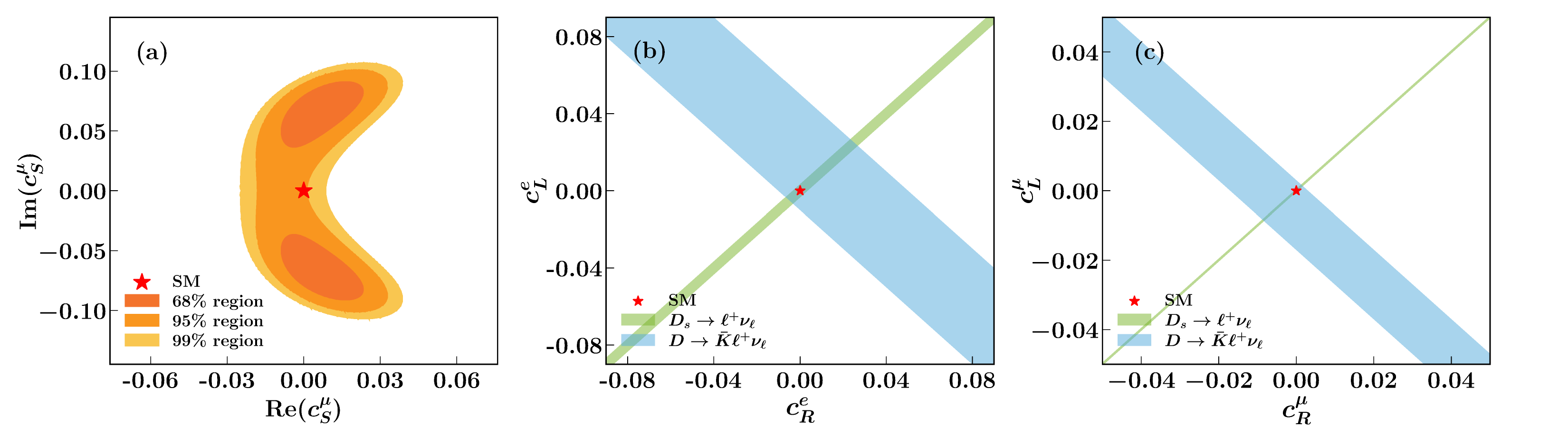}
	\caption{The confidence regions of (a) the scalar combination
	of complex Wilson coefficients $c_S^\mu$ with probabilities of
	68\%, 95\%, and 99\% from $D\to \bar K\ell^+\nu_\ell$, as well as (b)(c) the
	right- and left-handed real Wilson coefficients $c^\ell_{R}$
	and $c^\ell_{L}$ with probabilities of 68\% by combining with the
	$D_s^+\to\ell^+\nu_\ell$ decay rate~\cite{BESIII:2026uin,BESIII:2026ydr}. The red dot indicates the
	SM value~\cite{Fajfer:2015ixa}.}
	\label{fig:D_Klnu_constraint}
\end{figure*}

Based on the fit with a scalar current
contribution included, no significant deviation from the SM is found
in the positron channels. For the muon channels, $c^\mu_S$ is
found to deviate from zero as shown in Fig.~\ref{fig:D_Klnu_constraint}(a),
especially in its imaginary part. 

Further constraints on the right- and left-handed components of the scalar
current are obtained by combining with the decay rates of
$D_s^+\to\ell^+\nu_\ell$, which are sensitive to the pseudoscalar
combination of Wilson coefficients $c^\ell_P=c^{\ell}_R-c^{\ell}_L$
as~\cite{Fajfer:2015ixa}
\begin{equation}
\begin{split}
\mathcal{B}(D_s^+\to\ell^+\nu_\ell)&=\tau_{D_s}\frac{m_{D_s}}{8\pi}f^2_{D_s}\left(1-\frac{m_\ell^2}{m^2_{D_s}}\right)^2 G^2_F \\
&\times|V_{cs}|^2 m^2_{\ell} |1-c^\ell_P\frac{m^2_{D_s}}{(m_c+m_s)m_\ell}|^2.
\end{split}	
\end{equation}
Here, the measured branching fractions $\mathcal{B}(D_s^+\to \ell^+\nu_\ell)$ are
from the Particle Data Group (PDG)~\cite{ParticleDataGroup:2024cfk}, and
the inputs of $f_{D_s}=(249.9\pm0.5)$ MeV from an LQCD calculation~\cite{Bazavov:2017lyh} and
$|V_{cs}|=0.97349\pm0.00016$ from the global SM
fit~\cite{ParticleDataGroup:2024cfk} are used. Since the
$D_s^+\to\ell^+\nu_\ell$ decay cannot constrain the real and imaginary parts
of $c^\ell_P$ simultaneously, the Wilson coefficient $c^\ell_{R(L)}$
is assumed to be real with T-parity conserved here. As shown in
Figs.~\ref{fig:D_Klnu_constraint}(b) and~\ref{fig:D_Klnu_constraint}(c), both
the real parts of the right- and left-handed Wilson coefficients
$c^\ell_{R}$ and $c^\ell_{L}$ are consistent with zero.

A simultaneous fit to the differential decay rates, $d\Gamma/dq^2$, where $q^2=M^2(\ell^+\nu)$ is the
square of the four-momentum transfer between initial state $D$
and final state $\bar K$, leads to the product of the hadronic transition form factor at
$q^2=0$ and the $c \to s$ CKM matrix element $f_{+}^{D\to K}(0)|V_{cs}|=0.7183\pm 0.0007\pm0.0014$.
Using the value of $|V_{cs}|=0.97349\pm0.0016$ given by the SM-constrained fit~\cite{ParticleDataGroup:2024cfk}, the $f_{+}^{D\to K}(0)$ is measured to be $0.7378\pm0.0007\pm0.0014$.
Alternatively, using the value of $f_{+}^{D\to K}(0)=0.7452\pm0.0031$ from a recent LQCD calculation~\cite{FermilabLattice:2022gku} gives $|V_{cs}|=0.9639\pm0.0009\pm0.0018\pm0.0040$.
Based on a simultaneous fit to the first measured
forward-backward asymmetries and precisely determined partial decay
rates, the parameters of the scalar current are determined to be ${\rm
	Re}(c_S^{\mu})=0.017\pm0.008\pm0.006$ and
${\rm Im}(c_S^{\mu})=\pm(0.077\pm0.011\pm0.009)$, which deviates from the SM prediction by a significance of
$2.3\sigma$. The significance is estimated by comparing the fit quality with and without the scalar current in the muon channel.
The branching fractions of $D^{0(+)}\to \bar
K\ell^+\nu_\ell$, the hadronic transition form factor $f_{+}^{D\to K}(0)$, and the modulus
of the $c\to s$ CKM matrix element $|V_{cs}|$ are also determined with
improved precision. In addition, no evidence of lepton flavor universality violation is found
via the ratios of the decay rates between semimuonic and
semielectronic decays in the full momentum
transfer range and in subranges.

\begin{figure*}[htbp]
	\centering
	\includegraphics[width=0.8\linewidth]{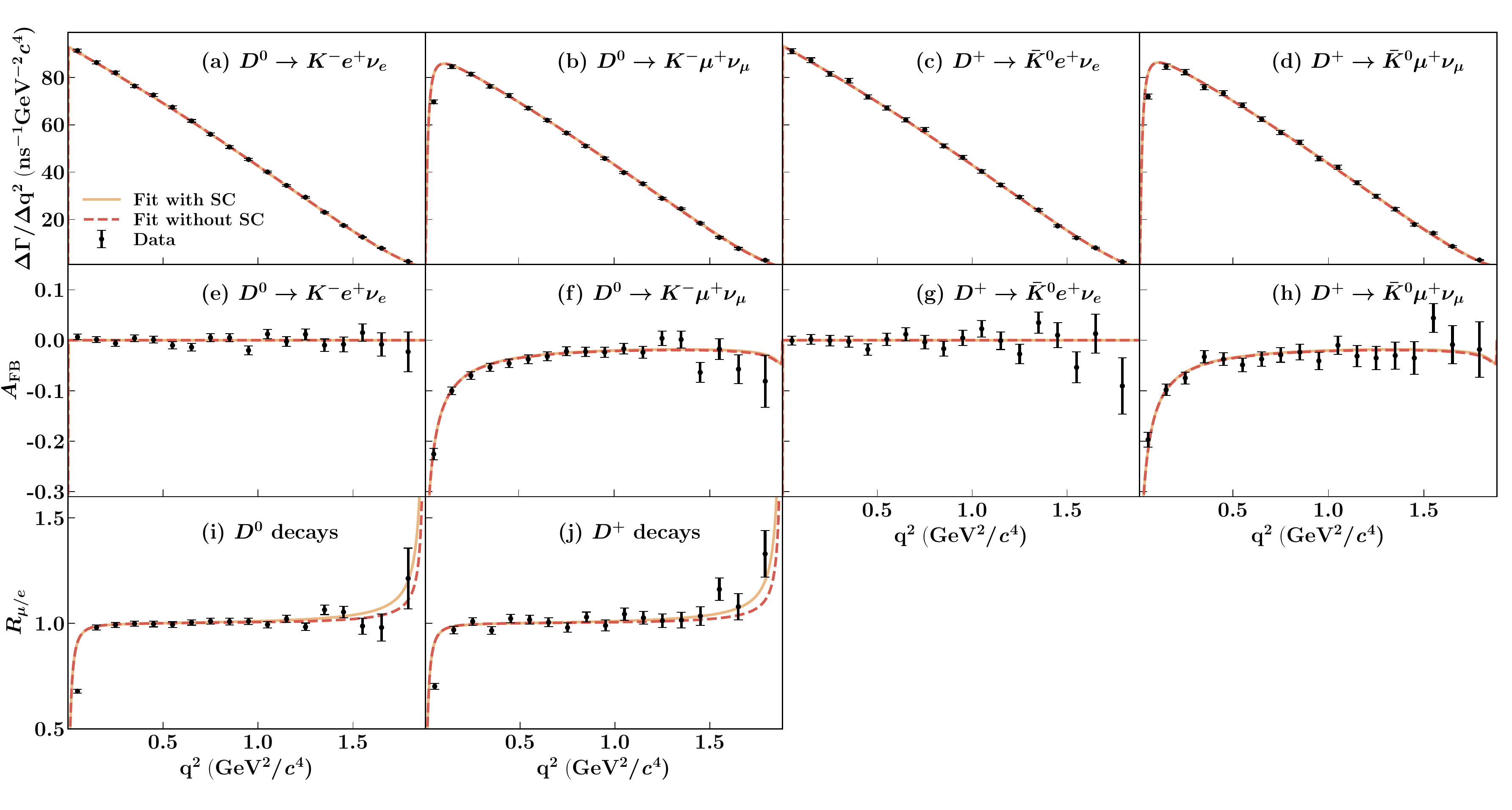}
	\caption{Simultaneous fit to the measured partial decay
	rates~(top) and forward-backward asymmetries~(middle) of
	$D\to \bar K\ell^+\nu_\ell$, and the fit projections on the ratios of differential
	decay rates $\mathcal{R}_{\mu/e}$~(bottom)~\cite{BESIII:2026uin,BESIII:2026ydr}. The black points with red error bars are data, the solid lines are the fit projections with scalar current (SC) contribution included, and the dashed lines are the fit projections without scalar current contribution.}
	\label{fig:D_Klnu_fit}
\end{figure*}

The measured partial decay rates of $D\to \bar K \ell^+\nu_\ell$, $\frac{\Delta\Gamma}{\Delta q^2}$,
the measured forward-backward asymmetries, $A_{\rm FB}$, and the calculated ratios $\mathcal{R}_{\mu/e}\equiv\frac{\Delta\Gamma_\mu}{\Delta\Gamma_e}$ in various $q^2$ intervals with and without scalar current contribution included are
shown in the top, middle, and bottom plots of Fig.~\ref{fig:D_Klnu_fit},
respectively.

\begin{figure}[htbp]
  \centering
  \includegraphics[width=0.4\textwidth]{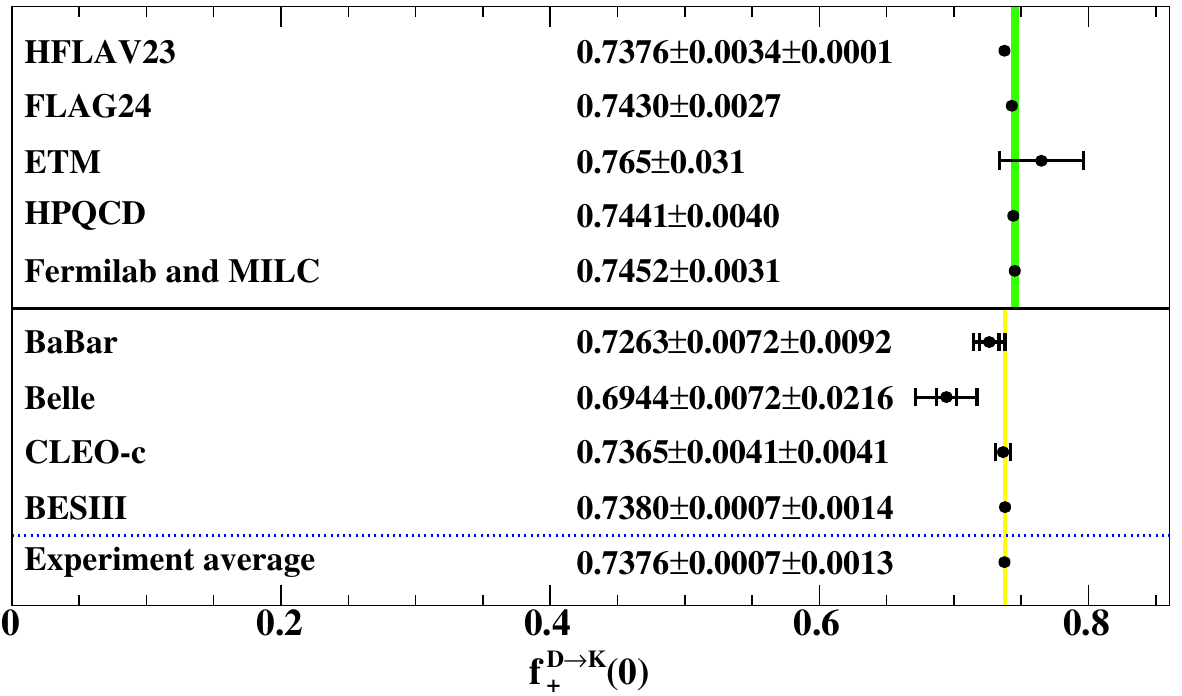}
  \put(-146,110){\tiny~\cite{HeavyFlavorAveragingGroupHFLAV:2024ctg}}
  \put(-146,100.5){\tiny~\cite{FlavourLatticeAveragingGroupFLAG:2024oxs}}
  \put(-146,91){\tiny~\cite{Lubicz:2017syv}}
  \put(-146,81.5){\tiny~\cite{Parrott:2022rgu}}
  \put(-146,72){\tiny~\cite{FermilabLattice:2022gku}}
  \put(-146,59.5){\tiny~\cite{BaBar:2007zgf}}
  \put(-146,50){\tiny~\cite{Belle:2006idb}}
  \put(-146,40.5){\tiny~\cite{CLEO:2009svp}}
  \put(-146,31){\tiny~\cite{BESIII:2026uin,BESIII:2026ydr,BESIII:2015jmz}}
   \caption{Comparison of $f^{D\to K}_+(0)$ from experimental measurements of
   BaBar~\cite{BaBar:2007zgf}, Belle~\cite{Belle:2006idb}, CLEO-c~\cite{CLEO:2009svp}, and BESIII~\cite{BESIII:2026uin,BESIII:2026ydr,BESIII:2015jmz}
   as well as recent LQCD calculations
   of ETM~\cite{Lubicz:2017syv}, HPQCD~\cite{Parrott:2022rgu}, and Fermilab and MILC~\cite{FermilabLattice:2022gku}  as well as HFLAV23~\cite{HeavyFlavorAveragingGroupHFLAV:2024ctg} and FLAG24~\cite{FlavourLatticeAveragingGroupFLAG:2024oxs}. The green band is the $\pm 1\sigma$ region of the result of Fermilab and MILC~\cite{FermilabLattice:2022gku} and the yellow band denotes the $\pm 1\sigma$ region of the result averaged over all measurements of $D\to \bar K\ell^+\nu_\ell$.
}
  \label{fig:f0_DK}
\end{figure}

The comparison of the $f^{D\to K}_+(0)$ measured by different experiments and those predicted by recent LQCD calculations~\cite{FermilabLattice:2022gku} is presented in Fig.~\ref{fig:f0_DK}. 
The measured  $f_{+}^{D\to K}(0)$ is consistent with the LQCD calculation~\cite{FermilabLattice:2022gku} within 3$\sigma$.
Since LQCD calculations of the $D \to K$ hadronic transition form factor have reached sub-percent precision, we do not include results from other theoretical approaches, such as the quark model or QCD sum rule, when comparing the experimental measurements with the theoretical predictions.

\begin{figure}[htbp]\centering
\includegraphics[width=0.475\textwidth]{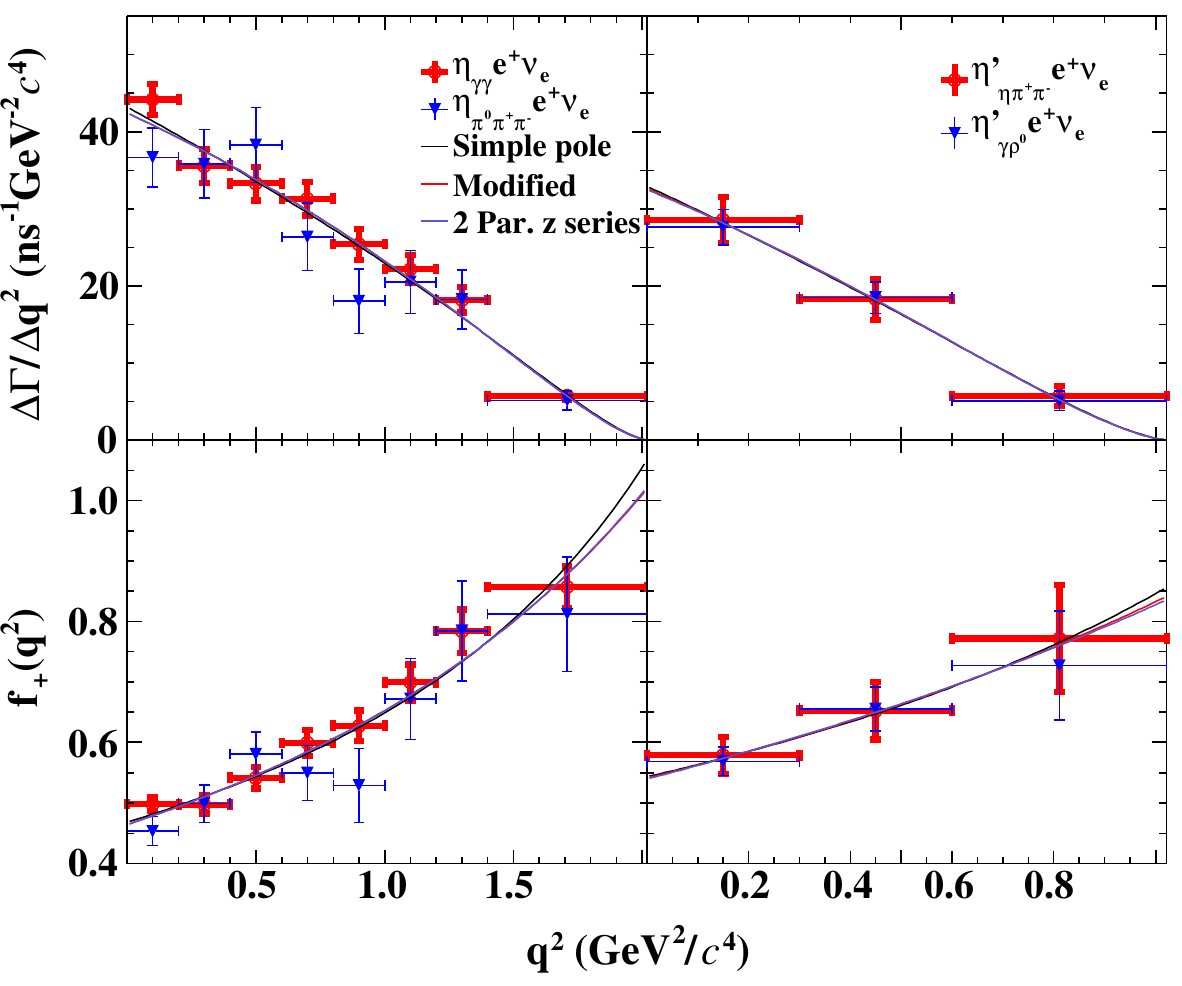}
\caption{(Top row) Simultaneous fits to the differential decay rates of (left) $D^+_s\to \eta_{\gamma\gamma} e^+\nu_e$ and  $D^+_s\to \eta_{\pi^0\pi^+\pi^-} e^+\nu_e$ and (right) $D^+_s\to \eta^\prime_{\eta\pi^+\pi^-} e^+\nu_e$ and  $D^+_s\to \eta^\prime_{\rho(770)^0\gamma} e^+\nu_e$,
and (bottom row)  projection on the hadronic transition form factors as a function of $q^2$~\cite{BESIII:2023ajr}. The red circles and blue triangles with error bars are (top row) the measured differential decay rates for two $\eta^{(\prime)}$ channels and (bottom row) the hadronic transition form factors. 
\label{fig:Ds_etaetapenu_FFs}}
\end{figure}

\begin{figure}
\centering
  \includegraphics[width=0.475\textwidth]{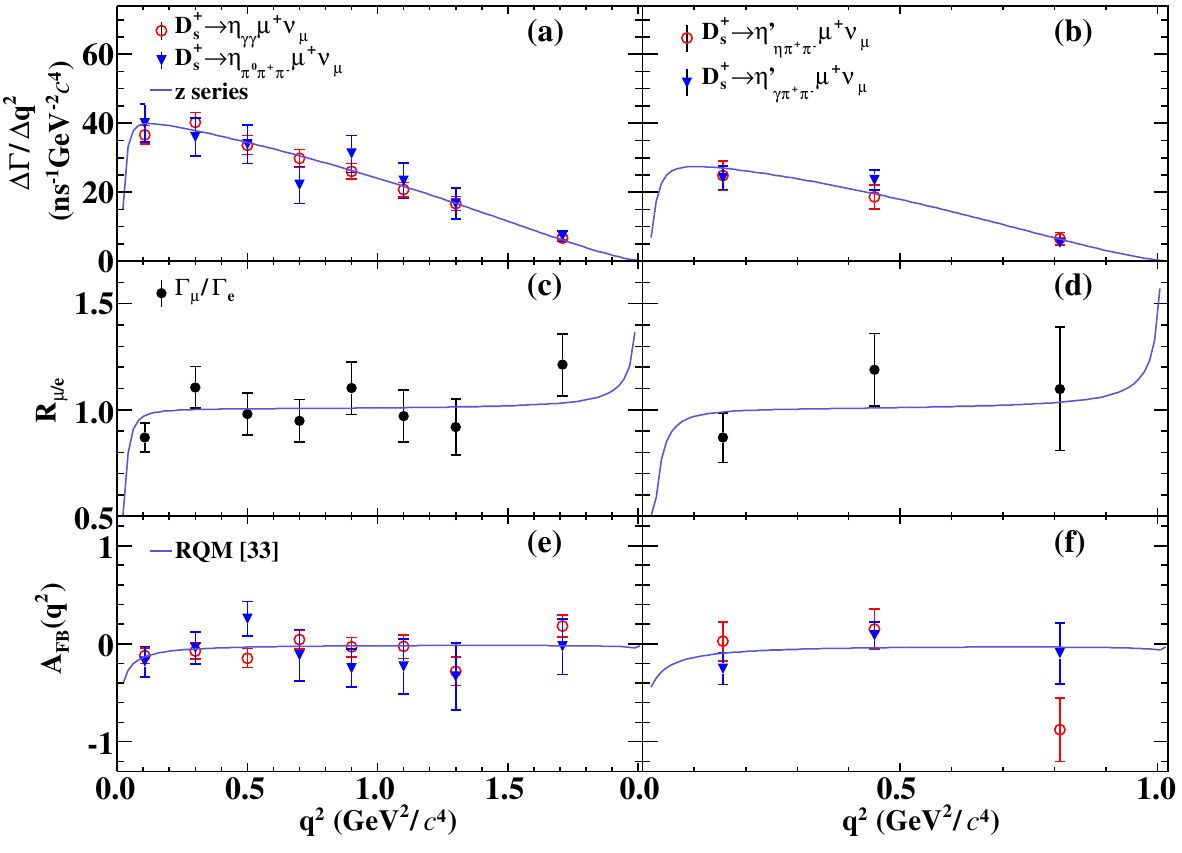}
\caption{\small
(a,b) Fits to measured partial decay rates.
 (c,d) The measured $\mathcal R_{\mu/e}$ combining the two signal channels in each $q^2$ interval~\cite{BESIII:2023gbn}.
  (e,f) Comparisons of the measured $A_{\rm FB}$ and theoretical predictions~\cite{Faustov:2019mqr}.
}
  \label{fig:Ds_etaetapmunu_FFs}
\end{figure}

The earliest measurements of the semileptonic decays $D^+_s\to \eta \ell^+\nu_\ell$, $D^+_s\to \eta^\prime \ell^+\nu_\ell$,
and $D^+_s\to \phi \ell^+\nu_\ell$~($\ell =e$ or $\mu$) were performed by using 13k of tagged $D_s^-$ mesons
with $e^+e^-\to D^+_sD^-_s$ from 482 pb$^{-1}$ of data at 4.009~GeV~\cite{BESIII:2016ult,BESIII:2017ikf}.
In these measurements, up to dozens of signal events for each decay were observed,
and thus only branching fractions were reported.
Using 3.19~fb$^{-1}$ of data at 4.178 GeV,
BESIII reported the studies of $D^+_s\to \eta^{(\prime)} e^+\nu_{e}$~\cite{BESIII:2019qci},
in which the branching fractions with improved precision and the first measurements of the hadronic transition form factors, $f^{D_s\to \eta^{(\prime)}}(0)$,
are reported.
In 2023, BESIII reported the improved measurements of ${\cal B}(D^+_s\to \eta^{(\prime)} e^+\nu_e)$ and $f^{D_s\to \eta^{(\prime)}}(0)$
based on 7.3~fb$^{-1}$ of data at 4.128-4.226 GeV~\cite{BESIII:2023ajr}. Also in 2023,
with the same data sample and analysis strategy, Ref.~\cite{BESIII:2023gbn} reported the first observation of $D^+_s\to \eta^\prime\mu^+\nu_\mu$
and the first study of the decay dynamics of $D^+_s\to \eta^{(\prime)} \mu^+\nu_{\mu}$.
In these studies, the $\eta$ meson is reconstructed using the decay channels $\eta\to\gamma\gamma$ and $\eta\to \pi^0\pi^+\pi^-$, while the $\eta^\prime$ is reconstructed via $\eta^\prime\to\eta_{\gamma\gamma}\pi^+\pi^-$ and  $\eta^\prime\to\rho(770)^0\gamma$. The branching fractions and hadronic transition form factors are determined by performing a simultaneous fit to the two reconstructed decay modes, with the relevant parameters constrained to common values.
From a sample of 0.82 million tagged $D^-_s$ mesons,
4.0k $D^+_s\to \eta e^+\nu_e$,
675 $D^+_s\to \eta^\prime e^+\nu_e$,
3.1k $D^+_s\to \eta\mu^+\nu_\mu$,
and 387 $D^+_s\to \eta^\prime\mu^+\nu_\mu$ signal events are observed. 
The obtained branching fractions are
${\cal B}(D^+_s\to \eta e^+\nu_e)=(2.255\pm 0.039\pm0.051)\%$,
${\cal B}(D^+_s\to \eta^\prime e^+\nu_e)=(0.810\pm 0.038\pm0.024)\%$,
${\cal B}(D^+_s\to \eta \mu^+\nu_\mu)=(2.235\pm 0.051\pm0.052)\%$, and
${\cal B}(D^+_s\to \eta^\prime \mu^+\nu_\mu)=(0.801\pm 0.055\pm0.028)\%$.
Separate fits to differential decay rates of these four decays give
$f_{+}^{D_s\to \eta}(0)|V_{cs}|_e=0.452\pm 0.007\pm0.007$,
$f_{+}^{D_s\to \eta^\prime}(0)|V_{cs}|_e=0.525\pm 0.024\pm0.009$,
$f_{+}^{D_s\to \eta}(0)|V_{cs}|_\mu=0.452\pm 0.010\pm0.007$,
$f_{+}^{D_s\to \eta^\prime}(0)|V_{cs}|_\mu=0.504\pm 0.037\pm0.012$. 
Using the value of $|V_{cs}|$ given by the SM-constrained fit~\cite{ParticleDataGroup:2024cfk},
one obtains
$f_{+}^{D_s\to \eta}(0)_e=0.464\pm0.007\pm0.007$,
$f_{+}^{D_s\to \eta^\prime}(0)_e=0.540\pm0.025\pm0.009$,
$f_{+}^{D_s\to \eta}(0)_\mu=0.465\pm0.010\pm0.007$, and
$f_{+}^{D_s\to \eta^\prime}(0)_\mu=0.518\pm0.038\pm0.012$. 
The forward-backward asymmetries are determined to be $\langle A_{\rm FB}^\eta\rangle=-0.059\pm0.031\pm0.005$ and $\langle A_{\rm FB}^{\eta^\prime}\rangle=-0.064\pm0.079\pm0.006$ for the first time, which are consistent with the theoretical calculation~\cite{Faustov:2019mqr}.
Figure~\ref{fig:Ds_etaetapenu_FFs} shows
the simultaneous fits to the partial decay rates of $D_s^+\to\eta e^+\nu_e$ or $D_s^+\to\eta^\prime e^+\nu_e$ reconstructed with two different decay modes and the hadronic transition form factors as a function of $q^2$.
Figures~\ref{fig:Ds_etaetapmunu_FFs}(a,b) show the fits to the differential decay rates; the ${\mathcal R}^{\eta}_{\mu/e}$ and ${\mathcal R}^{\eta^\prime}_{\mu/e}$
in different $q^2$ intervals after considering the correlated uncertainties, are shown in Figs.~\ref{fig:Ds_etaetapmunu_FFs}(c,d); these are also consistent with the SM predictions.
The forward-backward asymmetry parameter $A_{\rm FB}$ is defined as $A_{\rm FB}(q^2)=\frac{\int_0^1d\cos\theta_\ell d\Gamma/d\cos\theta_\ell-\int_{-1}^0d\cos\theta_\ell d\Gamma/d\cos\theta_\ell}{\int_0^1d\cos\theta_\ell d\Gamma/d\cos\theta_\ell+\int_{-1}^0d\cos\theta_\ell d\Gamma/d\cos\theta_\ell}$, where $\theta_\ell$ is the angle between the momentum of the $\mu^+$ in the rest frame of the  $W$-boson and the
direction to the $W$-boson momentum in the rest frame of $D_s^+$.
The measured $A_{\rm FB}$ in various $q^2$  intervals are shown in Figs.~\ref{fig:Ds_etaetapmunu_FFs}(e,f).

In 2024, Ref.~\cite{BESIII:2024mot} reported other independent measurements of branching fractions of semielectronic $D^+_s$
decays by using 124k of tagged $D^{*-}_s$ mesons with $e^+e^-\to D^{*+}_sD^{*-}_s$ from 10.64 fb$^{-1}$ of data at 4.237-4.699~GeV.
With up to hundreds of signal events,
the branching fractions of each decay are determined,
and the hadronic transition form factors of $D^+_s\to \eta$ and $D^+_s\to \eta^\prime$
are determined to be
$f^{D_s\to\eta}_+(0) = 0.442\pm 0.022\pm 0.017$ and
$f^{D_s\to\eta^{\prime}}_+(0) = 0.557\pm 0.062\pm0.024$.
The precisions of the obtained branching fractions  are not better than those measured via $e^+e^-\to
D_s^{*\pm}D_s^{\mp}$ with 7.33 fb$^{-1}$ of $e^+e^-$ data
taken between 4.128 and 4.226 GeV at BESIII, but better than those measured by CLEO-c via $e^+e^-\to
D_s^{*\pm}D_s^{\mp}$ with 0.6 fb$^{-1}$ of $e^+e^-$ data
taken at 4.17~GeV.

\begin{figure}[htbp]
  \centering
  \includegraphics[width=0.4\textwidth]{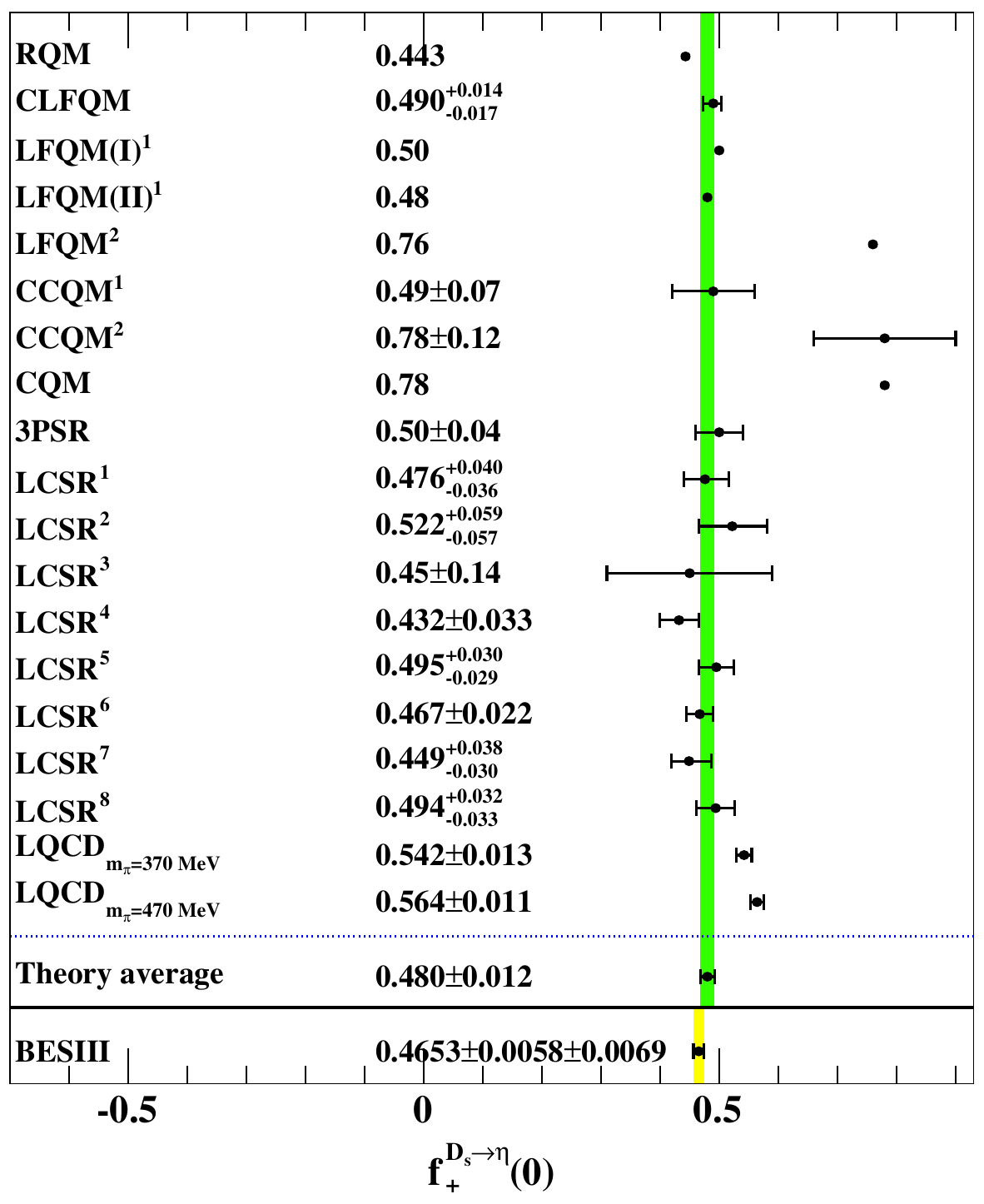}
 \put(-158,236)  {\tiny~\cite{Faustov:2019mqr}}
 \put(-158,226.278)  {\tiny~\cite{Yang:2025gfz}}
 \put(-158,216.556){\tiny~\cite{Wei:2009nc}}
 \put(-158,206.833)  {\tiny~\cite{Wei:2009nc}} 
 \put(-158,197.111){\tiny~\cite{Verma:2011yw}}
 \put(-158,187.389)  {\tiny~\cite{Ivanov:2019nqd}}       
 \put(-158,177.667){\tiny~\cite{Soni:2018adu}}         
 \put(-158,167.944){\tiny~\cite{Melikhov:2000yu}}      
 \put(-158,158.222){\tiny~\cite{Colangelo:2001cv}}     
 \put(-158,148.5){\tiny~\cite{Hu:2021zmy}}           
 \put(-158,138.777){\tiny~\cite{Zhang:2025yeu}}        
 \put(-158,129.056){\tiny~\cite{Azizi:2010zj}}         
 \put(-158,119.333){\tiny~\cite{Offen:2013nma}}        
 \put(-158,109.611){\tiny~\cite{Duplancic:2015zna}}    
 \put(-158,99.889){\tiny~\cite{Melic:2025uha}}        
 \put(-158,90.167) {\tiny~\cite{Huang:2026hvs}}        
 \put(-158,80.444) {\tiny~\cite{Zhou:2026llv}}         
 \put(-158,70.722) {\tiny~\cite{Bali:2014pva}}         
 \put(-158,61) {\tiny~\cite{Bali:2014pva}}         
 \put(-158,30.5){\tiny~\cite{BESIII:2023ajr,BESIII:2023gbn,BESIII:2024mot}}
    \caption{Comparison of $f^{D_s\to \eta}_+(0)$ from experimental measurements of
   BESIII~\cite{BESIII:2023ajr,BESIII:2023gbn,BESIII:2024mot} and those predicted by theoretical calculations
   of RQM~\cite{Faustov:2019mqr},
CLFQM~\cite{Yang:2025gfz},
LFQM$^1$~\cite{Wei:2009nc},
LFQM$^2$~\cite{Verma:2011yw},
CCQM$^1$~\cite{Ivanov:2019nqd},
CCQM$^2$~\cite{Soni:2018adu},
CQM~\cite{Melikhov:2000yu},
3PSR~\cite{Colangelo:2001cv},
LCSR$^1$~\cite{Hu:2021zmy},
LCSR$^2$~\cite{Zhang:2025yeu},
LCSR$^3$~\cite{Azizi:2010zj},
LCSR$^4$~\cite{Offen:2013nma},
LCSR$^5$~\cite{Duplancic:2015zna},
LCSR$^6$~\cite{Melic:2025uha},
LCSR$^7$~\cite{Huang:2026hvs},
LCSR$^8$~\cite{Zhou:2026llv}, and
LQCD~\cite{Bali:2014pva}. The green band is the $\pm 1\sigma$ region of averaged theoretical calculations~\cite{notef0Deta} and the yellow band denotes the $\pm 1\sigma$ region of the result averaged over all measurements of $D_s^+\to \eta\ell^+\nu_\ell$. 
}
  \label{fig:f0_Dseta}
\end{figure}

\begin{figure}[htbp]
  \centering
  \includegraphics[width=0.4\textwidth]{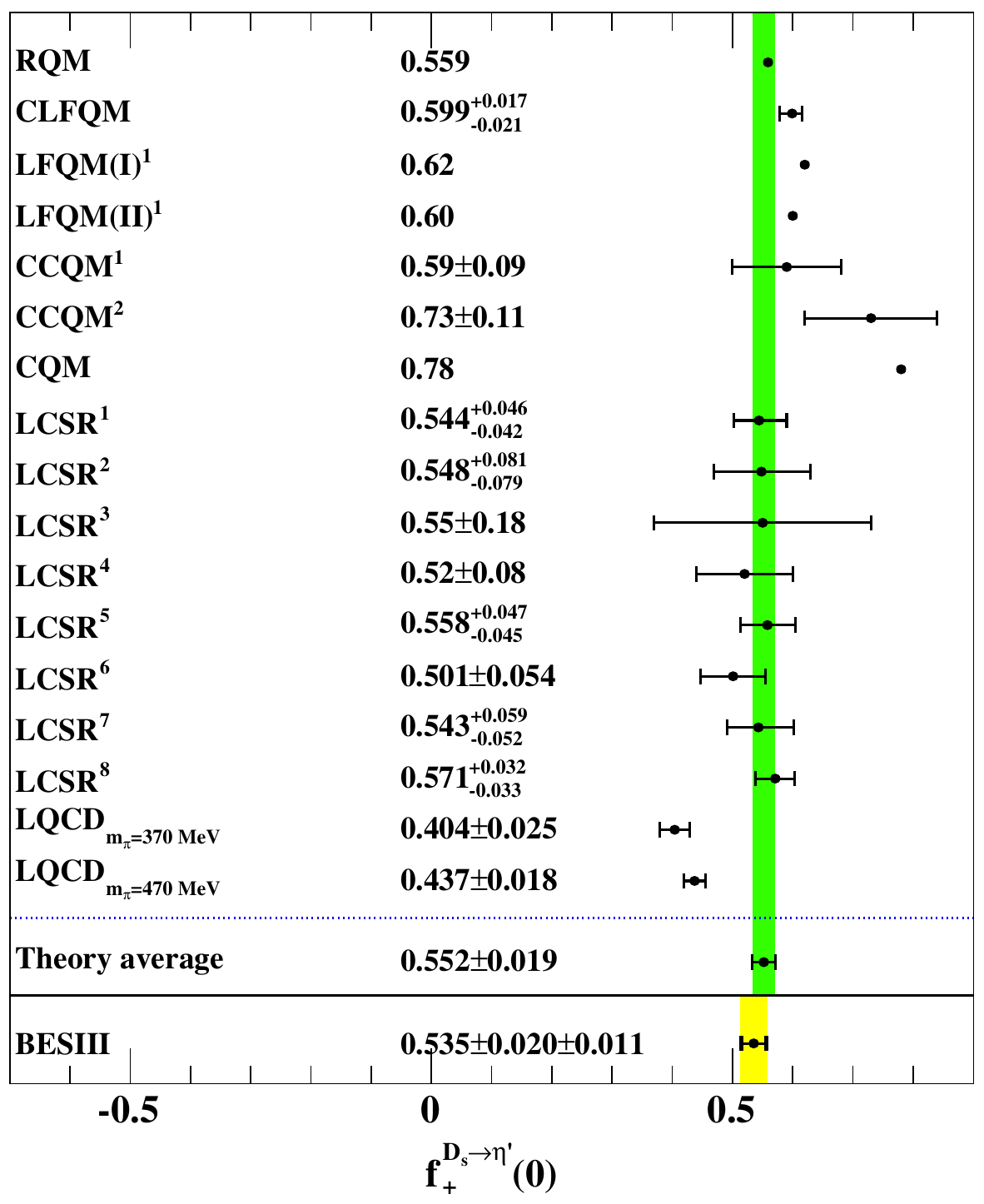}
 \put(-158,235.5)  {\tiny~\cite{Faustov:2019mqr}}
 \put(-158,224.875)  {\tiny~\cite{Yang:2025gfz}}
 \put(-158,214.25){\tiny~\cite{Wei:2009nc}}
 \put(-158,203.625)  {\tiny~\cite{Wei:2009nc}}
 \put(-158,193)  {\tiny~\cite{Ivanov:2019nqd}}
 \put(-158,182.375){\tiny~\cite{Soni:2018adu}}
 \put(-158,171.75){\tiny~\cite{Melikhov:2000yu}}
 \put(-158,161.125){\tiny~\cite{Hu:2021zmy}}
 \put(-158,150.5){\tiny~\cite{Zhang:2025yeu}}
 \put(-158,139.875){\tiny~\cite{Azizi:2010zj}}
 \put(-158,129.25){\tiny~\cite{Offen:2013nma}}
 \put(-158,118.625){\tiny~\cite{Duplancic:2015zna}}
 \put(-158,108){\tiny~\cite{Melic:2025uha}}
 \put(-158,97.375) {\tiny~\cite{Huang:2026hvs}}
 \put(-158,86.75) {\tiny~\cite{Zhou:2026llv}}
 \put(-158,76.125) {\tiny~\cite{Bali:2014pva}}
 \put(-158,65.5) {\tiny~\cite{Bali:2014pva}}
 \put(-158,32){\tiny~\cite{BESIII:2023ajr,BESIII:2023gbn,BESIII:2024mot}}
   \caption{
Comparison of $f^{D_s\to \eta^\prime}_+(0)$ from experimental measurements of BESIII~\cite{BESIII:2023ajr,BESIII:2023gbn,BESIII:2024mot} and those predicted by theoretical calculations
   of RQM~\cite{Faustov:2019mqr},
  CLFQM~\cite{Yang:2025gfz},
LFQM$^{1}$~\cite{Wei:2009nc},
CCQM$^1$~\cite{Ivanov:2019nqd},
CCQM$^2$~\cite{Soni:2018adu},
CQM~\cite{Melikhov:2000yu},
LCSR$^1$~\cite{Hu:2021zmy},
LCSR$^2$~\cite{Zhang:2025yeu},
LCSR$^3$~\cite{Azizi:2010zj},
LCSR$^4$~\cite{Offen:2013nma},
LCSR$^5$~\cite{Duplancic:2015zna},
LCSR$^6$~\cite{Melic:2025uha},
LCSR$^7$~\cite{Huang:2026hvs},
LCSR$^8$~\cite{Zhou:2026llv}, and
LQCD~\cite{Bali:2014pva}. The green band is the $\pm 1\sigma$ region of averaged theoretical calculations~\cite{notef0Deta} and the yellow band denotes the $\pm 1\sigma$ region of the result averaged over all measurements of $D_s^+\to \eta^\prime\ell^+\nu_\ell$.
}
  \label{fig:f0_Dsetap}
\end{figure}

Following the combination of their uncorrelated statistical uncertainties and fully correlated systematic uncertainties, the averaged hadronic transition form factors from Refs.~\cite{BESIII:2023ajr,BESIII:2023gbn,BESIII:2024mot} are determined as $f_+^{D_s\to\eta}(0)=0.4653\pm0.0058\pm0.0069$ and $f_+^{D_s\to\eta^\prime}(0)=0.535\pm0.020\pm0.011$, respectively.
The obtained $f_+^{D_s\to\eta^{(\prime)}}(0)$ are consistent with the theoretical calculations of 
RQM~\cite{Faustov:2019mqr},
CLFQM~\cite{Yang:2025gfz},
LFQM$^1$~\cite{Wei:2009nc},
CCQM$^1$~\cite{Ivanov:2019nqd},
3PSR~\cite{Colangelo:2001cv},
LCSR$^1$~\cite{Hu:2021zmy},
LCSR$^2$~\cite{Zhang:2025yeu},
LCSR$^3$~\cite{Azizi:2010zj},
LCSR$^4$~\cite{Offen:2013nma},
LCSR$^5$~\cite{Duplancic:2015zna},
LCSR$^6$~\cite{Melic:2025uha},
LCSR$^7$~\cite{Huang:2026hvs}, and
LCSR$^8$~\cite{Zhou:2026llv}.
They disfavor the 
LFQM$^2$~\cite{Verma:2011yw},
CCQM$^2$~\cite{Soni:2018adu},
CQM~\cite{Melikhov:2000yu}, and
LQCD~\cite{Bali:2014pva} by more than $4\sigma$.
Due to the lack of precise LQCD calculations for the hadronic transition form factors $f_{+}^{D_s\to \eta^{(\prime)}}(0)$, and because existing quark models cannot simultaneously reproduce the experimental measurements for both $f_{+}^{D_s\to \eta^{(\prime)}}(0)$ and $f_{+}^{D^+\to \eta^{(\prime)}}(0)$, we average the theoretical calculations by excluding those values that lack error estimates or exhibit excessively large deviations~\cite{notef0Deta}. 
Comparisons of the $f^{D_s\to \eta}_+(0)$ and $f^{D_s\to \eta^\prime}_+(0)$ measured by different experiments and those predicted by different theoretical calculations are  presented in Table~\ref{tab:f0:Deta}, Figs.~\ref{fig:f0_Dseta} and~\ref{fig:f0_Dsetap}. The measured $f_{+}^{D_s\to \eta}(0)$ and $f_{+}^{D^+\to \eta^{\prime}}(0)$ are consistent with the averaged theoretical calculations within 1$\sigma$.

\subsubsection{Cabibbo-suppressed decays}

To date, the measurements of $D\to \pi\ell^+\nu_\ell$ were only performed based on 2.93~fb$^{-1}$ of data.
From this sample, up to 2.5 million $\bar D^0$ and 1.7 million $D^-$ pairs are tagged,
about 6.3k $D^0\to \pi^- e^+\nu_e$~\cite{Ablikim:2015ixa},
 3.4k $D^+\to \pi^0 e^+ \nu_e$~\cite{BESIII:2017ylw},
 2.3k $D^0\to \pi^-\mu^+\nu_\mu$~\cite{Ablikim:2018frk}, and
 1.3k $D^+\to \pi^0\mu^+\nu_\mu$~\cite{Ablikim:2018frk} signal events are observed.
These correspond to absolute branching fractions of
${\cal B}(D^0 \to \pi^-e^+\nu_e)=(0.295\pm 0.004\pm0.003)\%$,
${\cal B}(D^+ \to \pi^0e^+\nu_e)=(0.363\pm 0.008\pm0.005)\%$,
${\cal B}(D^0 \to \pi^-\mu^+\nu_\mu)=(0.272 \pm 0.008\pm0.006)\%$, and
${\cal B}(D^+ \to \pi^0\mu^+\nu_\mu)=(0.350 \pm 0.011\pm0.010)\%$.
The former three decays are measured with significantly improved precision compared with previous measurements;
while the last one is the first measurement.
In Refs.~\cite{Ablikim:2015ixa,BESIII:2017ylw},
separate fits to the partial decay rates of $D^0\to \pi^- e^+\nu_e$ and $D^+\to \pi^0 e^+ \nu_e$
give the products of the hadronic transition form factor at
$q^2=0$ and the $c \to d$ CKM matrix element to be
$f_{+}^{D\to \pi}(0)|V_{cd}|_{D^0}=0.1435\pm 0.0018\pm0.0009$ and $f_{+}^{D\to \pi}(0)|V_{cd}|_{D^+}=0.1400\pm0.0026\pm0.0007$, respectively.
Using the value of $|V_{cd}|=0.22487\pm0.00068$ given by the SM-constrained fit~\cite{ParticleDataGroup:2024cfk}
leads to $f_{+}^{D\to \pi}(0)_{D^0}=0.623\pm0.012\pm0.003$ and $f_{+}^{D\to \pi}(0)_{D^+}=0.638\pm0.008\pm0.004$.
Alternatively, using the value of $f_{+}^{D\to \pi}(0)=0.6300\pm0.0051$ from a recent LQCD calculation~\cite{FermilabLattice:2022gku}
leads to $|V_{cd}|_{D^0}=0.2222\pm0.0041\pm0.0011\pm0.0018$ and $|V_{cd}|_{D^+}=0.2278\pm0.0029\pm0.0014\pm0.0018$.

\begin{figure*}[htbp]
  \centering
  \includegraphics[width=0.4\textwidth]{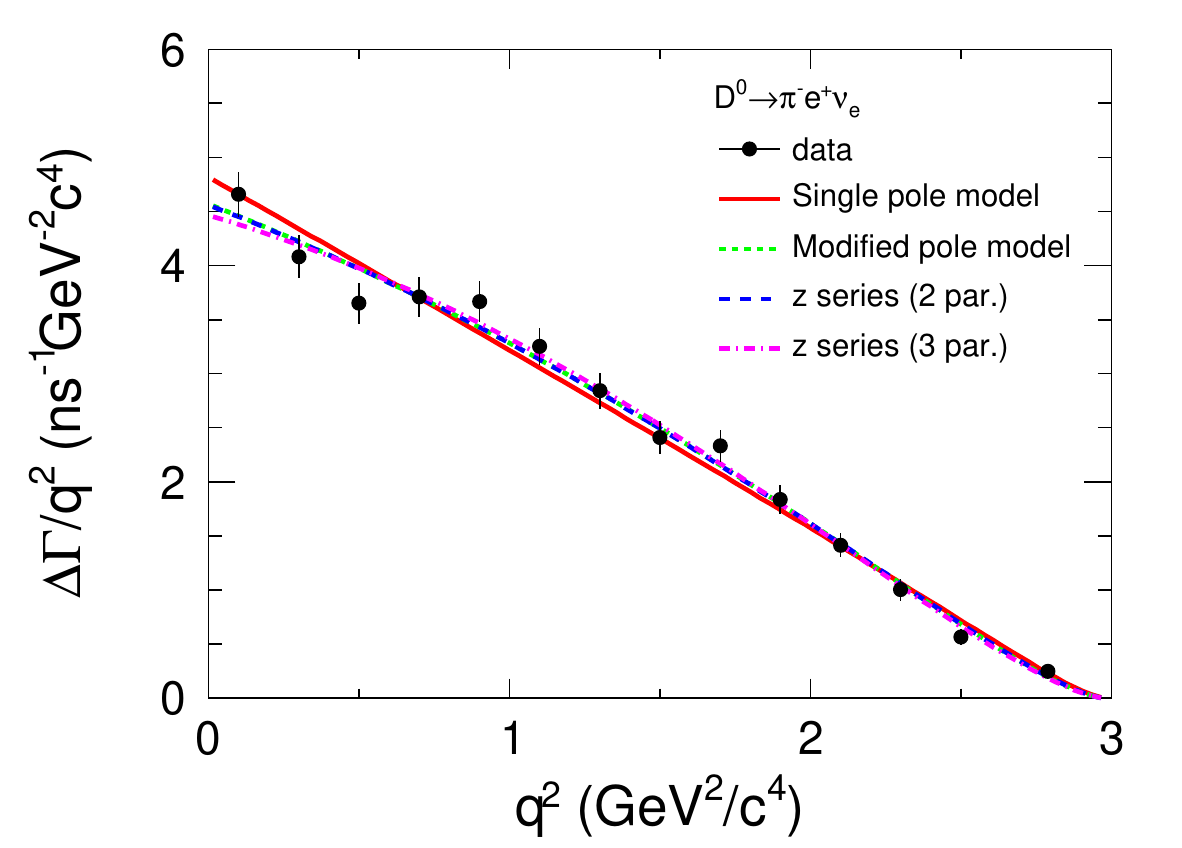}
  \includegraphics[width=0.4\textwidth,height=0.28\textwidth]{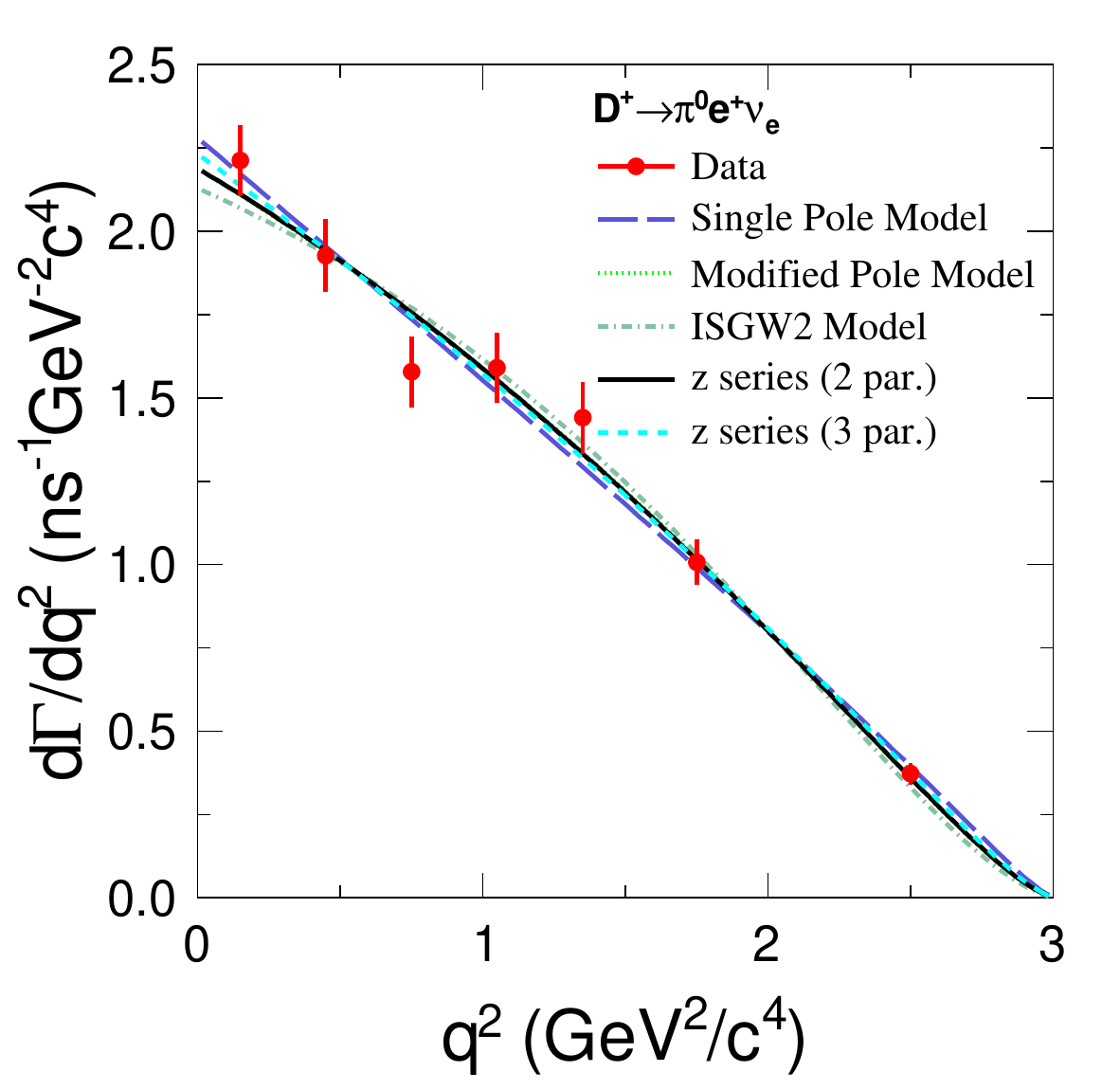}
   \caption{Fits to the partial decay rates of (left) $D^0\to \pi^-e^+\nu_e$~\cite{Ablikim:2015ixa} and (right) $D^+\to \pi^0e^+\nu_e$~\cite{BESIII:2017ylw}.
}
  \label{fig:D_pilnu}
\end{figure*}

\begin{figure}[htp]
  \centering
  \includegraphics[width=0.8\columnwidth]{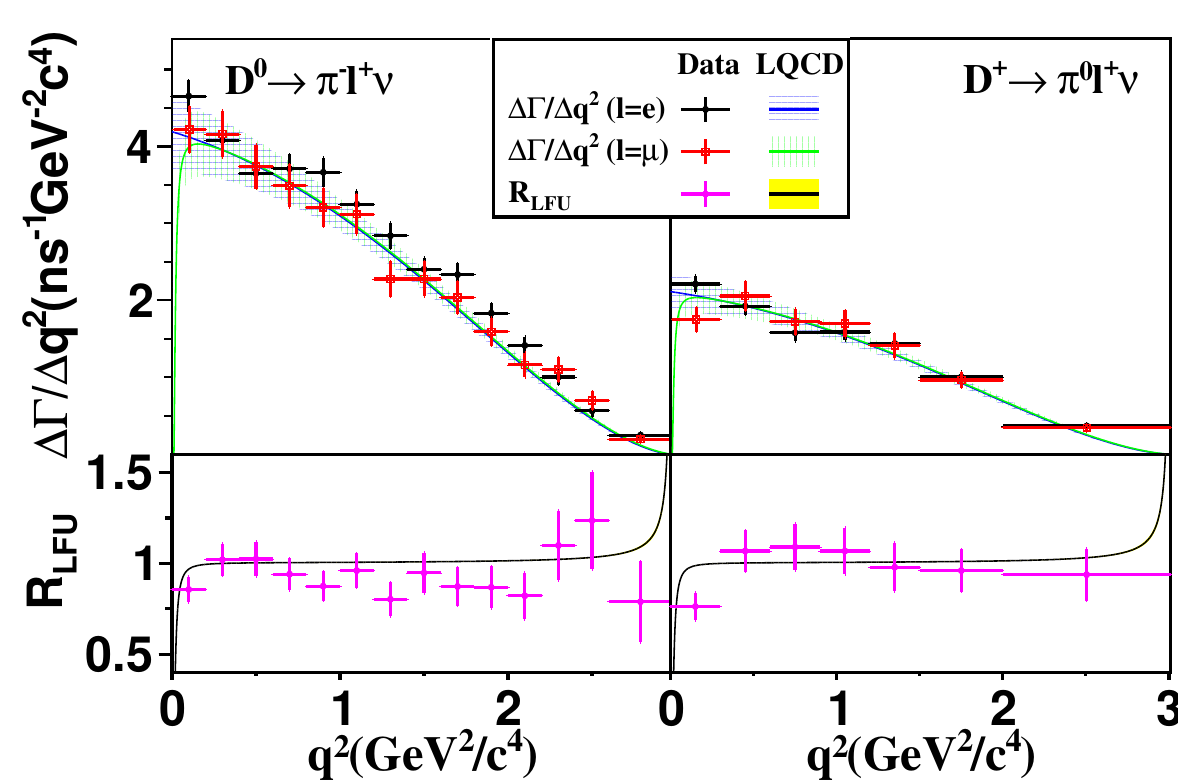}
  \caption{
$\Delta \Gamma^{0(+)}_{i}/\Delta q^2$ of $D^{0(+)}\to \pi^{-(0)}\ell^+\nu_\ell$ (top)
and ${\mathcal R}^{0(+)}_{\rm LFU}$ (bottom) in various $q^2$ bins~\cite{Ablikim:2018frk,BESIII:2015tql,BESIII:2017ylw}. The blue, green and black curves with bands show the LQCD predictions with uncertainties~\cite{Riggio:2017zwh,Lubicz:2017syv}.
}
\label{fig:D_pilnu_Rmue}
\end{figure}

\begin{figure}[htbp]
  \centering
  \includegraphics[width=0.4\textwidth]{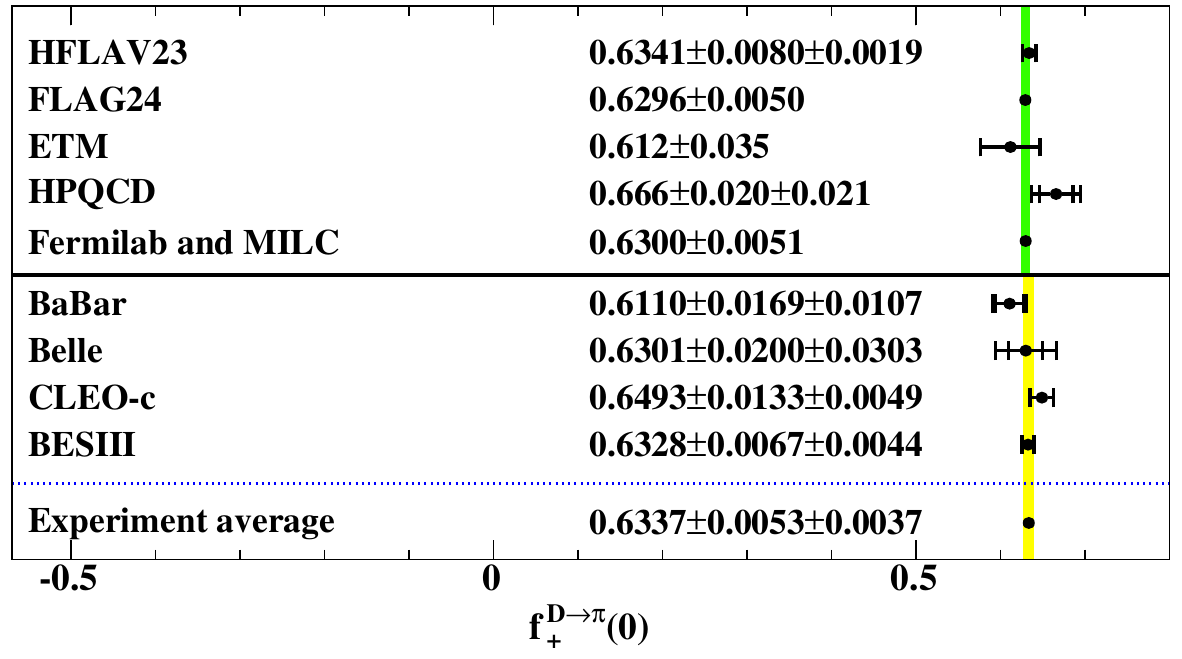}
 \put(-138,103){\tiny~\cite{HeavyFlavorAveragingGroupHFLAV:2024ctg}}
  \put(-138,95){\tiny~\cite{FlavourLatticeAveragingGroupFLAG:2024oxs}}
  \put(-138,87){\tiny~\cite{Lubicz:2017syv}}
  \put(-138,79){\tiny~\cite{Na:2011mc}}
  \put(-138,71){\tiny~\cite{FermilabLattice:2022gku}}
  \put(-138,60){\tiny~\cite{BaBar:2014xzf}}
  \put(-138,52){\tiny~\cite{Belle:2006idb}}
  \put(-138,44){\tiny~\cite{CLEO:2009svp}}
  \put(-138,36){\tiny~\cite{Ablikim:2015ixa,BESIII:2017ylw}}
   \caption{Comparison of $f^{D\to \pi}_+(0)$ from experimental measurements of
   BaBar~\cite{BaBar:2014xzf}, Belle~\cite{Belle:2006idb}, CLEO-c~\cite{CLEO:2009svp}, and BESIII~\cite{Ablikim:2015ixa,BESIII:2017ylw}
    and LQCD calculations of ETM~\cite{Lubicz:2017syv}, HPQCD~\cite{Na:2011mc}, and Fermilab and MILC~\cite{FermilabLattice:2022gku}  as well as HFLAV23~\cite{HeavyFlavorAveragingGroupHFLAV:2024ctg} and FLAG24~\cite{FlavourLatticeAveragingGroupFLAG:2024oxs}. The green band is the $\pm 1\sigma$ region of the result of Fermilab and MILC~\cite{FermilabLattice:2022gku} and the yellow band denotes the $\pm 1\sigma$ region of the result averaged over all measurements of $D\to \pi\ell^+\nu_\ell$.
}
  \label{fig:f0_Dpi}
\end{figure}

Figure~\ref{fig:D_pilnu} shows the fits  to the partial decay rates of $D^0\to \pi^-e^+\nu_e$ and $D^+\to \pi^0e^+\nu_e$.
Figure~\ref{fig:D_pilnu_Rmue} shows $\Delta \Gamma^{0(+)}_{i}/\Delta q^2$ and ${\mathcal R}^{0(+)}_{\rm LFU}$
in various $q^2$ bins, as well as the LQCD predictions for comparison.
The measured values are consistent with the SM predictions within $2\sigma$ in most of the $q^2$ regions.
The comparison of the $f^{D\to \pi}_+(0)$ measured by different experiments and those predicted by recent LQCD calculations is presented in Fig.~\ref{fig:f0_Dpi}. The measured  $f_{+}^{D\to \pi}(0)$ is consistent with the LQCD calculation~\cite{FermilabLattice:2022gku} within 1$\sigma$.
Since LQCD calculations of the $D \to \pi$ hadronic transition form factor have reached sub-percent precision, we do not include results from other theoretical approaches, such as the quark model or QCD sum rule, when comparing the experimental measurements with the theoretical predictions.

\begin{figure}[tp!]
\begin{center}
   \flushleft
   \includegraphics[width=\linewidth]{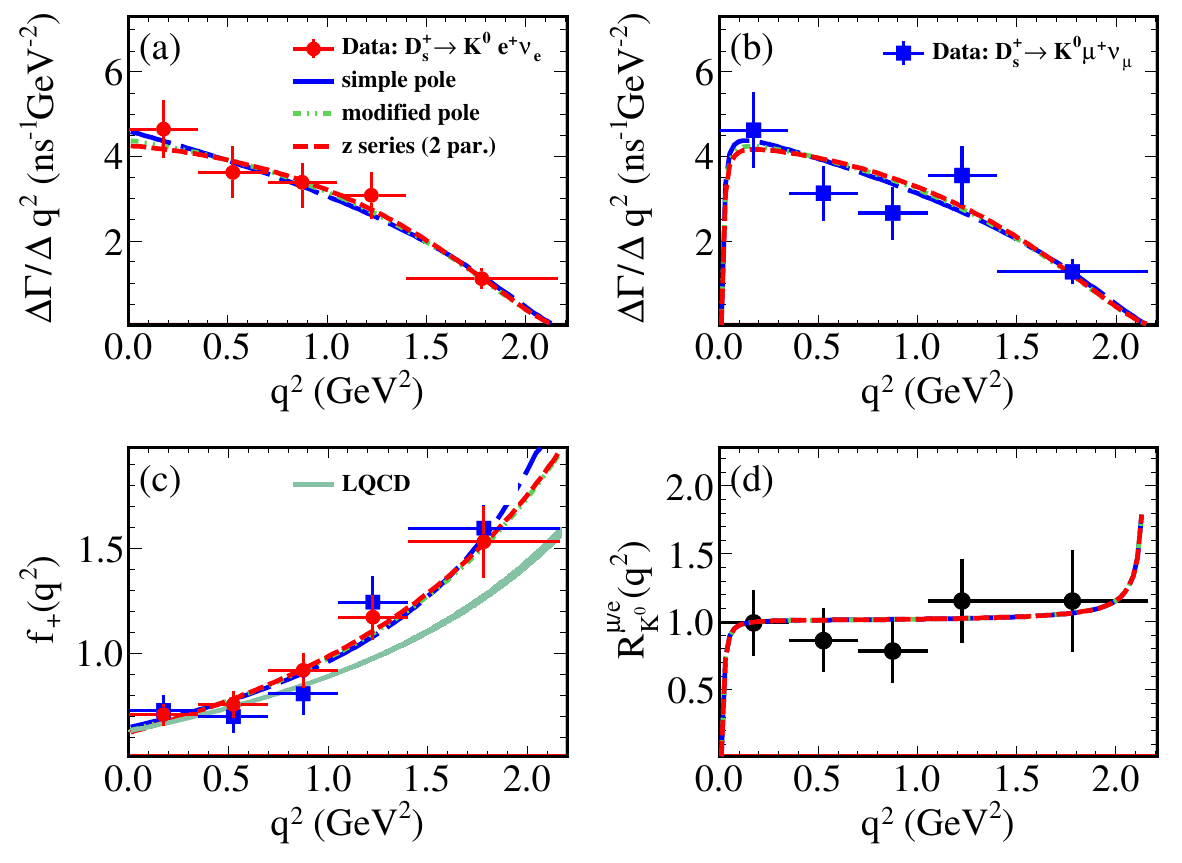}
   \caption{ Fits to the partial decay rates for (a) $D_s^+\to K^0e^+\nu_e$ and (b) $D_s^+\to K^0\mu^+\nu_{\mu}$~\cite{BESIII:2025gov}. (c) Projection onto $f^{D_s\to K^0}_+(q^2)$, and the hatched curves show the LQCD prediction~\cite{FermilabLattice:2022gku}. (d) The measured $\mathcal{R}^{\mu/e}_{K^0}(q^2)$ in various $q^2$ intervals. 
   }
\label{fig:Ds_K0lnu_FFs}
\end{center}
\end{figure}

The earliest study of $D^+_s\to K^0 e^+\nu_e$ was made by analyzing 3.19~fb$^{-1}$ of data at
4.178 GeV~\cite{BESIII:2018xre}. Its branching fraction is reported with improved precision and the hadronic transition form factor of
$D^+_s\to K^0$ at $q^2=0$ ($f_{+}^{D_s\to K^0}(0)$) is determined for the first time.
In 2025, BESIII reported improved measurements of ${\cal B}(D^+_s\to K^0 e^+\nu_e)$ and $f_{+}^{D_s\to K^0}(0)$ based on 7.3~fb$^{-1}$ of
data at 4.128-4.226 GeV~\cite{BESIII:2024zft}. Also in 2025,
with the same data sample and analysis strategy, Ref.~\cite{BESIII:2025gov} reported the first observation of $D^+_s\to K^0\mu^+\nu_\mu$
and a simultaneous fit to the differential decay rates of $D^+_s\to K^0 e^+\nu_e$ and $D^+_s\to K^0\mu^+\nu_\mu$.
From a sample of 0.78 million tagged $D^-_s$ mesons,
225 $D^+_s\to K^0 e^+\nu_e$ and 147 $D^+_s\to K^0\mu^+\nu_\mu$ signal
events are observed.
The obtained branching fractions are
${\cal B}(D^+_s\to K^0 e^+\nu_e)=(2.98\pm 0.23\pm0.12)\times 10^{-3}$ and
${\cal B}(D^+_s\to K^0\mu^+\nu_\mu)=(2.89\pm 0.27\pm0.12)\times 10^{-3}$.
The simultaneous fit to differential decay rates of these two decays  gives $f_{+}^{D_s\to K^0}(0)|V_{cd}|=0.140\pm 0.008\pm0.003$.
Using the value of $|V_{cd}|=0.22487\pm0.00068$ given by the SM-constrained fit~\cite{ParticleDataGroup:2024cfk}, the $f_{+}^{D_s\to K^0}(0)$ is measured to be $0.623\pm0.036\pm0.0011$.
Taking $f^{D_s\to K^0}_{+}(0)=0.6307\pm0.0020$ from LQCD calculations as an input, one obtains $|V_{cd}|=0.222\pm0.013\pm0.005\pm0.001$. 
Reference~\cite{BESIII:2024mot} also reported the independent measurement of semielectronic $D^+_s\to K^0 e^+\nu_e$ by using 124k of tagged $D^{*-}_s$ mesons with $e^+e^-\to D^{*+}_sD^{*-}_s$ from 10.64 fb$^{-1}$ of data at 4.237-4.699~GeV.
The branching fractions and the hadronic transition form factor
are determined to be
${\cal B}(D^+_s\to K^0 e^+\nu_e)=(2.4\pm 0.4\pm0.1)\times 10^{-3}$ and
$f^{D_s\to K^0}_+(0) = 0.677\pm0.098\pm0.023$.
Figure~\ref{fig:Ds_K0lnu_FFs} shows the fits to the partial decay rates and the projections of $f^{D_s\to K^0}_+(q^2)$ for $D_s^+\to K^0e^+\nu_e$ and $D_s^+\to K^0\mu^+\nu_{\mu}$. In addition, the $\mathcal{R}^{\mu/e}_{K^0}(q^2)$ results
in various $q^2$ intervals are shown in Fig.~\ref{fig:Ds_K0lnu_FFs}(d), which are also consistent with the SM predictions.

\begin{figure}[htbp]
  \centering
  \includegraphics[width=0.4\textwidth]{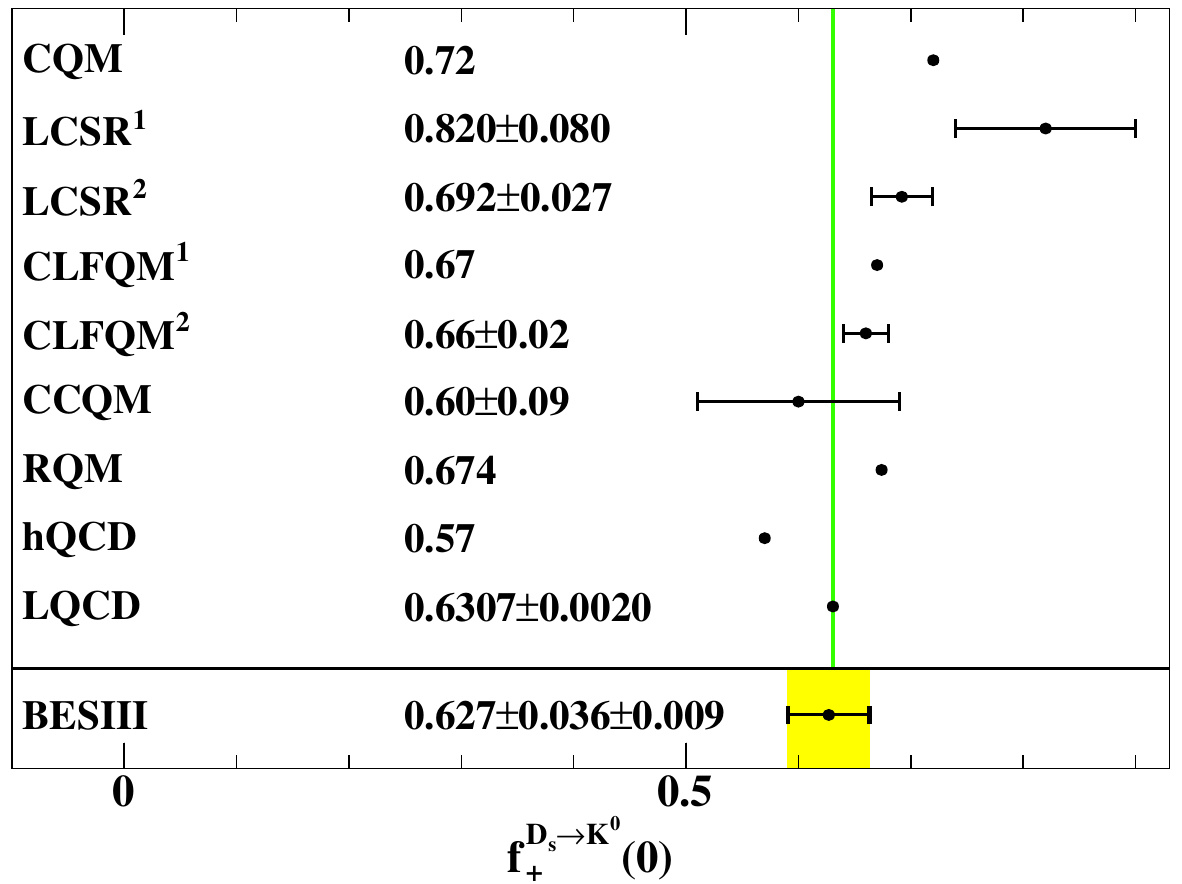}
  \put(-168,141){\tiny~\cite{Melikhov:2000yu}}
  \put(-168,129.125){\tiny~\cite{Wu:2006rd}}
  \put(-168,117.25){\tiny~\cite{Tian:2024lrn}}
  \put(-168,105.375){\tiny~\cite{Fajfer:2004mv}}
  \put(-168,93.5){\tiny~\cite{Cheng:2017pcq}}
  \put(-168,81.625){\tiny~\cite{Ivanov:2019nqd}}
  \put(-168,69.75){\tiny~\cite{Faustov:2019mqr}}
  \put(-168,57.875){\tiny~\cite{Ahmed:2023pod}}
  \put(-168,46){\tiny~\cite{FermilabLattice:2022gku}}
  \put(-168,27.5){\tiny~\cite{BESIII:2024mot,BESIII:2025gov}}
     \caption{Comparison of $f^{D_s\to K^0}_+(0)$ from experimental measurements of BESIII~\cite{BESIII:2025gov,BESIII:2024mot} and those predicted by theoretical calculations
   of CQM~\cite{Melikhov:2000yu},
   LCSR$^1$~\cite{Wu:2006rd},
   LCSR$^2$~\cite{Tian:2024lrn}.
   CLFQM$^1$~\cite{Fajfer:2004mv},
   CLFQM$^2$~\cite{Cheng:2017pcq},
   CCQM~\cite{Ivanov:2019nqd},
   RQM~\cite{Faustov:2019mqr},
   hQCD~\cite{Ahmed:2023pod}, and
   LQCD~\cite{FermilabLattice:2022gku}. The green band is the $\pm 1\sigma$ region of the LQCD result~\cite{FermilabLattice:2022gku} and the yellow band denotes the $\pm 1\sigma$ region of the result averaged over all measurements of $D_s^+\to K^0\ell^+\nu_\ell$.
}
  \label{fig:f0_DsK0}
\end{figure}

The comparison of the $f^{D_s\to K^0}_+(0)$ measured by different experiments and those predicted by recent LQCD calculations is presented in Fig.~\ref{fig:f0_DsK0}.

\begin{figure*}[htbp]
	\begin{center}
		\includegraphics[width=0.8\textwidth]{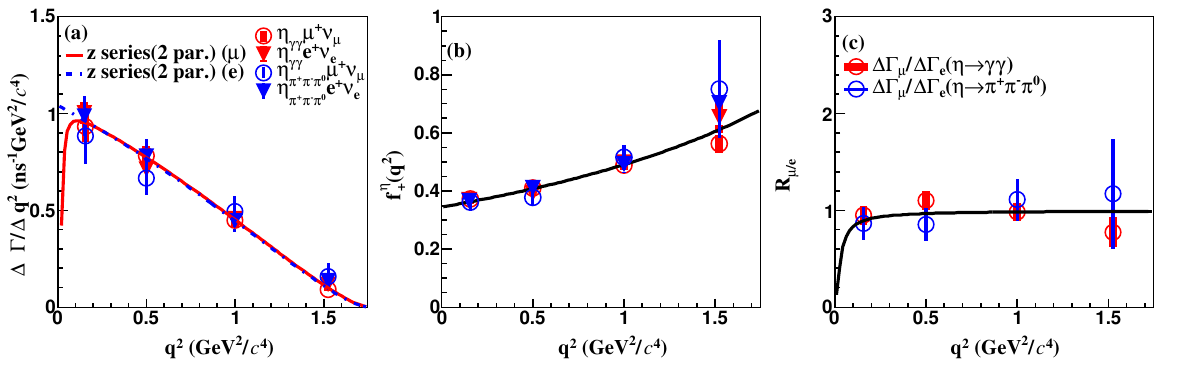}
		\caption{(a) Simultaneous fit to partial decay rates of $D^+\to \eta \ell^+\nu_\ell$ and 
(b) projection to the hadronic transition form factor as a function of $q^2$ of $D^+\to \eta \ell^+\nu_\ell$~\cite{BESIII:2025hjc}.
(c) The measured $R_{\mu/e}$ in each $q^2$ interval, and the black curve is the SM prediction~\cite{Becher:2005bg,Faustov:2019mqr}.
			\label{fig:Dp_etalnu_FFs}}
	\end{center}
\end{figure*}

Using 2.93~fb$^{-1}$ of data at 3.773 GeV, the study of $D^+\to \eta e^+\nu_e$ and  the first observation of $D^+\to \eta\mu^+\nu_\mu$ were reported in Ref.~\cite{BESIII:2018xre} and Ref.~\cite{Ablikim:2020hsc}, respectively, in which both branching fractions and hadronic transition form factors are presented.
In 2025, Ref.~\cite{BESIII:2025hjc} reported a joint analysis of $D^+\to \eta e^+\nu_e$  and  $D^+\to \eta\mu^+\nu_\mu$
by using full 20.3~fb$^{-1}$ of data at 3.773 GeV.
From a sample of 10.7 million tagged $D^-$ mesons,
2.0k $D^+\to \eta e^+\nu_e$ and 1.8k $D^+\to \eta\mu^+\nu_\mu$ signal
events are observed.
The obtained branching fractions are
${\cal B}(D^+\to \eta e^+\nu_e)=(9.75\pm0.29\pm0.28)\times10^{-4}$ and
${\cal B}(D^+\to \eta\mu^+\nu_\mu)=(9.08\pm0.35\pm0.23)\times10^{-4}$.
The simultaneous fit to their differential decay rates gives $f_{+}^{D\to\eta}(0)|V_{cd}|=0.078\pm0.002\pm0.001$.
Using the value of $|V_{cd}|=0.22487\pm0.00068$ given by the SM-constrained fit~\cite{ParticleDataGroup:2024cfk}, the $f_{+}^{D\to\eta}(0)$ is measured to be $0.345\pm0.008\pm0.003$.
Figure~\ref{fig:Dp_etalnu_FFs}(a) exhibits the fit result,
Fig.~\ref{fig:Dp_etalnu_FFs}(b) shows the extracted hadronic transition form factor, and
Fig.~\ref{fig:Dp_etalnu_FFs}(c) displays the measured ratios of partial decay rates of $D^+\to\eta\mu^+\nu_{\mu}$ and $D^+\to\eta e^+\nu_e$, $R_{\mu/e}=\Delta\Gamma_\mu/\Delta\Gamma_e$, in each $q^2$ interval.

\begin{figure}
\centering
  \includegraphics[width=0.475\textwidth]{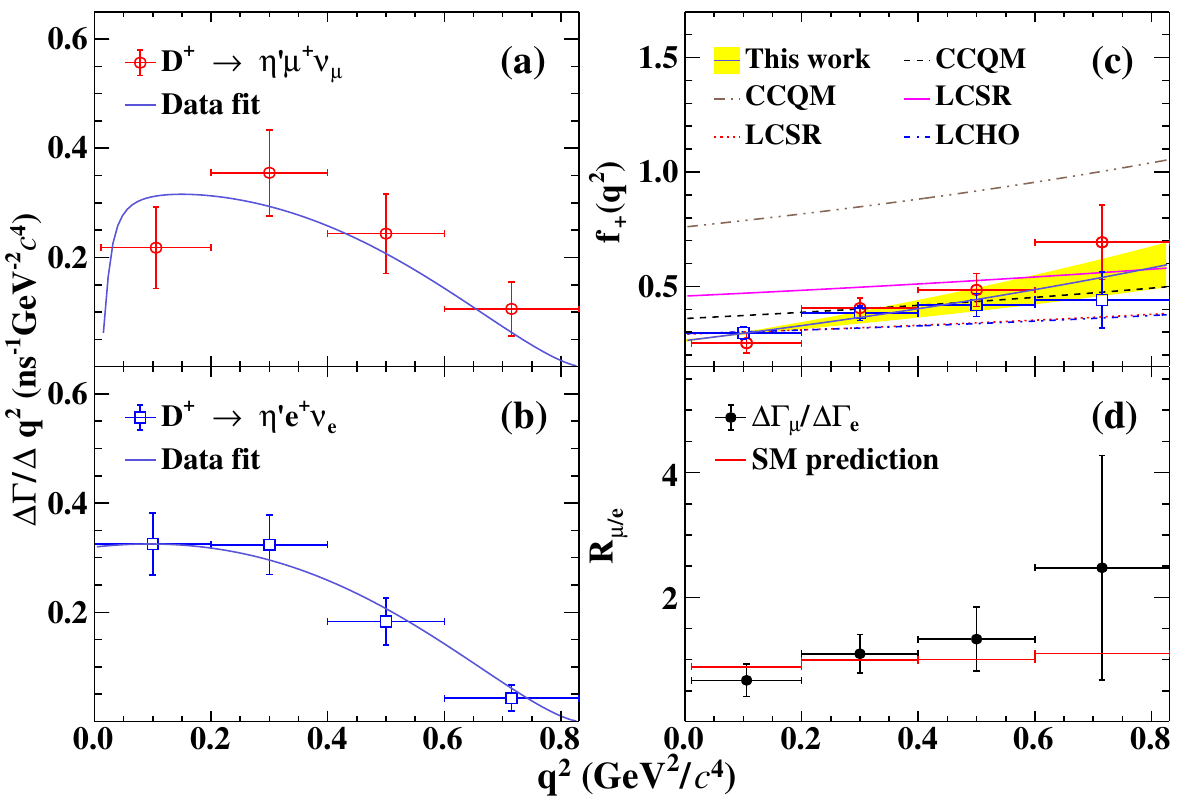}
\caption{\small (a, b) Fits to measured partial decay rates, (c) projections to $f_+^{\eta^{\prime}}(q^2)$, and (d) the measured $\mathcal R_{\mu/e}$ in each $q^2$ interval.
The yellow bands are
the $\pm1\sigma$ intervals of the fitted parameters~\cite{BESIII:2024njj}.}
  \label{fig:Dp_etaplnu_FFs}
\end{figure}

Reference~\cite{BESIII:2018xre} also reported a measurement of the branching fraction of $D^+\to \eta^\prime e^+\nu_e$.
In 2025, Ref.~\cite{BESIII:2024njj} reported the first observation of $D^+\to \eta^\prime\mu^+\nu_\mu$ and
a joint analysis of $D^+\to \eta^\prime\ell^+\nu_\ell$~($\ell=e$ or $\mu$), 
by using full 20.3~fb$^{-1}$ of data at 3.773 GeV.
From a sample of 10.7 million tagged $D^-$ mesons,
151 $D^+\to \eta^\prime e^+\nu_e$ and 90 $D^+\to \eta^\prime\mu^+\nu_\mu$ signal
events are observed.
The obtained branching fractions are
${\cal B}(D^+\to \eta^\prime e^+\nu_e)=(1.92\pm0.28\pm0.08)\times10^{-4}$ and
${\cal B}(D^+\to \eta^\prime\mu^+\nu_\mu)=(1.79\pm0.19\pm0.07)\times10^{-4}$.
The simultaneous fit to their differential decay rates gives $f_{+}^{D\to\eta^\prime}(0)|V_{cd}|=0.0592\pm0.0056\pm0.0013$.
Using the value of $|V_{cd}|=0.22487\pm0.00068$ given by the SM-constrained fit~\cite{ParticleDataGroup:2024cfk}, the $f_{+}^{D\to\eta^\prime}(0)$ is measured to be $0.263\pm0.025\pm0.006$.
The $\eta-\eta^\prime$ mixing angle in the quark flavor basis is determined to be $\phi_P=(39.8\pm0.8\pm0.3)^\circ$.
Figures~\ref{fig:Dp_etaplnu_FFs}(a) and~\ref{fig:Dp_etaplnu_FFs}(b) exhibit the fit results and Fig.~\ref{fig:Dp_etaplnu_FFs}(c) shows the extracted hadronic transition form factors.
In addition, we examine ${\mathcal R}_{\mu/e}$
in different $q^2$ intervals after considering the correlated uncertainties, with results shown in Fig.~\ref{fig:Dp_etaplnu_FFs}(d); these are also consistent with the SM predictions.

\begin{figure}[htbp]
  \centering
  \includegraphics[width=0.4\textwidth]{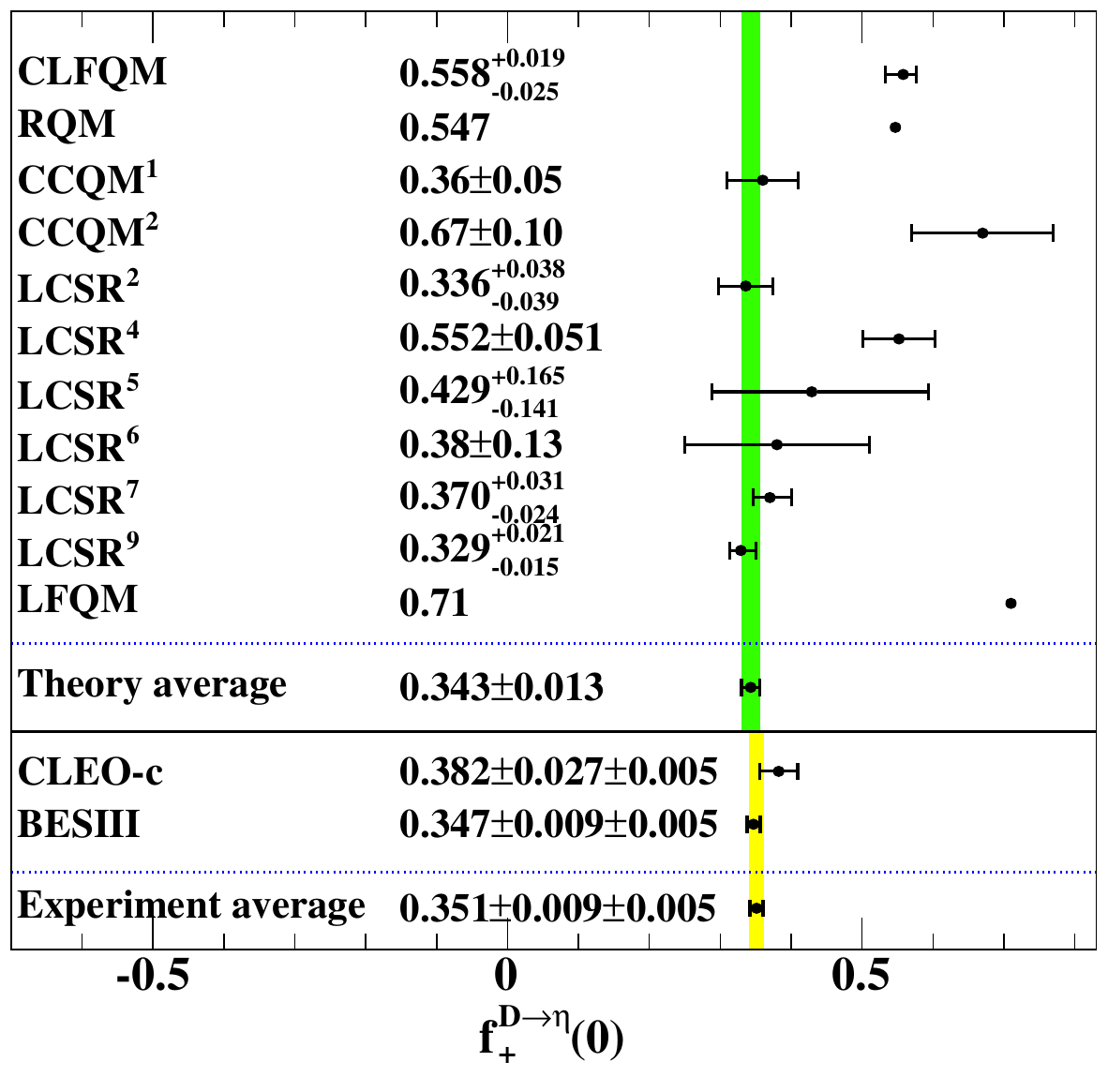}
  \put(-165,187){\tiny~\cite{Yang:2025gfz}}
  \put(-165,177.2){\tiny~\cite{Faustov:2019mqr}}
  \put(-165,167.4){\tiny~\cite{Ivanov:2019nqd}}
  \put(-165,157.6){\tiny~\cite{Soni:2018adu}}
  \put(-165,147.8){\tiny~\cite{Zhang:2025yeu}}
  \put(-165,138){\tiny~\cite{Offen:2013nma}}
  \put(-165,128.2){\tiny~\cite{Duplancic:2015zna}}
  \put(-165,118.4){\tiny~\cite{Melic:2025uha}}
  \put(-165,108.6){\tiny~\cite{Huang:2026hvs}}
  \put(-165,98.8){\tiny~\cite{Hu:2023pdl}}
  \put(-165,89){\tiny~\cite{Verma:2011yw}}
  \put(-165,57){\tiny~\cite{CLEO:2010pjh}}
  \put(-165,47.1){\tiny~\cite{BESIII:2025hjc}}
   \caption{Comparison of $f^{D\to \eta}_+(0)$ from experimental measurements of CLEO-c~\cite{CLEO:2010pjh} and BESIII~\cite{BESIII:2025gov}
   as well as theoretical calculations of
   CLFQM~\cite{Yang:2025gfz},
   RQM~\cite{Faustov:2019mqr},
   CCQM$^1$~\cite{Ivanov:2019nqd},
   CCQM$^2$~\cite{Soni:2018adu},
   LCSR$^2$~\cite{Zhang:2025yeu},
   LCSR$^4$~\cite{Offen:2013nma},
   LCSR$^5$~\cite{Duplancic:2015zna},
   LCSR$^6$~\cite{Melic:2025uha},
   LCSR$^7$~\cite{Huang:2026hvs},
   LCSR$^9$~\cite{Hu:2023pdl},
   LFQM$^2$~\cite{Verma:2011yw}. The green band is the $\pm 1\sigma$ region of the averaged result of theoretical calculations~\cite{notef0Deta} and the yellow band denotes the $\pm 1\sigma$ region of the result averaged over all measurements of $D^+\to \eta\ell^+\nu_\ell$. 
}
  \label{fig:f0_Deta}
\end{figure}

\begin{figure}[htbp]
  \centering
  \includegraphics[width=0.4\textwidth]{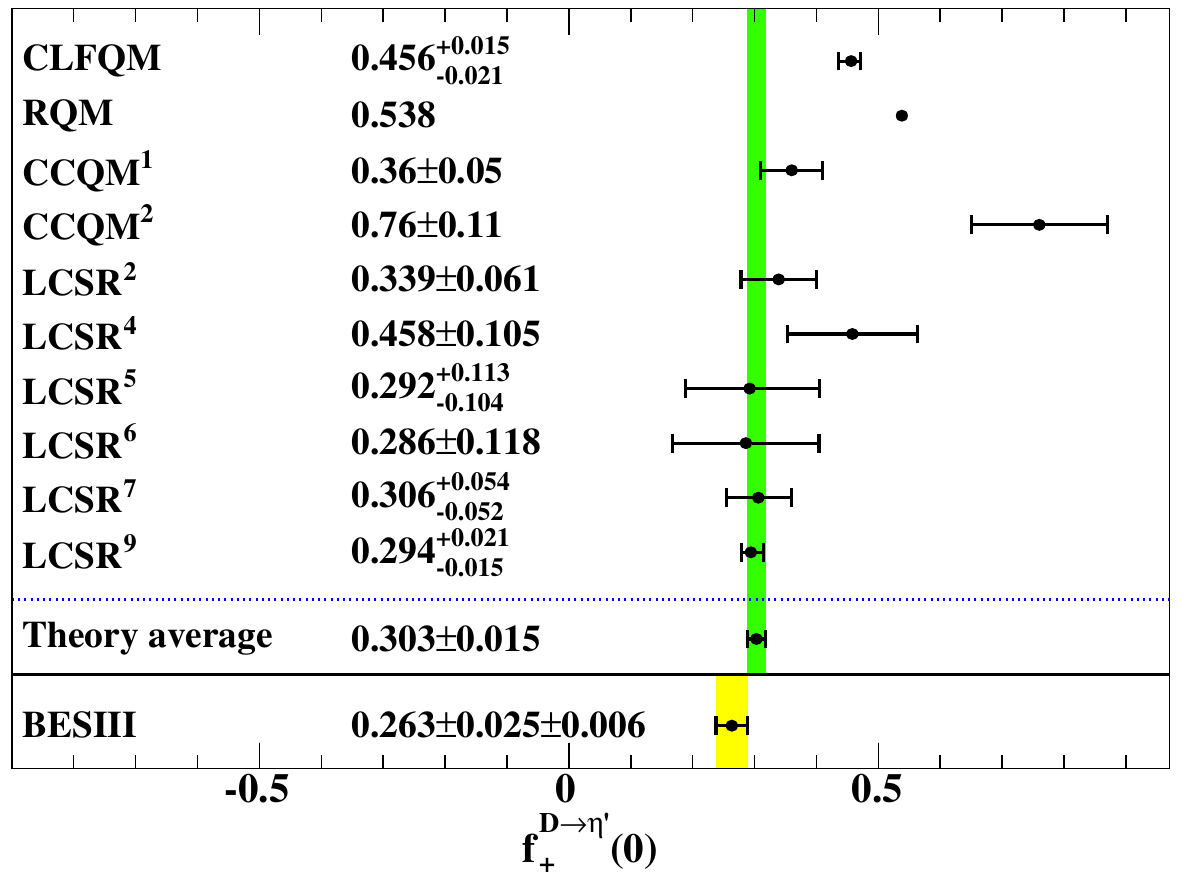}
 \put(-170,141){\tiny~\cite{Yang:2025gfz}}
  \put(-170,131.555){\tiny~\cite{Faustov:2019mqr}}
  \put(-170,122.111){\tiny~\cite{Ivanov:2019nqd}}
  \put(-170,112.666){\tiny~\cite{Soni:2018adu}}
  \put(-170,103.222){\tiny~\cite{Zhang:2025yeu}}
  \put(-170,93.777){\tiny~\cite{Offen:2013nma}}
  \put(-170,84.333){\tiny~\cite{Duplancic:2015zna}}
  \put(-170,74.889){\tiny~\cite{Melic:2025uha}}
  \put(-170,65.444){\tiny~\cite{Huang:2026hvs}}
  \put(-170,56){\tiny~\cite{Hu:2023pdl}}
  \put(-170,26){\tiny~\cite{BESIII:2024njj}}
   \caption{Comparison of $f^{D\to \eta^\prime}_+(0)$ from experimental measurements of BESIII~\cite{BESIII:2024njj} as well as theoretical calculations
    of
   CLFQM~\cite{Yang:2025gfz},
   RQM~\cite{Faustov:2019mqr},
   CCQM$^1$~\cite{Ivanov:2019nqd},
   CCQM$^2$~\cite{Soni:2018adu},
   LCSR$^2$~\cite{Zhang:2025yeu},
   LCSR$^4$~\cite{Offen:2013nma},
   LCSR$^5$~\cite{Duplancic:2015zna},
   LCSR$^6$~\cite{Melic:2025uha},
   LCSR$^7$~\cite{Huang:2026hvs}, and
   LCSR$^9$~\cite{Hu:2023pdl}. The green band is the $\pm 1\sigma$ region of the averaged result of theoretical calculations~\cite{notef0Deta} and the yellow band denotes the $\pm 1\sigma$ region of the BESIII result~\cite{BESIII:2024njj}. 
}
  \label{fig:f0_Detap}
\end{figure}

The obtained $f_+^{D\to\eta^{(\prime)}}(0)$ are consistent with the theoretical calculations of 
     CCQM$^1$~\cite{Ivanov:2019nqd},
   LCSR$^2$~\cite{Zhang:2025yeu},
   LCSR$^4$~\cite{Offen:2013nma},
   LCSR$^5$~\cite{Duplancic:2015zna},
   LCSR$^6$~\cite{Melic:2025uha},
   LCSR$^7$~\cite{Huang:2026hvs}, and
   LCSR$^9$~\cite{Hu:2023pdl}.
They disfavor the 
 CLFQM~\cite{Yang:2025gfz},
   RQM~\cite{Faustov:2019mqr}, and
   CCQM$^2$~\cite{Soni:2018adu}
 by more than $4\sigma$.
Due to the lack of precise LQCD calculations for the hadronic transition form factors $f_{+}^{D\to \eta^{(\prime)}}(0)$, and because existing quark models cannot simultaneously reproduce the experimental measurements for both $f_{+}^{D_s\to \eta^{(\prime)}}(0)$ and $f_{+}^{D\to \eta^{(\prime)}}(0)$, we average the theoretical calculations by excluding those values that lack error estimates or exhibit excessively large deviations~\cite{notef0Deta}. 
Comparisons of the $f^{D\to \eta^{(\prime)}}_+(0)$ measured by different experiments and those predicted by different theoretical calculations are  presented in Table~\ref{tab:f0:Deta} and Figs.~\ref{fig:f0_Deta} and ~\ref{fig:f0_Detap}. The measured $f_{+}^{D\to \eta}(0)$ and $f_{+}^{D\to \eta^{\prime}}(0)$ are consistent with the averaged theoretical calculations within 1$\sigma$.

\begin{table*}[htbp]
\centering
\caption{Results for $f^{D_{(s)}\to\eta^{(\prime)}}_+(0)$ from various theoretical calculations and their averages~\cite{notef0Deta}. 
}
\label{tab:f0:Deta}
\begin{tabular}{l c c c c}
\hline\hline
Model  &$f_+^{D_s\to\eta}(0)$ &$f_+^{D_s\to\eta^\prime}(0)$ &$f_+^{D\to\eta}(0)$&$f_+^{D\to\eta^\prime}(0)$ \\
\hline
RQM~\cite{Faustov:2019mqr}&0.443&0.559&0.547&0.538\\
CLFQM~\cite{Yang:2025gfz}&$0.490^{+0.014}_{-0.017}$&$0.599^{+0.017}_{-0.021}$&$0.558^{+0.019}_{-0.025}$&$0.456^{+0.015}_{-0.021}$\\
LFQM(I)$^{1}$~\cite{Wei:2009nc}&0.50&0.62&---&---\\
LFQM(I)$^{1}$~\cite{Wei:2009nc}&0.48&0.60&---&---\\
LFQM$^{2}$~\cite{Verma:2011yw}&0.76&---&0.71&---\\
CCQM$^{1}$~\cite{Ivanov:2019nqd}&$0.49\pm0.07$&$0.59\pm0.09$&$0.36\pm0.05$&$0.36\pm0.05$\\
CCQM$^{2}$~\cite{Soni:2018adu}&$0.78\pm0.12$&$0.73\pm0.11$&$0.67\pm0.10$&$0.76\pm0.11$\\
CQM~\cite{Melikhov:2000yu}&$0.78$&0.78&---&---\\
3PSR~\cite{Colangelo:2001cv}&$0.50\pm0.04$&---&---&---\\
LCSR$^{1}$~\cite{Hu:2021zmy}&$0.476^{+0.040}_{-0.036}$&$0.544^{+0.046}_{-0.042}$&---&---\\
LCSR$^{2}$~\cite{Zhang:2025yeu}&$0.522^{+0.059}_{-0.057}$&$0.548^{+0.081}_{-0.079}$&$0.336^{+0.038}_{-0.039}$&$0.339\pm0.061$\\
LCSR$^{3}$~\cite{Azizi:2010zj}&$0.45\pm0.14$&$0.55\pm0.18$&---&---\\
LCSR$^{4}$~\cite{Offen:2013nma}&$0.432\pm0.033$&$0.52\pm0.08$&$0.552\pm0.051$&$0.458\pm0.105$\\
LCSR$^{5}$~\cite{Duplancic:2015zna}&$0.495^{+0.030}_{-0.029}$&$0.558^{+0.047}_{-0.045}$&$0.429^{+0.165}_{-0.141}$&$0.292^{+0.113}_{-0.104}$\\
LCSR$^{6}$~\cite{Melic:2025uha}&$0.467\pm0.022$&$0.501\pm0.054$&$0.38\pm0.13$&$0.286\pm0.118$\\
LCSR$^{7}$~\cite{Huang:2026hvs}&$0.449^{+0.038}_{-0.030}$&$0.543^{+0.059}_{-0.052}$&$0.370^{+0.031}_{-0.024}$&$0.306^{+0.054}_{-0.052}$\\
LCSR$^{8}$~\cite{Zhou:2026llv}&$0.494^{+0.032}_{-0.033}$&$0.571^{+0.032}_{-0.033}$&---&---\\
LCSR$^{9}$~\cite{Hu:2023pdl}&---&---&$0.329^{+0.021}_{-0.015}$&$0.294^{+0.021}_{-0.015}$\\
LQCD$_{m_\pi=370 \rm MeV}$~\cite{Bali:2014pva}&$0.542\pm0.013$&$0.404\pm0.025$&---&---\\
LQCD$_{m_\pi=470 \rm MeV}$~\cite{Bali:2014pva}&$0.564\pm0.011$&$0.437\pm0.018$&---&---\\
\hline
Theory average&$0.480\pm0.012$&$0.552\pm0.019$&$0.343\pm0.013$&$0.303\pm0.015$\\
\hline\hline
\end{tabular}
\end{table*}

In addition, using 6.3 fb$^{-1}$ of data recorded at 4.178-4.226~GeV, the decay
$D_s^+\to \pi^0 e^+\nu_e$, which is expected to be sensitive to $\pi^0$--$\eta$ mixing,
is searched for the first time~\cite{BESIII:2022jcm}.
No significant signal is observed and an upper limit on the branching fraction of
$D_{s}^{+}\to \pi^0 e^+\nu_e$ at the $90\%$ confidence level
is set as $6.4 \times 10^{-5}$.

\begin{table*}[htbp]
\centering
\caption{Results for $f_+^{D\to P}(0)|V_{cs(d)}|$ from various experiments. The input $|V_{cs}| = 0.97296 \pm 0.00024$ and $|V_{cd}| = 0.2271 \pm 0.0010$ for the results of BaBar$^{a}$, Belle$^{b}$, and Belle$^{c}$ are taken from PDG2006. The systematic uncertainties of BESIII results are fully correlated while averaging. Specifically, the systematic uncertainties for the measurements of $D\to \bar K\ell^+\nu_\ell$~\cite{BESIII:2026uin,BESIII:2026ydr} and $D^{+} \to K_L^0e^+ \nu_e$~\cite{BESIII:2015jmz}  by BESIII are treated as uncorrelated due to the different analysis methods.
While calculating the $|V_{cs(d)}|$, the input hadronic transition form factors of $f_+^{D\to K}(0)=0.7452\pm0.0031$,  $f_+^{D\to \pi}(0)=0.6300\pm0.0051$, and $f_+^{D_s\to K^0}(0)=0.6307\pm0.0020$ are taken from LQCD calculations~\cite{FermilabLattice:2022gku}, and the hadronic transition form factor of  $f_+^{D_s\to \eta}(0)=0.473\pm0.012$ is averaged based on Refs.~\cite{Ivanov:2019nqd,Colangelo:2001cv,Hu:2021zmy,Azizi:2010zj,Offen:2013nma,Duplancic:2015zna,Melic:2025uha,Huang:2026hvs,Zhou:2026llv}, $f_+^{D_s\to \eta^\prime}(0)=0.551\pm0.019$ is averaged based on Refs.~\cite{Ivanov:2019nqd,Hu:2021zmy,Azizi:2010zj,Offen:2013nma,Duplancic:2015zna,Melic:2025uha,Huang:2026hvs,Zhou:2026llv}, $f_+^{D\to \eta}(0)=0.343\pm0.013$ and $f_+^{D\to \eta^\prime}(0)=0.303\pm0.015$ are averaged based on Refs.~\cite{Ivanov:2019nqd,Zhang:2025yeu,Duplancic:2015zna,Melic:2025uha,Huang:2026hvs,Hu:2023pdl}. Conversely, the input $|V_{cs}|=0.97349\pm0.00016$ and $|V_{cd}|=0.22487\pm0.00068$ are taken from PDG2024~\cite{ParticleDataGroup:2024cfk} for the hadronic transition form factor calculations.
}
\label{tab:fplus-vcd}
\begin{tabular}{l l c cc}
\hline\hline
Experiment & Decay &$f_+^{D\to K}(0)|V_{cs}|$&   $f_+^{D\to K}(0)$&$|V_{cs}|$\\
\hline 
BESIII~\cite{BESIII:2026uin,BESIII:2026ydr}  & $D^{0, +} \to \bar K\ell^+ \nu_\ell$ & $0.7183\pm0.0007\pm0.0014$&$0.7378\pm0.0007\pm0.0014$&$0.9639\pm0.0009\pm0.0044$ \\
BESIII~\cite{BESIII:2015jmz}  & $D^{+} \to K_L^0e^+ \nu_e$ & $0.728\pm0.006\pm0.011$&$0.7478\pm0.0062\pm0.0113$ &$0.9769\pm0.0081\pm0.0153$ \\
CLEO-c~\cite{CLEO:2009svp}  & $D^{0, +} \to \bar K e^+ \nu_e$ & $0.717\pm0.004\pm0.004$& $0.7365\pm0.0041\pm0.0041$&$0.9622\pm0.0054\pm0.0067$ \\
BaBar$^{a}$~\cite{BaBar:2007zgf}    & $D^0 \to K^- e^+ \nu_e$& $0.707\pm0.007\pm0.009$ &$0.7263\pm0.0072\pm0.0092$  &$0.9487\pm0.0094\pm0.0127$ \\
Belle$^{b}$~\cite{Belle:2006idb}  & $D^{0} \to K^- \ell^+ \nu_\ell$& $0.676\pm0.007\pm0.021$ &$0.6944\pm0.0072\pm0.0216$ &$0.9071\pm0.0094\pm0.0284$  \\
\hline
\multicolumn{2}{l}{BESIII average} & $0.7184\pm0.0007\pm0.0014$ &$0.7380\pm0.0007\pm0.0014$&$0.9640\pm0.0009\pm0.0044$\\
\multicolumn{2}{l}{Average} & $0.7180\pm0.0007\pm0.0013$ &$0.7376\pm0.0007\pm0.0013$ &$0.9635\pm0.0009\pm0.0044$\\
\hline 
 Experiment & Decay&$f_+^{D_s\to\eta}(0)|V_{cs}|$& $f_+^{D_s\to\eta}(0)$& $|V_{cs}|$\\
\hline
BESIII~\cite{BESIII:2023ajr}  & $D^+_s\to \eta e^+ \nu_e$ & $0.4519\pm0.0071\pm0.0065$&$0.4642\pm0.0073\pm0.0067$&$0.9415\pm0.0148\pm0.0272$\\
BESIII~\cite{BESIII:2024mot}  & $D^+_s\to \eta e^+ \nu_e$ & $0.430\pm0.021\pm0.016$&$0.4417\pm0.0216\pm0.0164$& $0.8958\pm0.0438\pm0.0402$\\
BESIII~\cite{BESIII:2023gbn}  & $D^+_s\to \eta \mu^+ \nu_\mu$ & $0.452\pm0.010\pm0.007$&$0.4643\pm0.0103\pm0.0072$&$0.9417\pm0.0208\pm0.0277$\\
\hline
\multicolumn{2}{l}{BESIII average} & $0.4530\pm0.0056\pm0.0067$ &$0.4653\pm0.0058\pm0.0069$&$0.9438\pm0.0117\pm0.0274$ \\
\hline
Experiment & Decay &$f_+^{D_s\to\eta^\prime}(0)|V_{cs}|$& $f_+^{D_s\to\eta^\prime}(0)$&$|V_{cs}|$ \\
\hline
BESIII~\cite{BESIII:2023ajr}  & $D^+_s\to \eta^\prime e^+ \nu_e$ & $0.525\pm0.024\pm0.009$&  $0.540\pm0.025\pm0.009$&$0.9511\pm0.0435\pm0.0366$ \\
BESIII~\cite{BESIII:2024mot}  & $D^+_s\to \eta^\prime e^+ \nu_e$ & $0.542\pm0.062\pm0.023$& $0.557\pm0.064\pm0.024$ &$0.9819\pm0.1123\pm0.0537$\\
BESIII~\cite{BESIII:2023gbn}  & $D^+_s\to \eta^\prime \mu^+ \nu_\mu$ & $0.504\pm0.037\pm0.012$&$0.518\pm0.038\pm0.012$&$0.9130\pm0.0670\pm0.0382$ \\
\hline
\multicolumn{2}{l}{BESIII average} & $0.521\pm0.019\pm0.011$ & $0.535\pm0.020\pm0.011$&$0.9438\pm0.0344\pm0.0381$\\
\hline
\multicolumn{3}{l}{$|V_{cs}|_{D_{(s)}\to P\ell^+\nu_\ell}$ average}&&$0.9629\pm0.0009\pm0.0043$\\
\multicolumn{3}{l}{$|V_{cs}|_{D_{(s)}\to P\ell^+\nu_\ell+D_s\to \ell^+\nu_\ell}$ average}&&$0.9686\pm0.0009\pm0.0038$\\
\hline\hline
Experiment & Decay  &$f_+^{D\to\pi}(0)|V_{cd}|$& $f_+^{D\to\pi}(0)$&$|V_{cd}|$ \\
\hline
BESIII~\cite{BESIII:2017ylw}  & $D^+ \to \pi^0 e^+ \nu_e$ & $0.1400\pm0.0026\pm0.0007$& $0.6226\pm0.0116\pm0.0036$&$0.2222\pm0.0041\pm0.0021$\\
BESIII~\cite{Ablikim:2015ixa}  & $D^0 \to \pi^- e^+ \nu_e$&$0.1435\pm0.0018\pm0.0010$ &$0.6381\pm0.0080\pm0.0044$ &$0.2278\pm0.0071\pm0.0109$\\
CLEO-c~\cite{CLEO:2009svp}   &$D^{0,+} \to \pi^{-,0} e^{+} \nu_e$& $0.146\pm0.003\pm0.001$ &$0.6493\pm0.0133\pm0.0049$&$0.2317\pm0.0048\pm0.0025$ \\
BaBar~\cite{BaBar:2014xzf}    & $D^0 \to \pi^- e^+ \nu_e$& $0.1374\pm0.0038\pm0.0024$ & $0.6110\pm0.0169\pm0.0107$&$0.2181\pm0.0060\pm0.0042$\\
Belle$^{c}$~\cite{Belle:2006idb}  & $D^0\to \pi^-\ell^+ \nu_\ell$& $0.1417\pm0.0045\pm0.0068$ & $0.6301\pm0.0200\pm0.0303$&$0.2249\pm0.0071\pm0.0109$\\
\hline
\multicolumn{2}{l}{BESIII average} & $0.1423\pm0.0015\pm0.0009$ &$0.6328\pm0.0067\pm0.0044$&$0.2259\pm0.0024\pm0.0023$\\
\multicolumn{2}{l}{Average} & $0.1425\pm0.0012\pm0.0007$ &$0.6337\pm0.0053\pm0.0037$ &$0.2262\pm0.0019\pm0.0021$\\
\hline
 Experiment & Decay  &$f_+^{D\to \eta}(0)|V_{cd}|$&$f_+^{D\to \eta}(0)$ & $|V_{cd}|$ \\
\hline
BESIII~\cite{BESIII:2025hjc}  & $D^+\to \eta\ell^+ \nu_\ell$ & $0.078\pm0.002\pm0.001$&$0.347\pm0.009\pm0.005$&$0.2274\pm0.0058\pm0.0091$\\
CLEO-c~\cite{CLEO:2010pjh}  & $D^+\to\eta e^+ \nu_e$ & $0.086\pm0.006\pm0.001$& $0.382\pm0.027\pm0.005$&$0.2507\pm0.0175\pm0.0099$ \\
\hline
\multicolumn{2}{l}{Average} & $0.079\pm0.002\pm0.001$ &$0.351\pm0.009\pm0.005$ &$0.2303\pm0.0058\pm0.0092$\\
\hline

Experiment & Decay &$f_+^{D\to \eta^\prime}(0)|V_{cd}|$&$f_+^{D\to \eta^\prime}(0)$&$|V_{cd}|$   \\
\hline
BESIII~\cite{BESIII:2024njj}  & $D^+\to \eta^\prime\ell^+ \nu_\ell$ & $0.0592\pm0.0056\pm0.0013$&$0.263\pm0.025\pm0.006$ &$0.1954\pm0.0185\pm0.0106$  \\
\hline

Experiment & Decay  &$f_+^{D_s\to K^0}(0)|V_{cd}|$&$f_+^{D_s\to K^0}(0)$ &$|V_{cd}|$  \\
\hline
BESIII~\cite{BESIII:2024mot}  & $D^+_s\to K^0 e^+ \nu_e$ & $0.152\pm0.022\pm0.005$& $0.677\pm0.098\pm0.022$&$0.2410\pm0.0349\pm0.0080$\\
BESIII~\cite{BESIII:2025gov}  & $D^+_s\to K^0 \ell^+ \nu_\ell$ & $0.140\pm0.008\pm0.003$&  $0.623\pm0.036\pm0.011$&$0.222\pm0.013\pm0.005$\\
\hline
\multicolumn{2}{l}{BESIII average} & $0.141\pm0.008\pm0.002$ & $0.627\pm0.036\pm0.009$&$0.2236\pm0.0127\pm0.0032$\\
\hline
\multicolumn{3}{l}{$|V_{cd}|_{D_{(s)}\to P\ell^+\nu_\ell}$ average}&&$0.2259\pm0.0018\pm0.0020$\\
\multicolumn{3}{l}{$|V_{cd}|_{D_{(s)}\to P\ell^+\nu_\ell+D\to \ell^+\nu_\ell}$ average}&&$0.2259\pm0.0014\pm0.0013$\\
\hline\hline
\end{tabular}
\end{table*}

\subsection{Results of $D\to V\ell^+\nu_\ell$ at BESIII}

\subsubsection{Cabibbo-favored decays}

\begin{figure*}[htp]
  \begin{center}
  \includegraphics[width=0.25\linewidth]{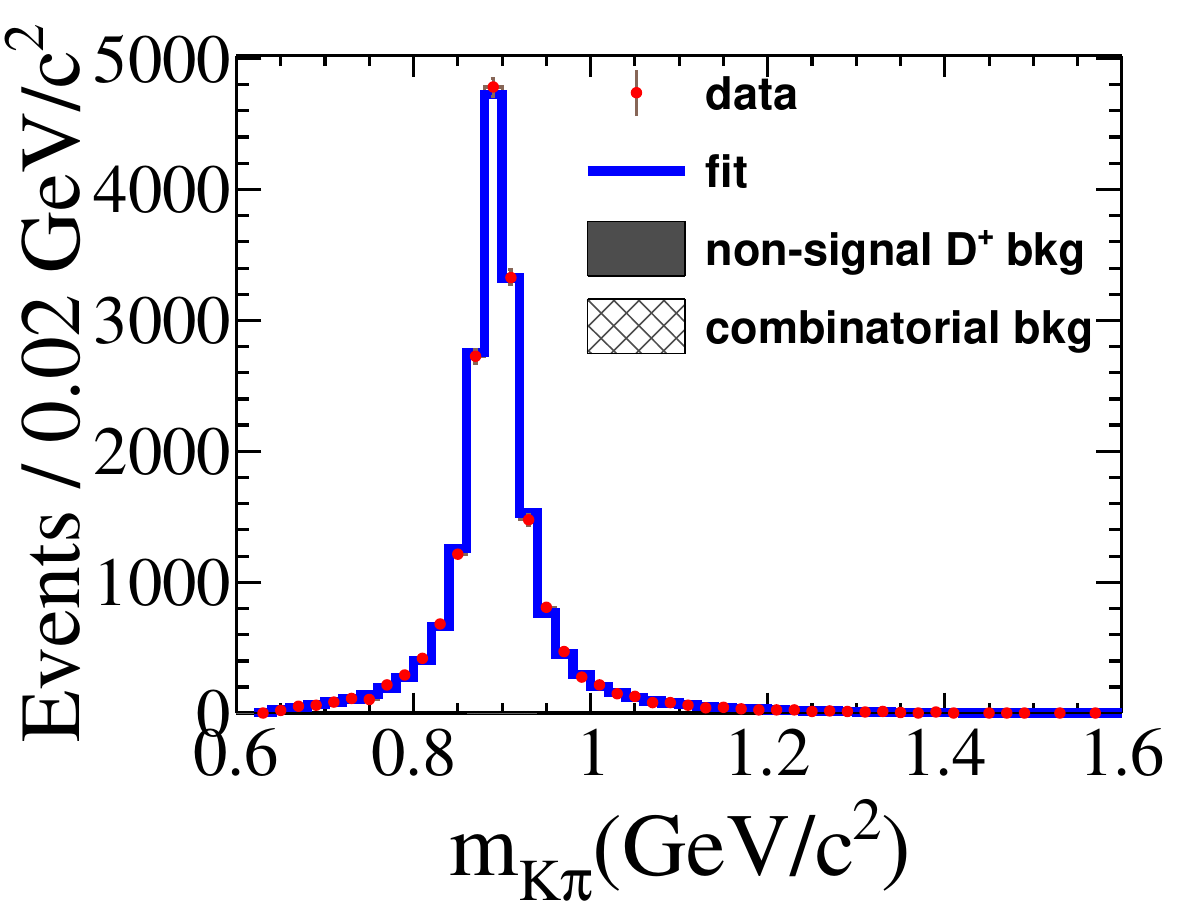}
  \includegraphics[width=0.25\linewidth]{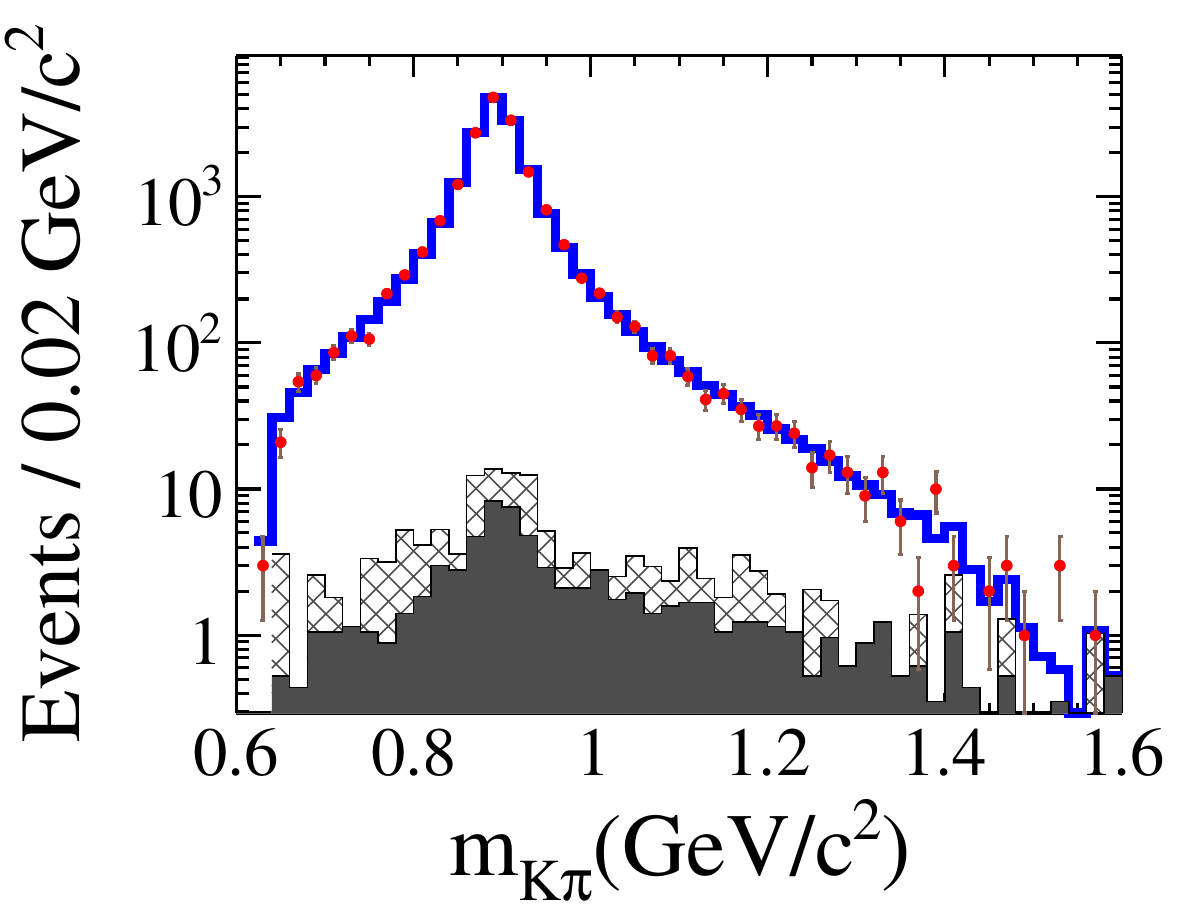}
  \includegraphics[width=0.25\linewidth]{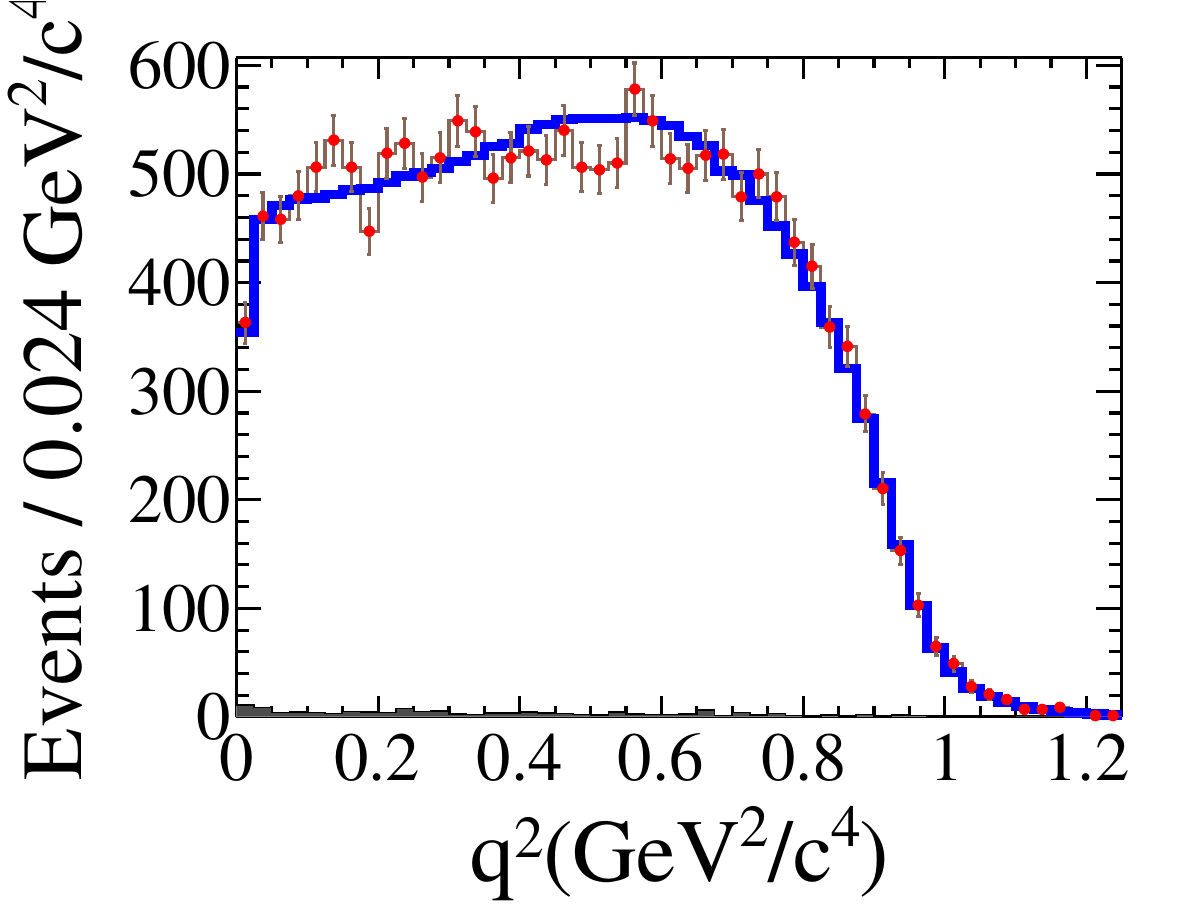}
  \includegraphics[width=0.25\linewidth]{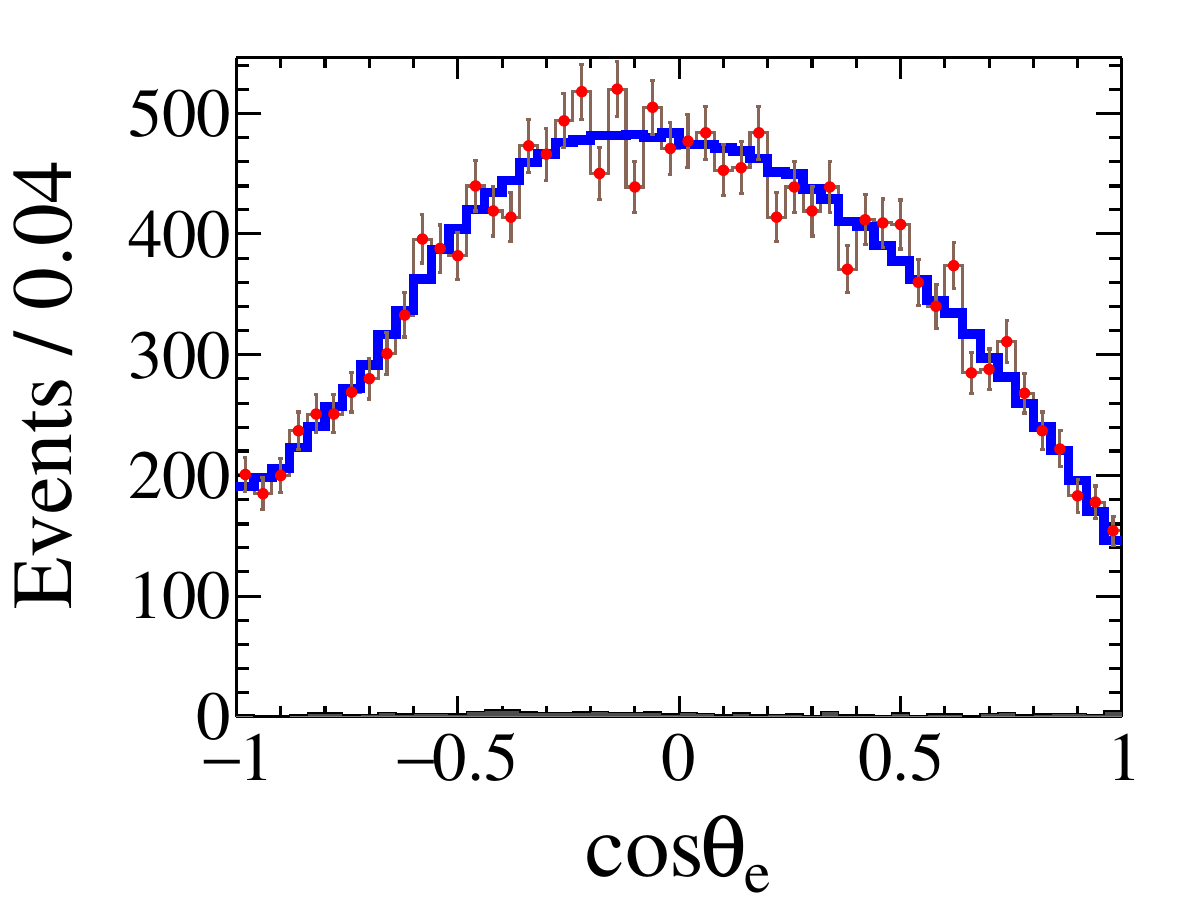}
  \includegraphics[width=0.25\linewidth]{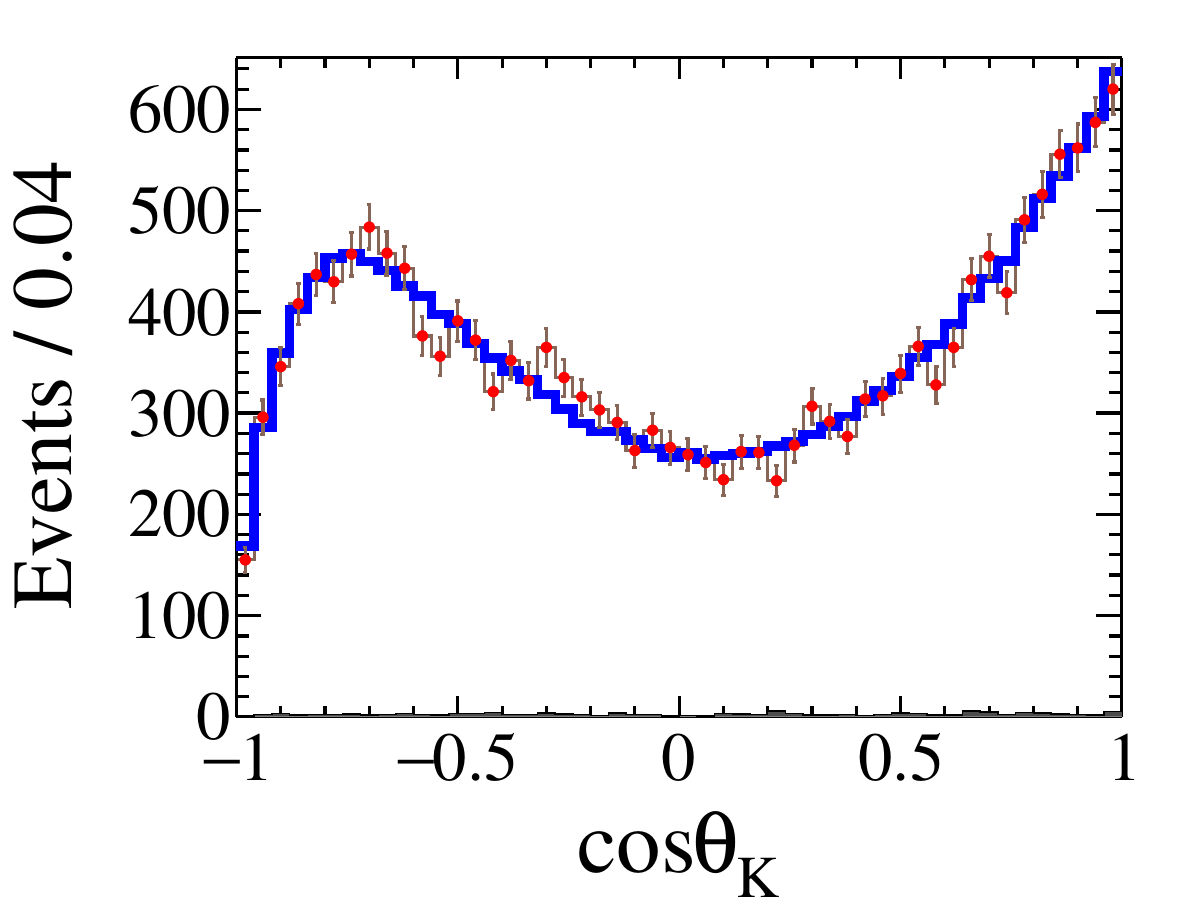}
  \includegraphics[width=0.25\linewidth]{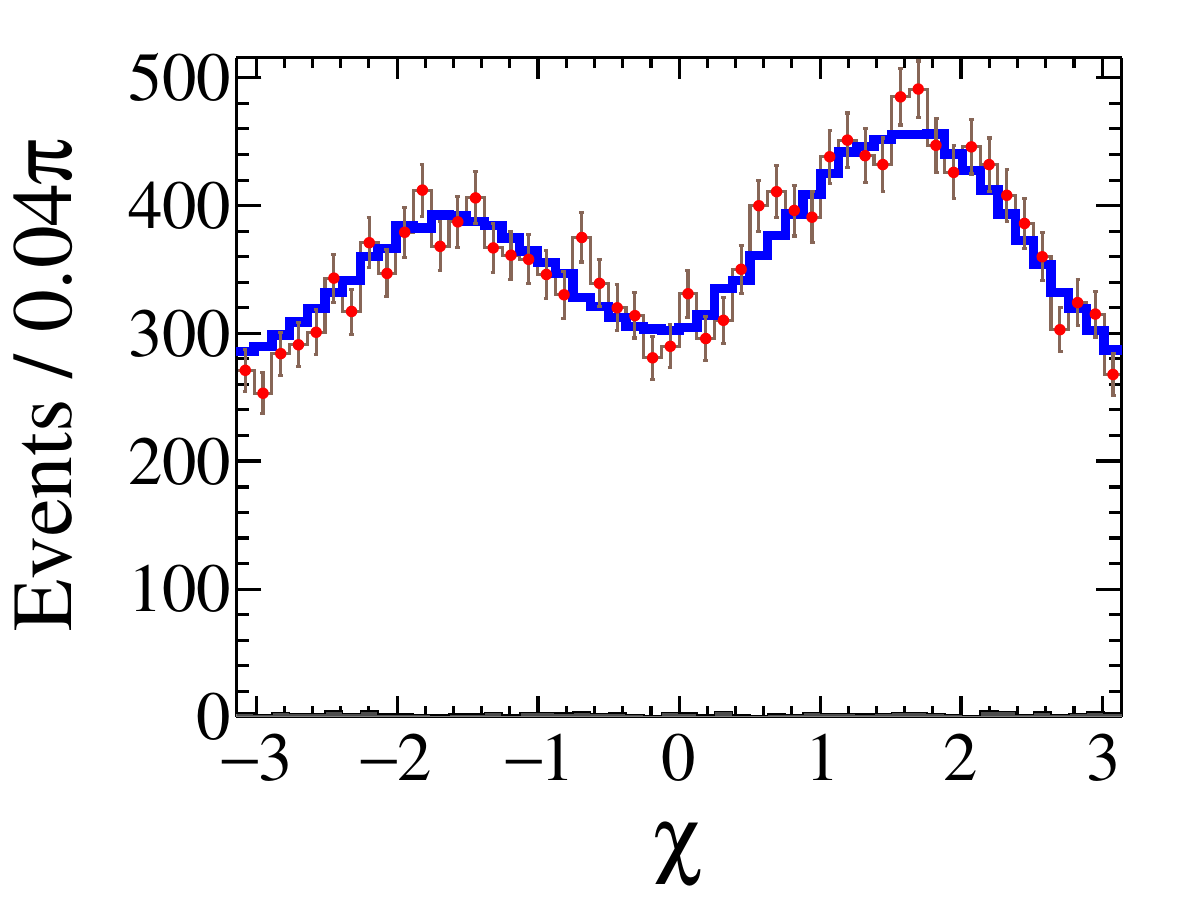}
  \caption{
The projections of amplitude analysis for $D^+\to K^-\pi^+e^+\nu_e$ on $M_{K^-\pi^+}$, $q^2$, $\cos\theta_{K^-}$, $\cos\theta_\ell$, and $\chi$~\cite{BESIII:2015hty}.
 The shadowed histogram
  shows the non-signal $D^+$ background estimated from MC simulation
  and the hatched area shows the combinatorial background
  estimated from the $M_{\rm BC}$ sideband of data.}
  \label{fig:Dp_Kpienu}
  \end{center}
\end{figure*}

Using 2.93 fb$^{-1}$ of data at 3.773~GeV, the decay $D^+ \to K^- \pi^+ e^{+} \nu_e$ is analyzed by using 18.3k signal events~\cite{BESIII:2015hty}. The branching fraction is measured to be $\mathcal{B}(D^{+} \to K^{-} \pi^+ e^+ \nu_e) = (3.77 \pm 0.03 \pm 0.08)\%$, with a partial branching fraction of $\mathcal{B}(D^{+} \to K^{-} \pi^+ e^+ \nu_e){[0.8,1.0]} = (3.39 \pm 0.03 \pm 0.08)\%$ for $0.8<M_{\bar K\pi}<1.0$ GeV/$c^{2}$. An amplitude analysis reveals a $\bar K\pi$ $\cal S$-wave component with a fraction of $(6.05\pm0.22\pm0.18)\%$, in addition to the dominant process $D^{+} \to \bar{K}^{*}(892)^{0} e^{+} \nu_e$.
The obtained hadronic transition form factor or their ratios at $q^2=0$ are $r_V=1.411 \pm 0.058 \pm 0.007$, $r_2=0.788 \pm 0.042 \pm 0.008$,
and $A_1(0)=0.589\pm0.010\pm0.012$. A model-independent measurement of the $\cal S$-wave phase variation with $M_{K^-\pi^+}$ is also performed. The projections of amplitude analysis for $D^+\to K^-\pi^+e^+\nu_e$ on $M_{K^-\pi^+}$, $q^2$, $\cos\theta_{K^-}$, $\cos\theta_\ell$, and $\chi$ are shown in Fig.~\ref{fig:Dp_Kpienu}.

\begin{figure*}[htbp]\centering
    \includegraphics[width=0.8\textwidth]{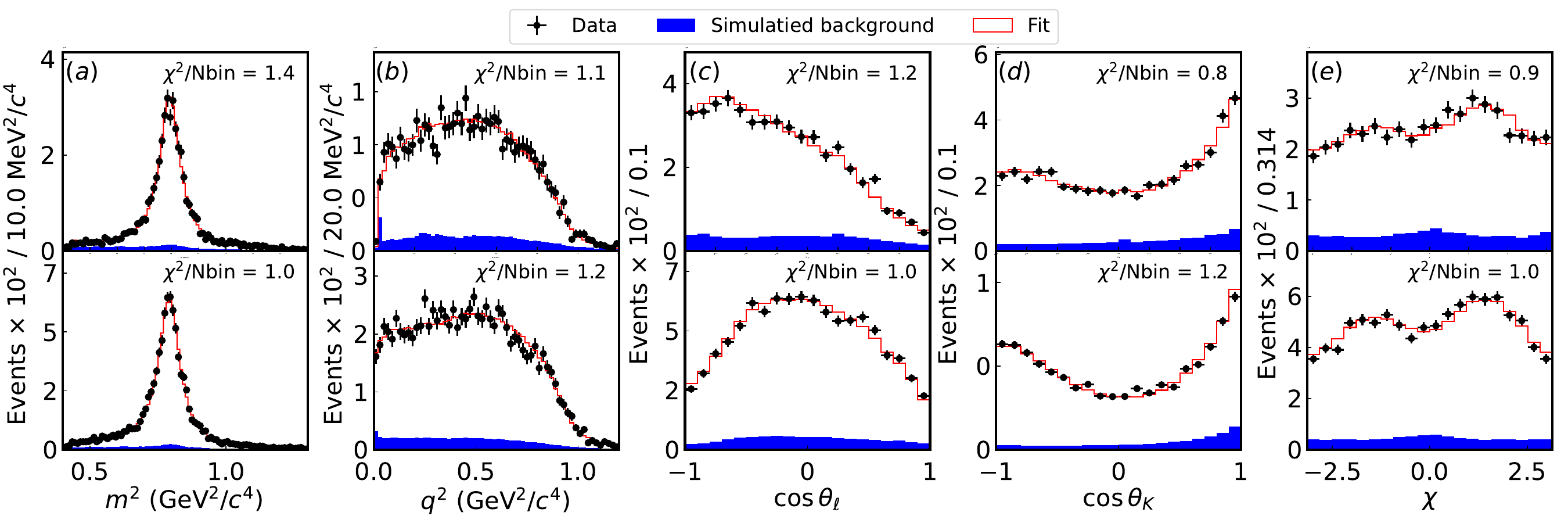}
    \caption{
 The projections of amplitude analysis for (top) $D^+\to K_{S}^{0} \pi^{0}\mu^+\nu_\mu$ and (bottom) $D^+\to K_{S}^{0} \pi^{0}e^+\nu_e$ on $M^2_{K_S^0\pi^0}$, $q^2$, $\cos\theta_{K_S^0}$, $\cos\theta_\ell$, and $\chi$~\cite{BESIII:2025fso}.}
    \label{fig:Dp_KSpi0lnu}
\end{figure*}

In 2024, the branching fraction of $D^+\to K_{S}^{0} \pi^{0}e^+\nu_e$ and the hadronic transition form factor ratios were measured for the first time~\cite{BESIII:2024awg}, using $7.9~\mathrm{fb}^{-1}$ of data at 3.773~GeV. These results were later superseded by those in Ref.~\cite{BESIII:2025fso}, based on 20.3~$\mathrm{fb}^{-1}$ of 3.773~GeV~data. The later article reports the first observation of $D^+\to K_{S}^{0} \pi^{0}\mu^+\nu_\mu$ and an improved measurement of $D^+\to K_{S}^{0} \pi^{0}e^+\nu_e$, using 6.8k and 11.1k signal events, respectively. The branching fractions of $D^+\to K_S^0\pi^0\mu^+\nu_\mu$ and $D^+\to K_S^0\pi^0e^+\nu_e$ are determined to be $(0.896\pm0.017\pm0.008)\%$ and $(0.943\pm0.012\pm0.010)\%$, respectively. Figure~\ref{fig:Dp_KSpi0lnu} shows the projections of the nominal fit result.
 A dynamics analysis reveals that the dominant $\bar{K}^\ast(892)^0$ component is accompanied by an $\cal S$-wave contribution, which accounts for $(7.10 \pm 0.68 \pm 0.41)\%$ of the total decay rate in the $\mu^+$ channel and $(6.39 \pm 0.17 \pm 0.14)\%$ in the $e^+$ channel. Assuming a single-pole dominance parametrization, the hadronic transition form factor ratios are extracted as $r_V=1.42 \pm 0.03 \pm 0.02$ and $r_2=0.75 \pm 0.03\pm 0.01$. A comprehensive angular and decay-rate $CP$ asymmetry analysis yields the first full set of averaged angular and $CP$ asymmetry observables as functions of the momentum-transfer squared; these are consistent with
the SM expectations. No evidence of $\mu-e$ lepton-flavor universality violation is observed, either in the full range or in five bins of momentum-transfer squared.

\begin{figure*}[htbp]\centering
    \includegraphics[width=0.8\textwidth]{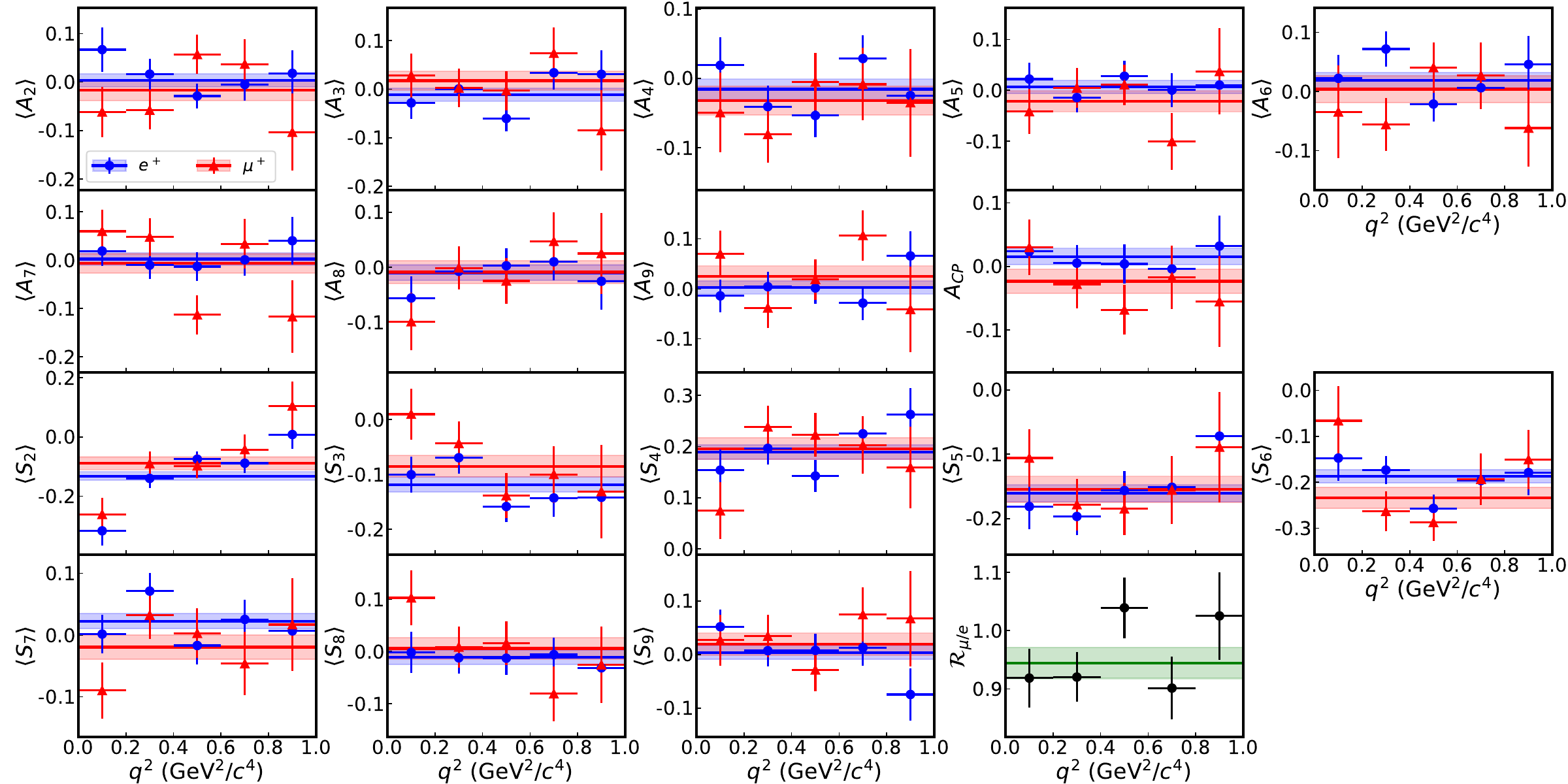}
    \caption{Measured observables $\langle A_i \rangle$, $\langle S_i \rangle$, $\mathcal{R}_{\mu/e}$ and the $CP$ asymmetry for $D^+ \to \bar{K}^{*}(892)^0\ell^+\nu_\ell$ in different $q^2$ regions with $0.8<M_{K_S^0\pi^0}<1.0$ GeV$/c^2$. The $e^+$ ($\mu^+$) results are shown in blue (red). The horizontal lines and shaded bands represent the mean value and 1 standard deviation across the full $q^2$ range~\cite{BESIII:2025fso}.}
    \label{fig:Dp_KSpi0lnu_asyAS}
\end{figure*}

Each angular observable, and $A_{CP}$, is determined from the partial decay rate $d\Gamma_i/dq^2_i\,(\text{Tag})$, where ``Tag'' refers to the angular and flavor tags defined by the piecewise integration.
The partial decay rate depends on the yield in each Tag and $q^2$ bin, and the yields are measured independently through the same fit methods used for the branching fraction measurement but with the candidates split by Tag and $q^2$ region. The fits use the same modeling described earlier, including the same treatment of floating components, but with shapes determined independently in each Tag and $q^2$ region. The results for $\langle A_i \rangle$, $\langle S_i \rangle$, and $A_{CP}$, including both statistical and systematic uncertainties added in quadrature, are shown in Fig.~\ref{fig:Dp_KSpi0lnu_asyAS}.

\begin{figure}[tp!]
\begin{center}
   \includegraphics[width=0.8\linewidth]{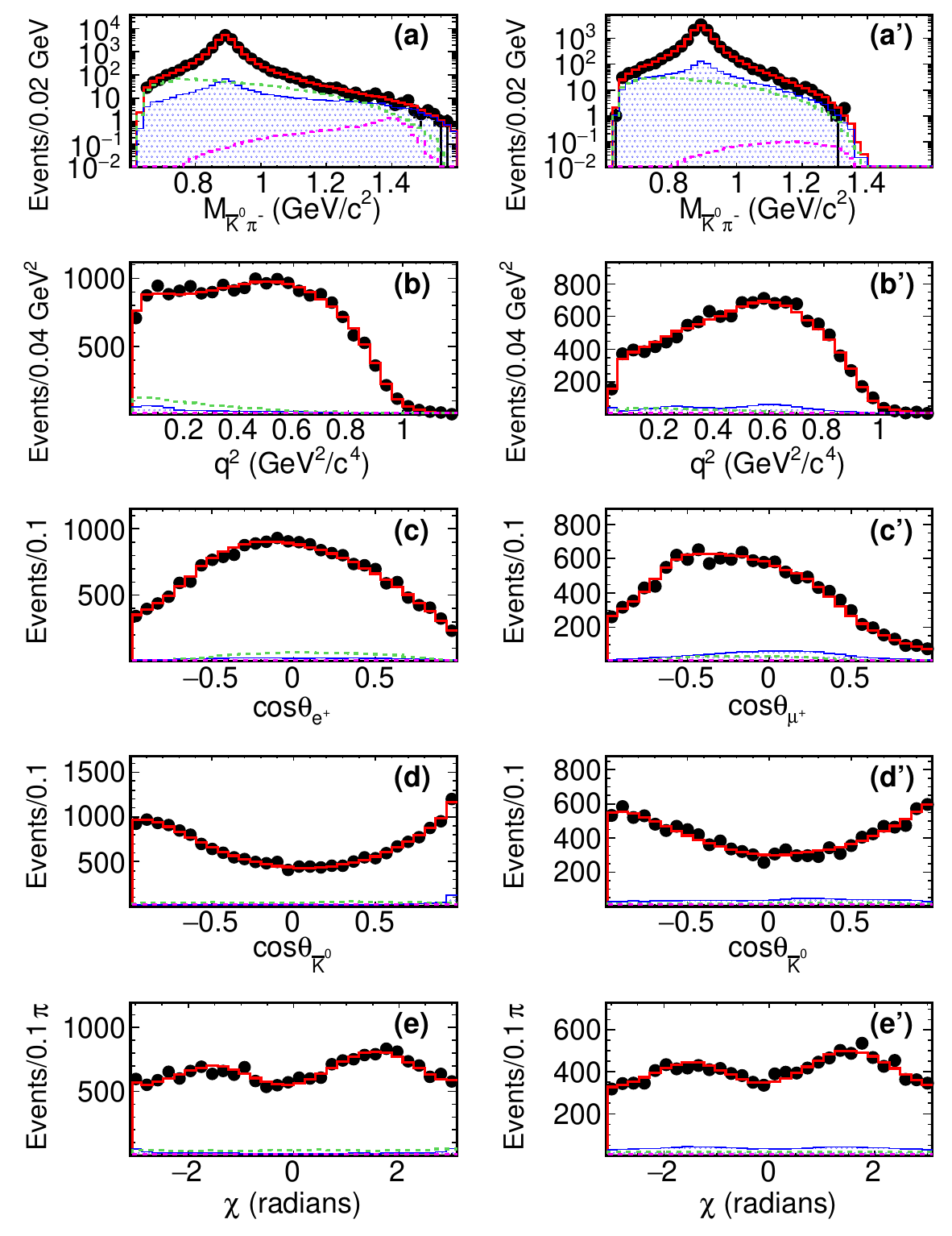}
   \caption{
   The projections of amplitude analysis for (left) $D^0\to \bar K^0\pi^-e^+\nu_e$ and (right) $D^0\to \bar K^0\pi^-\mu^+\nu_\mu$ on $M_{\bar K^0\pi^-}$, $q^2$, $\cos\theta_{\bar K^0}$, $\cos\theta_\ell$, and $\chi$~\cite{BESIII:2026txt}.
The dashed green and dashed pink curves show the contribution of the $\mathcal{S}-$wave and $\mathcal{D}-$wave components. The dots with error bars are data, the red histograms are the fit results, and the shaded histograms are the simulated background. }
\label{fig:D0_KSpilnu_FF}
\end{center}
\end{figure}

In 2019, a study of $D^0 \to \bar{K}^0\pi^-e^+\nu_e$ was first reported using 2.93 fb$^{-1}$ of data at 3.773~GeV~\cite{BESIII:2018jjm}, providing improved precision on the branching fraction and the first determination of the hadronic transition form factor ratios for this decay. These results were subsequently superseded by those obtained with a larger data sample of 7.9 fb$^{-1}$ at the same energy~\cite{BESIII:2024xjf}, and then by those based on a simultaneous study of $D^0 \to \bar{K}^0\pi^-e^+\nu_e$ and $D^0 \to \bar{K}^0\pi^-\mu^+\nu_\mu$ with 36.4k signal events using the 7.9 fb$^{-1}$ dataset~\cite{BESIII:2025wex}. More recently, a combined analysis of these two decay modes was performed with the full 20.3 fb$^{-1}$ of data sample accumulated at 3.773~GeV~\cite{BESIII:2026txt}.
The projected distributions of the fit onto the fitted variables are shown in Fig.~\ref{fig:D0_KSpilnu_FF}.
Based on an investigation of the decay dynamics in $D^0 \to \bar{K}^0\pi^-\ell^+\nu_{\ell}$, a $\mathcal{D}$-wave component from $D^0\to K_2^*(1430)^-\ell^+\nu_{\ell}$ is observed for the first time with a statistical significance of $8.0\sigma$, alongside the dominant $K^*(892)^-$ and $\mathcal{S}$-wave contributions. The $\mathcal{D}$-wave component accounts for $(0.092 \pm 0.028\pm 0.018)\%$ of the total decay rate. The branching fractions of the dominant $K^*(892)^-$ components are measured as $\mathcal{B}(D^0\to K^{*}(892)^-e^+\nu_e) = (2.043 \pm 0.018 \pm 0.012)\%$ and $\mathcal{B}(D^0\to K^{*}(892)^-\mu^+\nu_{\mu}) = (1.964 \pm 0.018\pm 0.012)\%$, representing the most precise measurements to date and significant improvements over previous world averages. The hadronic transition form factor ratios are measured to be $r_{V} = 1.444 \pm 0.026 \pm 0.010$, $r_{2}= 0.752 \pm 0.020\pm 0.004$, and $A_1(0)=0.618\pm0.002\pm0.004$. This is the most precise determination of the hadronic transition form factor ratios in a $D\to \bar K^*(892)$ transition to date. In addition, the first model-independent measurement of the $\mathcal{S}$-wave phase shift in the hadronic $\bar{K}^0\pi^-$ system is reported.

\begin{figure*}[tp!]
    \centering
      \includegraphics[width=0.25\linewidth]{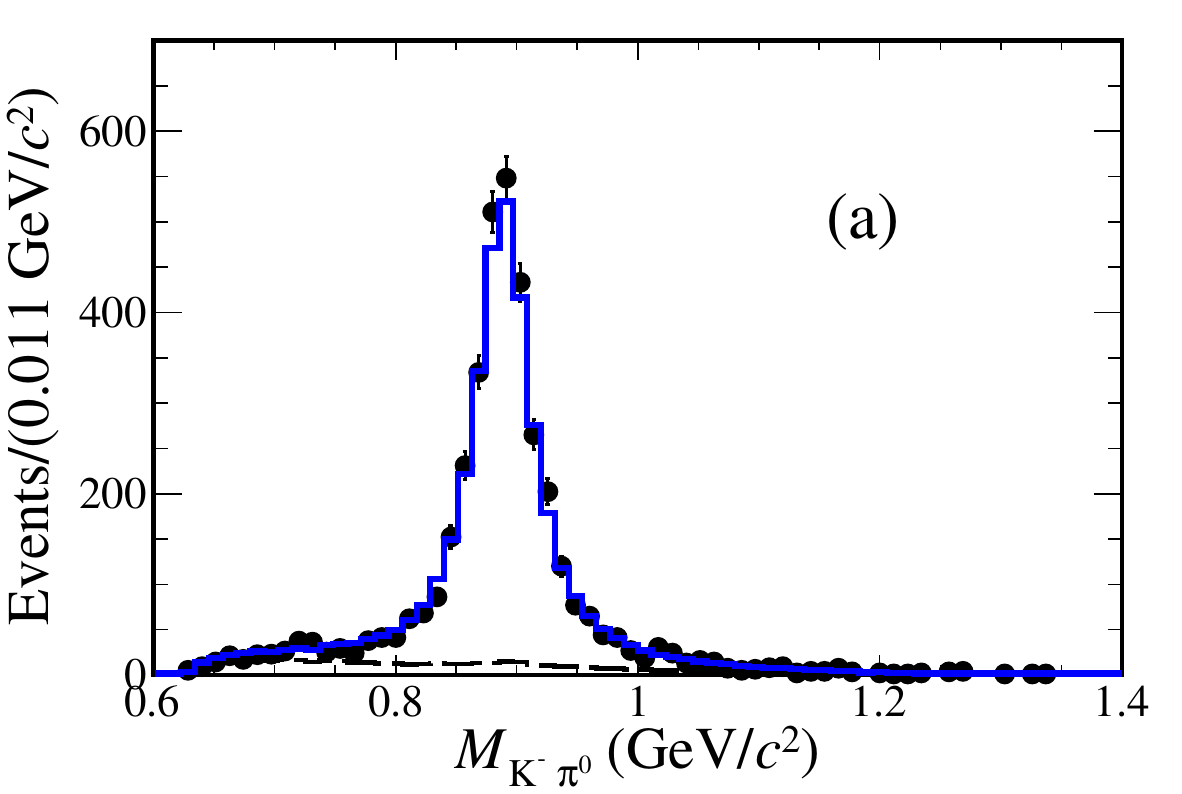}
      \includegraphics[width=0.25\linewidth]{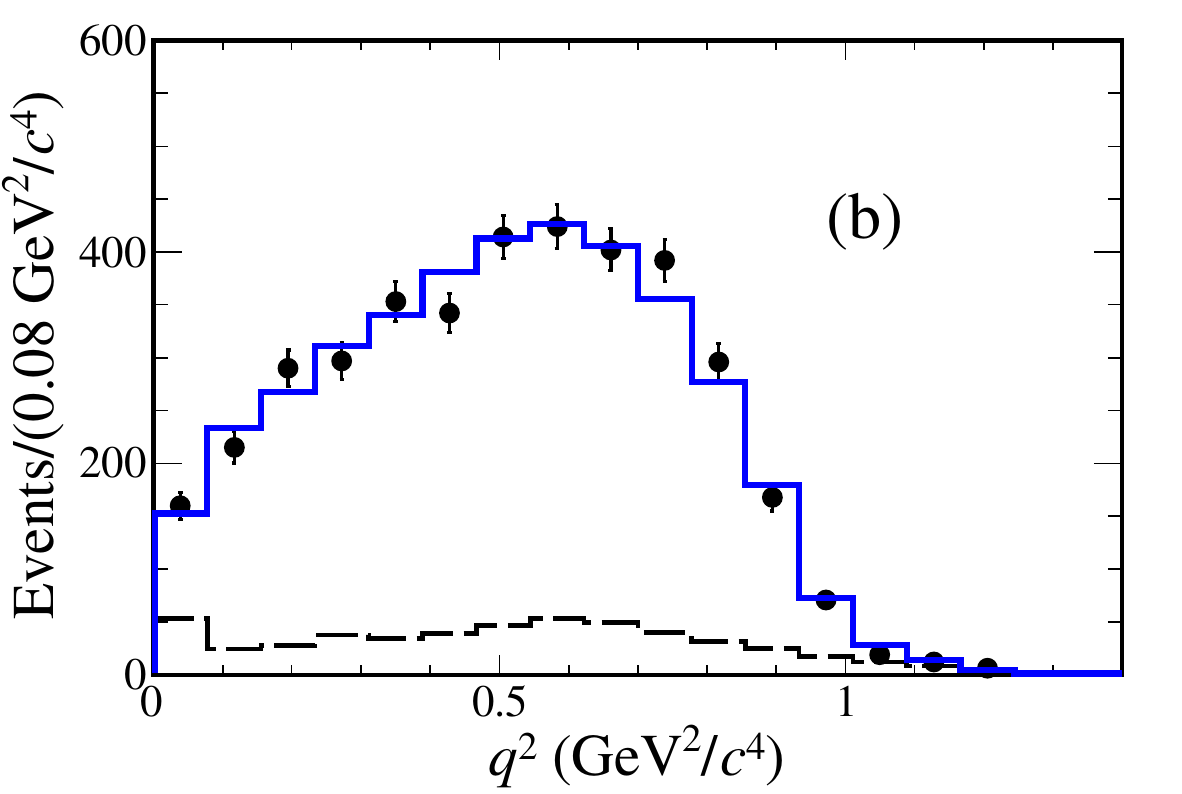}\\
       \includegraphics[width=0.25\linewidth]{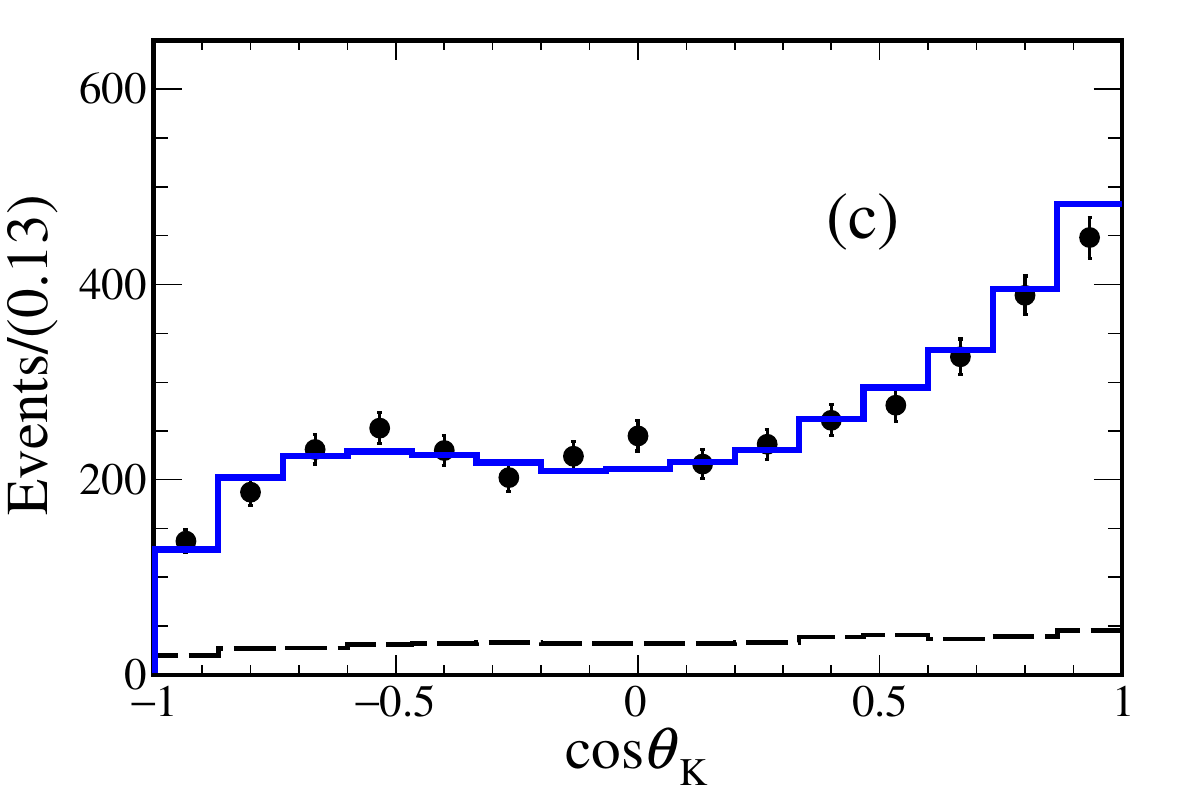}
      \includegraphics[width=0.25\linewidth]{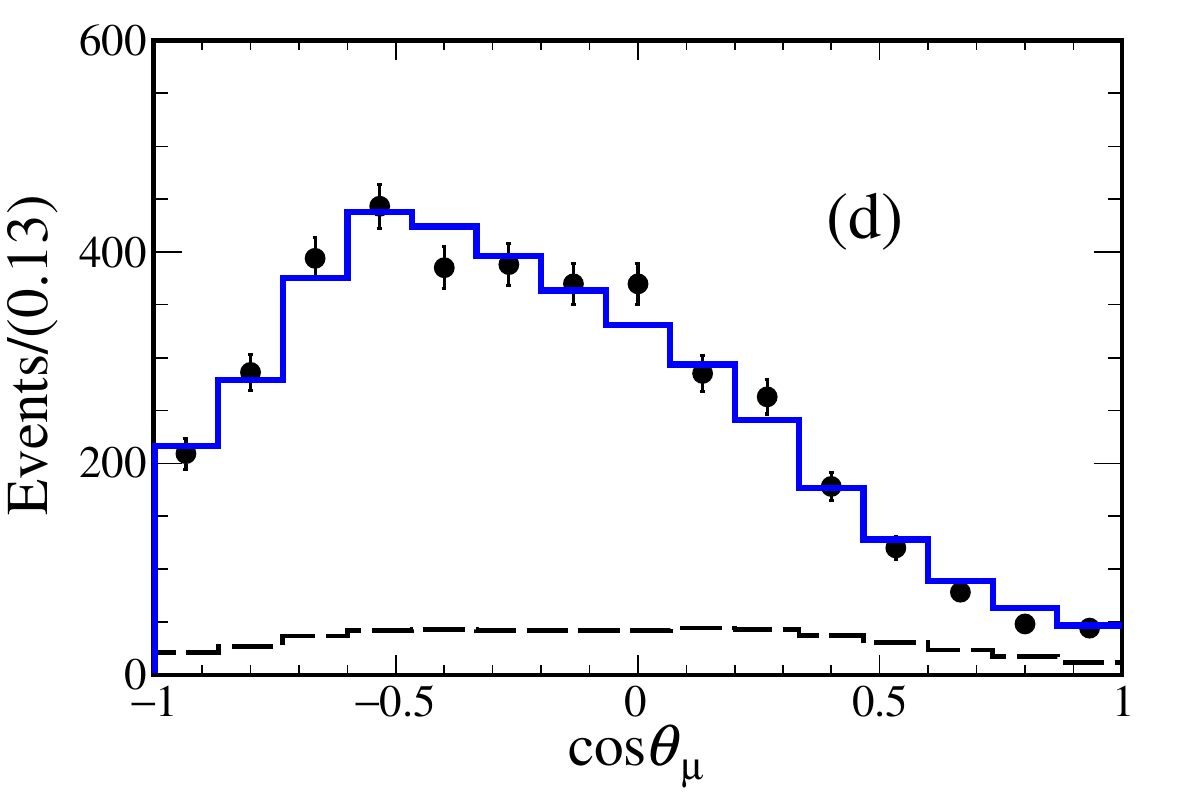}
      \includegraphics[width=0.25\linewidth]{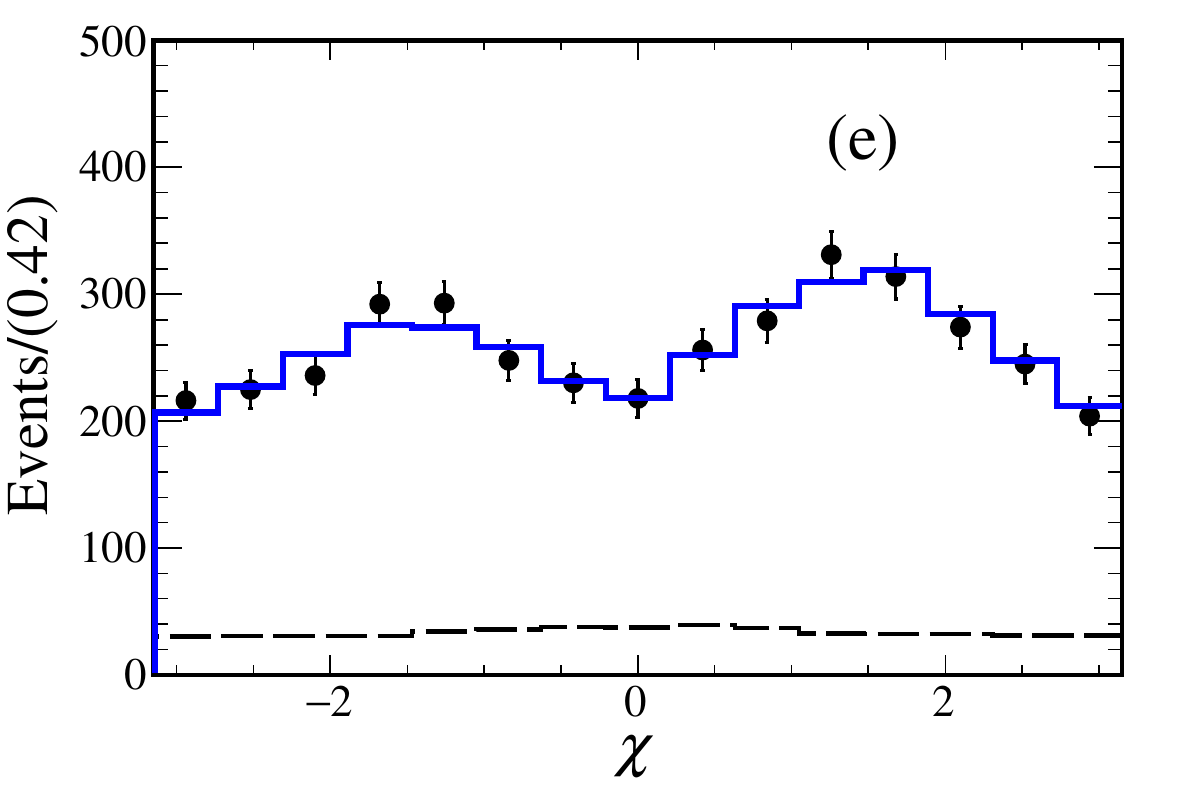}
   \caption{
 The projections of amplitude analysis for $D^0\to K^{-}\pi^0\mu^+\nu_\mu$ on $M_{K^-\pi^0}$, $q^2$, $\cos\theta_{K^-}$, $\cos\theta_\ell$, and $\chi$~\cite{BESIII:2024qnx}.
The dots with error bars are data,  the blue lines are the fit results, and the dashed lines show the sum of the simulated background contributions.}
\label{fig:fig:D0_Kpi0munu}
\end{figure*}

In 2025, the decay $D^0\to K^-\pi^0\mu^+\nu_{\mu}$ was studied for the first time with 6.4k signal events using $7.9~\mathrm{fb}^{-1}$ of data at 3.773~GeV~\cite{BESIII:2024qnx}.
The projected distributions of the fit onto the fitted variables are shown in Fig.~\ref{fig:fig:D0_Kpi0munu}.
The first amplitude analysis is performed, and an $\cal S$-wave component is observed with a fraction $f_{\cal S {\rm-wave}}=(5.76 \pm 0.35 \pm 0.29)\%$, yielding $\mathcal{B}(D^0\to (K^-\pi^0)_{\cal S {\rm-wave}}\mu^+\nu_\mu)=(4.223 \pm 0.268 \pm 0.222)\times 10^{-4}$. The dominant ${\cal P}$-wave component is observed with a fraction $f_{K^{*}(892)^-}=(94.24 \pm 0.35\pm 0.29)\%$, leading to $\mathcal{B}(D^0\to K^{*}(892)^-\mu^+\nu_\mu)=(2.073\pm0.039 \pm 0.032)\%$ after accounting for $\mathcal{B}(K^{*}(892)^-\to K^-\pi^0)=1/3$. This result is consistent with previous measurements and improves the precision by a factor of 5 compared to the world average. 
The hadronic transition form factor ratios are determined to be $r_V=1.37 \pm 0.09 \pm 0.03$ and $r_2=0.76 \pm 0.06 \pm 0.02$.

\begin{figure*}[htp]
    \centering
    \includegraphics[width=0.25\textwidth]{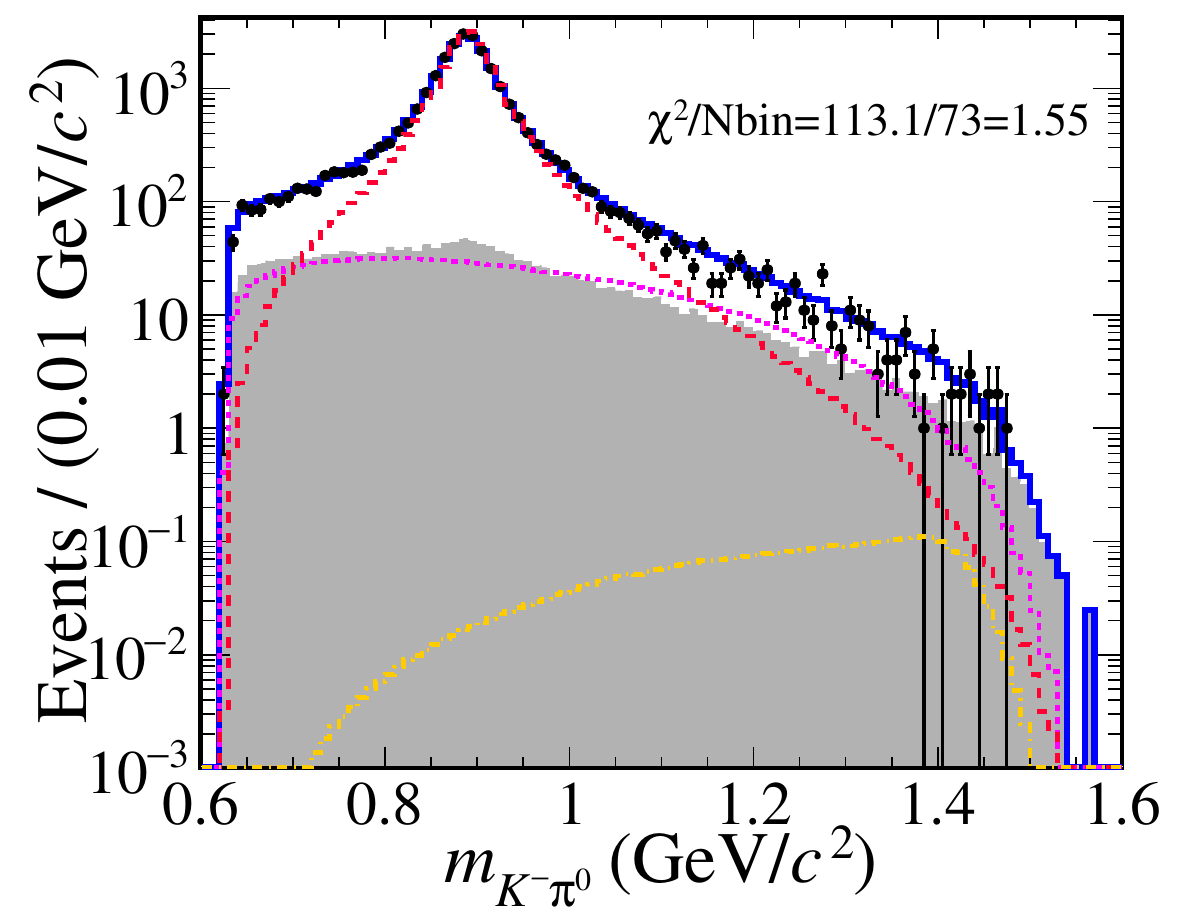}
    \includegraphics[width=0.25\textwidth]{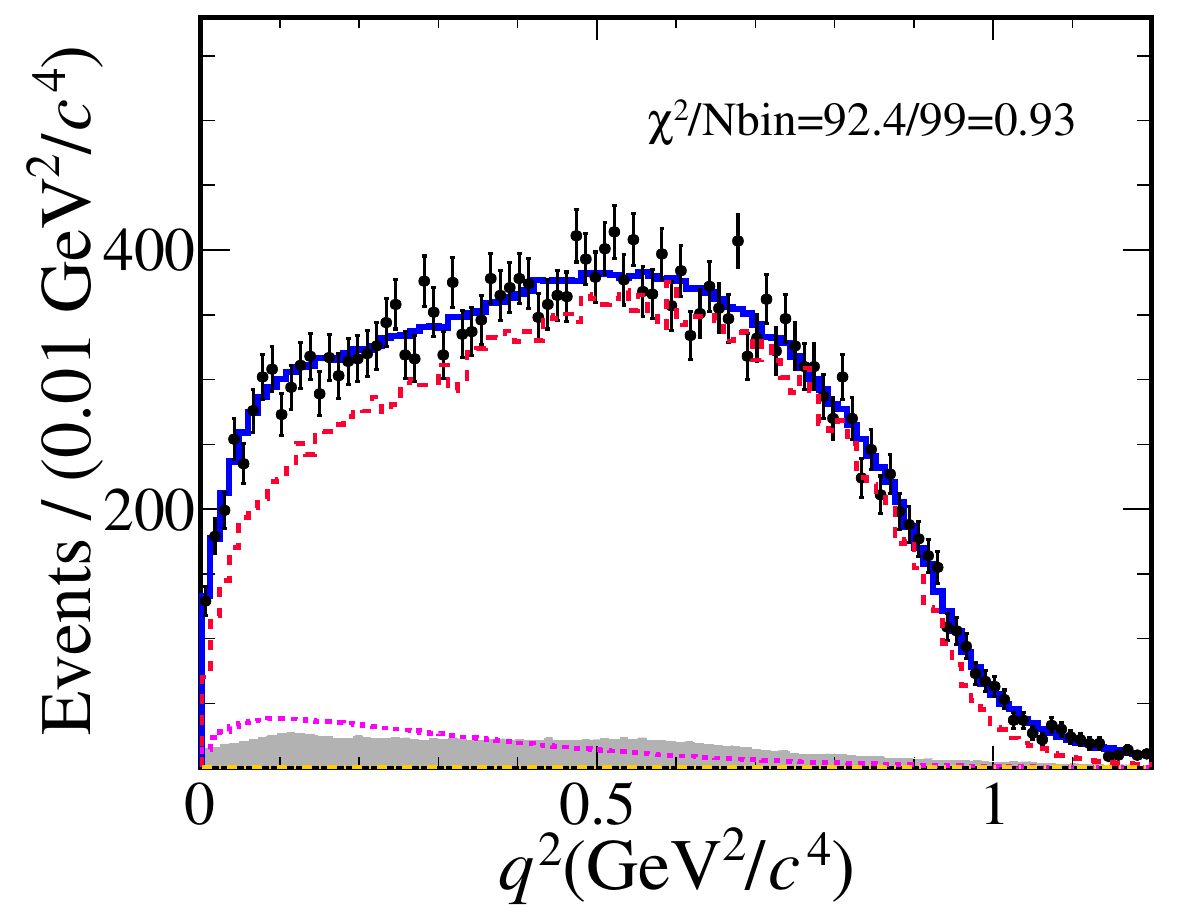}
    \includegraphics[width=0.25\textwidth]{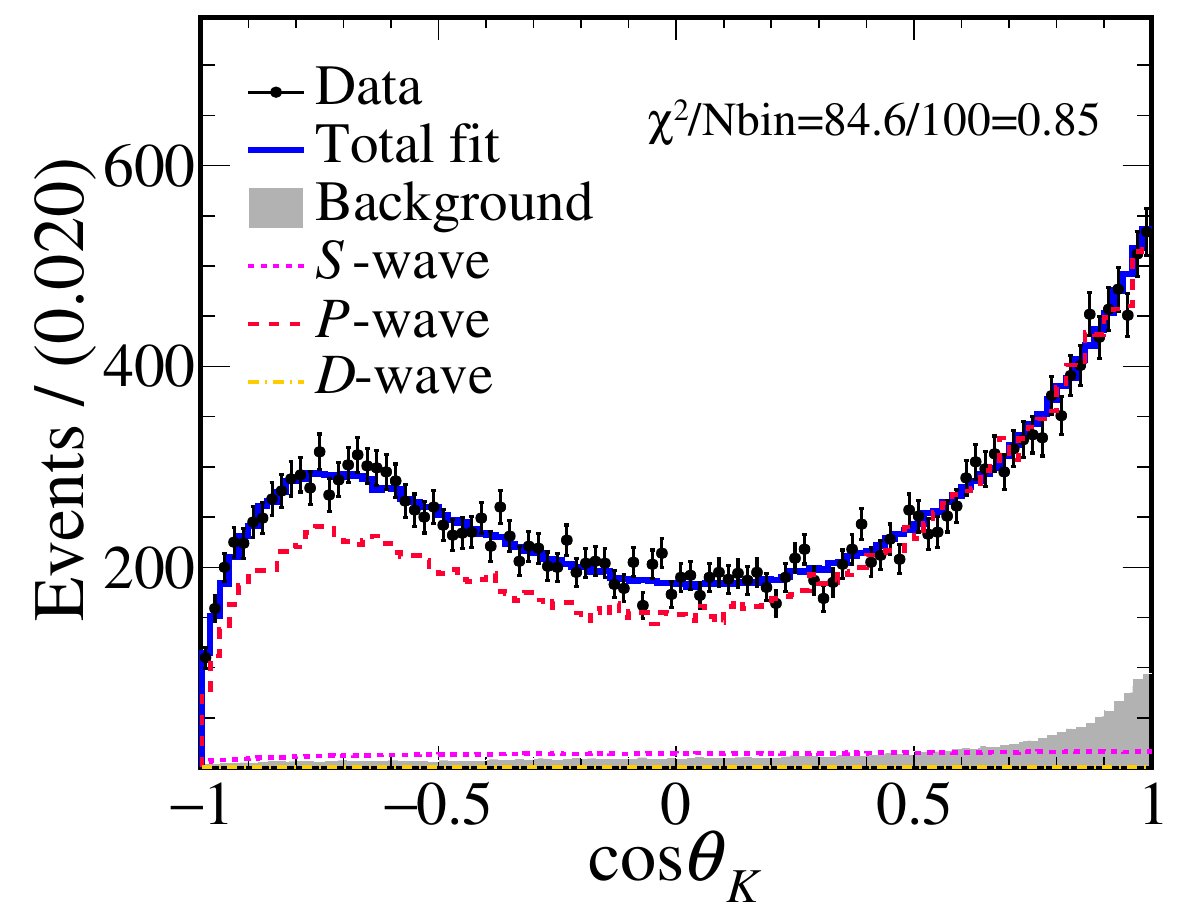}
    \includegraphics[width=0.25\textwidth]{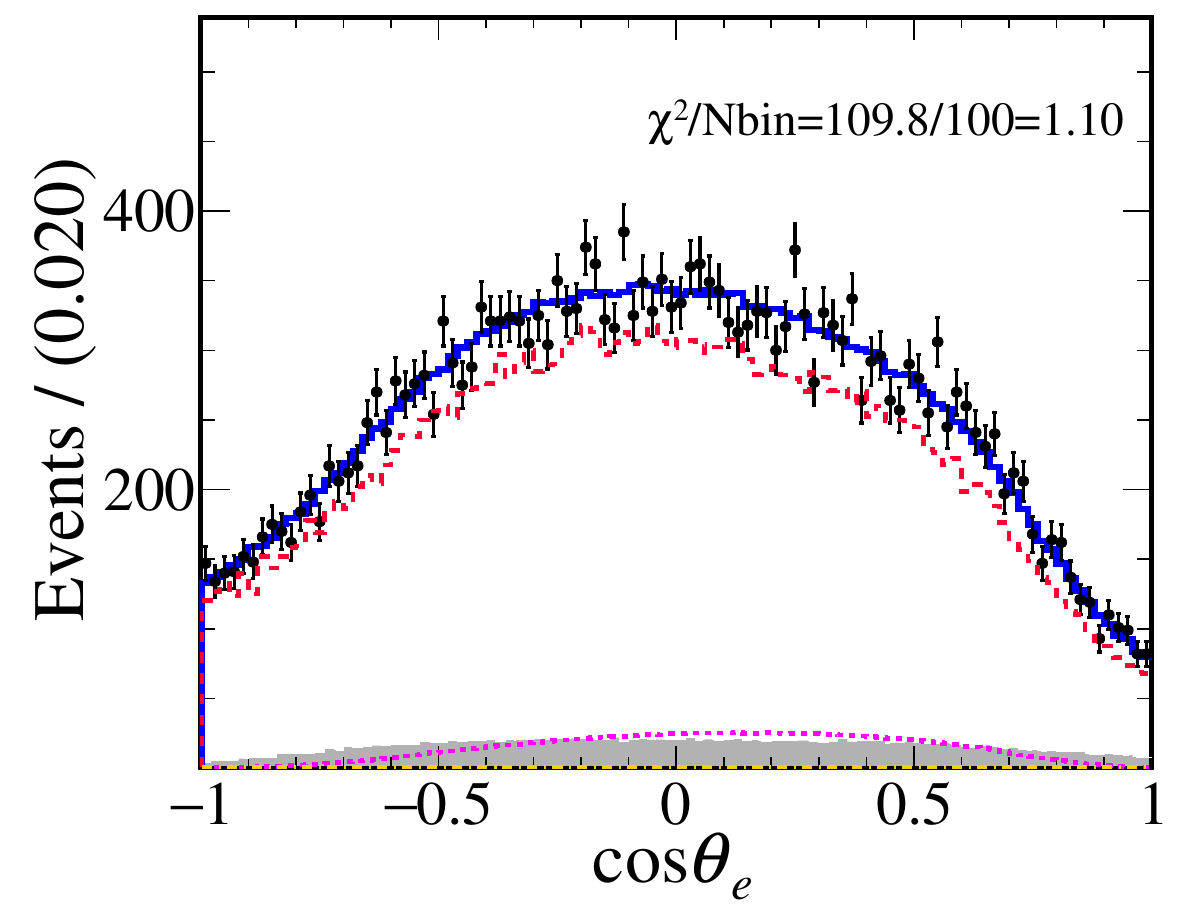}
    \includegraphics[width=0.25\textwidth]{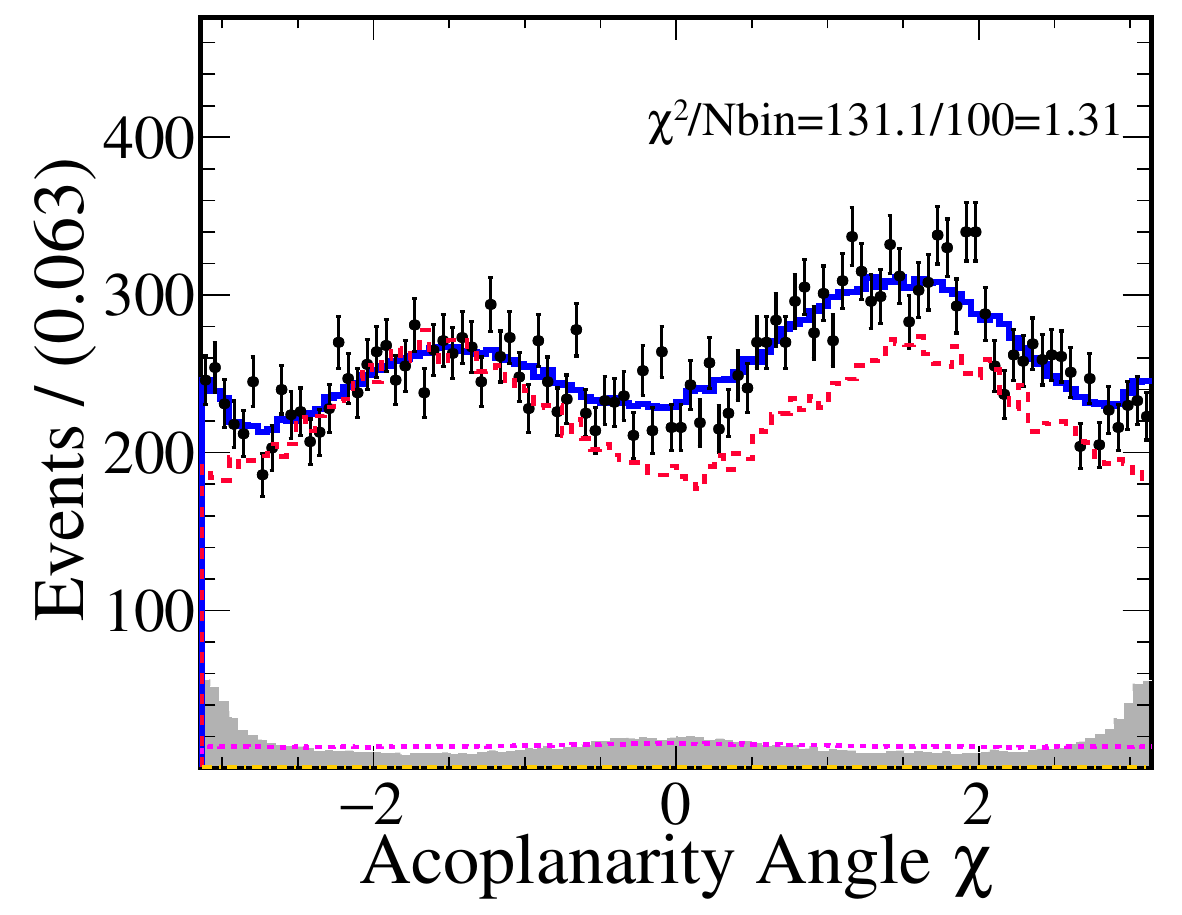}
    \caption{
 The projections of amplitude analysis for $D^0\to K^-\pi^0 e^+\nu_e $ on $M_{K^-\pi^0}$, $q^2$, $\cos\theta_{K^-}$, $\cos\theta_\ell$, and $\chi$~\cite{BESIII:2026ssp}.
    }
    \label{fit:pwa:kpienu}
    \end{figure*}

Very recently, the first amplitude analysis of $D^0\to K^-\pi^0 e^{+}\nu_e$ was reported with 28.9k signal events using the full 20.3~fb$^{-1}$ of data at 3.773~GeV~\cite{BESIII:2026ssp}.  A tiny $\cal D$-wave component from $K^*_2(1430)^-$, accounting for $(0.16\pm0.05\pm0.02)\%$ of the $K^-\pi^0$ system, is observed for the first time with a significance of $7.9\sigma$, alongside the dominant $K^*(892)^-$ ${\cal P}$-wave and the sub-dominant $\cal S$-wave. The hadronic transition form factors of the $D^0 \to K^*(892)^-$ transition are measured precisely as $r_V=1.41\pm0.05\pm0.01$ and $r_2=0.77\pm0.04\pm0.02$. The branching fraction of $D^0\to K^*(892)^-e^+\nu_e$ with $K^*(892)^-\to K^-\pi^0$ is measured to be $(7.403\pm0.061\pm0.048)\times10^{-3}$. Combining measurements of $D^0\to K^*(892)^-\ell^+ \nu_\ell, K^*(892)^-\to K^-\pi^0$~\cite{BESIII:2026ssp,BESIII:2024qnx}, lepton flavor universality is tested with the ratio $\mathcal{R}_{\rm LFU}=\mathcal{B}(D^0\to K^*(892)^-\mu^+ \nu_\mu)/\mathcal{B}(D^0\to K^*(892)^-e^+\nu_e)=0.933\pm0.019\pm0.012$, achieving unprecedented precision and showing no violation. Isospin symmetry in $\bar K^*(892) \to \bar K\pi$ decays is tested for the first time via $\mathcal R_{K^{-}} =\mathcal{B}(K^*(892)^-\to K^- \pi^0)/\mathcal{B}(K^*(892)^-\to K_S^0 \pi^-)= 1.09\pm0.02\pm0.02$, using previous measurements of $D^0\to K^*(892)^-e^+\nu_e$ with $K^*(892)^-\to K^0_S\pi^-$. Finally, the phase shift of the $K^-\pi^0$ $\cal S$-wave is extracted in a model-independent way, providing insights into the nature of the lightest strange scalar meson, the $K^*_0(700)$. The projections of amplitude analysis for $D^0\to K^-\pi^0 e^+\nu_e $ on $M_{K^-\pi^0}$, $q^2$, $\cos\theta_{K^-}$, $\cos\theta_\ell$, and $\chi$ are shown in Fig.~\ref{fit:pwa:kpienu}.

\begin{figure*}[htbp]
  \centering
  \includegraphics[width=0.3\textwidth]{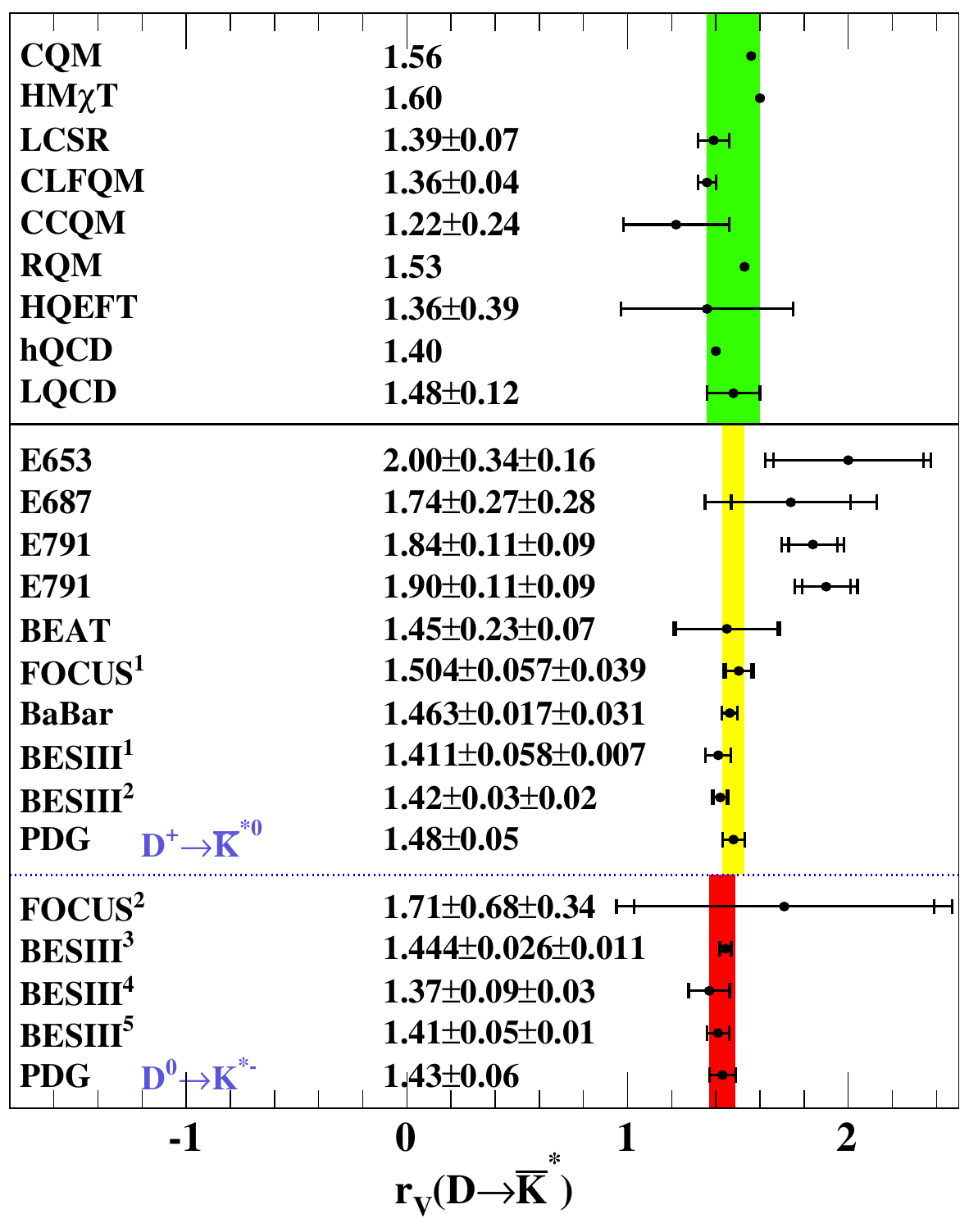}
  \put(-112,184.5){\tiny~\cite{Soni:2017eug}}
  \put(-112,178.0){\tiny~\cite{Fajfer:2005ug}}
  \put(-112,171.0){\tiny~\cite{Wu:2006rd}}
  \put(-112,165.){\tiny~\cite{Cheng:2017pcq}}
  \put(-112,158.){\tiny~\cite{Ivanov:2019nqd}}
  \put(-112,151.5){\tiny~\cite{Faustov:2019mqr}}
  \put(-112,144.5){\tiny~\cite{Wang:2002zba}}
  \put(-112,137.5){\tiny~\cite{Ahmed:2023pod}}
  \put(-112,131.5){\tiny~\cite{Abada:2002ie}}
  \put(-112,120.5){\tiny~\cite{FermilabE653:1992wyb}}
  \put(-112,114.){\tiny~\cite{E687:1993qiw}}
  \put(-112,107.5){\tiny~\cite{E791:1998wih}}
  \put(-112,101.){\tiny~\cite{E791:1997bnl}}
  \put(-112,94.0){\tiny~\cite{BEATRICE:1998hbq}}
  \put(-112,88.0){\tiny~\cite{FOCUS:2002lsy}}
  \put(-112,81.0){\tiny~\cite{BaBar:2010vmf}}
  \put(-112,74.5){\tiny~\cite{BESIII:2015hty}}
  \put(-112,67.5){\tiny~\cite{BESIII:2025fso}}
  \put(-112,61.0){\tiny~\cite{ParticleDataGroup:2024cfk}}
  \put(-112,49.5){\tiny~\cite{FOCUS:2004zbs}}
  \put(-112,43){\tiny~\cite{BESIII:2026txt}}
  \put(-112,36.5){\tiny~\cite{BESIII:2024qnx}}
  \put(-112,30){\tiny~\cite{BESIII:2026ssp}}
  \put(-112,23.5){\tiny~\cite{ParticleDataGroup:2024cfk}} 
  \includegraphics[width=0.3\textwidth]{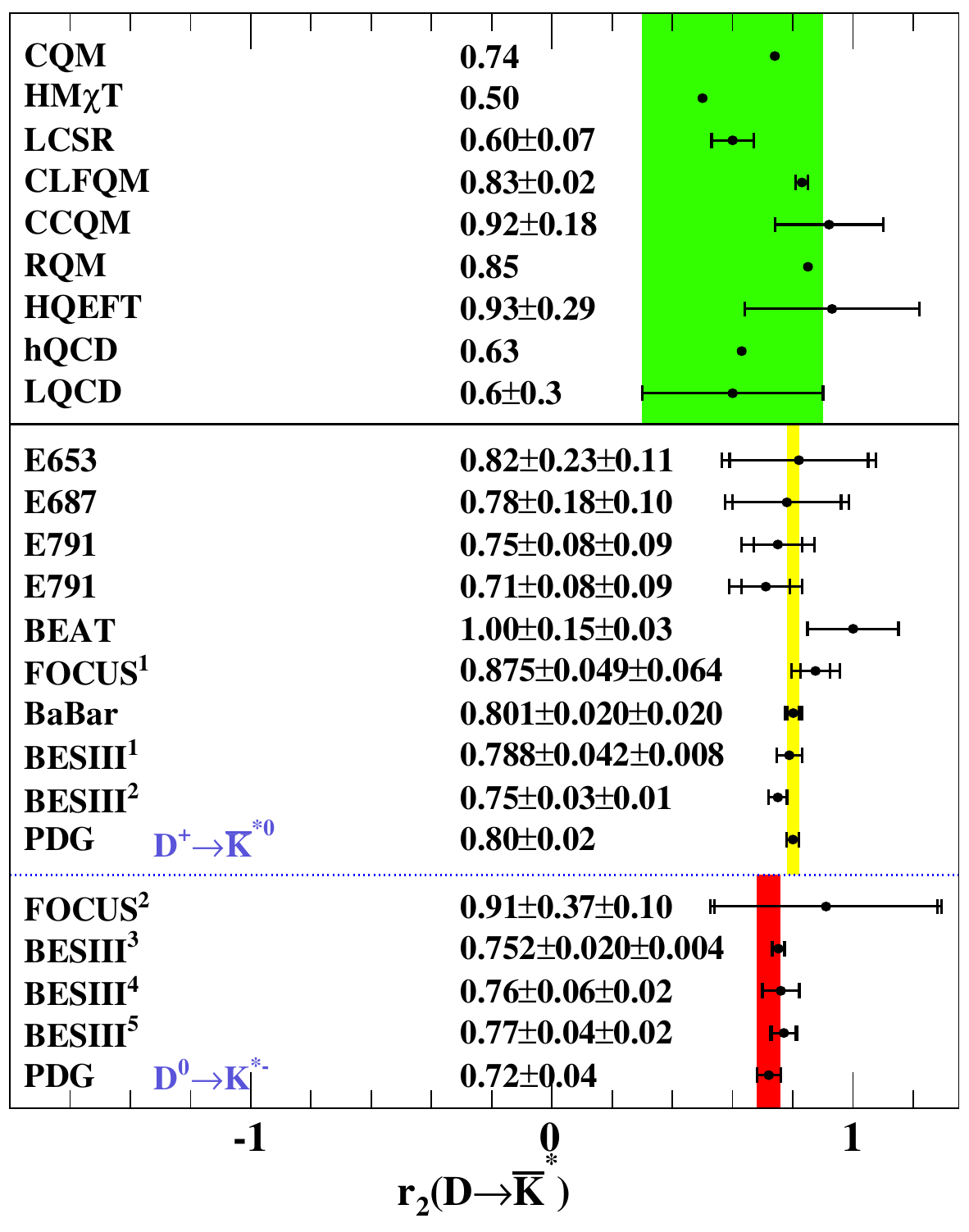}
  \put(-108,184.5){\tiny~\cite{Soni:2017eug}}
  \put(-108,178.0){\tiny~\cite{Fajfer:2005ug}}
  \put(-108,171.0){\tiny~\cite{Wu:2006rd}}
  \put(-108,165.){\tiny~\cite{Cheng:2017pcq}}
  \put(-108,158.){\tiny~\cite{Ivanov:2019nqd}}
  \put(-108,151.5){\tiny~\cite{Faustov:2019mqr}}
  \put(-108,144.5){\tiny~\cite{Wang:2002zba}}
  \put(-108,137.5){\tiny~\cite{Ahmed:2023pod}}
  \put(-108,131.5){\tiny~\cite{Abada:2002ie}}
  \put(-108,120.5){\tiny~\cite{FermilabE653:1992wyb}}
  \put(-108,114.){\tiny~\cite{E687:1993qiw}}
  \put(-108,107.5){\tiny~\cite{E791:1998wih}}
  \put(-108,101.){\tiny~\cite{E791:1997bnl}}
  \put(-108,94.0){\tiny~\cite{BEATRICE:1998hbq}}
  \put(-108,88.0){\tiny~\cite{FOCUS:2002lsy}}
  \put(-108,81.0){\tiny~\cite{BaBar:2010vmf}}
  \put(-108,74.5){\tiny~\cite{BESIII:2015hty}}
  \put(-108,67.5){\tiny~\cite{BESIII:2025fso}}
  \put(-108,61.0){\tiny~\cite{ParticleDataGroup:2024cfk}}
  \put(-108,49.5){\tiny~\cite{FOCUS:2004zbs}}
  \put(-108,43){\tiny~\cite{BESIII:2026txt}}
  \put(-108,36.5){\tiny~\cite{BESIII:2024qnx}}
  \put(-108,30){\tiny~\cite{BESIII:2026ssp}}
  \put(-108,23.5){\tiny~\cite{ParticleDataGroup:2024cfk}} 
  \includegraphics[width=0.3\textwidth]{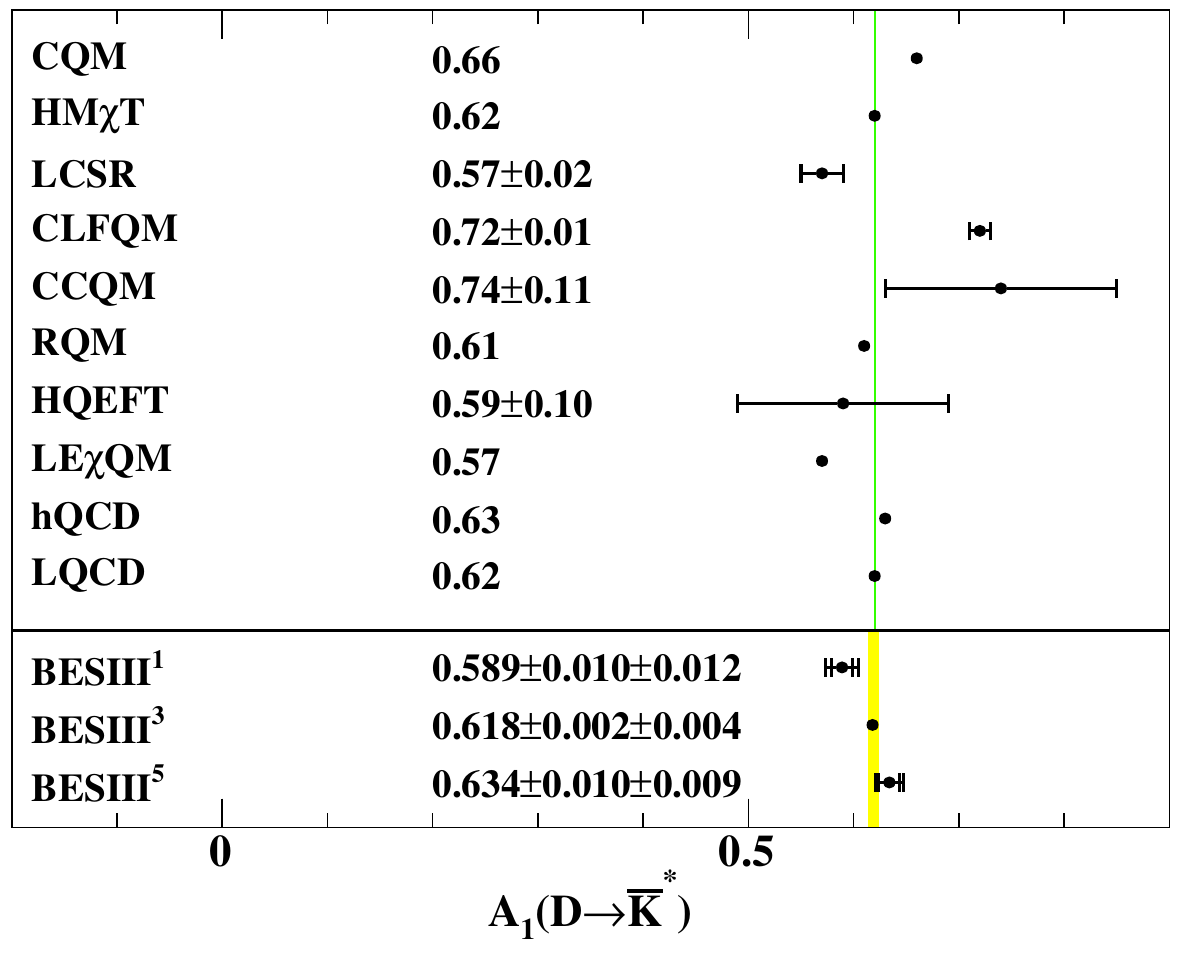}
  \put(-127,118.0){\tiny~\cite{Soni:2017eug}}
  \put(-127,110.5){\tiny~\cite{Fajfer:2005ug}}
  \put(-127,102.0){\tiny~\cite{Wu:2006rd}}
  \put(-127,94.5){\tiny~\cite{Cheng:2017pcq}}
  \put(-127,87.0){\tiny~\cite{Ivanov:2019nqd}}
  \put(-127,79.5){\tiny~\cite{Faustov:2019mqr}}
  \put(-127,72.0){\tiny~\cite{Wang:2002zba}}
  \put(-127,64.0){\tiny~\cite{Palmer:2013yia}}
  \put(-127,56.5){\tiny~\cite{Ahmed:2023pod}}
  \put(-127,49.0){\tiny~\cite{Abada:2002ie}}
  \put(-127,37.5){\tiny~\cite{BESIII:2015hty}}
  \put(-127,30.0){\tiny~\cite{BESIII:2026txt}}
  \put(-127,22.5){\tiny~\cite{BESIII:2026ssp}}
   \caption{Comparisons of the (left) $r_V$, (middle) $r_2$, and (right) $A_1(0)$ for $D^{0(+)}\to \bar K^*$
   measured by experimental measurements
($D^+\to\bar K^*$ from
  E653~\cite{FermilabE653:1992wyb},
  E687~\cite{E687:1993qiw},
  E791~\cite{E791:1998wih},
  E791~\cite{E791:1997bnl},
  BEAT~\cite{BEATRICE:1998hbq},
  FOCUS$^1$~\cite{FOCUS:2002lsy},
  BaBar~\cite{BaBar:2010vmf},
  BESIII$^1$ ($D^+\to K^-\pi^+e^+\nu_e$ based on 2.93 fb$^{-1}$ of data at 3.773 GeV)~\cite{BESIII:2015hty},
  BESIII$^2$ ($D^+\to K^0_S\pi^0\ell^+\nu_\ell$ based on 20.3 fb$^{-1}$ of data at 3.773 GeV)~\cite{BESIII:2025fso};
  and
$D^0\to K^{*}(892)^-$ from
FOCUS$^2$~\cite{FOCUS:2004zbs},
BESIII$^3$ ($D^0\to \bar K^0\pi^-\ell^+\nu_\ell$ based on 20.3 fb$^{-1}$ of data at 3.773 GeV)~\cite{BESIII:2026txt},
BESIII$^4$ ($D^0\to K^-\pi^0\mu^+\nu_\mu$ based on 7.9 fb$^{-1}$ of data at 3.773 GeV)~\cite{BESIII:2024qnx}, and
BESIII$^5$ ($D^0\to K^-\pi^0e^+\nu_e$ based on 20.3 fb$^{-1}$ of data at 3.773 GeV)~\cite{BESIII:2026ssp}.)
   and those predicted by theoretical calculations of
  CQM~\cite{Soni:2017eug},
  HM$\chi$T~\cite{Fajfer:2005ug},
  LCSR~\cite{Wu:2006rd},
  CLFQM~\cite{Cheng:2017pcq},
  CCQM~\cite{Ivanov:2019nqd},
  RQM~\cite{Faustov:2019mqr},
  HQEFT~\cite{Wang:2002zba},
  LE$\chi$QM~\cite{Palmer:2013yia},
  hQCD~\cite{Ahmed:2023pod}, and
  LQCD~\cite{Abada:2002ie}. The green bands are the $\pm 1\sigma$ region of the LQCD result~\cite{Abada:2002ie}. The yellow and red bands for $r_V$ and $r_2$ denote the $\pm 1\sigma$ region of the PDG averaged results for $D^+\to \bar K^{*}(892)^0$ and $D^0\to K^{*}(892)^-$, respectively. The yellow band for $A_1$ is the $\pm 1\sigma$ region of the BESIII result~\cite{BESIII:2026txt}.
}
  \label{fig:fD2Kst}
\end{figure*}

Figure~\ref{fig:fD2Kst} shows comparisons of the $r_V$, $r_2$, and $A_1(0)$  of $D^{0(+)} \to\bar K^*$ measured by different experiments and those predicted by theoretical calculations. In some measurements of the $D \to V \ell^+ \nu$ hadronic transition form factors, certain parameters such as $m_{\bar K^*}$, $\gamma_{\bar K^*}$, $r_{\text{BW}}$, $m_V$, and $m_A$, or a subset of them are fixed, thereby achieving better precision with the same statistical sample size. Therefore, we do not apply any weighting to the $D \to V$ hadronic transition form factors.

\begin{figure*}[htp]
  \centering
\includegraphics[width=0.25\textwidth]{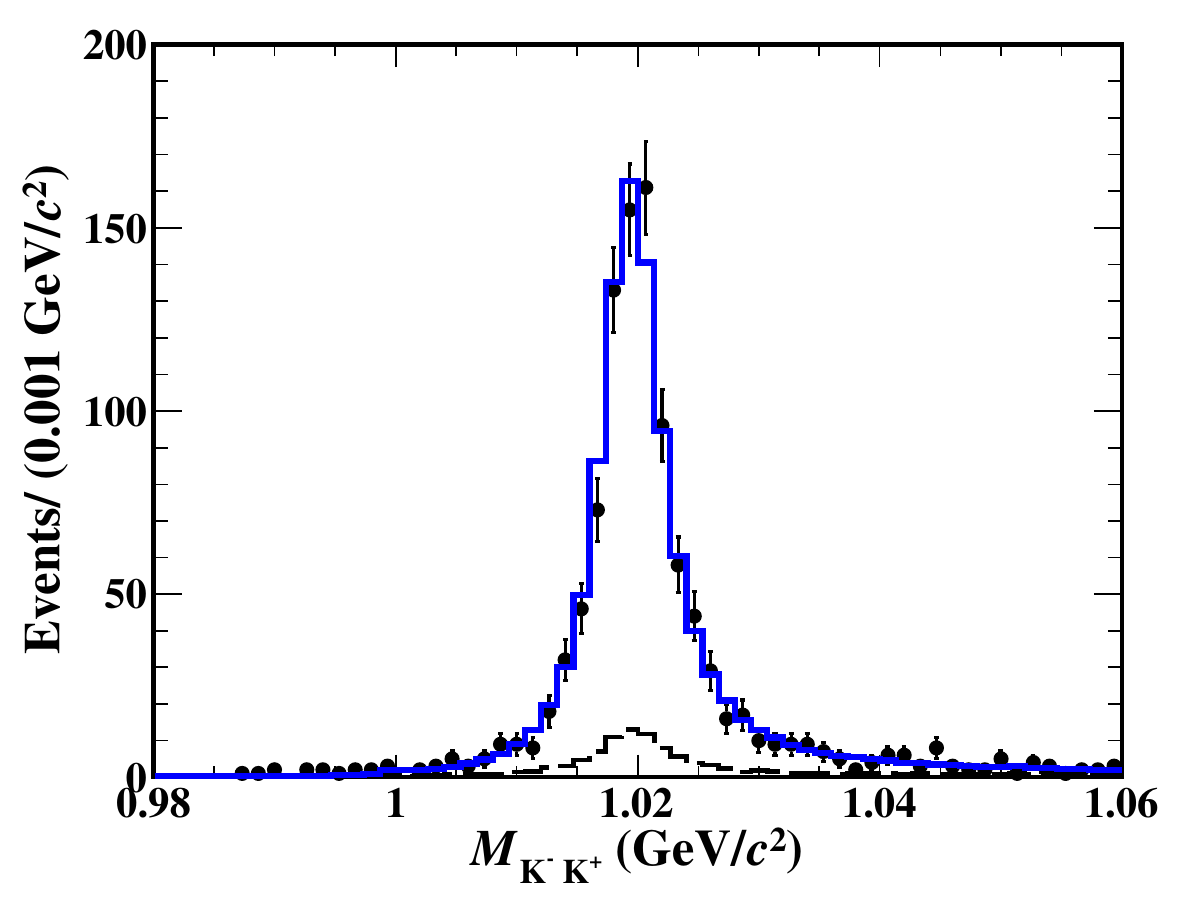}
\includegraphics[width=0.25\textwidth]{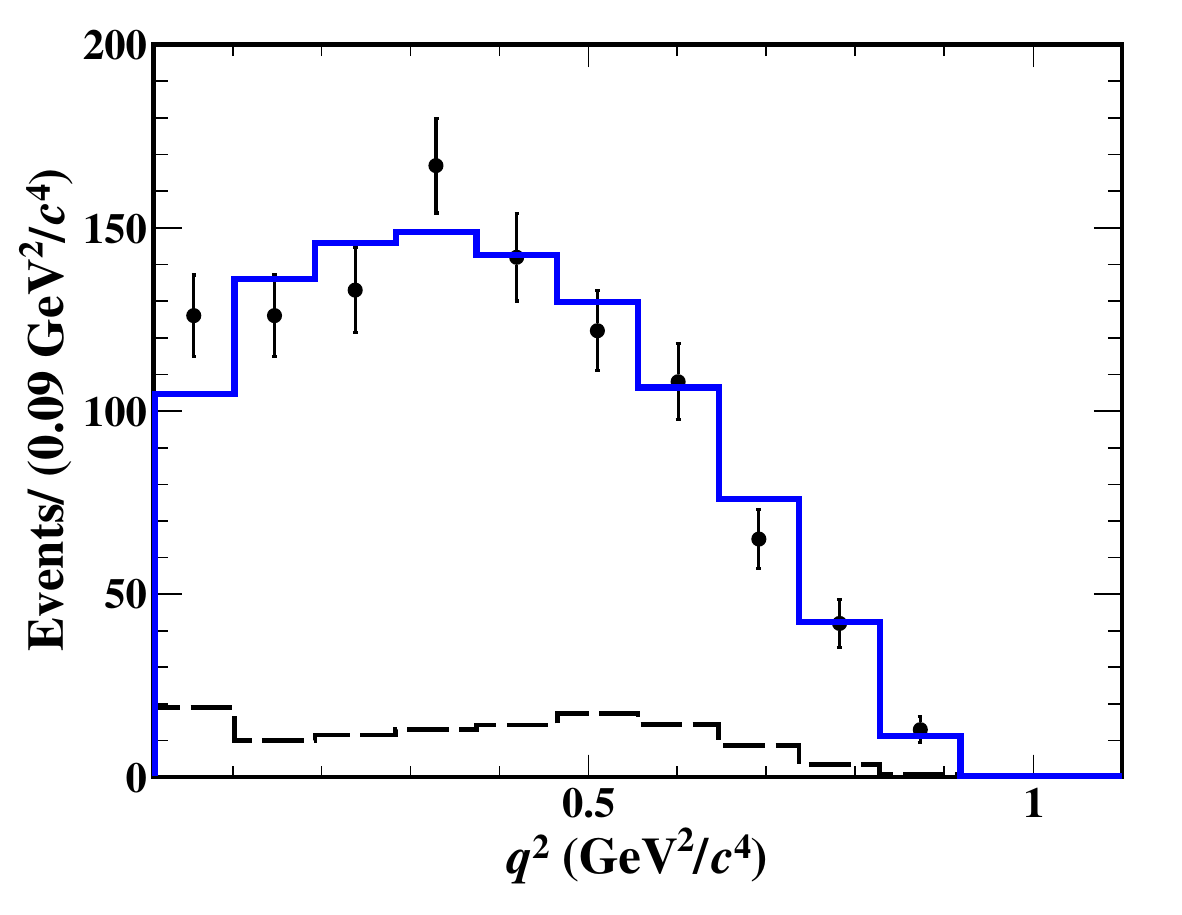} \\
\includegraphics[width=0.25\textwidth]{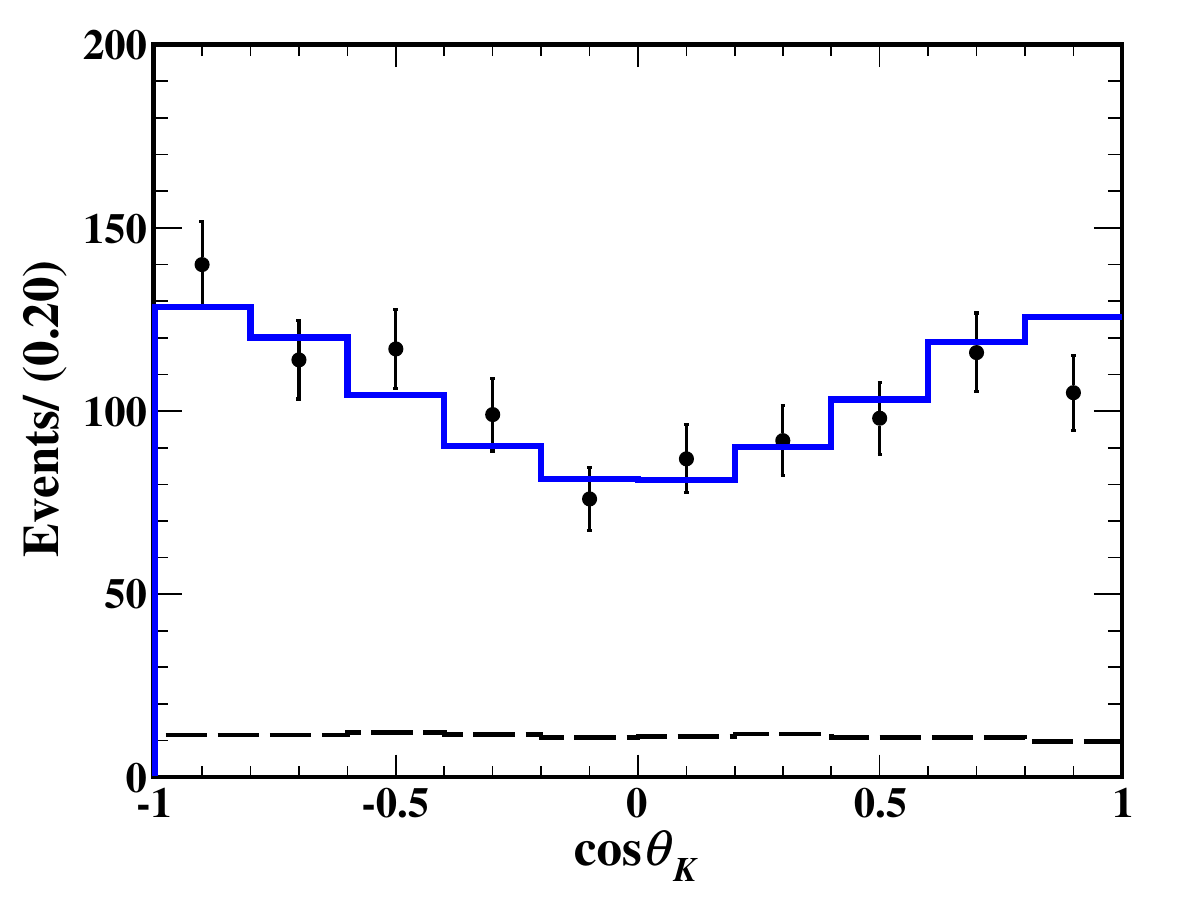}
\includegraphics[width=0.25\textwidth]{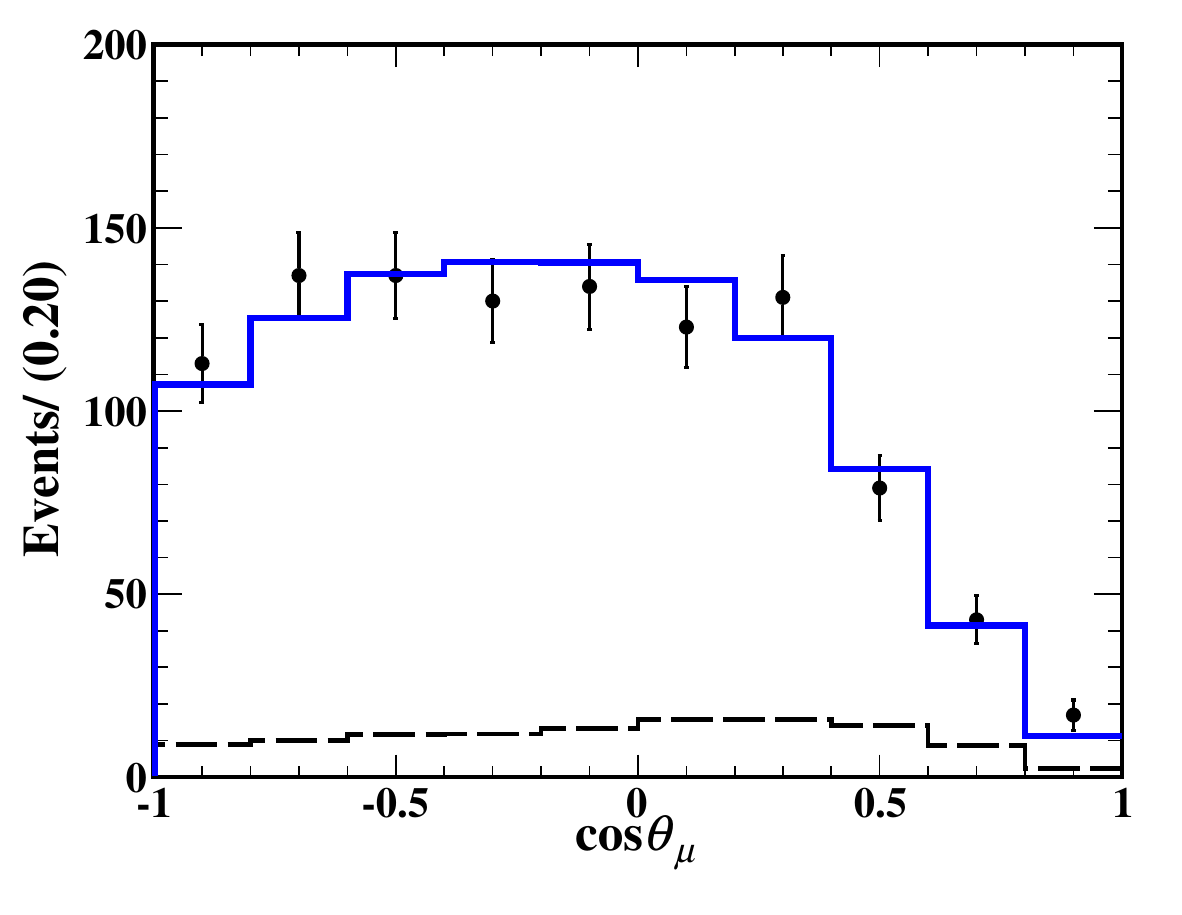}
\includegraphics[width=0.25\textwidth]{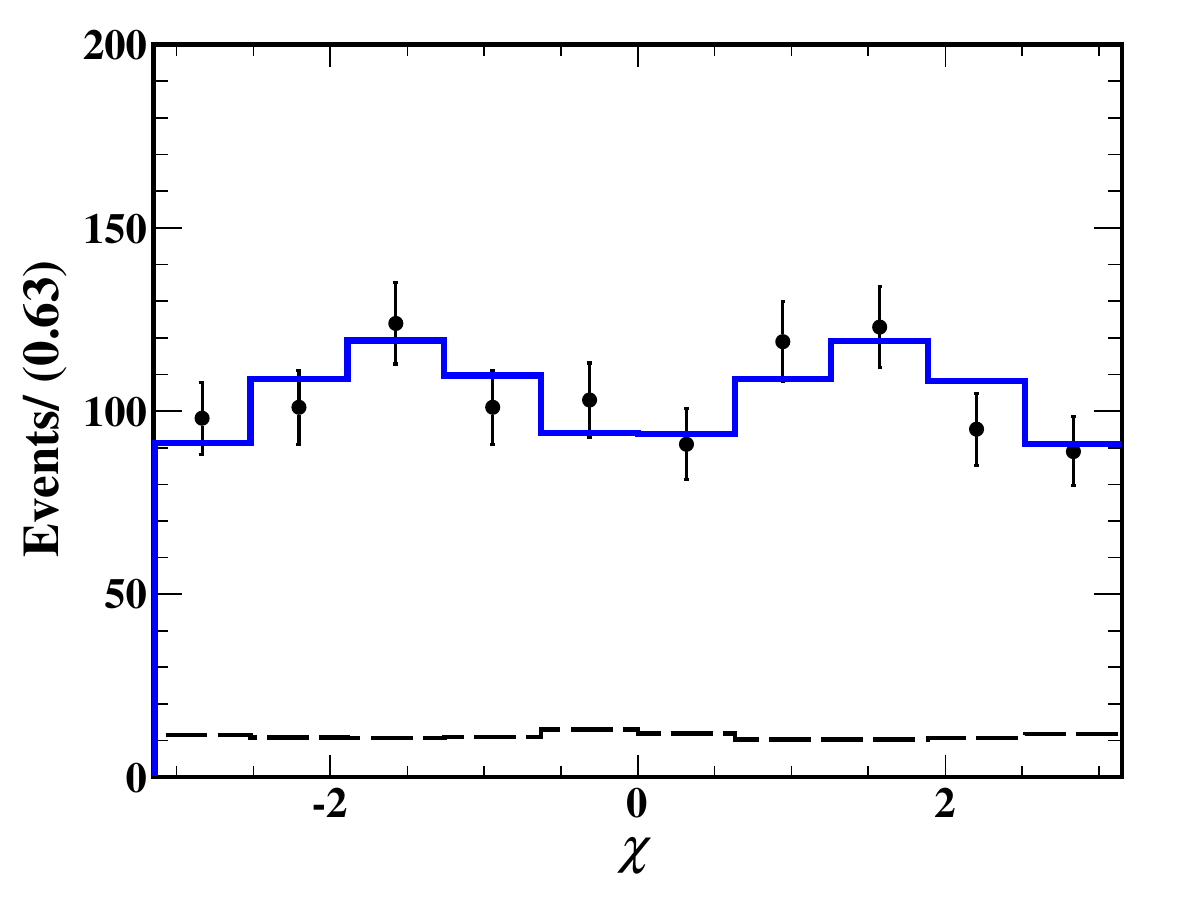}\\
\caption{
 The projections of amplitude analysis for  $D^+_s\to K^+K^-\mu^+\nu_\mu$ on $M_{K^+K^-}$, $q^2$, $\cos\theta_{K}$, $\cos\theta_\ell$, and $\chi$~\cite{BESIII:2023opt}.
The dots with error bars  are data, the blue lines are the best fit, and the dashed lines show the sum of the simulated background contributions.
}
\label{fig:Ds_KKmunu}
\end{figure*}

Using 7.33 fb$^{-1}$ of data at 4.128-4.226~GeV, the decay $D^+_s\to K^+K^-\mu^+\nu_\mu$ is studied with 1.7k signal events~\cite{BESIII:2023opt}. Its branching fraction is measured as ${\cal B}(D^+_s\to \phi \mu^+\nu_\mu) = (2.25\pm 0.09 \pm 0.07) \times10^{-2}$, the most precise measurement to date. Combining with the world average of ${\cal B}(D^+_s\to \phi e^+\nu_e)$ yields the ratio $\frac{{\cal B}(D^+_s\to \phi \mu^+\nu_\mu)}{{\cal B}(D^+_s\to \phi e^+\nu_e)} = 0.94\pm0.08$, consistent with lepton flavor universality.
An amplitude analysis, in which the projections of the five kinematic variables for the data are shown in Fig.~\ref{fig:Ds_KKmunu},
extracts the hadronic transition form factor ratios as $r_{V}=1.58\pm0.17\pm0.02$ and $r_{2}=0.71\pm0.14\pm0.02$. No significant $\cal S$-wave contribution from $f_0(980)\to K^+K^-$ is found, and an upper limit at the 90\% confidence level is set as $\mathcal{B}(D_s^+\to f_0(980)\mu^{+}{\nu}_{\mu}) \cdot{\cal B}(f_0(980)\to K^+K^-) < 5.45 \times 10^{-4}$.

\begin{figure*}[htbp]
  \centering
  \includegraphics[width=0.4\textwidth]{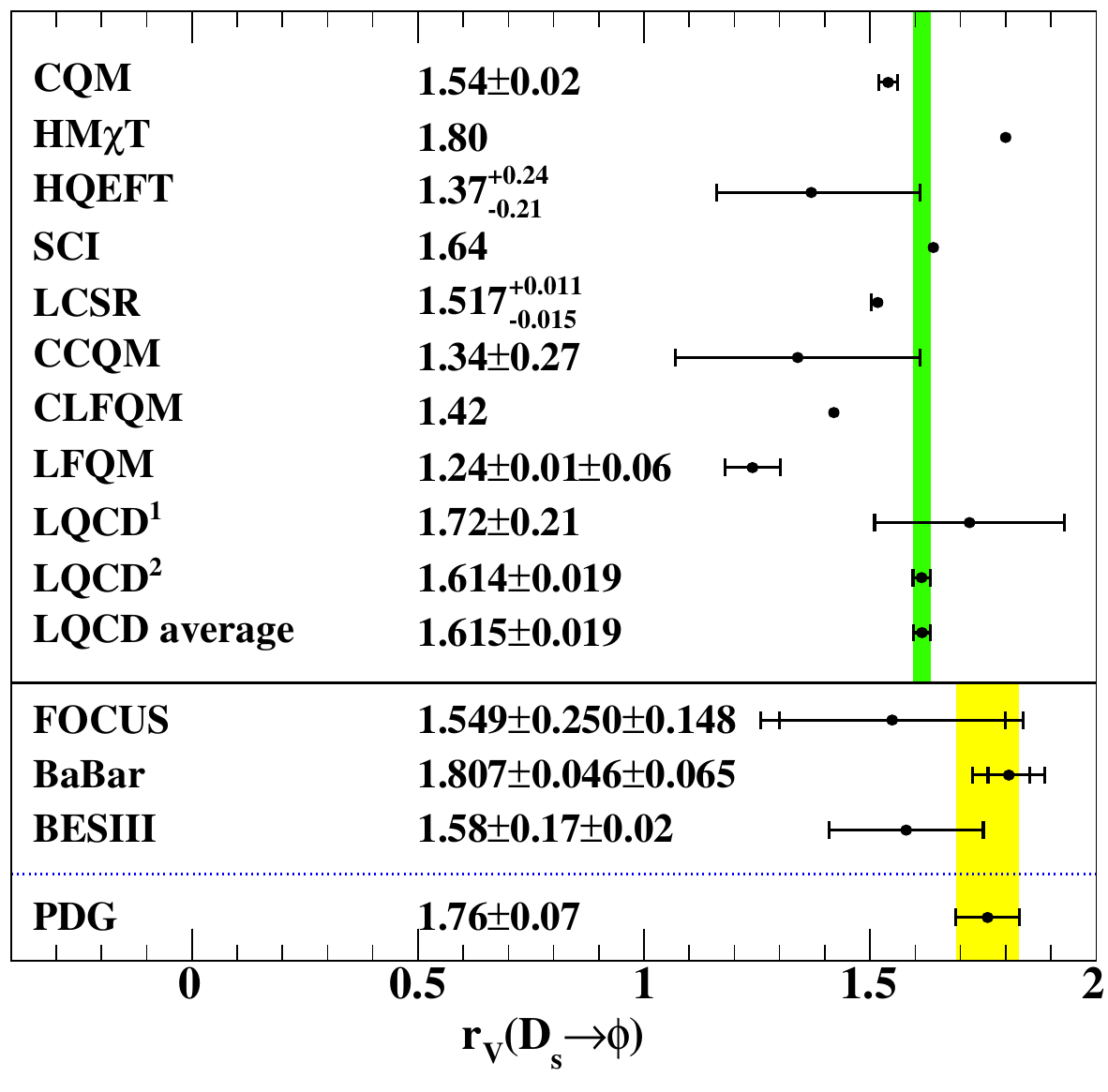}
  \put(-165,185){\tiny~\cite{Yang:2025gfz}}
  \put(-165,174.778){\tiny~\cite{Fajfer:2005ug}}
  \put(-165,164.556){\tiny~\cite{Wu:2006rd}}
  \put(-165,154.333){\tiny~\cite{Xing:2022sor}}
  \put(-165,144.111){\tiny~\cite{Wang:2025oix}}
  \put(-165,133.889){\tiny~\cite{Soni:2018adu}}
  \put(-165,123.667){\tiny~\cite{Verma:2011yw}}
  \put(-165,113.444){\tiny~\cite{Chang:2019mmh}}
  \put(-165,103.222){\tiny~\cite{Donald:2013pea}}
  \put(-165,93){\tiny~\cite{Fan:2025qgj}}
  \put(-165,67){\tiny~\cite{FOCUS:2004gfa}}
  \put(-165,56.778){\tiny~\cite{BaBar:2008gpr}}
  \put(-165,46.556){\tiny~\cite{BESIII:2023opt}}
  \put(-165,32){\tiny~\cite{ParticleDataGroup:2024cfk}}
  \includegraphics[width=0.4\textwidth]{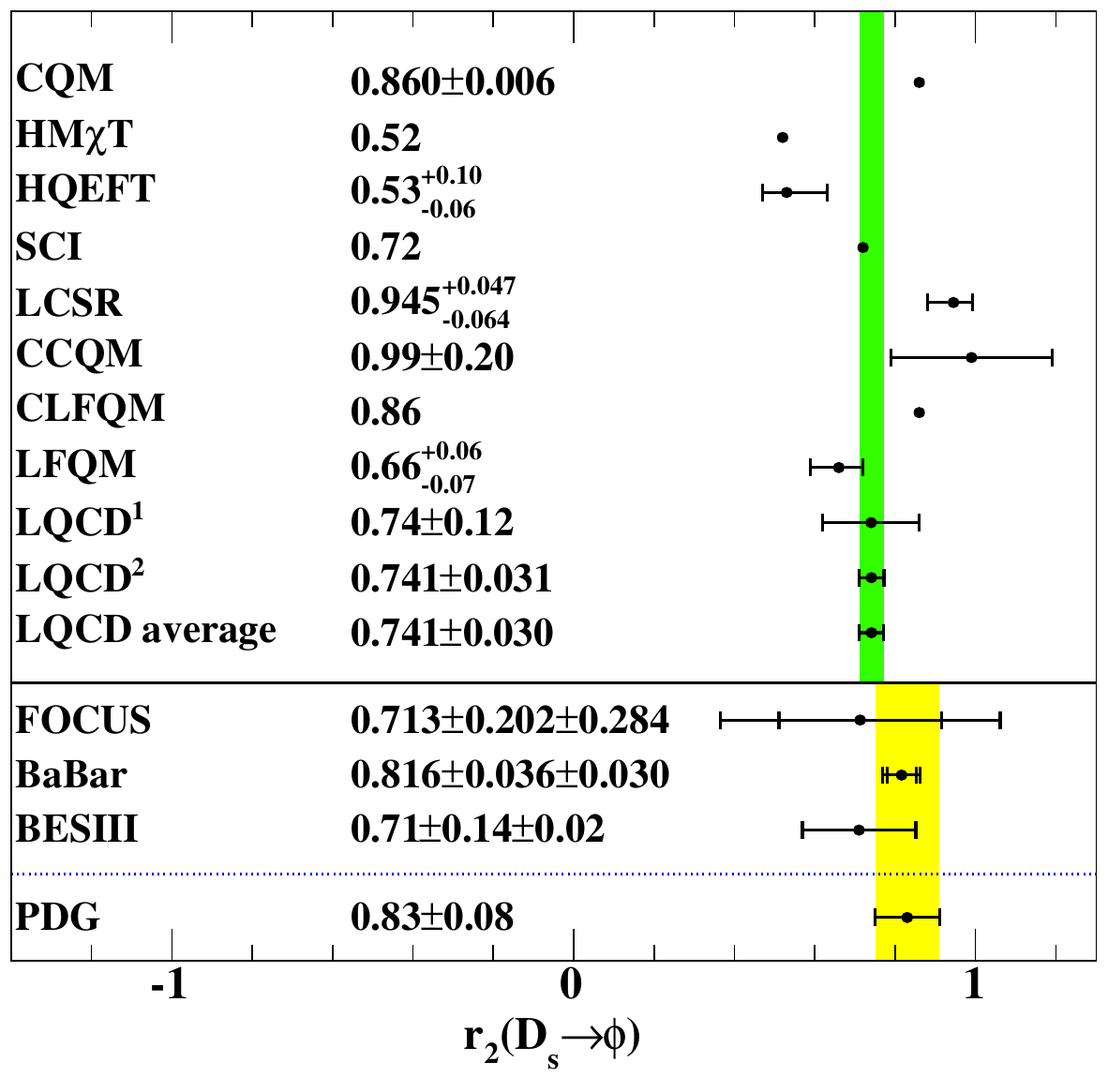}
  \put(-165,185){\tiny~\cite{Yang:2025gfz}}
  \put(-165,174.778){\tiny~\cite{Fajfer:2005ug}}
  \put(-165,164.556){\tiny~\cite{Wu:2006rd}}
  \put(-165,154.333){\tiny~\cite{Xing:2022sor}}
  \put(-165,144.111){\tiny~\cite{Wang:2025oix}}
  \put(-165,133.889){\tiny~\cite{Soni:2018adu}}
  \put(-165,123.667){\tiny~\cite{Verma:2011yw}}
  \put(-165,113.444){\tiny~\cite{Chang:2019mmh}}
  \put(-165,103.222){\tiny~\cite{Donald:2013pea}}
  \put(-165,93){\tiny~\cite{Fan:2025qgj}}
  \put(-165,67){\tiny~\cite{FOCUS:2004gfa}}
  \put(-165,56.778){\tiny~\cite{BaBar:2008gpr}}
  \put(-165,46.556){\tiny~\cite{BESIII:2023opt}}
  \put(-165,32){\tiny~\cite{ParticleDataGroup:2024cfk}}
   \caption{Comparisons of the (left) $r_V$ and (middle) $r_2$
   for $D^+_s\to \phi$ measured  by experimental measurements of FOCUS~\cite{FOCUS:2004gfa}, BaBar~\cite{BaBar:2008gpr}, and BESIII~\cite{BESIII:2023opt} as well as theoretical calculations
  of
  CQM~\cite{Yang:2025gfz},
   HM$\chi$T~\cite{Fajfer:2005ug},
   HQEFT~\cite{Wu:2006rd},
   SCI~\cite{Xing:2022sor},
   LCSR~\cite{Wang:2025oix},
   CCQM~\cite{Soni:2018adu},
   CLFQM~\cite{Verma:2011yw},
   LFQM~\cite{Chang:2019mmh},
   LQCD$^{1}$~\cite{Donald:2013pea}, and LQCD$^{2}$~\cite{Fan:2025qgj}. The green band is the $\pm 1\sigma$ region of the averaged LQCD$^{1,2}$ result~\cite{Donald:2013pea,Fan:2025qgj} and the yellow band denotes the $\pm 1\sigma$ region of the PDG averaged result~\cite{ParticleDataGroup:2024cfk}.
 }
  \label{fig:fDsphi}
\end{figure*}

Figure~\ref{fig:fDsphi} shows comparisons of the $r_V$ and  $r_2$  of $D^+_s\to \phi$ measured by different experiments and those predicted by theoretical calculations.

\subsubsection{Cabibbo-suppressed decays}

\begin{figure*}[htbp]
\centering
\includegraphics[width=0.8\textwidth]{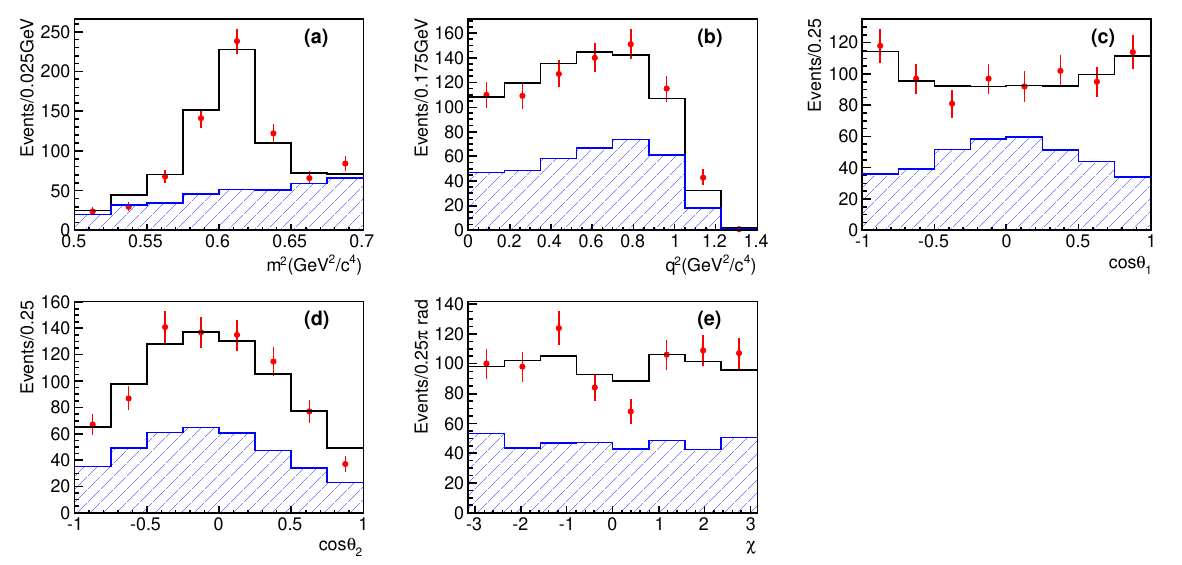}
\caption{
 The projections of amplitude analysis for $D^+\to \omega e^+\nu_e$ on $m^2=m^2_{\pi^+\pi^-\pi^0}$, $q^2$, $\cos\theta_1=\cos\theta_\pi$, $\cos\theta_2=\cos\theta_e$, and $\chi$~\cite{BESIII:2015kin}.
The points with  error bars are data, the solid histograms are the fit results, and
the filled histogram curves are the simulated backgrounds.}
\label{fig:Dp_omegaenu}
\end{figure*}

\begin{figure*}[htbp]
  \centering
  \includegraphics[width=0.4\textwidth]{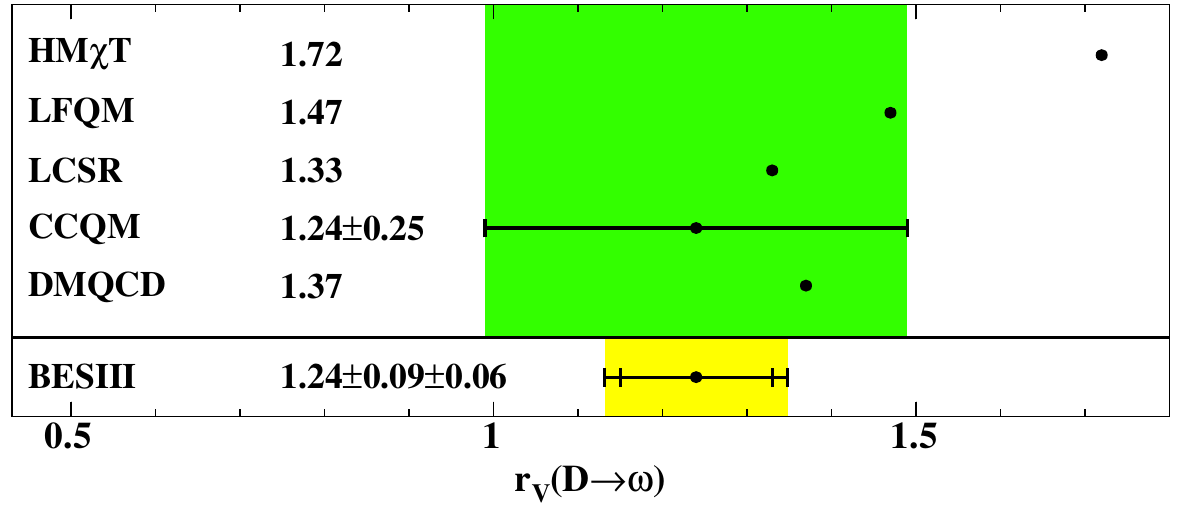}
   \put(-175,77){\tiny~\cite{Fajfer:2005ug}}
  \put(-175,67){\tiny~\cite{Verma:2011yw}}
  \put(-175,57){\tiny~\cite{Ball:2006yd}}
  \put(-175,47){\tiny~\cite{Ivanov:2019nqd}}
  \put(-175,37){\tiny~\cite{Voronin:2024ngw}}
  \put(-175,21){\tiny~\cite{BESIII:2015kin}}
  \includegraphics[width=0.4\textwidth]{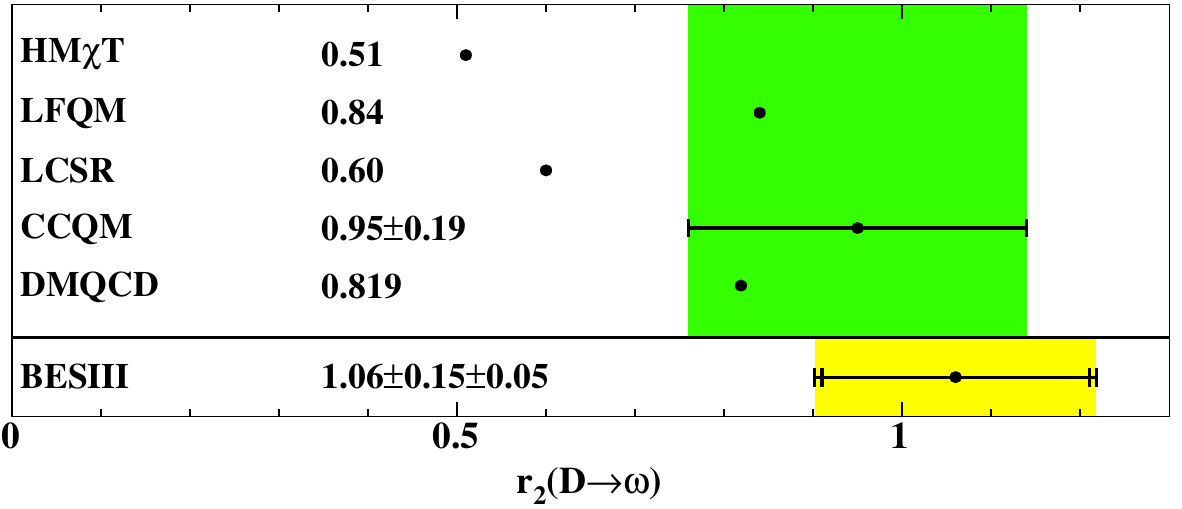}
 \put(-175,77){\tiny~\cite{Fajfer:2005ug}}
  \put(-175,67){\tiny~\cite{Verma:2011yw}}
  \put(-175,57){\tiny~\cite{Ball:2006yd}}
  \put(-175,47){\tiny~\cite{Ivanov:2019nqd}}
  \put(-175,37){\tiny~\cite{Voronin:2024ngw}}
  \put(-175,21){\tiny~\cite{BESIII:2015kin}}
   \caption{Comparisons of the (left) $r_V$ and (right) $r_2$ for $D^+\to \omega$ measured 
   by BESIII~\cite{BESIII:2015kin} and those predicted by theoretical calculations of
   HM$\chi$T~\cite{Fajfer:2005ug},
   LFQM~\cite{Verma:2011yw},
   LCSR~\cite{Ball:2006yd},
   CCQM~\cite{Ivanov:2019nqd}, and
   DMQCD~\cite{Voronin:2024ngw}.  The green band is the $\pm 1\sigma$ region of the CCQM result~\cite{Ivanov:2019nqd} and the yellow band denotes the $\pm 1\sigma$ region of the BESIII result~\cite{BESIII:2015kin}.
}
  \label{fig:fDomega}
\end{figure*}

Using 2.93 fb$^{-1}$ of data at 3.773~GeV, an improved measurement of the branching fraction $\mathcal{B}(D^+ \to \omega e^+ \nu_e) = (1.63\pm0.11\pm0.08)\times 10^{-3}$ is obtained based on 491 signal events~\cite{BESIII:2015kin}. The hadronic transition form factor ratios at zero momentum transfer are determined for the first time to be $r_V = 1.24\pm0.09\pm0.06$ and $r_2 = 1.06\pm0.15 \pm 0.05$.
Figure~\ref{fig:Dp_omegaenu} shows the $m^2=m^2_{\pi^+\pi^-\pi^0}$, $q^2$, $\cos\theta_1=\cos\theta_\pi$, $\cos\theta_2=\cos\theta_e$
and $\chi$ projections from the final fit to data.
BESIII also searched for $D^+ \to \phi e^+ \nu_e$, but no significant signal is observed; an improved upper limit at the 90\% confidence level is set as $\mathcal{B}(D^+ \to \phi e^+ \nu_e) < 1.3 \times 10^{-5}$. Using the same data sample, the first observation of $D^+\to \omega \mu^+\nu_\mu$ with 194 signal events is reported~\cite{Ablikim:2020tmg}, with a measured branching fraction of ${\cal B}(D^+\to \omega \mu^+\nu_\mu)=(1.77\pm0.18\pm0.11)\times 10^{-3}$. The ratio relative to the world average value of the branching fraction of $D^+\to \omega e^+\nu_e$, which probes lepton flavor universality, is determined to be ${\cal B}(D^+\to \omega \mu^+\nu_\mu)/{\cal B}^{\rm PDG}(D^+\to \omega e^+\nu_e)=1.05\pm0.14$, consistent with the SM expectation within one standard deviation.
Figure~\ref{fig:fDomega} shows comparisons of $r_V$ and  $r_2$ of $D\to \omega$ measured by BESIII and different theoretical calculations.

\begin{figure*}[htp]
\includegraphics[width=0.195\textwidth]{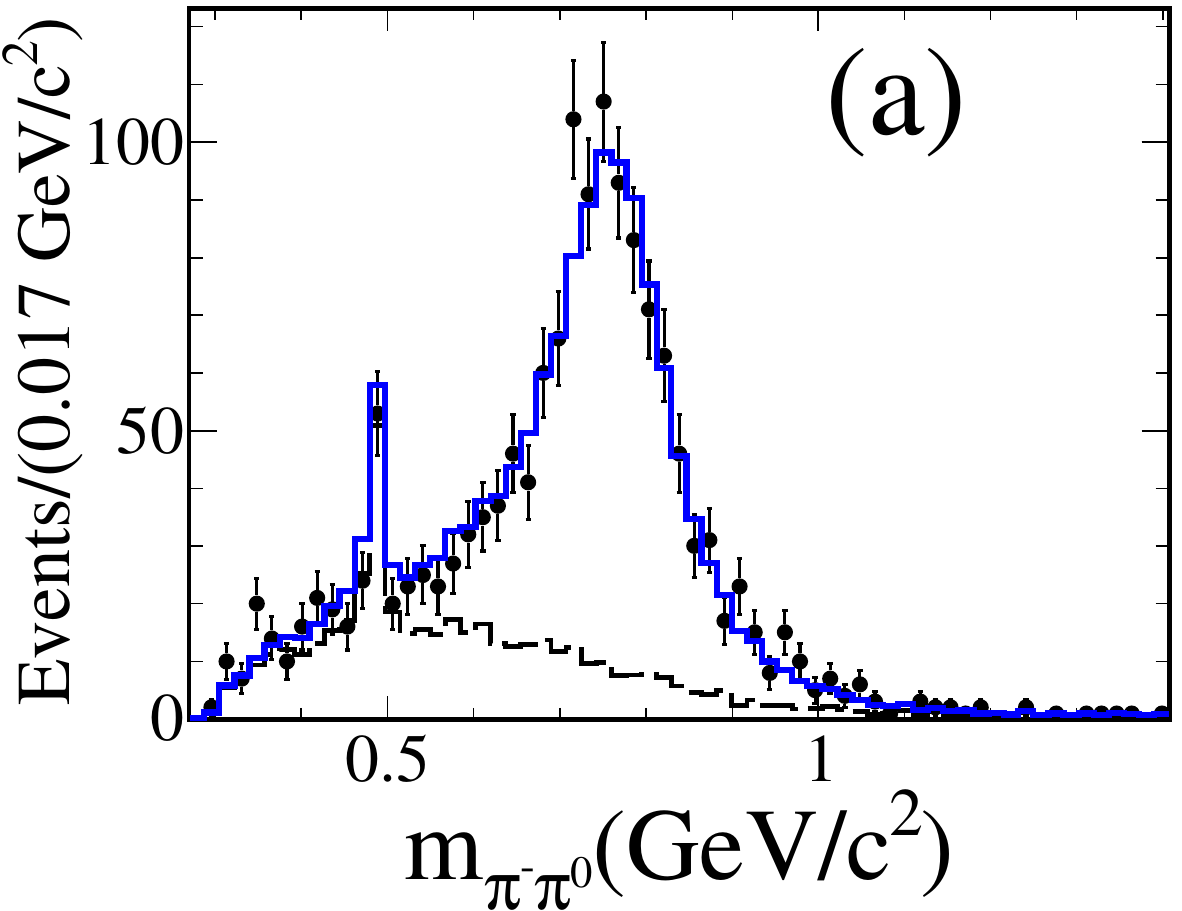}
\includegraphics[width=0.195\textwidth]{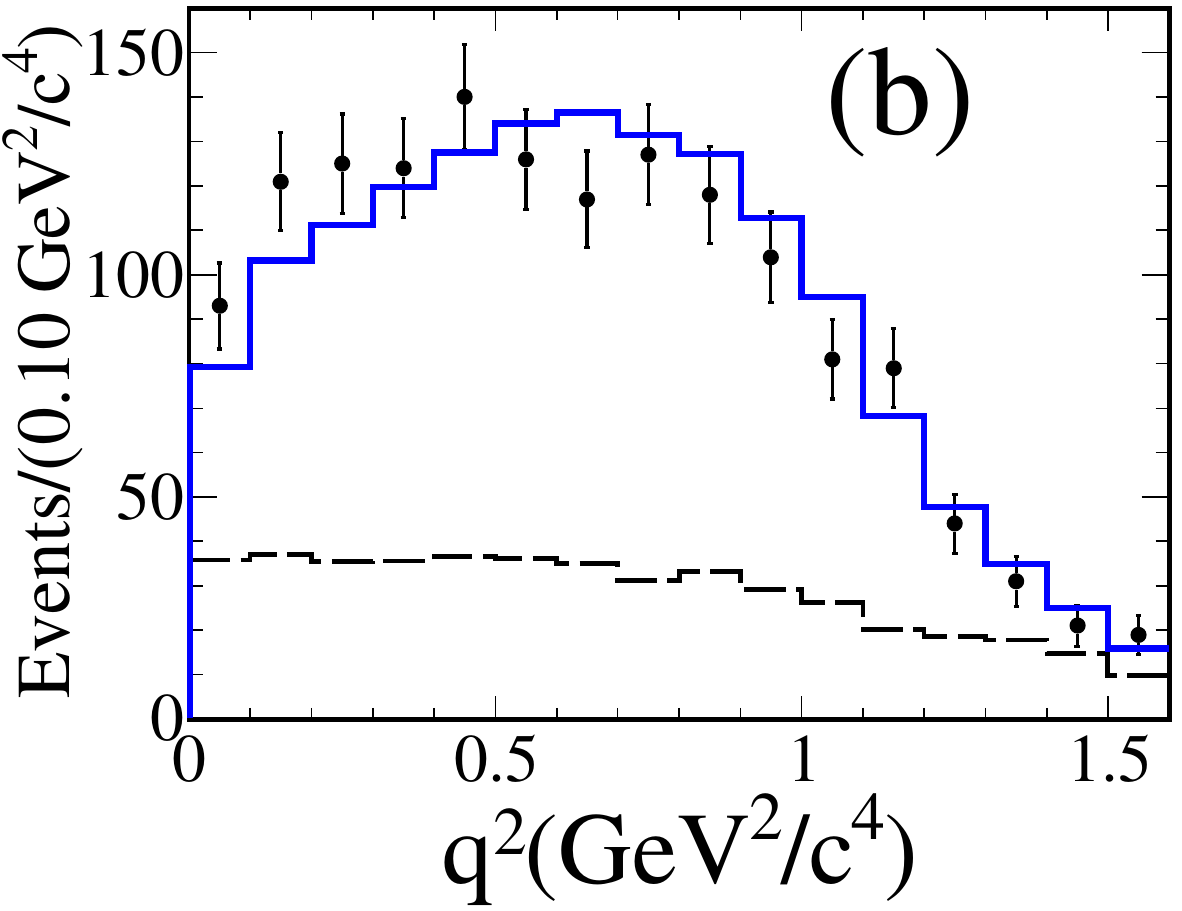}
\includegraphics[width=0.195\textwidth]{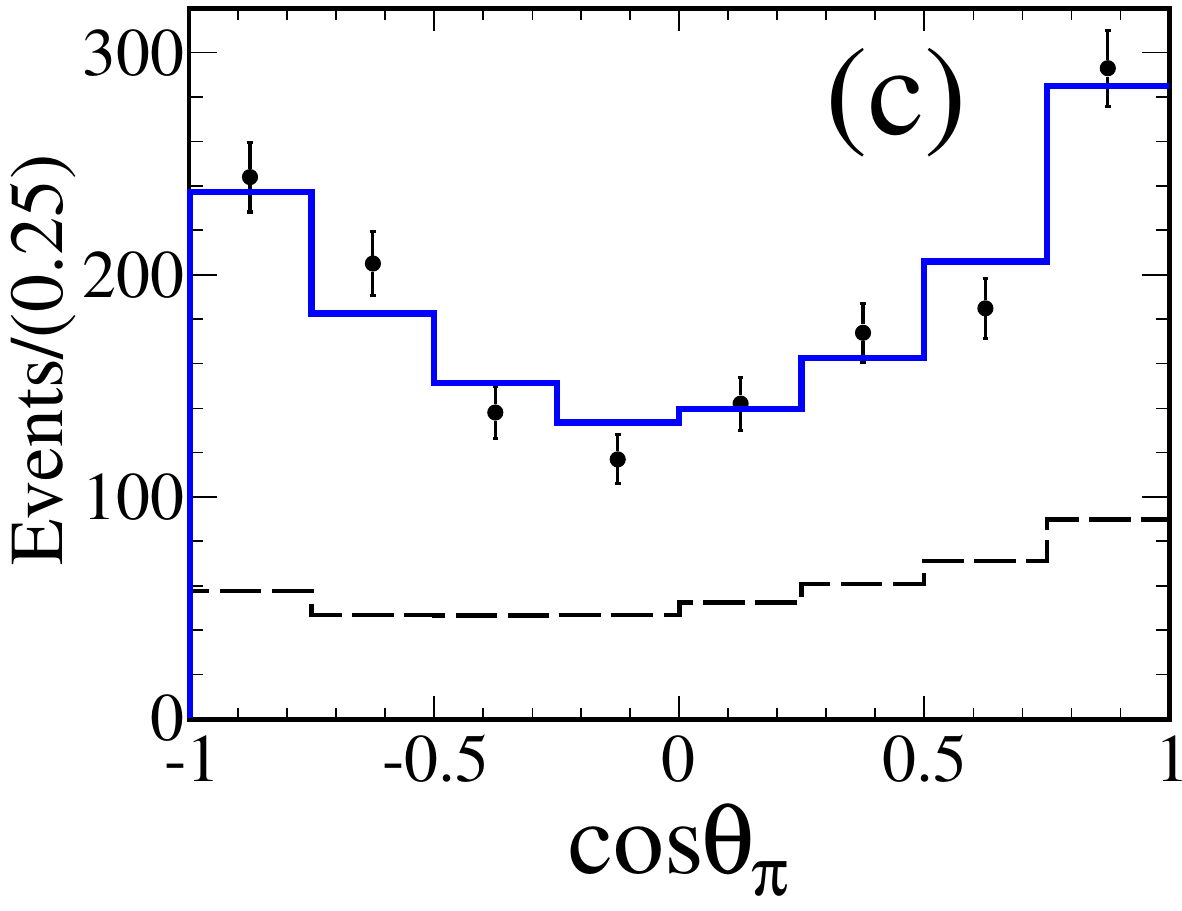}
\includegraphics[width=0.195\textwidth]{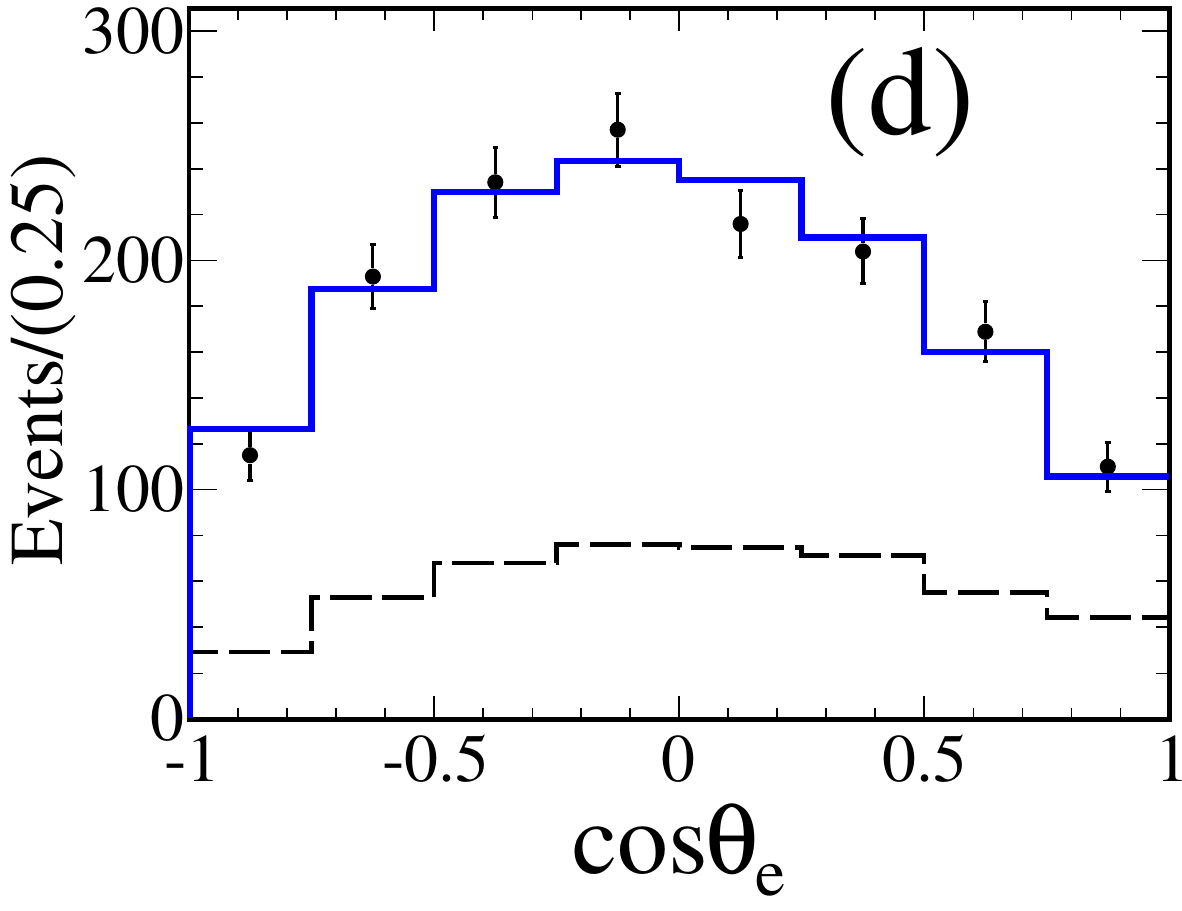}
\includegraphics[width=0.195\textwidth]{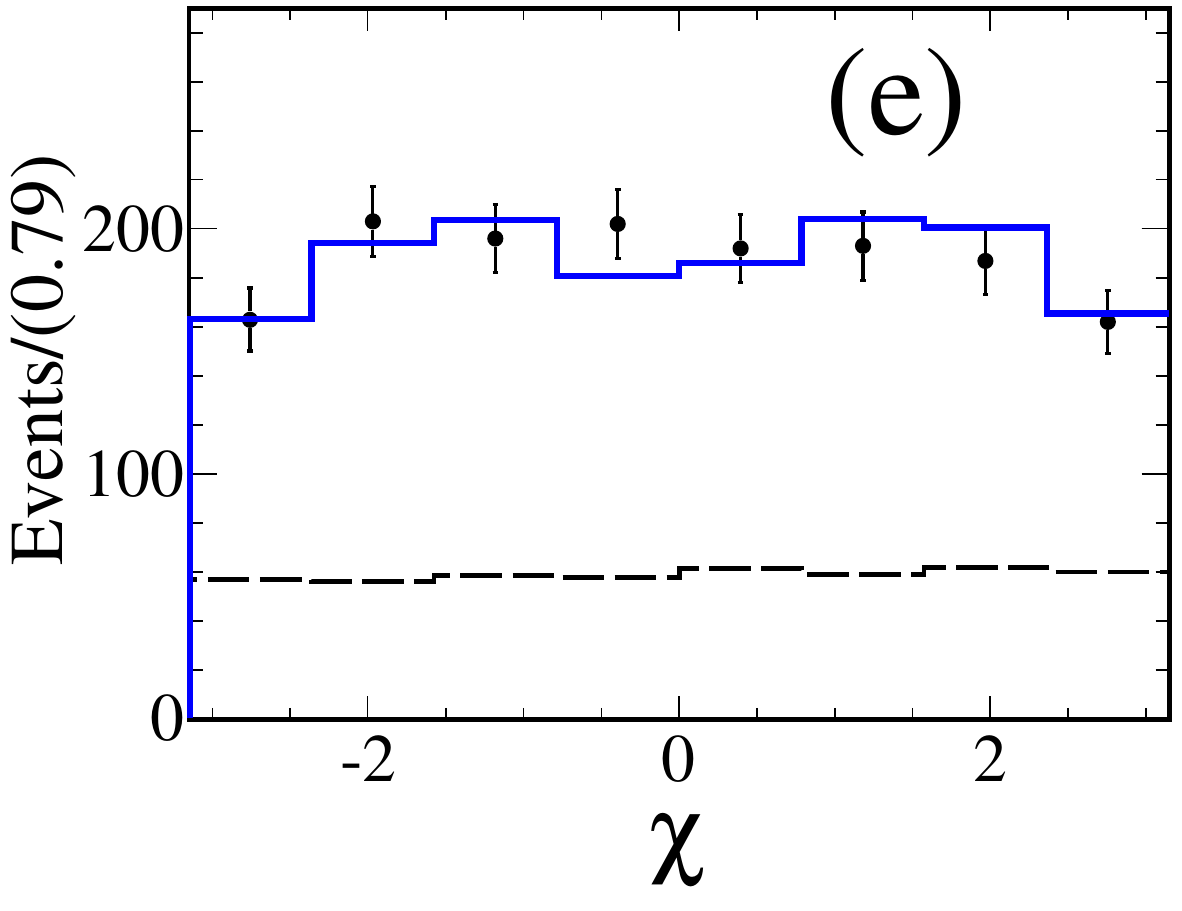}\\
\includegraphics[width=0.195\textwidth]{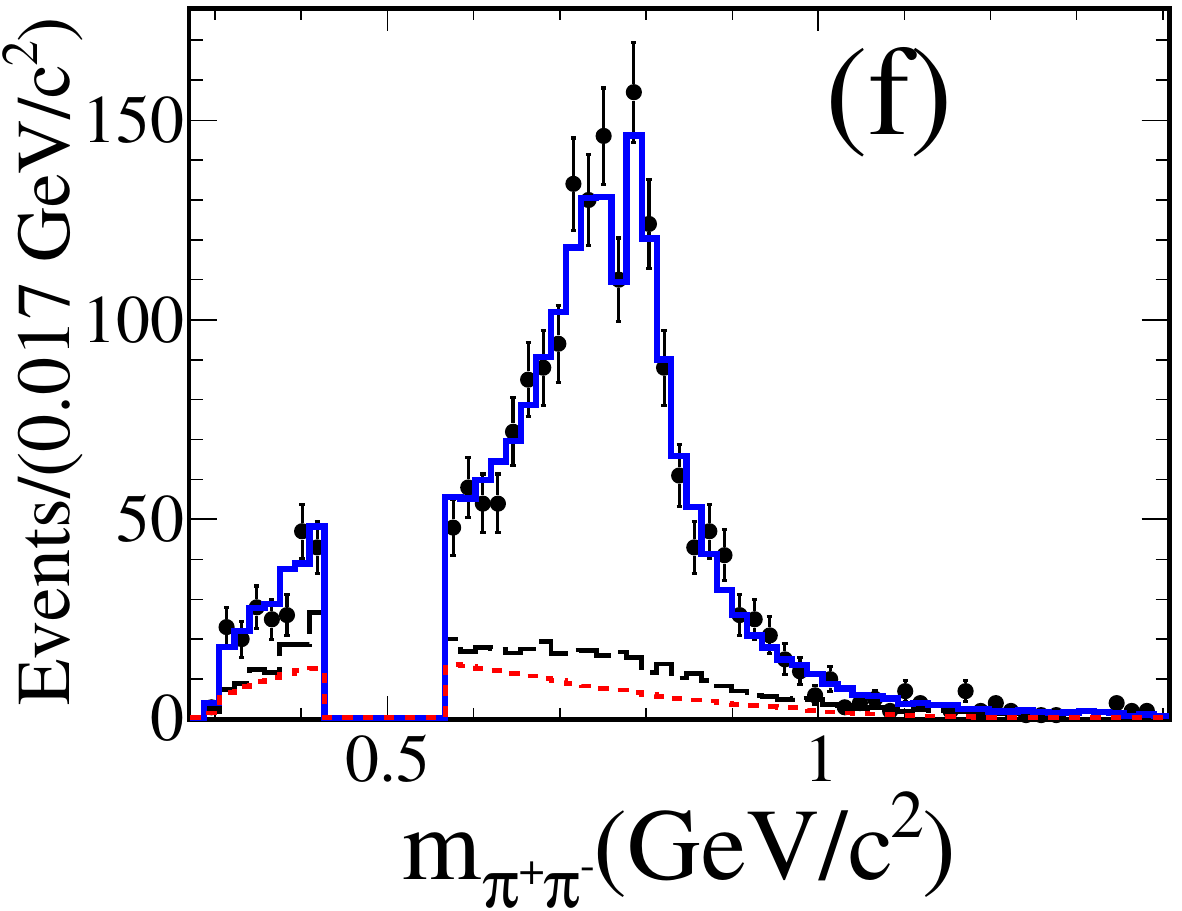}
\includegraphics[width=0.195\textwidth]{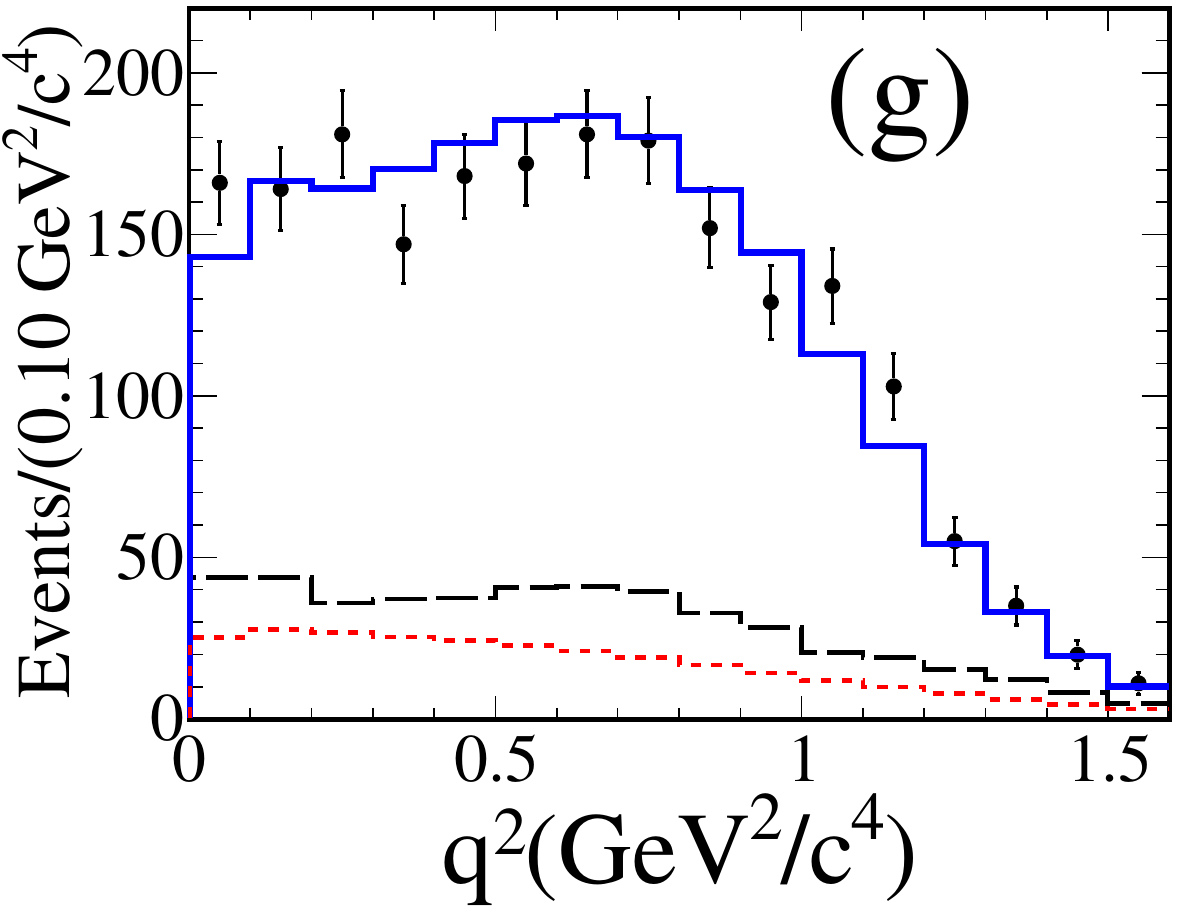}
\includegraphics[width=0.195\textwidth]{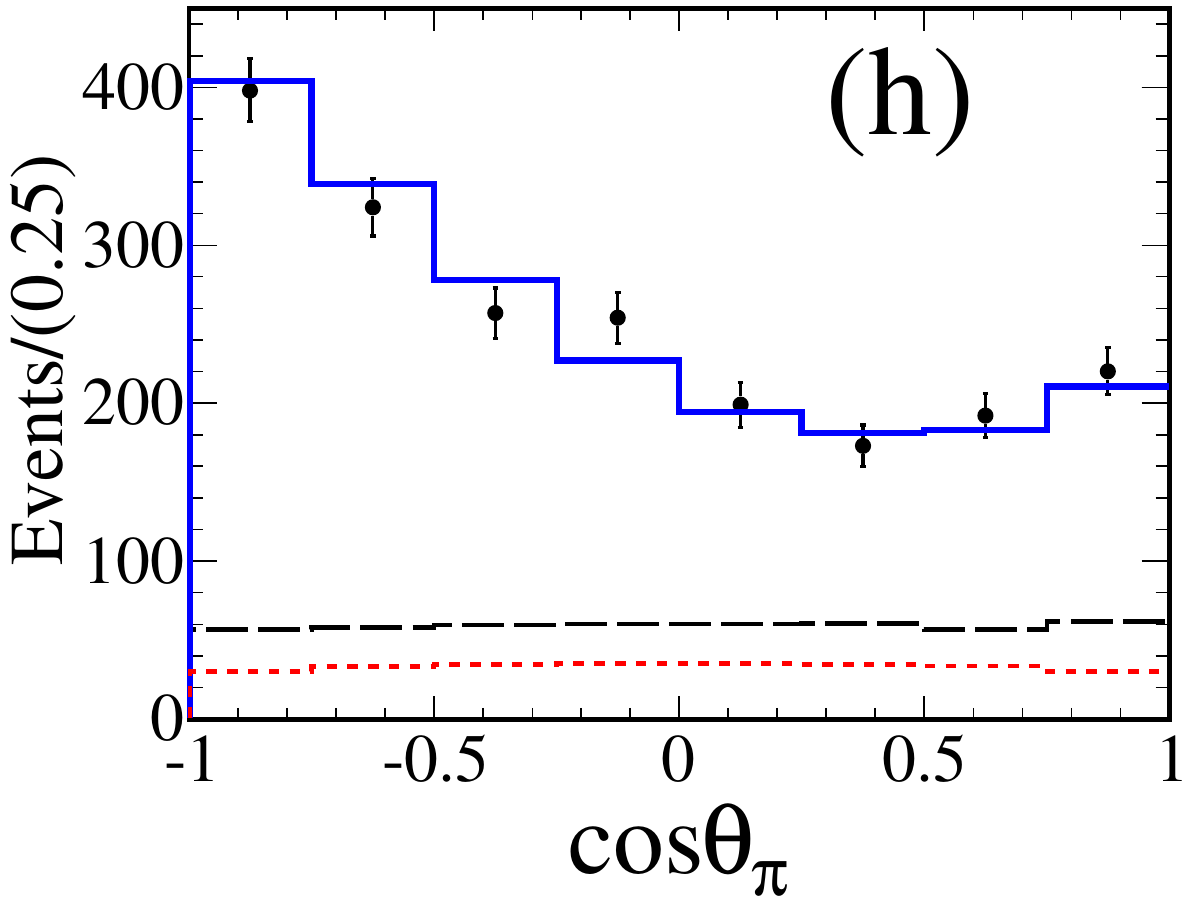}
\includegraphics[width=0.195\textwidth]{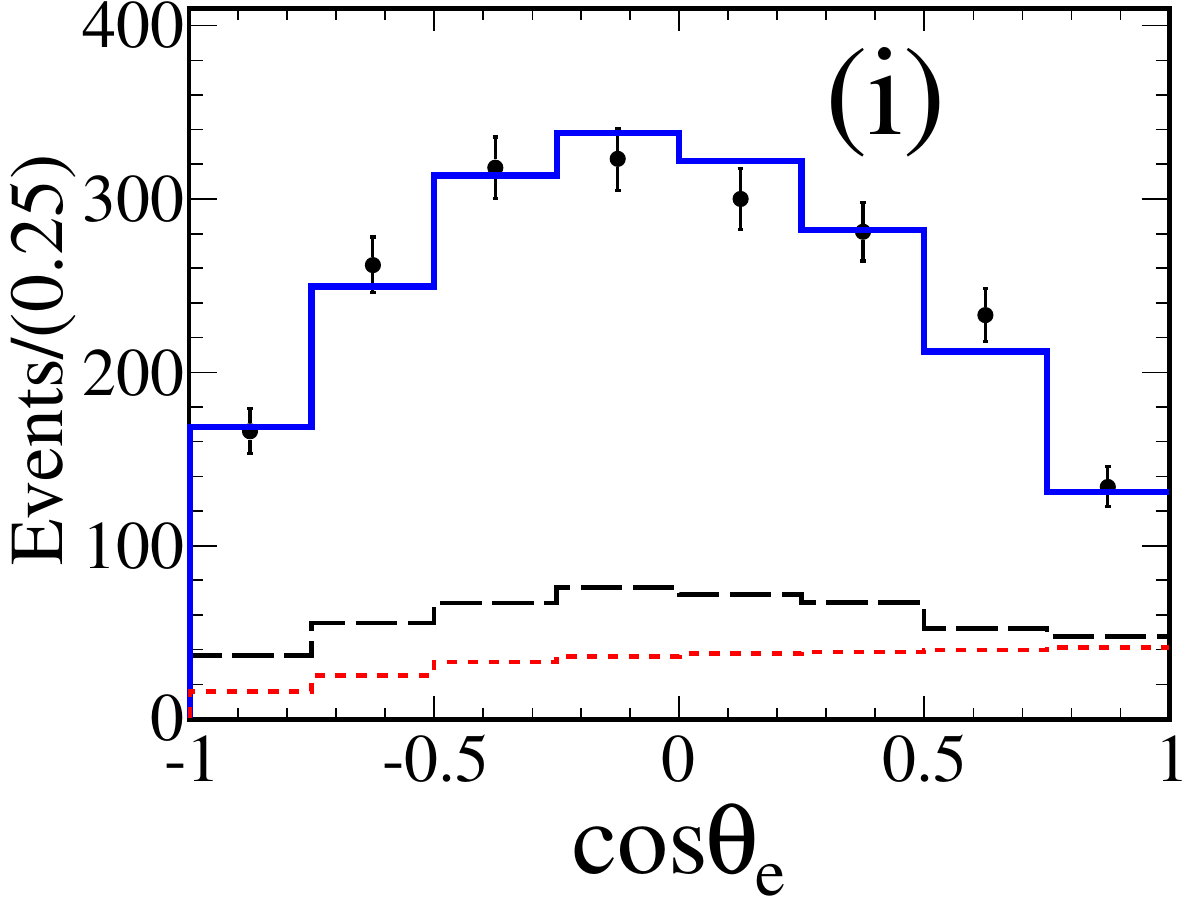}
\includegraphics[width=0.195\textwidth]{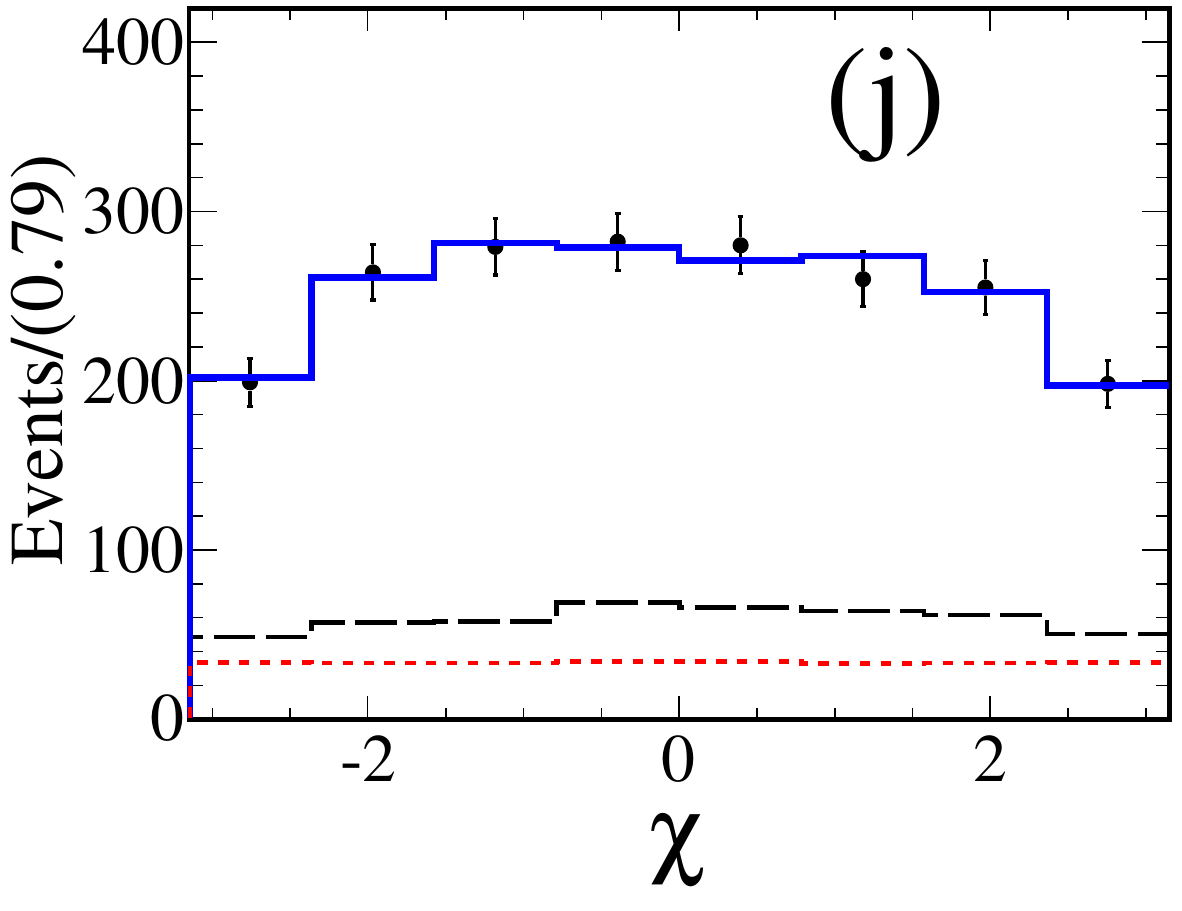}
\caption{
 The projections of amplitude analysis for (top) $D^{0} \to \pi^-\pi^0 e^+ \nu_e$ and (bottom) $D^{+} \to \pi^-\pi^+ e^+ \nu_e$ on $M_{\pi\pi}$, $q^2$, $\cos\theta_{\bar K}$, $\cos\theta_\ell$, and $\chi$~\cite{BESIII:2018qmf}.
The dots with error bars are data, the solid lines are the fits, the dashed lines show the MC simulated backgrounds,
and the short-dashed lines in (f)-(j) show the component of $D^+\to f_0(500)e^+\nu_e$.
}
\label{fit:D_pipienu_FFs}
\end{figure*}

Using 2.93~fb$^{-1}$ of 3.773~GeV~data, an analysis of the decays $D^{0} \to \pi^-\pi^0 e^+ \nu_e$ (1.7k signal events) and $D^{+} \to \pi^-\pi^+ e^+ \nu_e$ (1.1k signal events) was reported~\cite{BESIII:2018qmf}.
The projections of the five kinematic variables for the data are shown in Fig.~\ref{fit:D_pipienu_FFs}.
An amplitude analysis determines the $\pi^+\pi^-$ $\cal S$-wave contribution to $D^{+} \to \pi^-\pi^+ e^+ \nu_e$ to be $(25.7\pm1.6\pm1.1)\%$ with a statistical significance greater than 10$\sigma$, marking the first observation of the $\cal S$-wave contribution alongside the dominant ${\cal P}$-wave. The branching fractions are measured as $\mathcal{B}(D^{0} \to \rho(770)^- e^+ \nu_e) = (1.445\pm 0.058 \pm 0.039) \times10^{-3}$ and $\mathcal{B}(D^{+} \to \rho(770)^0 e^+ \nu_e) = (1.860\pm 0.070 \pm 0.061) \times10^{-3}$.
The hadronic transition form factor ratios of $D\to \rho$ at $q^{2}=0$ are determined to be $r_{V}=1.695\pm0.083\pm0.051$ and $r_{2}=0.845\pm0.056\pm0.039$.
Using $2.93~\mathrm{fb}^{-1}$ of 3.773~GeV~data, the decay $D^0\to \rho(770)^- \mu^+\nu_\mu$ is observed with 570 signal events for the first time~\cite{BESIII:2021pvy},
and the $D^+\to \rho(770)^0\mu^+\nu_\mu$ decay is measured with improved precision based on 496 signal events~\cite{BESIII:2024lnh}.
The obtained branching fractions are
${\cal B}(D^0\to \rho(770)^- \mu^+\nu_\mu)=(1.35\pm0.09\pm0.09)\times 10^{-3}$ and
${\cal B}(D^+\to \rho(770)^0 \mu^+\nu_\mu)=(1.64\pm0.13\pm0.10)\times 10^{-3}$.
Combining with the world average of ${\cal B}(D\to \rho(770) e^+\nu_e)$, the ratios
${\cal B}(D^0\to \rho(770)^- \mu^+\nu_\mu)/{\cal B}(D^0\to \rho(770)^- e^+\nu_e)=0.90\pm0.11$
and
${\cal B}(D^+\to \rho(770)^0 \mu^+\nu_\mu)/{\cal B}(D^+\to \rho(770)^0 e^+\nu_e)=0.88\pm0.10$
are obtained, consistent with the theoretical expectation of lepton flavor universality within uncertainty.
Combining the world average of ${\cal B}(D^+\to \rho(770)^0 \mu^+ \nu_\mu)$ and the lifetimes of $D^{0(+)}$, the partial decay width ratio ${\Gamma}(D^0\to \rho(770)^- \mu^+ \nu_{\mu})/(2{\Gamma}(D^+\to \rho(770)^0 \mu^+ \nu_{\mu}) = 0.71\pm0.14$ is obtained, which agrees with the isospin symmetry expectation of unity within $2.1\sigma$.

\begin{figure*}[htbp]
  \centering
  \includegraphics[width=0.4\textwidth]{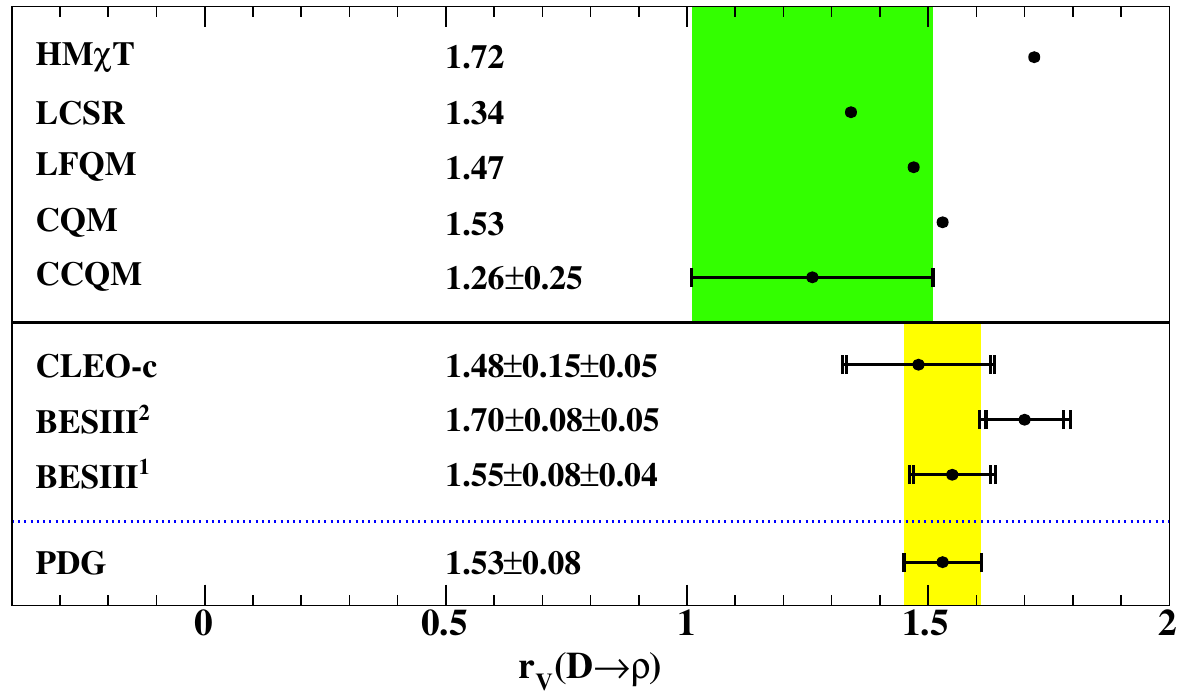}
 \put(-160,109.5){\tiny~\cite{Fajfer:2005ug}}
  \put(-160,99.875){\tiny~\cite{Wu:2006rd}}
  \put(-160,90.25){\tiny~\cite{Verma:2011yw}}
  \put(-160,80.625){\tiny~\cite{Ivanov:2019nqd}}
  \put(-160,71){\tiny~\cite{Melikhov:2000yu}}
  \put(-160,56){\tiny~\cite{CLEO:2011ab}}
  \put(-160,46.375){\tiny~\cite{BESIII:2024lxg}}
  \put(-160,36.75){\tiny~\cite{BESIII:2018qmf}}
  \put(-160,22){\tiny~\cite{ParticleDataGroup:2024cfk}}
  \includegraphics[width=0.4\textwidth]{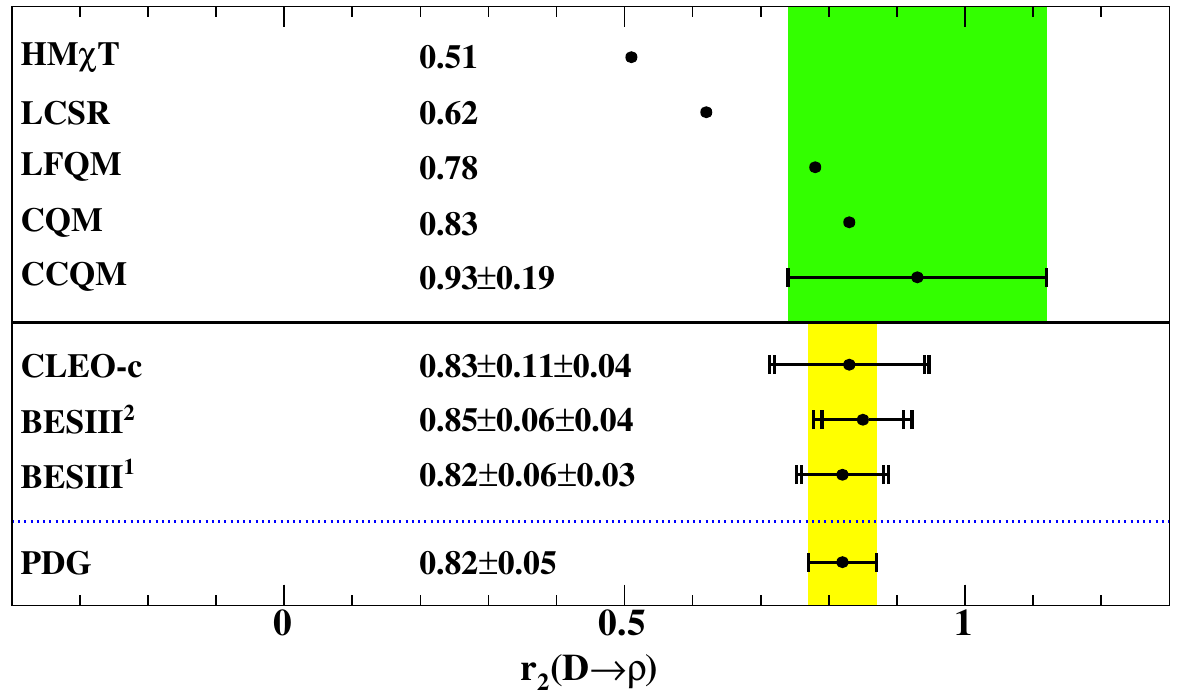}
 \put(-160,109.5){\tiny~\cite{Fajfer:2005ug}}
  \put(-160,99.875){\tiny~\cite{Wu:2006rd}}
  \put(-160,90.25){\tiny~\cite{Verma:2011yw}}
  \put(-160,80.625){\tiny~\cite{Ivanov:2019nqd}}
  \put(-160,71){\tiny~\cite{Melikhov:2000yu}}
  \put(-160,56){\tiny~\cite{CLEO:2011ab}}
  \put(-160,46.375){\tiny~\cite{BESIII:2024lxg}}
  \put(-160,36.75){\tiny~\cite{BESIII:2018qmf}}
  \put(-160,22){\tiny~\cite{ParticleDataGroup:2024cfk}}
   \caption{Comparisons of the (left) $r_V$ and (right) $r_2$ for $D\to \rho$ measured by CLEO-c~\cite{CLEO:2011ab}, BESIII$^1$~\cite{BESIII:2018qmf} ($D^0\to\pi^-\pi^0e^+\nu_e$
   and $D^+\to\pi^-\pi^+e^+\nu_e$ with 2.93 fb$^{-1}$ of data at 3.773 GeV),
   BESIII$^2$~\cite{BESIII:2024lxg} ($D^0\to\pi^-\pi^0e^+\nu_e$ with 7.9 fb$^{-1}$ of data at 3.773 GeV) as well as theoretical calculations
   of
   HM$\chi$T~\cite{Fajfer:2005ug},
   LCSR~\cite{Wu:2006rd},
   LFQM~\cite{Verma:2011yw},
   CQM~\cite{Ivanov:2019nqd}, and
   CCQM~\cite{Melikhov:2000yu}. The green band is the $\pm 1\sigma$ region of the  CCQM result~\cite{Melikhov:2000yu} and the yellow band denotes the $\pm 1\sigma$ region of the PDG averaged result~\cite{ParticleDataGroup:2024cfk}.
}
  \label{fig:fDrho}
\end{figure*}

In 2024, BESIII reported a study of $D^0\to \pi^-\pi^0e^{+}\nu_e$ based on 3.3k signal events using $7.9~\mathrm{fb}^{-1}$ of data at 3.773 GeV~\cite{BESIII:2024lxg}. The branching fraction of $D^0\to \rho(770)^-e^+\nu_e$ is measured to be $(1.439 \pm 0.033\pm 0.027) \times10^{-3}$, a factor of 1.6 more precise than previous measurements. From an amplitude analysis, the hadronic transition form factor ratios of $D\to \rho(770)$ are determined to be $r_{V}=1.548\pm0.079\pm0.041$ and $r_{2}=0.823\pm0.056\pm0.026$.
Figure~\ref{fig:fDrho} shows comparisons of  $r_V$ and $r_2$ of $D\to \rho$ measured by different experiments and those predicted by theoretical calculations.

\begin{figure*}[htbp]
  \centering
  \includegraphics[width=0.4\textwidth]{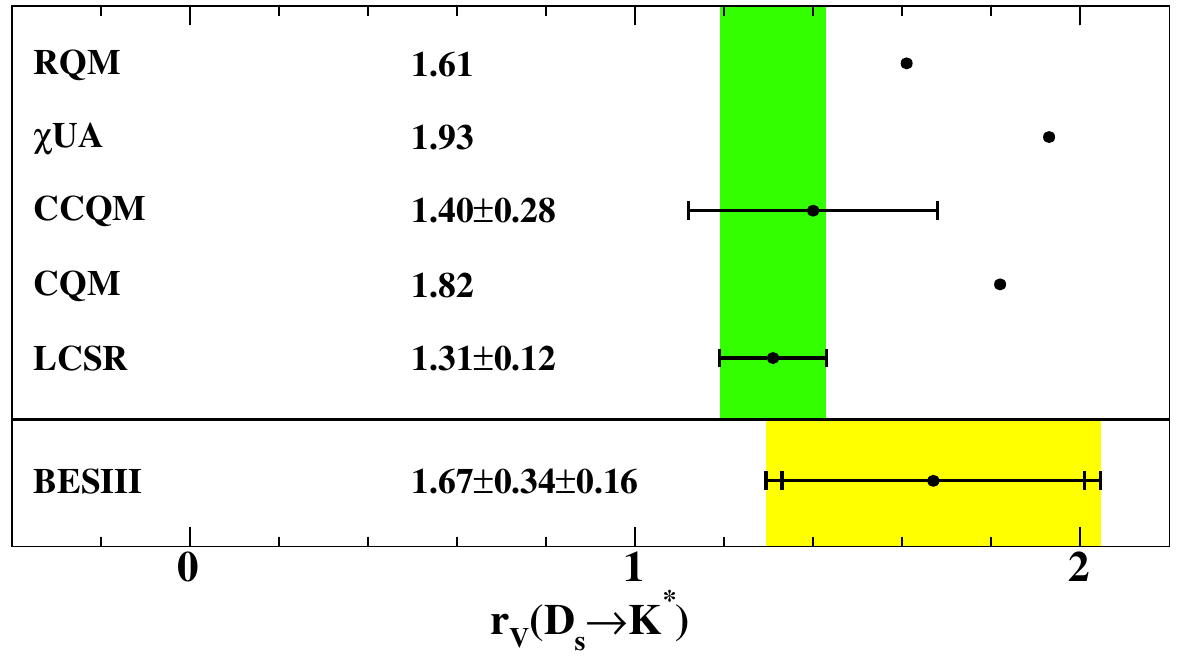}
  \put(-175,102){\tiny~\cite{Faustov:2019mqr}}
\put(-175,89.125){\tiny~\cite{Fajfer:2005ug}}
\put(-175,76.25){\tiny~\cite{Soni:2018adu,Ivanov:2019nqd}}
\put(-175,63.375){\tiny~\cite{Melikhov:2000yu}}
\put(-175,50.5){\tiny~\cite{Wu:2006rd}}
\put(-175,29.5){\tiny~\cite{BESIII:2018xre}}
  \includegraphics[width=0.4\textwidth]{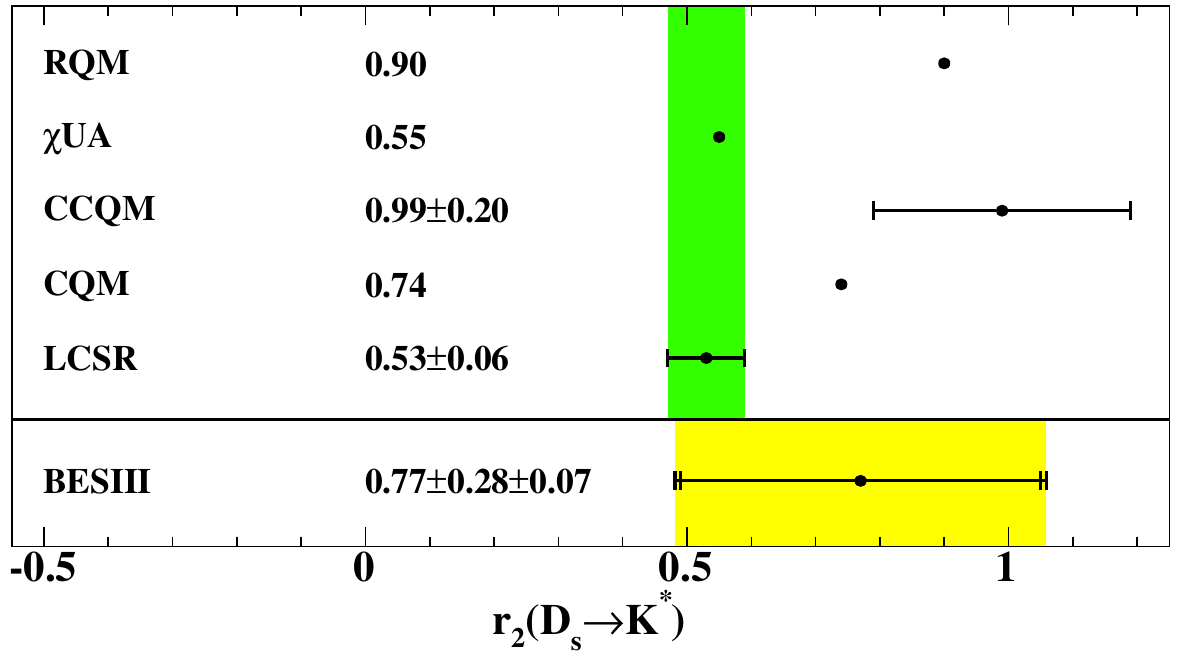}
  \put(-175,102){\tiny~\cite{Faustov:2019mqr}}
\put(-175,89.125){\tiny~\cite{Fajfer:2005ug}}
\put(-175,76.25){\tiny~\cite{Soni:2018adu,Ivanov:2019nqd}}
\put(-175,63.375){\tiny~\cite{Melikhov:2000yu}}
\put(-175,50.5){\tiny~\cite{Wu:2006rd}}
\put(-175,29.5){\tiny~\cite{BESIII:2018xre}}
   \caption{Comparisons of the (left) $r_V$ and (right) $r_2$ for $D^+_s\to K^*$ measured 
   by BESIII~\cite{BESIII:2018xre} and those predicted by theoretical calculations of
   RQM~\cite{Faustov:2019mqr},
   $\chi$UA~\cite{Fajfer:2005ug},
   CCQM~\cite{Soni:2018adu,Ivanov:2019nqd},
   CQM~\cite{Melikhov:2000yu}, and
   LCSR~\cite{Wu:2006rd}. The green band is the $\pm 1\sigma$ region of the  LCSR result~\cite{Wu:2006rd} and the yellow band denotes the $\pm 1\sigma$ region of the BESIII result~\cite{BESIII:2018xre}.
}
  \label{fig:fDsKstar0}
\end{figure*}

Using 3.19 fb$^{-1}$ of data at 4.178~GeV, the branching fraction of $D^+_s\to K^*(892)^0 e^+\nu_e$ is determined with improved precision, based on 155 signal events, to be
${\cal B}({D^+_s\to K^*(892)^0 e^+\nu_e})=(2.37\pm0.26\pm0.20)\times10^{-3}$~\cite{BESIII:2018xre}. The hadronic transition form factor ratios are measured for the first time as $r_V=1.67\pm0.34\pm0.16$ and $r_2=0.77\pm0.28\pm0.07$, where the $K^+\pi^-$ $\cal S$-wave component is treated as a source of systematic uncertainty due to limited sample size.
The comparisons of the $r_V$ and  $r_2$ of $D^+_s\to K^*(892)^0$ measured by BESIII and those predicted by different theoretical calculations are shown in Fig.~\ref{fig:fDsKstar0}.

\subsection{Results of $D\to S\ell^+\nu_\ell$ at BESIII}

\begin{figure}[htbp] \centering
\includegraphics[width=0.9\linewidth]{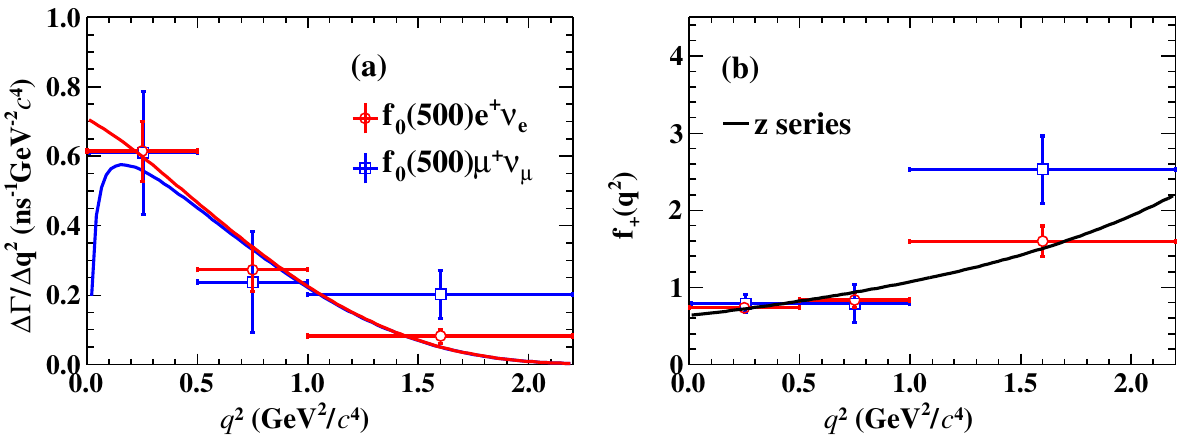}
\caption{ (a) The partial decay rates of $D^+\to f_0(500)\ell^+\nu_\ell$ and (b) projections to
  $f_+^{D\to f_0(500)}(q^2)$~\cite{BESIII:2024lnh}.
  }
\label{fig:Dp_f0500lnu_FFs}
\end{figure}

Based on a simultaneous amplitude analysis of $D^0\to \pi^-\pi^0e^+\nu_e$ and $D^+\to \pi^-\pi^+e^+\nu_e$
from 2.93 fb$^{-1}$ of data at 3.773 GeV, Ref.~\cite{BESIII:2018qmf} reported the first observation of the $\cal S$-wave contribution ($D^{+} \to f_0(500) e^+ \nu_e$)
with a fraction of  $(25.7\pm1.6\pm1.1)\%$, corresponding to a branching fraction of
$\mathcal{B}(D^{+} \to f_0(500) e^+ \nu_e, f_0(500)\to\pi^+\pi^-) = (6.30\pm 0.43 \pm 0.32) \times10^{-4}$,
in addition to the dominant ${\cal P}$-wave contribution of $D^+\to \rho(770)^0e^+\nu_e$.
Moreover, an upper limit of $\mathcal{B}(D^{+} \to f_0(980) e^+ \nu_e, f_0(980)\to\pi^+\pi^-) < 2.8 \times10^{-5}$ is set at the 90\% confidence level.
In 2024, Ref.~\cite{BESIII:2024lnh} reported the first observation of $D^+\to f_0(500)\mu^+\nu_\mu$ and
a joint analysis of $D^+\to f_0(500)\ell^+\nu_\ell$~($\ell=e$ or $\mu$),
by using the same data sample.
About 412 $D^+\to f_0(500) e^+\nu_e$ and 209 $D^+\to f_0(500)\mu^+\nu_\mu$ signal
events are observed.
The  branching fractions are determined to be ${\cal B}(D^+\to f_0(500)e^+\nu_e, f_0(500)\to\pi^+\pi^-)=(0.60\pm0.06\pm0.05)\times10^{-3}$ and 
${\cal B}(D^+\to f_0(500)\mu^+\nu_\mu, f_0(500)\to\pi^+\pi^-)=(0.72\pm0.13\pm0.08)\times10^{-3}$~\cite{BESIII:2024lnh}.
The simultaneous fit to their differential decay rates gives $f_{+}^{D\to f_0(500)}(0)|V_{cd}|=0.143\pm0.014\pm0.011$.
The fit results and the projections on $f^{D\to f_0(500)}_+(q^2)$ are shown in Fig.~\ref{fig:Dp_f0500lnu_FFs}.
Using the value of $|V_{cd}|=0.22487\pm0.00068$ given by the SM-constrained fit~\cite{ParticleDataGroup:2024cfk}, the $f_{+}^{D\to f_0(500)}(0)$ is measured to be $0.63\pm0.06\pm0.05$.

\begin{figure}[htp]
\begin{center}
\includegraphics[width=0.485\textwidth]{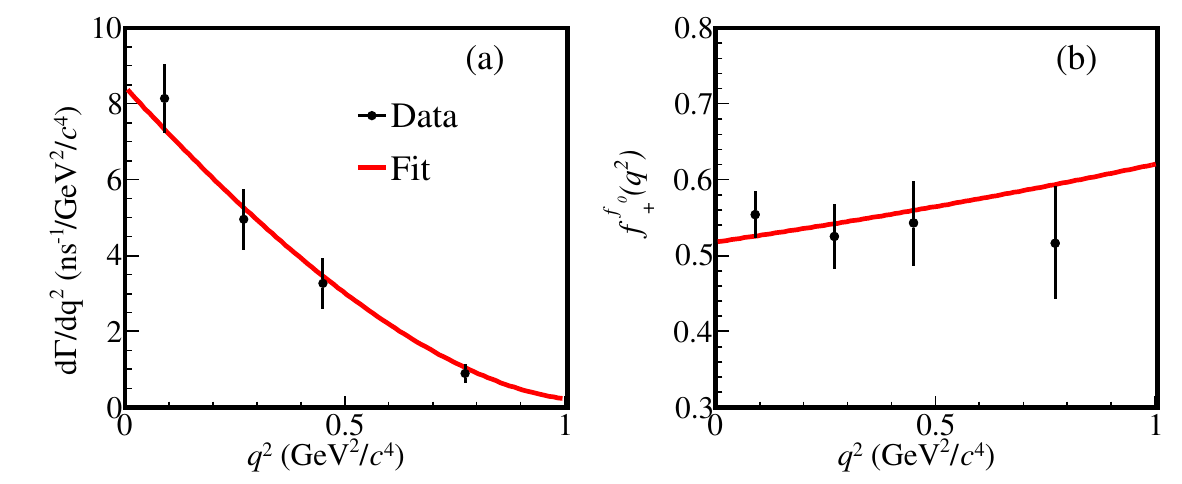}
\caption{(a) The fit to the differential decay rate as a function of $q^2$ and (b) projection on $f^{D_s\to f_{0}(980)}_{+}(q^2)$~\cite{BESIII:2023wgr}.
}
\label{fig:Ds_f0enu_FFs}
\end{center}
\end{figure}

Using 7.33~${\rm fb^{-1}}$ of data at 4.128-4.226~GeV,
an analysis of the decay $D_{s}^{+} \to \pi^{+}\pi^{-}e^{+}\nu_e$ with $f_{0}(980) \to \pi^{+}\pi^{-}$ was performed~\cite{BESIII:2023wgr},
where the $f_{0}(980)$ is observed in the $\pi^+\pi^-$ system.
The branching fraction of $D_{s}^{+} \to f_{0}(980)e^{+}\nu_e$ with $f_0(980)\to\pi^+\pi^-$ is measured to be $(1.72 \pm 0.13\pm 0.10) \times10^{-3}$.
From an analysis of its decay dynamics with the simple pole parameterization of the hadronic transition form factor
and the Flatt\'e formula describing the $f_0(980)$ in the differential decay rate,
the product of the hadronic transition form factor $f^{D_s\to f_0(980)}_{+}(0)$ and the $c\to s$ CKM matrix element $|V_{cs}|$
is determined for the first time to be $f^{D_s\to f_0(980)}_+(0)|V_{cs}|=0.504\pm0.017\pm0.035$.
Using the value of $|V_{cs}|=0.97349\pm0.0016$ given by the SM-constrained fit~\cite{ParticleDataGroup:2024cfk}, the $f_{+}^{D_s\to f_0(980)}(0)$ is measured to be
$0.518\pm0.018\pm0.036$.
The fit to the differential decay rate of the channel $D_{s}^{+}\to f_{0}(980)e^{+}\nu_{e}$ and the hadronic transition form factor projection are shown in Fig.~\ref{fig:Ds_f0enu_FFs}.

By analyzing 6.3 fb$^{-1}$ of data at 4.178-4.226 GeV, BESIII also studied
the semileptonic decay $D_{s}^{+} \to f_{0}(980)e^{+}\nu_e$ via $f_0(980)\to 2\pi^0$~\cite{BESIII:2021drk}.
Based on 55 signal events,
the product branching fraction of $D_s^+\to f_0(980)e^+\nu_e,\,f_0(980)\to 2\pi^0$ is measured
to be $(7.9\pm1.4 \pm0.4)\times 10^{-4}$.
According to isospin symmetry expectation
$\frac{\mathcal{B}(f_0\to 2\pi^0)}{\mathcal{B}(f_0\to\pi^+\pi^-)}=0.5$, this
result is consistent with the measurement of $D_s^+\to f_0(980) e^+\nu_e$ with
$f_0(980) \to \pi^+\pi^-$. No significant signal of $D_{s}^{+} \to f_{0}(500)e^{+}\nu_e$ via $f_0(980)\to 2\pi^0$
is observed. An upper limit on the product branching fraction at the 90\% confidence level
is set as
${\cal B}(D_s^+\to f_0(500)e^+\nu_e,\,f_0(500)\to 2\pi^0)<7.4\times 10^{-4}$.

\begin{figure}
\centering
\includegraphics[width=0.48\textwidth]{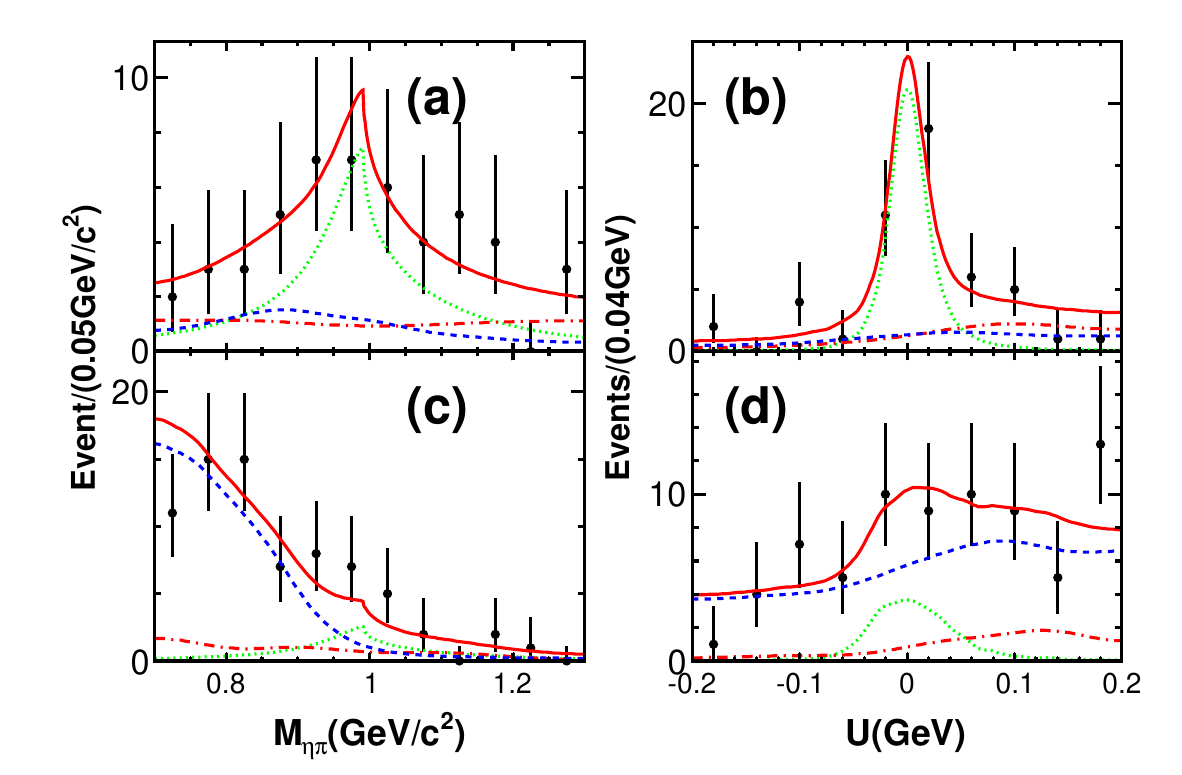}
\caption{Projections of the 2-D fit on (left)~$M_{\eta\pi}$ and (right)~$U$ for
    (a)(b)~$D^0\to \eta \pi^-e^+\nu_e$ and (c)(d)~$D^+\to \eta \pi^0e^+\nu_e$~\cite{BESIII:2018sjg}.
    }
    \label{fig:D_a0980enu}
\end{figure}

\begin{figure}[htbp]
    \begin{center}
        \mbox{
            \put(-130, 0){
            \includegraphics[width=0.25\textwidth]{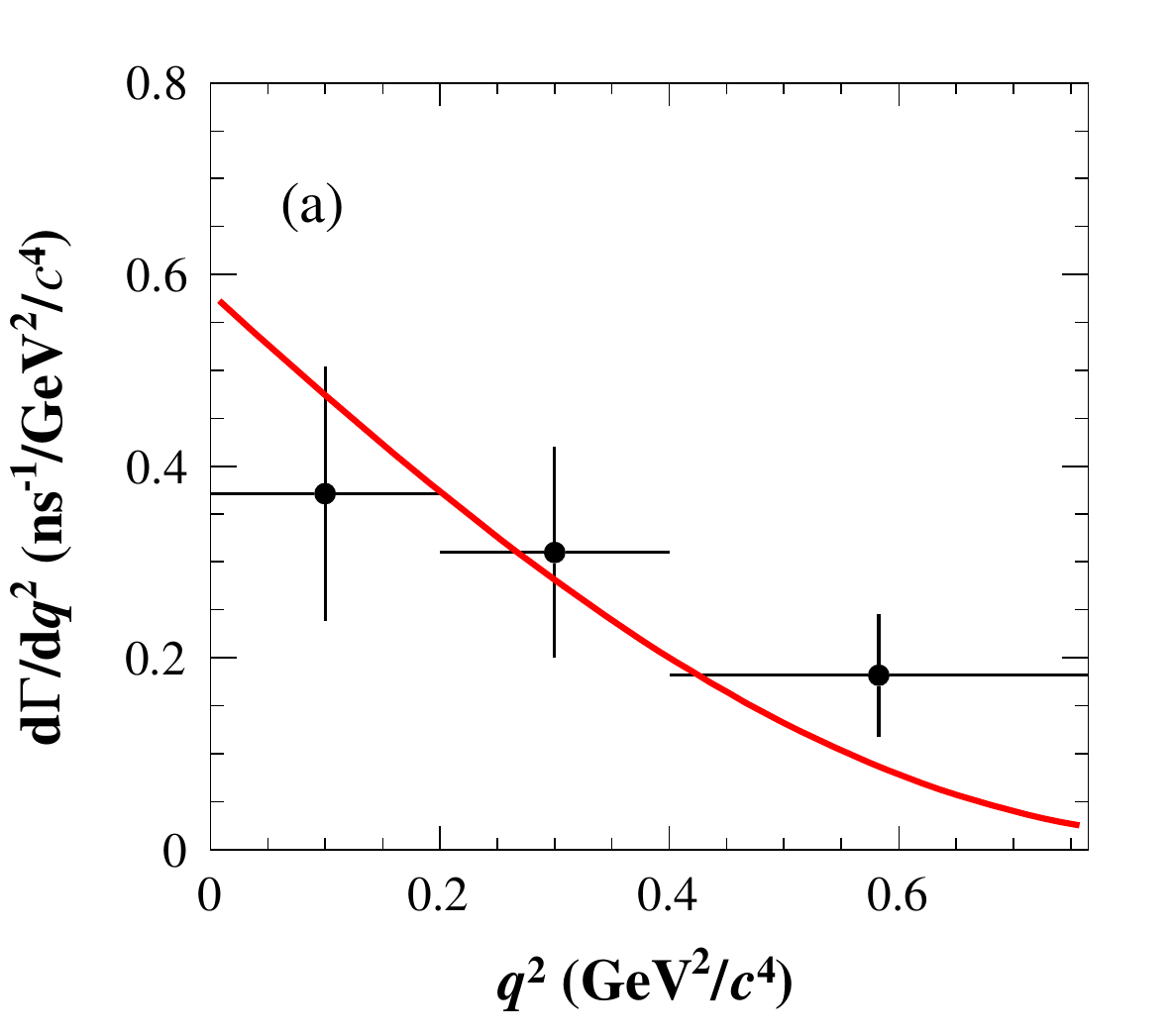}
            }
            \put(0, 0){
            \includegraphics[width=0.25\textwidth]{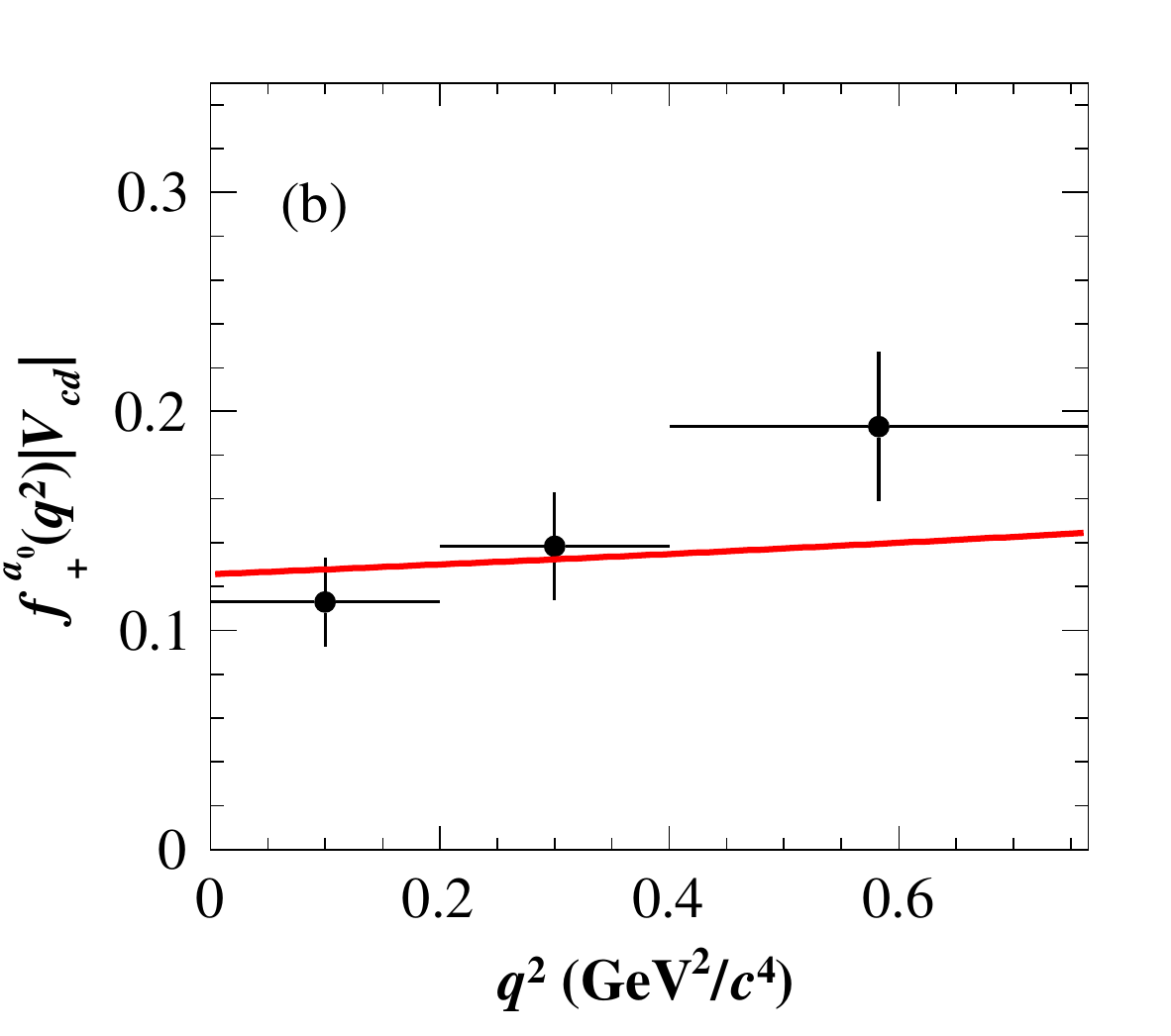}
            }
        }
    \end{center}
    \vspace{-0.5cm}
    \caption{
(a) The fit to the differential decay rate as a function of $q^2$ and (b) projection to $f^{D\to a_{0}(980)}_{+}(q^2)$~\cite{BESIII:2024zvp}.}
    \label{fig:DDR&&FF}
\end{figure}

Reference~\cite{BESIII:2018sjg} reported the first observation of $D^0\to a_0(980)^-e^+\nu_e$ 
and evidence for $D^+ \to a_0(980)^0 e^+ \nu_e$ with 2.93 fb$^{-1}$ of data at 3.773 GeV.
From the data sample, about 26 $D^0 \to a_0(980)^- e^+ \nu_e$ signal
events and 10 $D^+ \to a_0(980)^0 e^+ \nu_e$ signal
events are observed.
Projections of the 2-D unbinned
maximum likelihood fits to the $M_{\eta\pi}$ versus $U$ distributions are shown in Fig.~\ref{fig:D_a0980enu}.
The obtained products of branching fractions are
${\cal B}(D^0 \to a_0(980)^- e^+ \nu_e)\times {\cal B}(a_0(980)^-\to \eta\pi^-)=(1.33^{+0.33}_{-0.29}\pm0.09)\times 10^{-4}$
and
${\cal B}(D^+ \to a_0(980)^0 e^+ \nu_e)\times {\cal B}(a_0(980)^0\to \eta\pi^0)=(1.66^{+0.81}_{-0.66}\pm0.11)\times 10^{-4}$.
Using 7.9 fb$^{-1}$ of data at 3.773 GeV,
Ref.~\cite{BESIII:2024zvp} reported
an updated analysis of $D^0 \to a_0(980)^- e^+ \nu_e$.
From a sample of 6.3 million tagged $\bar D^0$ mesons,
52 $D^0 \to a_0(980)^- e^+ \nu_e$ signal events are obtained.
The obtained branching fraction is
${\cal B}(D^0 \to a_0(980)^- e^+ \nu_e, a_0(980)^- \to\eta\pi^-)=(0.86\pm0.17\pm0.05)\times10^{-4}$.
A fit is performed for the differential decay rate measured through the process $D^{0} \to a_0(980)^{-} e^{+} \nu_{e}$ and the results are presented in Figs.~\ref{fig:DDR&&FF}(a) and \ref{fig:DDR&&FF}(b), where the fitted decay rate and its projection onto the hadronic transition form factor are shown.
The fit to its differential decay rates gives $f_{+}^{D\to a_0(980)}(0)|V_{cd}|=0.126\pm0.013\pm0.003$.
Using the value of $|V_{cd}|=0.22487\pm0.00068$ given by the SM-constrained fit~\cite{ParticleDataGroup:2024cfk}, the $f_{+}^{D\to a_0(980)}(0)$ is measured to be $0.559\pm0.056\pm0.013$.

\begin{figure}[htbp]
  \centering
  \includegraphics[width=0.4\textwidth]{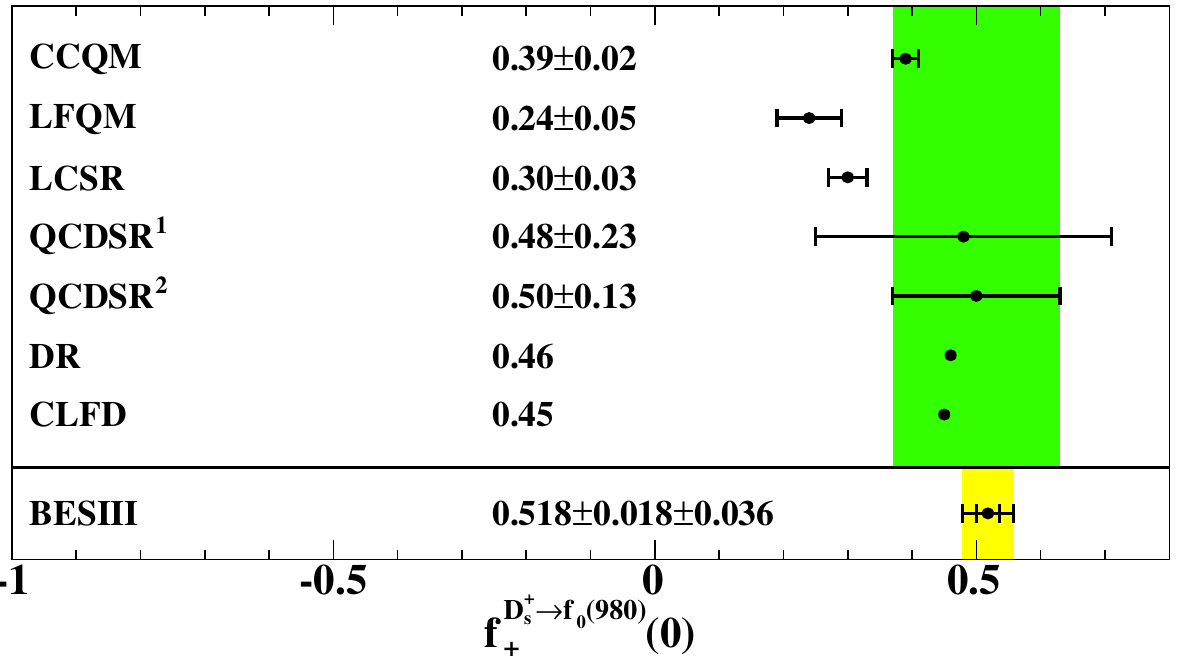}
  \put(-160,102){\scriptsize~\cite{Soni:2020sgn}}
  \put(-160,92){\scriptsize~\cite{Ke:2009ed}}
  \put(-160,82){\scriptsize~\cite{Colangelo:2010bg}}
  \put(-160,72){\scriptsize~\cite{Aliev:2007uu}}
  \put(-160,62){\scriptsize~\cite{Bediaga:2003zh}}
  \put(-160,51){\scriptsize~\cite{El-Bennich:2008rkp}}
  \put(-160,40){\scriptsize~\cite{El-Bennich:2008rkp}}
  \put(-160,22){\scriptsize~\cite{BESIII:2023wgr}}
   \caption{Comparison of the 
$f^{D_s\to f_0(980)}_+(0)$ measured by BESIII~\cite{BESIII:2023wgr} and those predicted by theoretical calculations of
CCQM~\cite{Soni:2020sgn}.
LFQM~\cite{Ke:2009ed},
LCSR~\cite{Colangelo:2010bg},
QCDSR$^{1}$~\cite{Aliev:2007uu},
QCDSR$^{2}$~\cite{Bediaga:2003zh},
DR~\cite{El-Bennich:2008rkp}, and
CLFD~\cite{El-Bennich:2008rkp}. The green band is the $\pm 1\sigma$ region of the  QCDSR result~\cite{Bediaga:2003zh} and the yellow band denotes the $\pm 1\sigma$ region of the BESIII result~\cite{BESIII:2023wgr}.
}
  \label{fig:fDsf0980}
\end{figure}

\begin{figure}[htbp]
  \centering
  \includegraphics[width=0.4\textwidth]{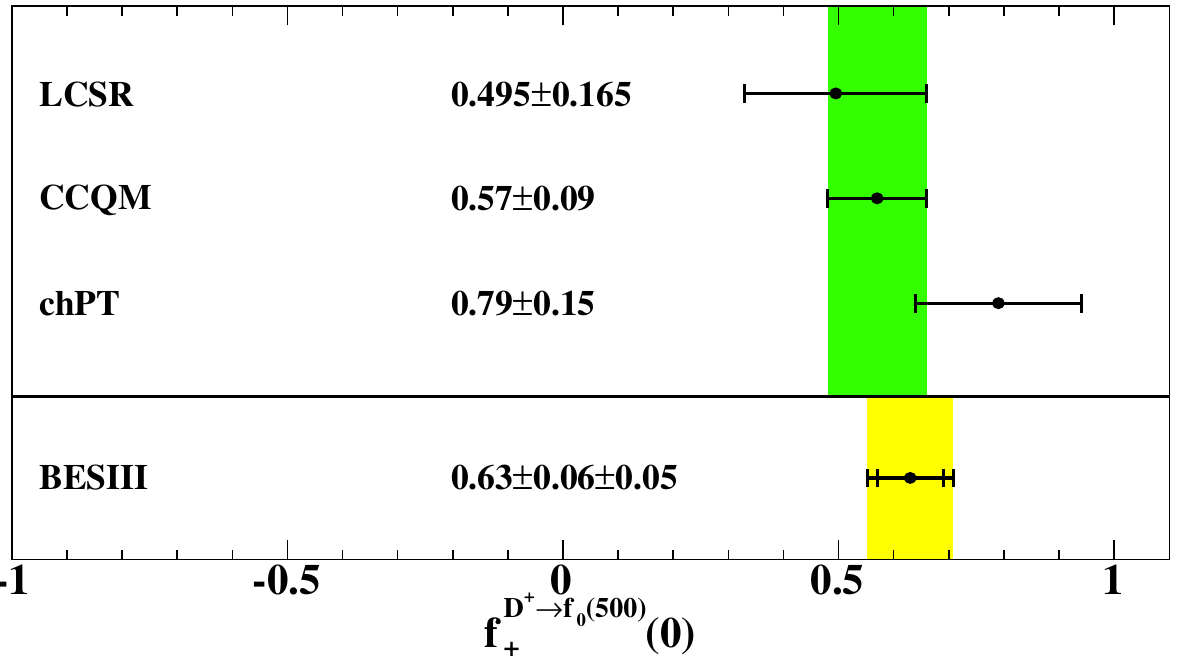}
  \put(-160,96){\scriptsize~\cite{Hsiao:2023qtk}}
  \put(-160,78){\scriptsize~\cite{Dosch:2002rh}}
  \put(-160,60){\scriptsize~\cite{Gatto:2000hj}}
  \put(-160,30){\scriptsize~\cite{BESIII:2024lnh}}
   \caption{Comparisons of hadronic transition form factors of $D$ transition into scalar mesons:
$f_+^{D\to f_0(500)}(0)$ measured by BESIII~\cite{BESIII:2024lnh} and those predicted by theoretical calculations
of LCSR~\cite{Hsiao:2023qtk}, CCQM~\cite{Dosch:2002rh}, and chPT~\cite{Gatto:2000hj}. The green band is the $\pm 1\sigma$ region of the  CCQM result~\cite{Dosch:2002rh} and the yellow band denotes the $\pm 1\sigma$ region of the BESIII result~\cite{BESIII:2024lnh}.
}
  \label{fig:fDf0500}
\end{figure}

\begin{figure}[htbp]
  \centering
  \includegraphics[width=0.4\textwidth]{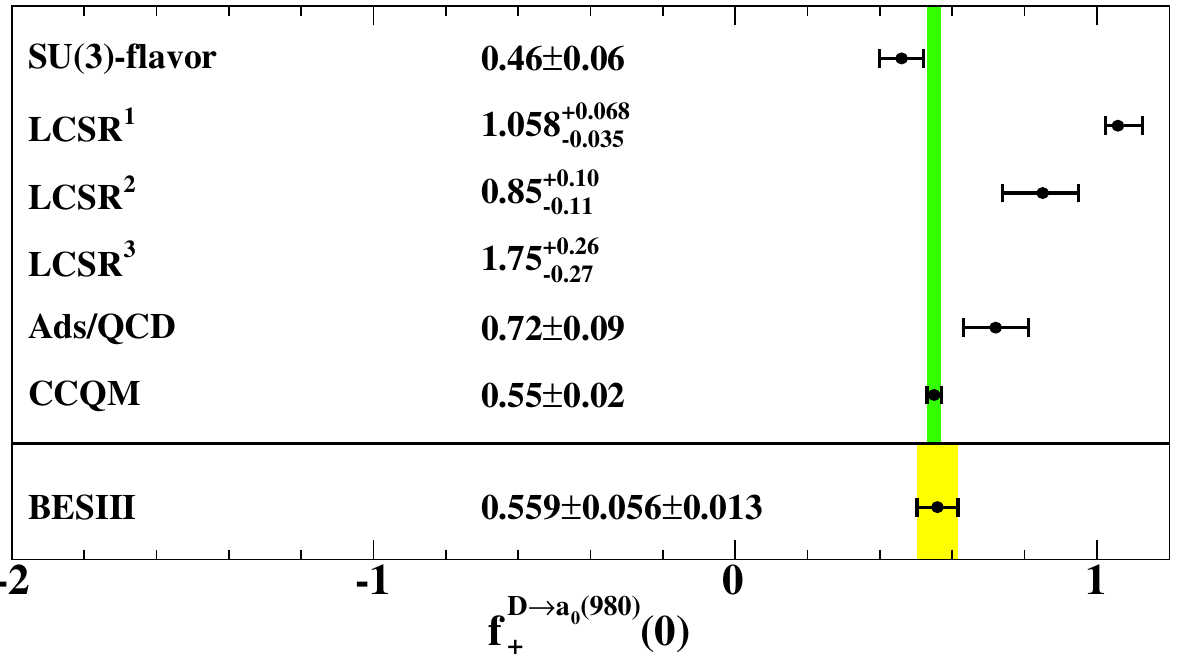}
  \put(-160,102){\scriptsize~\cite{Hsiao:2023qtk}}
  \put(-160,90){\scriptsize~\cite{Wu:2022qqx}}
  \put(-160,78){\scriptsize~\cite{Huang:2021owr}}
  \put(-160,66){\scriptsize~\cite{Cheng:2017fkw}}
  \put(-160,54){\scriptsize~\cite{Momeni:2022gqb}}
  \put(-160,42){\scriptsize~\cite{Soni:2020sgn}}
  \put(-160,25){\scriptsize~\cite{BESIII:2024zvp}}
   \caption{Comparisons of hadronic transition form factors of $D$ transition into scalar mesons:
  $f^{D\to a_0(980)}_+(0)$ measured by BESIII~\cite{BESIII:2024zvp} and those predicted by theoretical calculations
  of
  SU(3)-flavor~\cite{Hsiao:2023qtk},
  LCSR$^1$~\cite{Wu:2022qqx},
  LCSR$^2$~\cite{Huang:2021owr},
  LCSR$^3$~\cite{Cheng:2017fkw},
  Ads/QCD~\cite{Momeni:2022gqb}, and
  CCQM~\cite{Soni:2020sgn}. The green band is the $\pm 1\sigma$ region of the  CCQM result~\cite{Soni:2020sgn} and the yellow band denotes the $\pm 1\sigma$ region of the BESIII result~\cite{BESIII:2024zvp}.
  }
  \label{fig:fDa0980}
\end{figure}

Figures~\ref{fig:fDsf0980}, \ref{fig:fDf0500}, and \ref{fig:fDa0980} show comparisons of
hadronic transition form factors of $D$ transition into scalar mesons:
$f^{D_s\to f_0(980)}_+(0)$, $f^{D\to f_0(500)}_{+}(0)$, and $f^{D\to a_0(980)}_+(0)$,
from experimental measurements and different theoretical calculations, respectively.

Using 6.3 fb$^{-1}$ of data recorded at 4.178-4.226~GeV, the decay
$D_s^+\to a_0(980)^0 e^+\nu_e,\,a_0(980)^0\to \pi^0\eta$, which could proceed via $a_0(980)$-$f_0(980)$ mixing,
is searched for the first time~\cite{BESIII:2021tfk}.
No significant signal is observed and an upper limit on the product of the branching fractions of
$D_{s}^{+}\to a_0(980)^0 e^+\nu_e$ and $a_0(980)^0\to \pi^0\eta$ at the $90\%$ confidence level
is set as $1.2 \times 10^{-4}$.

\subsection{Results of $D\to A\ell^+\nu_\ell$ at BESIII}

\begin{figure*}[htbp]
	\centering
	\setlength{\abovecaptionskip}{0.0cm}
	\includegraphics[width=18.5cm]{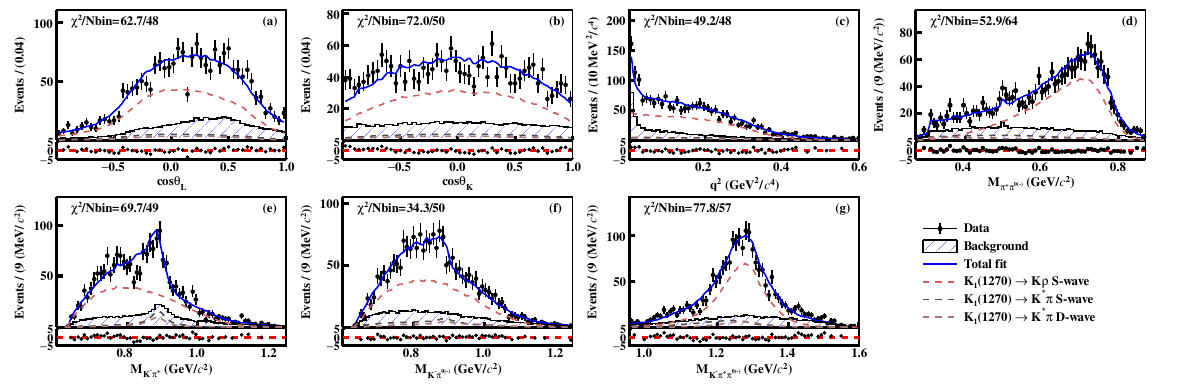}
	\caption{The projections of the amplitude analysis for $D^{+(0)}\to K^-\pi^+\pi^{0(-)}e^+\nu_e$ on (a) $\cos\theta_{L}$, (b) $\cos\theta_{\bar K}$, (c) $q^2$, (d) $M_{\pi^+\pi^{0(-)}}$, (e) $M_{K^-\pi^+}$, (f) $M_{K^-\pi^{0(-)}}$, and (g) $M_{K^-\pi^+\pi^{0(-)}}$~\cite{BESIII:2025hdt}.}
	\label{fig:D_K1enu_FF}
\end{figure*}

The semileptonic decays $D^+ \to \bar K_1(1270)^0 e^+ \nu_e$ and $D^0 \to K_1(1270)^- e^+ \nu_e$ were first observed by BESIII
using 2.93~fb$^{-1}$ of data at 3.773~GeV. With about one hundred signal events for each decay,
only branching fractions were measured~\cite{BESIII:2019eao,BESIII:2021uqr}.
In 2025, using 20.3~fb$^{-1}$ of 3.773~GeV~data, the decay dynamics of the semileptonic decays $D^{+(0)}\to K^-\pi^+\pi^{0(-)} e^+\nu_e$ have been studied for the first time, with about 2.1k signal events in total~\cite{BESIII:2025hdt}. An amplitude analysis provides the hadronic transition form factors for the semileptonic $D$ transitions into the axial-vector meson $\bar{K}_1(1270)$ to be $r_A=(-11.2\pm1.0\pm0.9)\times10^{-2}$ and $r_V = (-4.3\pm1.0\pm2.5)\times 10^{-2}$.
Figure~\ref{fig:D_K1enu_FF} shows the projections of the nominal fit result.
This is the first measurement of hadronic transition form factors in semileptonic decays of heavy mesons into axial-vector mesons. An angular analysis yields an up-down asymmetry $\mathcal{A}^\prime_{ud} = 0.01\pm0.11$, consistent with the SM prediction. Furthermore, the branching fractions of $D^+\to \bar K_1(1270)^0 e^+\nu_e$ and
$D^0\to K_1(1270)^- e^+\nu_e$ are determined with improved precision to be $(2.27\pm0.11\pm0.07\pm0.07)\times10^{-3}$ and $(1.02\pm0.06\pm0.06\pm0.03)\times10^{-3}$, respectively. The mass and width of $\bar K_1(1270)$ are determined to be $(1271\pm3\pm7)$~MeV/$c^2$ and $(168\pm10\pm20)$ MeV, respectively;
and the branching fraction ratio ${\cal B}(\bar K_1(1270)\to \bar K^*\pi)/{\cal B}(\bar K_1(1270)\to \bar K\rho(770))=(20.3\pm2.1\pm8.7)\%$ is obtained.
The $D^+\to \bar K_1(1400)^0 e^+\nu_e$ and
$D^0\to K_1(1400)^- e^+\nu_e$ decays are also searched for the first time, but no significant signals are observed; their branching fraction upper limits  at the 90\% confidence level are set as $1.4\times10^{-4}$ and $0.7\times10^{-4}$, respectively.

\begin{figure}[htbp]
	\vspace{-0.4cm}
	\setlength{\belowcaptionskip}{0cm}
	\centering
	\setlength{\abovecaptionskip}{-0.3cm}
	\includegraphics[width=8.5cm]{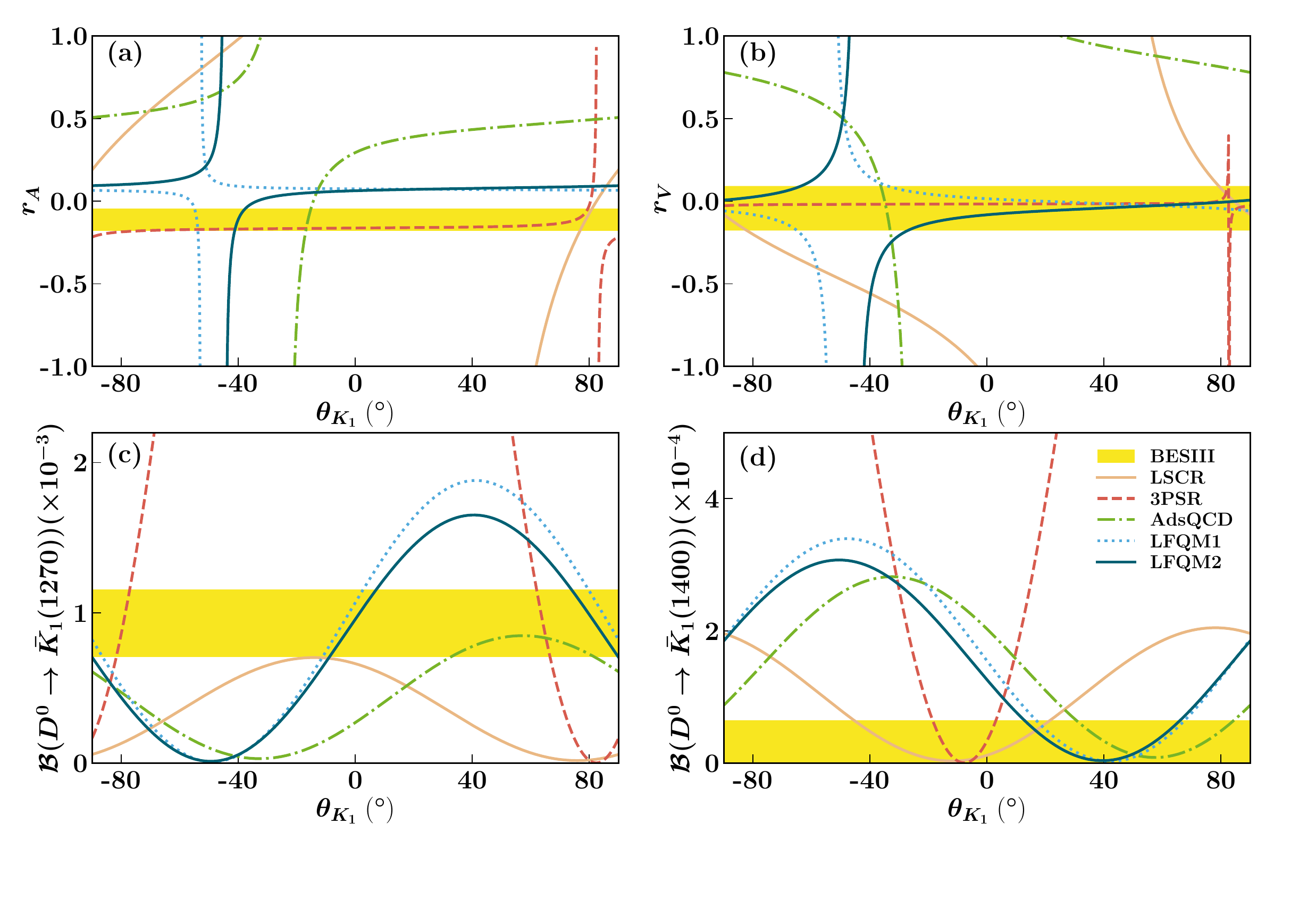}
	\caption{Comparisons of (a)~$r_A$, (b)~$r_V$, (c) $\mathcal{B}(D^0 \to K(1270)^-e^+\nu_e)$, and (d) $\mathcal{B}(D^0 \to K(1400)^- e^+\nu_e)$ measured
 by BESIII~\cite{BESIII:2025hdt} and predicted by various theoretical approaches (LSCR~\cite{Momeni:2019uag}, 3PSR~\cite{Khosravi:2008jw}, AdsQCD~\cite{Momeni:2022gqb}, LFQM1~\cite{Cheng:2003sm}, LFQM2~\cite{Verma:2011yw,Cheng:2017pcq}) as a function of $\theta_{\bar K_1}$. The yellow bands show $\pm 5\sigma$ limit for (a), (b), (c), and the upper limit at the 90\% confidence level for (d).}
	\label{fig:exp_th_compare}
\end{figure}

Figure~\ref{fig:exp_th_compare} shows comparisons of the measured hadronic transition form factors and branching fractions with different theoretical predictions~\cite{Momeni:2019uag,Momeni:2022gqb,Khosravi:2008jw,Cheng:2003sm,Verma:2011yw} as a function of $\theta_{\bar K_1}$. The obtained $r_A$, $r_V$, and $\mathcal{B}(D\to \bar K_1(1270) e^+\nu_e)$ are consistent with the 3PSR predictions~\cite{Khosravi:2008jw} with $\theta_{\bar K_1}\in(61,67)^\circ$,
and disfavor all other theoretical calculations by more than $5\sigma$. However, the upper limits on $\mathcal{B}(D\to \bar K_1(1400)e^+\nu_e)$ disfavor the corresponding 3PSR predictions. More universal theoretical calculations for all four variables are still desired. Additionally, the up-down asymmetry $\mathcal{A}^\prime_{ud}$ is extracted for the first time and no new physics effect is found with the current statistics. Forthcoming larger $B\to \bar K_1(1270)\gamma$~\cite{Belle-II:2022cgf,LHCb:2023hlw} and $D\to \bar K_1(1270) e^+\nu_e$~\cite{Cheng:2022tog} samples are expected to provide more effective restriction on the right-handed couplings in new physics models.

Using 2.93 fb$^{-1}$ of data at 3.773~GeV, the semileptonic decays $D^0\to K_S^0\pi^-\pi^0 e^+ \nu_e$ and $D^+\to K_S^0\pi^+\pi^- e^+ \nu_e$ were observed for the first time~\cite{BESIII:2024ieo},  with statistical significance of 5.4$\sigma$ and 5.6$\sigma$, respectively. When combined with the measurements of the $\bar K_1(1270)\to K^+\pi^-\pi$ decays, one finds the branching fractions $\mathcal{B}(D^0\to  K_1(1270)^-e^+\nu_e)=(1.08\substack{+0.14\\-0.13}\substack{+0.08\\-0.10}\pm0.21)\times10^{-3}$ and $\mathcal{B}(D^+\to \bar{K}_1(1270)^0e^+\nu_e)=(1.70^{+0.26}_{-0.23}\pm0.13\pm0.35)\times10^{-3}$.
Using 20.3~fb$^{-1}$ of data at 3.773~GeV, the semileptonic decays \mbox{$D^0\to K^-\omega e^+\nu_e$} and \mbox{$D^+\to K_S^0\omega e^+\nu_e$}
were observed for the first time~\cite{BESIII:2026zck}, with significances of $8.0\sigma$ and $5.8\sigma$, respectively.
Their decay branching fractions were measured to be ${\cal B}(D^0\to K^-\omega e^+\nu_e)=(9.4^{+2.0}_{-1.8}\pm0.6)\times10^{-5}$ and ${\cal B}(D^+\to K_S^0\omega e^+\nu_e)=(8.0^{+2.4}_{-2.1}\pm0.7)\times10^{-5}$.
Combining with the latest measurements of $D^{0(+)}\to  K^-\pi^+\pi^{-(0)} e^+\nu_e$ and assuming $\bar{K}_1(1270)$ to be the sole mediating resonance in all processes,
one finds the branching ratios to be
$\Gamma[K_1(1270)^-\to K^-\pi^+\pi^-]/\Gamma[K_1(1270)^-\to K^-\omega] = 3.4^{+0.8}_{-0.7}\pm0.3$ and
$\Gamma[\bar{K}_1(1270)^0\to K^-\pi^+\pi^0]/\Gamma[\bar{K}_1(1270)^0\to \bar{K}^0\omega] = 7.9^{+2.4}_{-2.1}\pm0.7$.
The combined branching fraction was determined to be
${\cal B}(\bar{K}_1(1270)\to \bar{K}\omega) = (8.4\pm 1.4\pm0.5)\%$, which is the most precise measurement from a collider experiment.

Analyzing 7.9~fb$^{-1}$ of data at 3.773~GeV, the semimuonic decays of $D^+\to \bar K_1(1270)^0\mu^+\nu_\mu$ and $D^0\to K_1(1270)^-\mu^+\nu_\mu$ were observed for the first time~\cite{BESIII:2025yot}, with statistical significances of $12.5\sigma$ and $6.0\sigma$, respectively.
Their decay branching fractions were determined to be  ${\cal B}(D^{+}\to \bar{K}_1(1270)^0 \mu^{+}\nu_{\mu})=(2.36\pm0.20^{+0.18}_{-0.27}\pm0.48)\times10^{-3}$ and ${\cal B}(D^{0}\to K_1(1270)^{-} \mu^{+}\nu_{\mu})=(0.78\pm0.11^{+0.05}_{-0.09}\pm0.15)\times10^{-3}$.
Combining these two branching fractions with the previous measurements of ${\cal B}(D^+\to \bar K_1(1270)^0e^+\nu_e)$ and ${\cal B}(D^0\to K_1(1270)^-e^+\nu_e)$ gives the branching fraction ratios to be
${\cal B}(D^+\to \bar K_1(1270)^0\mu^+\nu_{\mu})/{\cal B}(D^+\to \bar K_1(1270)^0e^+\nu_e)=1.03\pm0.14\substack{+0.11\\-0.15}$ and
${\cal B}(D^0\to K_1(1270)^-\mu^+\nu_{\mu})/{\cal B}(D^0\to K_1(1270)^-e^+\nu_e)=0.74\pm0.13\substack{+0.08\\-0.13}$.
Using the measured branching fractions and the world-average lifetimes of the $D^+$ and $D^0$ mesons, the semimuonic partial decay width ratio was determined to be $\Gamma [D^+\to \bar K_1(1270)^0 \mu^+\nu_\mu]/\Gamma [D^0\to K_1(1270)^- \mu^+\nu_\mu]=1.22\pm0.10\substack{+0.06\\-0.09}$, which is consistent with unity as predicted by isospin conservation.

\begin{figure}[htbp]
  \centering
  \includegraphics[width=0.45\textwidth]{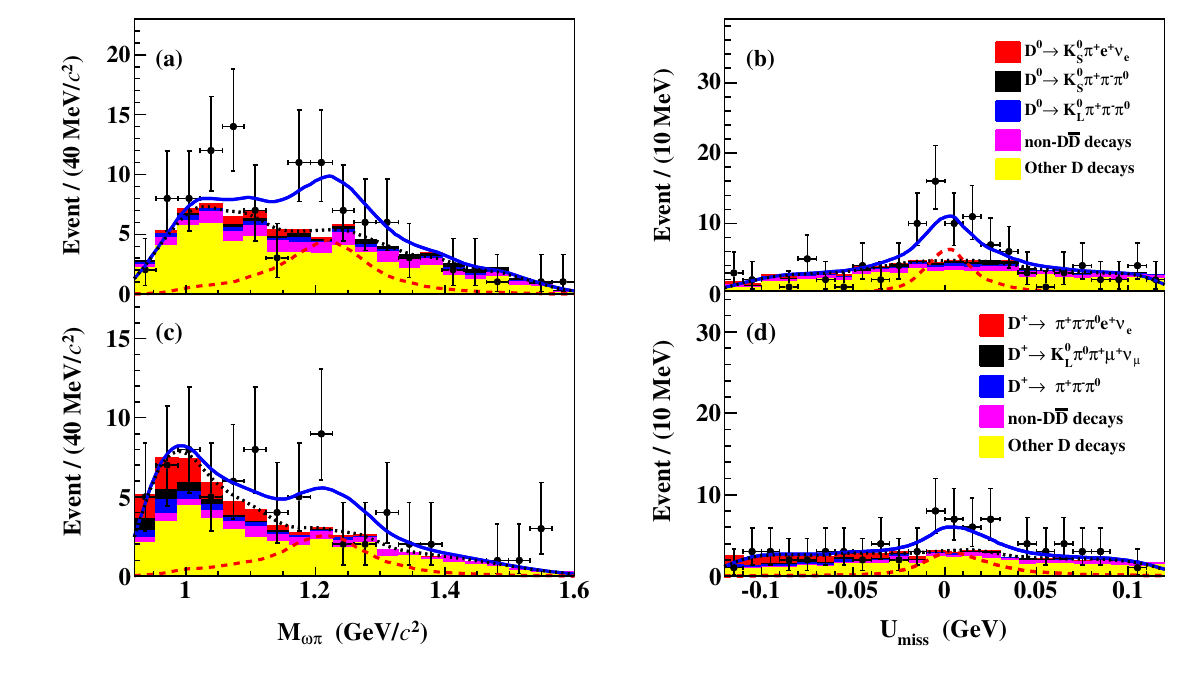}
  \caption{The 2-D fits to the accepted candidates for
  (top) $D^0\to b_1(1235)^-e^+\nu_e$ and (bottom) $D^+\to b_1(1235)^0e^+\nu_e$~\cite{BESIII:2024pwp}.
}
  \label{fig:D_b1enu}
\end{figure}

The earliest search for the semileptonic $D^{0(+)}$ decays into a $b_1(1235)^{-(0)}$ axial-vector meson was performed by
using 2.93~fb$^{-1}$ of data at 3.773~GeV; but no significant signal was observed for each decay~\cite{BESIII:2020jan}.
The upper limits on their product branching fractions at the 90\%  confidence level were set to be
${\cal B}(D^0\to b_1(1235)^- e^+\nu_e, b_1(1235)^-\to \omega\pi^-)<1.12\times 10^{-4}$ and ${\cal B}(D^+\to b_1(1235)^0 e^+\nu_e, b_1(1235)^0\to \omega\pi^0)<1.75\times 10^{-4}$.
In 2026, BESIII reported the first observation of $D^0\to b_1(1235)^-e^{+}\nu_e$ and the first evidence for $D^+\to b_1(1235)^0 e^+\nu_e$,
with significances of $5.2\sigma$  and $3.1\sigma$, respectively~\cite{BESIII:2024pwp}, from an analysis of 7.9~fb$^{-1}$ of data at 3.773~GeV.
Figure~\ref{fig:D_b1enu} shows the 2-D fit to the accepted candidates for $D^0\to \omega \pi^-e^+\nu_e$ and $D^+\to \omega \pi^0e^+\nu_e$.
Their product branching fractions are determined to be ${\cal B}(D^0\to b_{1}(1235)^-e^{+}\nu_e, b_1(1235)^-\to \omega \pi^-) = (0.72\pm0.18^{+0.06}_{-0.08})\times10^{-4}$  and ${\cal B}(D^+\to b_{1}(1235)^0e^{+}\nu_e, b_1(1235)^0~\to
\omega \pi^0) = (1.16\pm0.44\pm0.16)\times10^{-4}$. The ratio of their partial decay widths is determined to be
$\Gamma[D^0\to b_{1}(1235)^-e^{+}\nu_e]/2\Gamma[D^+\to b_{1}(1235)^0e^{+}\nu_e]=0.78\pm0.19^{+0.04}_{-0.05}$, which is consistent with unity, predicted by isospin invariance, within  uncertainties.

Using 7.33~fb$^{-1}$ of data at 4.128-4.226~GeV,
the semileptonic decays $D^+_s \to K_1(1270)^0 e^+\nu_e$ and $D^+_s \to b_1(1235)^0 e^+\nu_e$~\cite{BESIII:2023clm}
as well as
$D^+_s\to\ f_1(1420)e^+\nu_e$ and $D^+_s\to\ f_1(1285)e^+\nu_e$~\cite{BESIII:2026rsd} were also searched for the first time.
The first evidence for the decay $D^+_s\to\ f_1(1420)e^+\nu_e$ is found with a statistical significance of 3.4$\sigma$, and its product branching fraction
$\mathcal{B}(D^+_s\to\ f_1(1420)e^+\nu_e, f_1(1420)\to\ K^+K^-\pi^0)$ is determined to be $\rm (4.5^{+2.0}_{-1.7}\pm0.4) \times 10^{-4} $.
No significant signals of $D^+_s \to K_1(1270)^0 e^+\nu_e$ and  $D^+_s \to b_1(1235)^0 e^+\nu_e$, and $D^+_s\to\ f_1(1285)e^+\nu_e$ were observed and
the upper limit on their (product) branching fractions  at the 90\% confidence level was set to be
${\cal B}(D^+_s \to K_1(1270)^0 e^+\nu_e) < 4.1\times 10^{-4}$,
${\cal B}(D^+_s \to b_1(1235)^0 e^+\nu_e, b_1(1235)^0\to \omega\pi^0) < 6.4\times 10^{-4}$, and
$\mathcal{B}(D^+_s\to\ f_1(1285)e^+\nu_e, f_1(1285)\to\ \pi^+\pi^-\eta) < 1.7\times10^{-4}$, respectively.

\subsection{Other semileptonic $D$ decays at BESIII}

From an analysis of 2.93~fb$^{-1}$ of data at 3.773~GeV,
a search for the rare semileptonic decay $D^+\to D^0 e^+ \nu_e$ was performed~\cite{BESIII:2023exq}.
No obvious signal is observed, and its branching fraction upper limit  at the 90\% confidence level
was set to be $1.0\times10^{-4}$. In addition, BESIII also searched for
$D_{s}^{+} \to p \bar{p} e^{+} \nu_e$ for the first time~\cite{BESIII:2019kwo}, by analyzing 3.19 fb$^{-1}$ of data at 4.178 GeV.
With no obvious signal being observed, an upper limit on its decay branching fraction at the 90\% confidence level was set to be
$\mathcal{B}(D_{s}^{+} \to p\bar{p} e^{+}\nu_e)  < 2.0 \times 10^{-4}$.

Using 7.9 fb$^{-1}$ of data at 3.773 GeV, BESIII also searched for some other semileptonic decays of
$D\to PP' e^+\nu_e$~\cite{BESIII:2024jde}, where $P$ denotes a pseudoscalar meson.
The first evidence for $D^0\to K^-\eta e^+\nu_e$ with $11.1^{+4.5}_{-3.8}$ signal events (significance of $3.3\sigma$) was obtained,
corresponding to a branching fraction of
${\cal B}(D^0\to K^-\eta e^+\nu_e)=(0.84_{-0.34}^{+0.29}\pm0.22)\times 10^{-4}$.
No significant signals of
$D^0\to K^0_SK^-e^+\nu_e$,
$D^+\to K^0_SK^0_Se^+\nu_e$,
$D^+\to K^+K^-e^+\nu_e$,
$D^+\to K^0_S\eta e^+\nu_e$, and $D^+\to 2\eta e^+\nu_e$
were observed. Their branching fraction upper limits
at the 90\% confidence level were set to be
$2.1\times 10^{-5}$, $1.5\times 10^{-5}$, $2.1\times 10^{-5}$,
$2.0\times 10^{-4}$, and $1.0\times 10^{-4}$, respectively.

Based on an analysis of 6.3 fb$^{-1}$ of data at 4.178-4.226 GeV, BESIII also searched for
$D_s^+\to K_S^0 K_S^0 e^+ \nu_e$ but did not find any significant signal~\cite{BESIII:2021drk}.
Thus, an upper limit on its decay branching fraction at the 90\% confidence level is set to be $3.8\times 10^{-4}$.

By analyzing 6.3 fb$^{-1}$ of data at 4.178-4.226 GeV, about 1.7k inclusive $D_s^+ \to Xe^{+}\nu_e$
decay events are obtained with the only tag mode $D^-_s\to K^+K^-\pi^-$~\cite{BESIII:2021duu}.
Its branching fraction is determined to be
$\mathcal{B}(D_s^+ \to Xe^{+}\nu_e)=\left(6.30\pm0.13\pm 0.10\right)\%$, showing no evidence for unobserved
exclusive semielectronic modes. Combining it with external
data taken from literature determines the ratio of the $D_s^+$ and
$D^0$ semielectronic widths, $\frac{\Gamma(D_{s}^{+}\to
	Xe^+\nu_e)}{\Gamma(D^0\to Xe^+\nu_e)}=0.790\pm
0.016\pm0.020$. These results are
consistent with and more precise than previous measurements.

\subsection{Test of lepton flavor universality with semileptonic $D$ decays at BESIII}

Because the same hadronic transition form factors occur in the semimuonic and semielectronic
$D$ decays into the same particle, they cancel in the ratio of branching
fractions, thereby allowing for a model-independent test of lepton flavor universality.
Tantalizing indications of possible lepton flavor universality violations
were reported in some semileptonic $B$ decays, and these have stimulated
interest in more stringent tests in the charm sector.
These could provide useful constraints on models proposed
as explanations for the $B$ decay anomalies~\cite{Bifani:2018zmi}.
Table~\ref{tab:semi} gives a summary of
the branching fractions of $e$ and $\mu$ channels as well as their ratios,
the hadronic transition form factors for different semileptonic $D$ decays at BESIII.
No evidence for violation of lepton flavor universality was found.

The most precise results for CF and singly Cabibbo-suppressed (SCS) semileptonic $D$ decays
are from $D\to \bar K\ell^+\nu_\ell$~\cite{Ablikim:2018evp,Ablikim:2018frk,BESIII:2024slx,BESIII:2026uin},
$D\to \pi\ell^+\nu_\ell$~\cite{Ablikim:2018frk}, respectively.
As examples, the obtained branching fraction ratios for $D\to \bar K\ell^+\nu_\ell$:
 ${\mathcal R}_K^0  \equiv \frac{{\cal B}(D^0\to   K^-\mu^+\nu_\mu)}
 {{\cal B}(D^0\to   K^-e^+\nu_e)}=0.971\pm0.003\pm0.004$~\cite{BESIII:2026uin},
 and
 ${\mathcal R}_K^+  \equiv \frac{{\cal B}(D^+\to \bar{K}^0\mu^+\nu_\mu)}
 {{\cal B}(D^+\to \bar{K}^0e^+\nu_e)}=0.982\pm0.004\pm0.002$,
are in agreement with the lepton flavor universality expected value of
${\mathcal R}_K=0.975\pm 0.001$~\cite{Riggio:2017zwh,Soni:2017eug}, within
$0.1$ and $2.3$ standard deviations, respectively.
Likewise, the obtained results for $D\to \pi\ell^+\nu_\ell$:
 ${\mathcal R}_\pi^0\equiv \frac{{\cal B}(D^0\to \pi^-\mu^+\nu_\mu)}
 {{\cal B}(D^0\to \pi^-e^+\nu_e)}=0.922\pm0.030\pm0.022$,
 and
 ${\mathcal R}_\pi^+\equiv \frac{{\cal B}(D^+\to \pi^0\mu^+\nu_\mu)}
 {{\cal B}(D^+\to \pi^0e^+\nu_e)}=0.964\pm0.037\pm0.026$~\cite{Ablikim:2018frk},
are compatible with lepton flavor universality based theoretical expectations:
 ${\mathcal R}_\pi=0.985\pm 0.002$~\cite{Riggio:2017zwh,Soni:2017eug},
within $1.7$ and $0.5$ standard deviations, respectively.

In addition, investigations of the ratios of
differential branching fractions for different four-momentum
transfer regions were also performed for the semileptonic decays
$D\to \bar K\ell^+\nu_\ell$~\cite{Ablikim:2018evp,Ablikim:2018frk,BESIII:2024slx,BESIII:2026uin},
$D\to \pi\ell^+\nu_\ell$~\cite{Ablikim:2018frk},
$D\to \eta\ell^+\nu_\ell$~\cite{Ablikim:2020hsc,BESIII:2025hjc},
$D\to \eta^\prime\ell^+\nu_\ell$~\cite{BESIII:2024njj},
$D\to f_0(500)\ell^+\nu_\ell$~\cite{BESIII:2024lnh},
$D^+_s\to \eta\ell^+\nu_\ell$~\cite{BESIII:2023gbn,BESIII:2023ajr},
$D^+_s\to \eta^\prime\ell^+\nu_\ell$~\cite{BESIII:2023gbn,BESIII:2023ajr}, and
$D^+_s\to K^0\ell^+\nu_\ell$~\cite{BESIII:2025fso}.
No evidence for violation of lepton flavor universality was found.

\begin{sidewaystable*}[htbp]
	\centering
	\caption{Summary of the branching fractions of $e$ and $\mu$ channels as well as their ratios,
the hadronic transition form factors for different semileptonic $D$ decays.}
	\scalebox{0.55}{
	\begin{tabular}{ccccc}
		\hline\hline
		Decay & ${\cal B}_e$ & ${\cal B}_\mu$& ${\cal B}_\mu/{\cal B}_e$ & Form factor at $q^2=0$\\
		\hline
		$D^0 \to K^- \ell^+ \nu_\ell$~\cite{BESIII:2026uin} & $(3.548\pm0.006\pm0.017)\%$ & $(3.445\pm0.007\pm0.017)\%$ & $0.971\pm0.003\pm0.004$ & \multirow{2}{*}{$f_{+}=0.7378\pm0.0007\pm0.0014$} \\
		$D^+ \to \bar{K}^0 \ell^+ \nu_\ell$~\cite{BESIII:2026uin} & $(8.928\pm0.025\pm0.050)\%$ & $(8.770\pm0.029\pm0.053)\%$ & $0.982\pm0.004\pm0.002$ &  \\
		$D^0 \to \pi^- \ell^+ \nu_\ell$~\cite{BESIII:2015tql,BESIII:2018nzb} & $(0.295\pm0.004\pm0.003)\%$ & $(0.272\pm 0.008\pm0.006)\%$ & $0.922\pm0.030\pm0.022$ & $f_+=0.6372\pm0.0080\pm0.0044$ \\
		$D^+ \to \pi^0 \ell^+ \nu_\ell$~\cite{BESIII:2017ylw,BESIII:2018nzb} & $(0.363\pm0.008\pm0.005)\%$ & $(0.350\pm 0.011\pm0.010)\%$  & $0.964\pm0.037\pm0.026$ & $f_+ = 0.622\pm0.012\pm0.003$\\
		$D^+ \to \eta \ell^+ \nu_\ell$~\cite{BESIII:2025hjc} & $(9.75\pm0.29\pm0.28)\times10^{-4}$ & $(9.08\pm0.35\pm0.23)\times10^{-4}$ & $0.93\pm0.05\pm0.02$ & $f_+=0.345\pm0.008\pm0.003$ \\
		$D^+ \to \eta^\prime \ell^+ \nu_\ell$~\cite{BESIII:2024njj} & $(1.79\pm0.19\pm0.07)\times10^{-4}$ & $(1.92\pm0.28\pm0.08)\times10^{-4}$ & $1.07\pm0.19\pm0.03$ & $f_+=0.263\pm0.025\pm0.006$ \\ \hline
		
		$D^+ \to (K^-\pi^+)_{{\cal S}{\rm -wave}} \ell^+ \nu_\ell$~\cite{BESIII:2015hty} & $(0.228 \pm 0.008 \pm 0.008) \%$ & --- & --- & $r_S=-11.57\pm0.58\pm0.46$ \\
		$D^+ \to (K^0_S\pi^0)_{{\cal S}{\rm -wave}} \ell^+ \nu_\ell$~\cite{BESIII:2025fso} & $(6.28 \pm 0.56 \pm 0.37) \times 10^{-4}$ & $
		(5.73 \pm 0.16 \pm 0.13) \times 10^{-4}$ & $0.855 \pm 0.085 \pm 0.055$ & \makecell[c]{$r^{e}_S=-8.44\pm0.13\pm0.37$; $r^{\mu}_S=-9.59\pm0.46\pm0.58$} \\

		$D^0 \to (\bar{K}^0\pi^-)_{{\cal S}{\rm -wave}} \ell^+ \nu_\ell$~\cite{BESIII:2026txt} & $(8.4 \pm 0.3 \pm 0.1)\times10^{-4}$ & $(8.1 \pm 0.3 \pm 0.1)\times10^{-4} $ & $0.961\pm0.012\pm0.005$ & $r_S=-13.21\pm0.49\pm0.36$ \\
		$D^0 \to (K^-\pi^0)_{{\cal S}{\rm -wave}} \ell^+ \nu_\ell$~\cite{BESIII:2026ssp} & $(4.62 \pm 0.15 \pm 0.16) \times 10^{-4}$ & --- & --- & $r_S=-7.53\pm0.22\pm0.11$ \\
		$D^+ \to a_0(980)^0(\to\eta\pi^0) \ell^+ \nu_\ell$~\cite{BESIII:2018sjg} & $\mathcal{B}\times\mathcal{B}_{a_0(980)^0}=(1.66^{+0.81}_{-0.66}\pm0.11)\times 10^{-4}$ & --- & --- & --- \\
		$D^0 \to a_0(980)^-(\to\eta\pi^-) \ell^+ \nu_\ell$~\cite{BESIII:2024zvp} & $\mathcal{B}\times\mathcal{B}_{a_0(980)^-}=(0.86\pm0.17\pm0.05)\times10^{-4}$ & --- & --- & $f_+ = 0.559\pm0.056\pm0.013$ \\
		$D^+ \to f_0(500)(\to\pi^+\pi^-) \ell^+ \nu_\ell$~\cite{BESIII:2024lnh} & $\mathcal{B}\times\mathcal{B}_{f_0(500)}=(0.60\pm0.06\pm0.05)\times10^{-3}$ & $\mathcal{B}\times\mathcal{B}_{f_0(500)^0}=(0.72\pm0.13\pm0.08)\times10^{-3}$ & $1.14\pm0.26$ & $f_+=0.63\pm0.06\pm0.05$ \\
		$D^+ \to f_0(980)(\pi^+\pi^-) \ell^+ \nu_\ell$~\cite{BESIII:2018qmf} & $\mathcal{B}\times\mathcal{B}_{f_0(980)}<0.028\times10^{-3}$ & --- & --- & --- \\ \hline

	 	$D^+ \to \bar{K}^{*}(892)^{0}(\to K^-\pi^+) \ell^+ \nu_\ell$~\cite{BESIII:2015hty} & $\mathcal{B}\times\mathcal{B}_{\bar{K}^{*}(892)^{0}}=(3.54 \pm 0.03 \pm 0.08) \%$ & --- & --- & \makecell[c]{$r_V=1.411\pm0.058\pm0.007$;$r_2=0.788\pm0.042\pm0.008$} \\
		$D^+ \to \bar{K}^{*}(892)^{0}(\to K^0_S\pi^0)\ell^+ \nu_\ell$~\cite{BESIII:2025fso} & $\mathcal{B}\times\mathcal{B}_{\bar{K}^{*}(892)^{0}}=(8.82 \pm 0.11 \pm 0.14) \times 10^{-3}$ & $\mathcal{B}\times\mathcal{B}_{\bar{K}^{*}(892)^{0}}=(8.32 \pm 0.16 \pm 0.12) \times 10^{-3}$ & $0.9433 \pm 0.0216 \pm 0.0202$ & \makecell[c]{$r_V=1.42\pm0.03\pm0.02$;$r_2=0.75\pm0.03\pm0.01$} \\
		
		$D^0 \to {K}^{*}(892)^{-}(\to\bar{K}^0\pi^-) \ell^+ \nu_\ell$~\cite{BESIII:2026txt} & $\mathcal{B}\times\mathcal{B}_{{K}^{*}(892)^{-}}=(1.362 \pm 0.012 \pm 0.009)\%$ & $\mathcal{B}\times\mathcal{B}_{{K}^{*}(892)^{-}}=(1.309 \pm 0.013 \pm 0.008)\%$& $0.961\pm0.012\pm0.005$ & \makecell[c]{$r_V=1.444\pm0.026\pm0.011$;$r_2=0.752\pm0.020\pm0.004$} \\

		$D^0 \to {K}^{*}(892)^{-}(\to K^-\pi^0) \ell^+ \nu_\ell$~\cite{BESIII:2026ssp} & $\mathcal{B}\times\mathcal{B}_{{K}^{*}(892)^{-}}=(7.403 \pm 0.061 \pm 0.048) \times 10^{-3}$ & --- & --- & \makecell[c]{$r_V=1.41\pm0.05\pm0.01$;$r_2=0.77\pm0.04\pm0.02$} \\
		$D^+ \to \rho(770)^0 \ell^+ \nu_\ell$~\cite{BESIII:2018qmf} & $(1.860\pm0.070\pm0.061)\times10^{-3}$ & --- & --- & \makecell[c]{$r_V=1.695\pm0.083\pm0.051$;$r_2=0.845\pm0.056\pm0.039$} \\
		$D^0 \to \rho(770)^- \ell^+ \nu_\ell$~\cite{BESIII:2018qmf,BESIII:2024lxg}& $(1.445\pm0.058\pm0.039)\times10^{-3}$ & $(1.439\pm0.033\pm0.027)\times10^{-3}$ & $0.996 \pm 0.046 \pm 0.033$ & \makecell[c]{$r^{e}_V=1.695\pm0.083\pm0.051$;$r^{e}_2=0.845\pm0.056\pm0.039$\\$r^{\mu}_V=1.548\pm0.079\pm0.041$;$r^{\mu}_2=0.823\pm0.056\pm0.026$} \\
		$D^+ \to \omega \ell^+ \nu_\ell$~\cite{BESIII:2015kin,Ablikim:2020tmg} & $(1.63\pm0.11\pm0.08)\times10^{-3}$ & $(1.77\pm0.18\pm0.11)\times10^{-3}$ & $1.05\pm0.14$ & \makecell[c]{$r_V=1.24\pm0.09\pm0.06$;$r_2=1.06\pm0.15\pm0.05$} \\
		$D^+ \to \phi \ell^+ \nu_\ell$~\cite{BESIII:2015kin} & $<1.3\times10^{-5}$ & --- & --- & --- \\\hline

		$D^0 \to \bar{K}_1(1270)^-(\to K^-\pi^+\pi^-)\ell^+ \nu_\ell$~\cite{BESIII:2025vza,BESIII:2025yot} & $(1.02\pm0.06\pm0.06\pm0.03)\times10^{-3}$ & $(0.78\pm0.11^{+0.05}_{-0.09}\pm0.15)\times10^{-3}$ & $0.74\pm0.13^{+0.08}_{-0.13}$ & \multirow{2}{*}{\makecell[c]{$r_A=(-11.2\pm1.0\pm0.9)\times10^{-2}$;$r_V=(-4.3\pm1.0\pm2.5)\times10^{-2}$}} \\
		$D^+ \to \bar{K}_1(1270)^0(\to K^-\pi^+\pi^0)\ell^+ \nu_\ell$~\cite{BESIII:2025vza,BESIII:2025yot} & $(2.27\pm0.11\pm0.07\pm0.07)\times10^{-3}$ & $(2.36\pm0.20^{+0.18}_{-0.27}\pm0.48)\times10^{-3}$ & $1.03\pm0.14^{+0.11}_{-0.15}$ &  \\

		$D^0 \to \bar K_1(1270)^-\ell^+ \nu_\ell$~\cite{BESIII:2024ieo} & $(1.08{^{+0.14}_{-0.13}}^{+0.08}_{-0.10}\pm0.21)\times10^{-3}$ & --- & --- & ---\\
		$D^0 \to \bar K_1(1270)^-(\to K^-\omega)\ell^+ \nu_\ell$~\cite{BESIII:2026zck} & $\mathcal{B}\times\mathcal{B}_{K_1(1270)}=(9.4^{+2.0}_{-1.8}\pm0.6)\times10^{-5}$ & --- & --- & ---\\
		
		$D^+ \to \bar K_1(1270)^0\ell^+ \nu_\ell$~\cite{BESIII:2024ieo} & $(1.70^{+0.26}_{-0.23}\pm0.13\pm0.35)\times10^{-3}$ & --- & --- & ---\\
		$D^+ \to \bar K_1(1270)^0(\to K^0_s\omega)\ell^+ \nu_\ell$~\cite{BESIII:2026zck} & $\mathcal{B}\times\mathcal{B}_{K_1(1270)}=(8.0^{+2.4}_{-2.1}\pm0.7)\times10^{-5}$ & --- & --- & ---\\
		$D^0 \to b_1(1235)^-(\to \omega \pi^-) \ell^+ \nu_\ell$~\cite{BESIII:2024pwp} & $\mathcal{B}\times\mathcal{B}_{b_1(1235)}=(0.72\pm0.18^{+0.06}_{-0.08})\times10^{-4}$ & --- & --- & ---\\
		$D^+ \to b_1(1235)^-(\to \omega \pi^0) \ell^+ \nu_\ell$~\cite{BESIII:2024pwp} & $\mathcal{B}\times\mathcal{B}_{b_1(1235)}=(1.16\pm0.44\pm0.16)\times10^{-4}$ & --- & --- & ---\\\hline
		
		$D^0 \to K^{*}_2(1430)^-(\to \bar K^0\pi^-) \ell^+ \nu_\ell$~\cite{BESIII:2026txt} & $\mathcal{B}\times\mathcal{B}_{{K}^{*}(1430)^{-}}=(1.879 \pm 0.571 \pm 0.367) \times 10^{-4}$ & $\mathcal{B}\times\mathcal{B}_{{K}^{*}(1430)^{-}}=(1.807 \pm 0.549 \pm 0.353) \times 10^{-4}$ & $0.961\pm0.012\pm0.005$ & --- \\
		$D^0 \to K^{*}_2(1430)^-(\to K^-\pi^0) \ell^+ \nu_\ell$~\cite{BESIII:2026txt} & $\mathcal{B}\times\mathcal{B}_{{K}^{*}(1430)^{-}}=(1.260 \pm 0.424 \pm 0.169) \times 10^{-4}$ & --- & --- & --- \\
		\hline\hline

		$D^+_s \to K^0 \ell^+ \nu_\ell$~\cite{BESIII:2024zft,BESIII:2025gov} & $(2.98\pm0.23\pm0.12)\times10^{-3}$ & $(2.89\pm0.27\pm0.12)\times10^{-3}$ & $0.97 \pm 0.12 \pm 0.04$ & $f_{+}=0.623\pm0.036\pm0.011$ \\
		$D^+_s \to \pi^0 \ell^+ \nu_\ell$~\cite{BESIII:2022jcm} & $<6.4\times10^{-5}$ & ---& --- & ---\\
		$D^+_s \to \eta \ell^+ \nu_\ell$~\cite{BESIII:2023gbn,BESIII:2023ajr} & $(2.255\pm0.039\pm0.051)\%$ & $(2.235\pm0.051\pm0.052)\times10^{-4}$ & $0.991\pm0.029\pm0.016$ & $f_+=0.4642\pm0.0073\pm0.0066$ \\
		$D^+_s \to \eta^\prime \ell^+ \nu_\ell$~\cite{BESIII:2023gbn,BESIII:2023ajr} & $(0.810\pm0.038\pm0.024)\%$ & $(0.801\pm0.055\pm0.028)\times10^{-4}$ & $0.988\pm0.082\pm0.031$ & $f_+=0.540\pm0.025\pm0.009$ \\ \hline

		$D^+_s \to a^0(980)^0(\to \pi^0\eta) \ell^+ \nu_\ell$~\cite{BESIII:2021tfk} & $\mathcal{B}\times\mathcal{B}_{a^0(980)}<1.2\times10^{-4}$ & --- & --- & --- \\
		$D^+_s \to f_0(500)(\to \pi^+\pi^-)\ell^+ \nu_\ell$~\cite{BESIII:2023wgr} & $\mathcal{B}\times\mathcal{B}_{f_0(500)}<3.3\times10^{-4}$ & --- & --- & --- \\
		$D^+_s \to f_0(500)(\to 2\pi^0)\ell^+ \nu_\ell$~\cite{BESIII:2021drk} & $\mathcal{B}\times\mathcal{B}_{f_0(500)}<7.3\times10^{-4}$ & --- & --- & --- \\
		$D^+_s \to f_0(980)(\to \pi^+\pi^-)\ell^+ \nu_\ell$~\cite{BESIII:2023wgr} & $\mathcal{B}\times\mathcal{B}_{f_0(980)}=(1.72\pm0.13\pm0.10)\times10^{-3}$ & --- & --- & $f_+ = 0.518\pm0.018\pm0.036$ \\
		$D^+_s \to f_0(980)(\to 2\pi^0)\ell^+ \nu_\ell$~\cite{BESIII:2021drk} & $\mathcal{B}\times\mathcal{B}_{f_0(980)}=(7.9\pm1.4\pm0.4)\times10^{-4}$ & --- & --- & --- \\  \hline
		
		$D^+_s \to K^*(892)^0\ell^+ \nu_\ell$~\cite{BESIII:2018xre} &$(2.37\pm0.26\pm0.20)\times10^{-3}$ & --- & --- & \makecell[c]{$r_V=1.67\pm0.34\pm0.16$;$r_2=0.77\pm0.28\pm0.07$}\\
		$D^+_s \to \phi\ell^+ \nu_\ell$~\cite{BESIII:2017ikf,BESIII:2023opt} &$(2.26\pm0.45\pm0.09)\%$ & $(2.25\pm0.09\pm0.07)\%$ & $0.94\pm0.08$ & \makecell[c]{$r_V=1.58\pm0.17\pm0.02$;$r_2=0.71\pm0.14\pm0.02$}\\\hline
		
		$D^+_s \to K_1(1270)^0\ell^+ \nu_\ell$~\cite{BESIII:2023clm} & $<4.1\times10^{-4}$ & --- & --- & --- \\
		$D^+_s \to b_1(1235)^0(\to\omega\pi^0)\ell^+ \nu_\ell$~\cite{BESIII:2023clm} & $\mathcal{B}\times\mathcal{B}_{b_1(1235)}<6.4\times10^{-4}$ & --- & --- & --- \\
		$D^+_s \to f_1(1420)^0(\to K^+ K^- \pi^0)\ell^+ \nu_\ell$~\cite{BESIII:2026rsd} & $\mathcal{B}\times\mathcal{B}_{f_1(1420)^0}=(4.5^{+2.0}_{-1.7}\pm0.4)\times10^{-4}$ & --- & --- & --- \\
		
		$D^+_s \to f_1(1285)^0(\to \pi^+ \pi^- \eta)\ell^+ \nu_\ell$~\cite{BESIII:2026rsd} & $\mathcal{B}\times\mathcal{B}_{f_1(1285)^0}<1.7\times10^{-4}$ & --- & --- & --- \\
		\hline
		\hline
	\end{tabular}
	}
	\label{tab:semi}
\end{sidewaystable*}

\section{Strong phase in $D^0$ decays}
\label{sec:strongphase}

In the SM, the charged-weak interaction in the quark sector is described by the CKM matrix~\cite{Cabibbo:1963yz,Kobayashi:1973fv}. One of the primary goals of flavor physics experiments is to determine the angles $\alpha, \beta$~and~$\gamma$ (or  $\phi_{2},\phi_{1}$ and $\phi_{3}$) of the $b-d$ CKM UT precisely. Currently, the most precise measurements of $\gamma$ are extracted using tree-level $B^{-}\to DK^{-}$ decays~\cite{Gronau:1991dp,Gronau:1990ra}. Here and elsewhere in this section $D$ refers to either a $D^{0}$ or a $\bar{D}^{0}$ meson decaying into the same final state and charge conjugation is implicit, unless stated otherwise.
The sensitivity to $\gamma$ arises from the interference of two amplitudes: $b\to c\bar{u}s$ that results in the $B^{-}\to D^{0}K^{-}$ decay, and $b\to u\bar c s$ that leads to the $B^{-}\to \bar{D}^{0}K^{-}$ decay. The latter amplitude is both CKM- and color-suppressed relative to the former.
The value of $\gamma$ measured with such tree-level transitions is insensitive to loop-level contributions \cite{Brod:2013sga}. Therefore, tests for new physics that are made by comparing UT parameters measured using tree and loop processes can be improved by more precise determinations of $\gamma$ \cite{Blanke:2018cya,Lenz:2019lvd}.
Based on $D$ decay products, there are three methods of determining $\gamma$, including $CP$ eigenstates (GLW method)~\cite{Gronau:1991dp,Gronau:1990ra}, flavor-eigenstates (ADS method)~\cite{Atwood:1996ci,Atwood:2000ck}, and self-conjugate multibody states (BPGGSZ method)~\cite{Giri:2003ty,Belle:2012ftx}.

\subsection{$(c_i,s_i)$ in $D\to K^0_{S,L}h^+h^-$~($h=\pi$ or $K$)}

The most widely used $D$ decays for the BPGGSZ method are $D\to K^{0}_{\rm{S}}h^{+}h^{-}$, where $h = \pi, K$.
The leading channel for the direct measurement of the $CP$-violating angle
$\gamma$ is $B^\pm \to D K^\pm$, $D \to K_S^0 \pi^+ \pi^-$~\cite{Giri:2003ty},
where $D$ is a superposition of $D^0$ and $\bar D^0$.
To determine the $\gamma$, one needs to input the strong-phase difference
between $D^0$ and $\bar D^0 \to K_S^0 \pi^+ \pi^-$ decays, $i.e.$,
\mbox{$\Delta\delta_{D} = \delta_{D}(m_{+}^{2},
	m_{-}^{2})-\delta_{D}(m_{-}^{2}, m_{+}^{2})$}, in which
$m_{\pm}^{2}$ is the squared mass of $K_{S}^{0}\pi^{\pm}$. Benefiting from
abundant intermediate processes in $D \to K_S^0 \pi^+ \pi^-$,
the $\Delta\delta_{D}$ varies across phase space, making this channel the
most sensitive to the $\gamma$.
Quantum-correlated (QC) $D\bar{D}$ pairs produced at the $\psi(3770)$
resonance provide the best environment to determine the $\Delta\delta_{D}$~\cite{Giri:2003ty}, allowing for a
model-independent measurement of the $\gamma$~\cite{LHCb:2020yot}, where the uncertainty from the strong-phase
difference can be reliably estimated.
In addition, the measured $\Delta\delta_{D}$ between $D^0$ and $\bar D^0 \to K_S^0 \pi^+ \pi^-$ can provide important inputs for
the determination of the UT angle $\beta$~\cite{Belle:2016ckr},
the study of charm mixing and $C\!P$ violation phenomena~\cite{LHCb:2021ykz},
and the measurement of strong-phase differences in other hadronic $D^0$ decays~\cite{BESIII:2022qkh}.

The $D \to K^0_{S,L}\pi^+\pi^-$ phase space is
divided into bins according to the equal binning scheme,
the optimal binning scheme and the modified optimal binning
scheme.  The detailed information on the choice of these regions is
given in Ref.~\cite{CLEO:2010iul}. These bins in the Dalitz plot are
symmetric with respect to the $m_{+}^{2}= m_{-}^{2}$ axis and are
indexed by $i$ from $-8$ to $8$, excluding zero. Positive (negative)
bins are located in the $m_{+}^{2}>m_{-}^{2}$ ($m_{+}^{2}<m_{-}^{2}$)
region.
The strong-phase difference $\Delta\delta_{D}$, which quantifies the
interference between the amplitudes of $D^0$ and $\bar D^0$ decays, is
parameterized using the amplitude-weighted averages of
$\cos\Delta\delta_{D}$ and $\sin\Delta\delta_{D}$ in each bin. These
parameters are defined as
\begin{equation}
\label{eq_cidefine}
    \begin{aligned}
        c_{i} &= \frac{1}{\sqrt{F_{i}F_{-i}}}\int_{i} |f_{D^0}(m_{+}^{2}, m_{-}^{2})| |f_{D^0}(m_{-}^{2}, m_{+}^{2})| \\
        & \cos[\Delta\delta_{D}(m_{+}^{2}, m_{-}^{2})]\rm{d} \mathit{m}_{+}^{2}\rm{d}
        \mathit{m}_{-}^{2},  \\
         s_{i} &= \frac{1}{\sqrt{F_{i}F_{-i}}}\int_{i} |f_{D^0}(m_{+}^{2}, m_{-}^{2})| |f_{D^0}(m_{-}^{2}, m_{+}^{2})|\\
          &\sin[\Delta\delta_{D}(m_{+}^{2}, m_{-}^{2})]\rm{d} \mathit{m}_{+}^{2}\rm{d} \mathit{m}_{-}^{2},
    \end{aligned}
\end{equation}
where $F_{i}$ represents the fraction of events found in the $i^{\rm th}$ bin of the decay $D^0 \to K^0_{S,L}\pi^+\pi^-$.
These measurements are based on events with tag modes reconstructed against the signal
$K_{S,L}^0\pi^+\pi^-$ decays, with categories of flavor, $C\!P$ eigenstate or self-conjugate.

\begin{figure*}
\centering
 \includegraphics[width=0.8\textwidth]{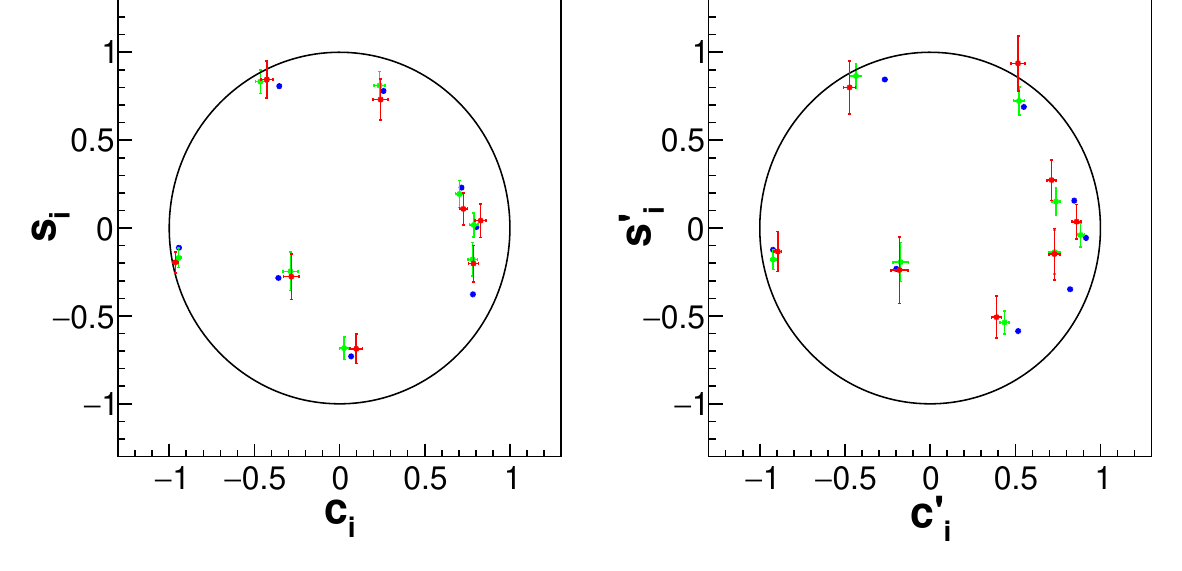}
\caption{The $c^{(\prime)}_{i}/s_i^{(\prime)}$ parameters determined without constraints (red), with constraints (green) and predicted by the amplitude models~\cite{BaBar:2018cka} (blue) under the modified optimal binning scheme. See Ref.~\cite{BESIII:2025nsp} for details.}
\label{fig:2Dci_nocs}
\end{figure*}

The strong-phase differences between $D^0$ and $\bar D^0\to K_{S}^0\pi^+\pi^-$ decays were determined for the first time
by the CLEO-c experiment using a dataset collected at the $\psi(3770)$ resonance with an integrated
luminosity of 818~fb$^{-1}$~\cite{CLEO:2010iul}.
BESIII reported improved measurements of these strong-phase parameters between $D^0$ and $\bar{D}^0$ decays to
$K^0_{S,L}\pi^+\pi^-$ by using 2.93 fb$^{-1}$ of data at 3.773 GeV~\cite{BESIII:2020hlg,BESIII:2020khq}.
The most precise measurements of the strong-phase parameters between $D^{0}$
and $\bar{D}^{0} \to K^{0}_{S,L}\pi^{+}\pi^{-}$ decays are reported by BESIII based on an analysis of
7.9 fb$^{-1}$ of data at 3.773 GeV~\cite{BESIII:2025nsp}.
The $c_{i}^{(\prime)}$ and $s_{i}^{(\prime)}$ parameters have been measured with or without the constraints of
model-predicted differences between the strong-phase parameters in the
$D^0 \to K_{S,L}^0 \pi^+ \pi^-$ decays, as shown in Fig.~\ref{fig:2Dci_nocs}.
The obtained results provide essential inputs for
the CKM angle $\gamma$ measurement at LHCb and Belle. The propagated
uncertainty from the constrained strong-phase inputs contributing to
the $\gamma$ measurement is found to be $0.9^{\circ}$ based on the optimal
binning scheme, where it is $1.5^{\circ}$ from the unconstrained
result. The constraints in the strong-phase parameters are expected to
be negligible on the current $\gamma$ measurement at LHCb and Belle, whose statistical uncertainty is about $5^{\circ}$~\cite{LHCb:2020yot}.

The first measurements of the strong-phase difference between $D^{0}$ and $\bar{D}^{0}$ decaying to the $K^{0}_{S,L}K^{+}K^{-}$ final state were reported by the CLEO-c Collaboration, using a data set corresponding to an integrated luminosity of 818~pb$^{-1}$ at the $\psi(3770)$ resonance~\cite{CLEO:2010iul}. Due to low statistics, fewer bins are used for $D \to K^0_SK^+K^-$.
Based on an analysis of 2.93 fb$^{-1}$ of data at 3.773 GeV, BESIII reported measurements of the strong-phase difference parameters for $D\to K^{0}_{S,L}K^{+}K^{-}$ decays,
which are the best to date.
The major systematic uncertainties are from the input strong-phase difference parameters of $D\to K^{0}_{S,L}\pi^{+}\pi^{-}$,
and the background parametrization of the partially reconstructed $D\to K^{0}_{L}X$ decay modes. Both of them depend on the size of the charm sample available.
The estimated uncertainty on the $\gamma$ measurement arising from the uncertainties on the measured values of $c_i$ and $s_i$  ranges between (1.3-2.3)$^\circ$ for $\mathcal{N}=2-4$ equal-$\Delta\delta_{D}$ binning.

Very recently, a joint LHCb and BESIII analysis reported a measurement of the CKM angle $\gamma$ and related strong-phase parameters with a novel, model-independent approach in
$B^\pm\to {D(\to K^0_S h^{\prime+}h^{\prime-}) h^{\pm}}$ decays, where $h^{(\prime)} \equiv \pi, K$~\cite{LHCb:2026hot,LHCb:2026npj}.
This analysis uses 7.9 fb$^{-1}$ of $e^+e^-$ data at $\psi(3770)$ by BESIII during 2010--2011 and 2021--2022
as well as 9 fb$^{-1}$ of $pp$ data collected by LHCb during 2011--2018.
The two datasets are analyzed simultaneously by applying per-event weights based on the amplitude variation over the $D$-decay phase space to enhance the sensitivity to $CP$-violating observables.
The CKM angle $\gamma$ is determined to be $\gamma = (71.3\pm 5.0)^\circ$, which constitutes the most precise single measurement to date
 and is consistent with previous measurements and world averages.

\subsection{Strong phase in  $D\to Kn\pi$~($n=1,2,3$)}

The decays $D \to K^- n \pi$ ($n \geq 1$)
can proceed via CF amplitudes, DCS amplitudes, or through their interference.
Measurements of strong phases in  $D\to Kn\pi$~($n=1,2,3)$ decays offer important inputs
for the determination of $\gamma$, and determinations of parameters of charm mixing and $C\!P$ violation.
Among these, the final states $D\to K^\pm\pi^\mp\pi^+\pi^-$ and $D\to K^\pm\pi^\mp\pi^0$ are important for the $\gamma$ measurement.
Their sensitivity to the $\gamma$ in analyses of $B \to Dh$ decays depends on knowledge of the hadronic parameters:
the coherence factor $R_S$, the amplitude ratio $r_D^{S}$, and the $C\!P$-conserving strong-phase difference $\delta_D^{S}$ between the CF and DCS amplitudes~\cite{Atwood:2003mj}. These parameters are usually defined as
\begin{equation}
R_{S}e^{-i\delta_{D}^{S}}=\frac{\int {\cal A}^{\star}_{S}(\textbf{x}){\cal A}_{\bar{S}}(\textbf{x})\rm d\textbf{x}}{A_{S}A_{\bar{S}}} {\;\;\;{\rm and}\;\;\;}
r^{S}_{D} = A_{\bar{S}} / A_{S}.
\label{eq:rds}
\end{equation}
Here, ${\cal A}_{S}(\textbf{x})$ represents the decay amplitude of $D^0 \to S$ at the point $\textbf{x}$ in the multi-body phase space, and $A^2_{S}=\int |{\cal A}_{S}(\textbf{x})|^2 {\rm d}\textbf{x}$, with an analogous expression for $\bar{S}$. The coherence factor, which lies between 0 and 1, quantifies the degree of interference and is influenced by the resonance structures of the final state.
The amplitude ratio parameter $r_D^{S}$, which characterizes the relative strength of the DCS to CF processes, is typically of order $\lambda^2 \approx 0.05$, where $\lambda$ is the Wolfenstein parameter defined in Ref.~\cite{Wolfenstein:1983yz}.

\begin{figure*}[!ht]
	\begin{center}
		\includegraphics[width=.4\textwidth]{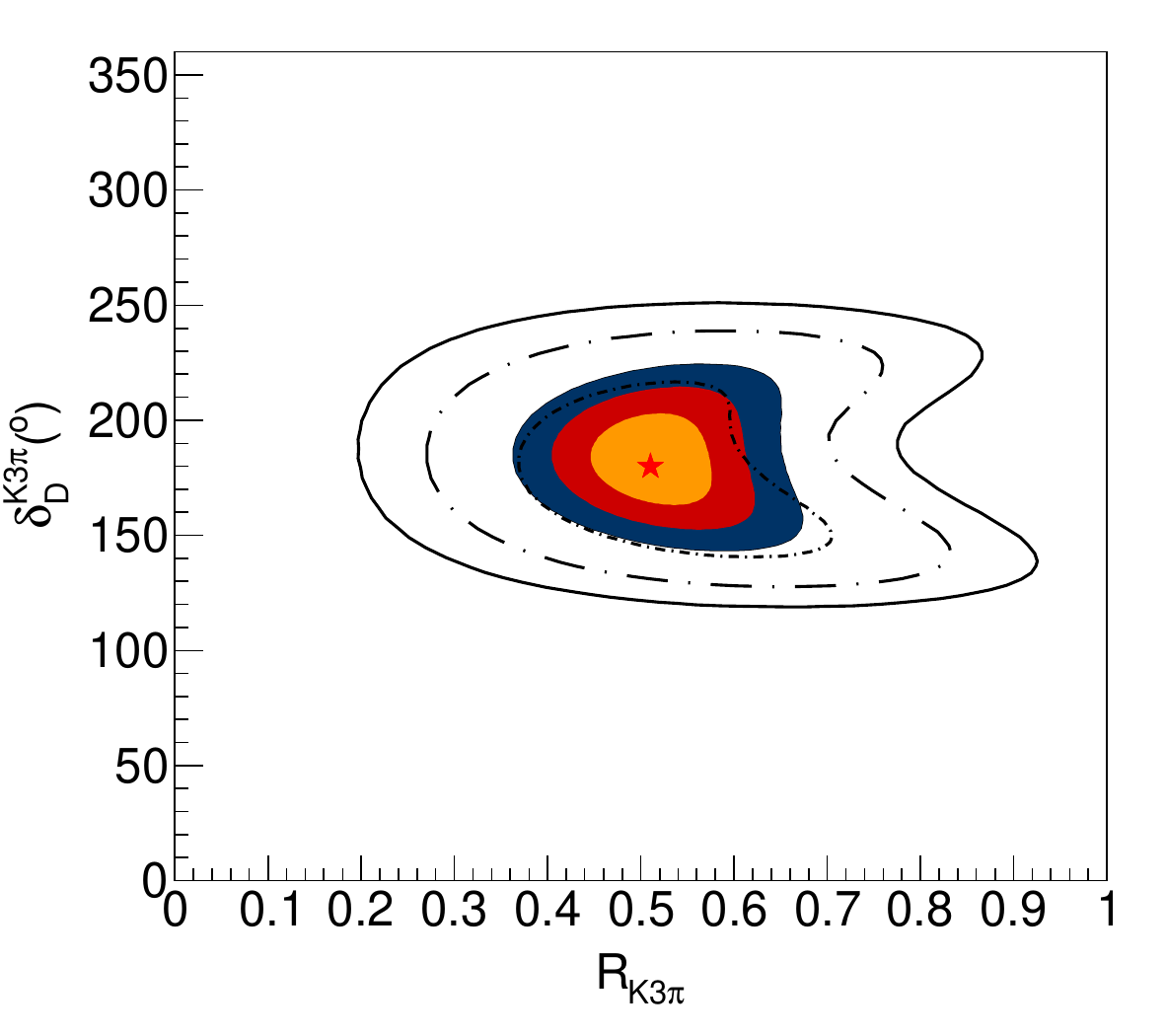}
		\includegraphics[width=.4\textwidth]{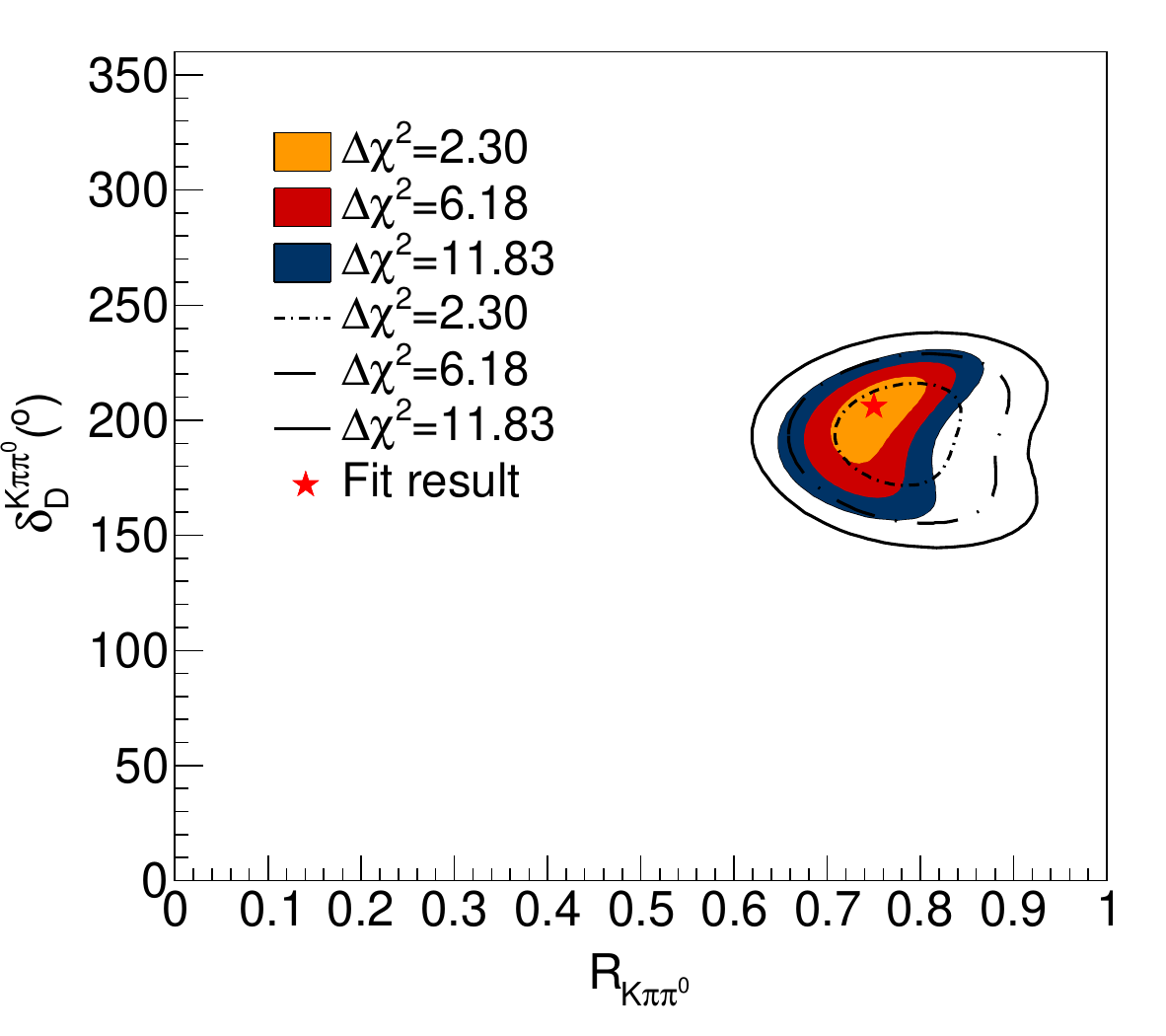}
		\caption{Scans of $\Delta \chi^2$ in the global ($R_{K3\pi}$, $\delta_D^{K3\pi}$) and ($R_{K\pi\pi^0}$, $\delta_D^{K\pi\pi^0}$) parameter space, where the filled color contours represent the current results and the various black lines indicate the previous BESIII measurements~\cite{BESIII:2021eud}. The $\Delta \chi^2$ = 2.30, 6.18, and 11.83 contours correspond to the 68.3\%, 95.4\%, and 99.7\% confidence levels in the 2-D parameter space, respectively.
See Ref.~\cite{BESIII:2026oav} for details.}
		\label{fig:R_Ddelta}
	\end{center}
\end{figure*}

The coherence factor and strong-phase difference in $D\to K^\pm\pi^\mp\pi^+\pi^-$ and $D \to K^\pm\pi^\mp\pi^0$  were extracted for the first time by CLEO-c with a dataset corresponding to 818~pb$^{-1}$ at 3.774 GeV~\cite{CLEO:2009ydx,Libby:2014rea,Evans:2016tlp};
and later by BESIII with 2.93~fb$^{-1}$ of data at 3.773 GeV~\cite{BESIII:2021eud}.
Complementary constraints on these parameters were also made from measurements of $D^0$-$\bar{D}^{0}$ oscillations above charm threshold~\cite{Harnew:2013wea,LHCb:2025zgk}.
Very recently, by analyzing 7.9\,fb$^{-1}$ of data at 3.773 GeV, BESIII reported updated measurements of the coherence factor and strong-phase difference
in $D\to K^\pm\pi^\mp\pi^+\pi^-$ and $D \to K^\pm\pi^\mp\pi^0$~\cite{BESIII:2026oav}.
The obtained results are $R_{K3\pi}=0.51\pm0.04$ and $R_{K\pi\pi^0}=0.75 \pm 0.03$, with values for the average strong-phase differences that are $\delta_D^{K3\pi}=\left(182^{+14}_{-13}\right)^\circ$ and  $\delta_D^{K\pi\pi^0}=\left(209^{+7}_{-8}\right)^\circ$,
where the uncertainties combine both statistical and systematic contributions.
In particular, it has been found that partitioning the phase space of the decay $D\to K^\pm\pi^\mp\pi^+\pi^-$ into regions will enhance its utility as one of the most sensitive channels for determining the angle $\gamma$~\cite{Evans:2019wza}.
For the measurement of the $\gamma$ with $D\to K^\pm\pi^\mp\pi^+\pi^-$, current BESIII inputs are expected to contribute an uncertainty of around 3.5$^\circ$~\cite{BESIII:2026oav}.
The $\Delta \chi^2$ contours in the ($R_{K3\pi}$, $\delta_D^{K3\pi}$) and ($R_{K\pi\pi^0}$, $\delta_D^{K\pi\pi^0}$)
parameter space are shown in Fig.~\ref{fig:R_Ddelta}.

The first measurement of the strong-phase difference between $D^{0}$ and $\bar{D}^{0}$ decaying to the $K^\pm\pi^\mp$ was also from
the CLEO-c experiment by using a data sample of 818~pb$^{-1}$ at 3.774 GeV~\cite{CLEO:2012fel};
and BESIII reported an improved measurement by analyzing a data sample of 2.93 fb$^{-1}$ at 3.773 GeV~\cite{BESIII:2014rtm}.
The BESIII result was later superseded by that in Ref.~\cite{BESIII:2022qkh} based on the same data sample,
but using more $CP$-even tag modes of $K^0_S2\pi^0$, $K^0_L\pi^0$, $K^0_L\omega$,
and the quasi-$CP$-even tag mode $\pi^+\pi^-\pi^0$
and more $CP$-odd tag modes $K^0_S\eta^\prime$, $K^0_S\phi$, and $K^0_L2\pi^0$.
The asymmetry between $C\!P$-odd and $C\!P$-even eigenstate decays into $K^-\pi^+$ is determined to be
${\cal A}_{K\pi} = 0.132 \pm 0.011 \pm 0.007$.
The branching fractions of the $K^0_L$ modes are determined as input to the analysis in a manner that is independent of any strong phase uncertainty,
which are
\begin{eqnarray}
{\cal B}({D^0 \to K^0_L \pi^0}) & = & (0.97 \pm 0.03 \pm 0.02)\,\%, \nonumber \\
{\cal B}({D^0 \to K^0_L \omega }) & = &  (1.09 \pm 0.06 \pm 0.03)\,\%, \nonumber \\
{\cal B}({D^0 \to K_L^0 2\pi^0}) & = & (1.26 \pm 0.05 \pm 0.03)\,\%.  \nonumber
\end{eqnarray}
Using the predominantly $C\!P$-even  tag $D\to \pi^+\pi^-\pi^0$ and the ensemble of $C\!P$-odd eigenstate tags, the observable ${\cal A}_{K\pi}^{\pi\pi\pi^0}$  is measured to be $0.130 \pm 0.012 \pm 0.008$.  The two asymmetries are sensitive to $r_D^{K\pi} \cos \delta_D^{K\pi}$, where $r_D^{K\pi}$ and $\delta_D^{K\pi}$ are the ratio of amplitudes and phase difference, respectively, between the DCS and CF decays.
In addition, events containing $D \to K^-\pi^+$  tagged by $D \to K^0_{S,L} \pi^+\pi^-$  are studied in bins of phase space of the three-body decays. This analysis has sensitivity to both $r_D^{K\pi} \cos\delta_D^{K\pi}$ and $r_D^{K\pi} \sin\delta_D^{K\pi}$.  A fit to ${\cal A}_{K\pi}$,  ${\cal A}_{K\pi}^{\pi\pi\pi^0}$ and the phase-space distribution of the $D \to K^0_{S,L} \pi^+\pi^-$ tags yields
$\delta_D^{K\pi}= \left( 187.6 {^{+8.9}_{-9.7}}{^{+5.4}_{-6.4}} \right)^\circ$, where external constraints are applied for $r_D^{K\pi}$ and other relevant parameters.   This is the most precise measurement of $\delta_D^{K\pi}$ in quantum-correlated $D\bar{D}$ decays.

\begin{figure*}
\includegraphics[width=0.8\linewidth]{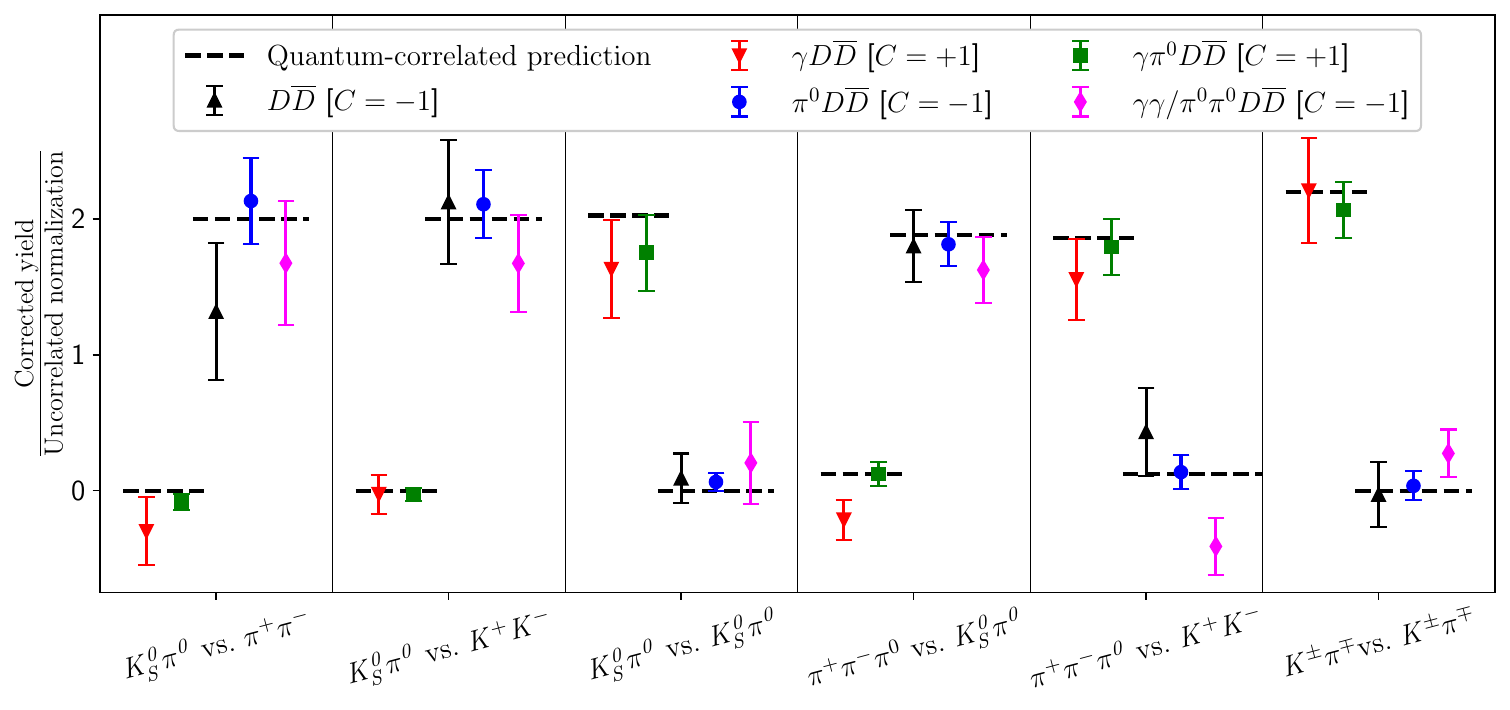}
\caption{The ratios of efficiency-corrected yields observed in data to those expected in the absence of correlations and mixing for each $D\bar D$ final state originating from each production mechanism~\cite{BESIII:2025xed,BESIII:2025pod}. The displayed errors include systematic uncertainties.}
\label{fig:DDstar_QC_plot}
\end{figure*}

Recently, BESIII reported the existence of quantum correlations in the $D\bar D$ system produced with $\cal C$ eigenvalues  through the processes $e^+e^-\to D\bar D$, $e^+e^- \to D^*\bar D$, and $e^+e^- \to D^*\bar D^*$~\cite{BESIII:2025xed,BESIII:2025pod}, where the lack of charge superscripts refers to an admixture of neutral-charm-meson particle and antiparticle states, using 7.13 fb$^{-1}$ of $e^+e^-$ data at $4.128-4.226$~GeV. Processes with either $\cal C$-even or $\cal C$-odd constraints are identified and separated. A procedure is presented that harnesses the entangled production process to enable measurements of $D^0$-meson hadronic parameters. This study provides the first confirmation of quantum correlations in  $e^+e^-\to X D\bar D$ processes   and the first observation of a $\cal C$-even constrained $D\bar D$ system.
Figure~\ref{fig:DDstar_QC_plot} shows the ratios of efficiency-corrected yields observed in data to those expected in the absence of correlations and mixing for each $D\bar D$ final state originating from each production mechanism.
The procedure is applied to measure $\delta_{K\pi}$, the strong phase between the $D^0\to K^-\pi^+$ and $\bar D^0\to K^-\pi^+$ decay amplitudes, which results in the determination of $\delta_{K\pi}=\left(192.8^{+11.0 + 1.9}_{-12.4 -2.4}\right)^\circ$.

\subsection{$CP+$ fraction}

Experimentally, the $\gamma$ can be extracted through the golden decay mode of $B^{\pm}\to DK^{\pm}$~\cite{Gronau:1991dp,Gronau:1990ra}.  The $D$ meson can be reconstructed in decays of mixed $CP$ content~\cite{Nayak:2014tea}, quantified by the $CP$-even fraction~($F_{+}$).
Quantum correlated $D\bar{D}$ pairs produced at the $\psi(3770)$ peak provide a unique platform to measure $CP$-even fractions.
Before BESIII, some measurements of $CP$-even fractions
of $D^0\to h^{+}h^{-}\pi^{0}$~($h=\pi$ or $K$),
$D^0\to \pi^+\pi^-\pi^{+}\pi^{-}$, and
$D^0\to K_{S}^{0}\pi^{+}\pi^{-}\pi^{0}$~\cite{Nayak:2014tea,Malde:2015mha,Harnew:2017tlp,Malde:2015xra}
were made based on 818~pb$^{-1}$ of data taken at 3.774 GeV by CLEO-c.

\begin{figure*}[htb]
    \centering
    \includegraphics[width=0.375\textwidth]{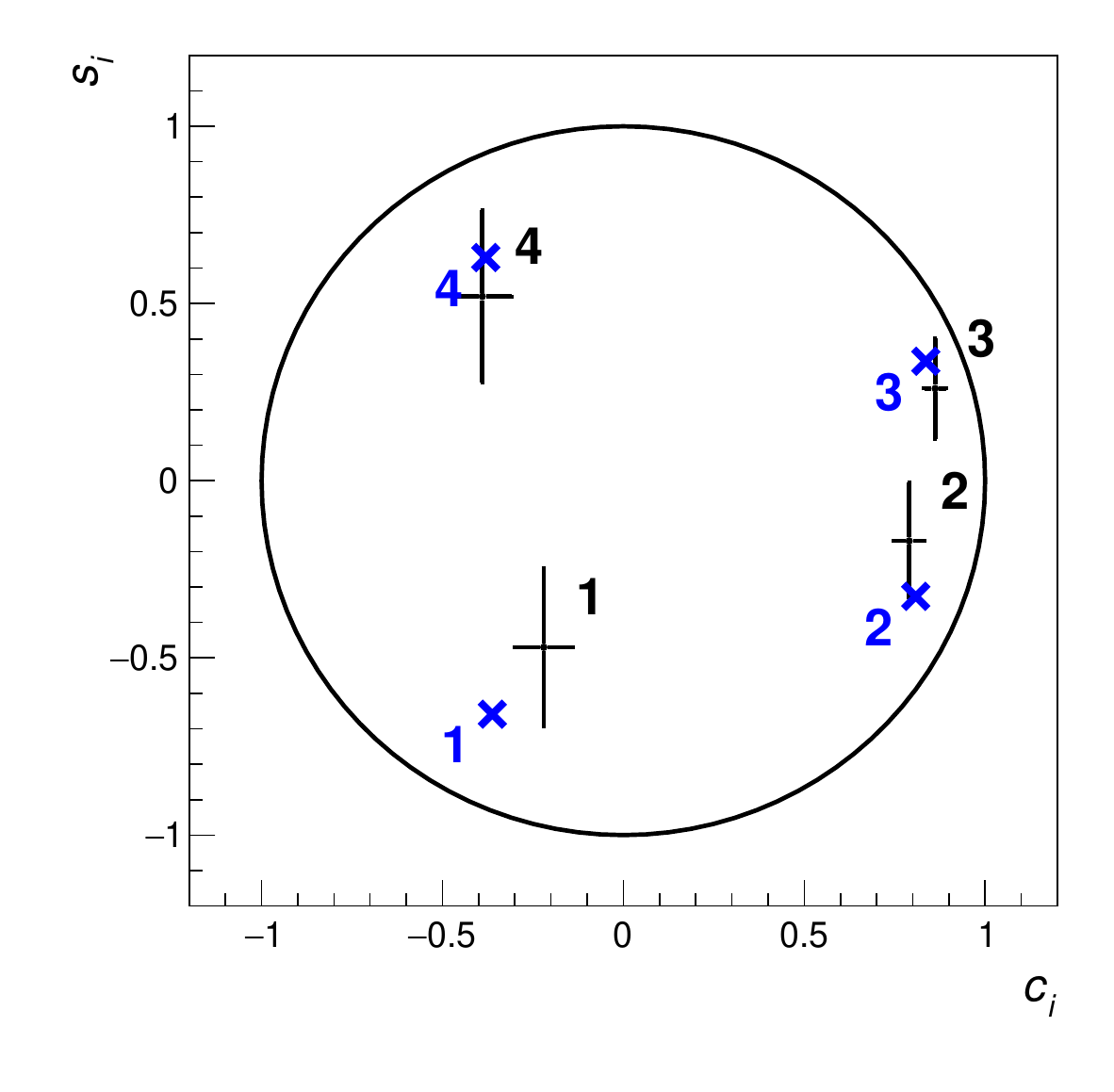}
    \includegraphics[width=0.375\textwidth]{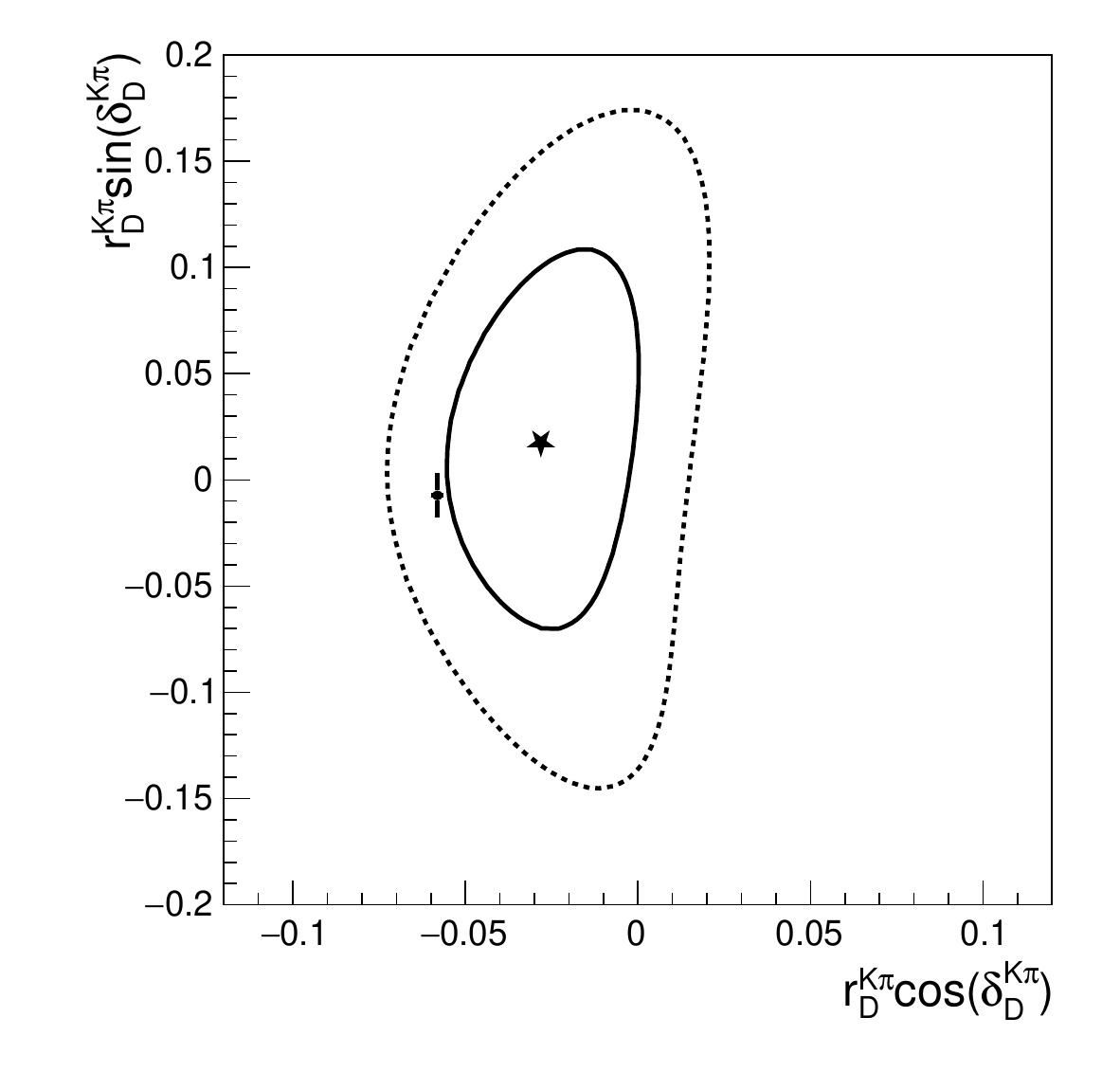}
    \caption{(Left) The fit results for $(c_i, s_i)$ in each phase-space bin~\cite{BESIII:2025wqu}, with error bars corresponding to $68\%$ confidence intervals. Only the statistical uncertainties are considered. The model predictions are shown as blue diagonal crosses, labelled with their bin numbers. The bin numbers are shown above and to the right of the measurements, and below and to the left of the model predictions.
    (right) The (solid) $\Delta\chi^2 = 2.30$ and (dashed) $\Delta\chi^2 = 6.18$ contours of $(r^{K\pi}_D\cos(\delta^{K\pi}_D), r^{K\pi}_D\sin(\delta^{K\pi}_D))$. The data point with error bars is the current average from HFLAV~\cite{HFLAV:2022esi} and the black asterisk is the fitted value.}
    \label{fig:D_KKpipi}
\end{figure*}

A determination of the $C\!P$-even fraction in the decay $D^0 \to K^+K^-\pi^+\pi^-$ is presented by using
2.93 fb$^{-1}$ of data at 3.773 GeV, yielding $F_+^{K^+K^-\pi^+\pi^-} = 0.730 \pm 0.037 \pm 0.021$~\cite{BESIII:2022ebh}.
This is the first model-independent measurement of $F_+$ in $D^0 \to K^+K^-\pi^+\pi^-$ decays.
A first determination of the strong-phase difference between $D^0$ and $\bar{D^0}\to K^+K^-\pi^+\pi^-$  is performed using $e^+e^-\to\psi(3770)\to D\bar{D}$ data collected by the BESIII detector corresponding to an integrated luminosity of $20.3$~fb$^{-1}$.  The measurements are made in four pairs of bins in phase space, which are chosen to provide optimal sensitivity to the angle $\gamma$ of the UT in $B^\pm \to DK^\pm$ decays. From these  measurements, it follows that the $C\!P$-even fraction of the decay is $F_+^{K^+K^-\pi^+\pi^-} = 0.754 \pm 0.010 \pm 0.008$.  In addition, the branching fraction of $D^0 \to K^+K^-\pi^+\pi^-$  is measured to be $(2.863 \pm 0.028 \pm 0.045)\times10^{-3}$~\cite{BESIII:2025wqu}, which is twice as precise as previous results obtained at other experiments.
The fit results of $(c_i, s_i)$, corresponding to a $68\%$ confidence interval, are plotted in Fig.~\ref{fig:D_KKpipi}(left) for each bin.
An alternative fit where $(r^{K\pi}_D\cos(\delta^{K\pi}_D))$ and $(r^{K\pi}_D\sin(\delta^{K\pi}_D))$ are allowed to vary freely is performed, and their $\Delta\chi^2 = 2.30$ and $\Delta\chi^2 = 6.18$ contours are shown in Fig.~\ref{fig:D_KKpipi}(right).

A measurement of the $C\!P$-even fraction of the decay $D^0\to 2(\pi^+\pi^-)$ is performed with 2.93 fb$^{-1}$ of data at 3.773 GeV~\cite{BESIII:2022wqs}.
The $C\!P$-even fraction is measured to be $F^{4\pi}_{+} = 0.735 \pm 0.015 \pm 0.005$.
Based on an analysis of the same data sample, measurements of the strong-phase difference between $D^0$ and $\bar{D}^0\to2(\pi^+\pi^-)$ are also performed in bins of phase space~\cite{BESIII:2024zco}.
The reported results are valuable for measurements of the $C\!P$-violating phase $\gamma$ (also denoted $\phi_3$) in $B^\pm \to DK^\pm$, $D \to 2(\pi^+\pi^-)$ decays, and the binning schemes are designed to provide good statistical sensitivity to this parameter. The expected uncertainty on $\gamma$ arising from the precision of the strong-phase measurements, when applied to very large samples of $B$-meson decays, is around $1.5^\circ$ or $2^\circ$, depending on the binning scheme.
The binned strong-phase parameters are combined to give a value of $F_+^{4\pi} = 0.746 \pm 0.010 \pm 0.004$ for the $C\!P$-even fraction of $D^0 \to 2(\pi^+\pi^-)$ decays, which is around 30\% more precise than the previous measurement~\cite{BESIII:2022wqs}.

In addition, by analyzing 2.93 fb$^{-1}$ and 7.9 fb$^{-1}$ of data at 3.773 GeV,
the $C\!P$-even fractions of $D^0\to K_{S}^{0}\pi^+\pi^-\pi^0$, $D^0 \to\pi^{+}\pi^{-}\pi^{0}$ and $D^0 \to K^{+}K^{-}\pi^{0}$ are measured
to be $F_+^{K_{S}^{0}\pi^+\pi^-\pi^0}=0.235\pm 0.010\pm 0.002$~\cite{BESIII:2023xgh}, ${F_+^{\pi^{+}\pi^{-}\pi^{0}}=0.9406\pm0.0036\pm0.0021}$~\cite{BESIII:2024nnf} and ${F_+^{K^{+}K^{-}\pi^{0}}=0.631\pm0.014\pm0.011}$~\cite{BESIII:2024nnf}, respectively.
They are consistent with the previous CLEO-c results~\cite{Nayak:2014tea,Malde:2015mha,Resmi:2017fuo} within $2\sigma$, with precision improved by factors of 1.7,
3.9 and 2.6 for $D^0\to K_{S}^{0}\pi^+\pi^-\pi^0$, ${D\to\pi^{+}\pi^{-}\pi^{0}}$ and ${D\to K^{+}K^{-}\pi^{0}}$, respectively.

\subsection{Other topics}

By taking advantage of quantum coherence between pairs of $D^0-\bar D^0$ mesons produced in $e^+e^-$
annihilations near threshold,
BESIII also determined the parameter in $D^0-\bar D^0$
oscillations to be $y_{CP}= (-2.0\pm1.3\pm0.7)\%$,
from an analysis of 2.93 fb$^{-1}$ of data at 3.773 GeV~\cite{BESIII:2015ado}. 
This result is compatible with the previous measurements~\cite{HFLAV:2014fzu,BaBar:2012bho,Staric:2012ta,LHCb:2013xvr} within about two standard deviations. However, the precision is still statistically limited and less precise than the current world average~\cite{ParticleDataGroup:2024cfk}.
Using 20.3 fb$^{-1}$ of data at 3.773 GeV, BESIII also searched for the
$e^+ e^- \to D^0\bar{D^0}\to (K^0_S\pi^0) (K^0_S\pi^0)$ process,
which is forbidden by $CP$ conservation, and only allowed via $D^0-\bar{D}^0$ mixing and $K^0-\bar K^0$ mixing mechanism in the secondary decays, and interference between them,
for the first time~\cite{BESIII:2025znu}. 

\section{Hadronic $D$ decays}
\label{sec:hadronBFs}

\subsection{Inclusive decays}

Before BESIII, the  branching fraction of the inclusive decay $D^0 \to K^- X$~($X$ denotes any possible particle combination) was measured by ACCMOR~\cite{ACCMOR:1992ypn}, MARKIII~\cite{MARK-III:1991qvo}, and BESII~\cite{BES:2007dfh};
the branching fractions of $D^0\to \phi X$ and $D^+\to \phi X$ were reported by BES~\cite{BES:1999tgo};
 while the MARKIII~\cite{MARK-III:1991qvo} and BESII~\cite{BES:2007dfh} experiments also reported the branching fraction of $D^0 \to K^+ X$. MARKIII~\cite{MARK-III:1991qvo} and BES~\cite{BES:2006wsn} reported the inclusive branching fractions of $D^0 \to K^0_S X$, and BESII~\cite{BES:2006wsn} additionally reported the branching fractions for inclusive $K^*(892)^-$ and $K^*(892)^+$ decays of $D^0$ and $D^+$ mesons.
CLEO-c~\cite{CLEO:2006hlm} reported the branching fractions for inclusive $\phi$, $\eta$, and $\eta^\prime$ decays of $D^0$, $D^+$, and $D_s^+$.
In a subsequent study, CLEO-c~\cite{CLEO:2009rrr} presented the branching fractions for inclusive decays of $D_s^+$ into $K^+$, $K^-$, $K_S^0$, $\pi^+$, $\pi^-$, $\pi^0$, $\eta$, $\eta^\prime$, $\phi$, $\omega$, $f_0(980)$, and $KK$, in which the results for the inclusive $\eta$, $\eta^\prime$ and $\phi$ decays of $D^+_s$ in Ref.~\cite{CLEO:2009rrr} supersede those of Ref.~\cite{CLEO:2006hlm}.

\begin{figure}
  \includegraphics[width=\columnwidth]{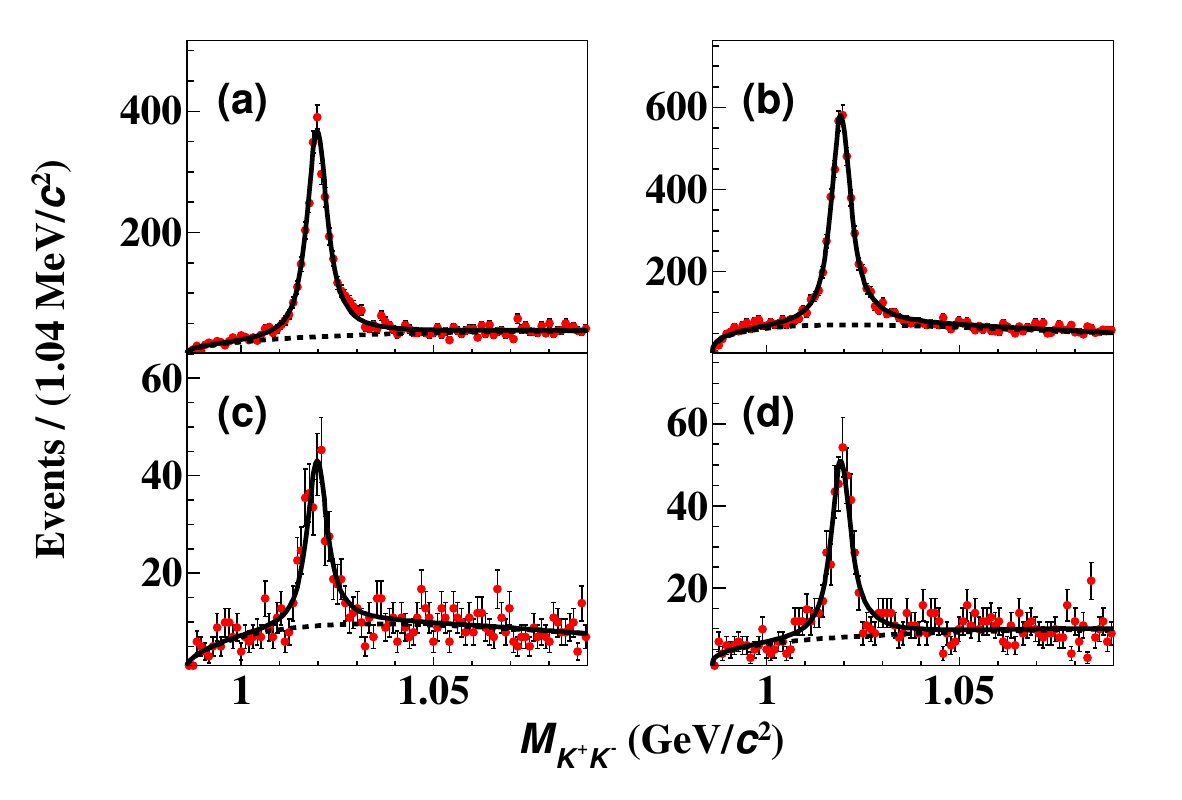}
  \caption{
Fits to the ${M}_{K^+K^-}$ spectra of the candidate events  for (a) $D^+\to\phi X$ and (b) $D^0\to\phi X$ in the $M_{\rm BC}$ signal region, (c) $D^+\to\phi X$ and (d) $D^0\to\phi X$ in the $M_{\rm BC}$ sideband region~\cite{BESIII:2019zan}.
}
\label{fig:D_phiX}
\end{figure}

By analyzing 2.93 fb$^{-1}$ of data collected at 3.773~GeV at BESIII, the branching fractions for the inclusive decays $D^+\to\phi X$ and $D^0\to\phi X$ were
measured to be
\[{\cal B}(D^+\to\phi X) = \left(1.135\pm0.034\pm0.031\right)\%\]
and
\[{\cal B}(D^0\to\phi X) = \left(1.091\pm0.027\pm0.035\right)\%,\] respectively~\cite{BESIII:2019zan}.
The results of fits to the $M_{K^+K^-}$ distributions are shown in Fig.~\ref{fig:D_phiX}.
These results are consistent with previous measurements but with significantly improved precision. They indicate that the nominal values for some known exclusive decays of the $D^+$ meson, such as $D^+\to\phi\pi^+\pi^0$, may be overestimated. Further precision measurements of exclusive $\phi X$ decays for both $D^+$ and $D^0$ mesons are needed to clarify this discrepancy.

\begin{figure}[htp]
  \centering
\includegraphics[width=0.9\linewidth]{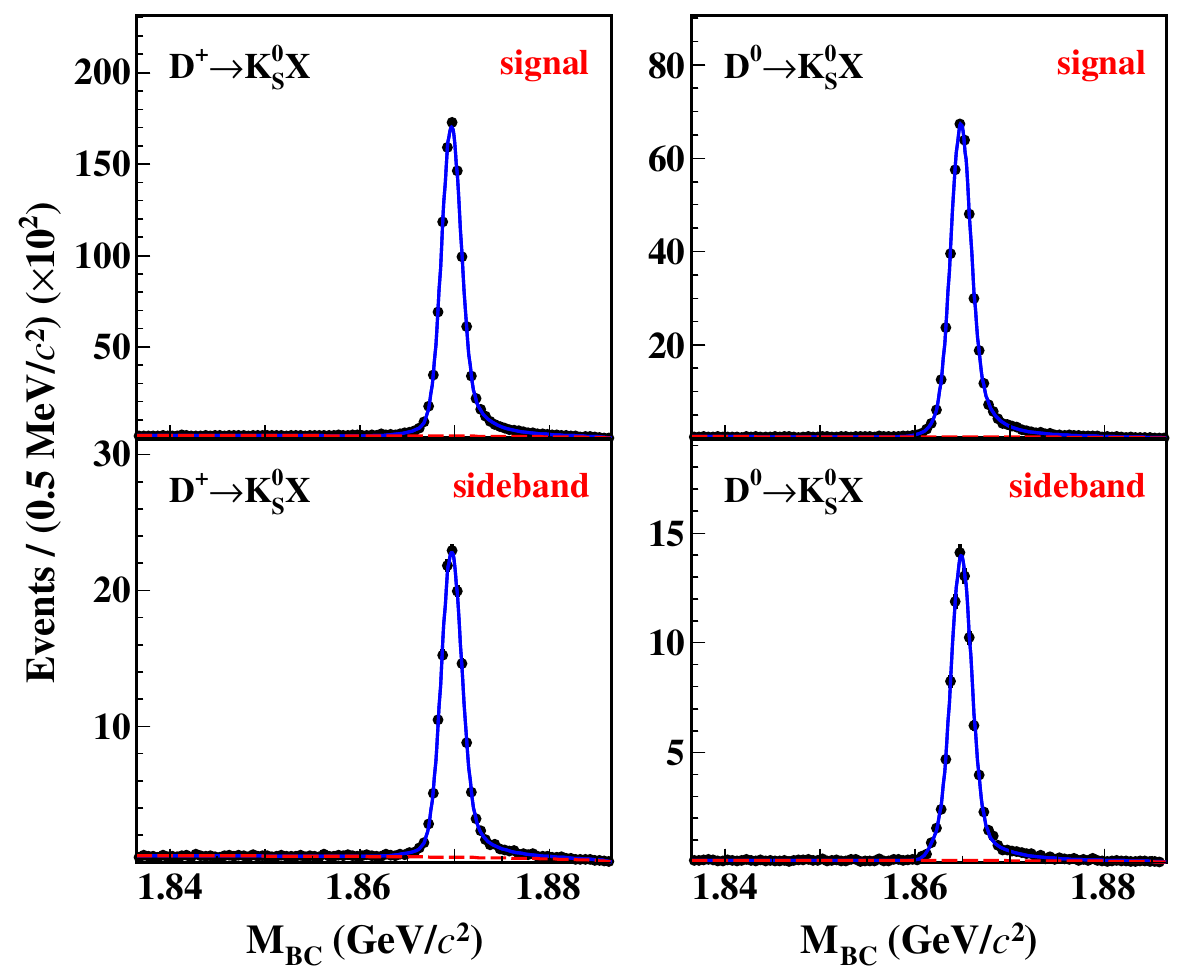}
  \caption{Fits to the $M_{\rm BC}$ distributions of the double-tag events for $D^+\to K^0_S X$ and $D^0\to K^0_S X$ in data~\cite{BESIII:2023has}.
The top and bottom plots correspond to events with $M_{\pi^+\pi^-}$ in the $K^0_S$ signal and sideband regions, respectively.
}
\label{fig:D0p_KSX}
\end{figure}

From an analysis of 2.93 fb$^{-1}$ of data at 3.773~GeV, the branching fractions for $D^+\to K^0_S X$ and $D^0\to K^0_S X$ are measured to be
\[{\cal B}(D^+\to K^0_S X)  = \left(33.11\pm 0.13\pm 0.36\right)\%\]
and
\[{\cal B}(D^0\to K^0_S X)  = \left(20.75\pm 0.12\pm 0.20\right)\%,\]
respectively~\cite{BESIII:2023has}.
The results of fits to the $M_{\rm BC}$ distributions are shown in Fig.~\ref{fig:D0p_KSX}.
Compared to the PDG averages~\cite{ParticleDataGroup:2022pth}, the precision of these measurements is improved by factors of 7.1 for $D^+$ and 7.6 for $D^0$. Summing the branching fractions of known $D^{+(0)}$ decay modes containing $K^0_S$ yields ${\cal B}^{\rm sum}_{\rm exclusive}(D^+\to K^0_SX)=(31.68\pm0.32)\%$ and ${\cal B}^{\rm sum}_{\rm exclusive}(D^0\to K^0_SX)=(18.16\pm0.72)\%$. The differences between inclusive and exclusive branching fractions are $(1.43\pm0.44)\%$ for $D^+$ and $(2.59\pm0.76)\%$ for $D^0$, suggesting that some decay modes involving $K_S^0$ may be missing for both $D^+$ and $D^0$.

Based on 3.19~fb$^{-1}$ of data at 4.178~GeV~\cite{BESIII:2022ydh} and 2.93~fb$^{-1}$ of data at 3.773~GeV~\cite{BESIII:2023sxm}, the branching fractions for the inclusive decays $D_s^+\to 2\pi^+\pi^- X$, $D^0\to 2\pi^+\pi^-X$, and $D^+\to 2\pi^+\pi^-X$ were measured for the first time to be
\[{\cal B}(D_s^+\to 2\pi^+\pi^- X) = \left(32.81\pm 0.35\pm 0.63\right)\%,\]
\[{\cal B}(D^0\to 2\pi^+\pi^- X)  = \left(17.60\pm 0.11\pm 0.22\right)\%,\]
and
\[{\cal B}(D^+\to 2\pi^+\pi^- X)  = \left(15.25\pm 0.09\pm 0.18\right)\%,\]
respectively.
The first value exceeds the sum of all observed exclusive branching fractions for $D^+_s$ decays by about 25\% based on PDG values~\cite{ParticleDataGroup:2022pth} and recent measurements, suggesting potentially unobserved decay modes with at least three charged pions in the final state.
The latter two are consistent with the sums of known decay modes within approximately $\pm 3\sigma$, indicating little room for possible missing $D^{0}$ or $D^+$ decays containing $2\pi^+\pi^-$.

In addition, using 482 pb$^{-1}$ of data at 4.009~GeV, the branching fraction for the inclusive decay $D_{s}^{+}\to \eta^\prime X$ is determined to be~\cite{BESIII:2015rrp}
\[{\cal B}(D_{s}^{+}\to \eta^\prime X)  = \left(8.8 \pm 1.8 \pm 0.5\right)\%.\]
This value is smaller than the sum of all known $D^{+}_s$	 decay modes containing an $\eta^\prime$, (12.0$\pm$0.2)\%~\cite{ParticleDataGroup:2022pth}, by 1.7$\sigma$, indicating little room for possible missing exclusive $D^+_s$ decays involving an $\eta^\prime$.

\subsection{Doubly-Cabibbo-suppressed decays}

DCS decays of $D$ mesons can provide unique insight into weak decay mechanisms of charmed hadrons.
The naive expectation for the DCS decay rate relative to its CF counterpart
is of the order $(0.5-2.0){\rm tan}^4\theta_C$\,$\sim$\,0.3\%, where $\theta_C$ is the Cabibbo mixing angle~\cite{Lipkin:2002za,Cheng:2010ry,Li:2012cfa,Qin:2013tje,Cheng:2024hdo}.
CLEOII~\cite{CLEO:1999vtg}, FOCUS~\cite{FOCUS:2004oyx}, Belle~\cite{Belle:2004ggv,Belle:2006ipk,Belle:2014yoi},
BaBar~\cite{BaBar:2007kib}, CDF~\cite{CDF:2007bdz,CDF:2013gvz}, and
LHCb~\cite{LHCb:2012zll,LHCb:2013zpr,LHCb:2016qsa,LHCb:2017uzt}
reported the branching fraction of $D^0\to K^+\pi^-$ relative to $D^0\to K^-\pi^+$;
CLEOII~\cite{CLEO:2001gxw}, Belle~\cite{Belle:2005xmv}, BaBar~\cite{BaBar:2006dln}, and CLEO-c~\cite{ Evans:2016tlp}
reported the branching fraction of
$D^0\to K^+\pi^-\pi^0$ relative to $D^0\to K^-\pi^+\pi^0$; while
E791~\cite{E791:1996ysd}, CLEOII~\cite{CLEO:2001fiv}, Belle~\cite{Belle:2005xmv,Belle:2013nfo}, and LHCb~\cite{LHCb:2016zmn}
reported the branching fraction of $D^0\to K^+2\pi^-\pi^+$ relative to $D^0\to K^-2\pi^+\pi^-$.
In addition, the branching ratio of the DCS decay $D^+ \to K^+\pi^+\pi^-$ to $D^+ \to K^-2\pi^+$ was measured by E687~\cite{E687:1995vdi}, E791~\cite{E791:1997rvr}, FOCUS~\cite{FOCUS:2004muk}, Belle~\cite{Belle:2009fyg}, and LHCb~\cite{LHCb:2018xff}. Amplitude analysis of $D^+ \to K^+\pi^+\pi^-$ was performed by E791~\cite{E791:1997rvr} and FOCUS~\cite{FOCUS:2004muk}. FOCUS~\cite{FOCUS:2002ame} and LHCb~\cite{LHCb:2018xff} reported the branching ratio of $D^+ \to 2K^+K^-$ to $D^+ \to K^-2\pi^+$. The branching fraction of $D^+ \to K^+\pi^0$ was reported by CLEO-c~\cite{CLEO:2006xrh} and BaBar~\cite{BaBar:2006jox}, while Belle~\cite{Belle:2011tmj} measured the branching fractions of $D^+ \to K^+\eta^{(\prime)}$ relative to $D^+ \to \pi^+\eta^{(\prime)}$. Moreover,
the branching fraction of $D_s^+ \to 2K^+\pi^-$ relative to $D_s^+ \to K^+K^-\pi^+$ was reported by FOCUS~\cite{FOCUS:2005wtq}, Belle~\cite{Belle:2009fyg}, BaBar~\cite{BaBar:2010wqe}, and LHCb~\cite{LHCb:2018xff}.
The known ratios of DCS and CF decay rates for $D^0\to K^+\pi^-$, $D^0\to K^+\pi^-\pi^0$,
$D^0\to K^+2\pi^-\pi^+$, and $D^+_s\to 2K^+\pi^-$ roughly support
this expectation.

\begin{figure}[tp]
  \centering
\includegraphics[width=1\linewidth]{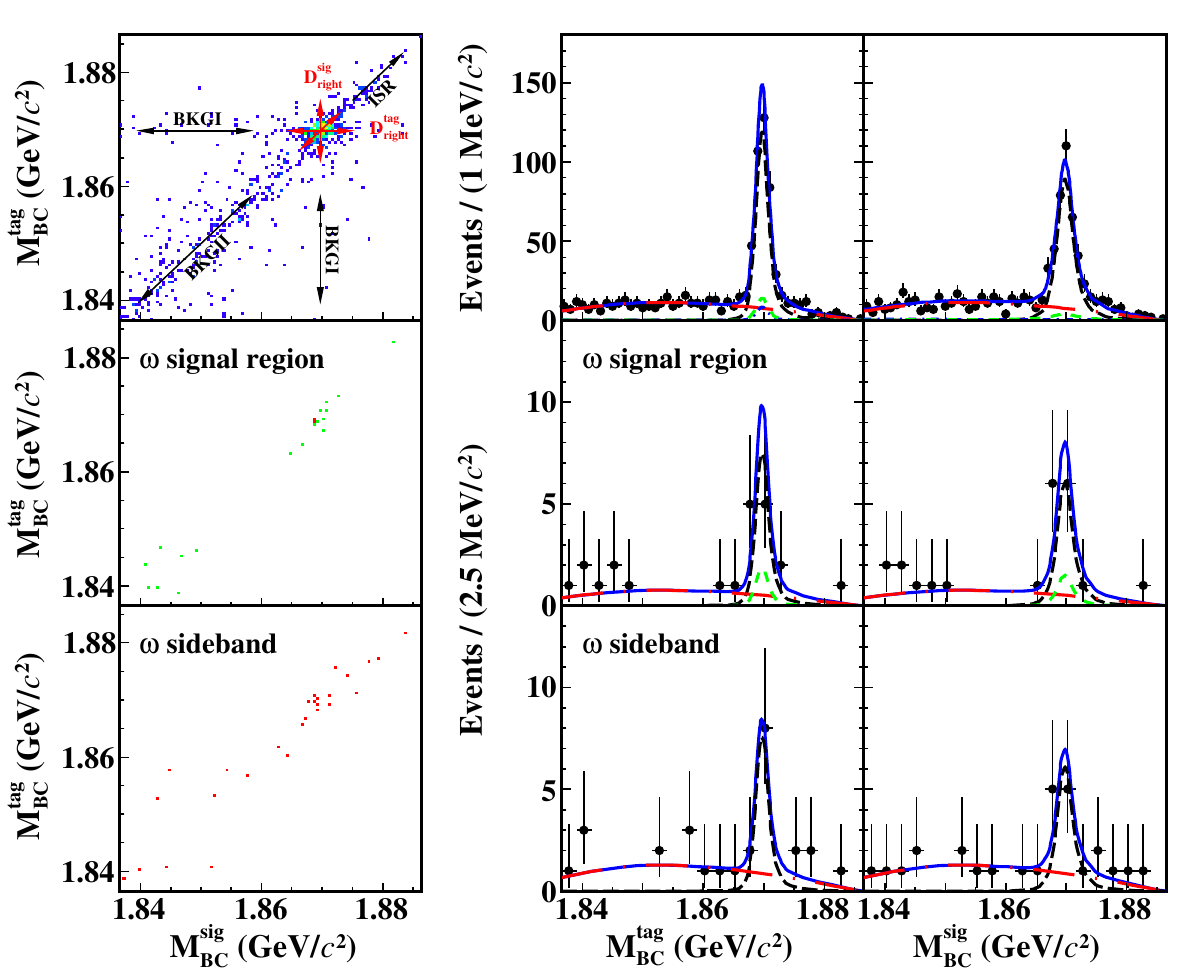}
  \caption{\small
Distributions of (left column) $M_{\rm BC}^{\rm tag}$ versus $M_{\rm BC}^{\rm sig}$, and the projections of the corresponding 2-D fits on (middle column) $M_{\rm BC}^{\rm tag}$ and (right column) $M_{\rm BC}^{\rm sig}$, for the double-tag candidate events of $D^-\to$ all tags versus $D^+\to K^+\pi^+\pi^-\pi^0$~\cite{BESIII:2020wnc}. The top, middle, and bottom rows correspond to all events, events lying in $\omega$ signal region, and those falling in $\omega$ sideband region, respectively.
}
\label{fig:Dp_DCS_K3pi}
\end{figure}

Using 2.93~fb$^{-1}$ of data collected at 3.773~GeV, the first observation of $D^+\to K^+\pi^+\pi^-\pi^0$ is reported based on 350 signal events~\cite{BESIII:2020wnc}.
After removing decays that contain narrow intermediate resonances, including $D^+\to K^+\eta$, $D^+\to K^+\omega$, and $D^+\to K^+\phi$, the branching fraction of the decay $D^+\to K^+\pi^+\pi^-\pi^0$ is measured to be $(1.13 \pm 0.08 \pm 0.03)\times 10^{-3}$.
The spectra in the middle and right columns in Fig.~\ref{fig:Dp_DCS_K3pi} show
the projections on $M_{\rm BC}^{\rm tag}$ and $M_{\rm BC}^{\rm sig}$ of the 2-D fits to data.
The ratio of branching fractions of $D^+\to K^+\pi^+\pi^-\pi^0$ over $D^+\to K^-2\pi^+\pi^0$ is found to be $(1.81\pm0.15)$\%, which corresponds to $(6.28\pm0.52)\tan^4\theta_C$, where $\theta_C$ is the Cabibbo mixing angle. This ratio is significantly larger than the corresponding ratios for other DCS decays. The asymmetry of the branching fractions of charge-conjugated decays $D^\pm\to K^\pm\pi^\pm\pi^\mp\pi^0$ is also determined, and no evidence for $C\!P$ violation is found.
In addition, the first evidence for the $D^+\to K^+\omega$ decay based on $9.2^{+4.0}_{-3.4}$ signal events, with a statistical significance of 3.3$\sigma$, is presented and the branching fraction is measured to be ${\cal B}(D^+ \to K^+ \omega)=({5.7^{+2.5}_{-2.1}}\pm0.2)\times10^{-5}$.
Later,
this decay is also confirmed by an independent measurement with a new semileptonic tagging method~\cite{BESIII:2021uix} using the same data sample.
Based on 112 signal events, tagged by $D^-\to K^0e^-\bar \nu_e$ and $D^-\to K^+\pi^-e^-\bar \nu_e$,
the branching fraction for $D^+\to K^+\pi^+\pi^-\pi^0$ is measured to be $(1.03\pm0.12\pm0.06)\times 10^{-3}$.

\begin{figure}[tp]
  \centering
\includegraphics[width=1.0\linewidth]{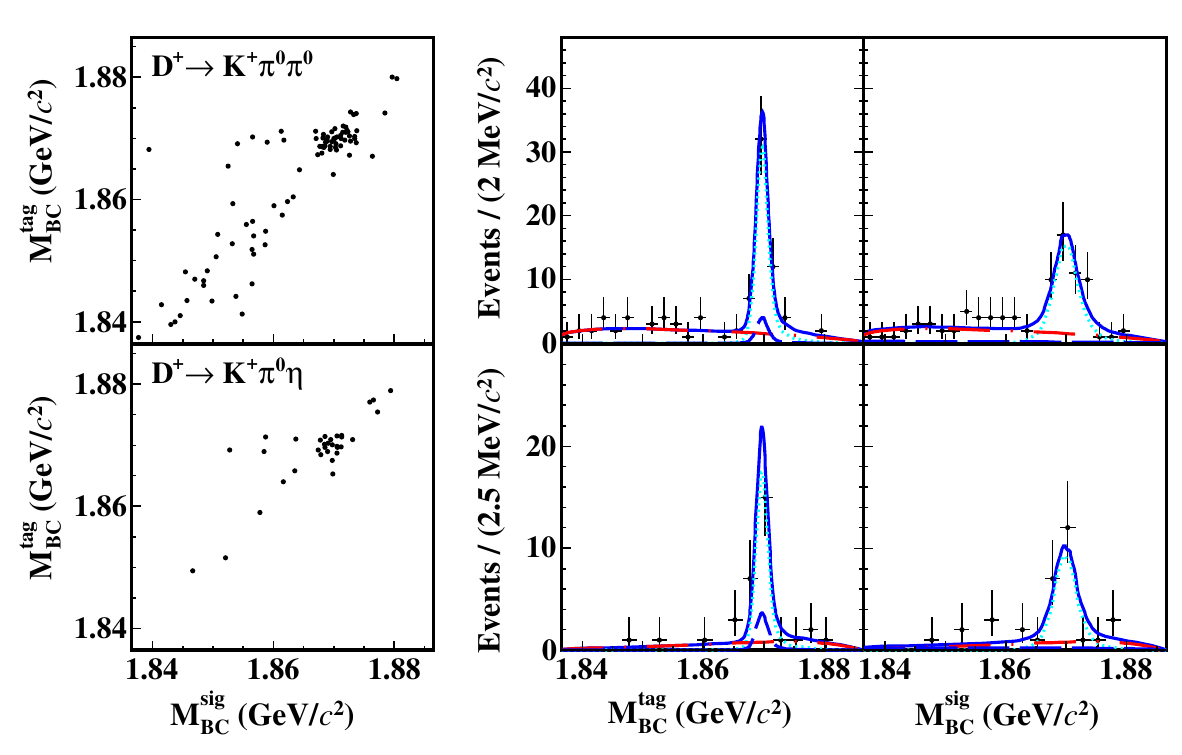}
  \caption{\small
Distributions of (left column) $M_{\rm BC}^{\rm tag}$ versus $M_{\rm BC}^{\rm sig}$  and
the projections on (middle column) $M_{\rm BC}^{\rm tag}$ and (right column) $M_{\rm BC}^{\rm sig}$ of the 2-D fits to the double-tag candidate events for (top row) $D^+\to K^{+}2\pi^0$ and (bottom row) $D^+\to K^{+}\pi^0\eta$~\cite{BESIII:2021qsa}.
}
\label{fig:Dp_DCS_K2pi0_Kpi0eta}
\end{figure}

By analyzing the same data sample, the decays $D^+\to K^+2\pi^0$ and $D^+\to K^+\pi^0\eta$ were observed for the first time~\cite{BESIII:2021qsa}.
The spectra of the middle and right columns in Fig.~\ref{fig:Dp_DCS_K2pi0_Kpi0eta} show
the projections on $M_{\rm BC}^{\rm tag}$ and $M_{\rm BC}^{\rm sig}$ of the 2-D fits to data.
Based on 43 and 19 signal events, the branching fractions of $D^+\to K^+2\pi^0$ and $D^+\to K^+\pi^0\eta$ are measured to be $(2.1 \pm 0.4 \pm 0.1)\times 10^{-4}$ and $(2.1 \pm 0.5 \pm 0.1)\times 10^{-4}$ with statistical significances of 8.8$\sigma$ and 5.5$\sigma$, respectively.
In addition, the subprocesses $D^+\to K^{*}(892)^{+}\pi^0$ and $D^+\to K^{*}(892)^{+}\eta$ with $K^{*}(892)^+\to K^+\pi^0$ are also investigated
by examining the $K^-\pi^0$ mass spectrum. The branching fraction of $D^+\to K^{*}(892)^{+}\eta$ is determined to be $({4.4^{+1.8}_{-1.5}}\pm0.2)\times10^{-4}$, with a statistical significance of 3.2$\sigma$. No significant signal for $D^+\to K^{*}(892)^{+}\pi^0$ is found
and its branching fraction upper limit at the 90\% confidence level is set to be $5.4\times10^{-4}$.

\begin{figure*}[htbp]
  \centering
  \includegraphics[width=0.8\textwidth]{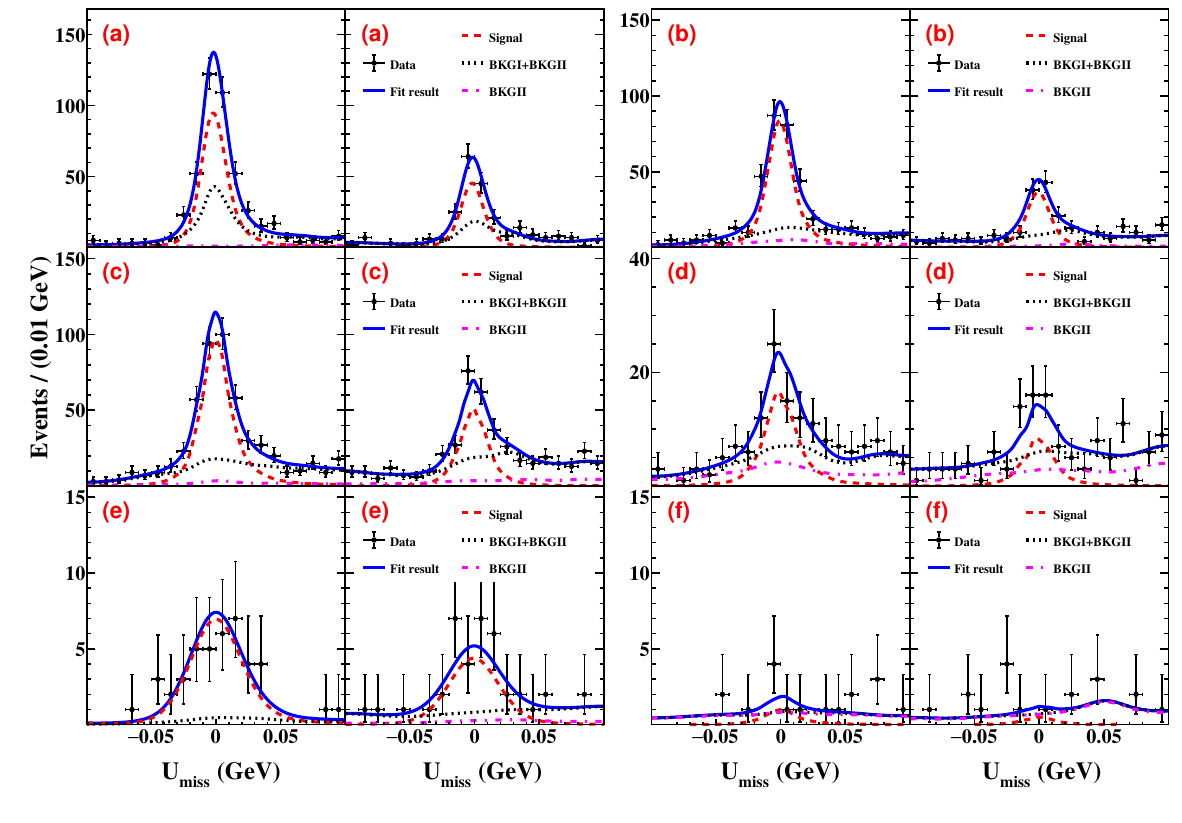}
  \caption{Fits to the $U_{\mathrm{miss}}$ distributions of the double-tag candidate events for (a) $D^0\to K^+\pi^-$, (b) $D^0\to K^+2\pi^-\pi^+$, (c) $D^0\to K^+\pi^-\pi^0$, (d) $D^0\to K^+\pi^-2\pi^0$, (e) $D^0\to K^+\pi^-\eta$ and (f) $D^0\to K^+\pi^-\pi^0\eta$ in data~\cite{BESIII:2025vbt}.  In each plot pair, the left panel is $\bar D^0\to K^+ e^-\bar \nu_e$ tags and the right one is $\bar D^0\to K^+ \mu^-\bar \nu_\mu$ tags.
  }
  \label{fig:D0_DCS_all}
\end{figure*}

\begin{figure*}[htbp]
  \centering
  \includegraphics[width=0.8\textwidth]{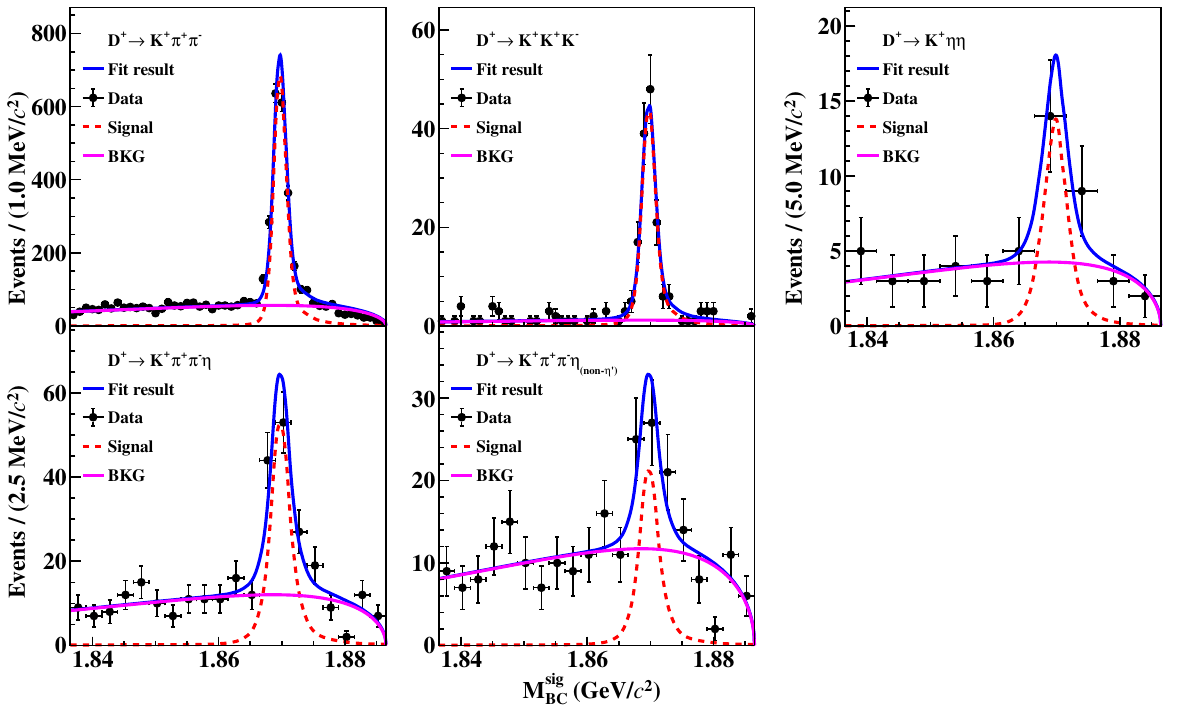}
  \caption{Fits to the $M_{\rm BC}^{\rm sig}$ distributions of the double-tag candidates for DCS $D^+$ decays in data~\cite{BESIII:2025vbt}.
  }
  \label{fig:Dp_DCS_all}
\end{figure*}

With the same data sample and the semileptonic tag of $\bar D^0\to K^+e^-\bar \nu_e$,
a measurement of the branching fraction of $D^0\to K^+\pi^-\pi^0$ and a search for $D^0\to K^+\pi^-2\pi^0$ were reported~\cite{BESIII:2022chs}.
The semileptonic tagging method avoids complex quantum correlation effects in measurements with the traditional hadronic tag method.
In 2025, these measurements were superseded by a study using 20.3~fb$^{-1}$ of data at 3.773~GeV, in which more decay modes of $D^0$ and $D^+$
are investigated~\cite{BESIII:2025vbt}. The fits to  the $U_{\rm miss}$ distributions of the accepted events
for DCS $D^0$ decays are shown in Fig.~\ref{fig:D0_DCS_all}; while
the fits to the $M_{\rm BC}^{\rm sig}$ distributions of the accepted events
for DCS $D^+$ decays are shown in Fig.~\ref{fig:Dp_DCS_all}.

Table~\ref{totresult} summarizes the obtained branching fractions, the PDG values of the corresponding DCS and CF decays, the individual DCS/CF ratios and the ratios in units of $\tan^{4}{\theta}_{C}$.
The decays
$D^0 \to K^+\pi^-\eta$,
$D^0 \to K^+\pi^-\pi^0\eta$,
$D^+\to K^+2\eta$,
$D^{+} \to  K^{+} \pi^{+} \pi^{-} \eta$,
and
$D^{+} \to  K^{+} \left(\pi^{+} \pi^{-} \eta\right)_{{\rm non}-\eta^{\prime}}$
are investigated for the first time.
For the known decays
$D^0 \to K^+ \pi^-$,
$D^0 \to K^+ 2\pi^- \pi^+$,
$D^+ \to K^+ \pi^+ \pi^-$,
and $D^+ \to 2K^+ K^-$, the measured branching fractions  are consistent with world average values.
In the future, amplitude analyses of the multi-body DCS $D$ decays with larger data samples will be able to extract the branching fractions of the intermediate two-body $D$ decays.  This will help to further explore quark SU(3)-flavor symmetry and its breaking effects, and potentially improve theoretical predictions of $\mathit{CP}$ violation in hadronic $D$ decays.

\begin{table*}[htbp]
	\centering
	\caption{The DCS branching fractions reported in Ref.~\cite{BESIII:2025vbt}, the PDG values of the corresponding DCS and CF decays, and the new DCS/CF branching fraction ratios, given both directly and in units of $\tan^{4}{\theta}_{C}$.
	}
	\renewcommand\arraystretch{1.3}
		\begin{tabular}{l|c|c|c|c|c}
			\hline \hline
			Signal decay                                               & ${\cal B}_{\rm DCS}~(\times10^{-4})$ & ${\cal B}_{\rm DCS}^{\rm PDG}~(\times10^{-4})$ & ${\cal B}_{\rm CF}^{\rm PDG}~(\times10^{-2})$ & ${\cal B}_{\rm DCS}/{\cal B}_{\rm CF}~(\%) $ & $\times \tan^{4}{\theta}_{C}$ \\ \hline
			$D^0 \to  K^{+} \pi^{-}$&$1.30 \pm 0.09 \pm 0.04$&$1.50 \pm 0.07$&$3.947 \pm 0.030$&$0.328 \pm 0.027$&$1.14 \pm 0.09$\\ 
			$D^0 \to  K^{+} 2\pi^{-} \pi^{+}$&$2.38 \pm 0.19 \pm 0.12$&$2.65 \pm 0.06$&$8.22 \pm 0.14$&$0.289 \pm 0.028$&$1.00 \pm 0.10$\\ 
			$D^0 \to  K^{+} \pi^{-} \pi^{0}$&$3.06 \pm 0.21 \pm 0.10$&$3.06 \pm 0.16$&$14.4 \pm 0.6$&$0.212 \pm 0.021$&$0.74 \pm 0.07$\\ 
			$D^0 \to  K^{+} \pi^{-} 2\pi^{0}$&$1.40 \pm 0.27 \pm 0.09$&$<3.6$&$8.86 \pm 0.23$&$0.158 \pm 0.036$&$0.55 \pm 0.12$\\ 
			$D^0 \to  K^{+} \pi^{-} \eta$&$1.04 \pm 0.16 \pm 0.08$&$-$&$1.88 \pm 0.05$&$0.555 \pm 0.092$&$1.93 \pm 0.32$\\ 
			$D^0 \to  K^{+} \pi^{-} \pi^{0}  \eta$&$<0.7$&$-$&$0.449 \pm 0.027$&$< 1.78 $&$< 6.19 $\\ 
			\hline
			$ D^{+} \to  K^{+} \pi^{+} \pi^{-}$&$4.50 \pm 0.12 \pm 0.35$&$4.91 \pm 0.09$&$9.38 \pm 0.16$&$0.480 \pm 0.019$&$1.67 \pm 0.07$\\ 
			$ D^{+} \to  K^{+} \pi^{+} \pi^{-} \eta$&$1.56 \pm 0.22 \pm 0.04$&$-$&$-$&$-$&$-$\\ 
			$ D^{+} \to  K^{+} \left(\pi^{+} \pi^{-} \eta\right)_{{\rm non}-\eta^\prime}$&$0.67 \pm 0.18 \pm 0.02$&$-$&$0.135 \pm 0.012$&$5.0 \pm 1.4$&$17.3 \pm 4.8$\\ 
			$ D^{+} \to 2K^{+} K^{-}$&$0.51 \pm 0.05 \pm 0.01$&$0.614 \pm 0.011$&$-$&$-$&$-$\\ 
			$ D^{+} \to  K^{+} 2\eta$&$0.59 \pm 0.23 \pm 0.02$&$-$&$-$&$-$&$-$\\ 
			\hline
			\hline
		\end{tabular}
	\label{totresult}
\end{table*}

Using 20.3~fb$^{-1}$ of data at 3.773~GeV, the measurements of the branching fractions of
$D^+\to K^+\pi^0$, $D^+\to K^+\eta$ and $ D^+ \to K^+ \eta^{\prime}$ with the double-tag method~\cite{BESIII:2025ykz} are performed with significantly improved precision compared to the previous measurements.
Their branching fractions are determined to be ${\cal B}(D^+\to K^+ \pi^0) = (1.45 \pm 0.06 \pm 0.08)\times 10^{-4}$, ${\cal B}(D^+\to K^+ \eta) = (1.17 \pm 0.10 \pm 0.03)\times 10^{-4}$ and ${\cal B}(D^+\to K^+ \eta^{\prime}) = (1.88 \pm 0.15 \pm 0.11)\times 10^{-4}$,
based on 629, 182, and 214 signal events, respectively.
The obtained branching fractions of $D^+\to K^+\eta$ and $ D^+ \to K^+ \eta^{\prime}$ are consistent with the world average values,
while the reported branching fraction of $D^+\to K^+\pi^0$ deviates from the world average value~\cite{ParticleDataGroup:2024cfk} by 3$\sigma$. The tension in $D^+\to K^+\pi^0$ is mainly due to the fact that the backgrounds in previous measurements were underestimated.

Based on 7.33 fb$^{-1}$  of data collected at 4.128-4.226 GeV, the studies of $D^+_s\to 2K^+\pi^-$ and $D^+_s\to 2K^+\pi^-\pi^0$ were reported~\cite{BESIII:2023coc}.
The absolute branching fraction of $D^+_s\to 2K^+\pi^-$ is determined to be $(1.23^{+0.28}_{-0.25}\pm0.06)\times 10^{-4}$.
No significant signal of $D^+_s\to 2K^+\pi^-\pi^0$ is observed and the upper limit on its decay branching fraction at the
90\% confidence level is set to be $1.7\times10^{-4}$.

\subsection{Other hadronic $D$ decays}

\subsubsection{Previous measurements}

\begin{figure}[htbp]
  \centering
  \includegraphics[width=0.4\textwidth]{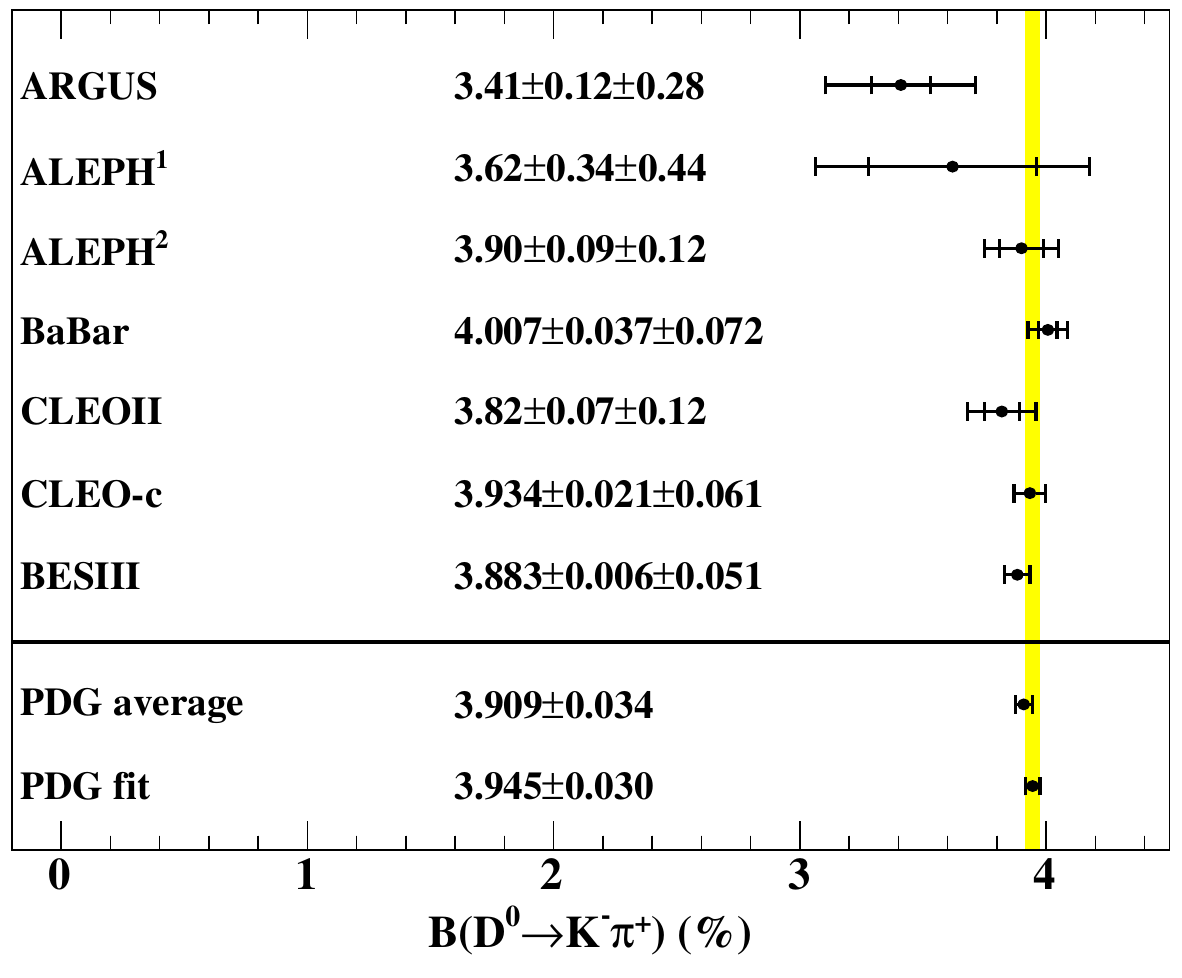}
  \put(-160,152.5){\tiny~\cite{ARGUS:1994onx}}
  \put(-160,138.417){\tiny~\cite{ALEPH:1991phy}}
  \put(-160,124.333){\tiny~\cite{ALEPH:1997jyv}}
  \put(-160,110.25){\tiny~\cite{BaBar:2007crr}}
  \put(-160,96.167){\tiny~\cite{CLEO:1997zun}}
  \put(-160,82.083){\tiny~\cite{CLEO:2013rjc}}
  \put(-160,68){\tiny~\cite{BESIII:2018apz}}
   \caption{Comparison of branching fraction of $D^0\to K^-\pi^+$ measured by BESIII~\cite{BESIII:2018apz}, CLEO-c~\cite{CLEO:2013rjc}, CLEOII~\cite{CLEO:1997zun}, BaBar~\cite{BaBar:2007crr}, ALEPH$^1$~\cite{ALEPH:1997jyv}, ALEPH$^2$~\cite{ALEPH:1991phy}, and ARGUS~\cite{ARGUS:1994onx}. The  yellow band denotes the $\pm 1\sigma$ region of the PDG global fit result~\cite{ParticleDataGroup:2024cfk}.
}
  \label{fig:BF:D0kpi}
\end{figure}

The hadronic decays of $D^0\to K^-\pi^+$, $D^0\to K^-\pi^+\pi^0$, and $D^0\to K^-2\pi^+\pi^-$ as well as
$D^+\to K^-2\pi^+$,
$D^+\to K^-2\pi^+\pi^0$,
$D^+\to K^0_S\pi^+$,
$D^+\to K^0_S\pi^+\pi^0$,
$D^+\to K^0_S2\pi^+\pi^-$, and
$D^+\to K^+K^-\pi^+$ are known as nine golden decay modes due to relatively high branching
fractions and low backgrounds. They were previously measured intensively in experiments.
Before the CLEO-c experiment, measurements of the branching fraction of $D^0\to K^-\pi^+$
were from
LGW~\cite{Peruzzi:1977ms},
MARKII~\cite{Schindler:1980ws},
MARKIII~\cite{MARK-III:1987jsm},
HRS~\cite{HRS:1988pmn},
CLEOII~\cite{CLEO:1993gww,CLEO:1997xir,CLEO:1997zun,CLEO:1997zun},
ARGUS~\cite{ARGUS:1993cew},
ALEPH~\cite{ALEPH:1991phy,ALEPH:1997jyv},
ARGUS~\cite{ARGUS:1994onx}, and
BaBar~\cite{BaBar:2007crr};
measurements for $D^0\to K^-\pi^+\pi^0$ were from
MARKII~\cite{Schindler:1980ws} and
MARKIII~\cite{MARK-III:1987jsm};
 measurements for $D^0\to K^-2\pi^+\pi^-$ were from
LGW~\cite{Peruzzi:1977ms},
MARKII~\cite{Schindler:1980ws},
MARKIII~\cite{MARK-III:1987jsm}, and
ARGUS~\cite{ARGUS:1994onx,ARGUS:1993cew};
measurements for $D^+\to K^-2\pi^+$ were from
MARKII~\cite{Schindler:1980ws},
MARKIII~\cite{MARK-III:1987jsm},
ACCMOR~\cite{ACCMOR:1992ypn} and CLEOII~\cite{CLEO:1994nfy};
while measurements for $D^0\to K^0_S\pi^+\pi^-$ were from
LGW~\cite{Peruzzi:1977ms},
MARKII~\cite{Schindler:1980ws},
MARKIII~\cite{MARK-III:1987qok}, and
ARGUS~\cite{ARGUS:1994onx}.
Later, CLEO-c~\cite{CLEO:2005pql,CLEO:2007rrw,CLEO:2013rjc} reported
measurements of the branching fractions of these nine golden decay modes. Figure~\ref{fig:BF:D0kpi} shows the comparisons of branching fraction of $D^0\to K^-\pi^+$ measured by different experiments.
To take into account correlated uncertainties among them,
the branching fractions of the other seven decay modes
relative to $D^+\to K^-2\pi^+$ or
$D^0\to K^-\pi^+$ were also presented.
In addition,
CLEO~\cite{CLEO:1990lrq}, ARGUS~\cite{ARGUS:1992gpk}, and CLEOII~\cite{CLEO:1996hzb} reported measurements of $D^0\to K^-\pi^+\pi^0$ relative to $D^0\to K^-\pi^+$;
MARKI~\cite{Piccolo:1977qr}, ACCMOR~\cite{ACCMOR:1983zhs,ACCMOR:1985jks}, ARGUS~\cite{ARGUS:1984cqw}, CLEO~\cite{CLEO:1988jcc}, NA142~\cite{NA142:1990qcz},
E691~\cite{FNAL-691:1992exu}, and SELEX~\cite{SELEX:1999xrm} presented
measurements of $D^0\to K^-2\pi^+\pi^-$ relative to $D^0\to K^-\pi^+$;
while E687~\cite{E687:1995jyc}, SELEX~\cite{SELEX:1999xrm}, BaBar~\cite{BaBar:2005enq}, BES~\cite{BES:2005irx}, and LHCb~\cite{LHCb:2018xff} made measurements of
$D^+\to K^+K^-\pi^-$ relative to $D^+\to K^-2\pi^+$.
Moreover,
ACCMOR~\cite{ACCMOR:1992ypn} and MARKIII~\cite{MARK-III:1991fvi} measured
the branching fraction of $D^0\to K^0_S\pi^+\pi^-\pi^0$;
CLEO~\cite{CLEO:1990lrq},
E691~\cite{FNAL-691:1992exu},
and ARGUS~\cite{ARGUS:1992gpk} reported
the branching fraction of $D^0\to K^0_S\pi^+\pi^-\pi^0$ relative to $D^0\to K^0_S\pi^+\pi^-$.

\begin{figure}[htbp]
  \centering
  \includegraphics[width=0.4\textwidth]{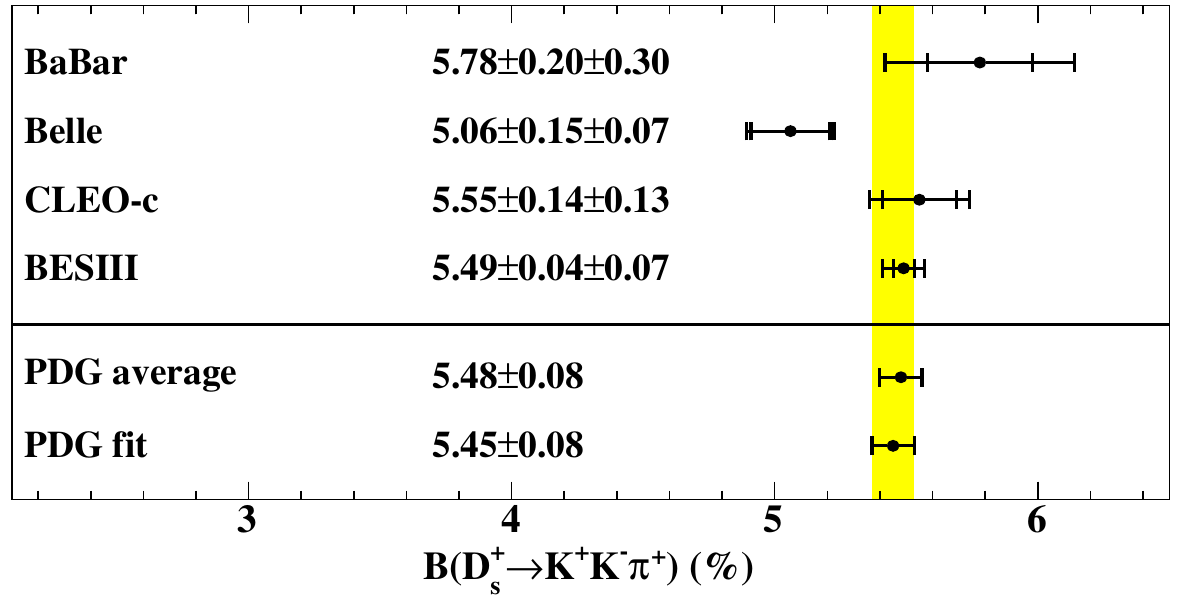}
  \put(-160,92){\tiny~\cite{BaBar:2010ixw}}
  \put(-160,80){\tiny~\cite{Belle:2013isi}}
  \put(-160,68){\tiny~\cite{CLEO:2013bae}}
  \put(-160,56){\tiny~\cite{BESIII:2024oth}}
   \caption{Comparison of branching fraction of $D^+_s\to K^+K^-\pi^+$ measured by BESIII~\cite{BESIII:2024oth}, CLEO-c~\cite{CLEO:2013bae},  BaBar~\cite{BaBar:2010ixw}, and Belle~\cite{Belle:2013isi}. The yellow band denotes the $\pm 1\sigma$ region of the PDG global fit result~\cite{ParticleDataGroup:2024cfk}.
}
  \label{fig:BF:Dskkpi}
\end{figure}

For hadronic $D^+_s$ decays,
CLEO-c~\cite{CLEO:2008hzo,CLEO:2013bae},
BaBar~\cite{BaBar:2010ixw}, and
Belle~\cite{Belle:2013isi} measured the branching fraction of $D^+_s\to K^+K^-\pi^+$. Figure~\ref{fig:BF:Dskkpi} shows the comparisons of branching fraction of $D^+_s\to K^+K^-\pi^+$ measured by different experiments.
In addition,
CLEO-c also reported the branching fractions of $D^+_s\to K^0_SK^+$,
$K^0_SK^+\pi^0$, $K^0_SK^0_S\pi^+$, $K^+K^-\pi^+\pi^0$,
$K^0_SK^+\pi^+\pi^-$,
$K^0_SK^-2\pi^+$,
$2\pi^+\pi^-$,
$\eta\pi^+$,
$\eta\pi^+\pi^0$,
$\eta^\prime\pi^+$,
$\eta^\prime\pi^+\pi^0$, and
$K^+\pi^+\pi^-$~\cite{CLEO:2008hzo,CLEO:2013bae},
$D^+_s\to p\bar n$~\cite{CLEO:2008aum},
as well as $D^+_s\to \pi^+2\pi^0$ and $D^+_s\to K^0\pi^+\pi^0$~\cite{CLEO:2009vke};
while Belle~\cite{Belle:2013isi} also reported the branching fraction of $D^+_s\to \eta\pi^+$.
FOCUS made a relative measurement of $D^+_s\to K^+\pi^+\pi^-$ to $D^+_s\to K^+K^-\pi^+$~\cite{FOCUS:2004muk};
E687~\cite{E687:1997jvh} and BaBar~\cite{BaBar:2008nlp} reported relative measurements of $D^+_s\to 2\pi^+\pi^-$ to $D^+_s\to K^+K^-\pi^+$;
while measurements of $D^+_s\to 3\pi^+2\pi^-$ relative to $D^+_s\to K^+K^-\pi^+$ were from E687~\cite{E687:1997dng} and FOCUS~\cite{FOCUS:2002psb}.
 
For two-body hadronic $D$ decays,
CLEO-c~\cite{CLEO:2009fiz} reported the branching fractions of
$D^0\to K^+K^-$, $K^0_SK^0_S$, $\pi^+\pi^-$, $2\pi^0$,
$K^0_S\pi^0$, $K^0_S\eta$, $\pi^0\eta$, $K^0_S\eta^\prime$, $\pi^0\eta^\prime$, $2\eta$, $\eta\eta^\prime$ relative to $D^0\to K^-\pi^+$,
those of $D^+\to K^0_SK^+$, $\pi^+\pi^0$, $K^0_S\pi^+$, $K^+\pi^0$, $\pi^+\eta$, and $\pi^+\eta^\prime$ relative to $D^+\to K^-2\pi^+$,
as well as those of $D^+_s\to K^0_S\pi^+$, $K^+\pi^0$, $K^+\eta$, $\pi^+\eta$, $K^+\eta^\prime$, $\pi^+\eta^\prime$ relative to $D^+_s\to K^0_SK^+$.
In addition,
E691~\cite{Anjos:1990nm}, CLEOII~\cite{CLEO:1997zxb} and  FOCUS~\cite{FOCUS:2001kwm}  reported the branching fraction of
$D^+\to K^0_S\pi^+$ relative to $D^+\to K^-2\pi^+$;
CLEO~\cite{CLEO:1990lrq}, ARGUS~\cite{ARGUS:1992gpk},
and CLEOII~\cite{CLEO:1992sni} reported the branching fraction of $D^0\to K^0_S\pi^0$ relative to $D^0\to K^0_S\pi^+\pi^-$,
and CLEOII~\cite{CLEO:1992sni} reported the branching fraction of $D^0\to K^0_S\eta$ relative to $D^0\to K^0_S\pi^0$ or $D^0\to K^0_S\pi^+\pi^-$.
Belle~\cite{Belle:2009ujf} reported the branching fraction of $D^+_s\to K^0_S\pi^+$ relative to $D^+_s\to K^0_SK^+$.
MARKIII~\cite{MARK-III:1991fvi}, CLEO~\cite{CLEO:1990lrq}, ARGUS~\cite{ARGUS:1992gpk}, and ARGUS~\cite{ARGUS:1988vgj}
reported the branching fraction of $D^0\to K^0_S\omega$ relative to $D^0\to K^0_S\pi^+\pi^-\pi^0$,
$D^0\to K^0_S\pi^+\pi^-$, $D^0\to K^0_S\pi^+\pi^-$, and $D^0\to K^-\pi^+$, respectively.
MARKII~\cite{Schindler:1980ws}, MARKIII~\cite{MARK-III:1985tfw}, E691~\cite{Anjos:1990nm}, E687~\cite{E687:1994qpa},
BES~\cite{BES:2005irx}, FOCUS~\cite{FOCUS:2001kwm}, CLEOII~\cite{CLEO:1997zxb,CLEO:2003rez}, and Belle~\cite{Belle:2009ujf}
reported the branching fraction of  $D^+\to K^0_SK^+$ relative to $D^+\to K^0_S\pi^+$.
FOCUS~\cite{FOCUS:2007hxo} and Belle~\cite{Belle:2009ujf} reported measurements of the branching fraction of
$D^+_s\to K^0_S\pi^+$ relative to $D^+_s\to K^0_SK^+$.
Belle reported measurements of $D^+_s\to h^+\pi^0/h^+\eta(h=K,\pi)$ relative to $D^+_s\to \phi\pi^+$~\cite{Belle:2021ygw}.
E687~\cite{E687:1995jyc},
CLEO-c~\cite{CLEO:2009nuz},
BaBar~\cite{BaBar:2010wqe}
reported the branching fractions of
$D^+_s\to \phi \pi^+$ and $D^+_s\to K^+\bar K^*(892)^0$ relative to $D^+_s\to K^+K^-\pi^+$.

In particular, CLEO-c~\cite{CLEO:2007rhw} reported the branching fractions of $D^0\to K^0_L\pi^0$ and $D^+\to K^0_S\pi^+$;
and it reported the branching fraction asymmetries
${\mathcal R}(D^0,\pi^0)=\frac{\mathcal{B}(D^0\to K_S^0 \pi^0)-\mathcal{B}(D^0\to K_L^0 \pi^0)}{\mathcal{B}(D^0\to K_S^0 \pi^0)+\mathcal{B}(D^0\to K_L^0 \pi^0)}=0.108\pm0.025\pm0.024$
and
${\mathcal R}(D^+,\pi^+)=\frac{\mathcal{B}(D^+\to K_S^0 \pi^+)-\mathcal{B}(D^+\to K_L^0 \pi^+)}{\mathcal{B}(D^+\to K_S^0 \pi^+)+\mathcal{B}(D^+\to K_L^0 \pi^+)}=0.022\pm0.016\pm0.018$. The $D^0$ asymmetry is consistent with the value based on the $U$-spin prediction
${\cal A}(D^0\to K^0\pi^0)/{\cal A}(D^0\to \bar K^0\pi^0)=-\tan^2\theta_C$~\cite{Rosner:2006bw}, where $\theta_C$ is the Cabibbo mixing angle.

For hadronic $D$ decays with pure multiple pions in the final states,
CLEO-c~\cite{CLEO:2005mti} reported the branching fractions of
$D^0\to \pi^+\pi^-$, $2\pi^0$, $\pi^+\pi^-\pi^0$, $2\pi^+2\pi^-$,
$\pi^+\pi^-2\pi^0$, $2\pi^+2\pi^-\pi^0$ to $D^0\to K^-\pi^+$.
$D^+\to \pi^+\pi^0$, $2\pi^+\pi^-$, $\pi^+2\pi^0$,  $2\pi^+\pi^-\pi^0$,  $3\pi^+2\pi^-$
relative to $D^+\to K^-2\pi^+$. The branching fractions of $D^+\to \eta\pi^+$, $\omega \pi^+$, $D^0\to \eta\pi^0$,
$\omega\pi^0$, $\eta\pi^+\pi^-$, and $\omega\pi^+\pi^-$ were also presented.
In addition,
CLEO~\cite{CLEO:1990dhi}, E691~\cite{Anjos:1991dr}, CLEOII~\cite{CLEO:1993pky,CLEO:2001lgl}, E687~\cite{E687:1993hjy},
BESII~\cite{BES:2005irx}, E791~\cite{E791:1997txw}, FOCUS~\cite{FOCUS:2002qvx}, and CDF~\cite{CDF:2004ohx}
 reported the branching fraction of
$D^0\to \pi^+\pi^-$ relative to $D^0\to K^-\pi^+$;
CLEOII~\cite{CLEO:2003rez} and BaBar~\cite{BaBar:2006jox}  reported the branching fraction of
$D^+\to \pi^+\pi^0$ relative to $D^+\to K^-2\pi^+$;
BaBar~\cite{BaBar:2006dxf} and Belle~\cite{Belle:2008qrk} reported the
branching fraction of $D^0\to \pi^+\pi^-\pi^0$ relative to $D^0\to K^-\pi^+\pi^0$.
E691~\cite{Anjos:1990se}, CLEO~\cite{CLEO:1991dma}, ARGUS~\cite{ARGUS:1994meq}, BESII~\cite{BES:2005irx},
E687~\cite{E687:1995spe}, E791~\cite{E791:1997btr}, and FOCUS~\cite{FOCUS:2004prc} reported the branching fraction of
$D^0\to 2\pi^+2\pi^-$ relative to $D^0\to K^-2\pi^+\pi^-$;
E691~\cite{Anjos:1988pu}, E687~\cite{E687:1997jvh}, E791~\cite{E791:2000vek}, and BESII~\cite{BES:2005irx}  reported the branching fraction of
$D^+\to 2\pi^+\pi^-$ relative to $D^+\to K^-2\pi^+$.
FOCUS reported a measurement of $D^+\to 3\pi^+2\pi^-$ relative to $D^+\to K^-3\pi^+\pi^-$~\cite{FOCUS:2002psb},
and a measurement of $D^0\to 3\pi^+3\pi^-$ relative to $D^0\to K^-2\pi^+\pi^-$~\cite{FOCUS:2004ypi}.

In addition, measurements of multi-body hadronic $D$ decays, e.g.,
$D\to \bar K\pi P (P=\omega,\eta,\eta^\prime)$,
$D\to \bar K n\pi (n=3,4,5)$, $D\to K\bar K n\pi (n=1,2,3)$, $D\to 3K$, and $D\to 3K\pi$,
were made by referring to various reference decay modes at different experiments,
as compiled in Table~\ref{tab:hadron_relative}.

\begin{table*}[htbp]
	\caption{Decay modes and reference modes in relative measurements of $D\to \bar K\pi P~(P=\omega,\eta,\eta^\prime)$,
$D\to \bar K n\pi~(n=3,4,5)$, $D\to K\bar K n\pi~(n=1,2,3)$, $D\to 3K$, and $D\to 3K\pi$.}
	\centering
		\centering
		\begin{tabular}{ccc}
			\hline
			\hline
Decay mode  & Reference mode & Experiment \\ \hline
$D^0\to K^-\pi^+\omega$ & $D^0\to K^-\pi^+$ & ARGUS~\cite{ARGUS:1992gpk} \\
$D^0\to K^-\pi^+\eta$ & $D^0\to K^-\pi^+$ & Belle~\cite{Belle:2020fbd} \\
$D^0\to K^-\pi^+\eta^\prime$ & $D^0\to K^-2\pi^+\pi^-$ & CLEOII~\cite{CLEO:1992sni} \\
$D^0\to K^0_S\pi^0\eta$ & $D^0\to K^0_S\pi^0$ & CLEOII~\cite{CLEO:2004umu} \\

$D^+_s\to K^+\pi^+\pi^-\pi^0$ & $D^+_s\to K^+K^-\pi^+\pi^0$ & Belle~\cite{Belle:2022aha}\\
$D^+_s\to K^0_S2\pi^+\pi^-$  & $D^+_s\to K^0_SK^-2\pi^+$ & FOCUS~\cite{FOCUS:2007hxo} \\ \hline
$D^0\to K^-2\pi^+\pi^-\pi^0$& $D^0\to K^-\pi^+$&ARGUS~\cite{ARGUS:1992gpk}\\
$D^0\to K^-2\pi^+\pi^-\pi^0$     & $D^0\to K^-2\pi^+\pi^-$&E691~\cite{Anjos:1990fv}\\
$D^+\to K^-3\pi^+\pi^-$     & $D^+\to K^-2\pi^+$&E691~\cite{Anjos:1990fv}\\
$D^0\to K^0_S2\pi^+2\pi^-$ &  $D^0\to K^0_S\pi^+\pi^-$& E691~\cite{Anjos:1990fv}, CLEO~\cite{CLEO:1991dma}, ARGUS~\cite{ARGUS:1992gpk}, FOCUS~\cite{FOCUS:2003pyu} \\
$D^+\to K^-3\pi^+2\pi^-$ & $D^+\to K^-2\pi^+$ & E691~\cite{Anjos:1990fv}, E687~\cite{E687:1997dng}, FOCUS~\cite{FOCUS:2002psb}  \\

$D^0\to K^+K^-\pi^0$   & $D^0\to K^-\pi^+\pi^0$   & CLEOII~\cite{CLEO:1996cav}, BaBar~\cite{BaBar:2006dxf}\\
$D^0\to K^0_SK^+\pi^-$ & $D^0\to K^0_S\pi^+\pi^-$ & CLEO~\cite{CLEO:1991dma} \\
$D^0\to K^0_SK^+\pi^-$ & $D^0\to K^0_SK^-\pi^+$   & CLEO-c~\cite{CLEO:2012obf}, LHCb~\cite{LHCb:2015lnk} \\
$D^0\to K^0_SK^\pm\pi^\mp$ & $D^0\to K^-\pi^+$    & E691~\cite{Anjos:1990se}\\

\multirow{2}{*}{$D^0\to K^+K^-\pi^+\pi^-$} & \multirow{2}{*}{$D^0\to K^-2\pi^+\pi^-$} &E691~\cite{Anjos:1990se}, CLEO~\cite{CLEO:1991dma}, ARGUS~\cite{ARGUS:1994meq}, BES~\cite{BES:2005irx}\\
  &  & E687~\cite{E687:1995spe}, E791~\cite{E791:1997btr}, FOCUS~\cite{FOCUS:2004prc} \\

$D^0\to 2K^0_S\pi^+\pi^-$  & $D^0\to K^0_S\pi^+\pi^-$ & ARGUS~\cite{ARGUS:1994meq}, FOCUS~\cite{FOCUS:2004met}, Belle~\cite{Belle:2022xof}\\
$D^+\to K^+K^-\pi^+\pi^0$  & $D^+\to K^-2\pi^+\pi^0$ & Belle~\cite{Belle:2022aha} \\
$D^+\to K^0_SK^-2\pi^+$    & $D^+\to K^0_S2\pi^+\pi^-$ & FOCUS~\cite{FOCUS:2001omf} \\
$D^+\to K^0_SK^+\pi^+\pi^-$& $D^+\to K^0_S2\pi^+\pi^-$ & FOCUS~\cite{FOCUS:2001omf} \\
$D^+_s\to K^0_SK^+\pi^+\pi^-$& $D^+_s\to K^0_SK^-2\pi^+$ & FOCUS~\cite{FOCUS:2001omf} \\
$D^0\to K^+K^-\pi^+\pi^-\pi^0$ & $D^0\to K^-2\pi^+\pi^-$ &  ACCMOR~\cite{ACCMOR:1992ypn} \\
$D^+_s\to 2K^0_S2\pi^+\pi^-$ & $D^+_s\to K^0_SK^-2\pi^+$ & FOCUS~\cite{FOCUS:2003pyu} \\  \hline

$D^0\to 3K^0_S$ & $D^0\to K^0_S\pi^+\pi^-$ &  ARGUS~\cite{ARGUS:1989ctp}, CLEO~\cite{CLEO:1991dma}, E687~\cite{E687:1994xwr}, CLEOII~\cite{CLEO:1996cav}, FOCUS~\cite{FOCUS:2004met}\\
$D^0\to K^0_SK^+K^-$ & $D^0\to K^0_S\pi^+\pi^-$ & ARGUS~\cite{ARGUS:1985hre}, CLEO~\cite{CLEO:1986pir,CLEO:1991dma}, E687~\cite{E-687:1992ogb}, BaBar~\cite{BaBar:2005vhe}\\
$D^+\to 2K^0_SK^+$ & $D^+\to K^-2\pi^+$ & CLEO~\cite{CLEO:1991dma}, FOCUS~\cite{ARGUS:1994meq}  \\
$D^+_s\to 2K^+K^-$ & $D^+_s\to K^+K^-\pi^+$ & FOCUS~\cite{FOCUS:2002ame}, BaBar~\cite{BaBar:2010wqe}\\

$D^0\to 2K^+K^-\pi^+$ & $D^0\to K^-2\pi^+\pi^-$ & E687~\cite{E687:1995spe}, E791~\cite{E791:2001qch}, FOCUS~\cite{FOCUS:2003gcs} \\
$D^0\to 2K^0_SK^\pm\pi^\mp$ & $D^0\to K^0_S\pi^+\pi^-$ & FOCUS~\cite{FOCUS:2004met} \\
$D^+\to K^0_SK^+K^-\pi^+$  & $D^+\to K^0_S2\pi^+\pi^-$ & FOCUS~\cite{FOCUS:2001omf} \\
$D^+_s\to K^0_SK^+K^-\pi^-$  & $D^+_s\to K^0_SK^+\pi^+\pi^-$ & Belle~\cite{Belle:2023bzn} \\
 \hline
			\hline
		\end{tabular}
\label{tab:hadron_relative}
\end{table*}

\subsubsection{Two-body $D$ decays at BESIII}

\begin{figure}
\begin{center}
\includegraphics[width=0.225\textwidth]{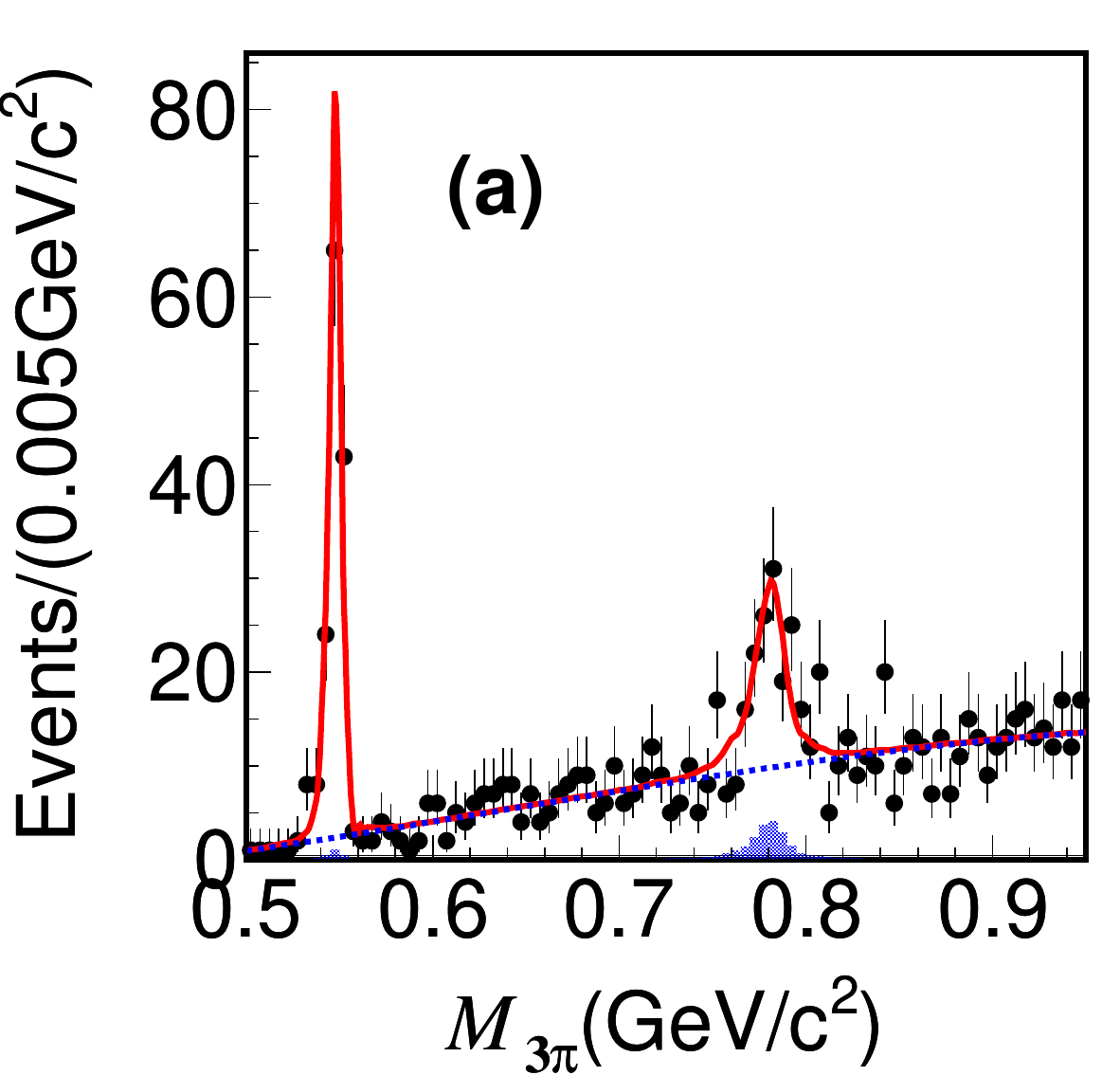}
\includegraphics[width=0.225\textwidth]{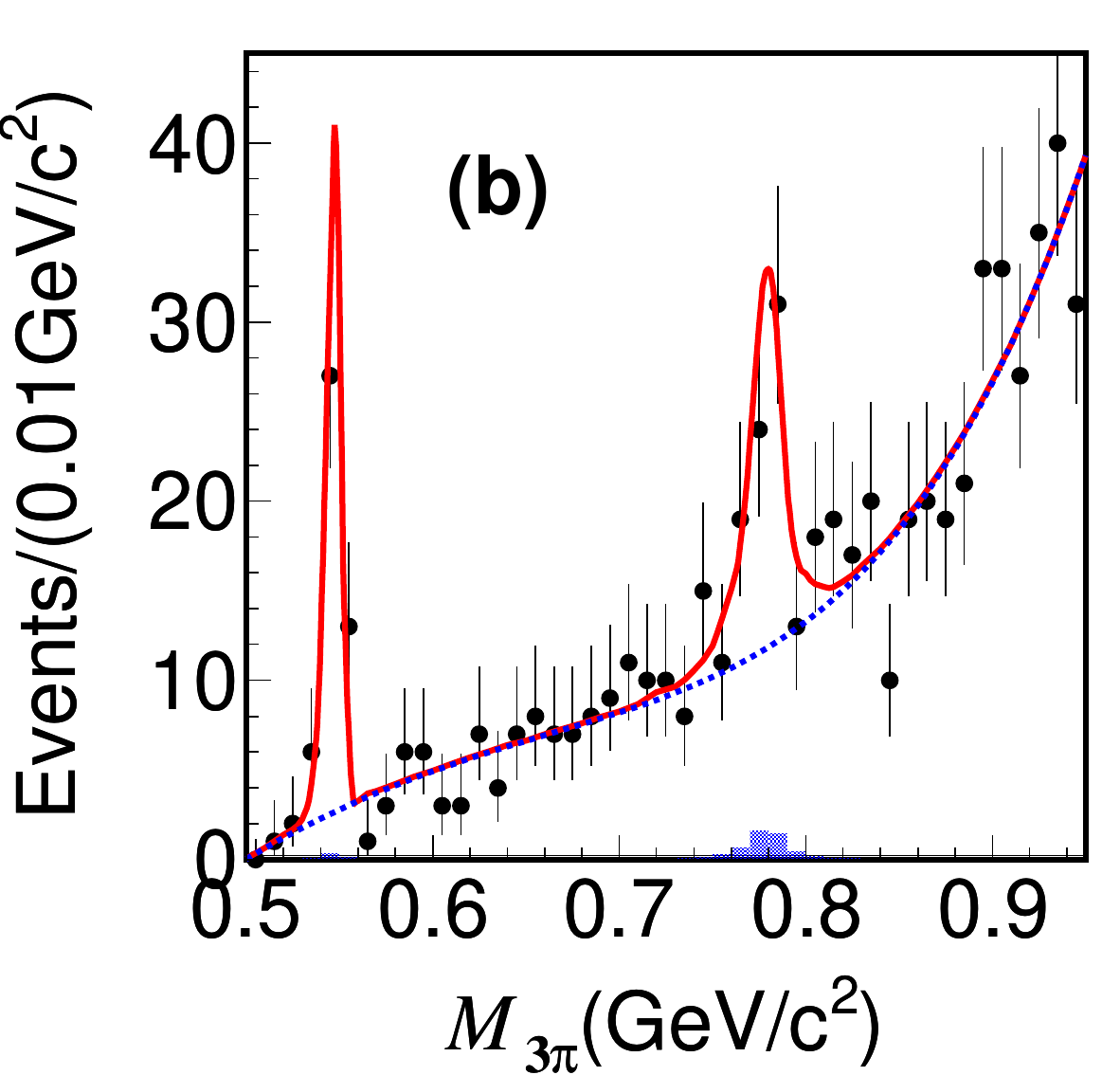}
\caption{Fits to the $3\pi$ mass spectra for (a) $D^+\to \pi^+\pi^-\pi^0\pi^+$
and (b) $D^0 \to \pi^+\pi^-2\pi^0$ in the signal region~\cite{BESIII:2015juf}.}
\label{fig:D_omegapi}
\end{center}
\end{figure}

\begin{figure}[htbp] \centering
\includegraphics[width=1.0\linewidth]{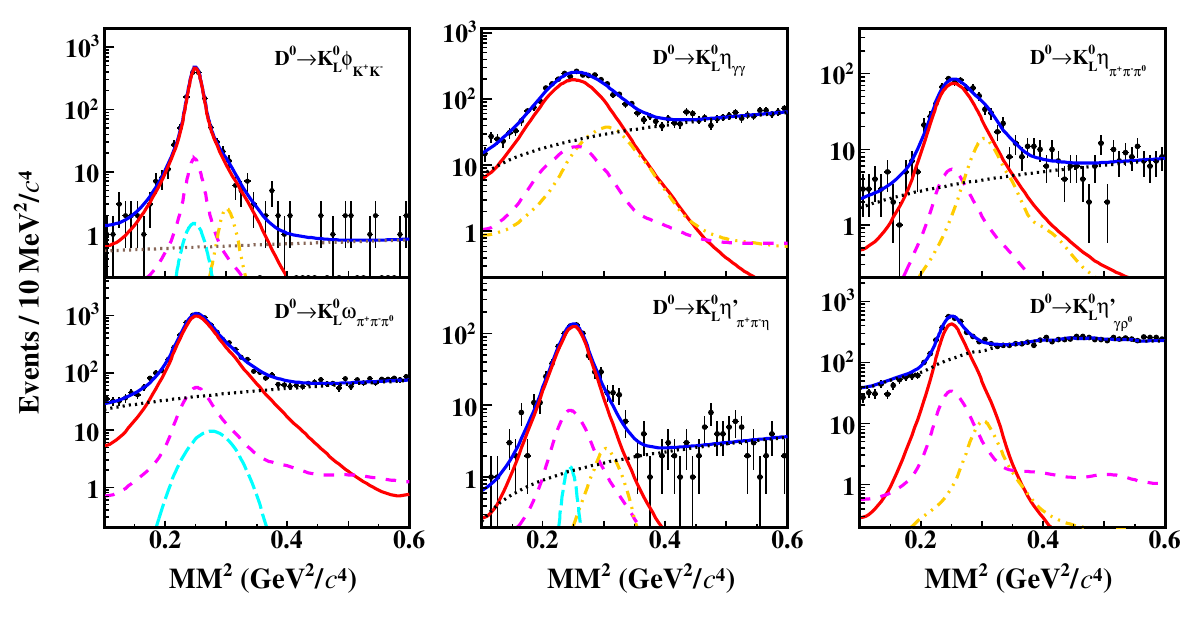}
\caption{
Fits to the MM$^2$ distributions of the accepted candidate events of $D^0\to K_L^0\phi$, $D^0\to K_L^0\eta$, $D^0\to K_L^0\omega$, and $D^0\to K_L^0\eta^{\prime}$ in data~\cite{BESIII:2022xhe}.}
\label{fig:D0_KLX}
\end{figure}

Using 2.93 ${\mbox{\,fb}^{-1}}$ of data at 3.773~GeV with the double-tag method,
the SCS decays $D^+\to\omega\pi^+$ and $D^0\to\omega\pi^0$
are measured with statistical significances of $5.5\sigma$ and $4.1\sigma$, respectively~\cite{BESIII:2015juf}.
The results of fits to the $M_{3\pi}$ distributions are shown in Fig.~\ref{fig:D_omegapi}.
Based on 79 and 45 signal events, their branching fractions are determined to be ${\cal B}(D^+\to\omega\pi^+)=(2.79\pm0.57\pm0.16)\times 10^{-4}$
and ${\cal B}(D^0\to\omega\pi^0)=(1.17\pm0.34\pm0.07)\times 10^{-4}$. In addition, the branching fractions
for $D^+\to\eta\pi^+$ and $D^0\to\eta\pi^0$ are also presented, which are ${\cal B}(D^+\to\eta\pi^+)=(3.07\pm0.22\pm0.13)\times 10^{-4}$
and ${\cal B}(D^0\to\omega\pi^0)=(0.65\pm0.09\pm0.04)\times 10^{-4}$.
Based on 9k, 2.7k, 6.1k, and 2.7k  double-tag signal events,
the  branching fractions of $D^0\to K_L^0\phi$, $D^0\to K_L^0\eta$, $D^0\to K_L^0\omega$, and $D^0\to K_L^0\eta^{\prime}$ are determined to be~\cite{BESIII:2022xhe}
$(0.414\pm0.021\pm0.010)\%$,
$(0.433\pm0.012\pm0.010)\%$,
$(1.164\pm0.022\pm0.028)\%$, and $(0.809\pm0.020\pm0.016)\%$, respectively.
Figure~\ref{fig:D0_KLX} shows the results of the fits to the $\rm MM^2$ distributions of the accepted candidates in data.
Combining these branching fractions with the known PDG values for $\mathcal{B}(D^0\to K_S^0 X)$ ($X=\phi$, $\eta$, $\omega$, and $\eta^\prime$), the asymmetries of $\mathcal{B}(D^0\to K_S^0 X)$ and $\mathcal{B}(D^0\to K_L^0 X)$,
\begin{eqnarray}
{\mathcal R}(D^0,X)&=&\frac{\mathcal{B}(D^0\to K_S^0 X)-\mathcal{B}(D^0\to K_L^0 X)}{\mathcal{B}(D^0\to K_S^0 X)+\mathcal{B}(D^0\to K_L^0 X)}\nonumber \\
\end{eqnarray}
are determined to be $(-0.1\pm4.7)\%$, $(8.0\pm2.2)\%$, $(-2.4\pm3.1)\%$, and $(8.0\pm2.3)\%$, respectively.
Clear asymmetries are found in $D^0\to K_{S,L}^0\eta, K_{S,L}^0\eta^{\prime}$, but none is found in the other two modes.
In addition, the asymmetries of the $C\!P$-conjugate branching fractions for these $D$ decays are determined
and no significant $C\!P$ violation is found.
In addition, based on 650 and 780 double-tag signal events,
Ref.~\cite{BESIII:2018pku} measured the branching fractions of
$D^+\to K_L^0K^+$ and
$D^+\to K_S^0K^+$ to be
$(3.02\pm0.09\pm0.08)\times 10^{-3}$ and
$(3.21\pm0.11\pm0.11)\times 10^{-3}$, respectively.

Using the same data sample but with the single-tag method,
the branching fractions of $D^0\to 2K^0_S$ and $D^0 \to 2\pi^0$ are measured as
$(1.67 \pm 0.11 \pm 0.11)\times 10^{-4}$~\cite{BESIII:2016nrs}
and $(8.24\pm0.21 \pm0.30)\times10^{-4}$~\cite{BESIII:2015kvk},
based on 576 and 6.3k single-tag signal events, respectively.
Meanwhile, with 17.5k, 3.3k, and 102 single-tag signal events,
the branching fractions of $D^{+}\to\phi\pi^{+}$, $D^{0}\to\phi\pi^{0}$, and $D^{0}\to\phi\eta$
are determined to be
${\cal B}(D^{0}\to\phi\pi^{0})=(1.168\pm0.028\pm0.028)\times10^{-3}$,
${\cal B}(D^{0}\to\phi\eta)=(1.81\pm0.46\pm0.06)\times10^{-4}$, and
${\cal B}(D^{+}\to\phi\pi^{+})=(5.70\pm0.05\pm0.13)\times10^{-3}$, respectively~\cite{BESIII:2019myc}.
There are three other measurements based on the same data sample and single-tag method.
Reference~\cite{BESIII:2018oqs} reported the first observation of $D^0\to \omega\eta$ 
based on 2.2k signal events, and improved measurements of $D^0\to\eta\pi^0$, $D^0\to\eta^\prime\pi^0$, $D^0\to2\eta$,
and $D^0\to\eta^\prime\eta$. The obtained branching fractions are
${\cal B}(D^0\to\omega\eta)=(2.15\pm0.17\pm0.15)\times 10^{-3}$,
${\cal B}(D^0\to\eta\pi^0)=(0.58\pm0.05\pm0.05)\times 10^{-3}$,
${\cal B}(D^0\to\eta^\prime\pi^0)=(0.93\pm0.11\pm0.09)\times 10^{-3}$,
${\cal B}(D^0\to2\eta)=(2.20\pm0.07\pm0.06)\times 10^{-3}$ and
${\cal B}(D^0\to\eta^\prime\eta)=(0.94\pm0.25\pm0.11)\times 10^{-3}$.
Based on 650 and 780 double-tag signal events,
Ref.~\cite{BESIII:2019kfh} measured the branching fractions of
$D^+_s\to K_L^0K^+$ and
$D^+_s\to K_S^0K^+$ to be
$(1.485\pm0.039\pm0.046)\%$ and
$(1.425\pm0.038\pm0.031)\%$, respectively;
which lead to the asymmetry of $\mathcal{B}(D^+_s\to K_S^0 K^+)$ and $\mathcal{B}(D^+_s\to K_L^0 K^+)$,
\begin{eqnarray}
{\mathcal R}(D^+_s,X)&=&\frac{\mathcal{B}(D^+_s\to K_S^0 K^+)-\mathcal{B}(D^+_s\to K_L^0 K^+)}{\mathcal{B}(D^+_s\to K_S^0 K^+)+\mathcal{B}(D^+_s\to K_L^0 K^+)}\nonumber \\
&=&(-2.1\pm1.9\pm1.6)\%.
\end{eqnarray}
The measured ${\mathcal R}(D^+_s,X)$ is consistent with different theoretical calculations~\cite{Bhattacharya:2009ps,Cheng:2010ry,Gao:2014ena,Muller:2015lua,Wang:2017ksn} within uncertainty.

\begin{figure*}[htbp]
\centering
\includegraphics[width=0.45\textwidth,height=0.275\textwidth]{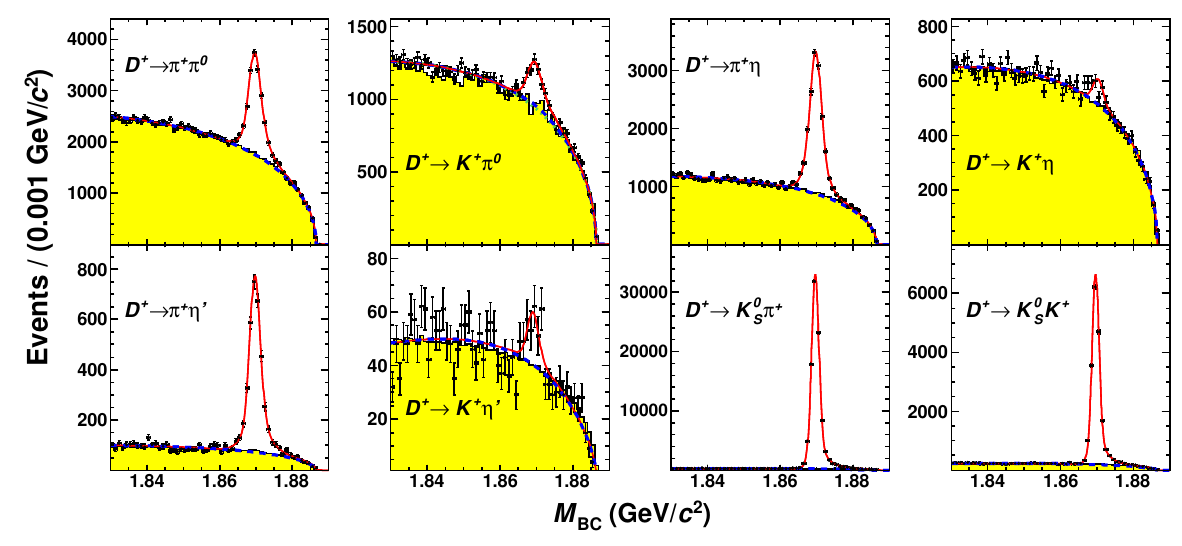}
\includegraphics[width=0.45\textwidth]{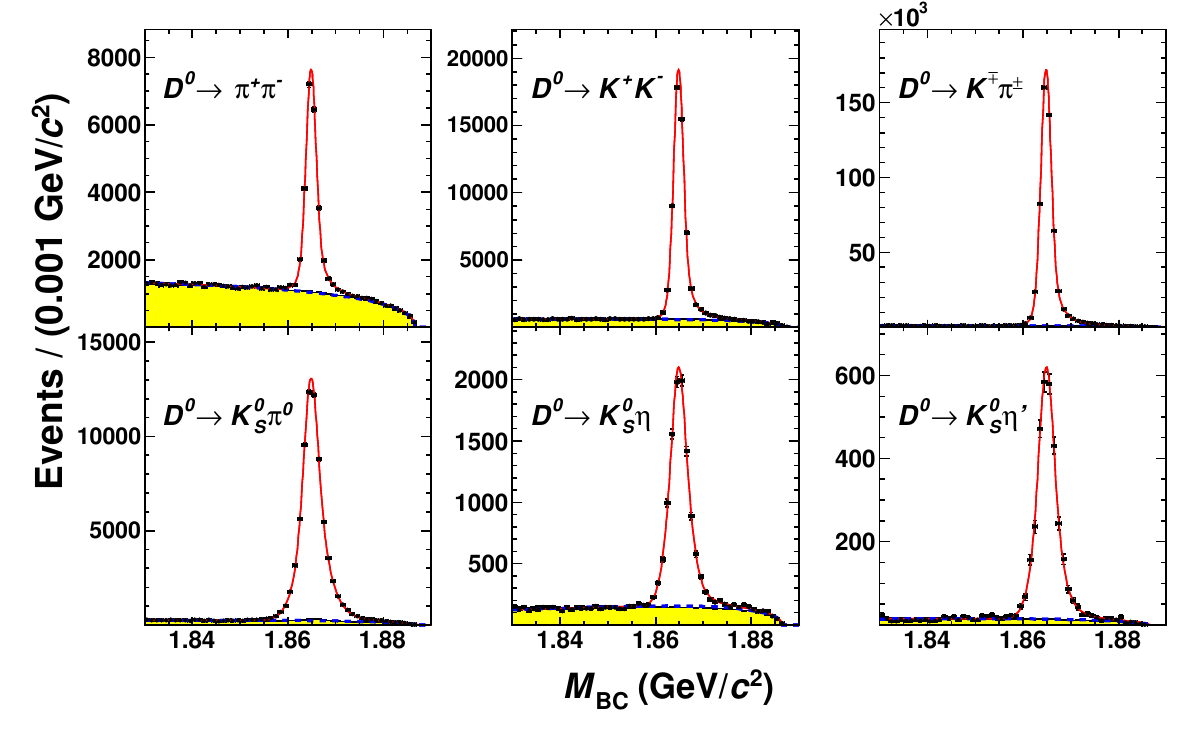}
\caption{
Fits to the $M_\mathrm{BC}$ distributions of the (left) $D^+$ and (right) $D^0$ candidate events~\cite{BESIII:2018apz}.}
\label{fig:D_PP}
\end{figure*}

Reference~\cite{BESIII:2018apz} presented improved measurements of the
branching fractions of more two-body hadronic decays
$D^+\to \pi^+ \pi^0$,
$K^+ \pi^0$, $\pi^+ \eta$, $K^+ \eta$, $\pi^+ \eta^\prime$, $K^+ \eta^\prime$,
$K_S^0 \pi^+ $, $K_S^0 K^+$, and $D^0\to \pi^+ \pi^-$,
$K^+ K^-$, $K^- \pi^+$, $K_S^0 \pi^0$, $K_S^0 \eta$, $K_S^0 \eta^\prime$.
Figure~\ref{fig:D_PP} shows the fits to the $M_{\rm BC}$ distributions of the single-tag $D^+$ and $D^0$ candidate events.
Especially, the measured branching fractions  for $D^+\to K^0_S\pi^+$ and $D^0\to K^-\pi^+$ are ${\cal B}(D^+\to K^0_S\pi^+)=(1.591\pm0.006\pm0.030)\%$ and ${\cal B}(D^0\to K^-\pi^+)=(3.898\pm0.006\pm0.051)\%$, respectively. 

\begin{figure}[htbp]
  \centering

 \includegraphics[width=0.5\textwidth]{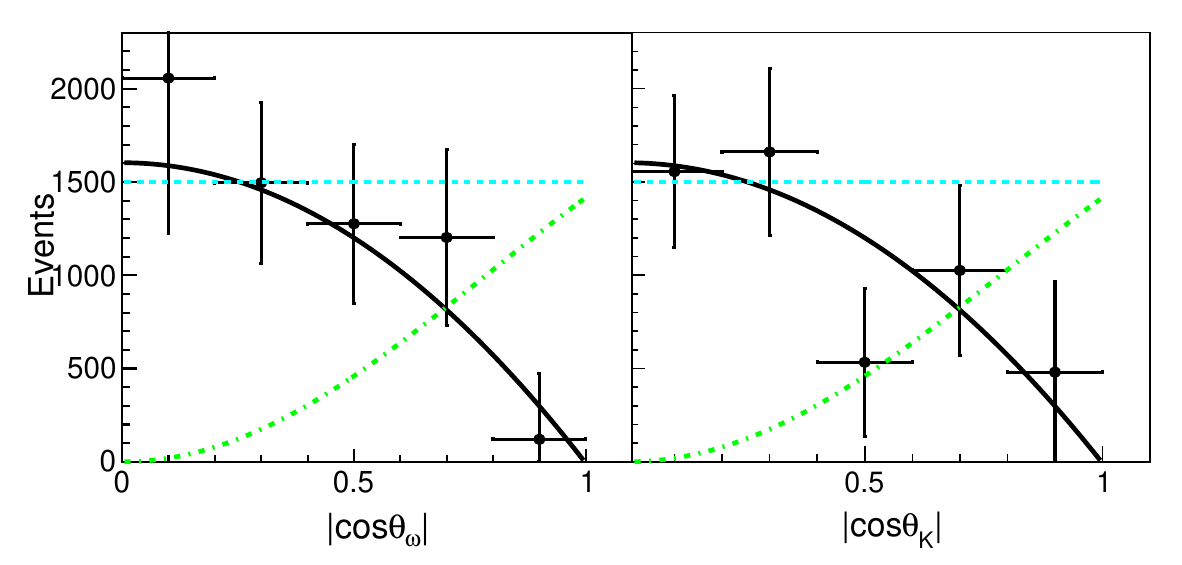}

  \caption{The distributions of the background-subtracted signal yields corrected by the efficiency versus $|\cos\theta_{\omega}|$ (left) and $|\cos\theta_K|$ (right)~\cite{BESIII:2021raf}.
  The black dots with error bars are data with both statistical and systematic uncertainties,
  and the solid black curves are the fit results.
  The distributions with the longitudinal polarization  and PHSP assumptions are  shown as the
  dotted dashed green and dashed cyan curves, respectively.
   }
  \label{fig:D0_omegaphi_Costheta}
\end{figure}

Reference~\cite{BESIII:2021raf} carried out the study of the SCS decay $D^{0}\to\omega\phi$.
This decay is observed for the first time based on $196\pm29$ signal events.
Its branching fraction is measured to be $(6.48 \pm 0.96 \pm 0.40)\times 10^{-4}$.
To study the polarization in the $D^0\to\omega\phi$ decay, the efficiency-corrected signal yields are evaluated in five equal bins of $|\cos\theta_{\omega}|$ and $|\cos\theta_K|$ as shown in Fig.~\ref{fig:D0_omegaphi_Costheta}. This reveals that the $\phi$ and $\omega$ mesons from the $D^{0} \to \omega \phi$ decay are transversely polarized.
The  upper limit on longitudinal polarization fraction is set to be less than $0.24$ at the $95\%$ confidence level, which
is inconsistent with current theoretical expectations~\cite{Cheng:2010rv,Hiller:2013cza} and challenges our understanding of the underlying dynamics in charm meson decays.

Using 3.19~fb$^{-1}$ of data at 4.178~GeV,
the observation of the $W$-annihilation decay $D^{+}_{s} \to \omega \pi^{+}$
and the evidence for $D_{s}^{+} \to \omega K^{+}$ were reported with the double-tag method~\cite{BESIII:2018mwk}.
Based on 65 and 29 signal events, their branching fractions are determined to be
$\mathcal{B}(D^{+}_{s} \to \omega \pi^{+})
= (1.77\pm0.32\pm0.11) \times 10^{-3}$
and $\mathcal{B}(D^{+}_{s} \to \omega K^{+})
= (0.87\pm0.24\pm0.07) \times 10^{-3}$,
respectively.
Moreover, based on 193 double-tag signal events~\cite{BESIII:2018cfe},
the branching fraction of $D^+_s \to p\bar{n}$ is determined to be $(1.21\pm0.10\pm0.05)\times10^{-3}$,
which confirms the previous CLEO-c measurement~\cite{CLEO:2008aum} with much improved precision.
The anomalously large branching fraction of $D^+_s \to p\bar{n}$ explicitly shows that the weak annihilation process featured as a short-distance dynamics is not the driving mechanism for this transition, while the hadronization process driven by non-perturbative dynamics determines the underlying physics~\cite{Pham:1980dc,Bediaga:1991eu,Chiesa:2025wnx,Chen:2008pf}.

Using the same data sample but with the single-tag method,
the branching fractions for seven two-body $D_{s}^{+}\to PP$ decays
are measured relative to the reference mode $D^+_s\to K^+K^-\pi^+$~\cite{BESIII:2020kim}.
Combining the world average of the branching fraction of $D_{s}^{+}\to K^{+}K^{-}\pi^{+}$ leads to
\begin{center}
	$\mathcal{B}(D_s^+\to K^+\eta^{\prime})=(2.68\pm0.17\pm0.17\pm0.08)\times10^{-3}$,\\
	$\mathcal{B}(D_s^+\to \eta^{\prime}\pi^+)=(37.8\pm0.4\pm2.1\pm1.2)\times10^{-3}$,\\
	$\mathcal{B}(D_s^+\to K^+\eta)=(1.62\pm0.10\pm0.03\pm0.05)\times10^{-3}$,\\
	$\mathcal{B}(D_s^+\to \eta\pi^+)=(17.41\pm0.18\pm0.27\pm0.54)\times10^{-3}$, \\
	$\mathcal{B}(D_s^+\to K^+K_S^0)=(15.02\pm0.10\pm0.27\pm0.47)\times10^{-3}$,\\
	$\mathcal{B}(D_s^+\to K_S^0\pi^+)=(1.109\pm0.034\pm0.023\pm0.035)\times10^{-3}$, \\
	$\mathcal{B}(D_s^+\to K^+\pi^0)=(0.748\pm0.049\pm0.018\pm0.023)\times10^{-3}$.
\end{center}

In addition, two-body hadronic $D\to VP$ decays
$D\to \rho(770) P(P=\pi,K,\eta^{(\prime)})$,
$D\to \omega P$,
$D\to K^*(892)P$, and
$D\to \phi P$
can be extracted from amplitude analyses of multiple-body hadronic $D$ decays
$D\to \pi\pi P$,
$D\to \pi^+\pi^-\pi^0 P$,
$D\to K^+K^- P$,
$D\to K\pi P$;
while two-body $D\to VV$ decays of
$K^*(892)\rho(770)$,
$K^*(892)K^*(892)$ and $\phi\rho(770)$,
$K^*(892)\omega$,
$\rho(770)\rho(770)$,
$\phi K^*$,
$\omega\rho(770)$,
can be extracted from amplitude analyses of multiple-body hadronic $D$ decays
$D\to K\pi\pi\pi$,
$D\to KK\pi\pi$,
$D\to K\pi\omega$,
$D\to \pi\pi\pi\pi$,
$D\to KKK\pi$, and
$\pi\pi\omega$, where $K$ includes charge conjugations.
From these amplitude analyses, rich information of two-body hadronic decays
of
$D\to PS$, $D\to TP$, $D\to TS$, and $D\to AP$ can also be obtained.
Throughout this article, $P$, $S$, $V$, $A$, and $T$ denote pseudoscalar, scalar,
vector, axial-vector, and tensor mesons, respectively. 
See Section~\ref{sec:amplitude} for more details.

\subsubsection{Multi-body $D$ decays at BESIII}

Based on a data sample of 2.93 fb$^{-1}$ at 3.773~GeV, BESIII has conducted a series of precision measurements
of multi-body hadronic $D$ meson decays using either single-tag or double-tag method.
With the single-tag method, Ref.~\cite{BESIII:2016nrs} reported the measurements of
$D^+\to 2K^0_SK^+$, $D^+\to 2K^0_S\pi^+$, and
$D^0\to 3K^0_S$; while Ref.~\cite{BESIII:2018exp} reported the studies of
$D^0\to K^-\pi^+\eta^\prime$, $D^0\to K^0_S\pi^0\eta^\prime$, and $D^+\to K^0_S\pi^+\eta^\prime$.
Among them, the decays $D^+\to 2K^0_S\pi^+$, $D^0\to K^0_S\pi^0\eta^\prime$ and
$D^+\to K^0_S\pi^+\eta^\prime$ are measured for the first time,
and the other decays are measured with significantly improved precision compared to
the previous measurements.

\begin{figure}[htp]
  \centering
\includegraphics[width=1.0\linewidth]{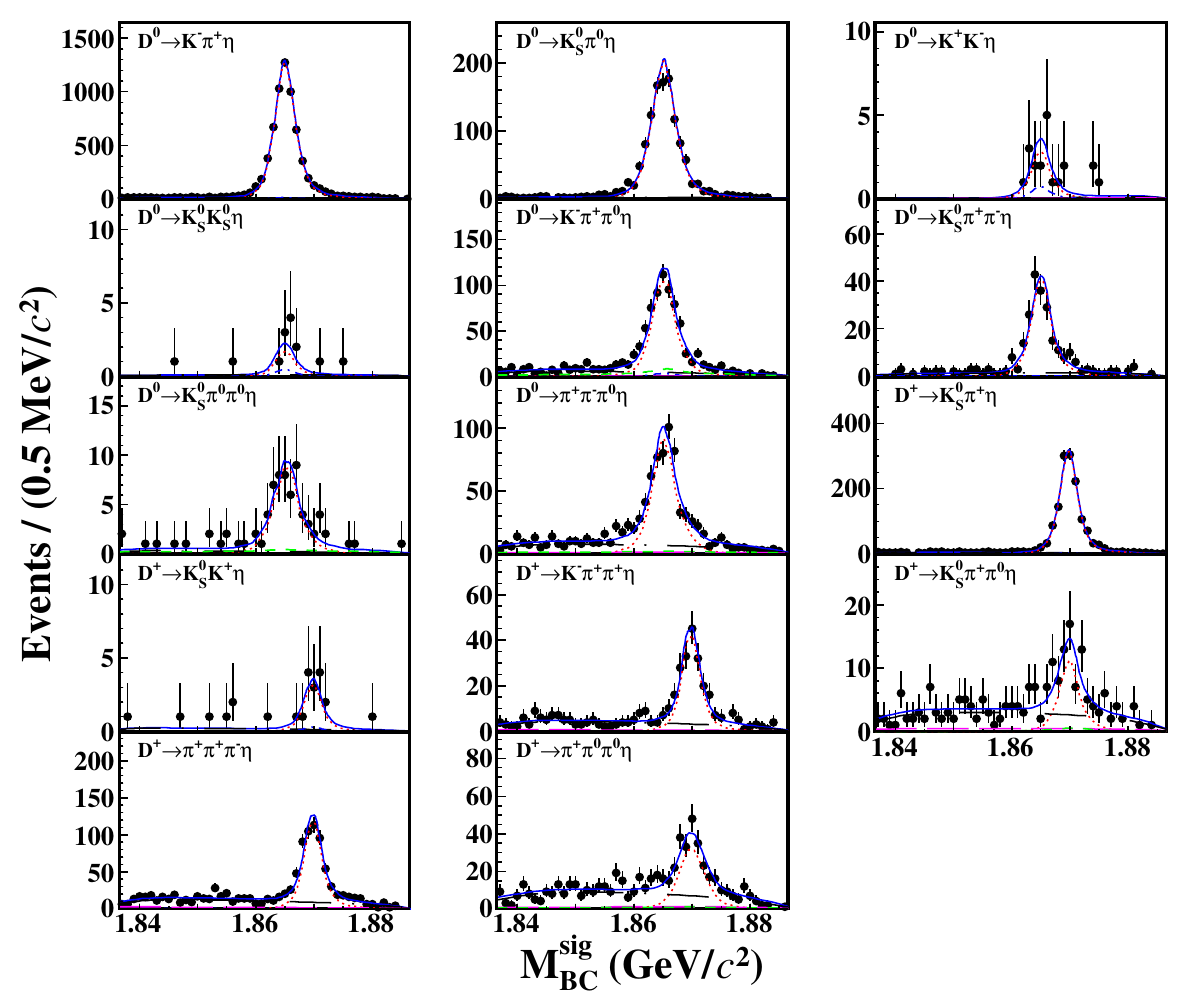}
  \caption{\small
Projections on
$M^{\rm sig}_{\rm BC}$ of the 2-D fits to the double-tag candidate events for $D^0\to K^-\pi^+\eta$, $K^0_S\pi^0\eta$, $K^+K^-\eta$, $K^0_SK^0_S\eta$, $K^-\pi^+\pi^0\eta$, $K^0_S\pi^+\pi^-\eta$, $K^0_S 2\pi^0\eta$, and $\pi^+\pi^-\pi^0\eta$; $D^+\to K^0_S\pi^+\eta$, $K^0_SK^+\eta$, $K^-2\pi^+\eta$, $K^0_S\pi^+\pi^0\eta$, $2\pi^+\pi^-\eta$, and $\pi^+2\pi^0\eta$ in data~\cite{BESIII:2020pxp}.}
\label{fig:D0p_etaX_exclusive}
\end{figure}

With  2.93 fb$^{-1}$ of data at 3.773~GeV and the double-tag method,
BESIII reported the first measurements of the absolute branching fractions of fourteen
hadronic $D^{0(+)}$ decays to exclusive final states with an $\eta$, e.g., $D^0\to K^-\pi^+\eta$, $K^0_S\pi^0\eta$, $K^+K^-\eta$, $K^0_SK^0_S\eta$, $K^-\pi^+\pi^0\eta$, $K^0_S\pi^+\pi^-\eta$, $K^0_S 2\pi^0\eta$, and $\pi^+\pi^-\pi^0\eta$; $D^+\to K^0_S\pi^+\eta$, $K^0_SK^+\eta$, $K^-2\pi^+\eta$, $K^0_S\pi^+\pi^0\eta$, $2\pi^+\pi^-\eta$, and $\pi^+2\pi^0\eta$~\cite{BESIII:2020pxp}.
Figure~\ref{fig:D0p_etaX_exclusive} shows the projections of 2-D fits on their $M_{\rm BC}^{\rm sig}$ distributions.
Among these decays, the $D^0\to K^-\pi^+\eta$ and $D^+\to K^0_S\pi^+\eta$ decays have the largest branching fractions, which are
$\mathcal{B} (D^0\to K^-\pi^+\eta )=(1.853\pm0.025\pm0.031)\%$ and
$\mathcal{B}(D^+\to K^0_S\pi^+\eta)=(1.309\pm0.037\pm0.031)\%$, respectively.
The $C\!P$ asymmetries for the six decays with highest event yields are determined to be percent level, and no statistically significant $C\!P$ violation is found.
In addition,
Ref.~\cite{BESIII:2021zma} presented the improved measurement of $D^0\to K^-\pi^+\omega$ as well
as the first measurements of $D^0\to K^0_S\pi^0\omega$ and
$D^+\to K^0_S\pi^+\omega$;
Ref.~\cite{BESIII:2022mji} presented the first measurements of the branching fractions of
the hadronic decays of
$D^0\to K^0_S3\pi^0$,
$D^0\to K^-\pi^+3\pi^0$,
$D^0\to K^0_S\pi^+\pi^-2\pi^0$,
$D^+\to K^0_S\pi^+2\pi^0$,
$D^+\to K^0_S\pi^+3\pi^0$,
$D^+\to K^-2\pi^+2\pi^0$, and
$D^+\to K^0_S2\pi^+\pi^-\pi^0$, in which significant non-($\eta,\omega$) contributions have been found
except for the decay $D^{+}\to K^{0}_{S}\pi^{+}3\pi^{0}$.
Moreover, Ref.~\cite{BESIII:2018pku} reported the branching fractions of
$D^+\to K_L^0K^+\pi^0$ and
$D^+\to K_S^0K^+\pi^0$ to be
$(5.24\pm0.22\pm0.22)\times 10^{-3}$ and
$(5.07\pm0.19\pm0.23)\times 10^{-3}$, respectively.

\begin{figure}[htbp]
  \centering
\includegraphics[width=1.0\linewidth]{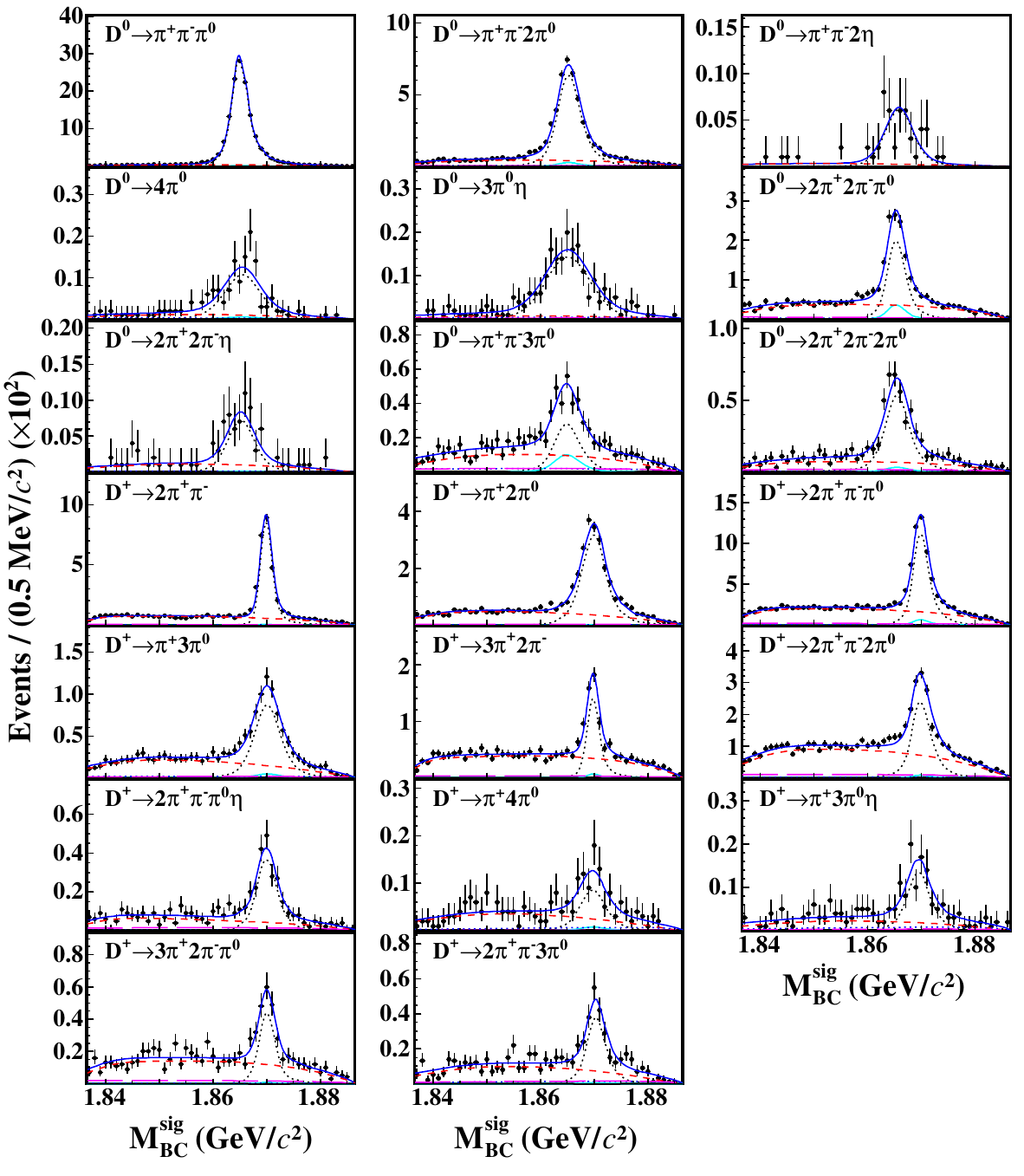}
  \caption{\small
Projections of
$M^{\rm sig}_{\rm BC}$ of the 2-D fits to the double-tag candidate events for
$D^0\to \pi^+\pi^-\pi^0$,
$\pi^+\pi^-2\pi^0$,
$\pi^+\pi^-2\eta$,
$4\pi^0$,
$3\pi^0\eta$,
$2\pi^+2\pi^-\pi^0$,
$2\pi^+2\pi^-\eta$,
$\pi^+\pi^-3\pi^0$,
$2\pi^+2\pi^-2\pi^0$,  and
$D^+\to 2\pi^+\pi^-$,
$\pi^+2\pi^0$,
$2\pi^+\pi^-\pi^0$,
$\pi^+3\pi^0$,
$3\pi^+2\pi^-$,
$2\pi^+\pi^-2\pi^0$,
$2\pi^+\pi^-\pi^0\eta$,
$\pi^+4\pi^0$,
$\pi^+3\pi^0\eta$,
$3\pi^+2\pi^-\pi^0$,
$2\pi^+\pi^-3\pi^0$ in data~\cite{BESIII:2022qrs}.
}
\label{fig:D0_pis}
\end{figure}

Also with 2.93 fb$^{-1}$ of data at 3.773~GeV and the double-tag method, more studies of SCS $D$ decays came from
Refs.~\cite{BESIII:2018hui,BESIII:2019xhl,BESIII:2020paw,BESIII:2022qrs}.
Reference~\cite{BESIII:2018hui} made the first observation of
$D^0\to\pi^02\eta$ as well as the first evidence for $D^0\to 3\pi^0$ and $2\pi^0\eta$;
Ref.~\cite{BESIII:2019xhl} reports the first observation of $D^+\to 2\eta\pi^+$
as well as the improved measurements of $D^+\to\eta\pi^+\pi^0$ and $D^0\to\eta\pi^+\pi^-$;
Ref.~\cite{BESIII:2020paw} provided the first observation of
$D^{+}\to \omega \pi^+\pi^0$ and the improved measurement of $D^{0} \to \omega \pi^+\pi^-$.
Particularly, Ref.~\cite{BESIII:2022qrs} presented the first absolute measurements of the branching fractions of
$D^0\to \pi^+\pi^-\pi^0$,
$\pi^+\pi^-2\pi^0$,
$\pi^+\pi^-2\eta$,
$4\pi^0$,
$3\pi^0\eta$,
$2\pi^+2\pi^-\pi^0$,
$2\pi^+2\pi^-\eta$,
$\pi^+\pi^-3\pi^0$,
$2\pi^+2\pi^-2\pi^0$,  and
$D^+\to 2\pi^+\pi^-$,
$\pi^+2\pi^0$,
$2\pi^+\pi^-\pi^0$,
$\pi^+3\pi^0$,
$3\pi^+2\pi^-$,
$2\pi^+\pi^-2\pi^0$,
$2\pi^+\pi^-\pi^0\eta$,
$\pi^+4\pi^0$,
$\pi^+3\pi^0\eta$,
$3\pi^+2\pi^-\pi^0$,
$2\pi^+\pi^-3\pi^0$;
 the projections of 2-D fits on individual $M_{\rm BC}^{\rm sig}$ distributions are shown in Fig.~\ref{fig:D0_pis}.
In these articles, the absolute branching fractions of each decay are presented.
The largest four branching fractions obtained are
$\mathcal{B}(D^0\to\pi^+\pi^-\pi^0)=   (1.343\pm0.013\pm0.016)\%$,
$\mathcal{B}(D^0\to\pi^+\pi^-2\pi^0)=  (0.998\pm0.019\pm0.024)\%$,
$\mathcal{B}(D^+\to 2\pi^+\pi^-\pi^0)= (1.174\pm0.021\pm0.021)\%$, and
$\mathcal{B}(D^+\to 2\pi^+\pi^-2\pi^0)=(1.074\pm0.040\pm0.030)\%$.
The precisions of the branching fractions of $D^0\to \pi^+\pi^-\pi^0$, $\pi^+\pi^-2\pi^0$, $2\pi^+2\pi^-\pi^0$, and $D^+\to 2\pi^+\pi^-$, $\pi^+2\pi^0$, $2\pi^+\pi^-\pi^0$,
$3\pi^+2\pi^-$ are improved by factors of 1.2-2.9 compared to the world average values~\cite{ParticleDataGroup:2022pth} based on relative measurements;
while the other 13 decay modes are measured for the first time. In addition, the $C\!P$ asymmetries for the six decays with highest event yields are determined to be percent level, and no significant violation is found.

From an analysis of 7.9~fb$^{-1}$ of data at 3.773 GeV with a double-tag method, Ref.~\cite{BESIII:2025qra}
reported the first observations of the hadronic decays $D^0\to K^- 2\pi^+\pi^-2\pi^0$ and $D^+\to K^- 3\pi^+\pi^-\pi^0$
as well as an improved measurement of $D^0\to K^- 3\pi^+2\pi^-$.
Using 20.3~fb$^{-1}$ of data at 3.773 GeV with a double-tag method,
Ref.~\cite{BESIII:2025qra} reported the improved measurements of
$D^+\to K^+K^-2\pi^+\pi^-$ and $D^+\to K^0_SK^+\eta$ as well as the first measurements of
$D^+\to \phi 2\pi^+\pi^-$,
$D^+\to K^0_SK^+\pi^+\pi^-\pi^0$, and $D^+\to K^0_SK^+\omega$;
Ref.~\cite{BESIII:2025eqp}
presented the first absolute measurements of the branching fractions of
$D^0\to K^0_S K^+K^-\pi^0$,
$D^0\to 2K^0_S K^-\pi^+$,
$D^0\to 2K^0_S K^+\pi^-$,
$D^0\to K^+2K^-\pi^+$, and
$D^+\to K^0_S K^+K^-\pi^+$.
The decays $D^0\to \phi K^0_S\pi^0$,
$D^0\to \phi K^-\pi^+$, and $D^+\to \phi K^0_S\pi^+$ are
found to be the main sub-processes in
$D^0\to K^0_S K^+K^-\pi^0$, $D^0\to K^+2K^-\pi^+$, and $D^+\to K^0_S K^+K^-\pi^+$,
respectively; and their branching fractions have also been determined.

Using 7.33~fb$^{-1}$ of data at 4.128-4.226 GeV and a double-tag method, the absolute branching fractions
of fifteen hadronic $D_s^+$ decays:
$D_{s}^{+} \to K^{0}_{S} K^{+}$,
$D_{s}^{+} \to K^{+} K^{-} \pi^{+}$,
$D_{s}^{+} \to K^{0}_{S} K^{+} \pi^{0}$,
$D_{s}^{+} \to 2 K^{0}_{S} \pi^{+}$,
$D_{s}^{+} \to K^{+} K^{-} \pi^{+} \pi^{0}$,
$D_{s}^{+} \to K^{0}_{S} K^{+} \pi^{+} \pi^{-}$,
$D_{s}^{+} \to K^{0}_{S} K^{-} 2 \pi^{+}$,
$D_{s}^{+} \to 2\pi^{+} \pi^{-}$,
$D_{s}^{+} \to \pi^{+} \eta$,
$D_{s}^{+} \to \pi^{+} \pi^{0} \eta$,
$D_{s}^{+} \to 2 \pi^{+} \pi^{-} \eta$,
$D_{s}^{+} \to \pi^{+} \eta^\prime$,
$D_{s}^{+} \to \pi^{+} \pi^{0} \eta^\prime$,
$D_{s}^{+} \to K^{0}_{S} \pi^{+} \pi^{0}$, and
$D_{s}^{+} \to K^{+} \pi^{+} \pi^{-}$
were determined by analyzing both single-tag and double-tag events~\cite{BESIII:2024oth}.
Additionally, the \emph{CP}-violating asymmetries of these hadronic $D_s^{\pm}$ decays have been measured to percent-level precision, and no significant asymmetries are observed.
The measured branching fractions or asymmetries of $D_{s}^{+}\to \pi^{+}\pi^{0}\eta$,
$D_{s}^{+}\to \pi^{+}\pi^{0}\eta^\prime$,
$D_{s}^{+} \to K^{+}K^{-}\pi^{+}$,
$D_{s}^{+} \to K^0_S\pi^+\pi^0$,
$D_{s}^{+} \to 2K^0_S\pi^+$,
$D_{s}^{+} \to K^0_SK^+\pi^0$,
$D_{s}^{+} \to K^+\pi^+\pi^-$,
$D^{+}_{s} \to 2\pi^{+}\pi^{-}\eta$,
$D^{+}_{s} \to K^0_SK^-2\pi^{+}$,
$D^{+}_{s} \to K^+K^-\pi^{+}\pi^{0}$ supersede those reported in
Refs.~\cite{BESIII:2019jjr,BESIII:2022ewq,BESIII:2020ctr,BESIII:2021xox,BESIII:2021anf,BESIII:2022npc,BESIII:2022vaf,BESIII:2021aza,BESIII:2021dot,BESIII:2021qfo},
in which the amplitude analyses and branching fraction measurements were performed based on 3.19 fb$^{-1}$ of data at 4.178 GeV
or 6.3 fb$^{-1}$ of data at 4.178-4.226 GeV.
As examples, the obtained branching fractions of $D_s^+ \to K^+ K^- \pi^+$,  $K_S^0 K^+$, and $K^+ K^- \pi^+ \pi^0$ are $\mathcal{B}(D_s^+ \to K^+ K^- \pi^+)=(5.49 \pm 0.04 \pm 0.07)\%$, $\mathcal{B}(D_s^+ \to K_S^0 K^+)=(1.50 \pm 0.01 \pm 0.01)\%$ and $\mathcal{B}(D_s^+ \to K^+ K^- \pi^+ \pi^0)=(5.50 \pm 0.05 \pm 0.11)\%$, respectively.  
In addition, the first observation of $D^+_s\to\omega\pi^+\eta$ is also reported with a statistical significance of $7.6\sigma$~\cite{BESIII:2023mie}.

\section{Amplitude analyses of hadronic $D$ decays}
\label{sec:amplitude}

Amplitude analyses of multibody hadronic $D$ decays are carried out by using an unbinned maximum-likelihood method. The probability density function is constructed as the sum of the signal amplitude and the background function with the corresponding fraction as the coefficient. The signal amplitude is parameterized with the isobar formulation in the covariant tensor formalism~\cite{Zou:2002ar}. The total signal amplitude $\mathcal{M}$ is a coherent sum of intermediate processes $\mathcal{M} = \sum \rho_n e^{i\phi_n}\mathcal{A}_n$, where $\rho_n e^{i\phi_n}$ is the coefficient of the $n^{\rm th}$ amplitude with magnitude $\rho_n$ and phase $\phi_n$. For $N$-body hadronic $D$ decays, the $n^{\rm th}$ amplitude $\mathcal{A}_n$ is usually given by the product of the Blatt-Weisskopf barrier factor of the $D$ meson $F_n^{D}$ and the intermediate state $F_n^i$~\cite{Blatt_Weisskopf_1973}, spin factor $S_n^i$~\cite{Zou:2002ar} and the propagator for the resonance $P_n^i$, $\mathcal{A}_n = F_n^{D}S^{D}\prod_{i=1}^{N-2} F_n^iS_n^iP_n^i$, where $i$ indicates the $i^{\rm th}$ intermediate process. The relativistic BW function~\cite{Jackson:1964zd} is used to describe the propagator for the resonances $\omega$, $K_0^*(700)$, $K^*(892)$, $\phi$, $a_1(1260)$, $b_1(1235)$, $K_1(1270)$, $K_1(1400)$, $K^*(1410)$, $K^*_0(1430)$, $K^*_2(1430)$, $K(1460)$, $K_1(1650)$, $K^*(1680)$, $K_2(1770)$, $K^*_3(1780)$, $f_1(1285)$, $f_1(1420)$, $f_2(1270)$, $\eta(1405)$, $\eta(1475)$, $f_0(1370)$, $f_0(1500)$, $f^\prime_2(1525)$, $f_0(1710)$, $a_0(1450)$, $a_1(1260)$, $a_2(1320)$, $a_2(1700)$, $\pi_1(1400)$, $\pi_1(1600)$, $S(1710)$, $a_0(1817)$.
The resonances $\rho(770)$, $\rho(1450)$ and $\rho(1700)$ are parametrized by the Gounaris-Sakurai line shape~\cite{Gounaris:1968mw}.
The resonances $a_0(980)$ and $f_0(980)$ are parameterized by a coupled Flatt\'{e} formula, and the parameters are fixed to the
values given in Ref.~\cite{BESIII:2016tqo} and Ref.~\cite{BES:2004twe},
respectively; while the $f_0(500)$ is parameterized by following Ref.~\cite{Bugg:1996ki}.
The $a_1(1260)$ is considered as a quasi-three-body decay and the width is determined by integrating the amplitude squared over phase space~\cite{BESIII:2023qgj};
and the masses and widths of other intermediate resonances used in the fits are taken from the PDG.
The background shape is estimated with the inclusive MC sample using the XGBoost package~\cite{Rogozhnikov:2016bdp,Liu:2019huh}.

Before BESIII, some amplitude analyses of charm meson decays were performed by various experiments.
For $D^+ \to K^- 2\pi^+$, amplitude analyses were carried out by MARKIII~\cite{MARK-III:1987qok}, NA14~\cite{NA142:1990qcz}, E691~\cite{E691:1992rwf}, E687~\cite{E687:1994wlh}, E791~\cite{E791:2002xlc,E791:2005gev}, and CLEO-c~\cite{CLEO:2008jus}. MARKIII also performed an amplitude analysis of $D^+ \to K_S^0 \pi^+ \pi^0$~\cite{MARK-III:1987qok}, while E691 reported amplitude analyses of $D^0 \to K^- \pi^+ \pi^0$, $D^0 \to \bar K^0 \pi^+ \pi^-$, and $D^+ \to K^- 2\pi^+$~\cite{E691:1992rwf}.
For $D^0$ decays, CLEOII and CLEO-c conducted amplitude analyses of $D^0 \to K^- \pi^+ \pi^0$~\cite{CLEO:2000fvk},
$D^0\to K^0_S\pi^0\eta$~\cite{CLEO:2004umu}, and $D^0 \to K_S^0 2\pi^0$~\cite{CLEO:2011cnt};
BaBar performed amplitude analyses of $D^0 \to K^+K^-\pi^0$~\cite{BaBar:2007soq} and $D^0 \to \bar K^0 K^+ K^-$~\cite{BaBar:2005vhe}; while both of them studied $D^0 \to \pi^+ \pi^- \pi^0$~\cite{CLEO:2005uoz,BaBar:2007dro}. Belle reported an amplitude analysis of $D^0 \to K^- \pi^+ \eta$~\cite{Belle:2020fbd}, and LHCb reported amplitude analyses of $D^0 \to K_S^0 K^- \pi^+$ and $D^0 \to K_S^0 K^+ \pi^-$~\cite{LHCb:2015lnk}. For $D_s^+$ decays, amplitude analyses of $D_s^+ \to K^+ K^- \pi^+$ were performed by E687~\cite{E687:1995jyc}, CLEO-c~\cite{CLEO:2009nuz}, and BaBar~\cite{BaBar:2010wqe}, while FOCUS reported an amplitude analysis of $D_s^+ \to K^+ \pi^+ \pi^-$~\cite{FOCUS:2004muk}. In the four-body sector, MARKIII~\cite{MARK-III:1991fvi} and E691~\cite{FNAL-691:1992exu} performed amplitude analyses of the four decays $K^- 2\pi^+ \pi^-$, $K_S^0 2\pi^+ \pi^-$, $K^- 2\pi^+ \pi^0$, and $K_S^0 \pi^+ \pi^- \pi^0$. FOCUS conducted an amplitude analysis of $D^0 \to K^+ K^- K^- \pi^+$~\cite{FOCUS:2003gcs}. Amplitude analyses of $D^0 \to K^+ K^- \pi^+ \pi^-$ were carried out by FOCUS~\cite{FOCUS:2004prc} and CLEO-c~\cite{CLEO:2012beo,dArgent:2017gzv}, with CLEO-c also studying $D^0 \to 2(\pi^+ \pi^-)$~\cite{dArgent:2017gzv}.
Finally, FOCUS reported a resonant substructure analysis of the five-body decays $D^+ \to K^+ K^- 2\pi^+ \pi^-$ and $D_s^+ \to K^+ K^- 2\pi^+ \pi^-$~\cite{FOCUS:2002psb}.

\subsection{Three-body decays}

Studies of three-body hadronic decays of $D$ and $D_s$ mesons offer a valuable window into their quasi-two-body intermediate channels, in particular those involving $VP$ and $SP$ final states.
The $D_{(s)}\to K\bar K\pi$ decays provide information on the $VP$ amplitudes $D_{(s)}\to \phi\pi$ and $D_{(s)}\to K^*K$, together with the $SP$ contributions $D_{(s)}\to f_0\pi$ and $D_{(s)}\to a_0\pi$. Similarly, analyses of $D_{(s)}\to K\pi\pi$ enable one to investigate the $VP$ transitions $D_{(s)}\to K\rho$ and $D_{(s)}\to K\pi$, as well as the $SP$ mode $D_{(s)}\to f_0 K$.
The decays $D_{(s)}\to \pi\pi\pi$ are probes of the $VP$ process $D_{(s)}\to \rho\pi$ and, in parallel, of the $SP$ reaction $D_{(s)}\to f_{0/2}\pi$.
Investigations of $D_{(s)}\to K\pi\eta$ shed light on the $VP$ channel $D_{(s)}\to K^*\eta$ and the $SP$ channel $D_{(s)}\to K a_0$.
The channels $D_{(s)}\to \pi\pi\eta^{(\prime)}$ allow one to explore the $VP$ amplitude $D_{(s)}\to \rho\eta^{(\prime)}$ together with the $SP$ mechanisms $D_{(s)}\to f_0\eta$ and $D_{(s)}\to a_0\pi$.
Finally, study of $D_{(s)}\to \pi2\eta$ yields insight into the $SP$ transition $D_{(s)}\to a_0\eta$.

Especially, the $D \to VP$ decays offer cleaner opportunities than other two-body processes for clarifying nonperturbative charm decays~\cite{Cheng:2016ejf,Cheng:2010ry}.
Unlike $D\to SP$, the $VP$ system has well-defined vector meson quark content, avoiding the controversies and rescattering effects of
scalar states~\cite{Hsiao:2019ait,Yu:2021euw,Hsiao:2023qtk}. It is also more favorably described than $PP$ and $VV$ systems~\cite{Cheng:2010ry,Cao:2023csx}.
Moreover, hadronic $D_{(s)}^{+}$ decays probe the interplay of short-distance weak-decay matrix elements and long-distance QCD interactions; the measured branching fractions provide valuable information on amplitudes and phases in decay processes~\cite{Bhattacharya:2008ke, Cheng:2010ry, Fu-Sheng:2011fji,Hsiao:2019ait}.
In addition, charmed-meson decays provide a versatile testing ground not only for investigating the nature of the scalar states $a_0$ and $f_0$, but also for scrutinizing the decay properties of the vector meson $\phi$.

For three-body hadronic $D^+_s$ decays, the studies of $D^+_s\to 2\pi^+\pi^-$, $D_{s}^{+}\to \pi^{+}\pi^{0}\eta$, and $D_{s}^{+} \to K^{+}K^{-}\pi^{+}$
were based on 3.19 fb$^{-1}$ of data at 4.178 GeV;
the investigation of $D^+_s\to K^0_SK^0_L\pi^+$ was based on 7.33 fb$^{-1}$ of data at 4.128-4.226 GeV;
while the analyses of other three-body hadronic $D^+_s$ decays were based on 6.3 fb$^{-1}$ of data at 4.178-4.226 GeV.
Amplitude analyses of $D^+_s\to K^0_S\pi^+\pi^0$,
$D^+\to K_S^0 K^+ \pi^0$, $D^0\to K^0_SK^+K^-$, $D^0 \to K_{L}^0\pi^+\pi^-$,
and $D^+ \to K^0_S\pi^+\eta$ were based on 2.93 fb$^{-1}$ of data at 3.773 GeV;
amplitude analyses of $D^{0(+)} \to \pi^{+} \pi^{-(0)} \eta$ and
$D^{+} \to K_{S}^{0}K_{S}^{0}\pi^{+}$ were performed by using 7.9 fb$^{-1}$ of data at 3.773 GeV;
while amplitude analyses of other three-body hadronic $D^{0(+)}$ decays were carried out
with full 20.3 fb$^{-1}$ of data at 3.773 GeV.
Moreover, the analyses of $D^+_s\to 2\pi^+\pi^-$ and $D^+\to K^0_S\pi^+\pi^0$ were based on single-tag signal candidates,
while all other analyses were based on double-tag signal candidates.

\subsubsection{Analyses of $D^+_{(s)}\to K\bar K\pi$}

\begin{figure}[htbp]
    \centering
    \mbox{
        \begin{overpic}[width=0.45\textwidth]{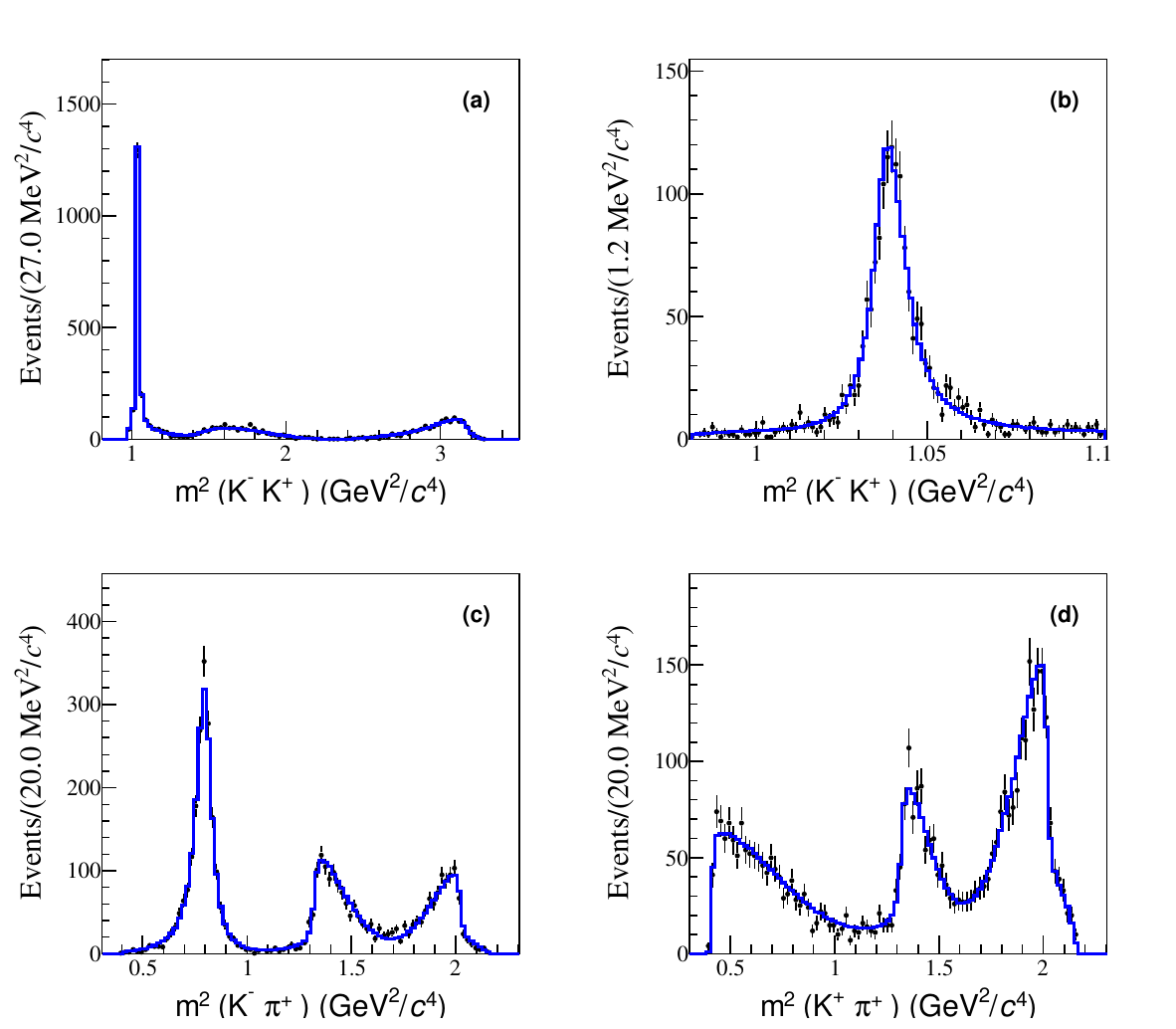}
        \end{overpic}
    }
    \caption{
    The projections of the amplitude analysis fit for $D^+_s\to K^+K^-\pi^+$ onto (a) $m^{2}(K^{+}K^{-})$, (b) $m^{2}(K^{+}K^{-})$ near the $\phi(1020)$ peak, (c) $m^{2}(K^{-}\pi^{+})$ and (d) $m^{2}(K^{+}\pi^{+})$~\cite{BESIII:2020ctr}.
    }
    \label{fig:Ds_KKpi}
\end{figure}

In experimental studies of $D_{s}^{+}$ decays, the CF decay $D_{s}^{+} \to K^{+}K^{-}\pi^{+}$ is a key reference mode due to its large branching fraction and low background.
Knowledge of its subresonances helps to reduce systematic uncertainties and improves precision in  branching fraction measurement.
Theoretical studies~\cite{Cheng:2016ejf} predict $\mathcal{B}(D_{s}^{+} \to \bar{K}^{*}(892)^{0}K^{+}) \sim (3.9-4.2)\%$ and $\mathcal{B}(D_{s}^{+} \to \phi(1020)\pi^{+}) \sim (3.4-4.51)\%$. Combining amplitude analysis results with branching fraction measurements yields branching fractions of intermediate processes,
therefore benefiting the refinement of theoretical models~\cite{Cheng:2016ejf}.
Previously, amplitude analyses of $D_{s}^{+} \to K^{+}K^{-}\pi^{+}$ were performed by E687~\cite{E687:1995jyc}, CLEO-c~\cite{CLEO:2009nuz}, and BaBar~\cite{BaBar:2010wqe}.
Based on 4.4k candidate events with a signal purity of 99.6\%, Ref.~\cite{BESIII:2020ctr} reported
the amplitude analysis of $D_{s}^{+} \to K^{+}K^{-}\pi^{+}$.
The Dalitz plot projections are shown in Fig.~\ref{fig:Ds_KKpi}.
From the amplitude analysis, one finds
            $D_{s}^{+} \to \bar K^{*}(892)^{0}K^{+}$ [(48.3$\pm$0.9$\pm$0.6)\%],
            $D_{s}^{+} \to \phi(1020)\pi^{+}$ [(40.5$\pm$0.7$\pm$0.9)\%],
            $D_{s}^{+} \to S(980)\pi^{+}$  [(19.3$\pm$1.7$\pm$2.0)\%],
            $D_{s}^{+} \to \bar K^{*}_{0}(1430)^{0}K^{+}$ [(3.0$\pm$0.6$\pm$0.5)\%],
            $D_{s}^{+} \to f_{0}(1710)\pi^{+}$ [(1.9$\pm$0.4$\pm$0.6)\%], and
            $D_{s}^{+} \to f_{0}(1370)\pi^{+}$ [(1.2$\pm$0.4$\pm$0.2)\%].
The results on fit fractions for $D_{s}^{+} \to f_{0}(1370)\pi^{+}$, $D_{s}^{+} \to f_{0}(1710)\pi^{+}$ and $D_{s}^{+} \to f_{0}(980)\pi^{+}/a_{0}(980)\pi^{+}$ are consistent with those of BaBar~\cite{BaBar:2010wqe} and E687~\cite{E687:1995jyc}.
In addition, the obtained fit fractions also agree with those of CLEO-c~\cite{CLEO:2009nuz}, except for $D_{s}^{+} \to f_{0}(980)\pi^{+}/a_{0}(980)\pi^{+}$ and  $D_{s}^{+} \to f_{0}(1370)\pi^{+}$ with 2.4$\sigma$ and 3.4$\sigma$ differences, respectively.
The branching fractions of each component are also presented by using the world average of ${\cal B}(D_{s}^{+} \to K^{+}K^{-}\pi^{+})$~\cite{ParticleDataGroup:2022pth}.

\begin{figure}[!htbp]
  \centering
  \includegraphics[width=0.225\textwidth]{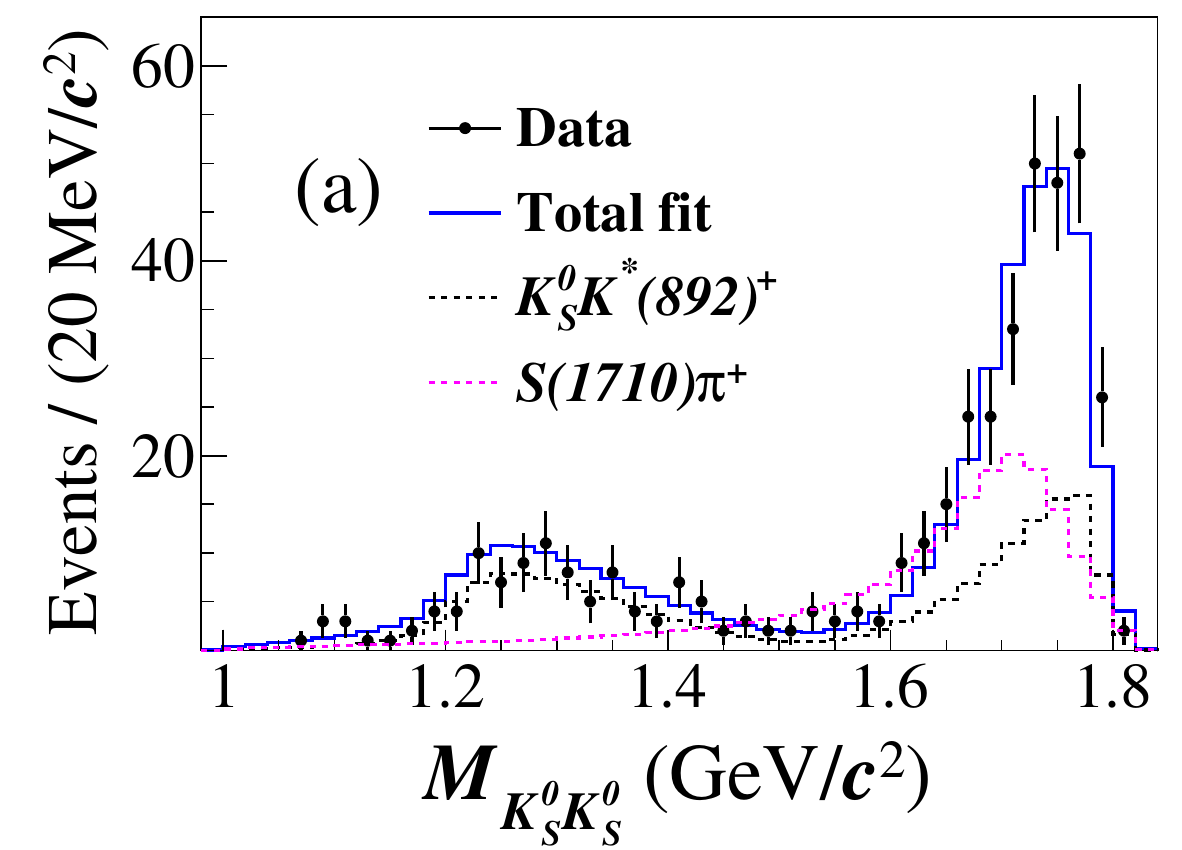}
  \includegraphics[width=0.225\textwidth]{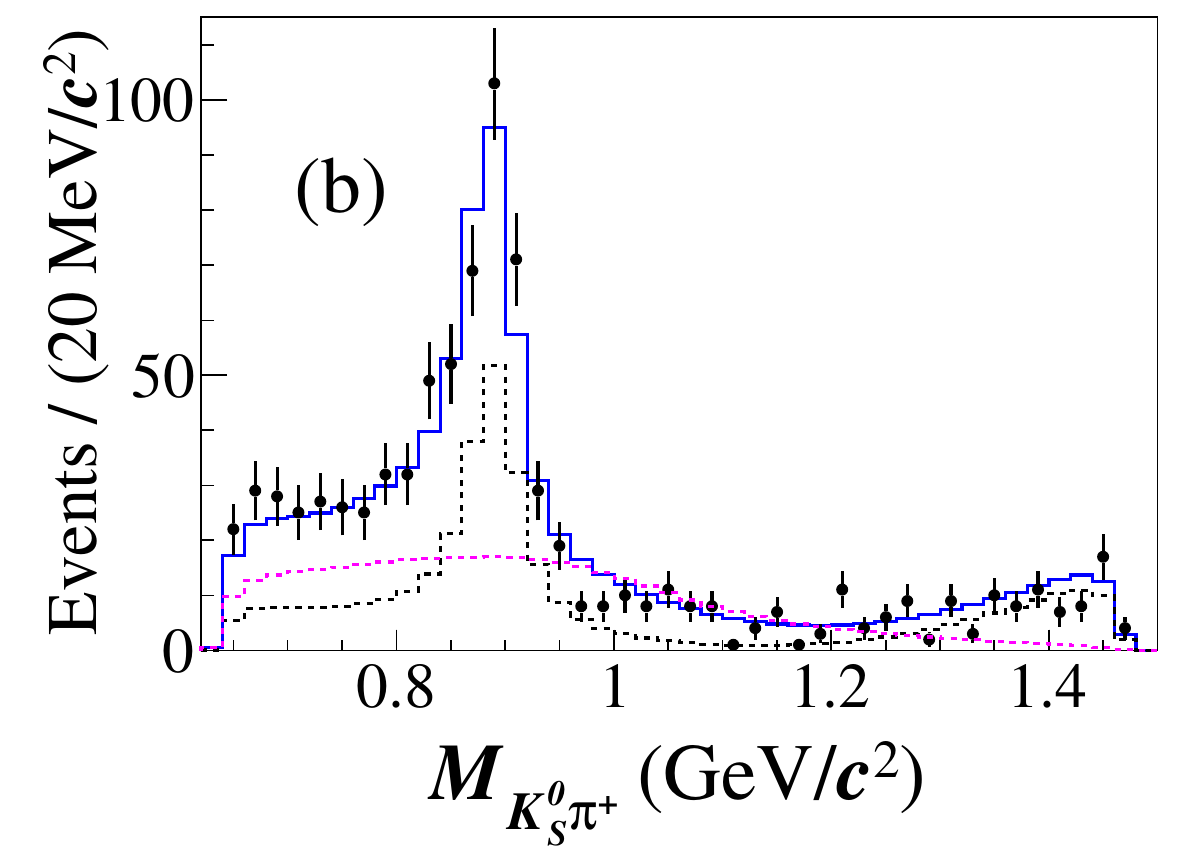}
  \caption{
    The projections of the amplitude analysis fit for  $D_{s}^{+} \to 2K^0_S\pi^+$  onto (a) $M_{K_S^0K_S^0}$ and (b) $M_{K_S^0\pi^+}$~\cite{BESIII:2021anf}.
    }
  \label{fig:Ds_KSKSpi}
\end{figure}

From the analysis of 412 candidates with a signal purity of 97\%,  Ref.~\cite{BESIII:2021anf}
reported the first amplitude analysis of $D_{s}^{+} \to 2K^0_S\pi^+$.
The fit projections of the amplitude analysis on two-body particle mass distributions are shown in Fig.~\ref{fig:Ds_KSKSpi}.
An enhancement is observed in the $2K_{S}^{0}$ mass spectrum near
1.7~GeV/$c^2$, named ${\cal S}(1710)$, which was not seen in an earlier analysis of $D_{s}^{+} \to K^+K^-\pi^{+}$, implying the existence of an isospin one partner
of the $f_0(1710)$. The amplitude analysis gives the fractions of $D_{s}^{+} \to K_{S}^{0}K^{*}(892)^{+}$
and $D_{s}^{+} \to S(1710)\pi^{+}$ to be
$(43.5\pm3.9\pm0.5)\%$ and $(46.3\pm4.0\pm1.2)\%$, respectively.
The mass and width of the $S(1710)$ are measured to be $(1.723 \pm 0.011 \pm 0.002)$~GeV/$c^2$ and $(0.140 \pm 0.014 \pm 0.004)$~GeV/$c^2$, consistent with the $f_0(1710)$ world averages within $1.2\sigma$ and $0.7\sigma$.
The measured branching fraction $\mathcal{B}(D_{s}^{+} \to K_{S}^0K_{S}^0\pi^{+}) = (0.68\pm0.04\pm0.01)\%$
agrees with the CLEO-c result~\cite{CLEO:2013bae} within $1.3\sigma$.
The branching fractions of two intermediate processes are also reported.
The obtained $\mathcal{B}(D_{s}^{+} \to \bar{K^0}K^{*}(892)^{+}) = (1.8\pm0.2\pm0.1)\%$
deviates from the CLEO value $(5.4\pm1.2)\%$~\cite{CLEO:1989zcg} by $2.9\sigma$.
A significant $D_{s}^{+} \to f_0(980)/a_0(980)^0\pi^+$ contribution is observed in $D^+_s\to K^+K^-\pi^+$~\cite{BESIII:2020ctr},
implying a sizeable contribution from $D_{s}^{+} \to f_0(980)/a_0(980)^0\pi^+$ with $f_0(980)/a_0(980)^0 \to K_S^0K_S^0$.
Yet almost no signal is observed below 1.1~GeV/$c^2$ in the $K_S^0K_S^0$ mass spectrum,
likely due to destructive interference between $a_0(980)^0$ and $f_0(980)$ in decays to two neutral kaons,
but constructive for charged kaons.
On the contrary, an enhancement appears near 1.7~GeV/$c^2$,
likely due to constructive interference in neutral kaon decays, but destructive in charged kaons.

\begin{figure*}[!htbp]
  \centering
  \includegraphics[width=0.225\textwidth]{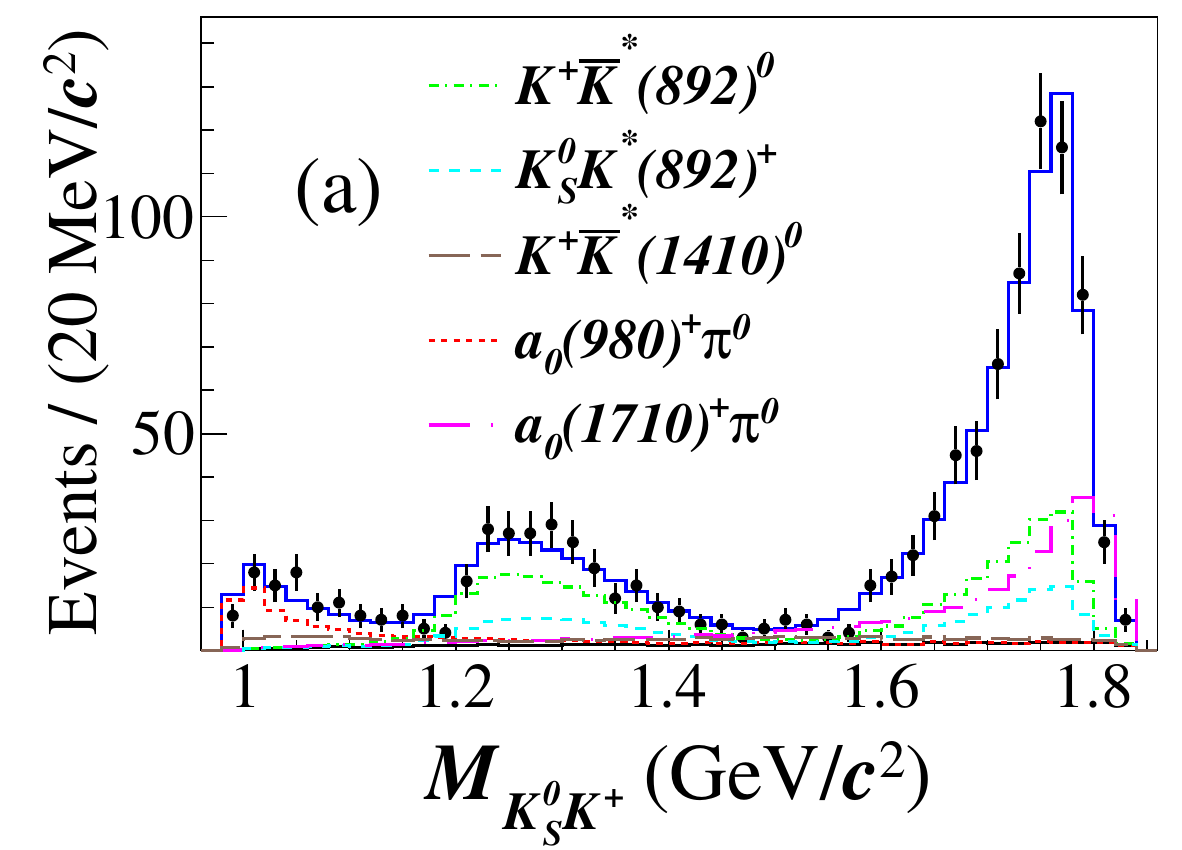}
  \includegraphics[width=0.225\textwidth]{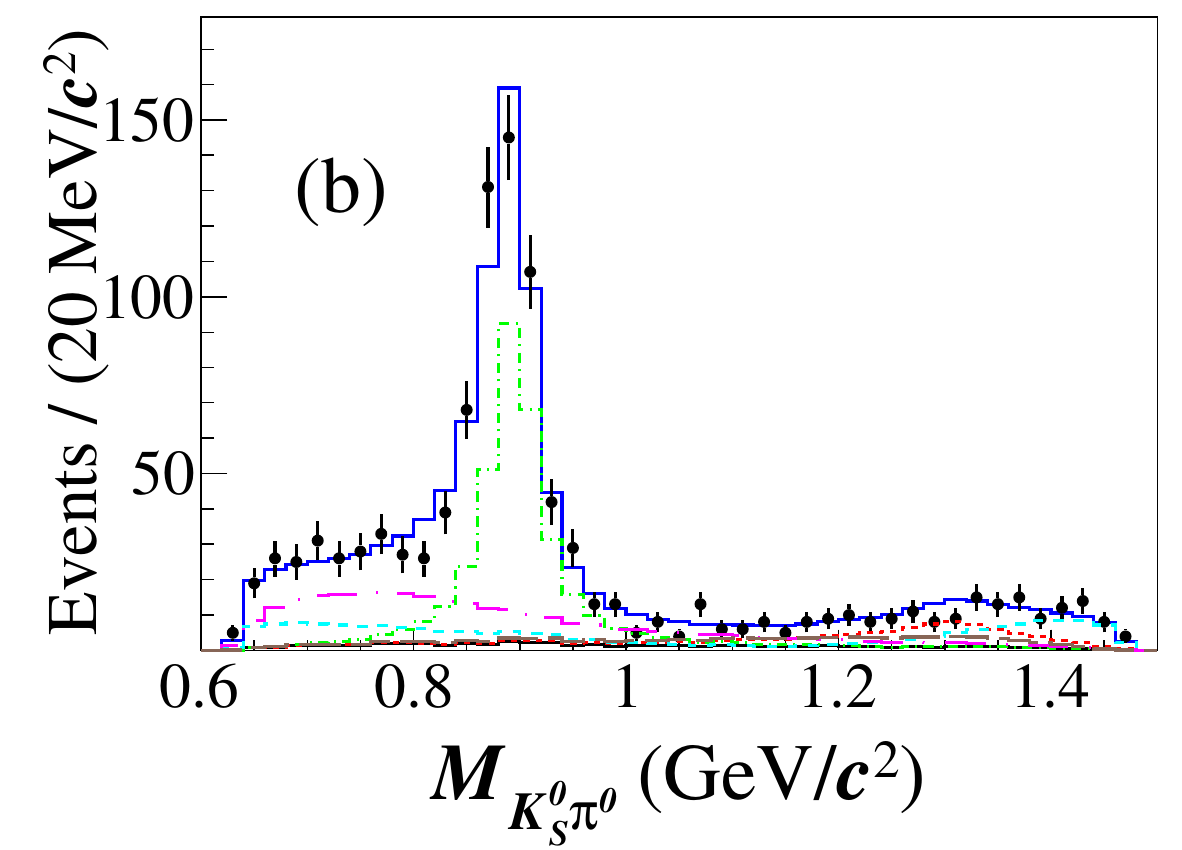}
  \includegraphics[width=0.225\textwidth]{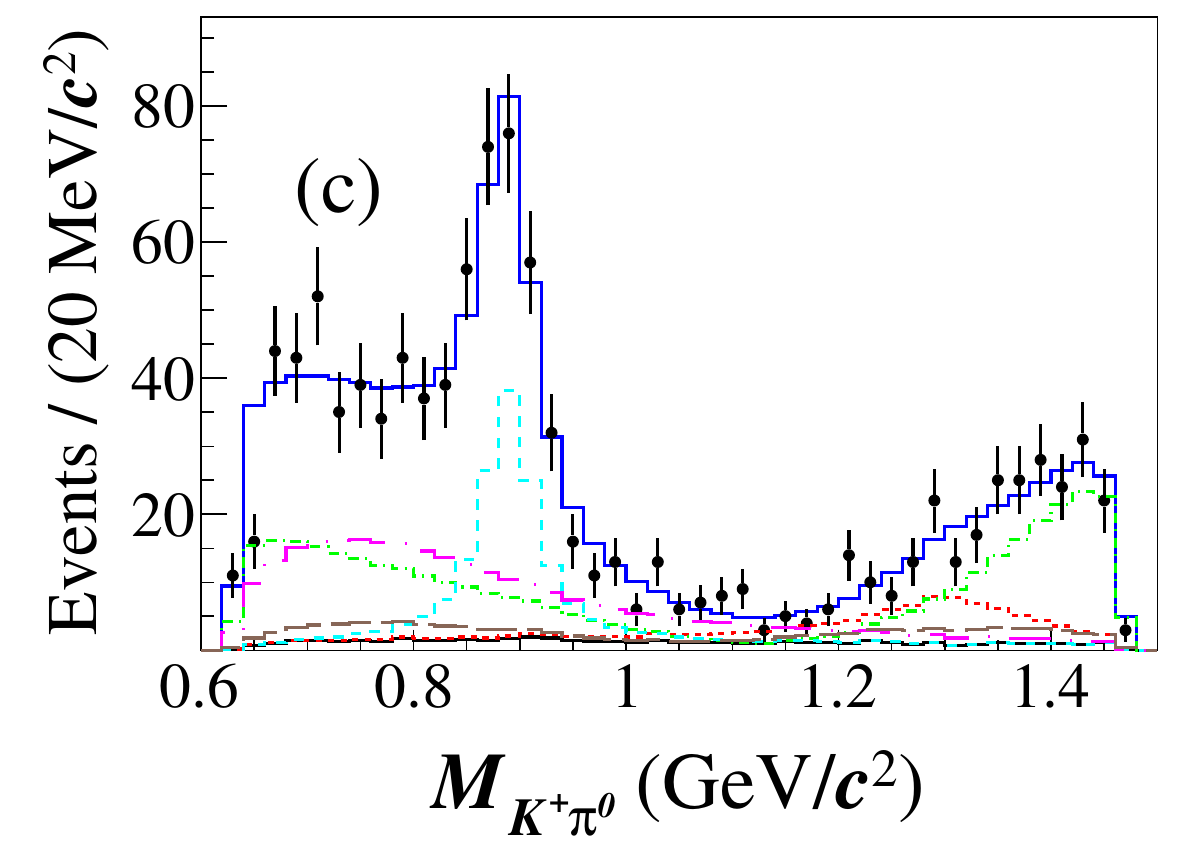}
  \caption{
  The projections of the amplitude analysis fit for $D_{s}^{+} \to K^0_SK^+\pi^0$ onto $M_{K_S^0K^+}$,
    $M_{K_S^0\pi^0}$, and $M_{K^+\pi^0}$~\cite{BESIII:2022npc}.
    }
  \label{fig:Ds_KSKpi0}
\end{figure*}

Reference~\cite{BESIII:2022npc} reported an amplitude analysis of the CF decay $D_{s}^{+} \to K^0_SK^+\pi^0$
based on 1.1k candidate events with a signal purity of 94.7\%.
The projections of the amplitude analysis fit on two-body particle mass distributions are shown in Fig.~\ref{fig:Ds_KSKpi0}.
An $a_0(1710)^+$, the isovector partner of the $f_0(1710)$ and $f_0(1770)$
mesons, named $a_0(1817)^+$, is observed in its decay to $K_S^0K^+$ for the first time.
From the amplitude analysis, the fractions of
      $D_s^+\to \bar K^{*}(892)^0K^+   $,
      $D_s^+\to K^{*}(892)^+K_S^0       $,
      $D_s^+\to a_{0}(980)^+\pi^0       $,
      $D_s^+\to \bar K^*(1410)^{0}K^+  $, and
      $D_s^+\to a_{0}(1817)^+\pi^0      $
are found to be
$(32.7\pm2.2\pm1.9)\%$,
$(13.9\pm1.7\pm1.3)\%$,
$(7.7\pm1.7\pm1.8)\%$,
$(6.0\pm1.4\pm1.3)\%$, and
$(23.6\pm3.4\pm2.0)\%$, respectively.
Along with the observed enhancement at the $K^0_SK^0_S$
mass threshold in $D^+_s\to K^0_SK^0_S\pi^+$~\cite{BESIII:2021anf}, this result supports the existence of a
new $a_0$ triplet. The branching fraction of $D^+_s\to a_0(1817)^+\pi^-$
 with $a_0(1817)^+\to K^0_SK^+$ is roughly consistent with the prediction~\cite{Dai:2021owu}
 assuming the $a_0(1817)^+$ meson is the candidate
isospin-one partner of the $f_0(1710)$ meson~\cite{Geng:2008gx,Dai:2021owu,Klempt:2021nuf,Sarantsev:2021ein,Wang:2022pin,Zhu:2022wzk}. However, the mass is about 100 MeV$/c^2$
greater than the predicted value. This higher mass may
imply instead that this $a_0$-like state is the isospin-one
partner of the $X(1812)$~\cite{Guo:2022xqu}. 
In addition, the ratio
$\frac{\mathcal{B}(D_{s}^{+} \to \bar K^{*}(892)^{0}K^{+})}{\mathcal{B}(D_{s}^{+} \to \bar K^{0}K^{*}(892)^{+})}$
is determined to be $2.35^{+0.42}_{-0.23}\pm 0.10$.
Combining with $\mathcal{B}(D_{s}^{+} \to a_0(980)^+\pi^0)$~\cite{BESIII:2019jjr}
leads to a ratio
$\frac{\mathcal{B}(a_0(980)^+ \to \bar K^{0}K^{+})}{\mathcal{B}(a_0(980)^+ \to \pi^+\eta)}=(13.7 \pm 3.6 \pm 4.2)\%$.
The branching fraction of $D_{s}^{+} \to K_{S}^0K^+\pi^{0}$ was determined to be
$(1.46\pm0.06\pm 0.05)\%$, which is consistent with
the CLEO-c result~\cite{CLEO:2013bae}, with precision improved by a factor of 2.8.

\begin{figure}[!htbp]
  \centering
  \includegraphics[width=0.235\textwidth]{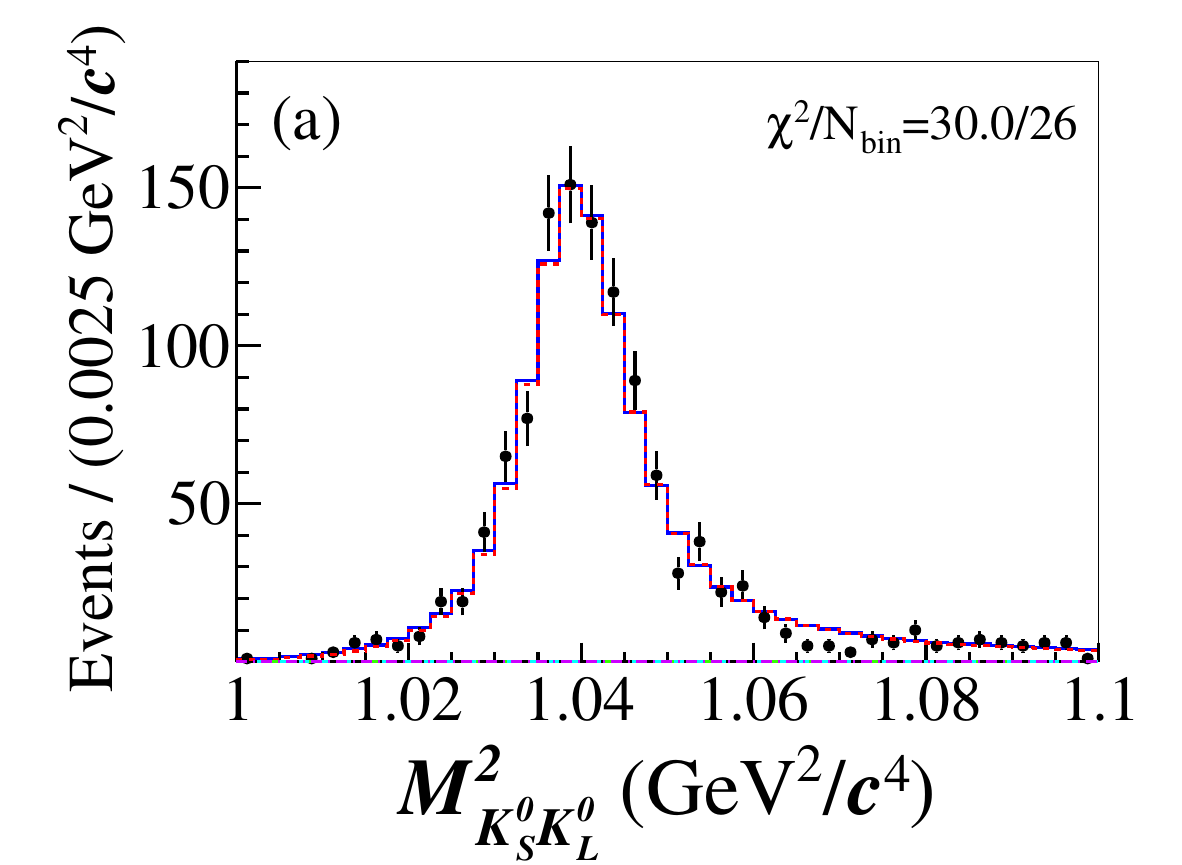}
  \includegraphics[width=0.235\textwidth]{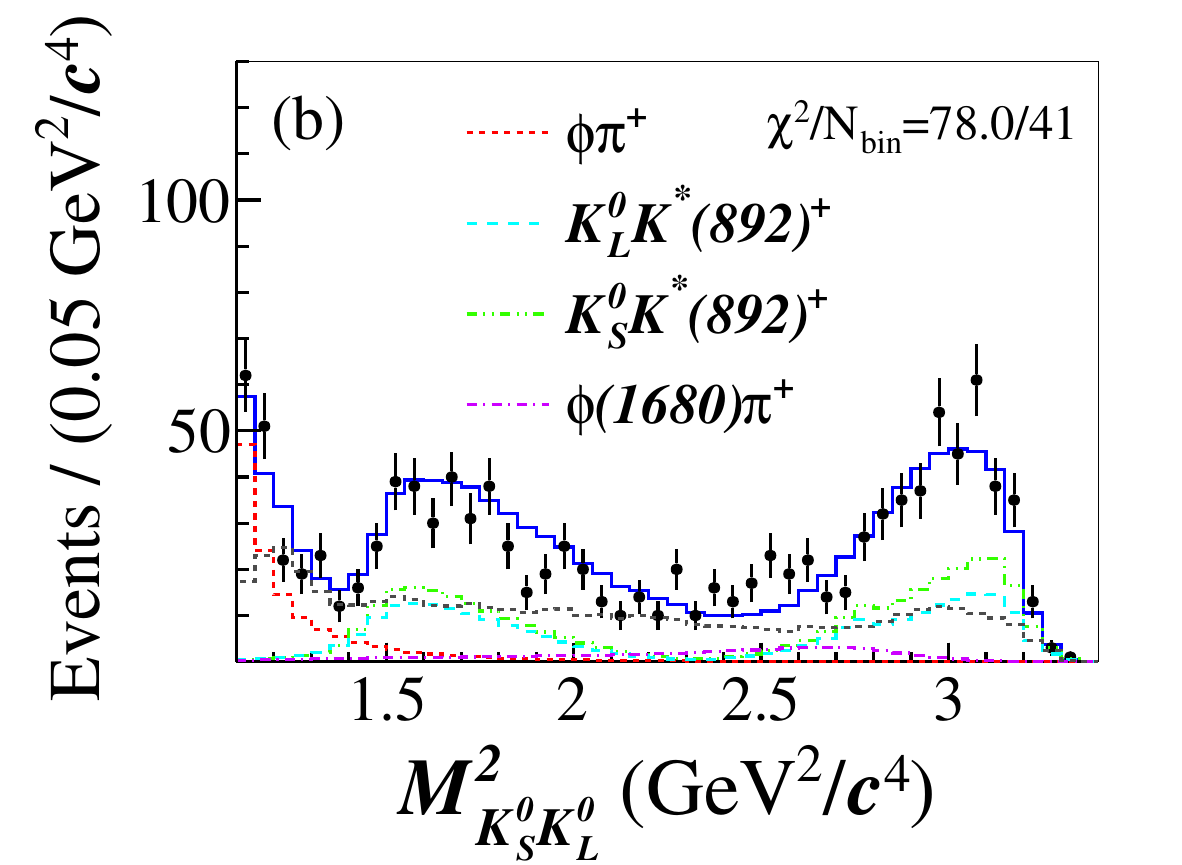}
  \includegraphics[width=0.235\textwidth]{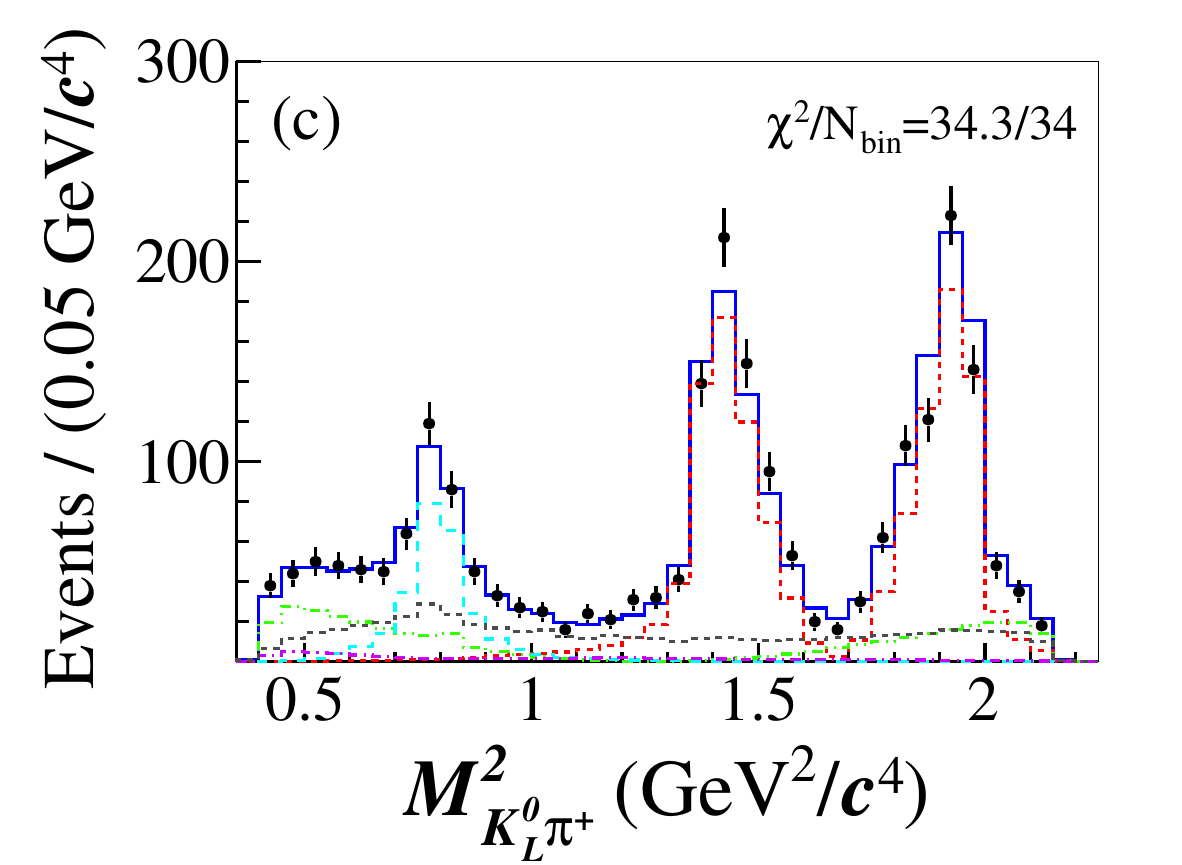}
  \includegraphics[width=0.235\textwidth]{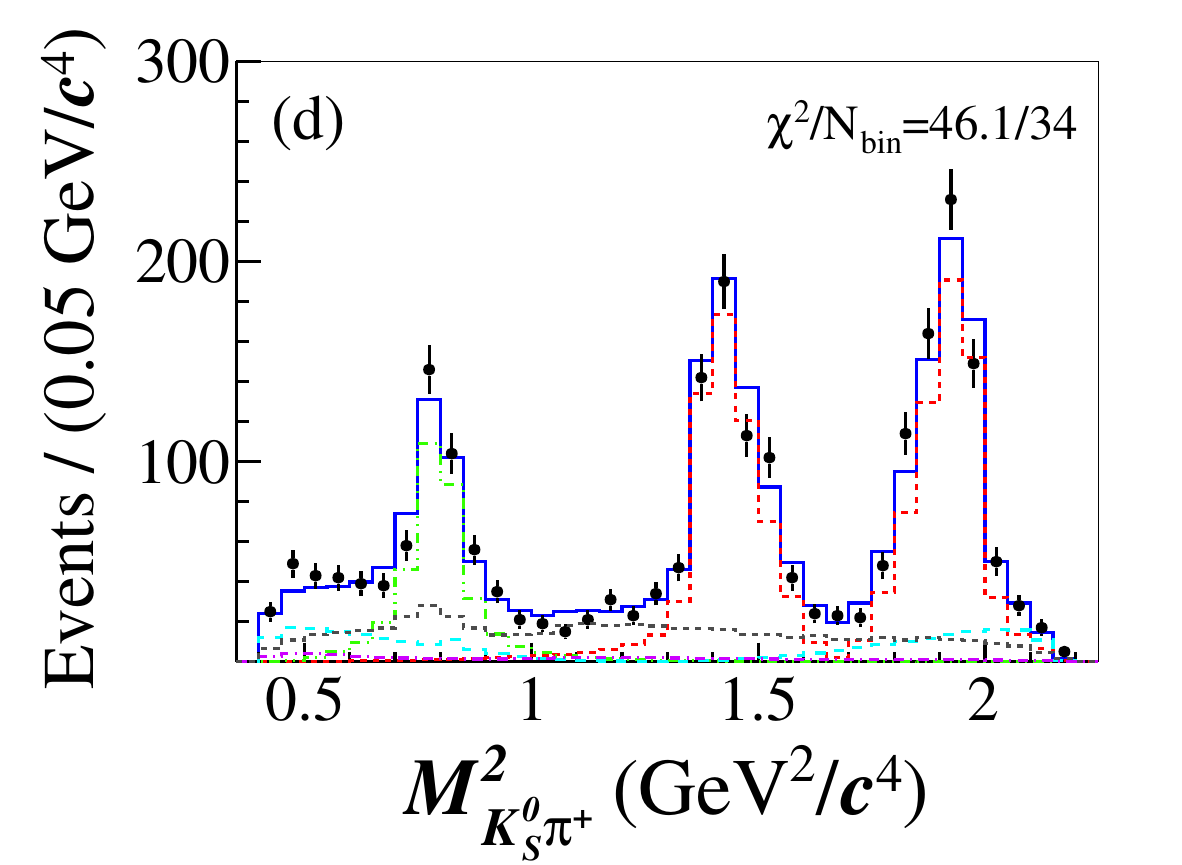}
  \caption{
     The projections of the amplitude analysis fit for $D_{s}^{+} \to K_{S}^0K_{L}^0\pi^{+}$  onto  (a) $M_{K_S^0K_L^0}^{2}$ less than 1.1~GeV$^{2}$/$c^4$, (b) $M_{K_S^0K_L^0}^{2}$ greater than 1.1~GeV$^{2}$/$c^4$, (c) $M_{K_L^0\pi^+}^{2}$ and (d) $M_{K_S^0\pi^+}^{2}$~\cite{BESIII:2025ygp}.
    }
  \label{fig:figure/Ds_KSKLpi}
\end{figure}

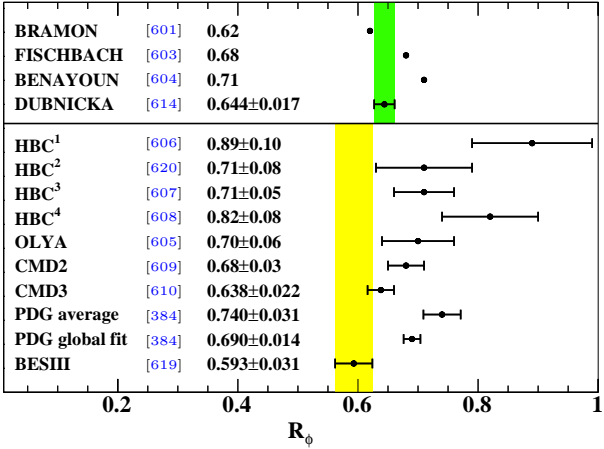
\begin{figure}[htp]
  \begin{center}
    \includegraphics[width=0.45\textwidth]{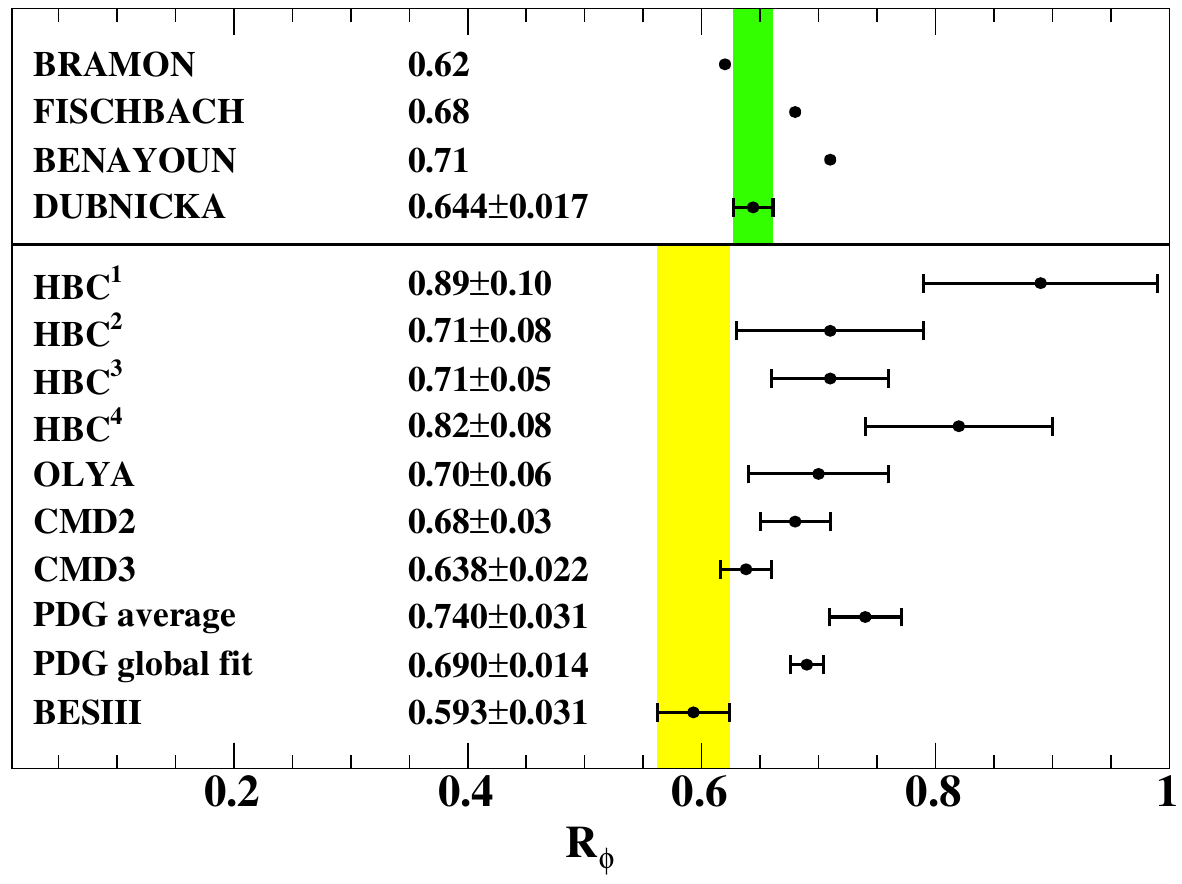}
         \put(-176,158){\tiny~\cite{Bramon:2000qe}}
      \put(-176,148.666){\tiny~\cite{Fischbach:2001ie}}
      \put(-176,139.333){\tiny~\cite{Benayoun:2012etq}}
      \put(-176,130){\tiny~\cite{Dubnicka:2024dta}}
      \put(-176,115){\tiny~\cite{Aguilar-Benitez:1972ngz}}
      \put(-176,105.666){\tiny~\cite{Lyons:1977rp}}
      \put(-176,96.333){\tiny~\cite{AACHEN-BERLIN-CERN-LONDON-VIENNA:1977sjv}}
      \put(-176,87){\tiny~\cite{Amsterdam-CERN-Nijmegen-Oxford:1977ctz}}
      \put(-176,77.666){\tiny~\cite{Bukin:1978wj}}
      \put(-176,68.333){\tiny~\cite{Akhmetshin:1995vz}}
      \put(-176,59){\tiny~\cite{Kozyrev:2017agm}}
      \put(-176,49.667){\tiny~\cite{ParticleDataGroup:2022pth}}
      \put(-176,40.333){\tiny~\cite{ParticleDataGroup:2022pth}}
      \put(-176,31){\tiny~\cite{BESIII:2025ygp}}
    \caption{    
    Comparison of the results for $R_{\phi}$ measured by BESIII~\cite{BESIII:2025ygp} and the HBC~\cite{Amsterdam-CERN-Nijmegen-Oxford:1977ctz,AACHEN-BERLIN-CERN-LONDON-VIENNA:1977sjv,Lyons:1977rp,Aguilar-Benitez:1972ngz}, OLYA~\cite{Bukin:1978wj}, CMD2~\cite{Akhmetshin:1995vz}, and CMD3~\cite{Kozyrev:2017agm} experiments. Above the dotted line are the theoretical calculations~\cite{Dubnicka:2024dta,Benayoun:2012etq,Fischbach:2001ie,Bramon:2000qe}; below are the experimental results. The green band is the $\pm 1\sigma$ region of the  DUBNICKA result~\cite{Dubnicka:2024dta} and the yellow band denotes the $\pm 1\sigma$ region of the BESIII result~\cite{BESIII:2025ygp}.
      }
    \label{fig:Ds_KSKLpi_R_phi}
  \end{center}
\end{figure}

Beyond the charm sector itself, hadronic $D$-meson decays provide a pristine laboratory for light-hadron spectroscopy. Among the most intriguing opportunities here is a measurement of the $\phi$ meson branching fraction.
About 80\% of $\phi$ mesons decay to $K\bar{K}$, and the ratio ${\cal R}_{\phi}\equiv\mathcal{B}(\phi \to K_S^0K_L^0)/\mathcal{B}(\phi \to K^+K^-)$ is expected to be 1.0 by isospin symmetry. Theoretical predictions range 0.62-0.71~\cite{Bramon:2000qe,Flores-Baez:2008owd, Fischbach:2001ie,Benayoun:2012etq},
while experimental measurements span 0.64-0.89~\cite{ParticleDataGroup:2024cfk,Bukin:1978wj,Aguilar-Benitez:1972ngz, AACHEN-BERLIN-CERN-LONDON-VIENNA:1977sjv,Amsterdam-CERN-Nijmegen-Oxford:1977ctz}, indicating overlap but need for further studies.
The PDG average ${\cal R}_{\phi}=0.740 \pm 0.031$~\cite{ParticleDataGroup:2024cfk} has not been updated for decades;
some measurements from CMD2~\cite{Akhmetshin:1995vz} and CMD3~\cite{Kozyrev:2017agm} are not included.
Previous experiments using $e^+e^-$ annihilation and $K-p$ scattering~\cite{ParticleDataGroup:2024cfk, Parrour:1975rt, Mattiuzzi:1995eze, Dolinsky:1991vq} suffer from high backgrounds. A recent study~\cite{Dubnicka:2024dta} obtained $0.644 \pm 0.017$, thereby challenging the PDG average~\cite{ParticleDataGroup:2024cfk}.
A cleaner determination of ${\cal R}_{\phi}$ can come from measuring $D_s^+ \to \phi(\to K_S^0 K_L^0) \pi^+$ alongside $D_s^+ \to \phi(\to K^+ K^-) \pi^+$~\cite{BESIII:2020ctr}.
In particular, BESIII observed a $>4\sigma$ deviation in the ratio $\mathcal{B}(\phi \to \pi^+ \pi^- \pi^0)/\mathcal{B}(\phi \to K^+K^-)$~\cite{BESIII:2024muy};
an amplitude analysis of $D_s^+ \to K_S^0K_L^0\pi^+$ helps to clarify this tension. In addition, the symmetry $\Gamma(\bar{K}^0)=2\Gamma(K_S^0)=2\Gamma(K_L^0)$ is often used for decays with neutral kaons,
but interference between CF and DCS transitions may induce a $K_{S}^{0}-K_{L}^{0}$ asymmetry~\cite{Bigi:1994aw, Rosner:2006bw, Bhattacharya:2009ps, Wang:2017ksn,Muller:2015lua,Cheng:2024hdo}, not yet observed in $D \to K^0_{S,L}V$~\cite{BESIII:2022xhe, Lipkin:1999qz, Gao:2006nb}. Models predict nonzero asymmetry for $D_s^+\to K^0_{S,L}V$~\cite{Wang:2017ksn,Cheng:2024hdo}. Measuring this asymmetry in $D_s^+ \to K_S^0K_L^0\pi^+$ via $D_{s}^{+} \to K_{S}^{0}K^{*}(892)^{+}$ and $D_{s}^{+} \to K_{L}^{0}K^{*}(892)^{+}$ will cancel most systematics and constrain decay dynamics.
Utilizing 2.3k candidates with a purity of 78\%,
an amplitude analysis of $D_{s}^{+} \to K_{S}^0K_{L}^0\pi^{+}$ was carried out~\cite{BESIII:2025ygp}.
The mass projections of the amplitude analysis fit are shown in Fig.~\ref{fig:figure/Ds_KSKLpi}.
The amplitude analysis finds four main components:
$D^+_s\to\phi\pi^+$,  $D^+_s\to K^0_LK^*(892)^+$, $D^+_s\to K^0_SK^*(892)^+$, and $D^+_s\to\phi(1680)\pi^+$,
with fractions of $(70.2\pm1.4\pm1.2)\%$, $(19.3\pm1.2\pm1.0)\%$, $(14.4\pm1.2\pm1.2)\%$,
 and $(3.3\pm0.9\pm0.7)\%$, respectively.
 The branching fraction of $D_{s}^{+}
\to K_{S}^{0}K_{L}^{0}\pi^{+}$ is determined to be $(1.86\pm0.06\pm0.03)\%$.
Combining the $\mathcal{B}(D_{s}^{+} \to \phi(\to K_{S}^0K_{L}^0) \pi^+)$ and the world average of $\mathcal{B}(D_{s}^{+} \to \phi(\to K^+K^-) \pi^+)$,
the relative branching fraction is determined as $\mathcal{B}(\phi \to K_S^0K_L^0)/\mathcal{B}(\phi \to K^+K^-)$=$0.593 \pm 0.023 \pm 0.014\pm 0.016$, where the third error is due to the uncertainty of the $\mathcal{B}(D_s^+ \to \phi \pi^+,\phi \to K^+K^-)$. Our result deviates from the PDG value~\cite{ParticleDataGroup:2024cfk} by more than 3$\sigma$.
Furthermore, the asymmetry of the branching fractions of $D^+_s\to K_{S}^0K^{*}(892)^{+}$ and $D^+_s\to K_{L}^0K^{*}(892)^{+}$,  $\frac{\mathcal{B}(D_{s}^{+} \to K_{S}^0K^{*}(892)^{+})-\mathcal{B}(D_{s}^{+} \to K_{L}^0K^{*}(892)^{+})}{\mathcal{B}(D_{s}^{+} \to K_{S}^0K^{*}(892)^{+})+\mathcal{B}(D_{s}^{+} \to K_{L}^0K^{*}(892)^{+})}$, is determined to be $(-14.5\pm 5.1\pm 1.8)\%$.
The obtained ${\cal R}_{\phi}$ is consistent with the theoretical expectation as reported in Ref.~\cite{Bramon:2000qe}.
However, it is $(1.0-2.8)\sigma$ below all previous measurements, see Fig.~\ref{fig:Ds_KSKLpi_R_phi}, and deviates from the PDG average (PDG fit~\cite{ParticleDataGroup:2024cfk}) by $3.2\sigma$ ($2.6\sigma$).

\begin{figure*}[thp]\centering
\includegraphics[trim = 9mm 0mm 0mm 0mm, width=0.225\textwidth]{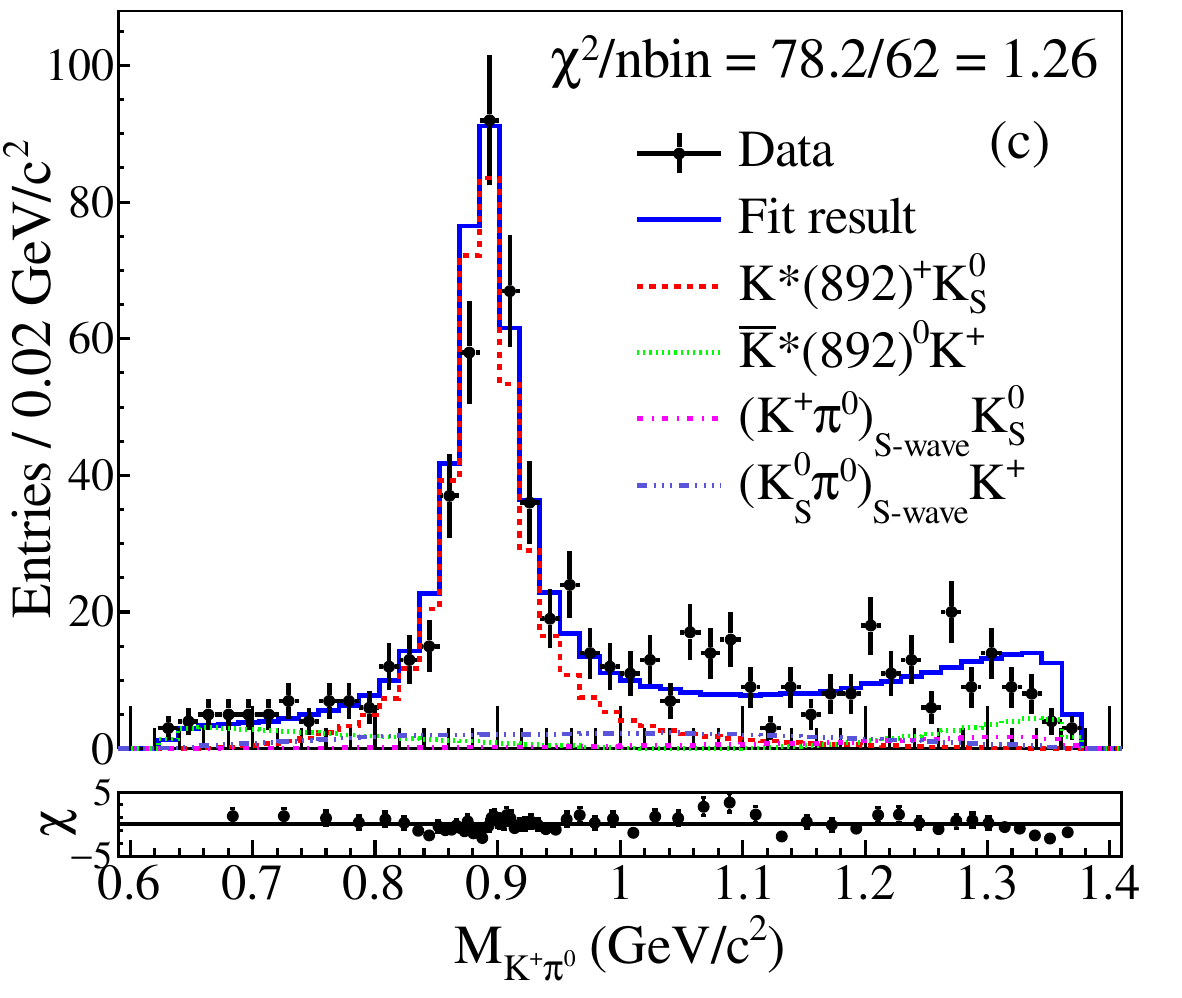}
\includegraphics[trim = 9mm 0mm 0mm 0mm, width=0.225\textwidth]{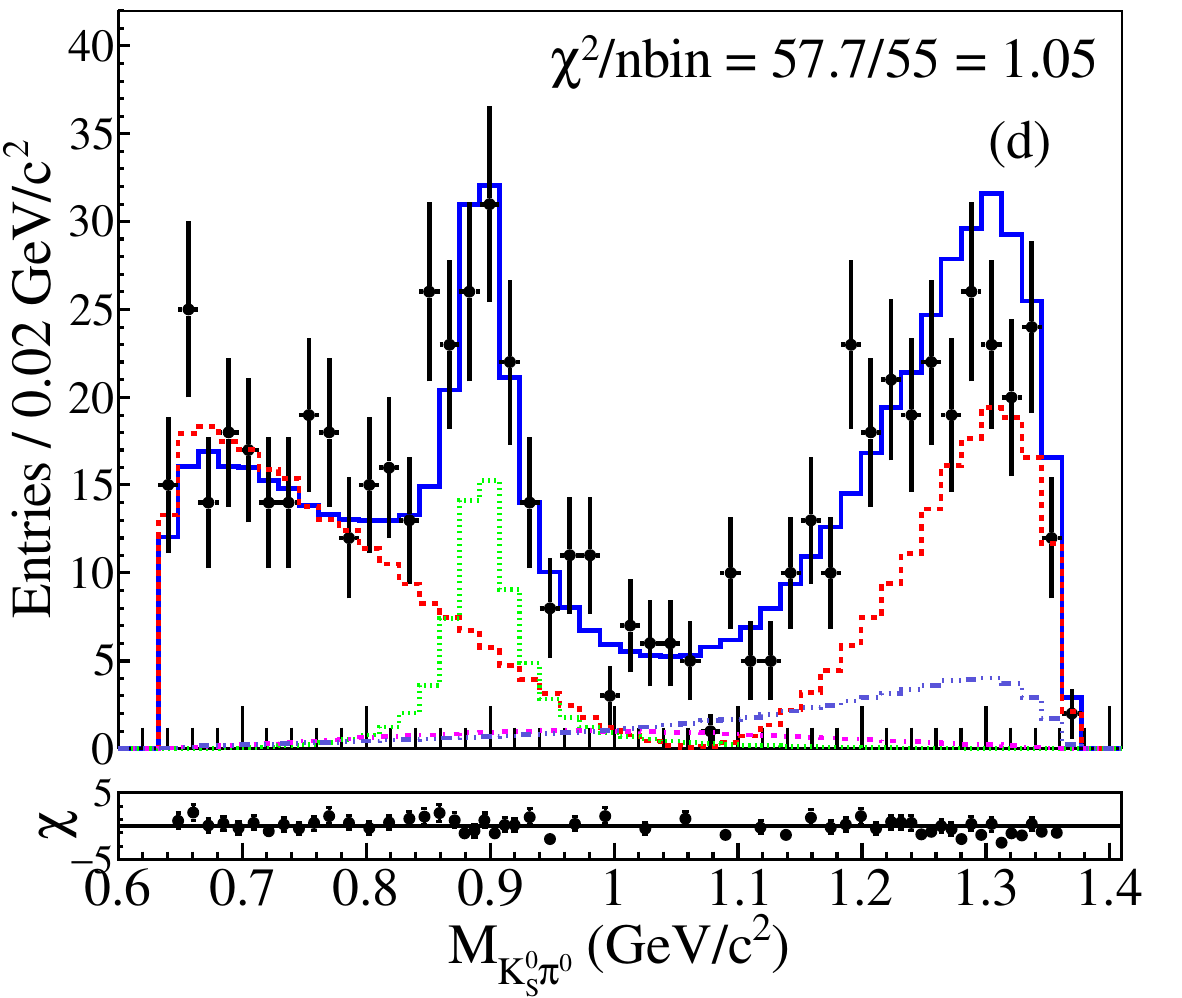}
\includegraphics[trim = 9mm 0mm 0mm 0mm, width=0.225\textwidth]{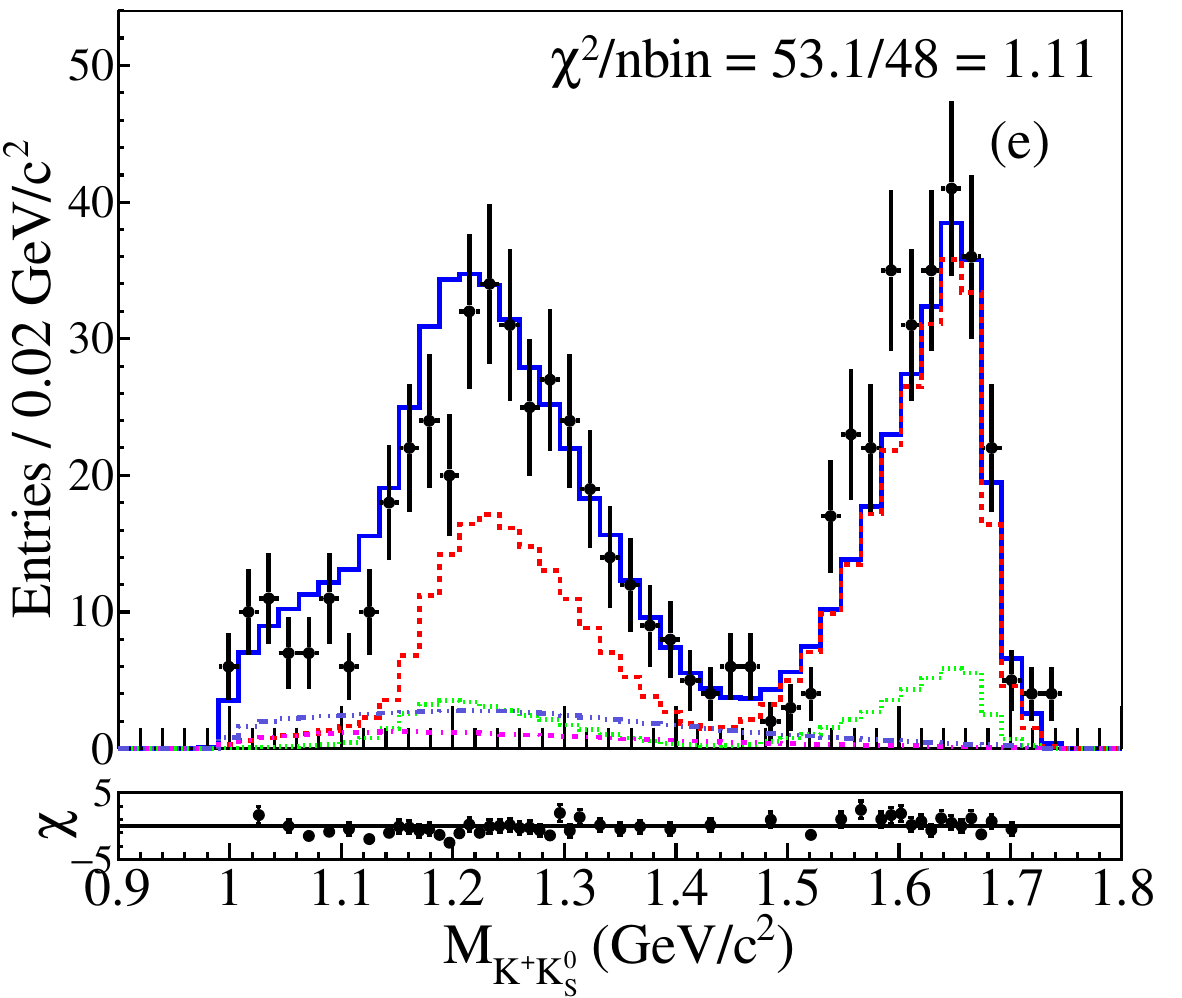}
\caption{The projections of the amplitude analysis fit
for $D^+\to K^+K^0_S\pi^0$ on two-body particle mass distributions~\cite{BESIII:2021dmo}.
}
\label{fig:Dp_KSKpi0}
\end{figure*}

Recent studies of $D\to PP$ and $D\to VP$ decays using QCD-derived models~\cite{Fu-Sheng:2011fji,Qin:2013tje,Cheng:2016ejf,Cheng:2019ggx} generally agree with experimental measurements,
except for the SCS decay $D^+\to K^*(892)^+K^0_S$, which involves color-favored tree, $W$-annihilation, and penguin diagrams.
Previously, E687 reported the branching fraction of $D^+\to K^*(892)^+K^0_S$ relative to $D^+\to K^0_S\pi^+$ to be $1.1\pm0.3\pm0.4$~\cite{E687:1994qpa},
which gives ${\cal B}(D^+\to K^*(892)^+K^0_S) = (17\pm8)\times 10^{-3}$ when combining with the world average of ${\cal B}(D^+\to K^0_S\pi^+)$.
In 2021, Ref.~\cite{BESIII:2021dmo} reported the first amplitude analysis of
the SCS decay $D^+\to K^+ K_S^0\pi^0$,
based on 692 candidates with a signal purity of 97.4\%.
The amplitude analysis fit projections on two-body particle mass distributions are shown in Fig.~\ref{fig:Dp_KSKpi0}.
The analysis reveals that
the $K^{*}(892)^+(\to K^+\pi^0)K_S^0$ component is found dominant with a fraction of $(57.1\pm2.6\pm4.2)\%$,
accompanied with the $\bar K^{*}(892)^0(\to K^0_S\pi^0)K^+$ component with a fraction of $(10.2\pm1.5\pm2.2)\%$.
Using ${\cal B}(D^+\to K^+K^0_S\pi^0)=(5.07\pm0.19\pm0.23)\times10^{-3}$ previously measured by BESIII~\cite{BESIII:2018pku}
gives ${\cal B} (D^+\to K^*(892)^+K^0_S)=(8.69\pm0.40\pm0.64\pm0.51)\times10^{-3}$.
This result is consistent with the previous results~\cite{E687:1994qpa}, with a precision improved by a factor of 4.6.
It is consistent with the prediction based on the pole model~\cite{Fu-Sheng:2011fji}, which suffers from large theoretical uncertainty;
but differs from the theoretical predictions in Refs.~\cite{Qin:2013tje,Cheng:2016ejf,Cheng:2019ggx} by about $4\sigma$.
This indicates that the QCD-derived models need further improvements, which may lead to variations in the predicted $CP$ violation effects.
In addition, ${\cal B}(D^+\to \bar K^{*}(892)^0K^+)=(3.10\pm0.46\pm0.68\pm0.18)\times10^{-3}$ is obtained,
which also agrees well with the previous measurements and theoretical predictions~\cite{Fu-Sheng:2011fji,Qin:2013tje,Cheng:2016ejf,Cheng:2019ggx}.

\begin{figure}[!htbp]
  \centering
  \hspace{-0.em}\includegraphics[width=0.225\textwidth]{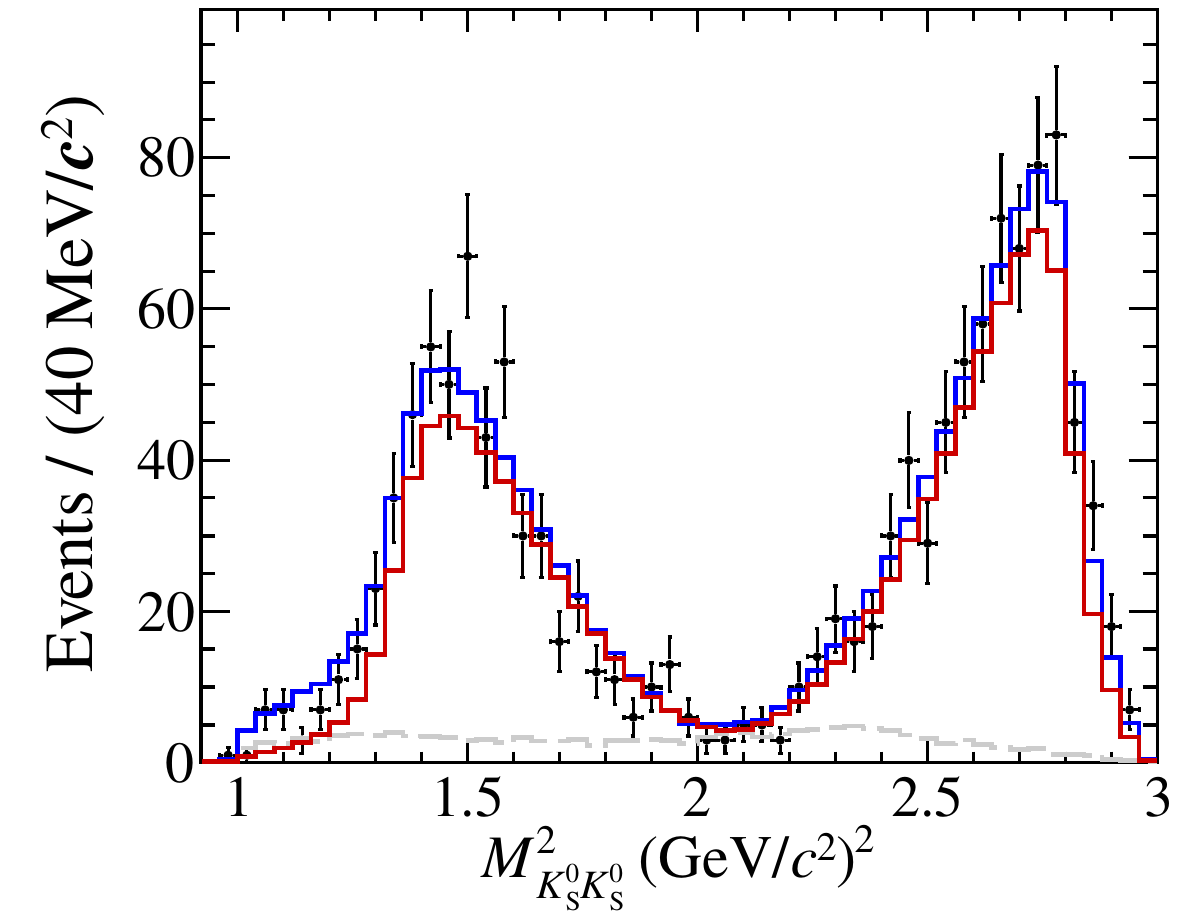}
  \hspace{-0.em}\includegraphics[width=0.225\textwidth]{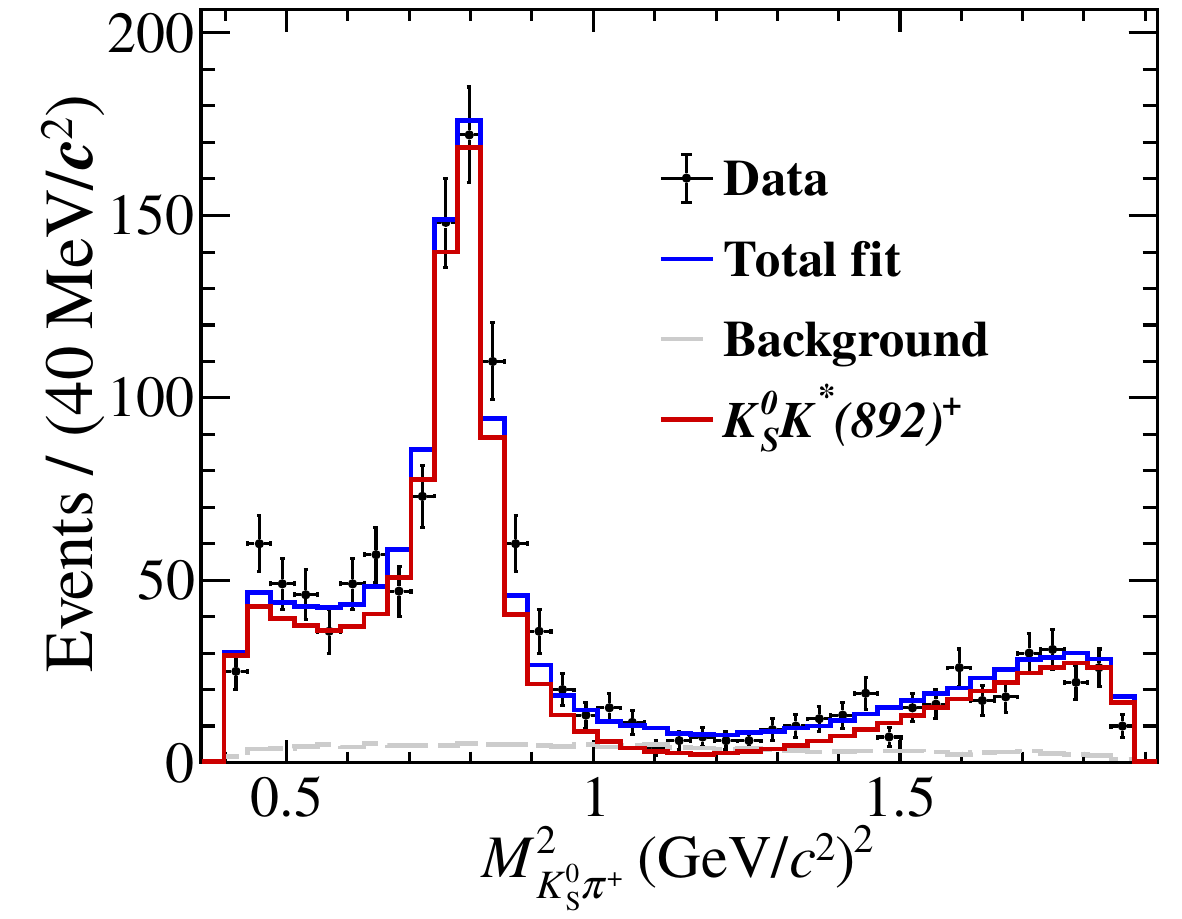}
  \caption{
 The projections of the amplitude analysis fits of $D^+\to 2K^0_S\pi^+$ on
  (left) $M^{2}_{K_S^0K_S^0}$ and (right) $M^{2}_{K_S^0\pi^+}$~\cite{BESIII:2024ncc}.
   }
  \label{fig:Dp_KSKSpi}
\end{figure}

Previously, BESIII measured $\mathcal{B}(D^+ \to K_S^0 K^*(892)^+) = (8.69 \pm 0.40\pm 0.64\pm 0.51)\times10^{-3}$ via
$D^+ \to K^+ K_S^0 \pi^{0}$~\cite{BESIII:2018pku,BESIII:2021dmo}, which differs from predictions in Refs.~\cite{Qin:2013tje,Cheng:2016ejf,Cheng:2019ggx} by about $4\sigma$.
The $D^{+} \to 2K_{S}^{0}\pi^{+}$ decay offers a cleaner environment, without interference from $D^+\to K^{+} K^*(892)^0$ present
in $D^{+} \to K^{+}K_{S}^{0}\pi^{0}$.
This decay was first observed by BESIII with $\mathcal{B}(D^+ \to 2K_S^0 \pi^{+}) = (2.70 \pm 0.05\pm 0.12) \times 10^{-3}$
based on the single-tag method~\cite{BESIII:2016nrs}.
The amplitude analysis of $D^{+} \to 2K_{S}^{0}\pi^{+}$ was performed by using 1.2k candidates with a signal purity of 89\%~\cite{BESIII:2024ncc}.
The amplitude analysis fit projections on two-body particle mass squared distributions are shown in Fig.~\ref{fig:Dp_KSKSpi}.
This decay is found to be dominated by $D^+\to K^0_SK^*(892)^+$
with a fraction of $(97.8\pm1.0\pm0.4)\%$, with the nonresonant decay
$D^+\to K^0_S(K^0_S\pi^+)_{{\cal S}{\rm -wave}}$
with a fraction of  $(4.4\pm1.0\pm0.5)\%$.
The branching fraction of $D^{+} \to 2K^{0}_{S}\pi^{+}$ is determined to be
$\mathcal{B}(D^{+} \to 2K^{0}_{S}\pi^{+})=(2.97\pm 0.09\pm 0.05) \times 10^{-3}$.
The branching fraction of $D^{+} \to K^{0}_{S}K^{*}(892)^{+}$ is calculated to be
$\mathcal{B}(D^{+} \to K^{0}_{S}K^{*}(892)^{+}) = (8.72 \pm 0.28 \pm 0.15) \times 10^{-3}$,
which is consistent with the previous BESIII measurement~\cite{BESIII:2021dmo} but with improved precision.
This result deviates from those in Refs.~\cite{Qin:2013tje,Cheng:2010ry} by more than five standard deviations,
while it is consistent with that based on the theoretical expectations~\cite{Fu-Sheng:2011fji,Cheng:2024hdo} within
$2\sigma$ and the predictions in Ref.~\cite{Cheng:2016ejf} within $3\sigma$.

\subsubsection{Analyses of $D^+_{(s)}\to K\pi\pi$}

\begin{figure*}[!htbp]
 \centering
 \includegraphics[width=0.225\textwidth]{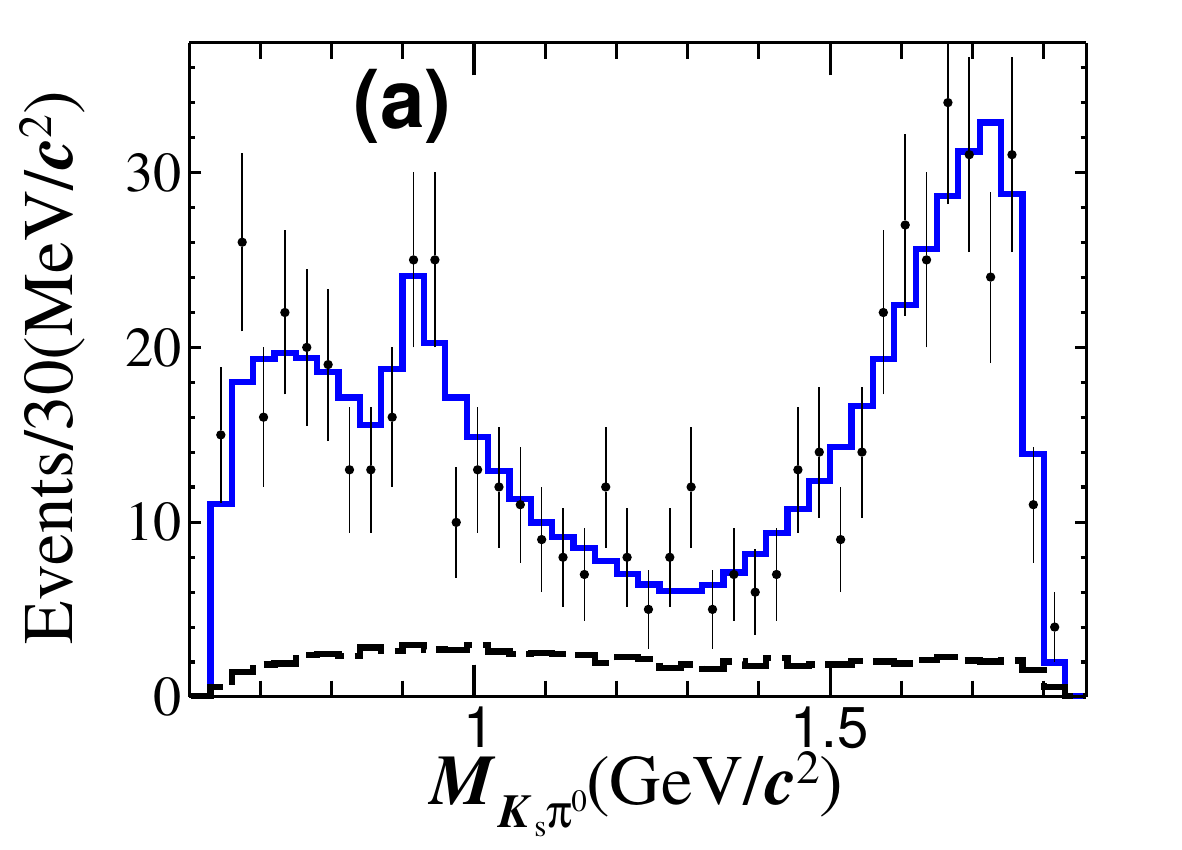}
 \includegraphics[width=0.225\textwidth]{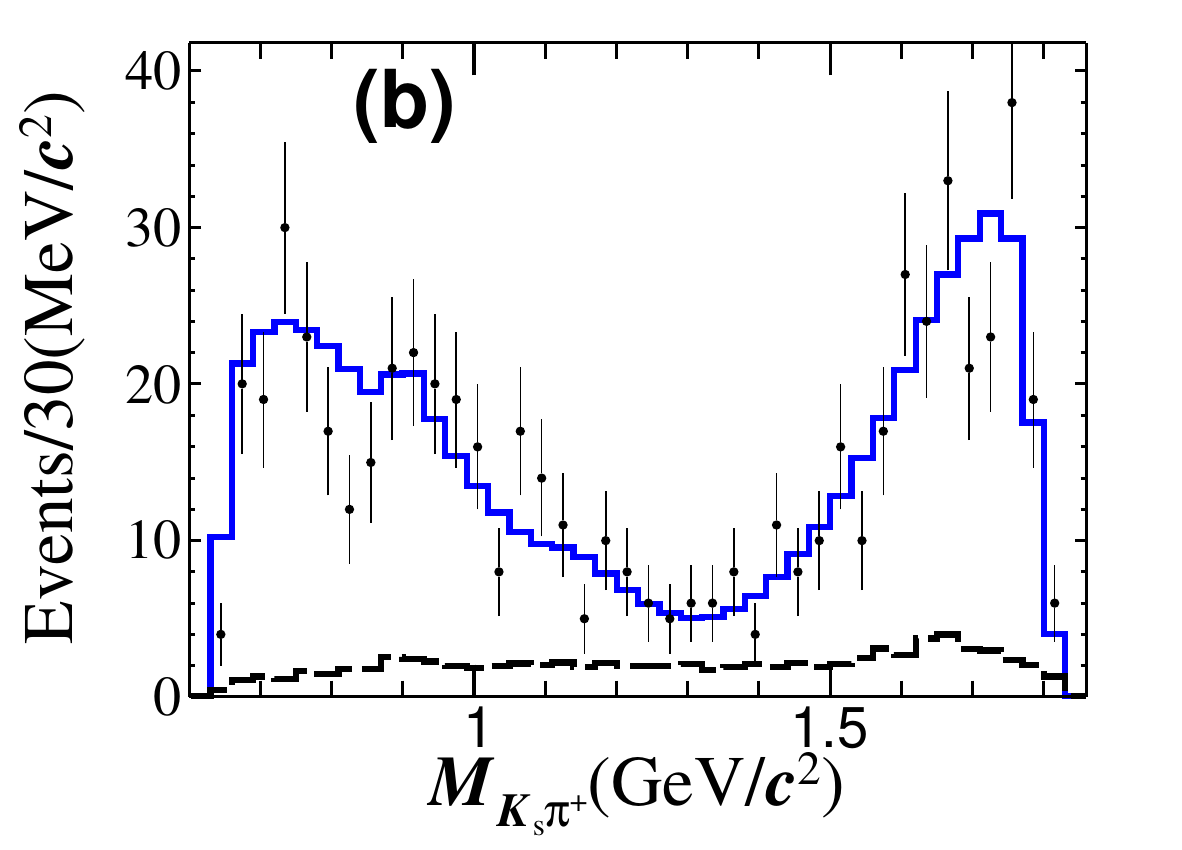}
 \includegraphics[width=0.225\textwidth]{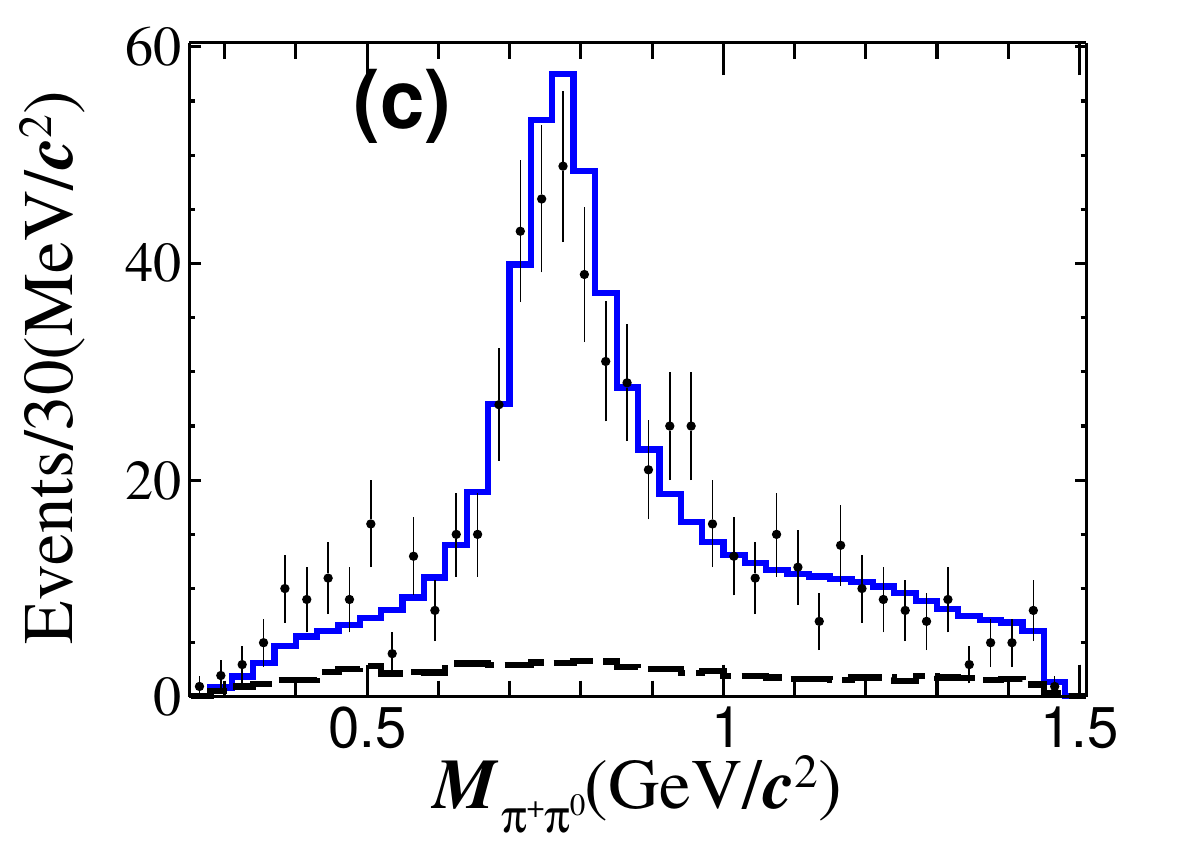}
 \caption{
   The projections of the amplitude analysis fit for $D^+_s\to K^0_S\pi^+\pi^0$ on (a) $M_{K_S^0\pi^0}$, (b) $K_{S}^0\pi^+$, and (c)
   $M_{\pi^+\pi^0}$~\cite{BESIII:2021xox}. The data are represented by points with error
   bars, the fit results by the solid blue lines, and the background
   estimated from generic MC samples by the black dashed lines.}
 \label{fig:Ds_KSpipi0}
\end{figure*}

Study of $D_{s}^{+} \to K^0_S\pi^+\pi^0$ is powerful to extract information about
different $D^+_s\to VP$ decays, e.g.,
$D^+_s\to K_{S}^{0}\rho(770)^+$, $K^{*}(892)^{0}\pi^+$, and $K^{*}(892)^{+}\pi^0$,
which were previously calculated in various theories~\cite{Wu:2004ht,Bhattacharya:2008ke,Cheng:2010ry,Cheng:2019ggx,Fu-Sheng:2011fji}.
Previously, only its branching fraction measurement was reported by CLEO-c,
which is $\mathcal{B}(D_{s}^{+} \to K^{0}\pi^{+}\pi^{0}) = (1.00\pm 0.18)\%$~\cite{CLEO:2009vke},
with 600~pb$^{-1}$ of data recorded around 4.17~GeV.
In Ref.~\cite{BESIII:2021xox}, an amplitude analysis of $D_{s}^{+} \to K^0_S\pi^+\pi^0$ was presented based on 609 candidate events with a signal purity of 80\%.
The amplitude analysis fit projections on two-body particle mass distributions are shown in Fig.~\ref{fig:Ds_KSpipi0}.
From this analysis, the main components (with fit fractions) are found to be
      $D_{s}^{+} \to K_{S}^{0}\rho(770)^{+}$ [$(50.2 \pm 7.2 \pm 3.9)\%$],
      $D_{s}^{+} \to K_{S}^{0}\rho(1450)^{+}$ [$(20.4 \pm 4.3 \pm 4.4)\%$],
      $D_{s}^{+} \to K^{*}(892)^{0}\pi^{+}$   [$(8.4  \pm 2.2 \pm 0.9)\%$],
      $D_{s}^{+} \to K^{*}(892)^{+}\pi^{0}$   [$(4.6  \pm 1.4 \pm 0.4)\%$], and
      $D_{s}^{+} \to K^{*}(1410)^{0}\pi^{+}$  [$(3.3  \pm 1.6 \pm 0.5)\%$].
The branching fraction of $D^+_s\to K^0_S\pi^+\pi^0$ is measured to be
$(5.43\pm0.30\pm0.15)\times 10^{-3}$ with
 precision improved by about a factor of 3 compared to the previous measurement~\cite{CLEO:2009vke}.
Assuming $\mathcal{B}(K^0\to K^0_S)=0.5$, the branching fractions for the intermediate processes are determined to be
$\mathcal{B}(D_{s}^{+} \to K^{0}\rho(770)^{+}) = (5.46\,\pm\,0.84\,\pm\,0.44)\times 10^{-3}$,
$\mathcal{B}(D_{s}^{+} \to K^{*}(892)^{0}\pi^{+}) = (2.71\,\pm\,0.72\,\pm\,0.30)\times 10^{-3}$, and
$\mathcal{B}(D_{s}^{+} \to K^{*}(892)^{+}\pi^{0}) = (0.75\,\pm\,0.24\,\pm\,0.06)\times 10^{-3}$.
These are valuable for a deeper understanding of quark flavor SU(3) symmetry, SU(3)
breaking effects, and other related theoretical issues.
The asymmetry for the branching fractions of the decays $D_{s}^{+} \to K_{S}^0\pi^{+}\pi^{0}$ and $D_{s}^{-} \to K_{S}^0\pi^{-}\pi^{0}$
is determined to be $(2.7\pm5.5\pm0.9)\%$. No evidence for $CP$ violation is found.

\begin{figure*}[htp]
          \centering
          \includegraphics[width=0.225\textwidth]{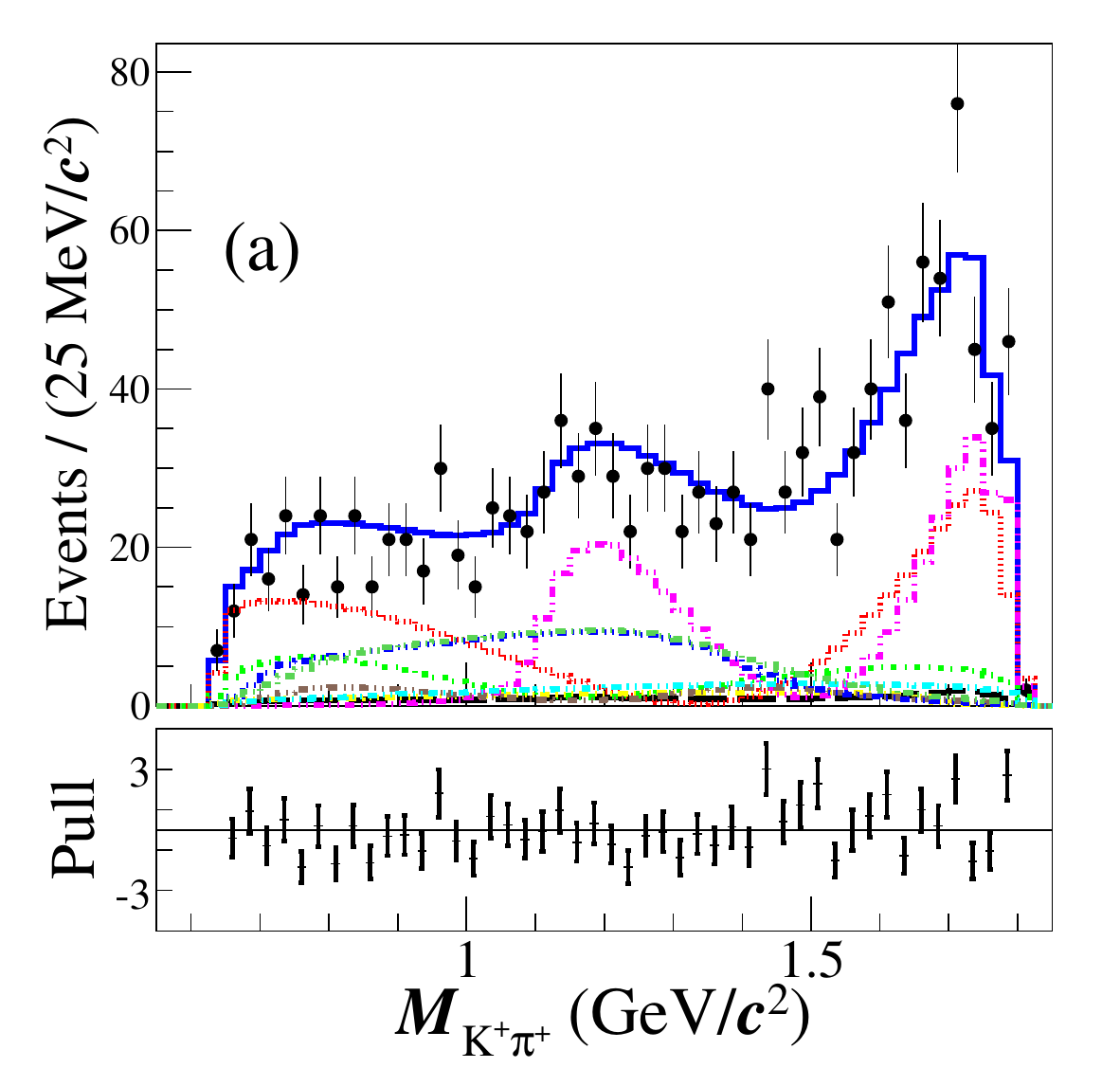}
          \includegraphics[width=0.225\textwidth]{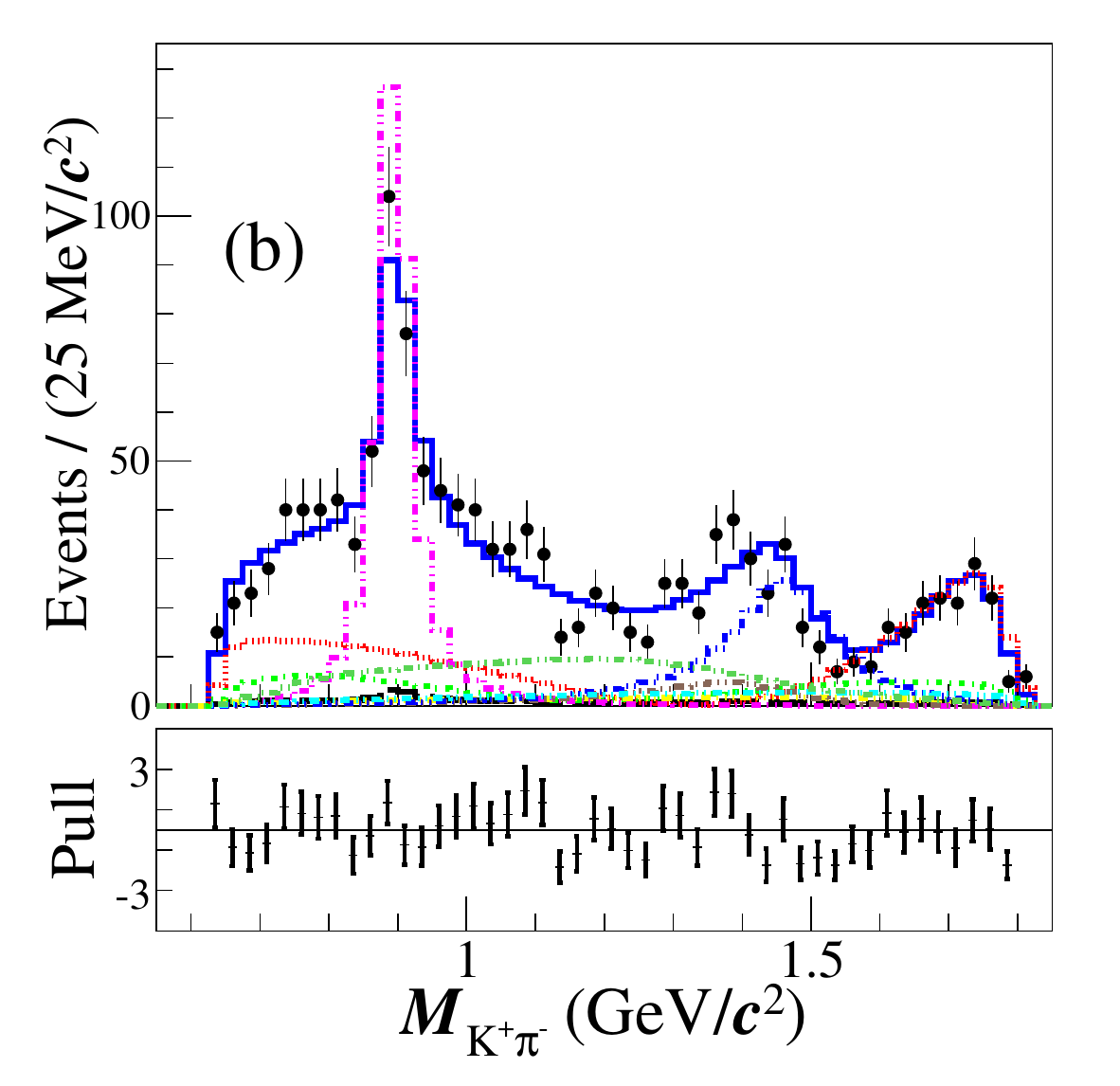}
          \includegraphics[width=0.225\textwidth]{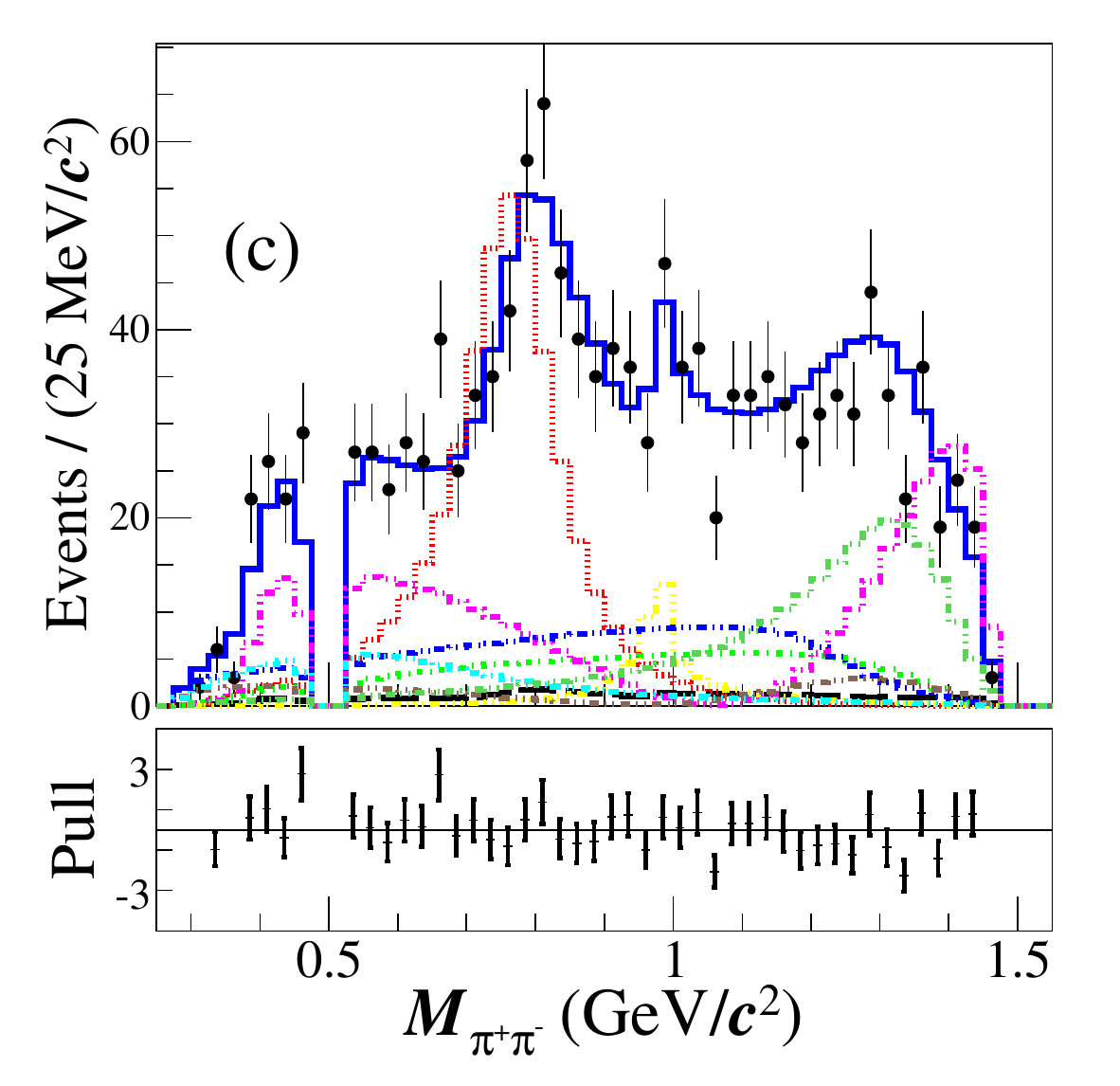}
          \includegraphics[width=0.225\textwidth]{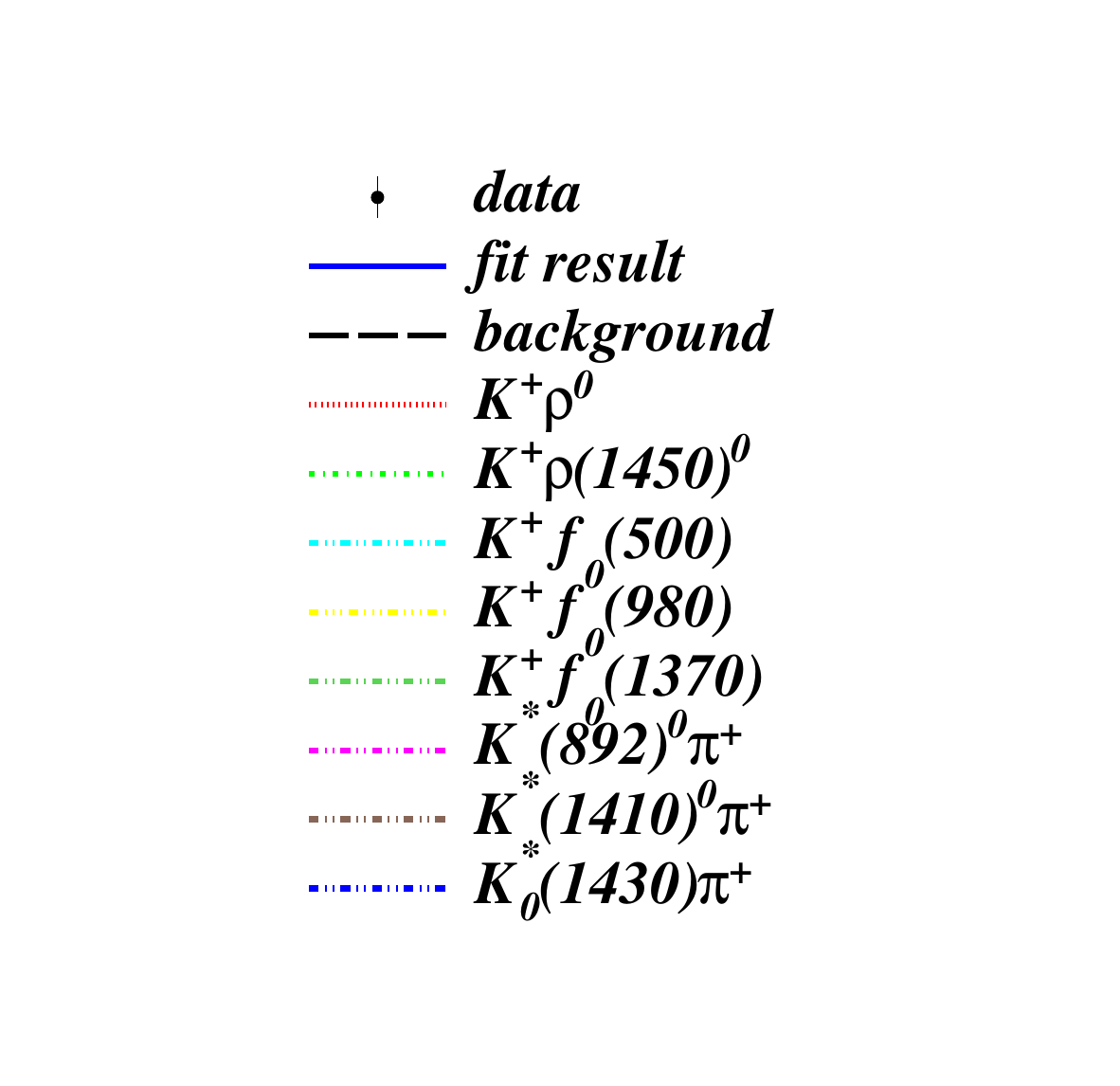}
          \caption{The projections of the amplitude analysis fit for $D^+_s\to K^+\pi^+\pi^-$ on (a) $M_{K^+\pi^+}$, (b) $M_{K^+\pi^-}$, and (c) $M_{\pi^+\pi^-}$~\cite{BESIII:2022vaf}.}
    \label{fig:Ds_Kpipi}
\end{figure*}

For $D_{s}^{+} \to K^+\pi^+\pi^-$,
CLEO-c previously reported its branching fraction to be $(0.654\pm0.033\pm0.025)\%$ with 600~pb$^{-1}$ of data recorded around 4.17~GeV~\cite{CLEO:2013bae} and
FOCUS reported its first amplitude analysis~\cite{FOCUS:2004muk}.
In 2022, based on 1.4k candidates with a signal purity of 95.2\%,
Ref.~\cite{BESIII:2022vaf} reported an amplitude analysis of $D_{s}^{+} \to K^+\pi^+\pi^-$.
The amplitude analysis fit projections on two-body particle mass distributions are shown in Fig.~\ref{fig:Ds_Kpipi}.
The intermediate resonances $f_0(500)$, $f_0(980)$, and $f_0(1370)$ are observed for the first time in this channel.
From the amplitude analysis, one finds that
the two-body decays $D_s^{+}\to K^+\rho(770)^0$ and $D_s^{+}\to K^*(892)^0\pi^+$ are dominated with
fractions of $(32.1\pm3.7\pm3.7)\%$ and $(30.2\pm1.8\pm2.0)\%$, respectively;
and fit fractions of other components of
		 $D_s^{+}\to K^+\rho(770)^0$,
		 $D_s^{+}\to K^+\rho(1450)^0$,
		 $D_s^{+}\to K^+f_0(500)$,
		 $D_s^{+}\to K^+f_0(980)$,
		 $D_s^{+}\to K^+f_0(1370)$,
		 $D_s^{+}\to K^*(1410)^0\pi^+$, and
		 $D_s^{+}\to K^*_0(1430)^0\pi^+$ range between $(4.5-19.9)\%$.
The branching fraction of $D^+_s\to K^+\pi^+\pi^-$ is measured to be $(6.11\pm0.18\pm0.11)\times 10^{-3}$,
with precision improved by about a factor of 2 compared to the world average value~\cite{ParticleDataGroup:2022pth}.
The branching fractions of the intermediate processes are also presented,
in which those of $D_s^+ \to K^+f_0(500)$, $D_s^+ \to K^+f_0(980)$, and $D_s^+ \to K^+f_0(1370)$ are reported for the first time.
The asymmetry of the branching fractions  of $D_s^+ \to K^+\pi^+\pi^-$ and $D_s^- \to K^-\pi^-\pi^+$ is determined to be $(3.3\pm{{3.0}}\pm1.3)\%$.
No indication of $\mathit{{CP}}$ violation is found.
The reported branching fraction of $D_s^+ \to K^+\rho(770)^0$ is in good agreement with the predictions in Ref.~\cite{Qin:2013tje},
and the measured branching fraction of $D_s^+ \to K^*(892)^0\pi^+$ is consistent with the prediction in Ref.~\cite{Cheng:2016ejf}.
Meanwhile, our result deviates from the predictions of $D_s^+ \to K^+\rho(770)^0$ in Refs.~\cite{Cheng:2016ejf,Wu:2004ht} and $D_s^+ \to K^*(892)^0\pi^+$ in Refs.~\cite{Wu:2004ht,Qin:2013tje} over two standard deviations. {{Moreover,
		Ref.~\cite{Qin:2013tje} predicts the ratio of the branching fraction of $D_s^+ \to K^+\rho(770)^0$ to that of $D_s^+ \to K^+\omega$ is far greater than one, while Ref.~\cite{Cheng:2016ejf} calculates that it should be close to one. The ratio is determined to be about two by taking the results in this analysis and in Ref.~\cite{BESIII:2022bvv}.}}

\begin{figure}
  \centering
  \includegraphics[width=0.9\linewidth]{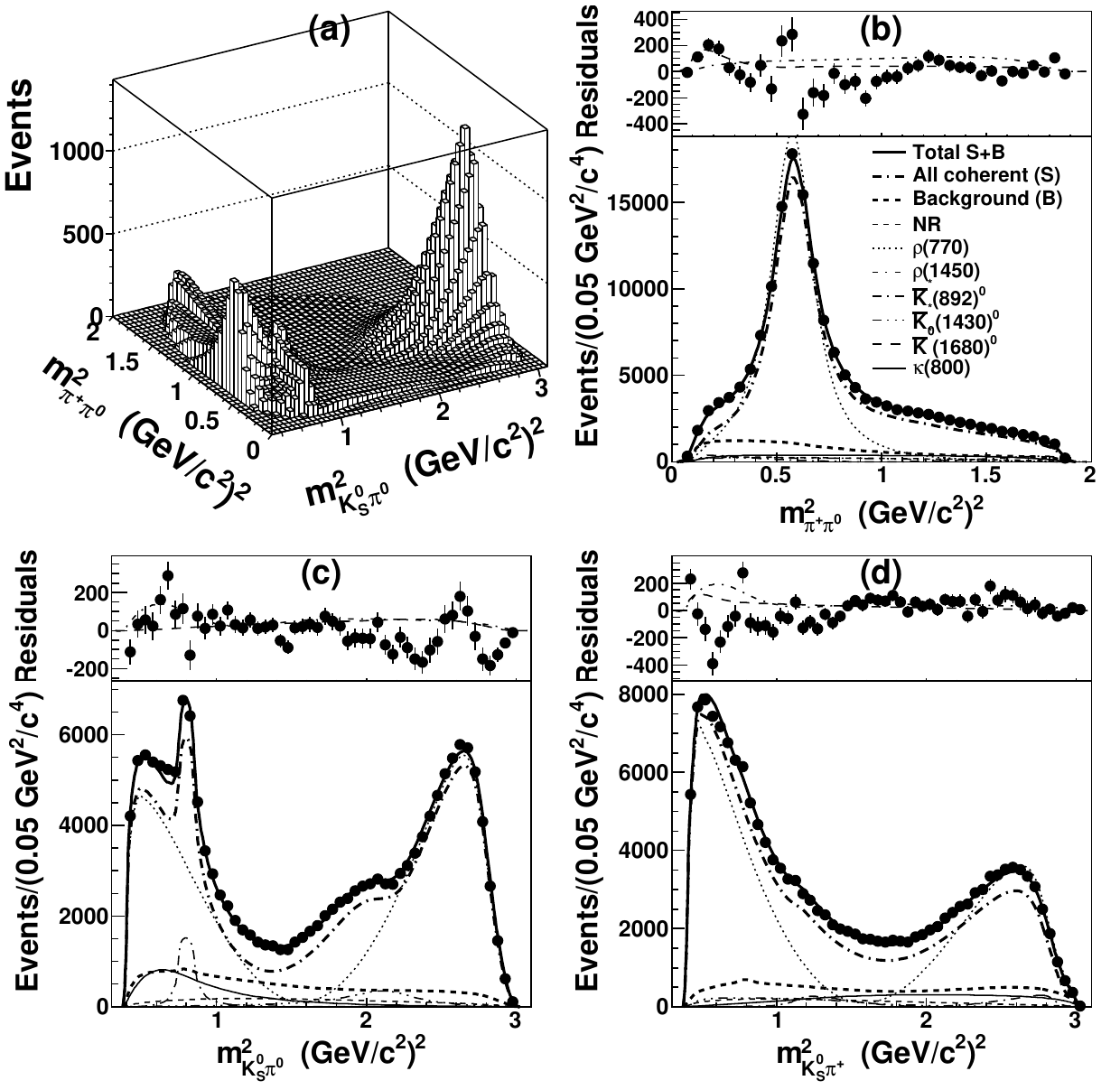}
  \caption{The projections of the amplitude analysis fit for $D^+\to K^0_S \pi^+\pi^0$ on (b) $m^2_{\pi^+\pi^0}$, (c) $m^2_{K^0_S\pi^0}$, and (d) $m^2_{K^0_S\pi^+}$~\cite{BESIII:2014oag}.
  }
  \label{fig:Dp_KSpipi0}
\end{figure}

A large contribution from a $\bar K\pi$ $\cal S$-wave intermediate state was observed in earlier experiments including MARKIII \cite{MARK-III:1987qok}, NA14 \cite{NA142:1990qcz}, E691 \cite{E691:1992rwf}, E687 \cite{E687:1994wlh}, E791 \cite{E791:2002xlc,E791:2005gev}, and CLEO-c \cite{CLEO:2008jus} in the $D^+ \to K^- 2\pi^+$ decay. Both E791 and CLEO-c interpreted their data with a Model-Independent Partial Wave Analysis (MIPWA) and found a phase shift at low $\bar K\pi$ mass to confirm the $\overline{\kappa}\pi$ component. Complementary to $D^+ \to K^- 2\pi^+$, the $D^+\to K^0_S \pi^+\pi^0$ decay is also a golden channel to study the $\bar K\pi$ $\cal S$-wave in $D$ decays.
The previous Dalitz plot analysis of $D^+\to K^0_S \pi^+\pi^0$ by MARKIII \cite{MARK-III:1987qok} included only two intermediate decay channels, $K^0_S\rho$ and $\overline{K}^{*0}\pi^+$, and was based on a small data set.
By analyzing 2.93 fb$^{-1}$ of data at 3.773 GeV, a Dalitz plot analysis of $D^+\to K^0_S \pi^+ \pi^0$ was performed~\cite{BESIII:2014oag}.
The analysis is based on 0.17 million candidate events with a background fraction of 15.1\%.
Figure~\ref{fig:Dp_KSpipi0} shows projections of the amplitude fit on $m^2_{\pi^+\pi^0}$, $m^2_{K^0_S\pi^0}$, and $m^2_{K^0_S\pi^+}$.
Its Dalitz plot is found to be well-represented by a combination of six quasi-two-body decay channels of $K^0_S\rho(770)^+$, $K^0_S\rho(1450)^+$, $\bar K^*(892)^0\pi^+$, $\bar K_0(1430)^0\pi^+$, $\bar K(1680)^0\pi^+$, and $\bar \kappa^0\pi^+$ plus a small non-resonant component.
The fit fractions of
  $D^+\to K^0_S\pi^+\pi^0_{NR}$,
  $D^+\to \rho(770)^+ K^0_S,\rho(770)^+\to \pi^+\pi^0$,
  $D^+\to \rho(1450)^+ K^0_S,\rho(1450)^+ \to \pi^+\pi^0$,
  $D^+\to \overline{K}^*(892)^0\pi^+,\bar{K}^*(892)^0\to K^0_S\pi^0$,
  $D^+\to \bar{K}^*_0(1430)^0\pi^+,\bar{K}^*_0(1430)^0\to K^0_S\pi^0$,
  $D^+\to \bar{K}^*(1680)^0\pi^+,\bar{K}^*(1680)^0\to K^0_S\pi^0$,
  $D^+\to \overline{\kappa}^0\pi^+,\overline{\kappa}^0\to K^0_S\pi^0$,
  $NR$+$\overline{\kappa}^0\pi^+$, and
  $K^0_S\pi^0$ $\cal S$-wave
  are
  $(4.6\pm0.7\pm 3.6^{+4.0}_{-3.6})\%$,
  $(83.4\pm2.2\pm2.8^{+6.5}_{-2.2})\%$,
  $(2.1\pm0.3\pm1.2^{+1.0}_{-1.5})\%$,
  $(3.6\pm0.2\pm0.2^{+0.4}_{-0.3})\%$,
  $(3.7\pm0.6\pm0.8\pm0.8)\%$,
  $(1.3\pm0.2\pm0.7^{+0.6}_{-1.1})\%$,
  $(7.7\pm1.2\pm4.0^{+5.1}_{-2.7})\%$,
  $(18.6\pm1.7\pm1.5^{+1.7}_{-4.4})\%$, and
  $(17.3\pm1.4\pm2.1^{+2.7}_{-3.8})\%$, respectively.
Using the fit fractions from this analysis, partial branching ratios are updated with higher precision than previous measurements.
Combining the obtained fit fractions and the world average value of ${\cal B}(D^+\to K^0_S\pi^+\pi^0)$,
partial branching ratios are reported with higher precision than previous measurements.
In addition, the $K^0_S\pi^0$ waves can be compared with the $K^-\pi^+$ waves in the $D^+\to K^-2\pi^+$ decay.

\begin{figure*}
	\centering
    \includegraphics[width=0.225\textwidth]{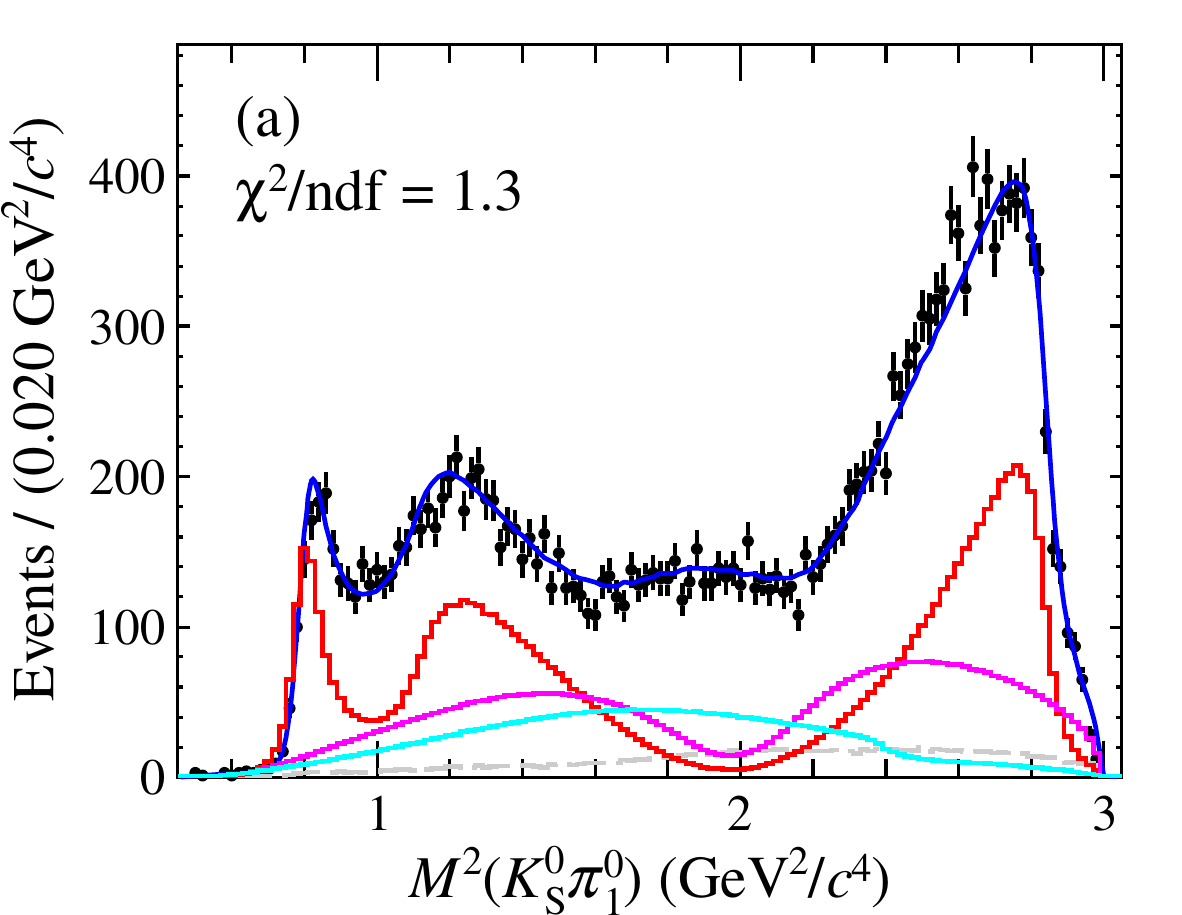}
    \includegraphics[width=0.225\textwidth]{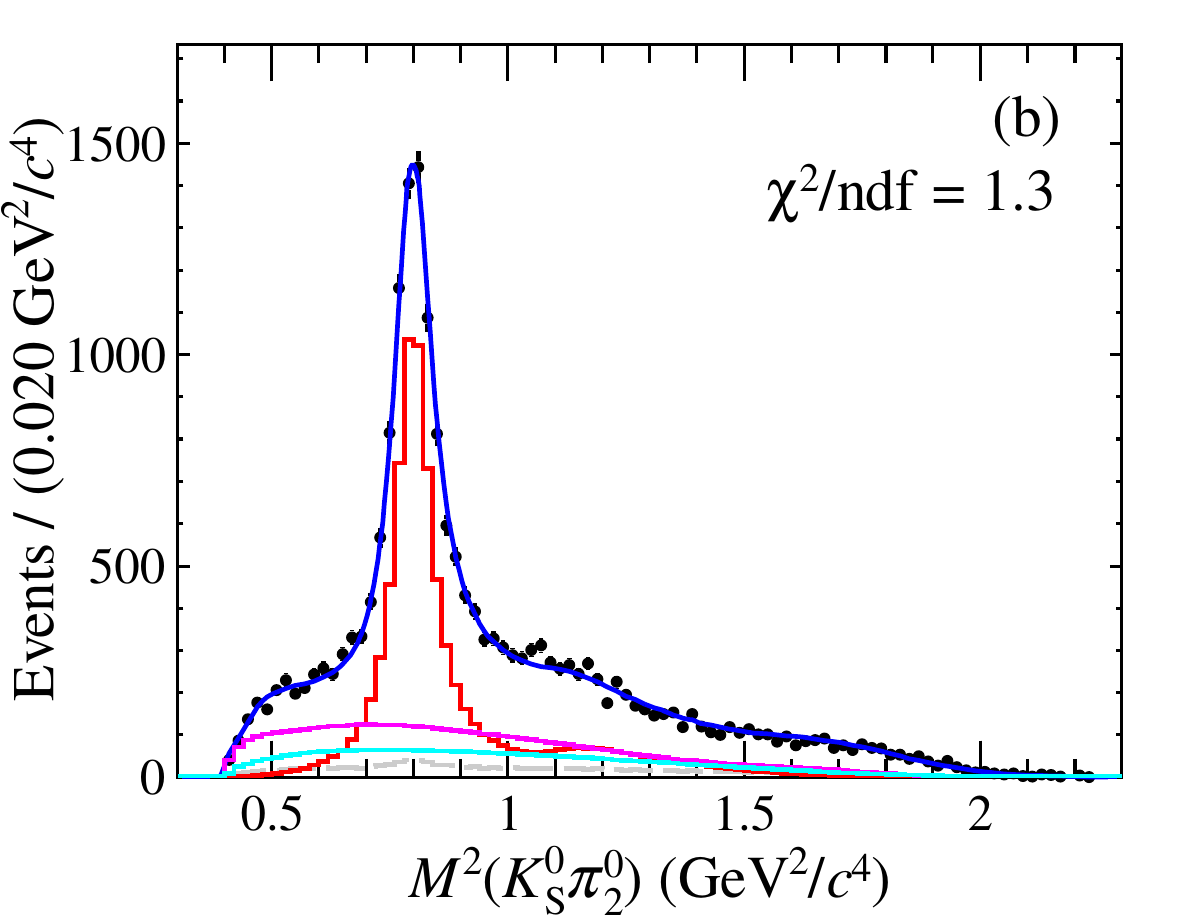}
    \includegraphics[width=0.225\textwidth]{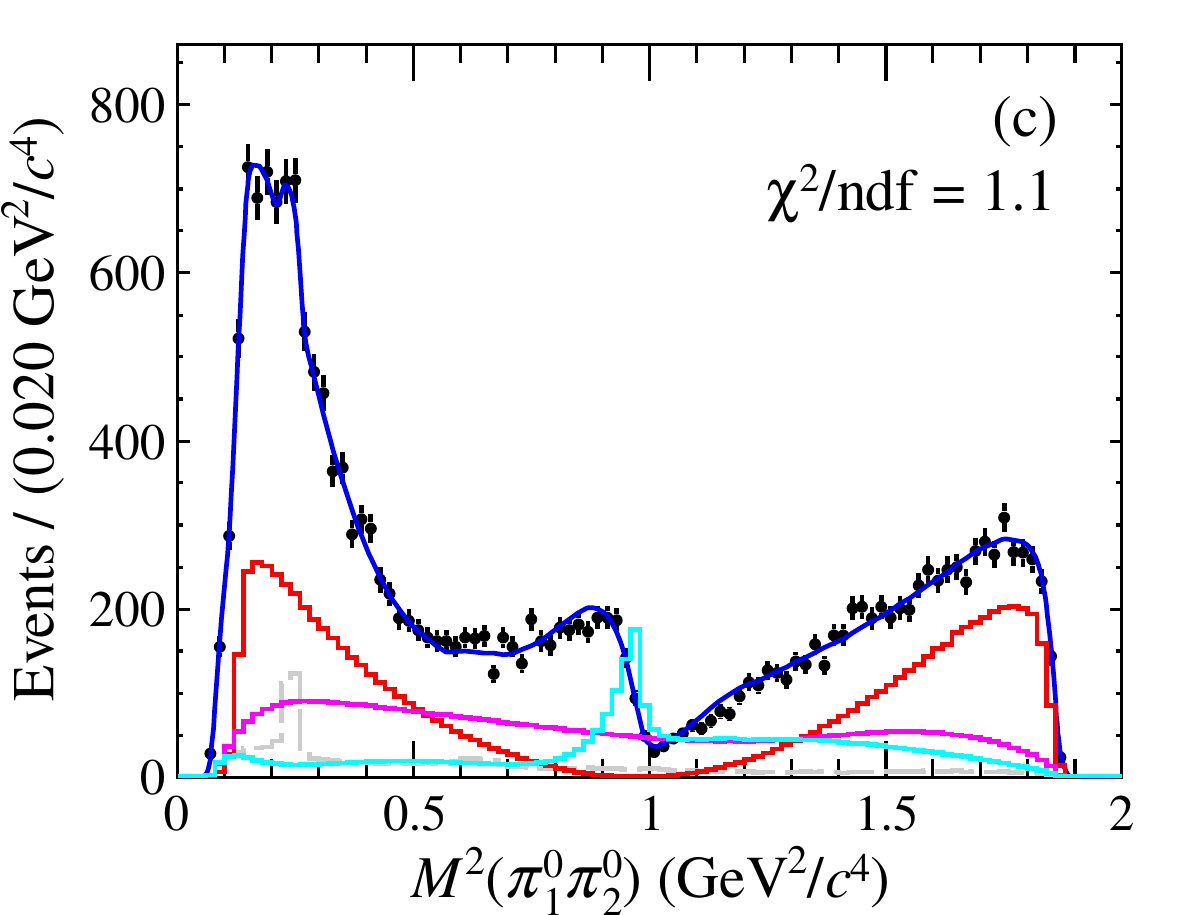}
    \includegraphics[width=0.225\textwidth]{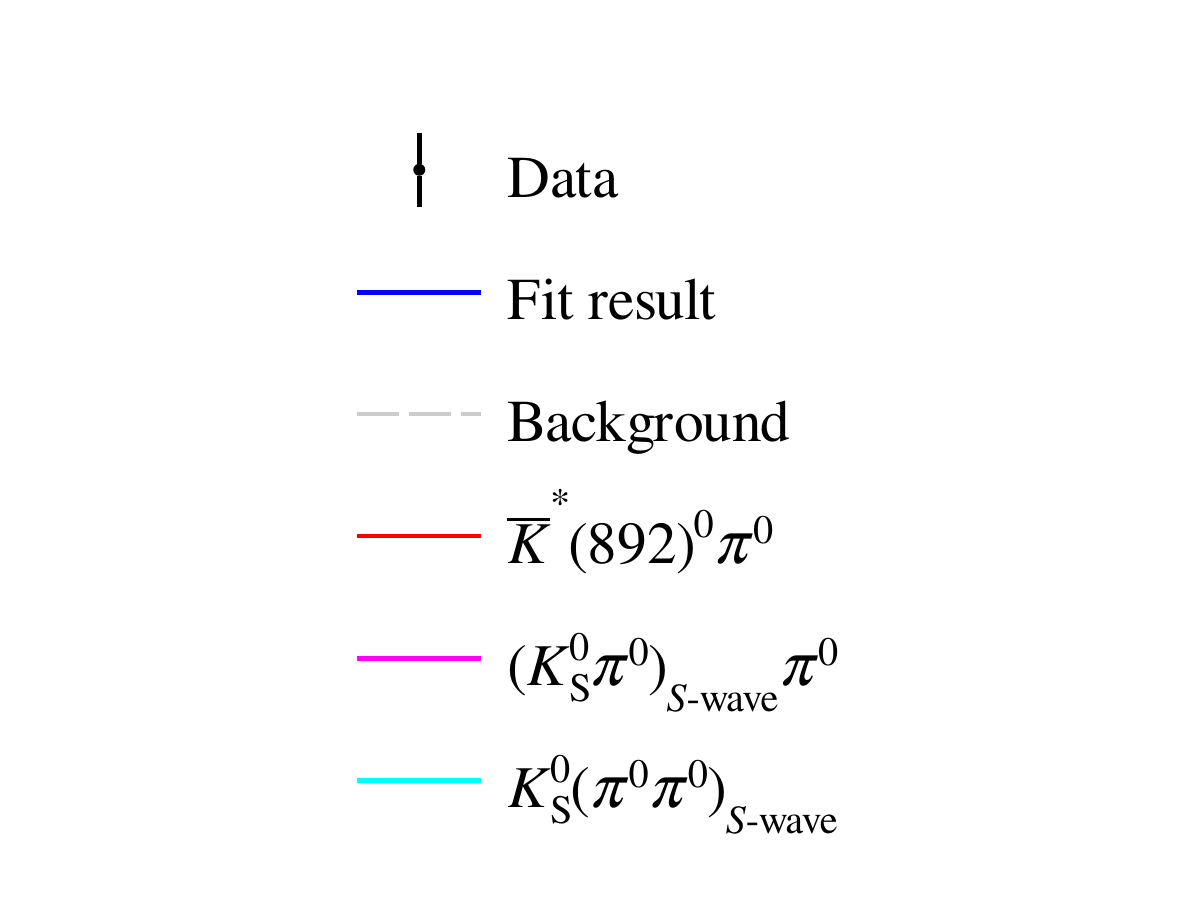}
	\caption{
The  projections of the amplitude analysis fit of $D^0\to K^0_S2\pi^0$ on (a) $M^2(K_S^0\pi^0_1)$, (b) $M^2(K_S^0\pi^0_2)$ and (c) $M^2(\pi_1^0\pi_2^0)$~\cite{BESIII:2025sea}.
 }

    \label{fig:D0_KSpi0pi0}
\end{figure*}

CF modes like $D^0 \to \bar K^{*}(892)^0\pi^0$ provide crucial constraints on topological amplitudes via color-suppressed internal $W$-emission and $W$-exchange diagrams~\cite{Cheng:2016ejf}. Theoretical predictions for this branching fraction span $(2.9 \pm 1.0)$\% to $(3.61 \pm 0.18)$\% across pole, FAT-mix, and TDA approaches~\cite{Fu-Sheng:2011fji,Qin:2013tje,Cheng:2024hdo}. However, experimental measurements from CLEO-c via amplitude analyses of $D^0 \to K^- \pi^+ \pi^0$~\cite{CLEO:2000fvk} and $D^0 \to K_S^0 2\pi^0$~\cite{CLEO:2011cnt} yield conflicting values of $(2.74 \pm 0.47)\%$ and $(4.16 \pm 0.49)\%$, respectively. 
With 20.3k candidate events with a signal purity of 93.5\%,
an amplitude analysis was performed on the $D^0 \to K_S^0 2\pi^0$ decay~\cite{BESIII:2025sea}.
The amplitude analysis fit projections on two-body particle mass squared distributions are shown in Fig.~\ref{fig:D0_KSpi0pi0}.
Five components $D^0\to\bar K^*(892)^0\pi^0$, $D^0\to\bar K^*_2(1430)^0\pi^0$,
$D^0\to\bar K^*(1680)^0\pi^0$, $D^0\to K^0_Sf_2(1270)$, and $D^0\to (K^0_S\pi^0)_{{\cal S}{\rm -wave}}\pi^0$
are observed, with fractions of
$(41.1\pm0.8\pm1.3)\%$, $(1.2\pm0.2\pm0.5)\%$, $(3.3\pm1.0\pm2.0)\%$,
$(1.8\pm0.4\pm1.0)\%$, and $(26.6\pm4.9\pm3.4)\%$, respectively.
Significant discrepancies are observed between these results and those reported by CLEO-c~\cite{CLEO:2011cnt}, mainly due to differences in both the amplitude model components and the propagator formalism. In the $M^2(K_S^0\pi^0)$ spectrum, an additional contribution from the $(K_S^0\pi^0)_{\cal S-{\rm wave}}$ component is accounted for in this analysis. For the $\cal S$-wave description of the $M^2(\pi^0\pi^0)$ spectrum,  the isobar model was utilized by CLEO-c, whereas the K-matrix formalism is adopted.
The branching fraction of $D^0 \to K^0_S2\pi^0$ is determined to be $(6.06 \pm 0.04\pm 0.07) \%$,
which is consistent with the CLEO-c measurement~\cite{CLEO:2011cnt}, with precision improved by a factor of 5.8.
The branching fractions of the intermediate processes are also presented.
The dominant intermediate process is $D^0 \to \bar{K}^{*}(892)^0\pi^0\to K_S^02\pi^0$.
Its branching fraction is determined as ${\cal B}(D^0 \to \bar{K}^{*}(892)^0\pi^0)=(2.54\pm0.05\pm0.08)\%$.
This result is significantly lower than the predicted results in Refs.~\cite{Fu-Sheng:2011fji,Qin:2013tje,Cheng:2024hdo} and the CLEO-c measurement from  the decay $D^0\to K_{S}^{0}2\pi^0$~\cite{CLEO:2011cnt} by about $3\sigma$. However, it aligns with the value $(2.74\pm0.23\pm0.41)\%$ obtained from the decay $D^0\to K^{-}\pi^+\pi^0$~\cite{CLEO:2000fvk} but with precision improved by a factor of 6.2.
Under the isospin symmetry with the predicted ratios $\frac{\mathcal{B}((\pi \pi)_{\cal S-{\rm wave}} \to \pi^+ \pi^-)}{\mathcal{B}((\pi \pi)_{\cal S-{\rm wave}} \to \pi^0 \pi^0)} = \frac{\mathcal{B}(f_{2}(1270) \to \pi^+ \pi^-)}{\mathcal{B}(f_{2}(1270) \to \pi^0 \pi^0)} = 2$, our measurements of $\mathcal{B}(D^0 \to K_S^0 (\pi \pi)_{\cal S-\rm wave})$ and $\mathcal{B}(D^0 \to K_S^0 f_2(1270))$ show excellent agreement with the $D^0 \to K_S^0 \pi^+ \pi^-$ results from Ref.~\cite{BaBar:2018cka}, providing robust validation of the isospin framework.

\subsubsection{Analyses of $D^{0,+}\to K^0_S\pi\eta$}

\begin{figure*}[!htbp]
  \centering
  \includegraphics[width=0.225\textwidth]{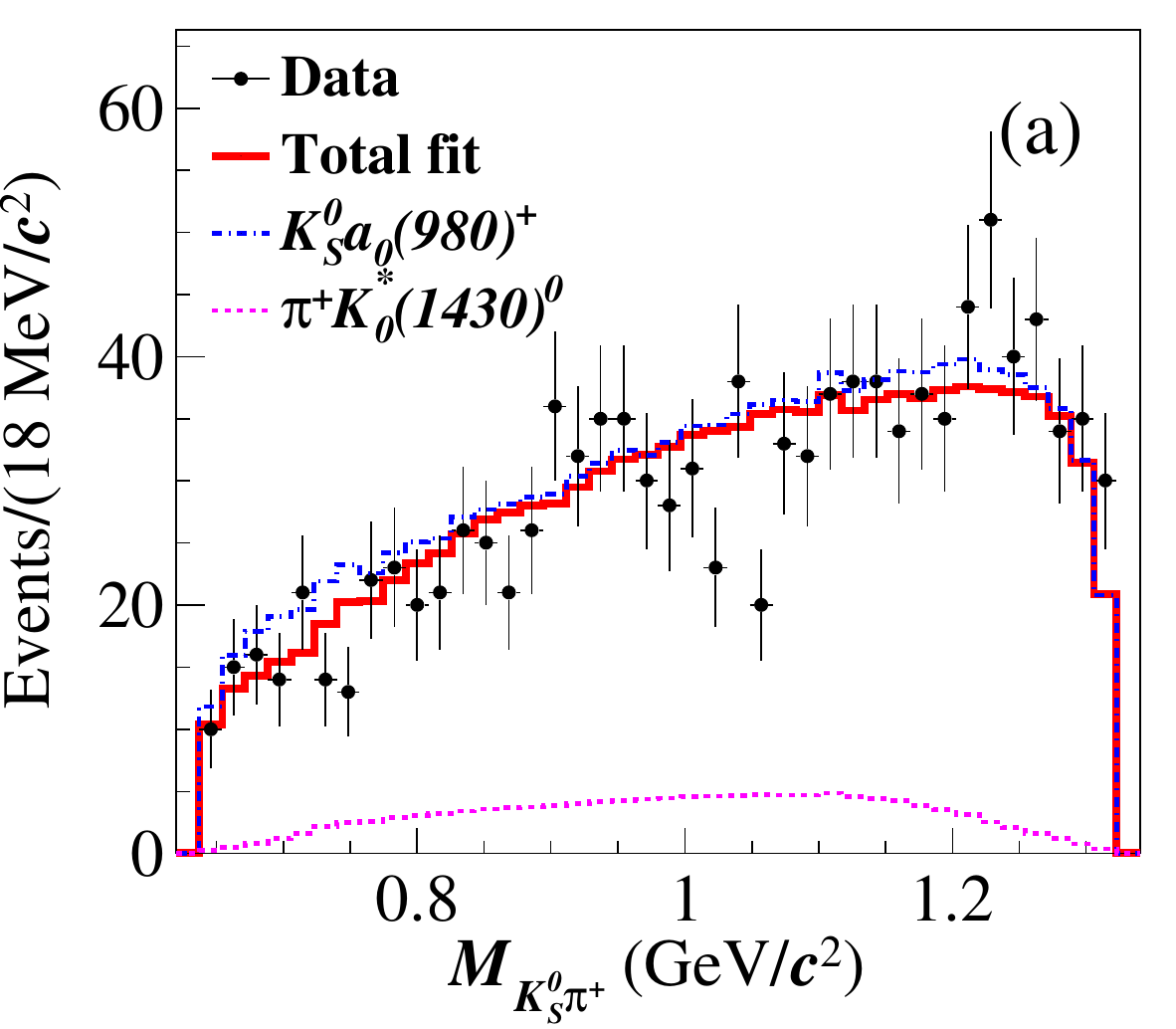}
  \includegraphics[width=0.225\textwidth]{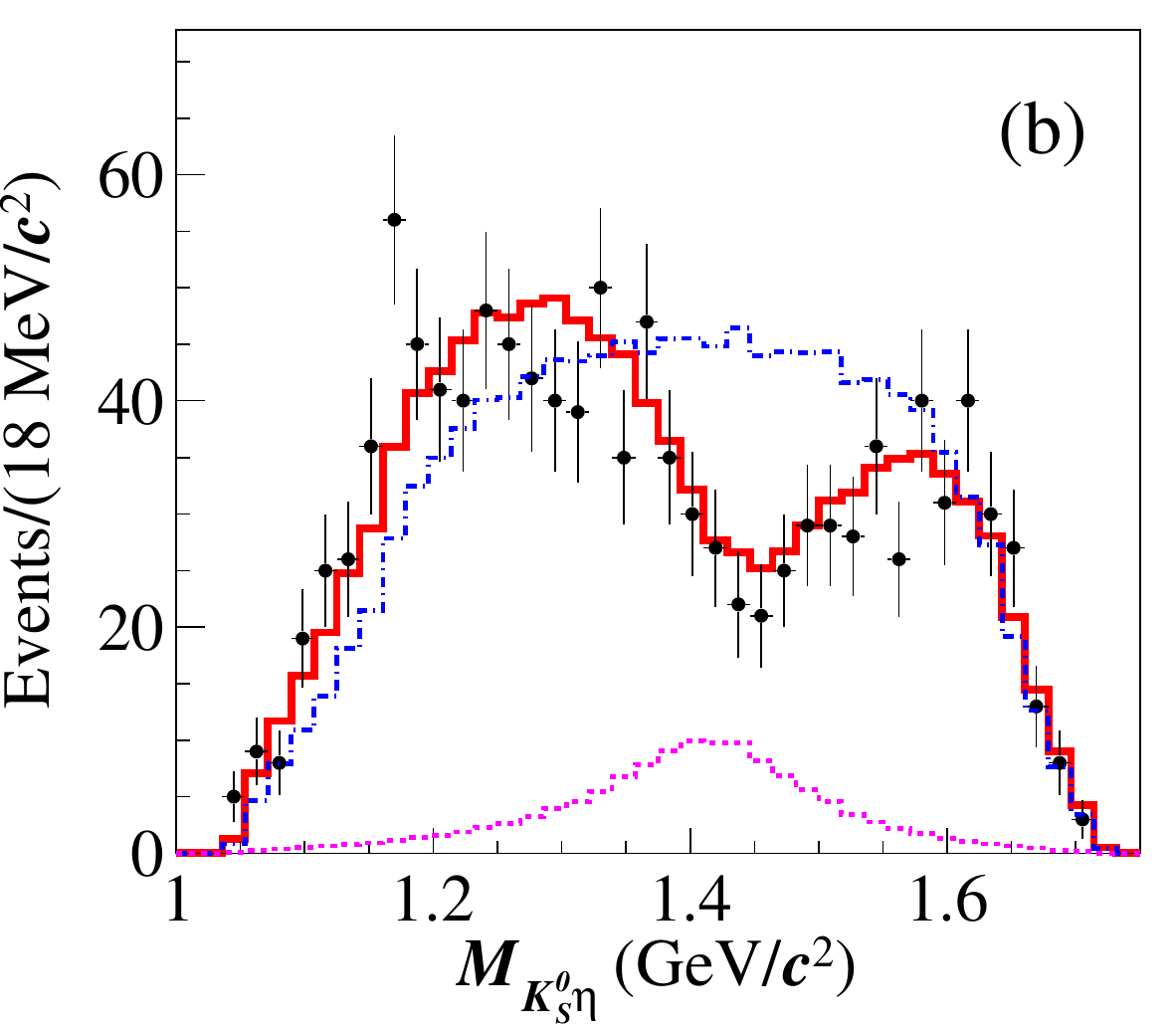}
  \includegraphics[width=0.225\textwidth]{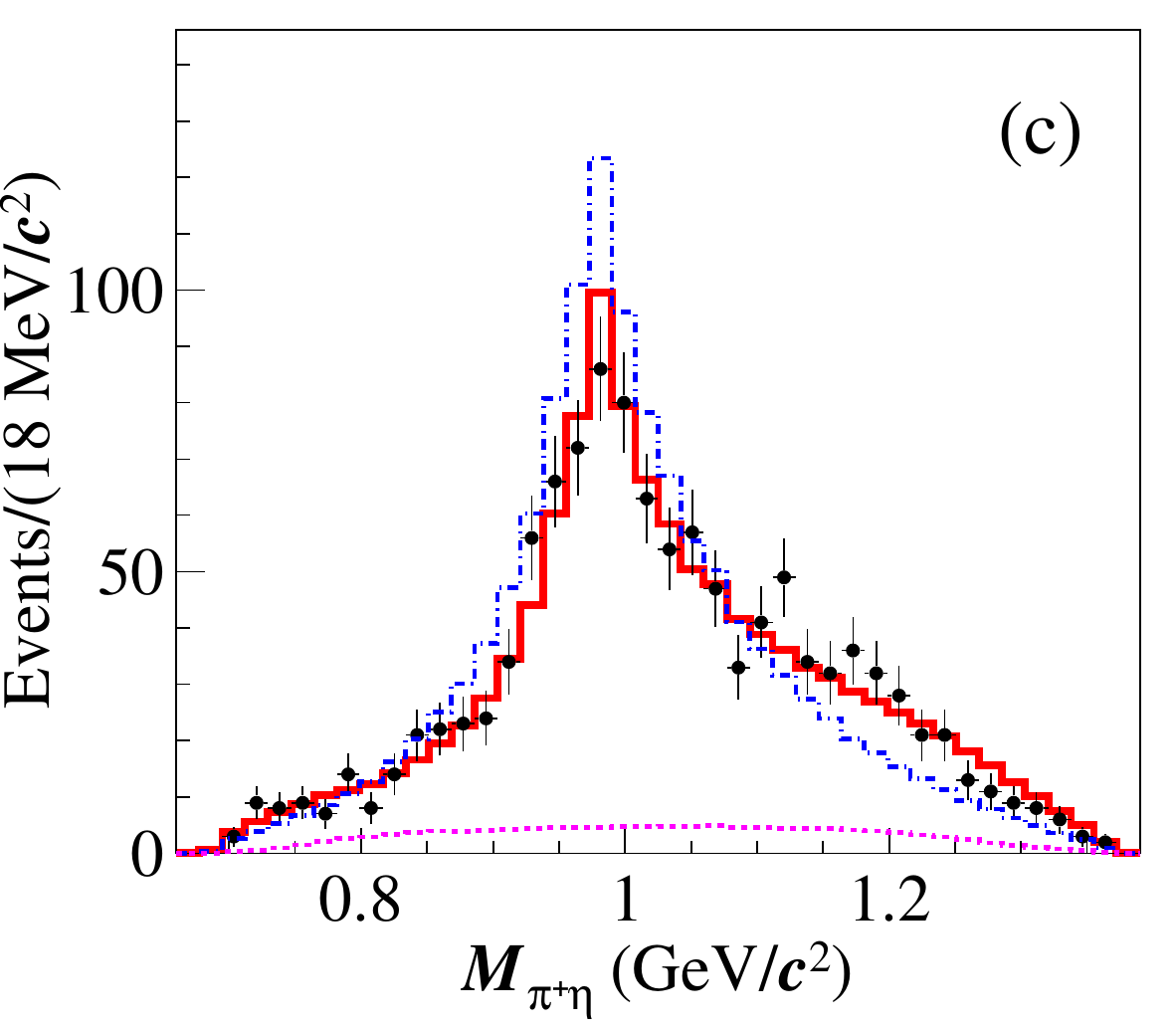}
  \caption{
  The projections of the amplitude analysis fit of $D^+\to K^0_S\pi^+\eta$ on
  (a) $M_{K_S^0\pi^+}$, (b)
    $M_{K_S^0\eta}$, and (c) $M_{\pi^+\eta}$~\cite{BESIII:2023htx}.
  }
  \label{fig:Dp_KSpieta}
\end{figure*}

The model-independent diagrammatic approach, based on SU(3) flavor symmetry,
 works well for $D\to PP, VP$~\cite{Cheng:2010ry,Cheng:2019ggx},
but experimental data for $D\to SP$ remain insufficient~\cite{Cheng:2002ai, Cheng:2010vk, Cheng:2022vbw, Hussain:1986xw, Katoch:1994ux, Fajfer:1995hp, Buccella:1996uy, El-Bennich:2008rkp, Boito:2008zk, Dedonder:2014xpa, Xie:2014tma, Dedonder:2021dmb}.
The decay $D^+ \to K_S^0a_0(980)^+$,  which could be accessed via $D^+\to K^0_S\pi^+\eta$,
provides a clean constraint on internal $W$-emission amplitudes involving $a_0(980)$~\cite{Cheng:2002ai, Cheng:2010vk, Cheng:2022vbw}.
Also, study of the $D^+ \to K_S^0a_0(980)^+$ decay helps to understand the nature of $a_0(980)$,
which is usually regarded as an exotic candidate (tetraquark, $K\bar K$ molecule, etc.),
and its production involves final-state interaction (FSI) effects such as
rescattering~\cite{Hsiao:2023qtk, Hsiao:2019ait, Yu:2021euw, Zhang:2022xpf, Achasov:2021dvt, Achasov:2017edm}.
Based on 1.1k signal events with a purity of 98.2\%, selected from 2.93 fb$^{-1}$ of data at 3.773 GeV,
Ref.~\cite{BESIII:2023htx} reported an amplitude analysis of $D^+ \to K^0_S\pi^+\eta$,
with fit projections on two-body mass distributions shown in Fig.~\ref{fig:Dp_KSpieta}.
Its amplitude is found to be dominated by $D^+\to K^0_Sa_0(980)^+$ [$(105.0\pm0.9\pm1.1)\%$] with a small fraction of
$\bar K^*_0(1430)^0\pi^+$ [$(10.8\pm1.5\pm1.3)\%$].
The branching fraction of $D^+ \to K_S^0\pi^+\eta$ is determined to be $(1.27\pm0.04\pm0.03)\%$,
which supersedes that reported in Ref.~\cite{BESIII:2020pxp}.
The obtained branching fractions of the intermediate processes are
$\mathcal{B}(D^+\to K_S^0a_0(980)^+, a_0(980)^+\to \pi^+\eta) = (1.33 \pm 0.05\pm 0.04)\%$
and
$\mathcal{B}(D^+\to \bar{K}_0^*(1430)^0\pi^+, \bar{K}_0^*(1430)^0\to K_S^0\eta) = (0.14 \pm 0.02\pm 0.02)\%$.
Theoretical studies insistently require experimental measurements of
$D^+ \to K_S^0a_0(980)^+$, as its observation can provide sensitive
constraints in the extraction of contributions from internal $W$-emission
diagrams of $D \to SP$~\cite{Cheng:2002ai, Cheng:2010vk, Cheng:2022vbw}. If
these measured branching fractions cannot be well described by the diagrammatic approach it
would indicate that significant FSIs must be
involved~\cite{Achasov:2021dvt}, which will provide important information on the
role of $a_0(980)$ in charmed meson decays and the nature of $a_0(980)$.
In addition, the obtained ${\cal B}(D^+\to \bar{K}_0^*(1430)^0\pi^+)=(3.26 \pm 0.47 \pm 0.47\,^{+1.02}_{-1.29\bar K_0^*})\%$
is consistent with that derived from $D^+\to K^-2\pi^+$ and
$D^+\to K_S^0\pi^+\pi^0$ by the PDG,
$(2.02 \pm 0.10\pm 0.22)\%$~\cite{ParticleDataGroup:2024cfk}.

\begin{figure*}[!htbp]
  \centering
  \includegraphics[width=0.225\textwidth]{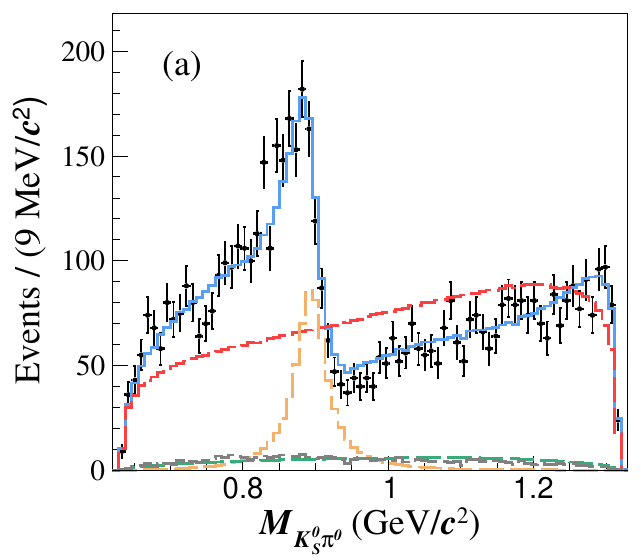}
  \includegraphics[width=0.225\textwidth]{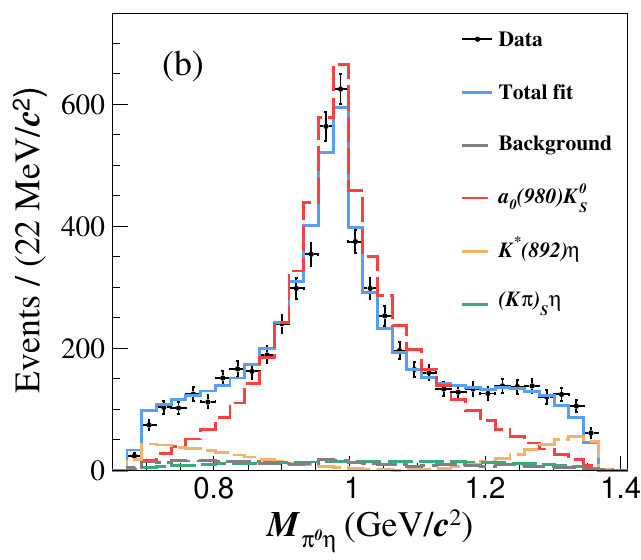}
  \includegraphics[width=0.225\textwidth]{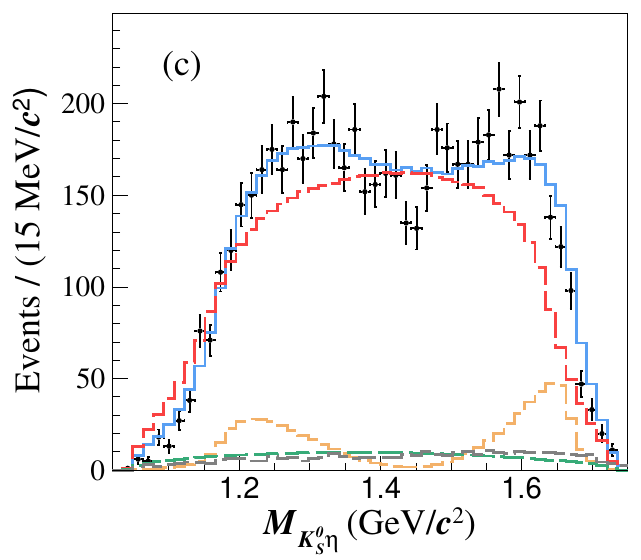}
  \caption{
  The  projections of the amplitude analysis fit of $D^0\to K^0_S\pi^0\eta$
  on (a) $M_{K_S^0\pi^0}$, (b)
    $M_{\pi^0\eta}$ and (c) $M_{K_S^0\eta}$~\cite{BESIII:2025wmd}.
  }
  \label{fig:D0_KSpi0eta}
\end{figure*}

In 2020, Belle measured the branching fraction of the CF decay $D^0\to \bar{K}^*(892)^{0}\eta$ ($\bar{K}^*(892)^{0}\to K^-\pi^+$) to be $(1.41^{+0.13}_{-0.12})\%$~\cite{Belle:2020fbd}, raising the world average by 35\%.
This larger branching fraction roughly doubles the estimated magnitude of the $W$-exchange amplitude $E_V$ and substantially alters predicted $CP$ asymmetries~\cite{Cheng:2021yrn,Cheng:2019ggx}.
To understand the origin of this issue,
an independent measurement of $D^0\to \bar K\pi\eta$ is highly desired.
In Ref.~\cite{BESIII:2025wmd}, using 6.1k candidates with a signal purity of 94.6\%,
an amplitude analysis was performed on the $D^0 \to K_S^0 \pi^0 \eta$ decay.
The amplitude analysis fit projections on two-body particle mass distributions are shown in Fig.~\ref{fig:D0_KSpi0eta}.
The most important component is $D^0\to K^0_Sa_0(980)^0$ with a fraction of $(93.2\pm3.3\pm3.2)\%$;
the secondary component is $D^0\to\bar K^*(892)^0\eta$ with a fraction of $(11.5\pm0.8\pm1.0)\%$,
while other components $D^0\to K^0_Sa_0(1320)^0$, $D^0\to\bar K^*(1410)^0\eta$,
$D^0\to (K^0_S\pi^0)_{{\cal S}{\rm -wave}}\eta$,
$D^0\to\bar K^*(1680)^0\eta$, and $D^0\to\bar K^*_2(1980)^0\eta$ are also observed with
significances greater than $5\sigma$ with fractions ranging between $(0.8-6.0)\%$.
The  branching fractions of $D^0 \to K_S^0 \pi^0 \eta$ as well as its two main subdecays $D^0\to K^0_Sa_0(980)^0$,
and $D^0\to\bar K^*(892)^0\eta$ are obtained to be
$\mathcal{B}(D^0 \to K_S^0 \pi^0 \eta)=(1.016 \pm 0.013\pm 0.014)\%$,
 $\mathcal{B}(D^0\to K_S^0a_0(980)^0, a_0(980)^0\to \pi^0\eta) = (9.88\pm 0.37\pm 0.42)\times10^{-3}$, and
$\mathcal{B}(D^0\to\bar K^*(892)^0\eta)=(0.73 \pm 0.05 \pm 0.03) \%$.
The ${\cal B}(D^0 \to\bar{K}^*(892)^0 \eta)$ measured by BESIII is $5 \sigma$ lower than the measurement by Belle~\cite{Belle:2020fbd},
$(1.41_{-0.12}^{+0.13} )\%$. It is consistent with
various theoretical predictions of $(0.51-0.93)\%$~\cite{Qin:2013tje,Fu-Sheng:2011fji, Cheng:2010ry,Cheng:2019ggx}.
As a result, the magnitudes of the $W$-exchange and QCD-penguin exchange amplitudes are found to be less than half of their current estimations~\cite{Cheng:2021yrn}.
This revision, as the major source, is expected to significantly impact
the predicted SM $CP$ violation for $D\to VP$ modes~\cite{Cheng:2021yrn}.
Moreover, the $\mathcal{B}(D^0 \to K_S^0 a_0(980)^0$, $a_0(980)^0 \to \pi^0 \eta)$
is measured with precision improved by a factor of 4.5 compared to the world average.
It is consistent with the prediction in Ref.~\cite{Cheng:2024zul} assuming a tetraquark scenario,
$(9.5-15)\times 10^{-3}$, but disfavors that in the diquark scenario, $6.5\times10^{-4}$.

\subsubsection{Analyses of $D^+_s\to \pi\pi\pi$}

\begin{figure*}
\centering
    \includegraphics[width=0.225\linewidth]{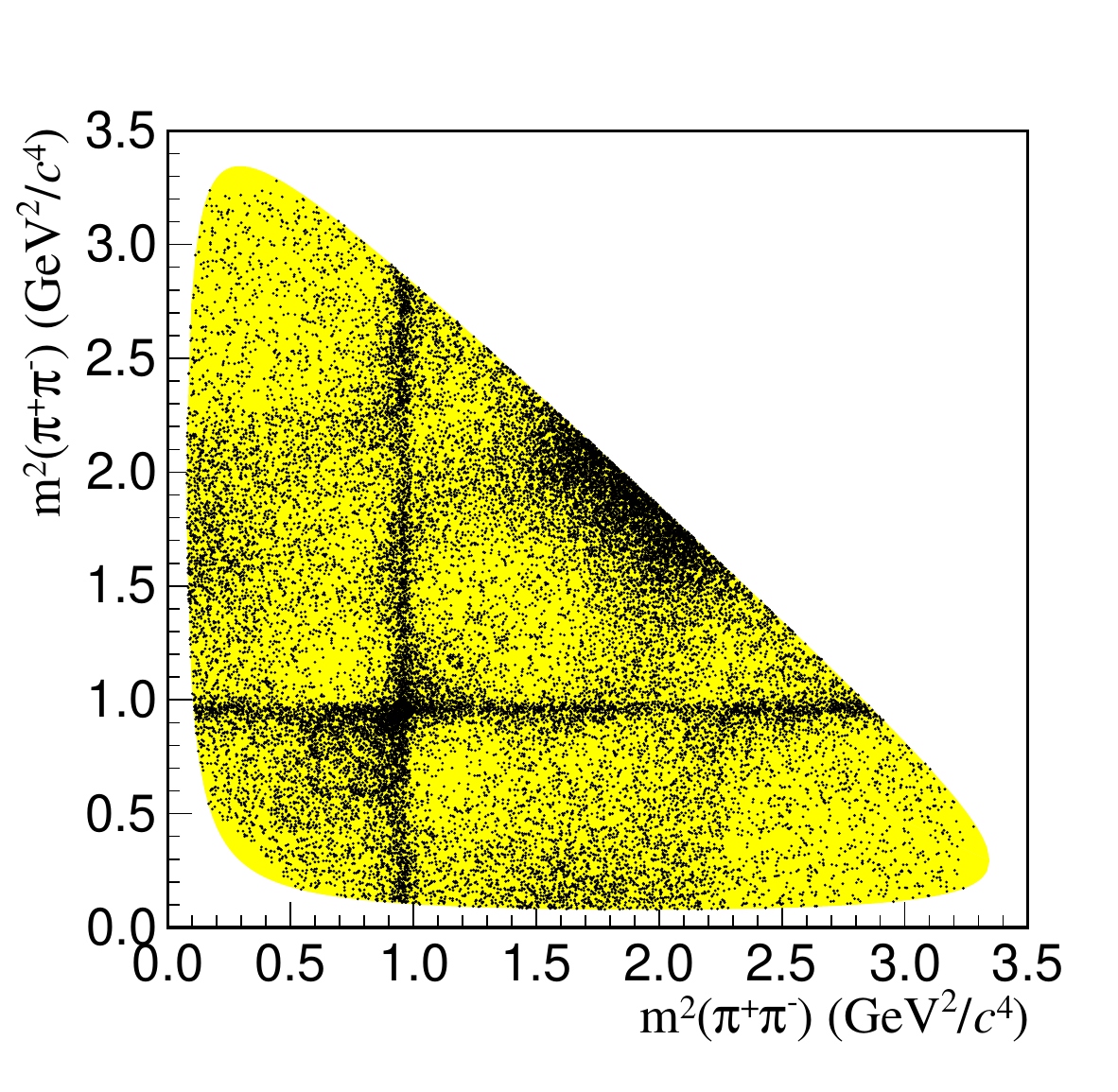}
    \includegraphics[width=0.225\linewidth,height=0.195\textwidth]{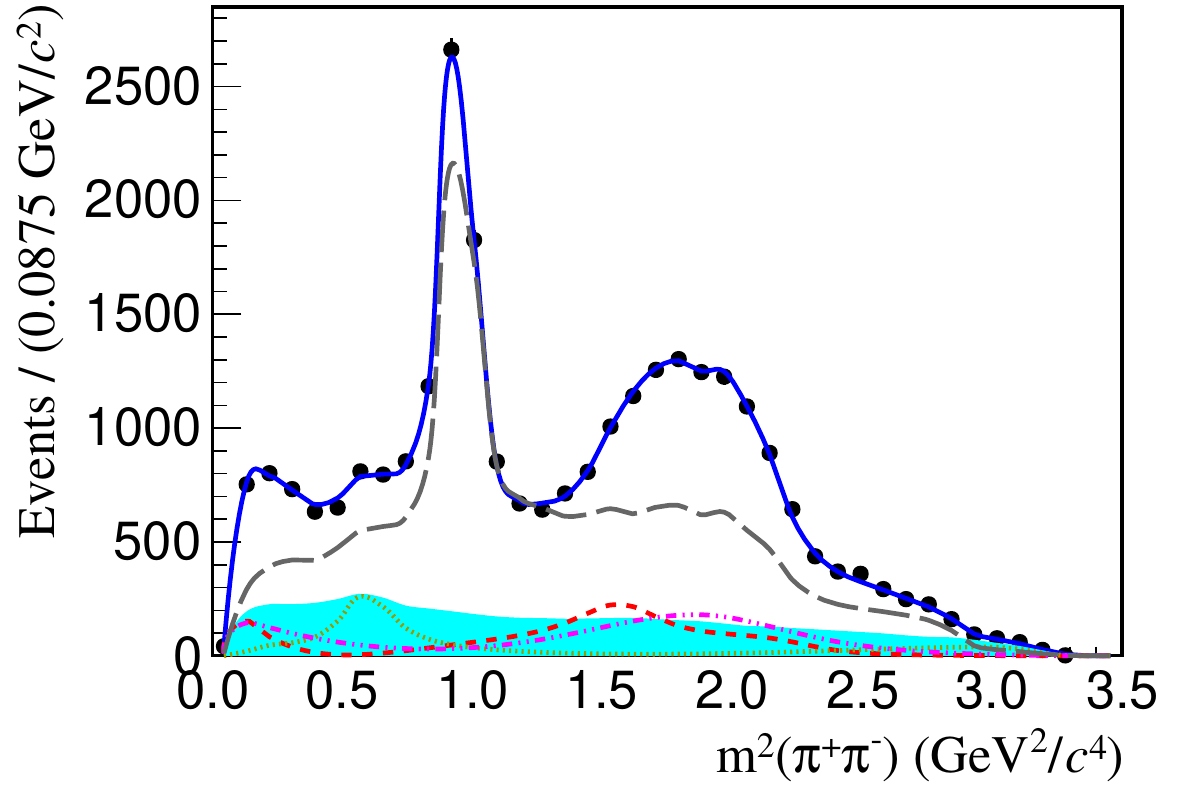}
    \includegraphics[width=0.225\linewidth,height=0.195\textwidth]{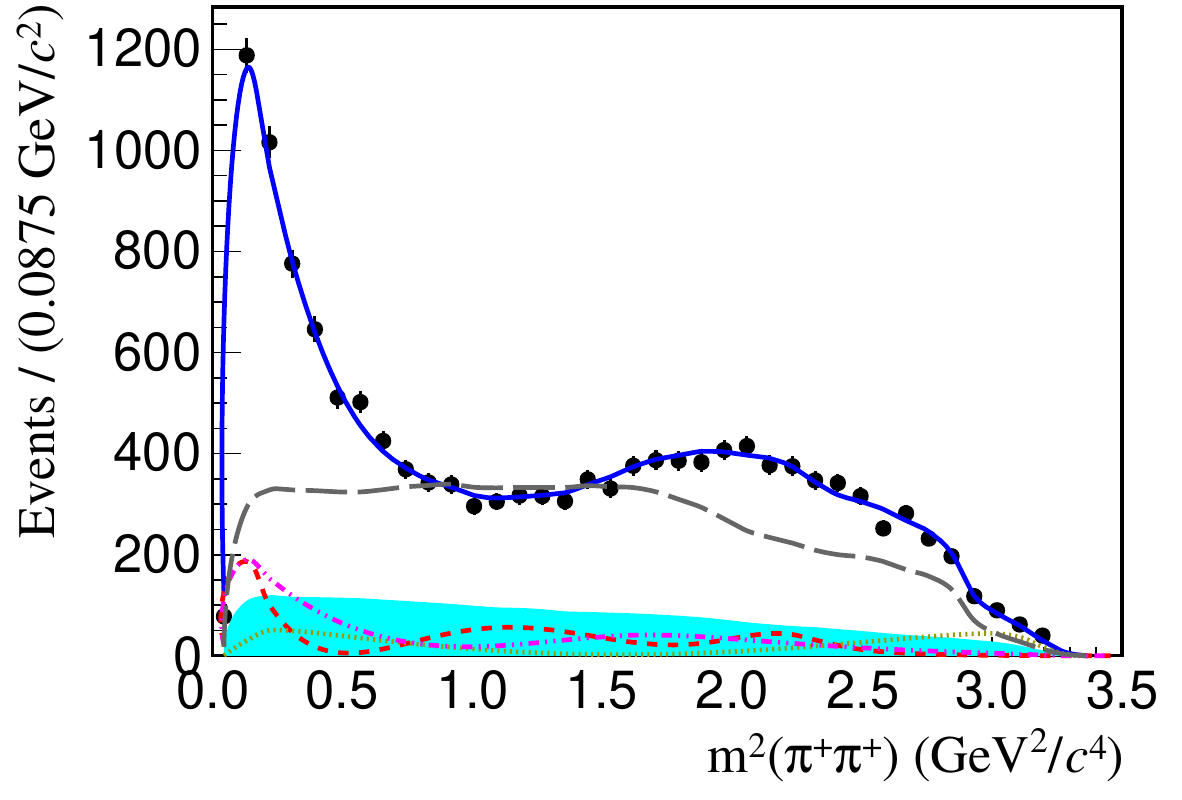}
    \caption{
(Left) The Dalitz plot  as well as
the projections of the amplitude analysis fit for $D_{s}^{+}\to 2\pi^{+}\pi^{-}$ on
 (middle) $m^2(\pi^+\pi^-)$ and (right) $m^2(\pi^+\pi^+)$~\cite{BESIII:2021jnf}.
      The shaded areas in
      cyan are the background contributions. Also shown in these figures
       are contributions from the $\cal S$-wave (gray long-dashed lines),
      $\rho(770)$ (yellow dotted lines, scaled by a factor of 10 for
      better visibility), $\rho(1450)$ (magenta dot-dashed line,
      scaled by a factor of 10 for better visibility), and $f_2(1270)$
      (red short-dashed lines).
      }
\label{fig:Ds_3pi}
\end{figure*}

The $D^+_s\to 2\pi^+\pi^-$ decay is dominated by $\cal S$-wave and has a relatively large branching fraction~\cite{FOCUS:2003tdy,BaBar:2008nlp,CLEO:2013bae},
enabling studies of the $\pi\pi$ $\cal S$-wave below 2~GeV as well as the light scalar mesons $f_0(980)$ and $f_0(1370)$.
The $f_0(980)$, produced via $s\bar{s}$ hadronization near $K\bar K$
threshold, can be probed through its couplings to $\pi\pi$ and $K\bar K$ in $D^+_s\to 2\pi^+\pi^-$ and $D^+_s\to K^+K^-\pi^+$~\cite{Dias:2016gou}.
Amplitude analysis also provides the branching fraction ${\cal B}(D^+_s\to\rho(770)^0\pi^+)$, which is unique as it involves the difference
(not sum) of $W$-annihilation amplitudes for $P$ and $V$ production~\cite{Cheng:2016ejf}.
This quantity is essential for determining the magnitudes and strong phases of $A_{P,V}$,
 and serves as a crucial input in global $D\to VP$ analyses~\cite{Cheng:2016ejf}.
Previous amplitude analyses of $D^+_s\to 2\pi^+\pi^-$  came from
E687~\cite{E687:1997jvh}, E791~\cite{E791:2000lzz}, FOCUS~\cite{FOCUS:2003tdy}, and
BaBar~\cite{BaBar:2008nlp}. BaBar also reported the first
quasi-model-independent partial wave analysis (QMIPWA) to model
the $\cal S$-wave amplitude in this channel using
a relatively large data sample of 13.2k $D^+_s$ candidates with a signal purity of 80\%.
Utilizing $3.19$ fb$^{-1}$ of data at 4.178 GeV, BESIII reported an amplitude analysis
of $D^+_s\to 2\pi^+\pi^-$ with 13.8k candidates with a signal purity of $\sim$80\%~\cite{BESIII:2021jnf}.
Figure~\ref{fig:Ds_3pi} shows projections of the amplitude analysis fit on $M^2(\pi^+\pi^-)$ and $M^2(\pi^+\pi^+)$.
The amplitude analysis shows that this decay is dominated by $\pi\pi$ $\cal S$-wave
with a fraction of $(84.2\pm0.8\pm1.2)\%$, accompanied with three two-body decays
$D^+_s\to f_2(1270)\pi^+$, $D^+_s\to \rho(770)^0\pi^+$, and $D^+_s\to \rho(1450)^0\pi^+$
with fractions of $(10.5\pm0.8\pm1.1)\%$, $(0.9\pm0.4\pm0.5)\%$, and $(1.3\pm0.4\pm0.5)\%$, respectively.
The obtained ${\cal F}(D^+_s \to \rho(770)^0\pi^+)
= (0.9\pm 0.4\pm 0.5)\%$ is compatible with the BaBar result.
Combining with the world average value of ${\cal B}(D^+_s\to 2\pi^+\pi^-)$
leads to ${\cal B}(D^+_s \to\rho(770)^0\pi^+) = {\cal B}(D^+_s\to 2\pi^+\pi^-)\times {\cal F}(D^+_s \to\rho(770)^0\pi^+)  = (0.009 \pm 0.007)$\% that agrees with the predictions
in Ref.~\cite{Qin:2013tje}.
In 2023, LHCb performed an amplitude analysis of $D^+_s\to 2\pi^+\pi^-$, based on
a sample containing over $7\times 10^5$ signal candidates with a signal purity of 95\%~\cite{LHCb:2022pjv},
in which additional components of
$D^+_s\to \omega\pi^+$,
$D^+_s\to \rho(1700)^0\pi^+$, and
$D^+_s\to f_2^\prime(1525)\pi^+$ are also observed.

\begin{figure}[!htbp]
 \centering
 \includegraphics[width=0.225\textwidth]{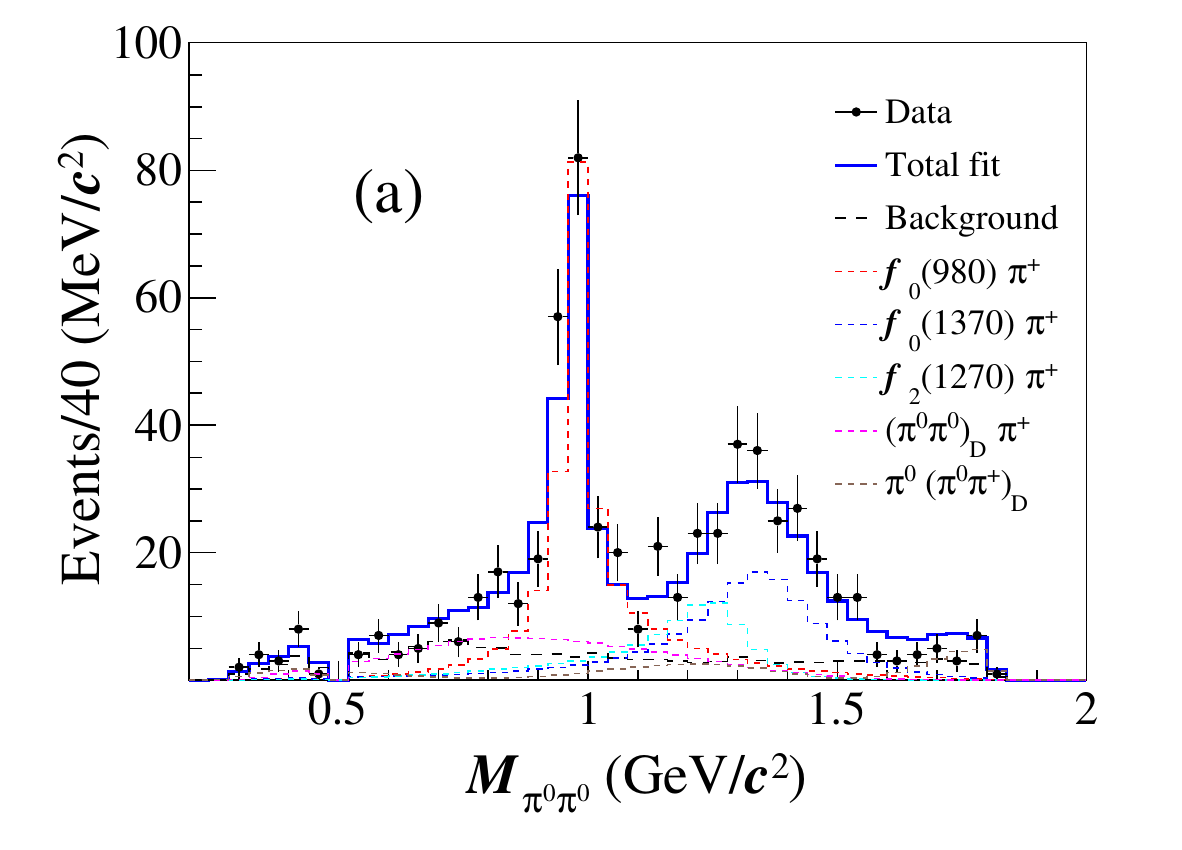}
 \includegraphics[width=0.225\textwidth]{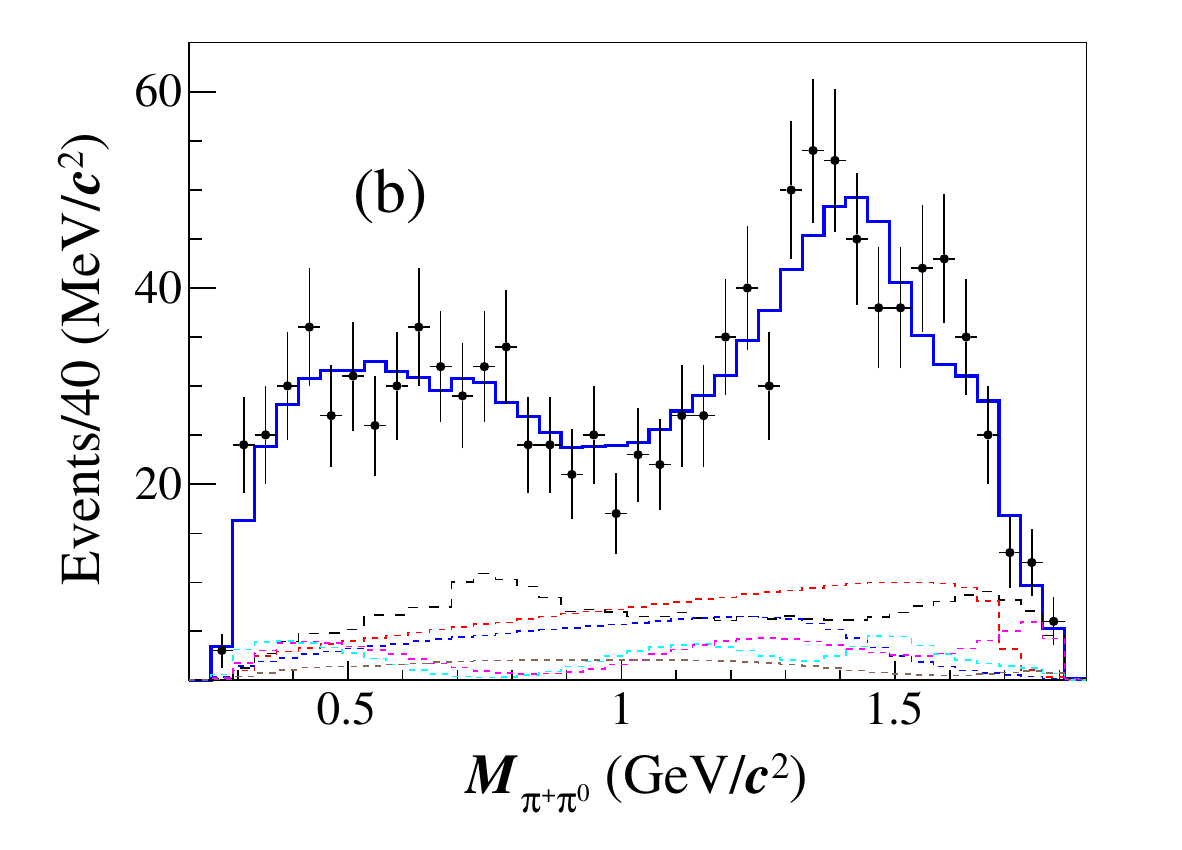}
 \caption{
   The projections of the amplitude analysis fit for $D^+_s\to \pi^+2\pi^0$ on (a) $M_{\pi^0\pi^0}$ and (b) $M_{\pi^+\pi^0}$~\cite{BESIII:2021eru}.
}
 \label{fig:Ds_pi2pi0}
\end{figure}

The $f_0(980)$ meson, a possible tetraquark candidate~\cite{Weinstein:1983gd,Weinstein:1990gu,Cheng:2017fkw}, can be studied via the hadronic decays $D_{s}^{+}\to\pi^{+}2\pi^{0}$, $D_{s}^{+} \to 2\pi^{+}\pi^{-}$ and $D_{s}^{+} \to K^{+}K^{-}\pi^{+}$.
The published branching fraction of $D^+_s\to f_{0(2)} \pi^+$ from $D_{s}^{+} \to 2\pi^{+}\pi^{-}$ shows large discrepancies~\cite{ParticleDataGroup:2024cfk,E687:1997jvh, E791:2000lzz} with that from $D_{s}^{+} \to K^{+}K^{-}\pi^{+}$. The $f_{0(2)}$ contributions may be contaminated by $a_{0}(980)\to K^+K^-$ or $\rho(770)^0 \to \pi^+ \pi^-$, while $D_{s}^{+} \to \pi^{+}2\pi^{0}$ offers a cleaner environment free of these backgrounds. 
Previously, the CLEO-c experiment only reported its branching fraction,
which is $\mathcal{B}(D_{s}^{+} \to \pi^{+}2\pi^{0}) = (0.65\pm 0.13)\%$~\cite{CLEO:2009vke},
by analyzing 600~pb$^{-1}$ of data taken around 4.17~GeV.
Based on the study of 572 candidate events with a signal purity of 77.5\%,
Ref.~\cite{BESIII:2021eru} reported an amplitude analysis of $D_{s}^{+} \to \pi^+2\pi^0$.
The amplitude analysis fit projections on two-body particle mass distributions are shown in Fig.~\ref{fig:Ds_pi2pi0}.
Except for the known decay modes of
$D^+_s\to f_0(980)\pi^+$, $D^+_s\to f_0(1370)\pi^+$, $D^+_s\to f_2(1270)\pi^+$,
non-resonant processes $D^+_s\to \pi^+(2\pi^0)_{\cal D}$, and $D^+_s\to (\pi^+\pi^0)_{\cal D}\pi^0$ are observed for the first time
in this channel. The fit fractions of each sub-process are
$(55.4\pm6.8\pm7.3)\%$,
$(25.5\pm5.1\pm9.3)\%$,
$( 9.7\pm2.9\pm6.0)\%$,
$(21.8\pm6.8\pm3.6)\%$, and
$( 5.7\pm2.6\pm2.0)\%$, respectively.
In addition, the $D^+_s\to \pi^+2\pi^0$ decay branching fraction is measured to be
$(0.50\pm 0.04\pm 0.02)\%$, with precision improved by about a factor of 2 compared to the CLEO-c value~\cite{CLEO:2009vke}.
The branching fractions for the intermediate processes are also presented,
in which that of $D^+_s\to f_0(980)\pi^+$ with $f_0(980)\to \pi^0\pi^0$ is measured for the first time.
In addition, no significant signal of $f_0(500)$ is observed. Assuming the branching fraction ratio between
$f_{0(2)}\to\pi^+\pi^-$ and $f_{0(2)}\to\pi^0\pi^0$ to be 2 based on isospin symmetry,
the results favor those from $D^+_s\to 2\pi^+\pi^-$ over those from $D_{s}^{+} \to K^{+}K^{-}\pi^{+}$.

\begin{figure*}[htp]
\centering
\begin{minipage}[t]{0.225\textwidth}
  \centering
  \includegraphics[width=\textwidth]{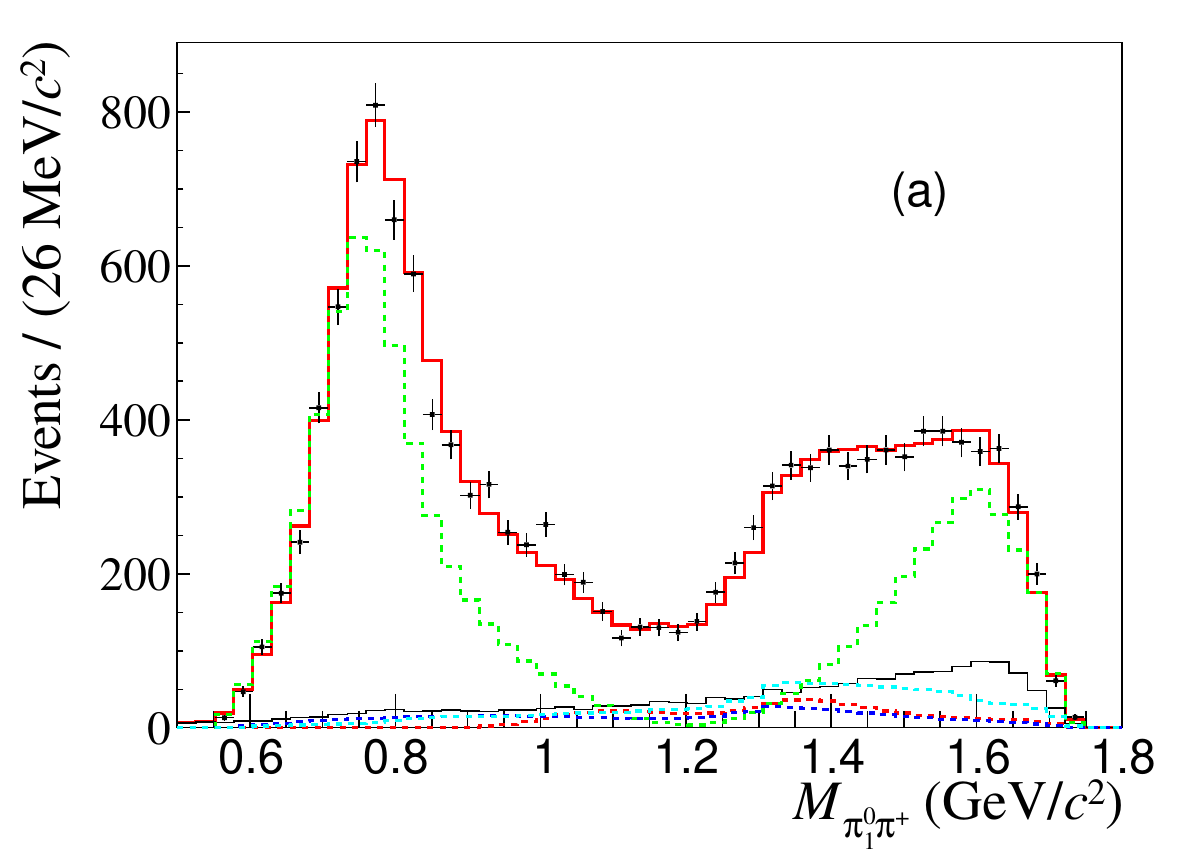}
\end{minipage}
\begin{minipage}[t]{0.225\textwidth}
  \centering
  \includegraphics[width=\textwidth]{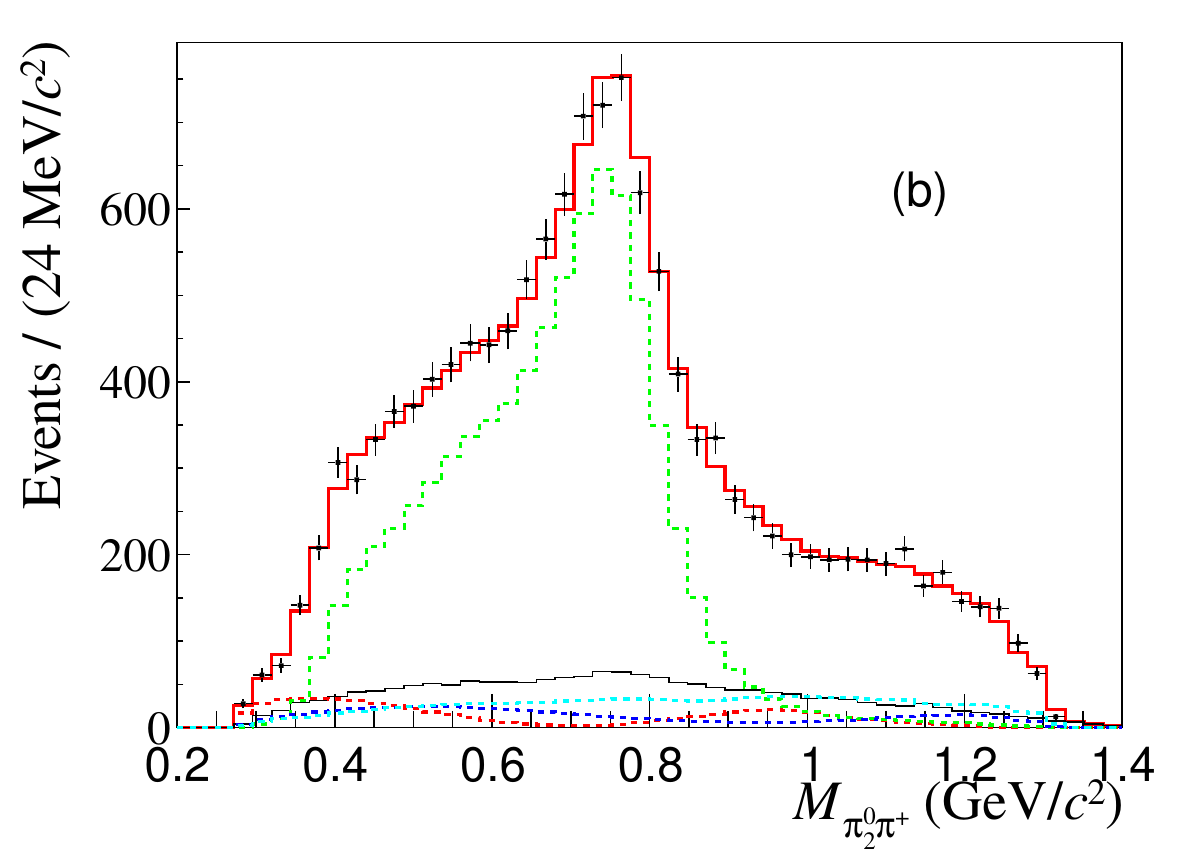}
\end{minipage}
\begin{minipage}[t]{0.225\textwidth}
  \centering
  \includegraphics[width=\textwidth]{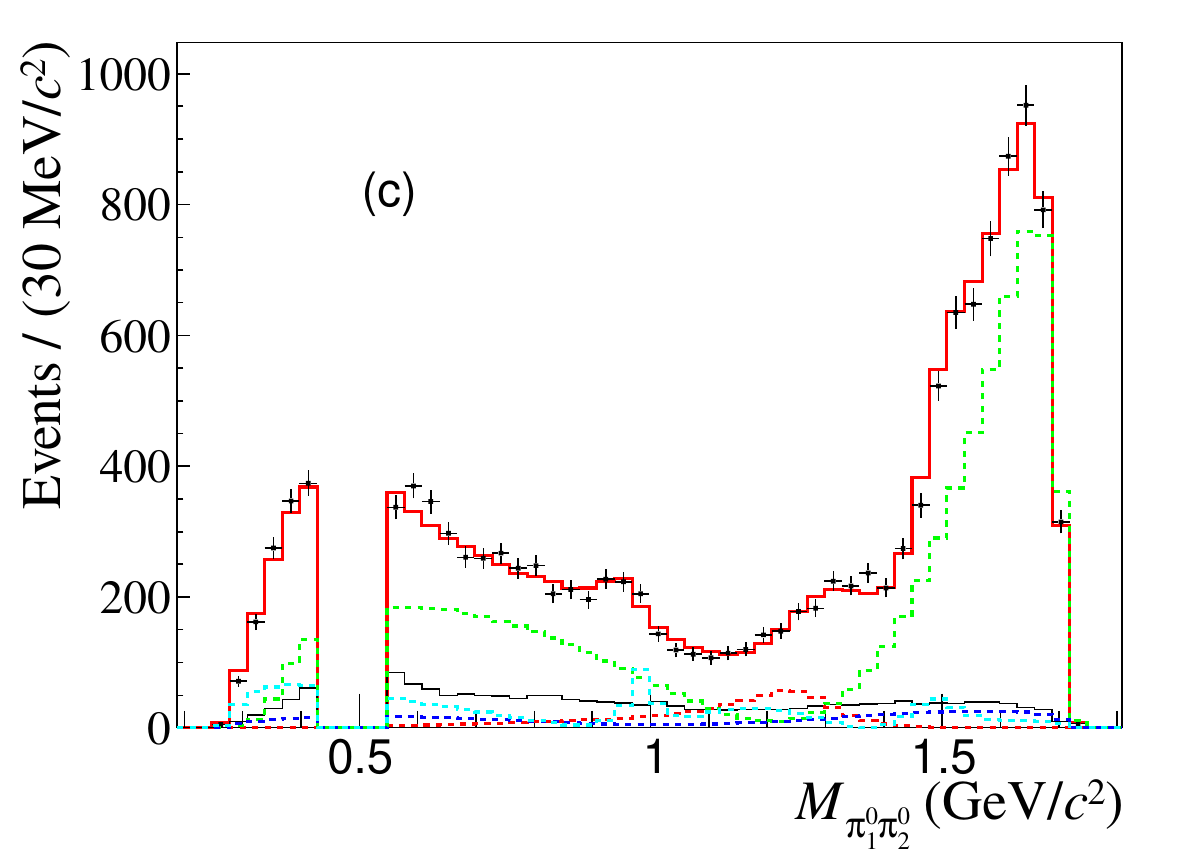}
\end{minipage}
\begin{minipage}[t]{0.225\textwidth}
  \centering
  \includegraphics[width=\textwidth]{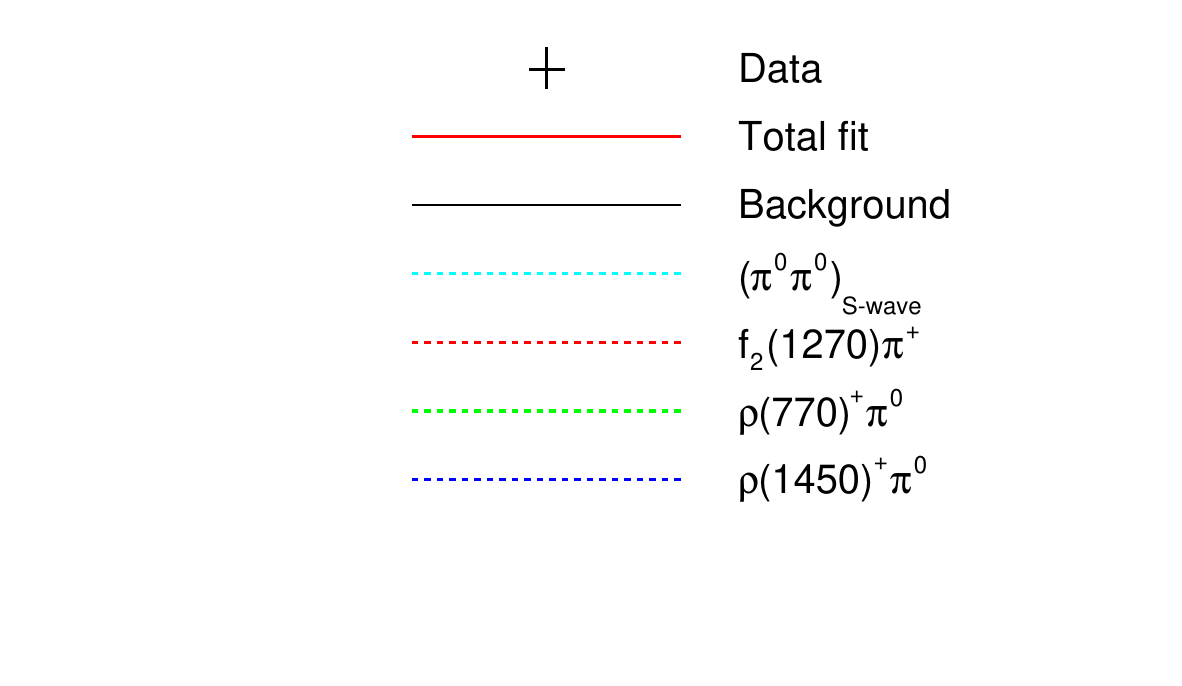}
\end{minipage}
\caption{The  projections of the amplitude analysis fit of $D^+\to \pi^+2\pi^0$ on (a) $M_{\pi^+ \pi_1^0}$, (b) $M_{\pi^+ \pi_2^0}$ and (c) $M_{\pi^0 \pi^0}$~\cite{BESIII:2025les}.
}
\label{fig:Dp_pipi0pi0}
\end{figure*}

The $D^+ \to \pi^+2\pi^0$ decay is expected to be dominated by the two-body channel $D^+\to\rho(770)^+(\to\pi^+\pi^0)\pi^0$,
featuring a color-allowed tree topology with possible contributions from color-suppressed internal emission, $W$-annihilation, and penguin diagrams~\cite{Cheng:2012wr}.
Theoretical predictions for the $D^+\to\rho(770)^+\pi^0$ decay branching fraction were
made via pole model, topological diagrammatic approaches, and factorization-assisted topological-amplitude
method~\cite{Qin:2013tje,Cheng:2015iom}. Precise measurements thus provide stringent constraints on amplitude compositions~\cite{Fu-Sheng:2011fji}.
In addition,
$CP$ violation in charm decays arises from interference between tree-level and higher-order processes in SCS
$c \to u \bar{d} d$ transitions. The SM predictions for such asymmetries lie at $10^{-4}$--$10^{-3}$~\cite{Grossman:2006jg}.
While LHCb observed direct $CP$ violation in $D^0$ decays at $\sim 0.15\%$~\cite{LHCb:2019hro},
no signal has yet been seen in $D^+$ decays. Predictions for $D\to \rho\pi$ give asymmetries of $(0.3\sim 5.0)\times10^{-4}$~\cite{Li:2012cfa,Cheng:2019ggx},
motivating searches in $D^+\to\pi^+2\pi^0$.
BESIII previously reported a hint of $\sim 2\sigma$ $CP$ asymmetry in $D^\pm\to\pi^\pm2\pi^0$ decays using 2.93~fb$^{-1}$ of data~\cite{BESIII:2022qrs}.
Amplitude analysis across the Dalitz plot may reveal larger localized $CP$ effects~\cite{Cheng:2021yrn},
making precision studies with larger samples crucial for probing new physics.
Reference~\cite{BESIII:2025les} conducted
the first amplitude analysis of $D^+ \to \pi^+2\pi^0$ based on 13.5k candidates with a purity of 87.4\%.
The amplitude analysis fit projections on two-body particle mass distributions are shown in Fig.~\ref{fig:Dp_pipi0pi0}.
 The amplitude analysis finds
four subprocesses $D^+\to \rho(770)^+\pi^0$, $D^+\to \rho(1450)^+\pi^0$, $D^+\to f_2(1270)\pi^+$,
and $D^+\to (2\pi^0)_{{\rm non-}f}$, corresponding to fractions of
$(63.5\pm2.0\pm0.9)\%$,
$( 5.2\pm0.8\pm0.7)\%$,
$( 4.5\pm0.3\pm0.2)\%$, and
$(11.6\pm0.9\pm0.5)\%$, respectively.
The branching fractions of this decay and its dominant component are determined to be
$\mathcal{B}(D^+ \to \pi^+2\pi^0)=(4.84\pm0.05\pm0.05)\times 10^{-3}$ and $\displaystyle\mathcal{B}(D^+ \to \rho(770)^+\pi^0)=(3.08\pm0.10\pm0.05)\times10^{-3}$.
The predictions in Ref.~\cite{Fu-Sheng:2011fji} are consistent with our results but have large uncertainties, whereas those in Ref.~\cite{Qin:2013tje} are given without uncertainties. In contrast, the predictions of Ref.~\cite{Cheng:2015iom} have smaller uncertainties and deviate from our measurement by about $2.3\sigma$.
The asymmetry of the branching fractions of $D^\pm \to \pi^\pm2\pi^0$ is determined to be
${\cal A}_{CP} = (-1.4\pm1.0\pm0.6)\%$, and no evidence for $CP$ violation is observed.

\subsubsection{Analyses of $D^+_{(s)}\to \pi\pi\eta^{(\prime)}$}

\begin{figure*}[htp]
\begin{center}
\begin{minipage}[b]{0.225\textwidth}
\epsfig{width=0.98\textwidth,clip=true,file=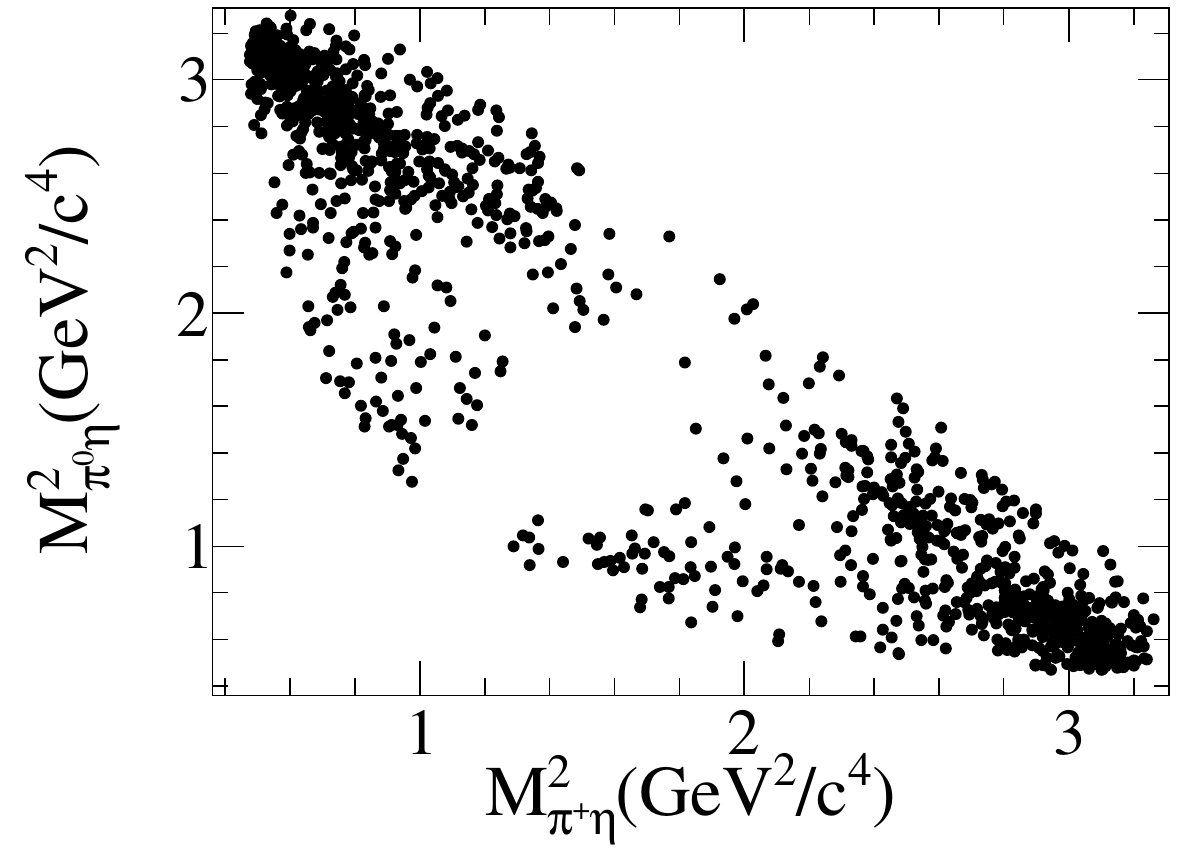}
\put(-35,65){(a)}
\end{minipage}
\begin{minipage}[b]{0.225\textwidth}
\epsfig{width=0.98\textwidth,clip=true,file=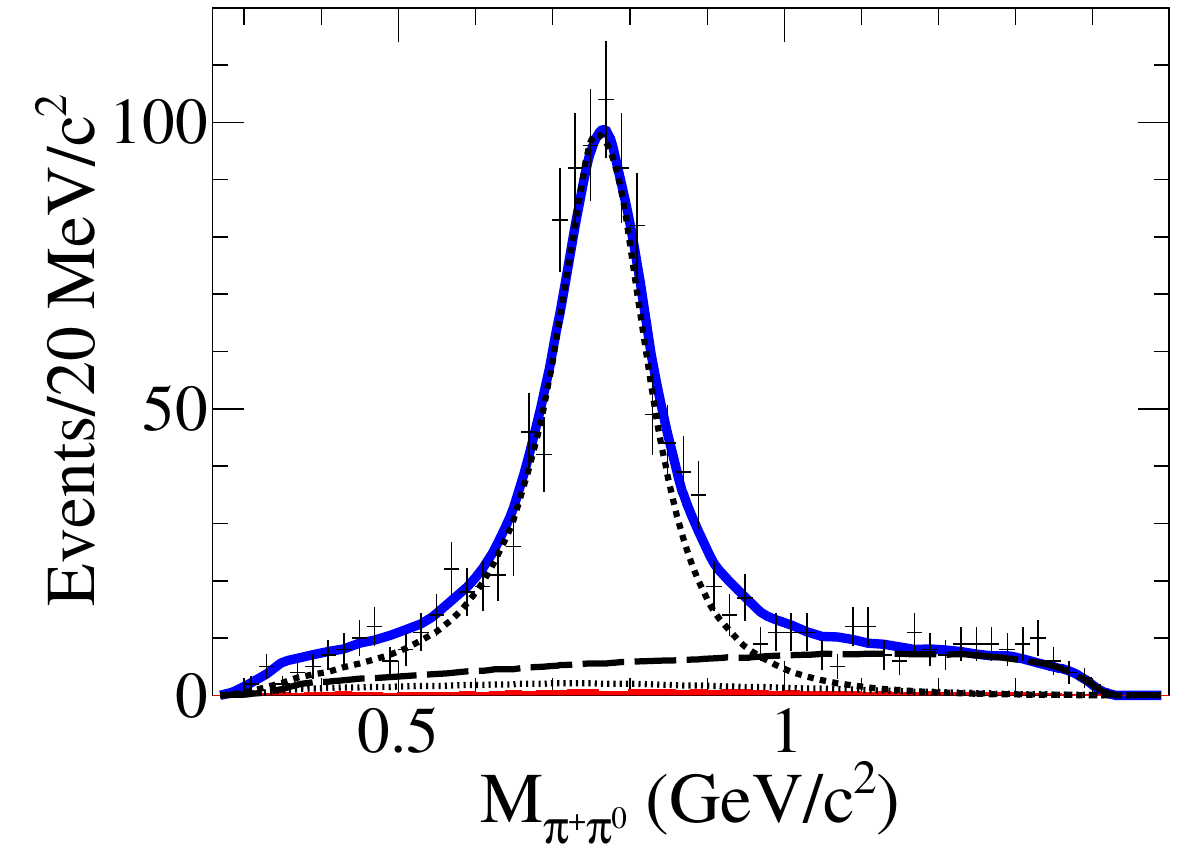}
\put(-35,65){(b)}
\end{minipage}
\begin{minipage}[b]{0.225\textwidth}
\epsfig{width=0.98\textwidth,clip=true,file=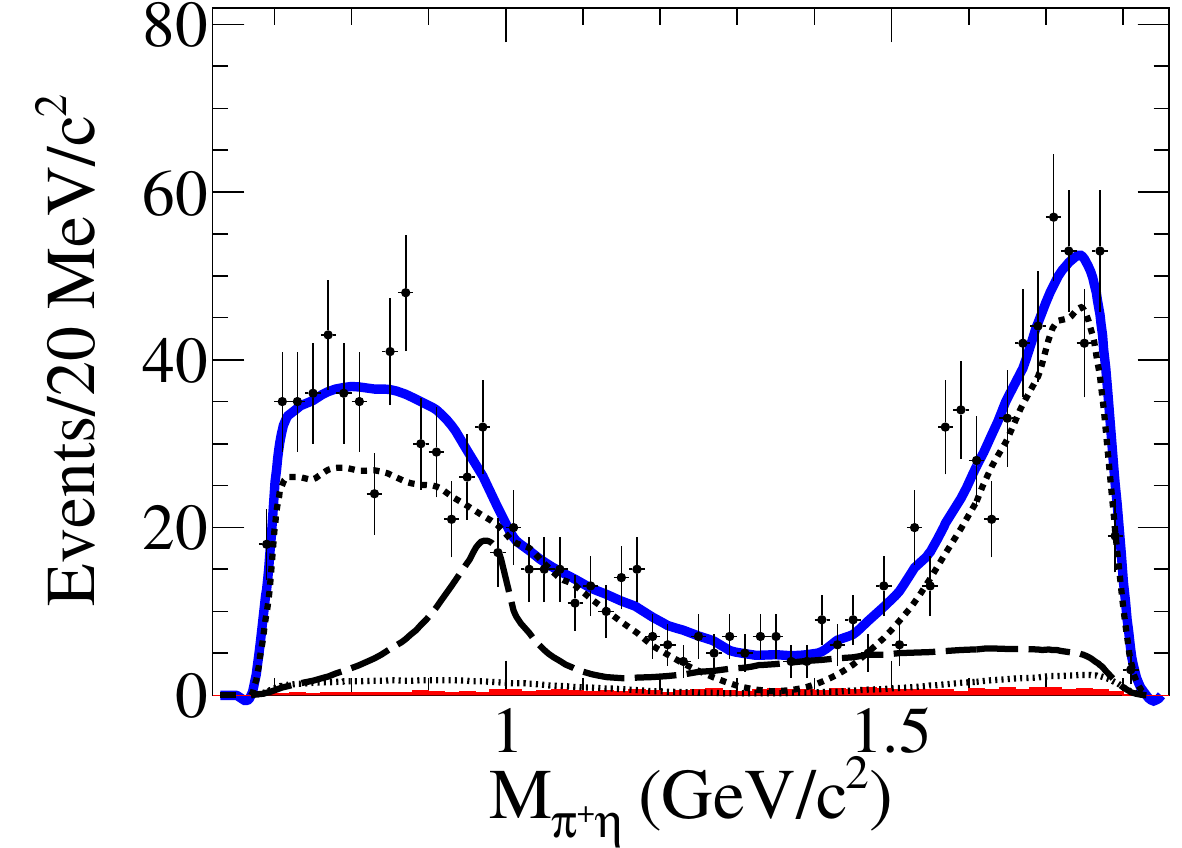}
\put(-35,65){(c)}
\end{minipage}
\begin{minipage}[b]{0.225\textwidth}
\epsfig{width=0.98\textwidth,clip=true,file=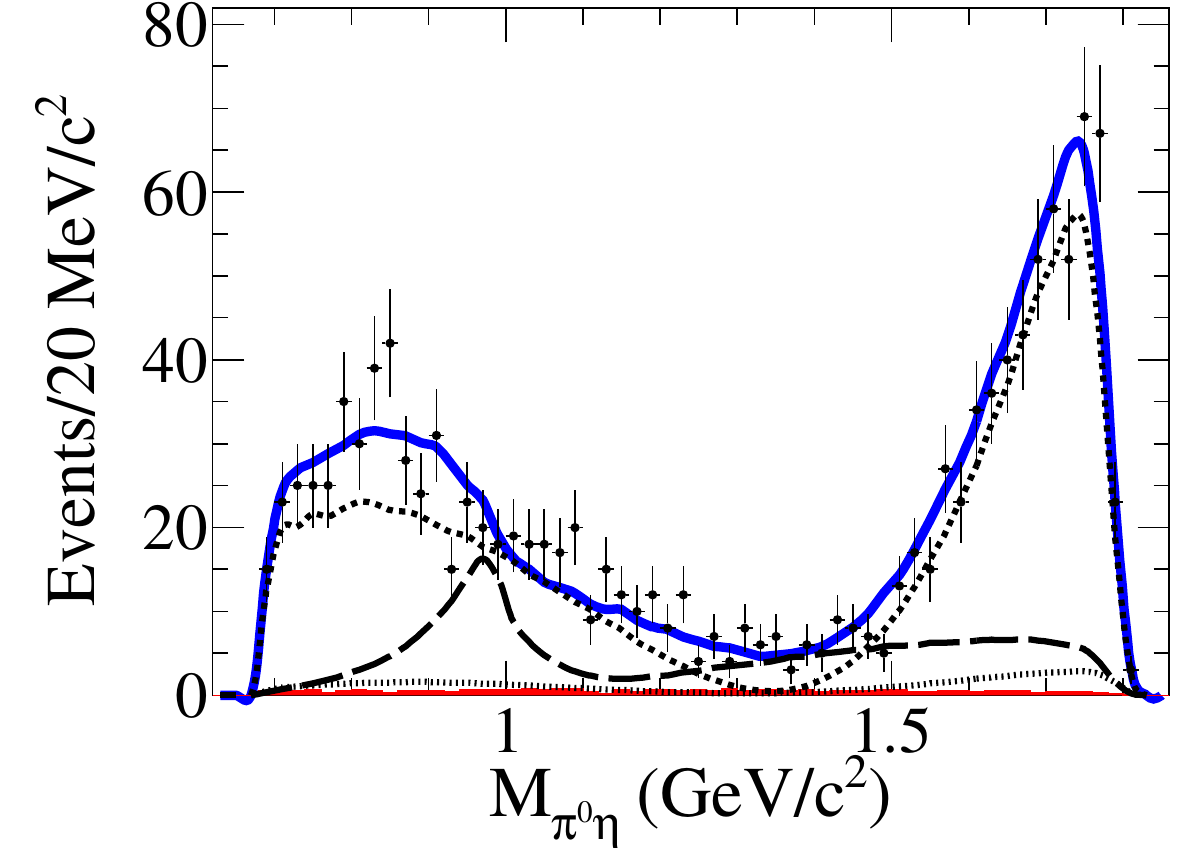}
\put(-35,65){(d)}
\end{minipage}
\caption{(a) The Dalitz plot  as well as
the projections of the amplitude analysis fit for $D_{s}^{+}\to \pi^{+}\pi^{0}\eta$ on (b) $M_{\pi^{-}\pi^{0}}$,
(c) $M_{\pi^{+}\eta}$ and (d) $M_{\pi^{0}\eta}$~\cite{BESIII:2019jjr}.
In (b-d), the dots with error bars and
the solid line are data and the total fit, respectively;
the dashed, dotted, and long-dashed  lines are the contributions from $D_{s}^{+} \to \rho(770)^{+}\eta$,
$D_{s}^{+} \to (\pi^{+}\pi^{0})_{V}\eta$, and
$D_{s}^{+} \to a_{0}(980)\pi$, respectively.  The (red) hatched histograms are the simulated background.}
\label{fig:Ds_etapipi0}
\end{center}
\end{figure*}

In 2019, the amplitude analysis of $D_{s}^{+}\to \pi^{+}\pi^{0}\eta$ was performed for the first time, with
1.2k candidate events with a signal purity of 97.7\%~\cite{BESIII:2019jjr}.
The Dalitz plot of $M^{2}_{\pi^{+}\eta}$ versus $M^{2}_{\pi^{0}\eta}$ for data is shown in Fig.~\ref{fig:Ds_etapipi0}(a).
The projections of the amplitude fit on $M_{\pi^{-}\pi^{0}}$, $M_{\pi^{+}\eta}$ and  $M_{\pi^{0}\eta}$
are shown in Figs.~\ref{fig:Ds_etapipi0}(b-d).
The corresponding projections on $M_{\pi^{+}\eta}$ and  $M_{\pi^{0}\eta}$ for events with $M_{\pi^{+}\pi^{0}}>1.0$ GeV$/c^{2}$
are shown in Figs.~\ref{fig:Ds_etapipi0}(e,f), in which $a_{0}(980)$ peaks are observed.
The analysis model includes the components of
$D_{s}^{+}\to \rho(770)^+\eta$,
$D_{s}^{+}\to (\pi^+\pi^0)_{{\rm non-}\rho}\eta$, and
$D_{s}^{+} \to a_{0}(980)^{+(0)}\pi^{0^(+)}, a_{0}(980)^{+(0)} \to \pi^{+(0)}\eta$,
with fractions of $(78.3\pm5.0\pm2.1)\%$, $(5.4\pm2.1\pm2.5)\%$, and $(23.2\pm2.3\pm3.3)\%$, respectively.
We obtain $\mathcal{B}(D_{s}^{+} \to \pi^{+}\pi^{0}\eta)$ to be
$(9.50\pm0.28\pm0.41)\%$.
Combining with the fit fractions of each intermediate process
leads to their branching fractions; those of
$D_{s}^{+} \to \rho(770)^{+}\eta$ and $D_{s}^{+} \to (\pi^{+}\pi^{0})_{V}\eta$
are calculated to be $(7.44\pm0.52\pm0.38)$\% and
$(0.51\pm0.20\pm0.25)$\%, respectively.
The obtained branching fraction of $D_{s}^{+} \to a_{0}(980)^{+(0)}\pi^{0(+)}$,
which is a pure $W$-annihilation decay,
is larger than other measured branching fractions of pure WA decays
$D_{s}^{+} \to \omega \pi^{+}$ and $D_{s}^{+} \to \rho(770)^{0} \pi^{+}$ by at least one order of magnitude.
Furthermore, when the measured $a_{0}(980)^{0}$-$f_{0}(980)$ mixing rate~\cite{BESIII:2018ozj} is considered,
the expected effect of $a_{0}(980)^{0}$-$f_{0}(980)$ mixing is lower than the WA contribution in $D_{s}^{+} \to a_{0}(980)^{0}\pi^{+}$ decay
by two orders of magnitude, which is negligible.
With the measured $\mathcal{B}(D_{s}^{+} \to a_{0}(980)^{+(0)}\pi^{0(+)})$,
the WA contribution with respect to the tree-external-emission contribution in $SP$ mode is estimated to be $0.84\pm0.23$,
which is significantly greater than that (0.1$\sim$0.2) in $VP$ and $PP$ modes~\cite{Li:2012cfa,Cheng:2016ejf}.
This measurement sheds light on FSI effects and non-perturbative strong interaction~\cite{Cheng:2010ry,Cheng:2016ejf}, and
provides a theoretical challenge to understand such a large WA contribution.
In addition, the result of this analysis is an essential input to determine the effect from $a_{0}(980)^{0}$ on the $K^{+}K^{-}$
$\cal S$-wave contribution to the model-dependent amplitude analysis of $D_{s}^{+} \to K^{+}K^{-}\pi^{+}$~\cite{CLEO:2009nuz,BaBar:2010wqe}.

\begin{figure*}[htbp]
	\centering
	\subfigure{
		\includegraphics[width=0.225\textwidth]{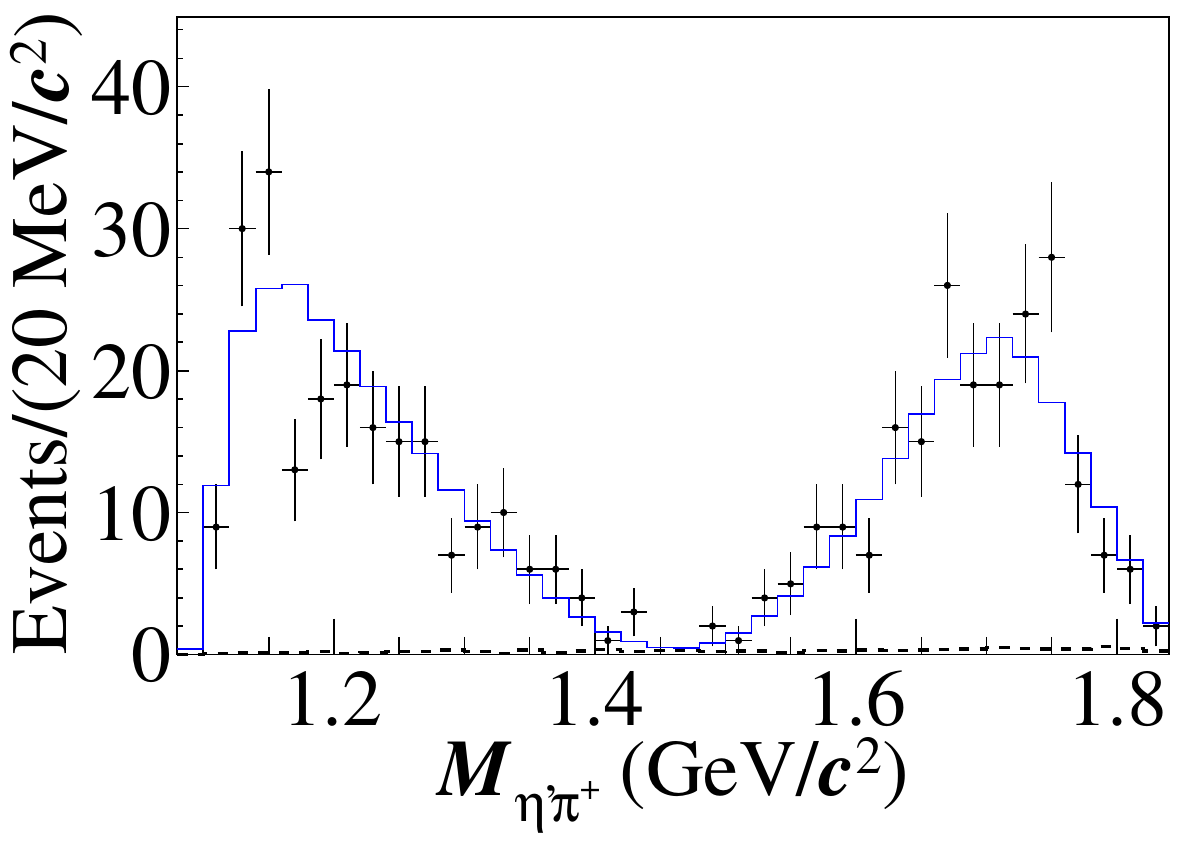}}
	\subfigure{
		\includegraphics[width=0.225\textwidth]{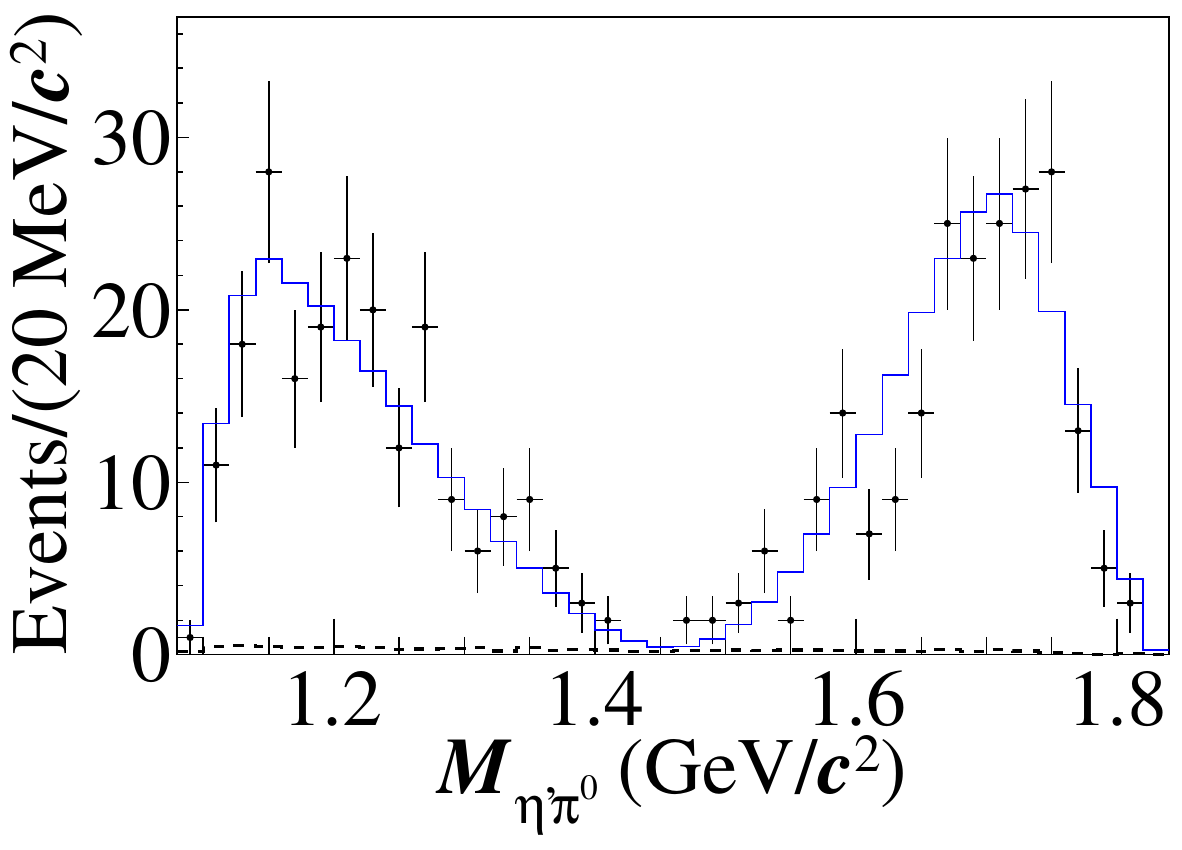}}
	\subfigure{
		\includegraphics[width=0.225\textwidth]{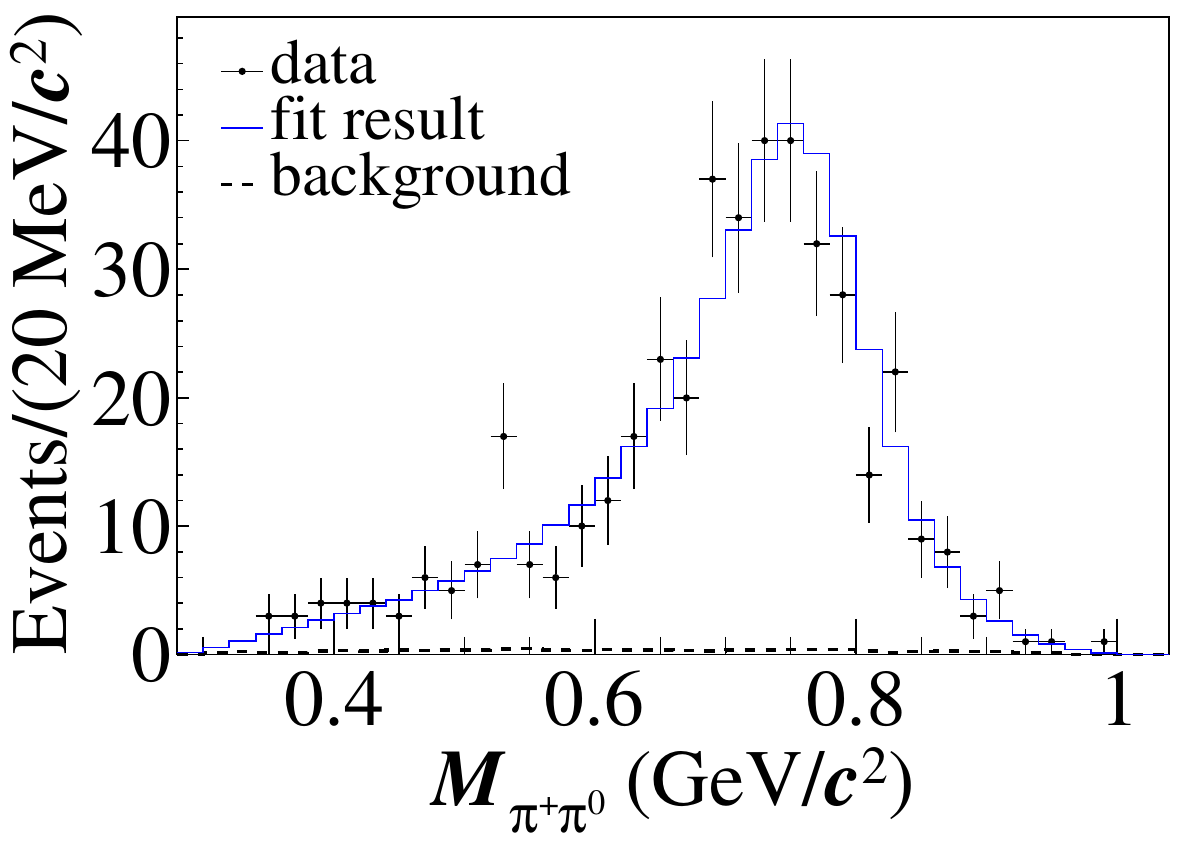}}
	\caption{
The projections of the amplitude analysis fit for $D^+_s\to \eta^\prime \pi^+\pi^0$ on two-body particle mass distributions~\cite{BESIII:2022ewq}.
}
	\label{fig:Ds_etappipi0}
\end{figure*}

Theoretically, the $D^+_s\to\rho(770)^+\eta^{\prime}$ decay is expected to be dominated by two topological diagrams: tree ($T$) and annihilation ($A$)~\cite{Cheng:2011qh}.
Based on these, the amplitudes for $D^+_s\to\rho(770)^+\eta$, $\rho(770)^+\eta^{\prime}$ and $\pi^+\omega$ satisfy~\cite{Cheng:2016ejf}:
\begin{eqnarray}
\label{1}
\frac{1}{\sin\phi}\mathcal{A}(D^+_s\to\pi^+\omega)&=&\frac{\cos\phi}{\sin\phi}\mathcal{A}(D^+_s\to\rho(770)^+\eta)\nonumber\\
&+&\mathcal{A}(D^+_s\to\rho(770)^+\eta^{\prime}),
\end{eqnarray}
where $\phi$ is the $\eta$-$\eta^{\prime}$ mixing angle.
The measured branching fractions of $D^+_s\to\pi^+\omega$ and $\rho(770)^+\eta$
together with the triangular inequality gives the bound $(2.19\pm0.27)\% < \mathcal{B}(D^+_s\to\rho(770)^+\eta^{\prime}) < (4.51\pm0.38)\%$~\cite{Cheng:2016ejf}.
However, theoretical predictions~\cite{Fu-Sheng:2011fji,Qin:2013tje} are lower than experimental measurements by about $2\sigma$,
suggesting possible missing QCD hairpin contributions~\cite{Cheng:2011qh}.
Previous measurements of $D^+_s\to\rho(770)^+\eta^{\prime}$ were from BESIII (482 pb$^{-1}$ at $4.009$ GeV)~\cite{BESIII:2015rrp} and
CLEO-c (586 pb$^{-1}$ at $4.17$ GeV)~\cite{CLEO:2013bae}.
In 2022, Ref.~\cite{BESIII:2022ewq} presented the first amplitude analysis of $D_{s}^{+}\to \pi^{+}\pi^{0}\eta^\prime$
with 411 signal events with a purity of 96\%, which reveals that this decay is dominated by $D^+_s \to\rho(770)^+ \eta^{\prime}$.
The upper limits of the branching fractions of $\cal S-$wave and $\cal P-$wave nonresonant components of $D^+_s\to \pi^+\pi^0\eta^\prime$
are set to be $\mathcal{B}(D^+_s\to (\pi^+\pi^0)_{S}\eta^{\prime}) < 0.10\%$ and $\mathcal{B}(D^+_s\to (\pi^+\pi^0)_{P}\eta^{\prime}) < 0.74\%$ at the 90\% confidence level, respectively.
Figure~\ref{fig:Ds_etappipi0} shows projections of the amplitude analysis fit on $M_{\eta^{\prime}\pi^+}$, $M_{\eta^{\prime}\pi^0}$ and $M_{\pi^+\pi^0}$.
The measured $\mathcal{B}(D^+_s\to\pi^+\pi^0\eta^{\prime})=(6.15\pm0.25\pm0.18)\%$ is consistent within 1$\sigma$ of the CLEO-c result~\cite{CLEO:2013bae} $\mathcal{B}(D^+_s\to\pi^+\pi^0\eta^{\prime})=(5.6\pm0.5\pm0.6)\%$, but with significantly improved precision.
The branching fraction of $D^+_s \to\rho(770)^+ \eta^{\prime}$ is $(6.15\pm0.25\pm0.18)\%$ based on the amplitude analysis results.
This result is more than $3\sigma$ above current theoretical predictions and suggests that other contributions, such as
QCD flavor-singlet hairpin amplitude~\cite{Cheng:2011qh}, should be taken into account.

\begin{figure*}[htbp]
\begin{center}
\begin{minipage}[b]{0.225\textwidth}
\epsfig{width=0.98\textwidth,clip=true,file=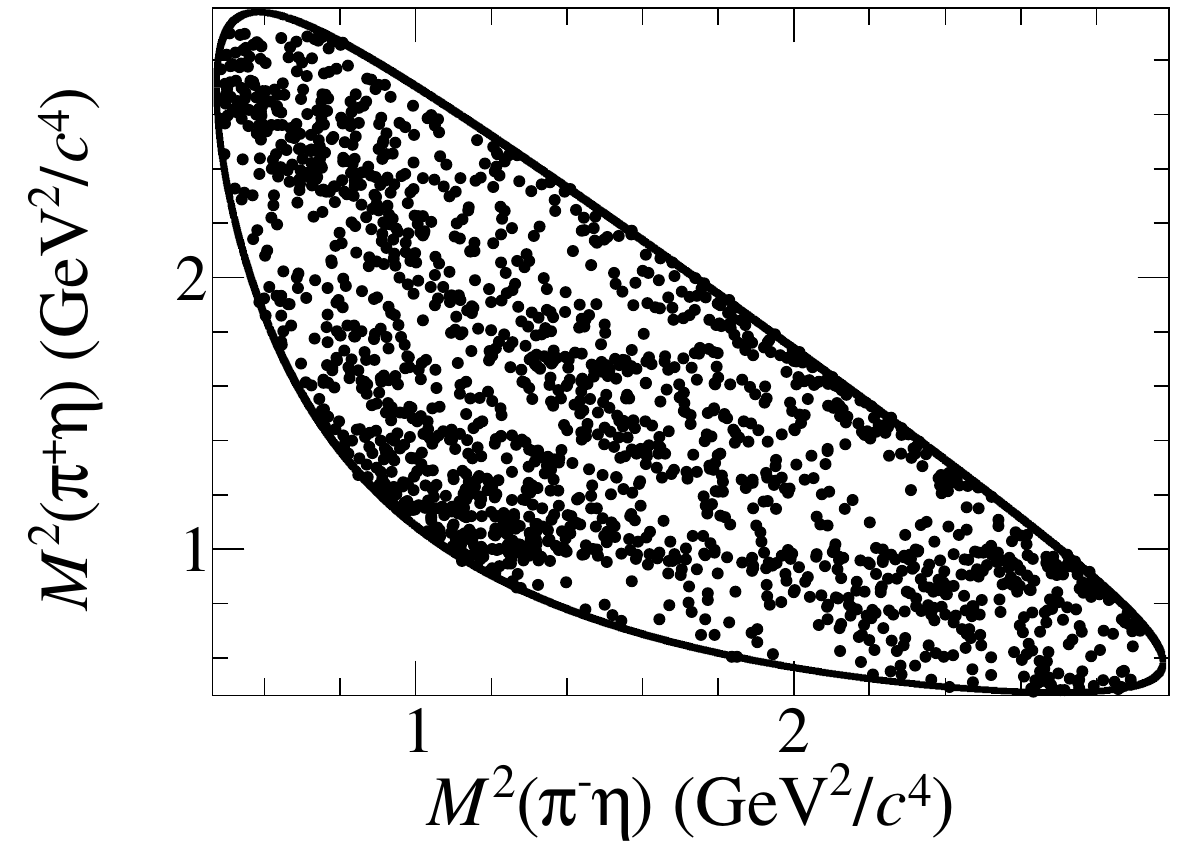}
\put(-25,65){(a)}
\end{minipage}
\begin{minipage}[b]{0.225\textwidth}
\epsfig{width=0.98\textwidth,clip=true,file=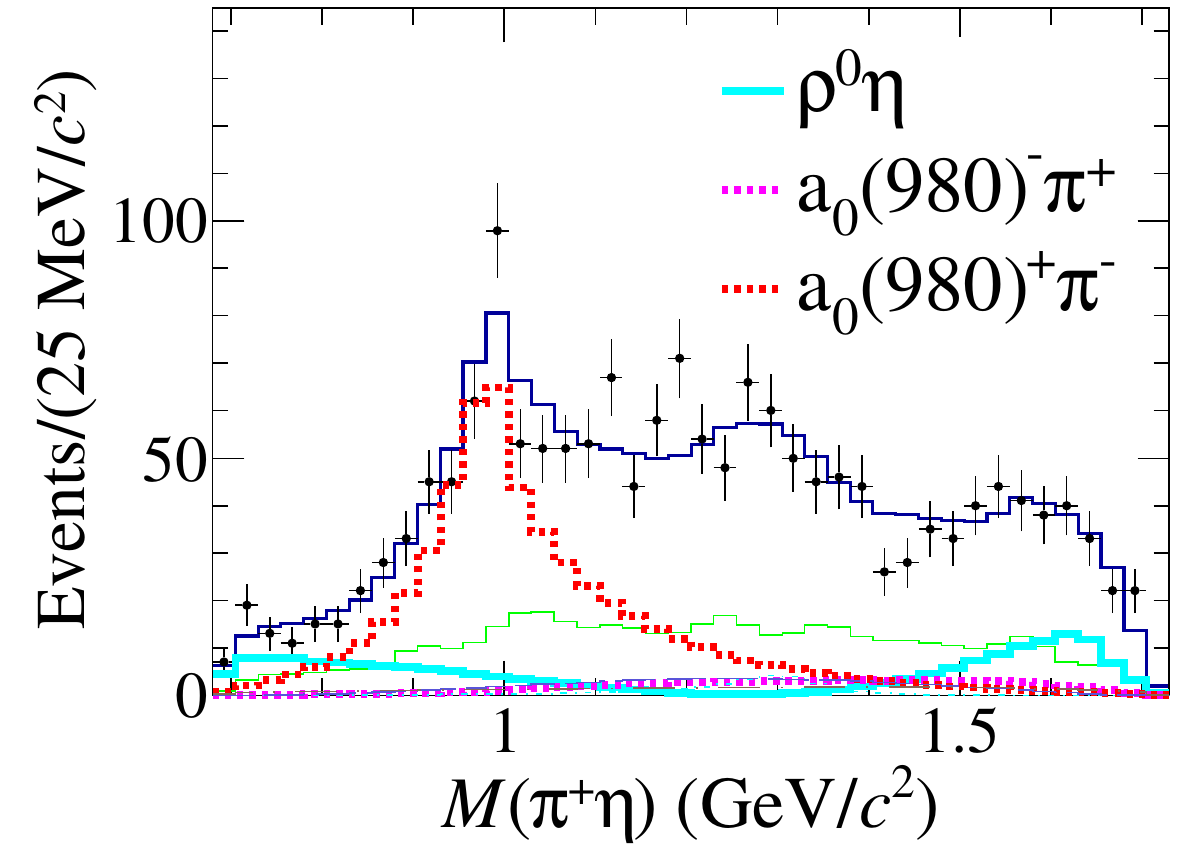}
\put(-85,65){(b)}
\end{minipage}
\begin{minipage}[b]{0.225\textwidth}
\epsfig{width=0.98\textwidth,clip=true,file=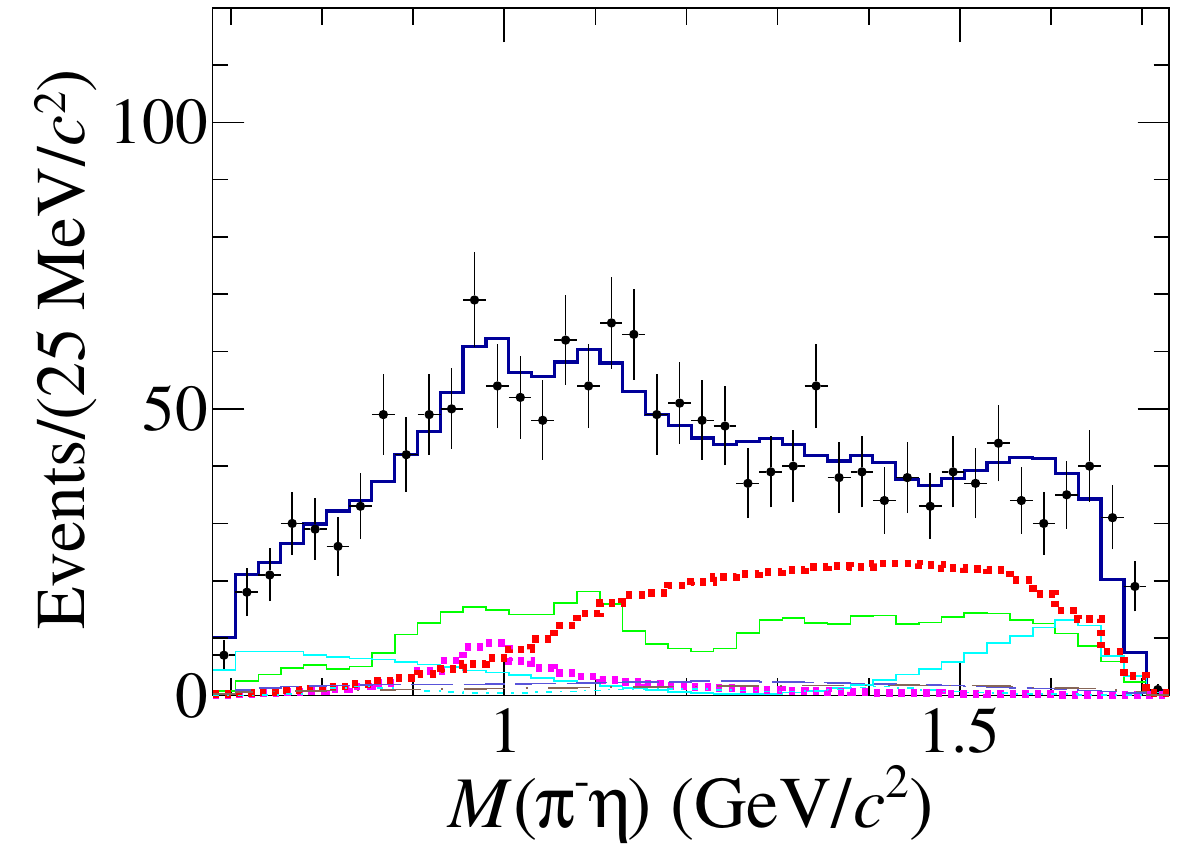}
\put(-85,65){(c)}
\end{minipage}
\begin{minipage}[b]{0.225\textwidth}
\epsfig{width=0.98\textwidth,clip=true,file=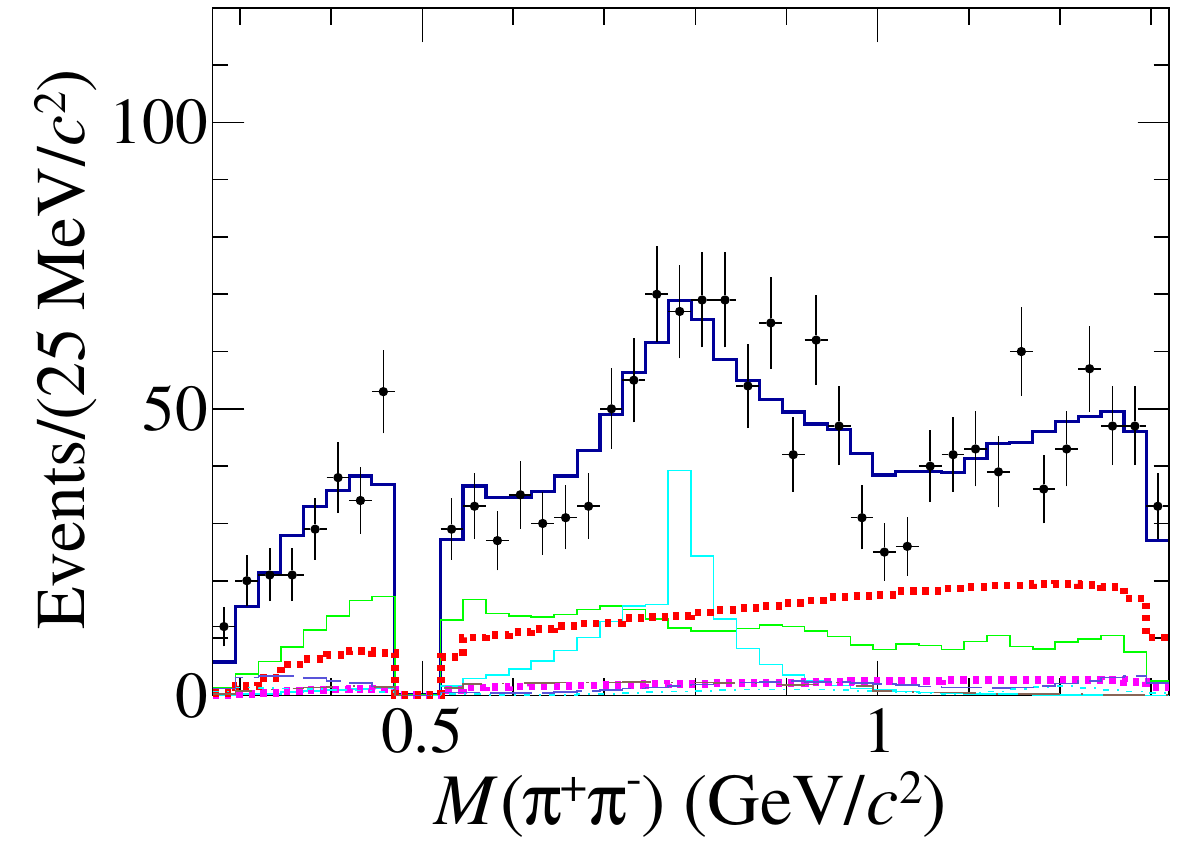}
\put(-85,65){(d)}
\end{minipage}
\begin{minipage}[b]{0.225\textwidth}
\epsfig{width=0.98\textwidth,clip=true,file=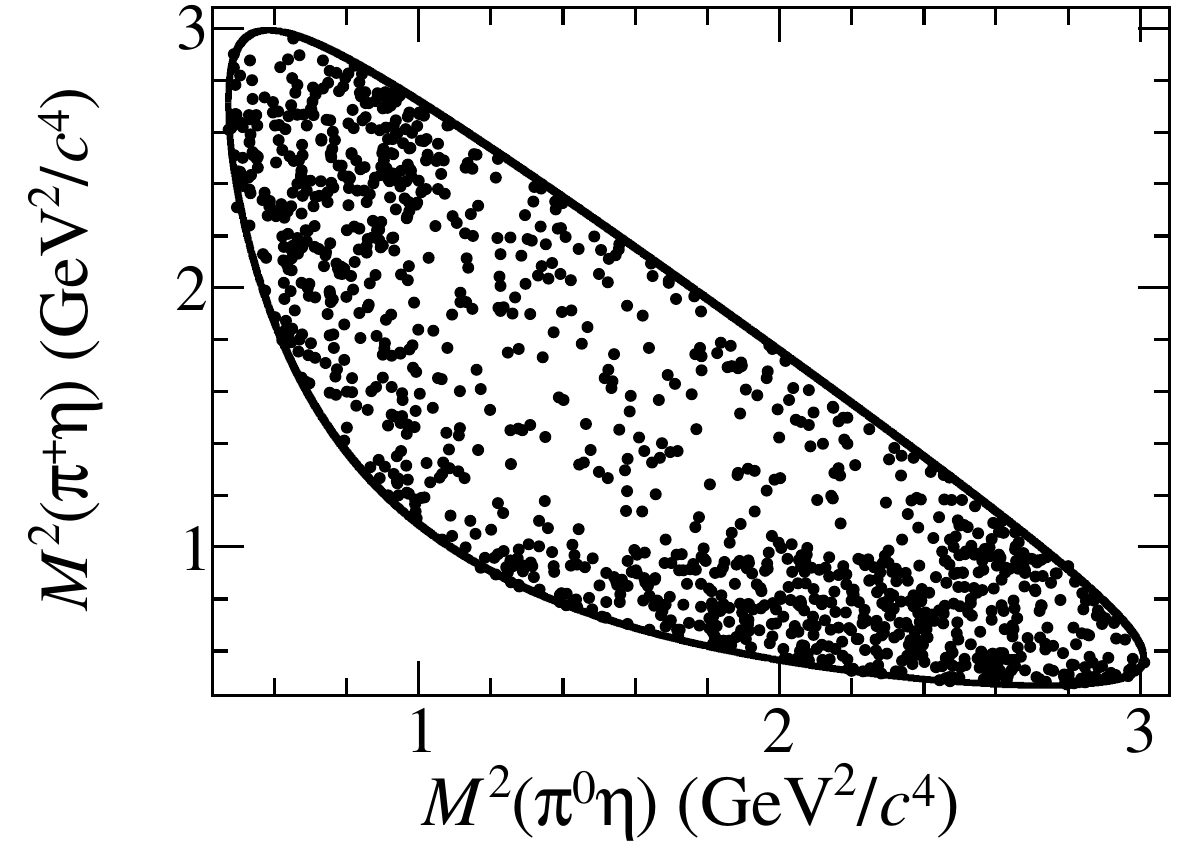}
\put(-25,65){(e)}
\end{minipage}
\begin{minipage}[b]{0.225\textwidth}
\epsfig{width=0.98\textwidth,clip=true,file=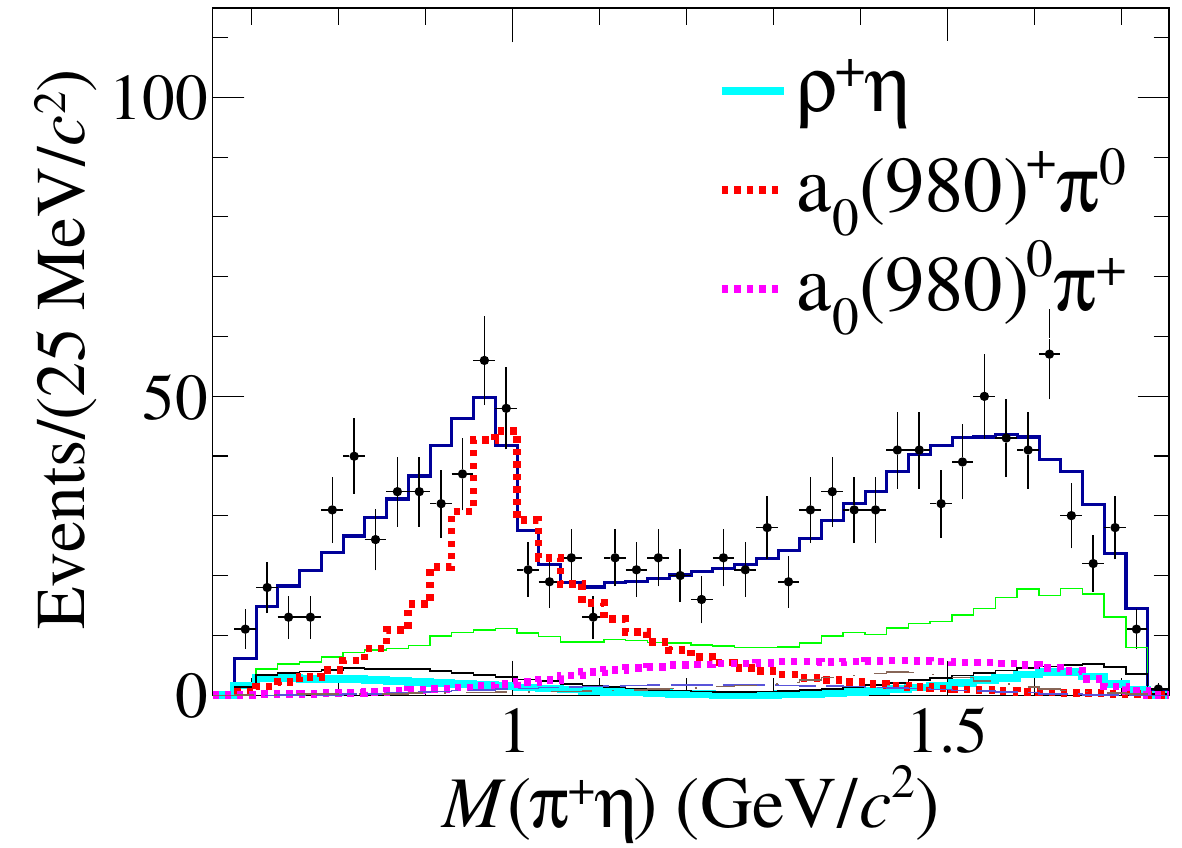}
\put(-85,65){(f)}
\end{minipage}
\begin{minipage}[b]{0.225\textwidth}
\epsfig{width=0.98\textwidth,clip=true,file=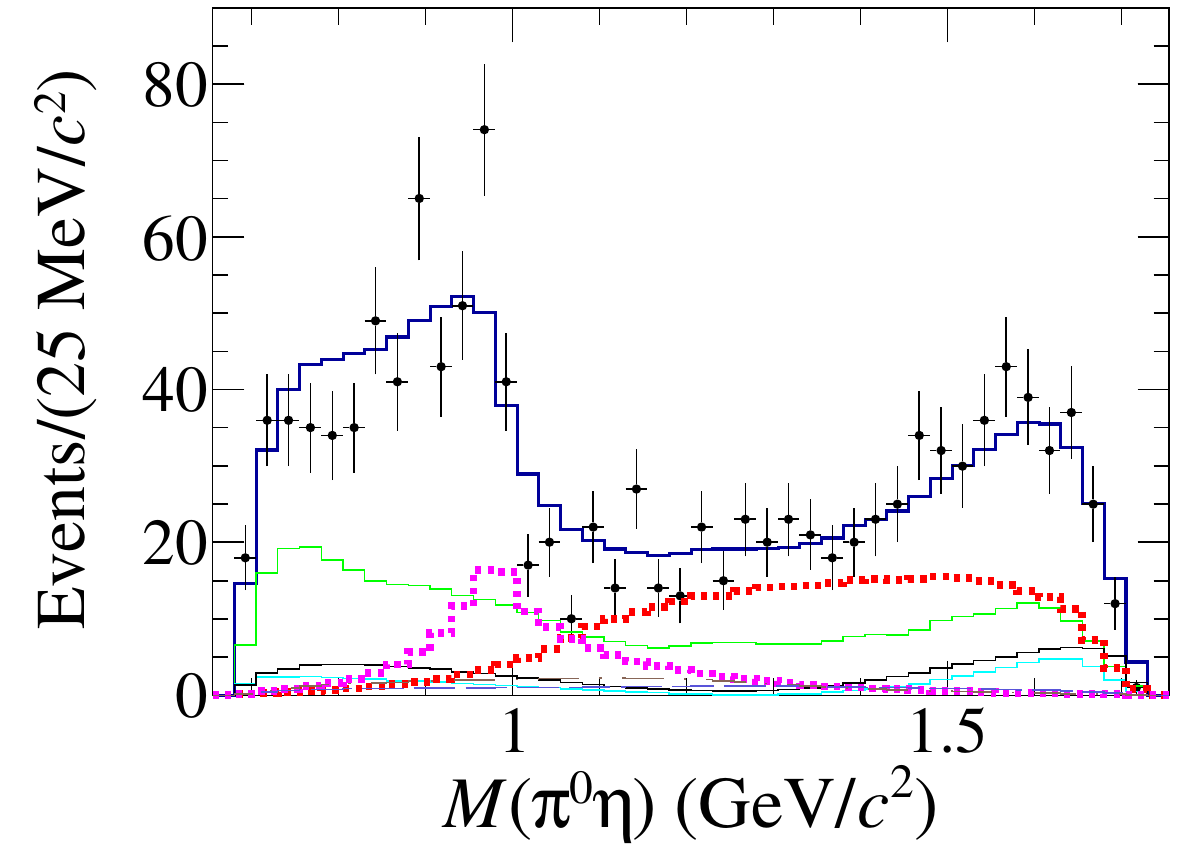}
\put(-85,65){(g)}
\end{minipage}
\begin{minipage}[b]{0.225\textwidth}
\epsfig{width=0.98\textwidth,clip=true,file=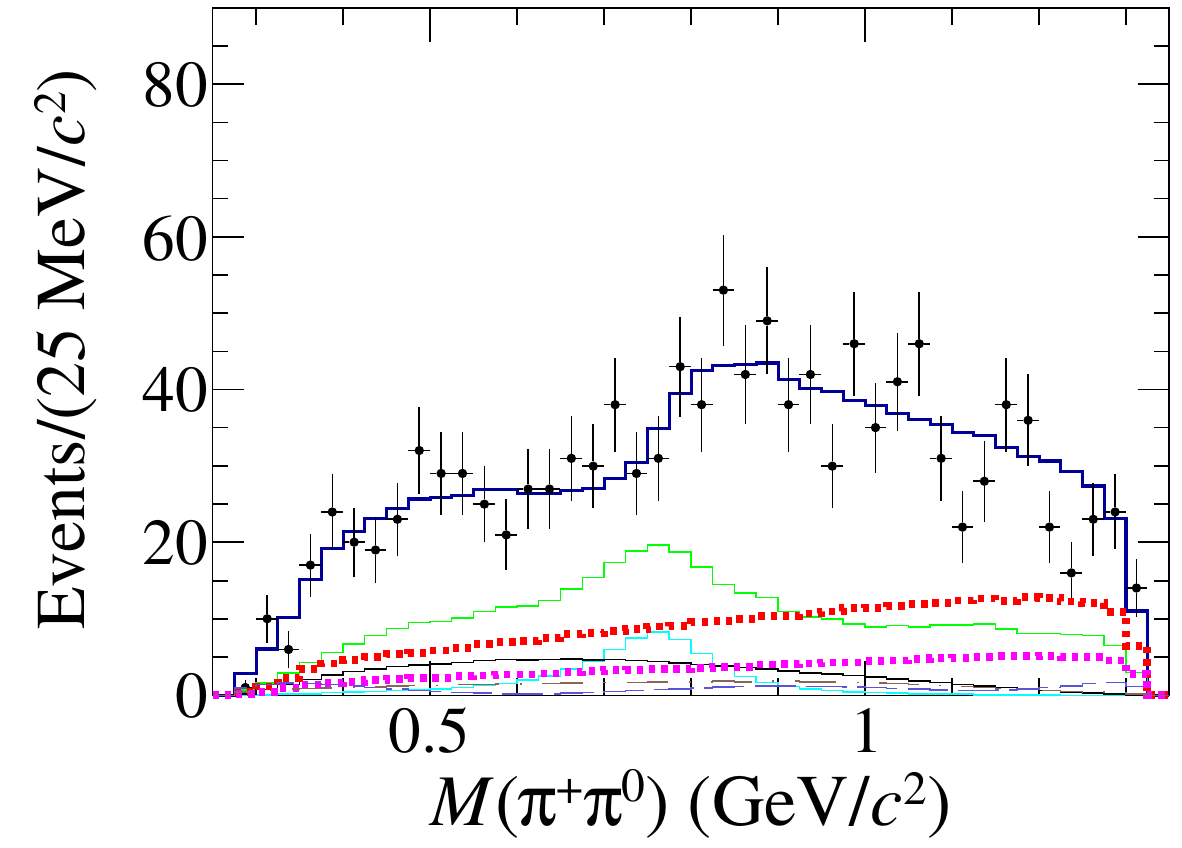}
\put(-85,65){(h)}
\end{minipage}
\caption{The Dalitz plots as well as the projections of the amplitude analysis fits
on two-body particle mass distributions of
(top) $D^{0} \to \pi^{+}\pi^{-}\eta$ and (bottom) $D^{+} \to \pi^{+}\pi^{0}\eta$~\cite{BESIII:2024tpv}.
}
\label{fig:D_pipieta}
\end{center}
\end{figure*}

The $W$-annihilation (WA) and $W$-exchange (WE) processes, suppressed in $B$ decays, can occur in $D$ decays via
FSI and are dominated by non-perturbative effects, causing large theoretical
uncertainties~\cite{Cheng:2010ry,Qin:2013tje}. Study of hadronic $D$ decays with significant WA/WE contributions
thus probes the decay dynamics of charmed mesons. Previously, BESIII observed pure WA decays $D_{s}^{+} \to a_{0}(980)^{+(0)} \pi^{0(+)}$~\cite{BESIII:2019jjr},
indicating sizable FSI in $D\to SP$. Theorists explained the large WA amplitude and the symmetry $A(D_{s}^{+} \to a_0(980)^{+}\pi^{0}) = -A(D_{s}^{+} \to a_0(980)^{0}\pi^{+})$
by considering contributions from the $D_{s}^{+} \to \rho(770)^{+}\eta$ and $D_{s}^{+} \to \bar{K}^{*}(892)^{0}K^{+}(K^{*}(892)^{+}\bar{K}^{0})$
decays~\cite{Hsiao:2019ait,Ling:2021qzl}, supporting $a_{0}(980)$ as a tetraquark or molecule.
Subsequent studies observed $a_{0}(1817)^{+(0)}$~\cite{BESIII:2021anf,BESIII:2022npc}, which is explained
as an excited state of $a_{0}(980)^{+(0)}$~\cite{Guo:2022xqu}, supporting
the interpretation of these two resonances as $K^{(*)}\bar{K}^{(*)}$ molecules~\cite{Wang:2021jub,Oset:2023hyt}.
In $D^{0}$ decays, the ratio $r_{+/-} = \mathcal{B}(D^{0} \to a_{0}(980)^{+} \pi^{-})/ \mathcal{B}(D^{0} \to a_{0}(980)^{-} \pi^{+})$ is expected to be $<0.05$ without WE contribution~\cite{Cheng:2022vbw}. Measurements from CLEO-c~\cite{CLEO:2012obf}, LHCb~\cite{LHCb:2015lnk} and Belle~\cite{Belle:2021dfa} remain inconclusive.
For $D^{+}$ decays, sizable FSI is expected to enhance WA contribution, but the symmetry observed in $D_{s}^{+}$ decays is violated due to short-distance contributions (color-allowed and color-suppressed external $W$-emission)~\cite{Cheng:2022vbw}.
Measurements of branching fractions and ratios $r_{+/0}$ ($r_{+/0} = \mathcal{B}(D^{+} \to a_{0}(980)^{+} \pi^{0})/ \mathcal{B}(D^{+} \to a_{0}(980)^{0} \pi^{+})$)
would effectively constrain WA/WE amplitudes and the role of $a_{0}(980)$ in charmed meson decays~\cite{Ikeno:2021kzf}.
From the analyses of 1.7k(1.2k) candidates for $D^{0(+)} \to \pi^{+} \pi^{-(0)} \eta$ with signal purities of 74(65)\%,
the first amplitude analyses of these two hadronic decays were conducted~\cite{BESIII:2024tpv}.
The amplitude analysis fit projections on two-body particle mass distributions are shown in Fig.~\ref{fig:D_pipieta}.
For $D^{0} \to \pi^{+} \pi^{-} \eta$, the components of
$D^0\to\rho(770)^0\eta$,  $D^0\to a_0(980)^-\pi^+$, $D^0\to a_0(980)^+\pi^-$,
$D^0\to a_2(1320)^+\pi^-$, $D^0\to a_2(1700)^+\pi^-$, and $D^0\to (\pi^+\pi^-)_{{\cal S}{\rm -wave}}\eta$
are observed with fractions of
$(15.2\pm1.7\pm1.0)\%$, $(5.9\pm1.3\pm1.0)\%$, $(44.0\pm4.0\pm5.3)\%$,
$(2.1\pm0.9\pm0.8)\%$, $(5.5\pm1.8\pm2.7)\%$, and $(3.9\pm1.8\pm2.1)\%$, respectively.
For $D^{+} \to \pi^{+} \pi^{0} \eta$,
 the components of
$D^+\to\rho(770)^+\eta$,  $D^+\to (\pi^+\pi^0)_{\rm non-\rho(770)}\eta$, $D^+\to a_0(980)^+\pi^0$, $D^+\to a_0(980)^0\pi^+$,
$D^0\to a_2(1700)^+\pi^0$, and $D^0\to  a_0(1450)^+\pi^0$
are observed with fractions of
$(9.3\pm3.0\pm2.1)\%$, $(15.8\pm4.8\pm5.2)\%$, $(43.7\pm5.6\pm1.9)\%$,
$(17.0\pm4.4\pm1.7)\%$, $(4.2\pm2.1\pm0.7)\%$, and $(7.0\pm2.8\pm0.7)\%$, respectively.
The decays $D^{0(+)} \to a_{0}(980)^{+} \pi^{-(0)}$ and $D^{0(+)} \to a_{0}(980)^{-(0)} \pi^{+}$ are observed for the first time
and are dominant intermediate resonances in both channels.
The contribution from the process $D^{0(+)} \to a_{0}(980)^{+} \pi^{-(0)}$ is significantly larger than the $D^{0(+)} \to a_{0}(980)^{-(0)} \pi^{+}$ contribution.
The ratios  $\mathcal{B}(D^{0} \to a_{0}(980)^{+}\pi^{-})/\mathcal{B}(D^{0} \to a_{0}(980)^{-}\pi^{+})$ and $\mathcal{B}(D^{+} \to a_{0}(980)^{+}\pi^{0})/\mathcal{B}(D^{+} \to a_{0}(980)^{0}\pi^{+})$ are measured to be $7.5^{+2.5}_{-0.8}\pm1.7$ and $2.6\pm0.6\pm0.3$, respectively.
The measured $D^{0}$ ratio disagrees with the theoretical predictions by orders of magnitude~\cite{Cheng:2022vbw,Cheng:2024hdo}, thus implying a substantial contribution from FSIs.

\subsubsection{Analysis of $D^+\to \pi^+2\eta$}

\begin{figure*}[htbp]
\begin{center}
\begin{minipage}[b]{0.225\textwidth}
\epsfig{width=0.98\textwidth,clip=true,file=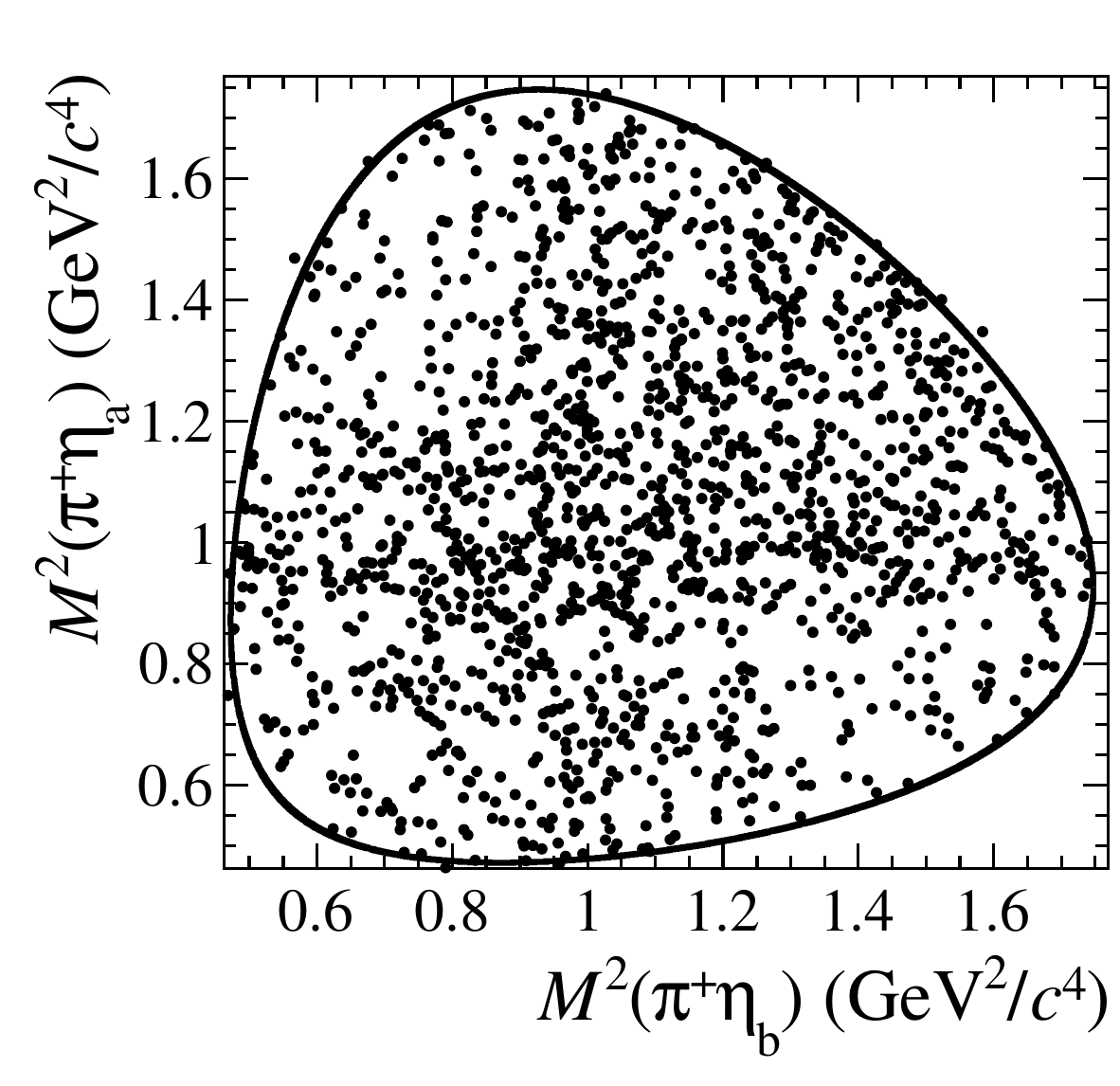}
\end{minipage}
\begin{minipage}[b]{0.225\textwidth}
\epsfig{width=0.98\textwidth,clip=true,file=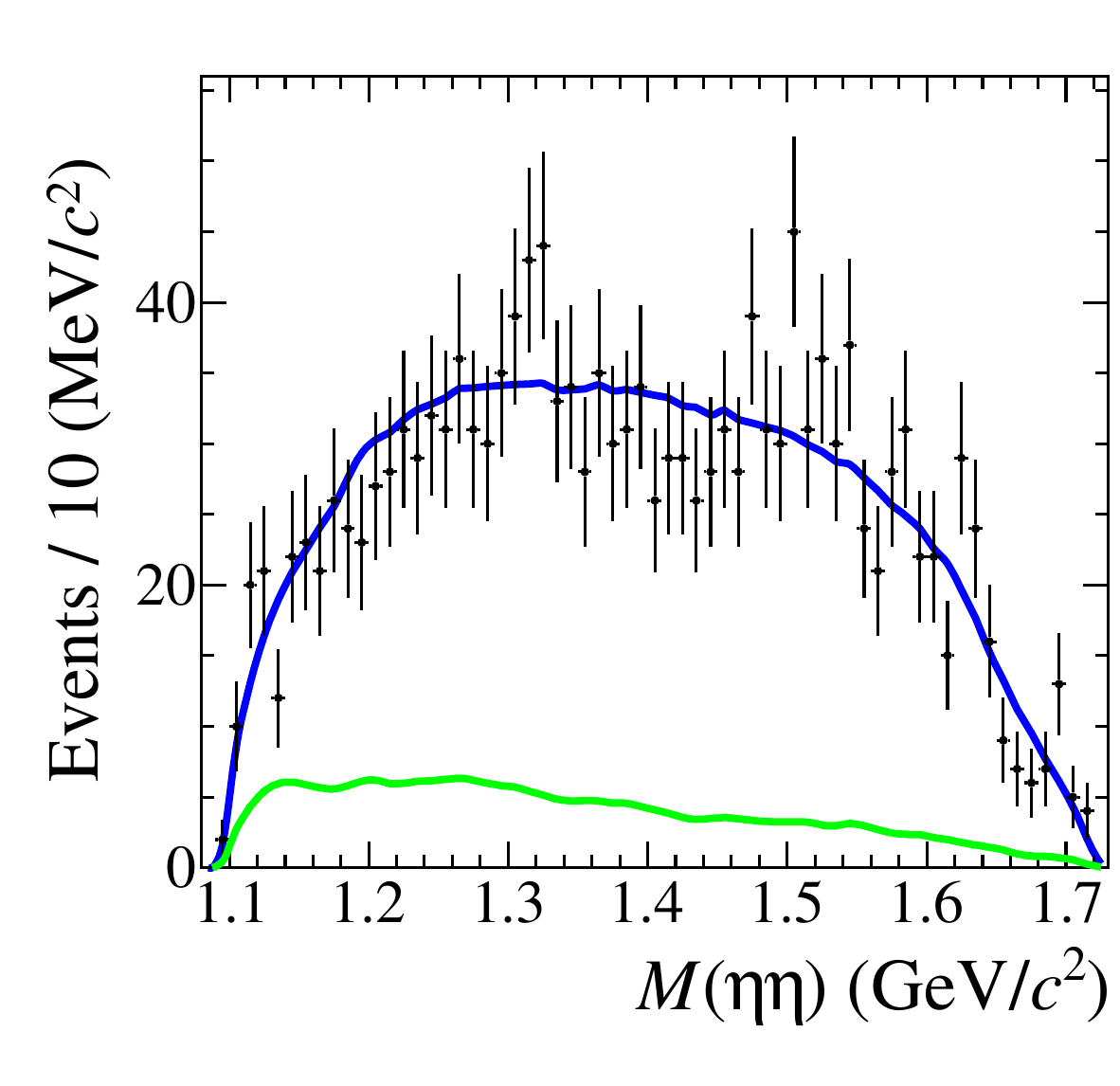}
\end{minipage}
\begin{minipage}[b]{0.225\textwidth}
\epsfig{width=0.98\textwidth,height=0.925\textwidth,clip=true,file=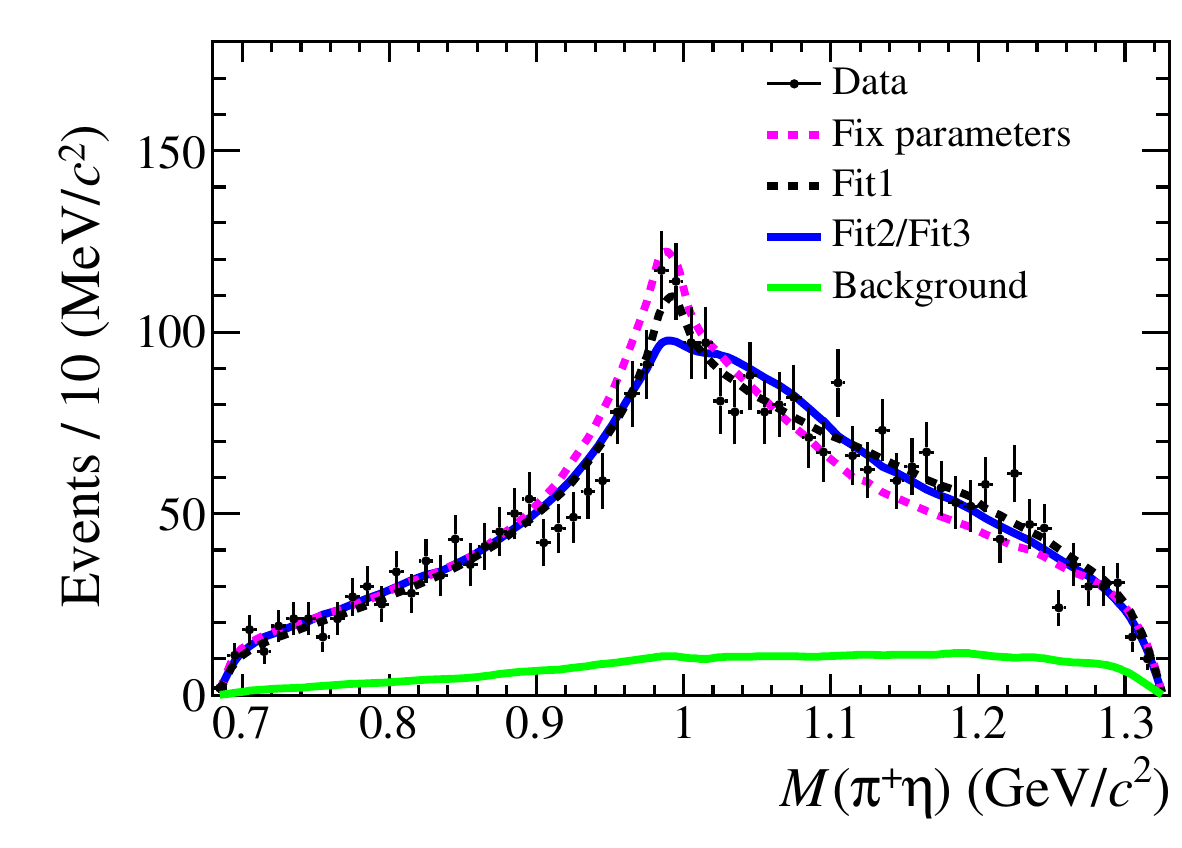}
\end{minipage}
\caption{(Left) The Dalitz plot of the data as well as
the projections of the amplitude analysis fit of $D^+\to \pi^+2\eta$
 on (middle) $M(\eta\eta)$ and (right) $M(\pi^{+}\eta)$~\cite{BESIII:2025yag}.
}
\label{fig:Dp_pietaeta}
\end{center}
\end{figure*}

QCD exhibits non-perturbative features in the low-energy regime, where hadronic loop contributions become essential. A striking example is the cusp in the $\pi^0\pi^0$ invariant mass spectrum of $K^\pm \to \pi^\pm2\pi^0$ at the $\pi^+\pi^-$ threshold, explicable only via a two-point loop diagram with an intermediate $\pi^+\pi^-$ pair~\cite{NA482:2005wht}. Similarly, triangle singularities arising from three-point loop diagrams are vital for interpreting processes such as the large isospin-breaking decay $\eta(1405/1475) \to \pi\pi\pi$ observed by BESIII~\cite{BESIII:2012aa}, where a $K^*\bar{K}K$ triangle loop enhances the effect \cite{Wu:2011yx,Aceti:2012dj,Wu:2012pg}.
More generally, when a parent particle couples comparably to two intermediate channels, triangle loops must be accounted for in data analysis to avoid biased resonance parameter extraction. While such diagrams are widely studied theoretically, especially for hadronic molecular candidates~\cite{Guo:2017jvc}, their incorporation into experimental amplitude analyses remains limited. Recent studies by COMPASS~\cite{COMPASS:2020yhb} and LHCb~\cite{LHCb:2019kea} have considered triangle contributions, yet direct experimental evidence for their decisive role is still lacking, partly because a clear methodology to identify such effects within an amplitude analysis framework has not yet been established. Using 1.6k candidates with a signal purity of 85\%,
the first amplitude analysis of $D^{+} \to \pi^{+}2\eta$ was carried out~\cite{BESIII:2025yag}.
The amplitude analysis fit projections on two-body particle mass distributions are shown in Fig.~\ref{fig:Dp_pietaeta}.
The intermediate process $D^{+} \to a_{0}(980)^{+}\eta, a_{0}(980)^{+} \to \pi^{+}\eta$ is observed and is found to be the only component and its branching fraction is measured to be $(3.67\pm0.12\pm 0.06)\times 10^{-3}$.
Unlike the $a_{0}(980)$ line-shape observed in the decays of charmed mesons to $a_{0}(980)\pi$~\cite{BESIII:2019jjr,BESIII:2024tpv} and in the decay $D^{0} \to a_{0}(980)^{-}e^{+}\nu_e$~\cite{BESIII:2024zvp},  where the low-mass side of the $a_0(980)$ is wider than the high-mass side,
the $a_{0}(980)$ line-shape in $D^{+} \to a_{0}(980)^{+}\eta$ is found to be significantly altered, with the high-mass side being wider than the low-mass side.

A systematic study is performed using several conventional descriptions of the $a_0(980)$ amplitude, including the Flatt\'e parameterization~\cite{Flatte:1976xu}, the dispersively modified Flatt\'e parameterization~\cite{Bugg:2008ig}, the $T$-matrix formalism~\cite{Ikeno:2021kzf}, and the $K$-matrix formalism~\cite{Anisovich:1997pe}. 
However, none of these descriptions reproduce the observed line shape satisfactorily. The observed line shape distortion cannot be explained solely by the choice of the $a_0(980)$ line shape parameterization or by interference with small additional conventional amplitudes. 
Additional dynamical effects beyond the conventional direct-production amplitude descriptions may be relevant for describing the $a_0(980)$ line shape in $D^+\to\pi^+\eta\eta$.

The triangle loop rescattering of $D^+ \to \bar K_0^*(1430)^0K^+ \to a_0(980)^+ \eta$ and $D^+ \to K_0^*(1430)^+\bar K^0 \to a_0(980)^+ \eta$ is established to describe the $a_0(980)$ line-shape. The significance of the triangle-loop rescattering contribution is determined to be 5.8$\sigma$, which constitutes the first experimental confirmation of the triangle-loop rescattering effect.

\subsubsection{Other results}

The  $D^0\to K^0_SK^+K^-$ decay  is a self-conjugate channel with a resonant substructure containing $CP$ eigenstates as well as non-$CP$ eigenstates.
An accurate measurement of the decay and its substructure has implications for various fields.
Its substructure  is dominated by the $K\bar K$ $\cal S$-wave which can be studied in an almost background-free environment.
In particular, light scalar mesons are of interest since their spectrum is not free of doubt.
The branching fractions of the resonant substructure and the total branching fraction are inputs to a better theoretical understanding of $D^0$-$\bar D^0$ mixing.
Furthermore, the strong phase difference between the decays $D^0\to K^0_SK^+K^-$ and $\bar D^0\to K^0_SK^+K^-$ can be determined from the amplitude model.
This phase is an input to a measurement of the CKM angle $\gamma$ of the CKM UT using the decay of $B^-\to D^0K^-$
 with $D^0\to K^0_SK^+K^-$~\cite{BaBar:2008inr}.
A previous analysis of $D^0\to K^0_SK^+K^-$ was performed by BaBar~\cite{BaBar:2005vhe},
with a branching fraction measured relative to $D^0\to K^0_S\pi^+\pi^-$.
Reference~\cite{BESIII:2020hfw} presented an amplitude analysis of $D^0\to K^0_SK^+K^-$,
based on about 1.9k signal events. Its Dalitz plot can be well described by six resonances:
$a_0(980)^0$, $a_0(980)^+$, $\phi(1020)$, $a_2(1320)^+$, $a_2(1320)^-$ and $a_0(1450)^-$.
Their magnitudes, phases and fit fractions are determined as well as the coupling of $a_0(980)$ to $K\bar K$, $g_{K \bar K}=(3.77\pm0.24\pm0.35)$~GeV.
The branching fraction of this decay is measured using about 11.7k untagged signal events to be
$(4.51\pm0.05\pm0.16)\times 10^{-3}$. Both measurements are limited by their systematic uncertainties.
A joint LHCb and BESIII measurement of the $\gamma$ via $B^-\to D^0K^-$
 with $D^0\to K^0_Sh^+h^-$ ($h=\pi$ and $K$) was recently presented in separate papers~\cite{LHCb:2026hot,LHCb:2026npj}.

A model-independent BESIII analysis measured the average sine and cosine of $\Delta \delta_D$ for $D\to K^0_S\pi^+\pi^-$ $\left(c_i,s_i\right)$ and $D\to K^0_L\pi^+\pi^-$ $\left(c_i^{\prime},s_i^{\prime}\right)$~\cite{BESIII:2020khq,BESIII:2020hlg}. Inclusion of the $D\to K^0_L\pi^+\pi^-$  mode provides a three-times-larger data sample at BESIII due to higher $K^0_LS$ reconstruction efficiency and combinatorics of $D\bar{D} \to (K_{S}^0\pi^+\pi^-)^2$ versus $D\bar{D} \to (K_{S}^0\pi^+\pi^-, K_{L}^0\pi^+\pi^-)$ decays. However, including these $D\to K^0_L\pi^+\pi^-$  decays introduces a systematic uncertainty related to assumptions about the values of complex {\it U-spin breaking} parameters $\hat{\rho}$ that separate the decay amplitudes of $D\to K^0_L\pi^+\pi^-$  and $D\to K^0_S\pi^+\pi^-$  modes. In previous analyses~\cite{BESIII:2020khq,BESIII:2020hlg,CLEO:2010iul}, the nominal value of the parameters was unity; and a systematic uncertainty on this assumption was derived by assuming  $\left|\hat{\rho}\right|$ had an uncertainty of $50\%$ and $\arg\left(\hat{\rho}\right)$ could have any value in the interval ($-180^{\circ}, 180^{\circ}$). These U-spin breaking parameters have never been experimentally determined and the only way to measure them is through an amplitude analysis of the $D\to K^0_L\pi^+\pi^-$  decay.
From the amplitude analysis of $D^0 \to K_{L}^0\pi^+\pi^-$ with tens of thousands of candidates,
the main components are
$\rho(770)$, $\omega(782)$, $f_2(1270)$, $\rho(1450)$,
$K^*(892)^-$, $K^*_2(1430)^-$, $K^*_2(1680)^-$, $K^*(1410)^-$,
$K^*(892)^+$, $K^*_2(1430)^+$, $K^*(1410)^+$, $K^*_0(1430)^-$, and $\pi\pi$ $\cal S$-wave~\cite{BESIII:2022qvy}.
This is the first amplitude analysis of a decay mode involving a $K_{L}^0$, which also results in the first measurement of the complex {\it U-spin breaking parameters} ($\hat{\rho}$) related to various $\mathit{CP}$-eigenstate resonant modes through which the three-body decay proceeds. The moduli of the $\hat{\rho}$ parameters have central values in a wide range from $0.4$ to $12.1$, which indicates substantial U-spin symmetry breaking. The fractional resonant contributions and average strong-phase parameters over regions of phase space for both $K_{S}^0\pi^+\pi^-$ and $K_{L}^0\pi^+\pi^-$ modes are presented. In addition, the ratio of the branching fractions between $D^0 \to K_{L}^0\pi^+\pi^-$ and $D^0 \to K_{S}^0\pi^+\pi^-$ decay modes and the $CP$-even fraction of $D^0 \to K_{L}^0\pi^+\pi^-$ calculated using the U-spin breaking parameters are also provided.

\subsection{Four-body decays}

Investigations of four-body hadronic decays of charmed mesons provide a rich framework for isolating a broad spectrum of intermediate quasi-two-body processes. Specifically, analyses of the $D_{(s)}\to K\bar K\pi\pi$ channel grant access to $VV$ amplitudes such as $D_{(s)}\to K^*(892)\rho(770)$, $AP$ contributions including $D_{(s)}\to \bar K a_{1}(1260)$, $K_1(1270)\pi$, and $K_1(1400)\pi$, as well as $SP$ modes exemplified by $D_{(s)}\to f_1 K$, and $PP$ configurations like $D_{(s)}\to \eta(1405) K$. Turning to the $D_{(s)}\to K\pi\pi\pi$ final state, one may extract $VV$ dynamics via $D_{(s)}\to K^*(892)K^*(892)$, additional $AP$ channels such as $D_{(s)}\to \bar K a_{1}(1260)$, $K_1(1270)K$, and $K_1(1400) K$, and $VP$ transitions represented by $D_s^+\to \omega K^+$. The $D_{(s)}\to \pi\pi\pi\pi$ decays serve as a versatile laboratory for examining a wide array of intermediate combinations. These include $SV$ pairs like $D_{(s)}\to f_{0(2)}\rho$ and $a_0(980)\rho$, $VV$ processes such as $D_{(s)}\to \rho\rho$, $VP$ modes involving $D_{(s)}\to \phi\pi$ and $D_{(s)}\to \omega\pi$, and $AP$ interactions with $D_{(s)}\to a_1(1320)\pi$. Moreover, the $D_{(s)}\to \pi\pi\pi\eta$ decays offer a window into $SS$ production via $D_{(s)}\to a_0f_0$, further $SP$ contributions in the form of $a_1(1260)\eta$ and $f_1\pi$, and additional $VP$ channels like $\omega\eta$ and $\phi\eta$. By systematically disentangling these resonant substructures, such multi-body decay studies will significantly enhance our understanding of both weak decay mechanisms and the spectroscopy of light-hadrons.

In order to improve signal purities, double-tag signal candidates are used in amplitude analyses of all four-body hadronic $D$ decays.
Among them,  the investigations of $D^{+}_{s}\to 2\pi^+\pi^-\pi^0$ and $D^+_s\to \pi^+2\pi^0\eta$ were performed by using 7.33 fb$^{-1}$ of data at 4.128-4.226 GeV,
while the analyses of all other four-body $D^+_s$ hadronic decays were carried out with 6.3 fb$^{-1}$ of data at 4.178-4.226 GeV.
Meanwhile,
the study of $D^+ \to K^-2\pi^+\pi^0$ was based on 7.9 fb$^{-1}$ of data at 3.773 GeV;
the studies of $D^0\to K^+K^-2\pi^0$ and $D^{+}\to \pi^+\pi^{+(0)}\pi^{-(0)}\eta$
were based on 20.3 fb$^{-1}$ of data at 3.773 GeV;
while the investigations of all other four-body hadronic $D^{0(+)}$ decays were conducted with
2.93 fb$^{-1}$ of data at 3.773 GeV.

\subsubsection{Analyses of $D^+_{(s)}\to K\bar K\pi\pi$}

\begin{figure*}[htb]
 \centering
 \begin{overpic}[width=0.18\textwidth]{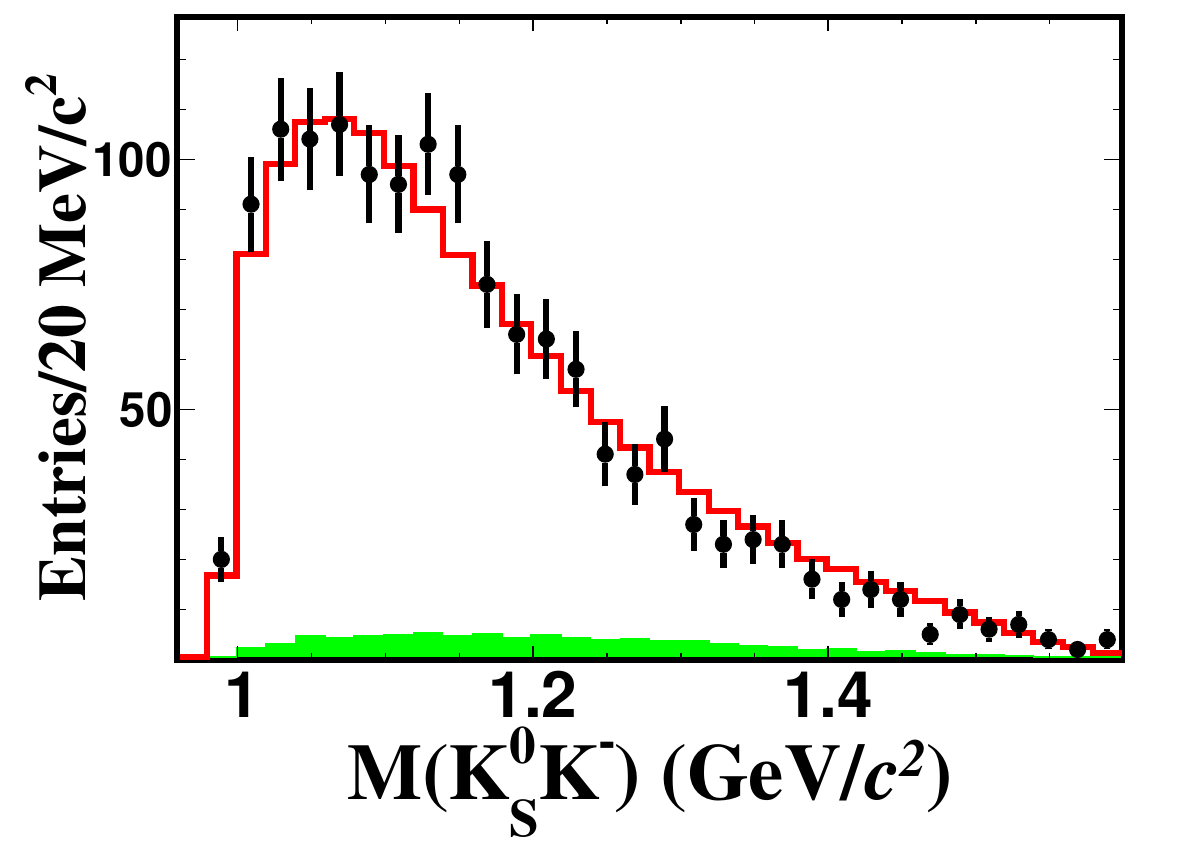}
 \put(65,60){$(a)$}
 \end{overpic}
 \begin{overpic}[width=0.19\textwidth]{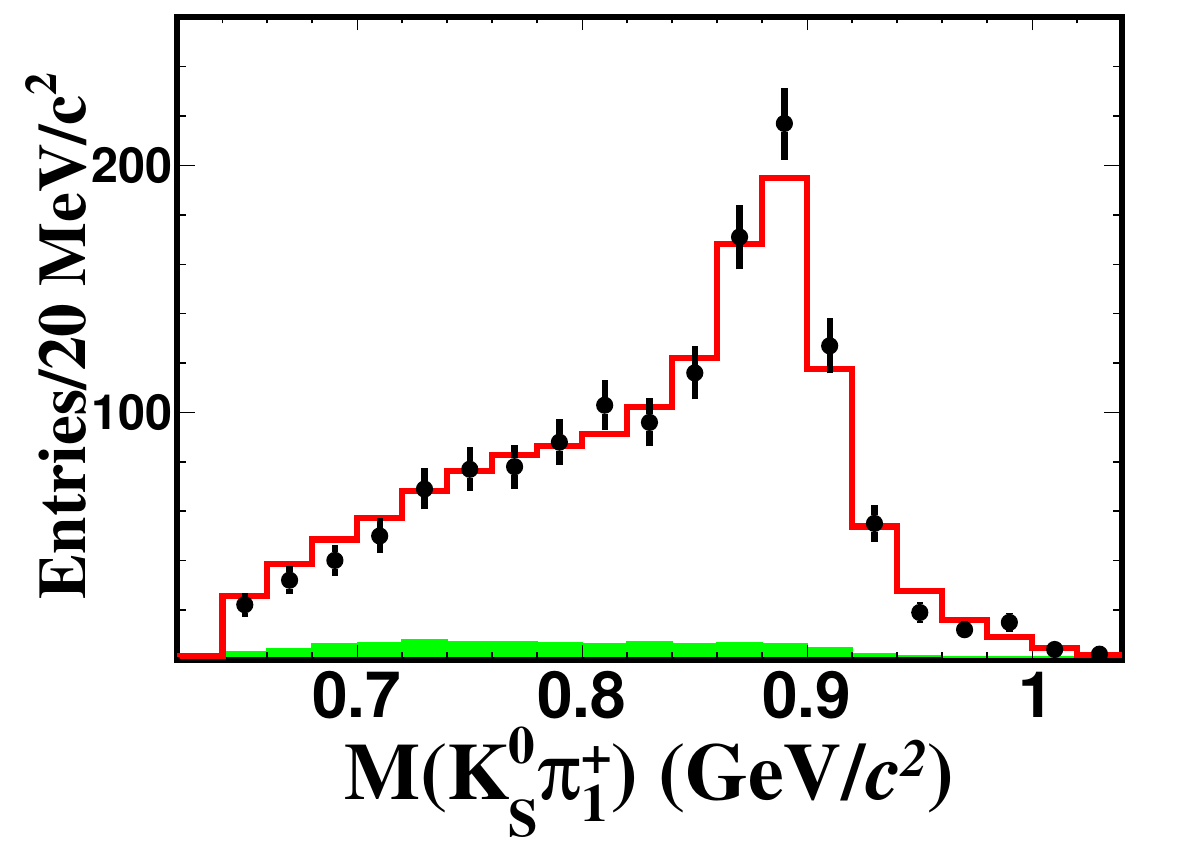}
 \put(65,60){$(b)$}
 \end{overpic}
 \begin{overpic}[width=0.19\textwidth]{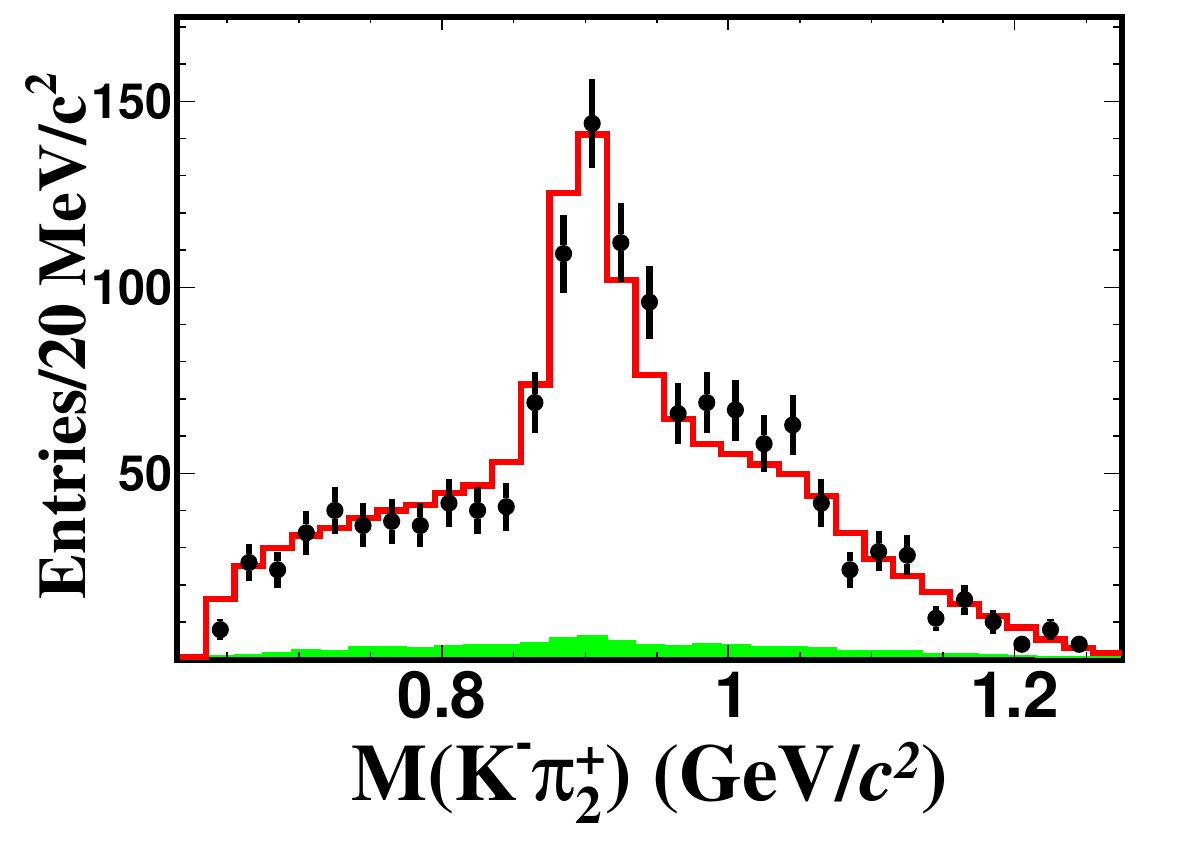}
 \put(65,60){$(c)$}
 \end{overpic}
 \begin{overpic}[width=0.19\textwidth]{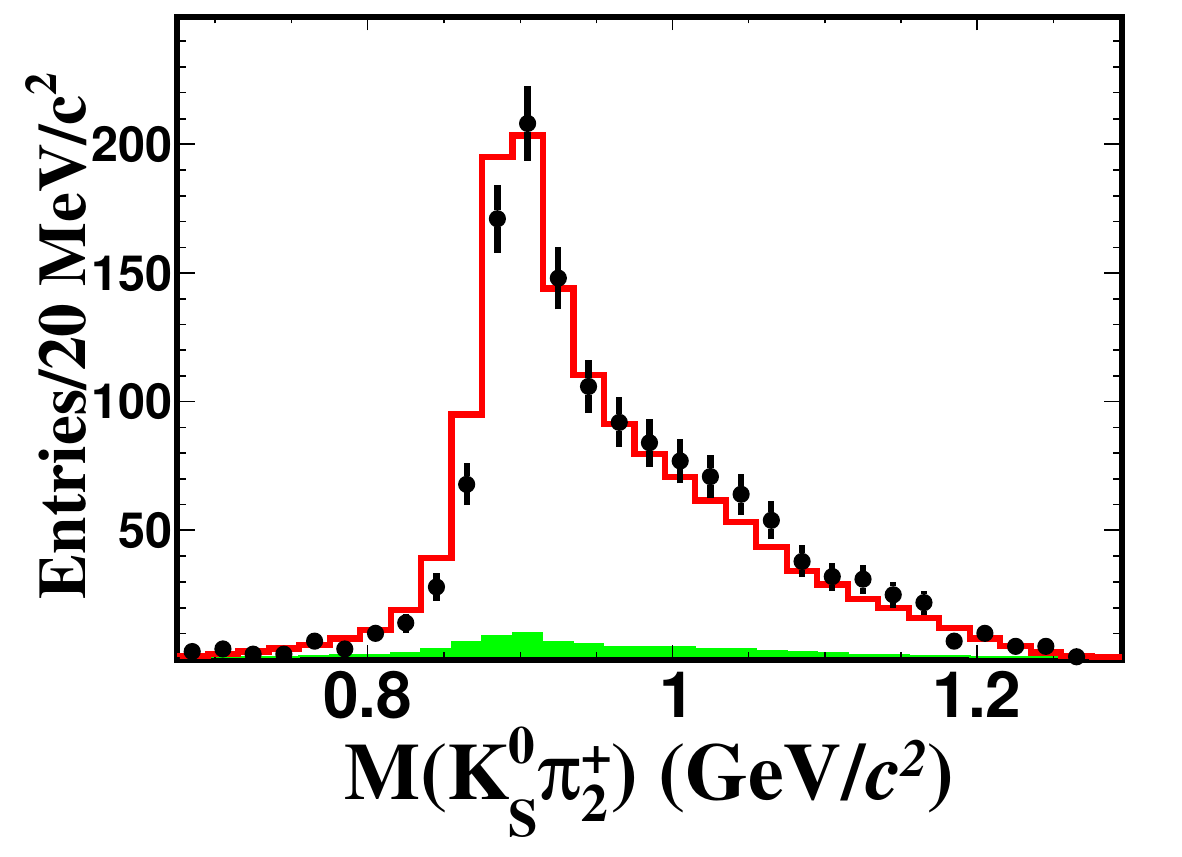}
 \put(65,60){$(d)$}
 \end{overpic}
 \begin{overpic}[width=0.19\textwidth]{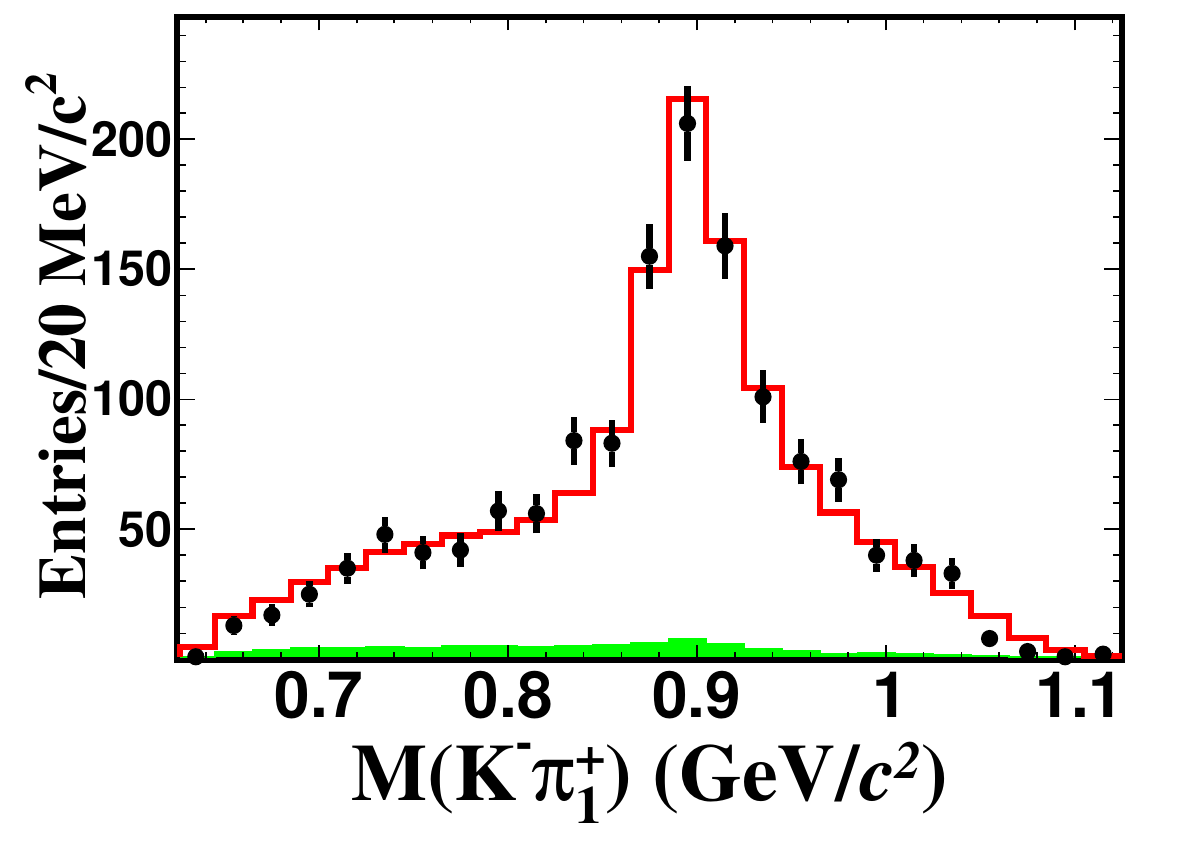}
 \put(65,60){$(e)$}
 \end{overpic}
 \begin{overpic}[width=0.19\textwidth]{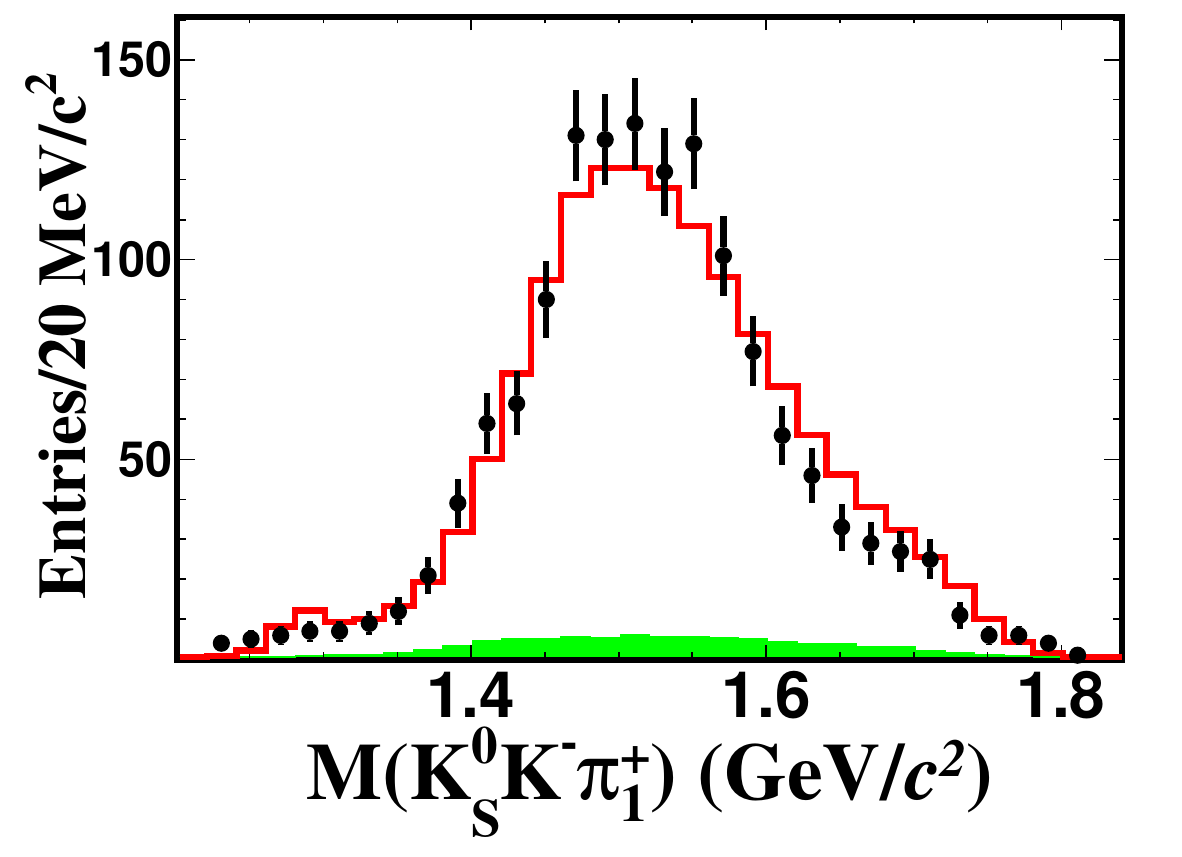}
 \put(65,60){$(f)$}
 \end{overpic}
 \begin{overpic}[width=0.19\textwidth]{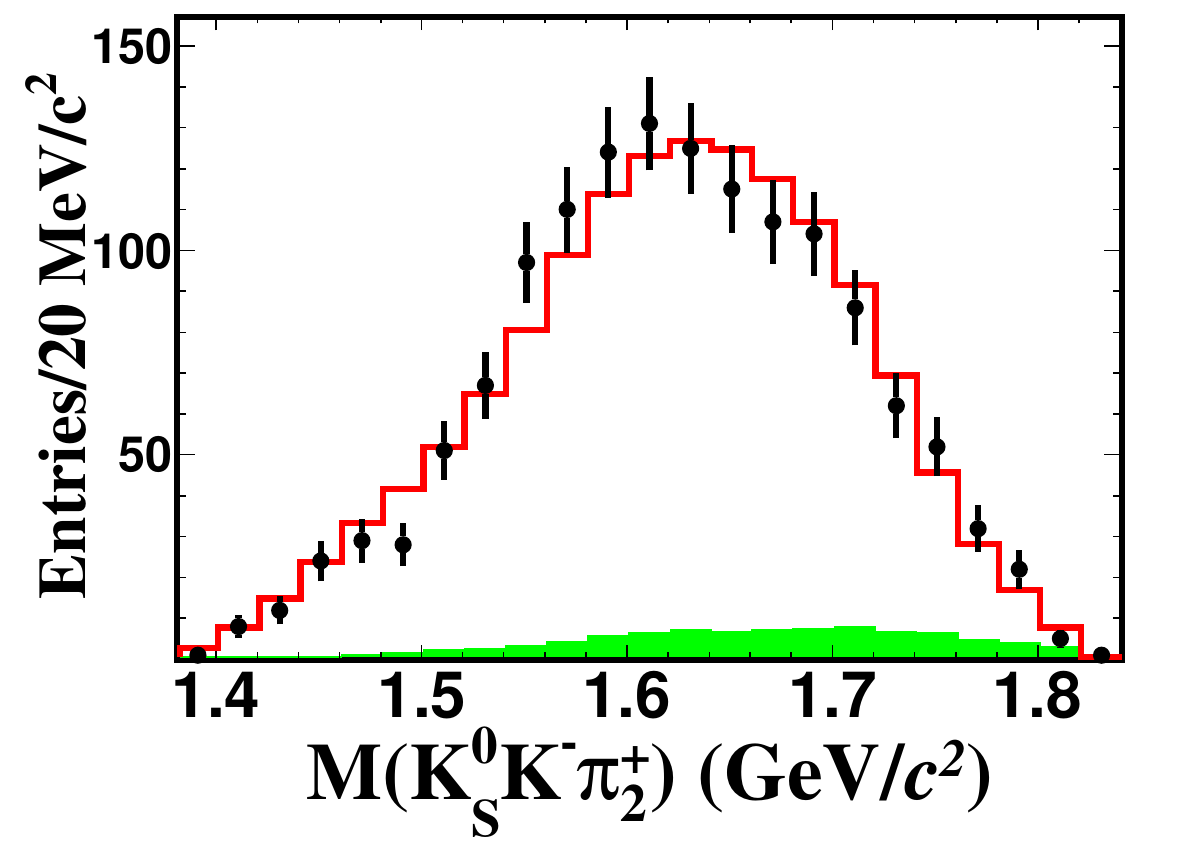}
 \put(65,60){$(g)$}
 \end{overpic}
 \begin{overpic}[width=0.19\textwidth]{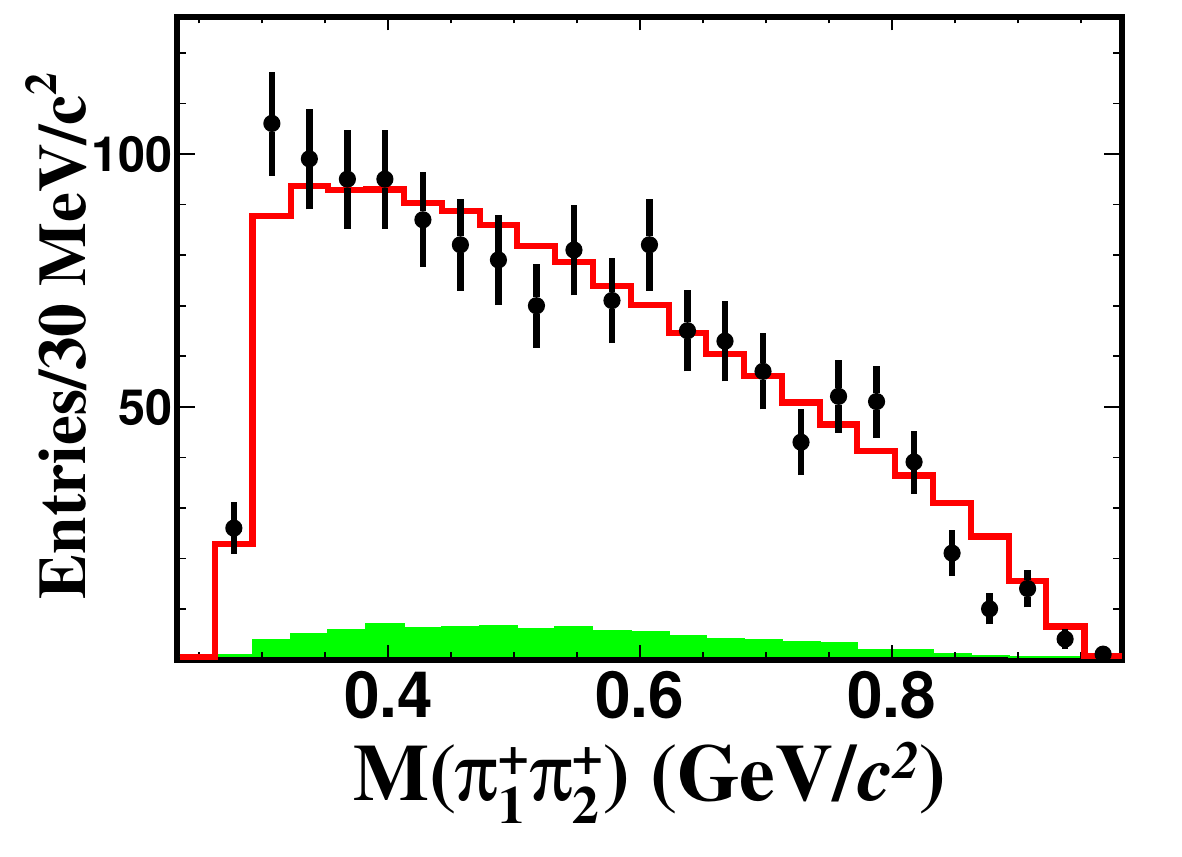}
 \put(65,60){$(h)$}
 \end{overpic}
 \begin{overpic}[width=0.19\textwidth]{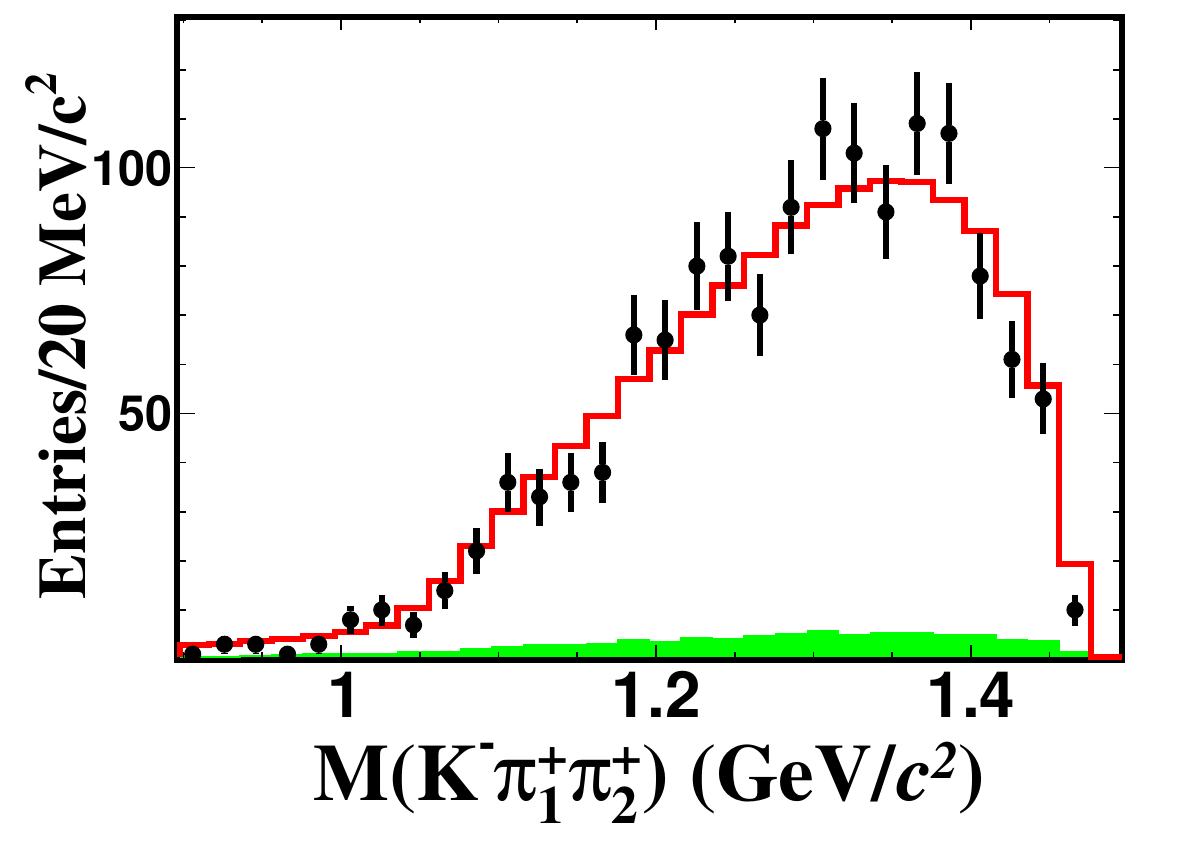}
 \put(65,60){$(i)$}
 \end{overpic}
   \caption{
The projections of the amplitude analysis fit of $D^{+}_{s}\to K^0_SK^-2\pi^+$ on
    two-body and three-body particle mass distributions~\cite{BESIII:2021dot}.
   }
\label{fig:Ds_KSKpipi}
\end{figure*}

Analysis of $D^{+}_{s}\to K^{0}_{S}K^{-}2\pi^{+}$ helps to explore the $D_{s}^+\to K^*(892)^+\bar{K}^*(892)^0$ decay,
thus providing valuable information on $CP$ violation with T-violating triple-products~\cite{Kang:2009iy}.
In addition, studies of $D^{+}_{s}\to K^{0}_{S}K^{-}2\pi^{+}$ also help to explore light hadrons with $J^{P} = 0^-,1^+$,
e.g., $\eta(1295)$, $\eta(1405)$, $\eta(1475)$, $f_1(1285)$, $f_1(1420)$ and $f_1(1510)$, which may decay into the $(K\bar{K}\pi)^0$ final state.
Previous investigations came from CLEO-c and FOCUS;
the former one only reported its branching fraction to be $(1.64\pm0.07\pm0.08)\%$,
based on 586~pb$^{-1}$ of data at 4.17~GeV~\cite{CLEO:2013bae};
while the latter one reported the branching fraction of $D^{+}_{s}\to K^*(892)^+\bar{K}^*(892)^0$ and claimed it is dominant in $D^{+}_{s}\to K^{0}_{S}K^{-}2\pi^{+}$,
but with large uncertainty~\cite{ARGUS:1991jlu}.
From the study of about 1.3k candidates with a signal purity of 95\%,
an amplitude analysis of $D^{+}_{s} \to K^0_SK^-2\pi^{+}$ is presented for the first time~\cite{BESIII:2021dot}.
The amplitude analysis fit projections on two-body or three-body particle mass distributions  are shown in Fig.~\ref{fig:Ds_KSKpipi}.
This analysis indicates that the decay $D^{+}_{s}\to K^*(892)^+\bar K^*(892)^0$ is dominant with a fit fraction of ($40.6\pm2.9\pm4.9)$\%.
In addition, significant contributions from $f_1(1285)$, $\eta(1475)$ and $(K^{*}(892)^{+}K^-)_P$ are observed in the mass spectrum of $K^{0}_{S}K^{-}\pi^{+}$. The $\eta(1475)$ meson decays to both $K^*K$ and $a_0(980)\pi$ final states, while the $f_1(1285)$ meson decays only to $a_0(980)\pi$.
The branching fractions of $D^+_s\to K^0_SK^-2\pi^+$ and $D^{+}_{s}\to K^*(892)^+ \bar K^*(892)^0$ are $(1.46\pm0.05\pm0.05$)\%
and $(5.34\pm0.39\pm0.64)$\%, respectively, which are consistent with the world averages but much more precise.

\begin{figure*}[htbp]
\centering
    \begin{overpic}[width=0.19\textwidth]{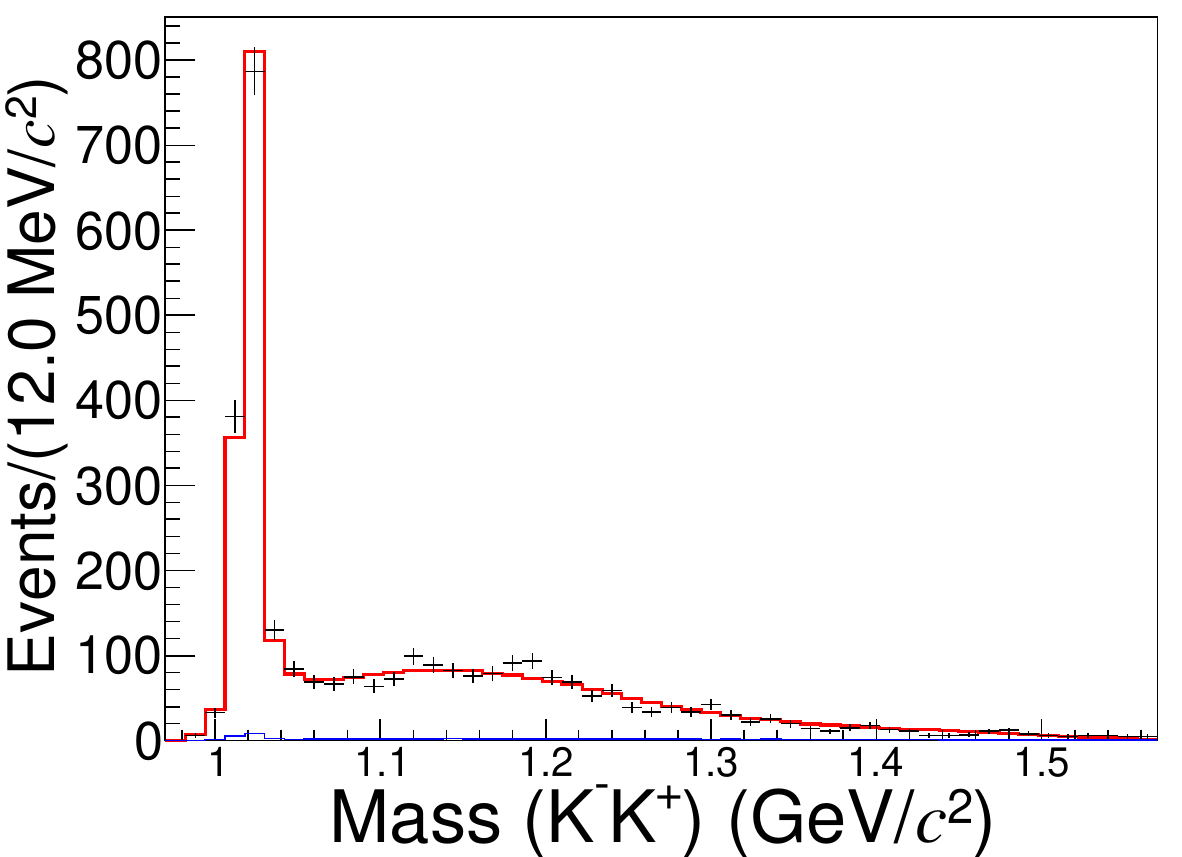}
    \put(80,50){{ (a) }}
    \end{overpic}
    \begin{overpic}[width=0.19\textwidth]{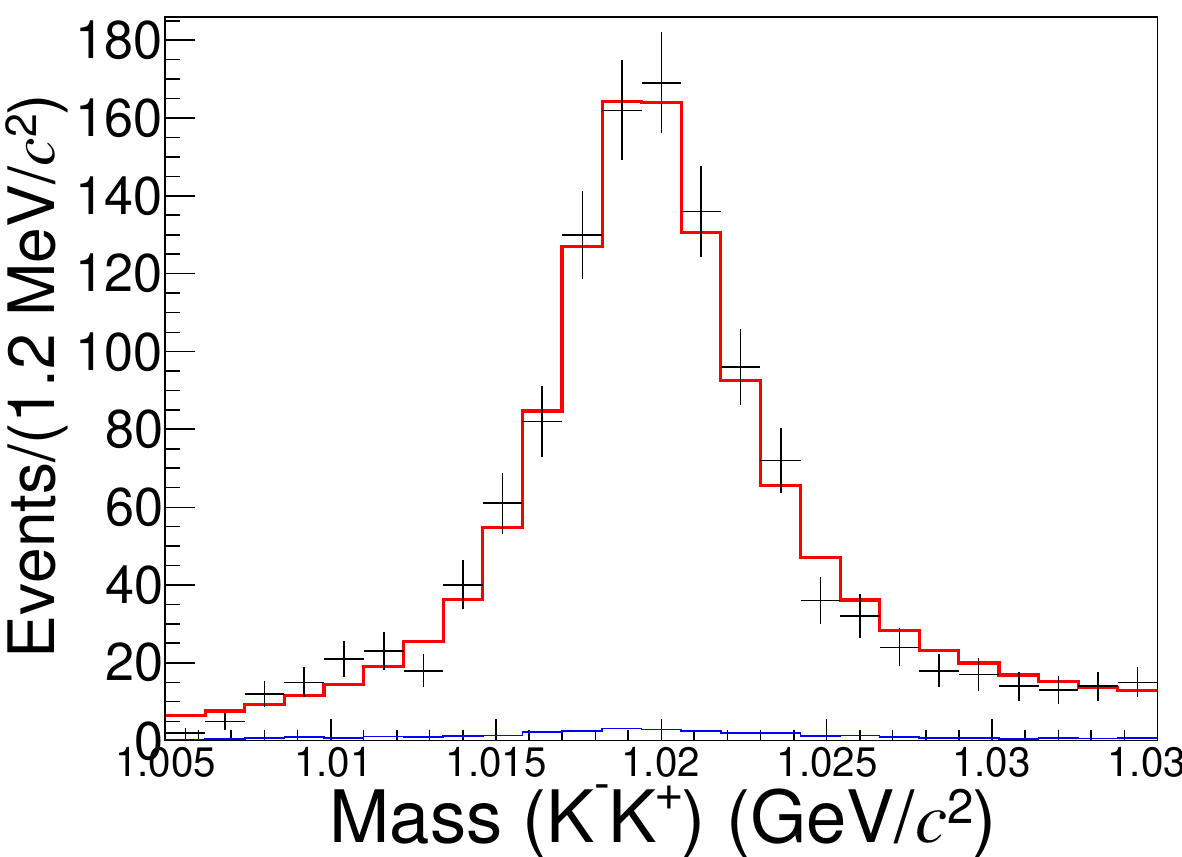}
    \put(80,50){{ (b) }}
    \end{overpic}
    \begin{overpic}[width=0.19\textwidth]{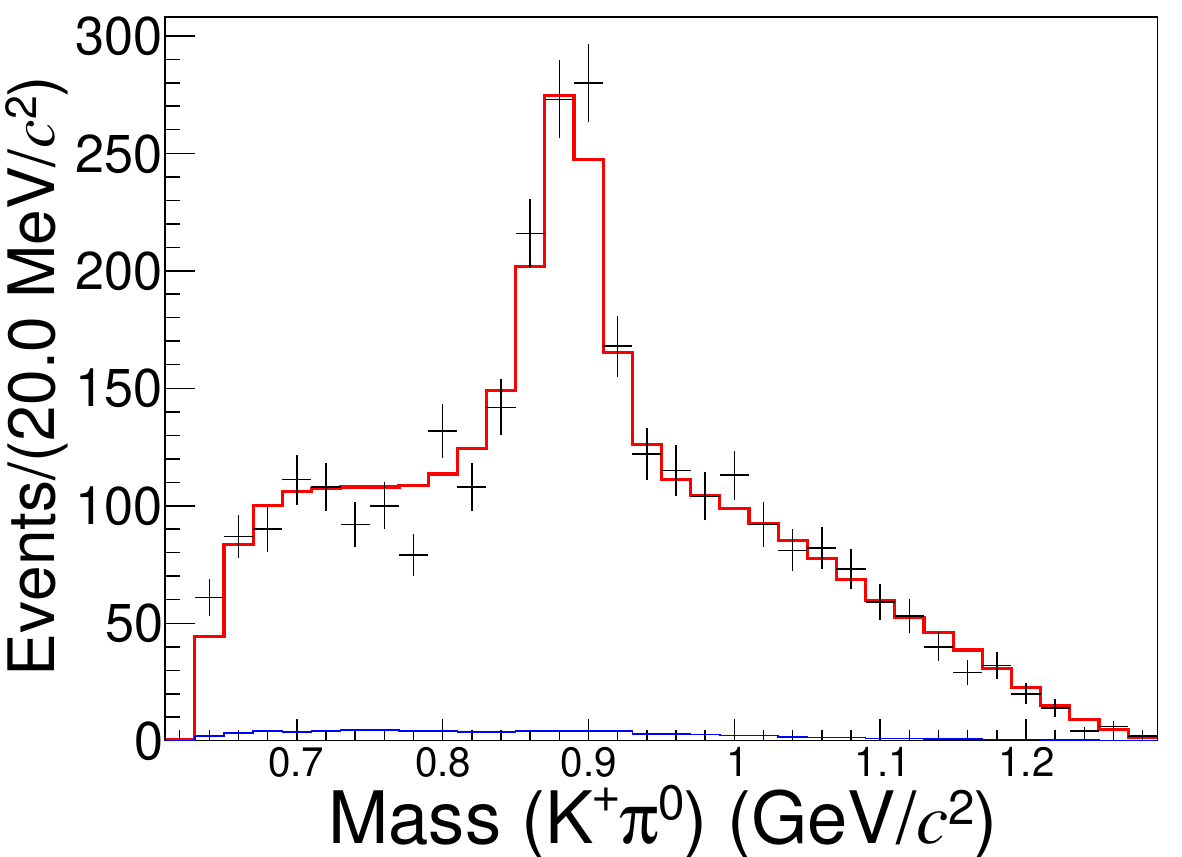}
    \put(80,50){{ (c) }}
    \end{overpic}
      \begin{overpic}[width=0.19\textwidth]{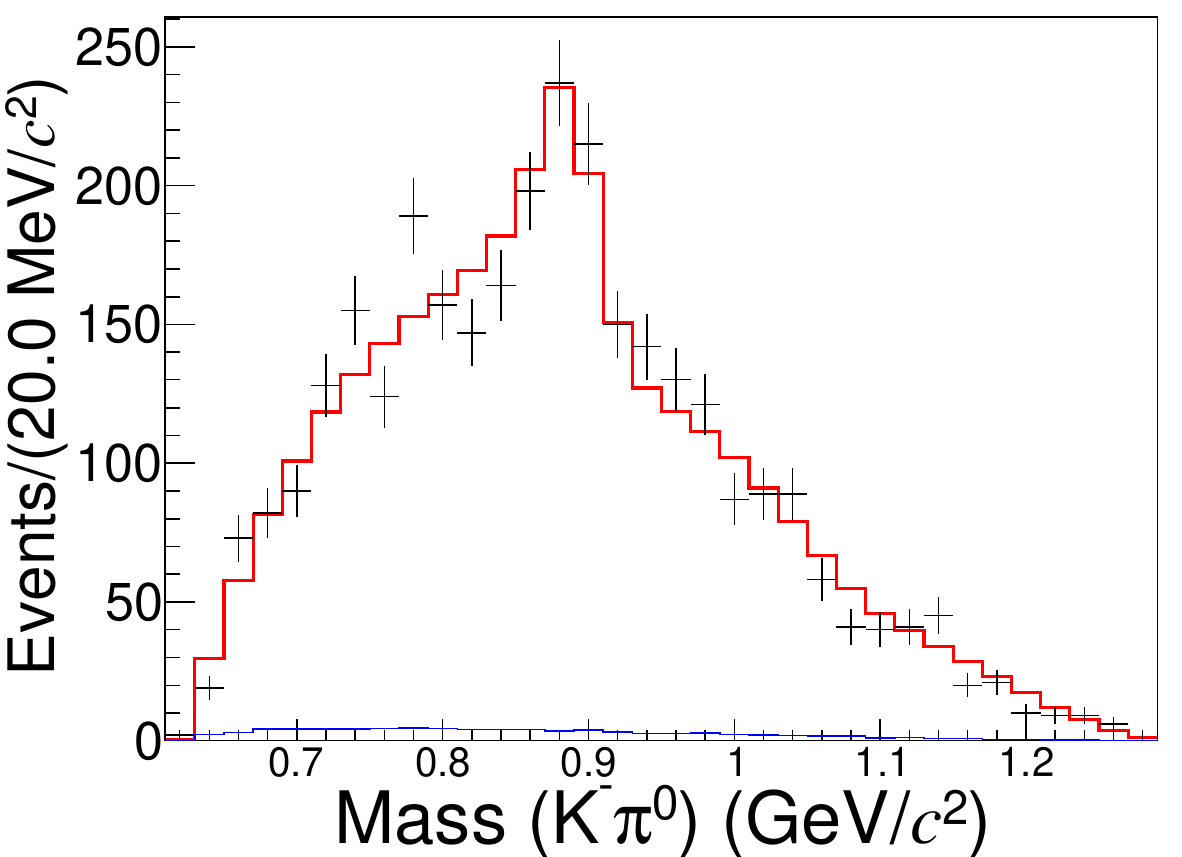}
    \put(80,50){{ (d) }}
    \end{overpic}
    \begin{overpic}[width=0.19\textwidth]{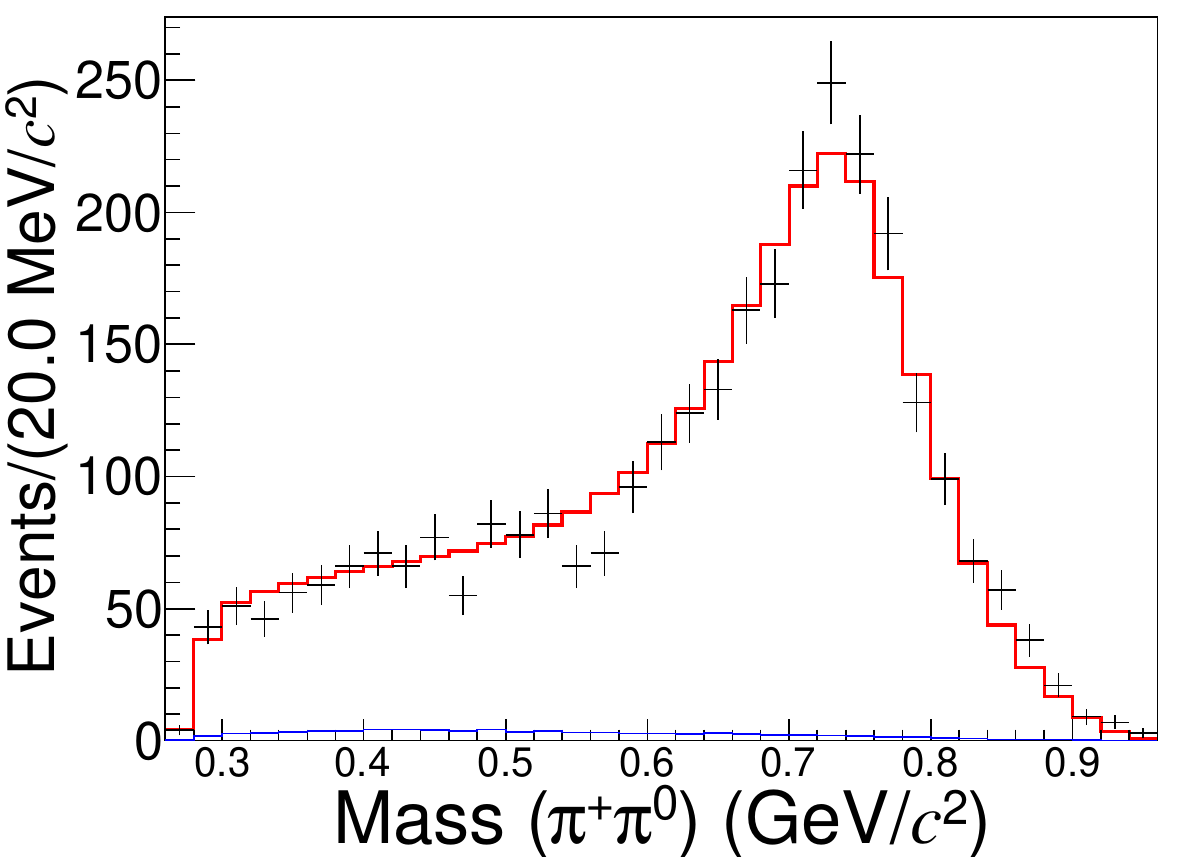}
    \put(80,50){{ (e) }}
    \end{overpic}
    \begin{overpic}[width=0.19\textwidth]{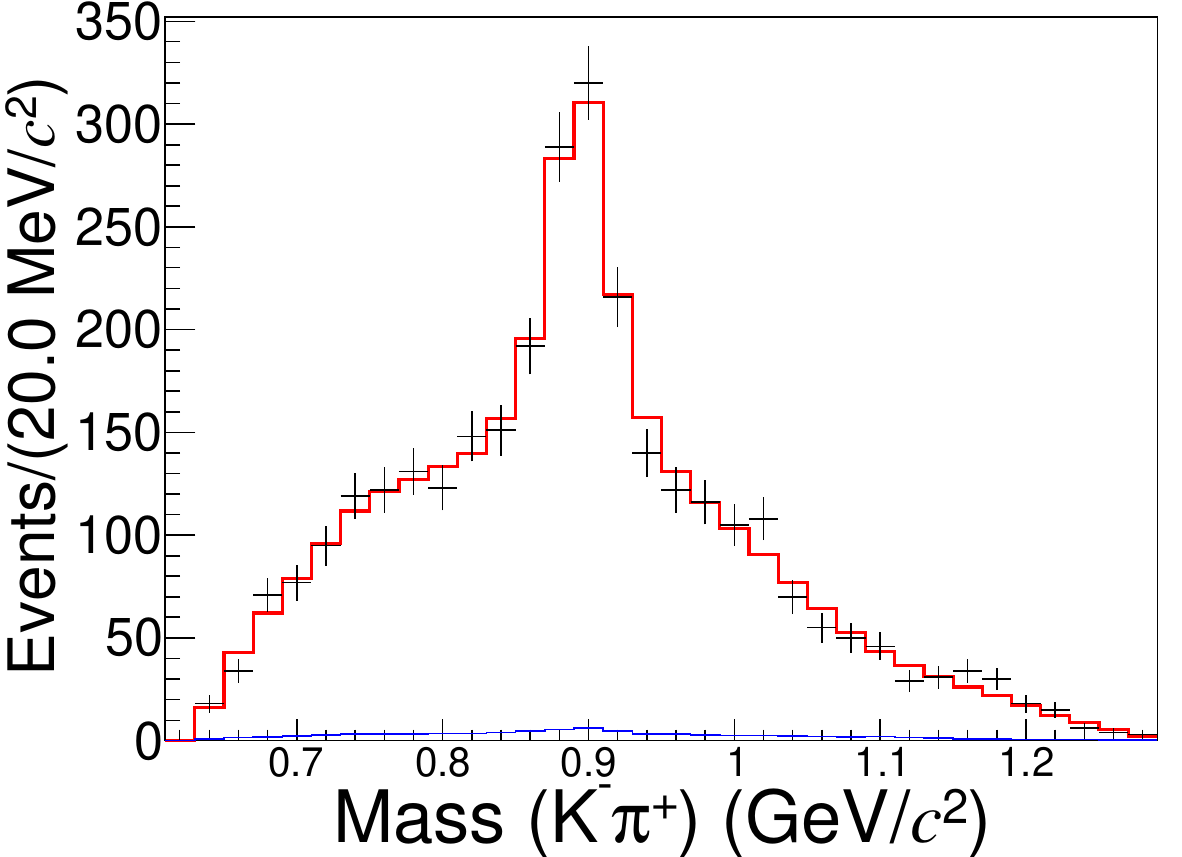}
    \put(80,50){{ (f) }}
    \end{overpic}
     \begin{overpic}[width=0.19\textwidth]{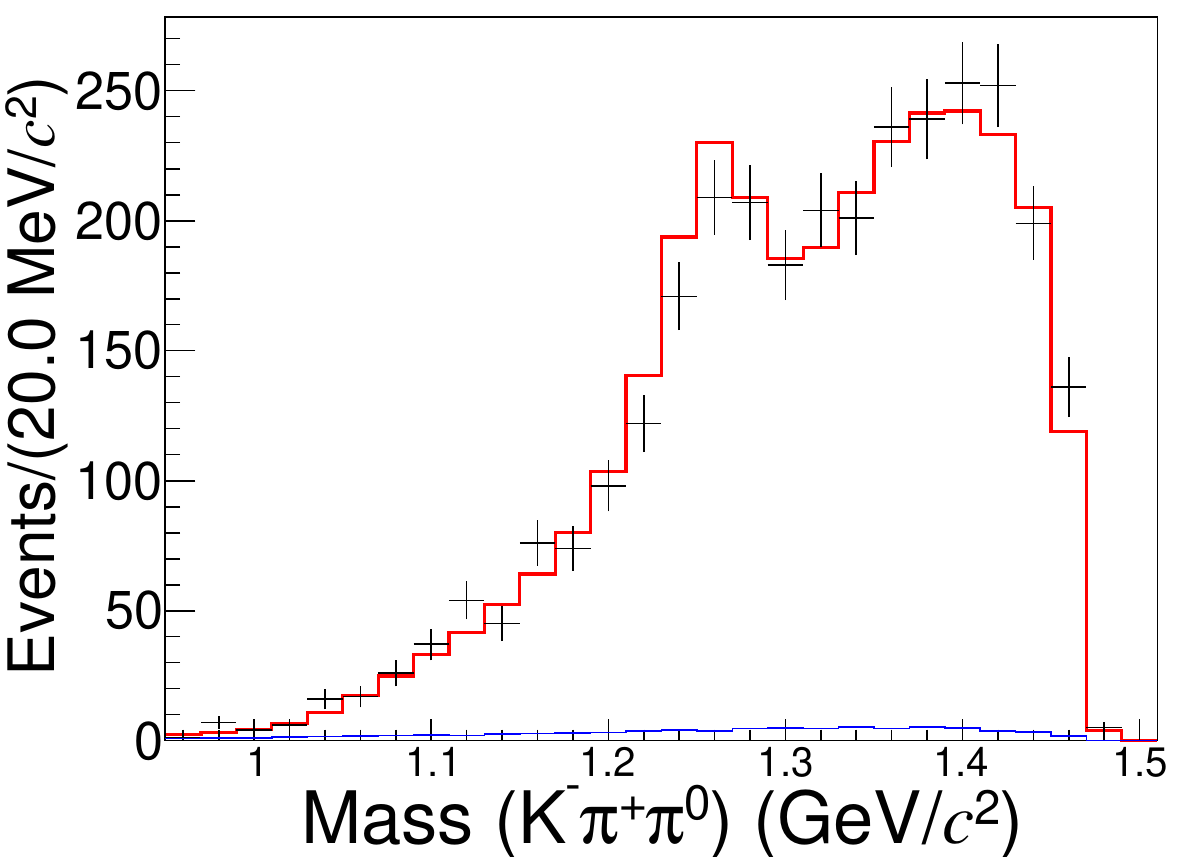}
    \put(30,50){{ (g) }}
    \end{overpic}
    \begin{overpic}[width=0.19\textwidth]{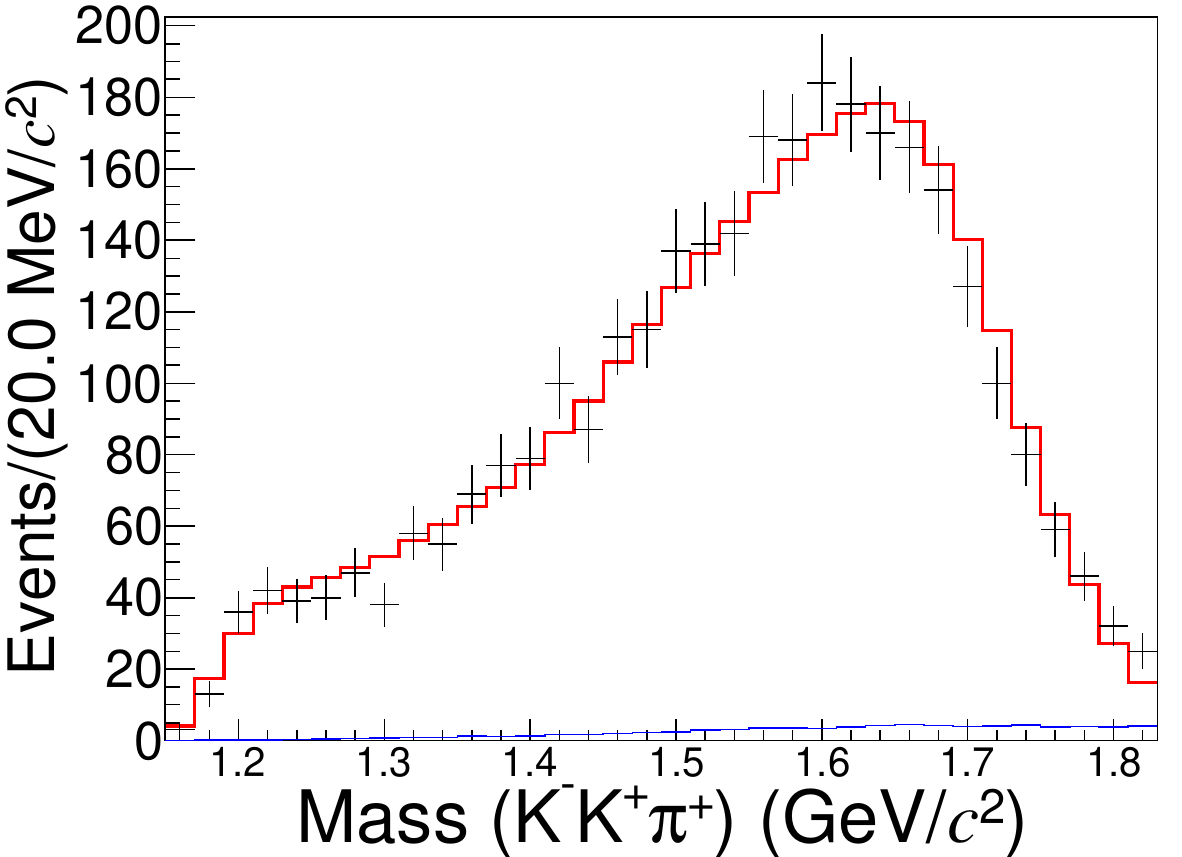}
    \put(30,50){{ (h) }}
    \end{overpic}
    \begin{overpic}[width=0.19\textwidth]{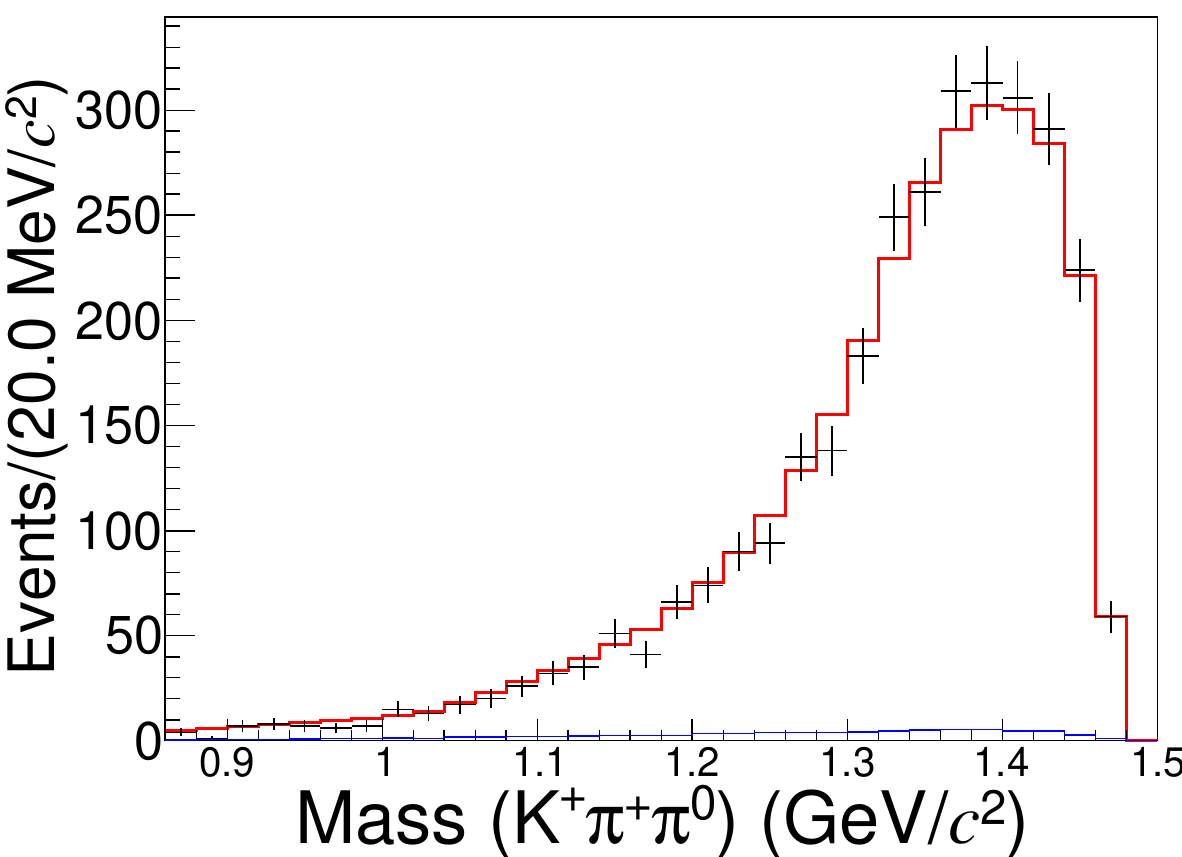}
    \put(30,50){{ (i) }}
    \end{overpic}
    \begin{overpic}[width=0.19\textwidth]{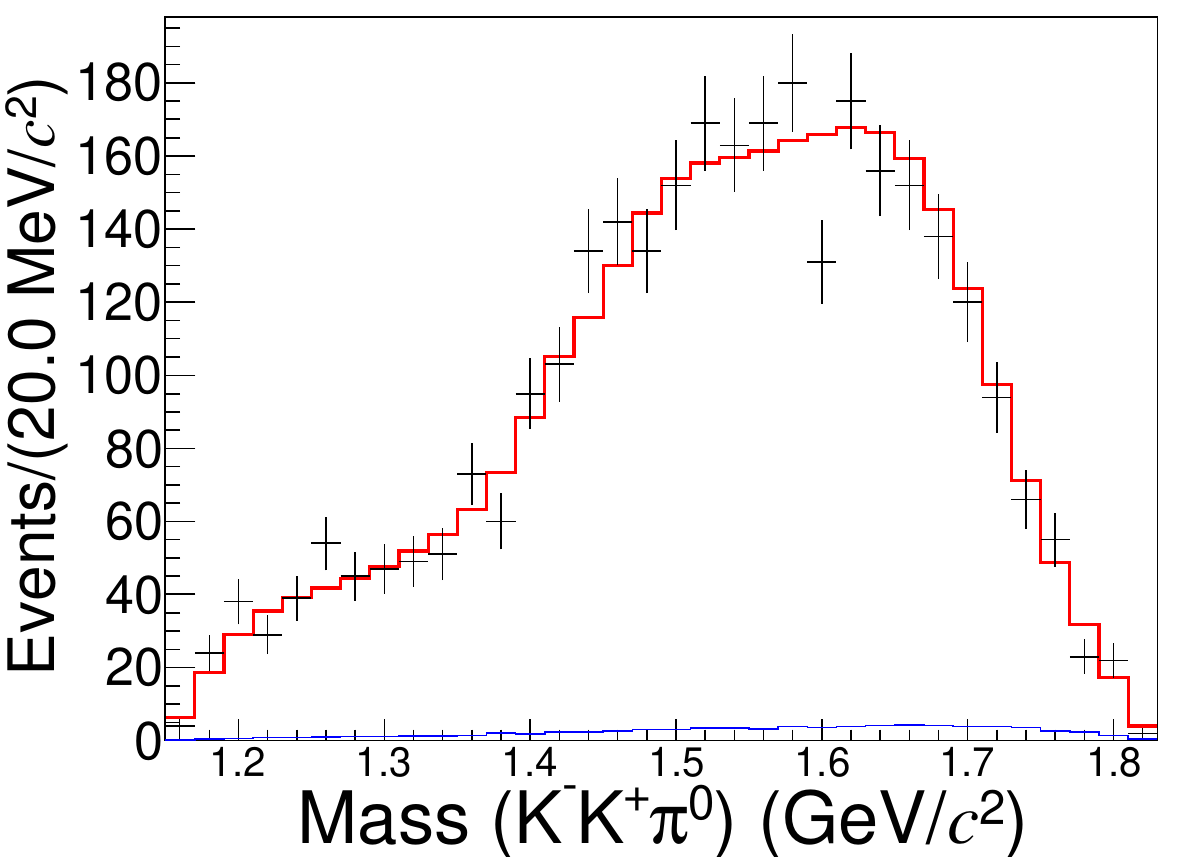}
     \put(30,50){{ (j) }}
    \end{overpic}
\caption{
The projections of the amplitude analysis fit of $D^{+}_{s}\to K^+K^-\pi^+\pi^0$ on
    two-body and three-body particle mass distributions~\cite{BESIII:2021qfo}.
}
\label{fig:Ds_KKpipi0}
\end{figure*}

Study of $D^{+}_{s} \to K^+K^-\pi^{+}\pi^{0}$ provides a valuable opportunity to access both $D_s^+\to AP$ ($A$ could be $K_1(1270)$, $K_1(1400)$, or $f_1(1420)$) and $D_s^+\to VV$ (e.g. $\phi\rho(770)^+$ or $K^*(892)K^*(892)$) processes,
knowledge of which is essential for testing theoretical models and probing $D_s^+$ decay mechanisms~\cite{Bauer:1986bm,Kamal:1990ky,Bedaque:1993fb,Hinchliffe:1995hz}. 
Information of $D_s^+\to AP$ decays helps to constrain the mixing of the $K_1$ axial-vector mesons~\cite{Cheng:2003bn,Cheng:2010vk,Cheng:2010rv,Cheng:2011pb,Cheng:2013cwa,Guo:2018orw};
it also offers insights into the ratio ${\cal R}_{K_1(1270)}=\frac{\mathcal{B}(K_1(1270)^0\to K^*\pi)}{\mathcal{B}(K_1(1270)^0\to K\rho)}$, which is complementary to understand the existing  discrepancies  between measured values across different experiments~\cite{LHCb:2018mzv,CLEO:2012beo,BESIII:2017jyh,LHCb:2017swu,Belle:2010wrf,ACCMOR:1981yww,dArgent:2017gzv},
which is unexpected in theory~\cite{Guo:2018orw}.
Previous measurements of $\mathcal{B}(D_s^+\to \phi\rho(770)^+)=(8.4^{+1.9}_{-2.3})\%$ by CLEO~\cite{CLEO:1992jqc}, and $\mathcal{B}(D_s^+\to \bar{K}^{0}K^{+})=(7.2\pm2.6)\%$ by ARGUS~\cite{ARGUS:1991jlu} were only based on limited data samples.
Reference~\cite{BESIII:2021qfo} reported an amplitude analysis of $D^{+}_{s} \to K^+K^-\pi^{+}\pi^{0}$ for the first time,
with about 3.1k candidates with a signal purity of 97.5\%.
The amplitude analysis fit projections on two-body or three-body particle mass distributions  are shown in Fig.~\ref{fig:Ds_KKpipi0}.
Its amplitude consists of $D_s^+\to \phi\rho(770)^+$,
$D_s^+\to \bar K^*(892)^0K^*(892)^+$,
$D_s^+\to a_0(980)^0 \rho(770)^+$,
$D_s^+\to \bar K _1(1270)^0K^+(\bar K_1(1270)^0\to K^-\rho(770)^+)$,
$D_s^+\to \bar K_1(1270)^0K^+(\bar K_1(1270)^0\to K^{*}(892)\pi)$,
$D_s^+\to \bar K_1(1400)^0 K^+(\bar K_1(1400)^0\to K^{*}(892)\pi)$,
$D_s^+\to f_1(1420)\pi^+(f_1(1420)\to K^{*}(892)^\mp K^{\pm})$,
$D_s^+\to f_1(1420)\pi^+(f_1(1420)\to a_0(980)^0\pi^0)$, and
$D_s^+\to \eta(1475)\pi^+(\eta(1475)\to a_0(980)^0\pi^0)$.
The $D_s^+\to \phi \rho(770)^+$ and $D_s^+\to \bar K^*(892) K^*(892)^+$ decays are found to be
dominant, and the decays involving $K_1(1270), K_1(1400), \eta(1475),
f_1(1420)$, and $a_0(980)^0$ mesons are also observed with significances
larger than $4\sigma$.
The obtained branching fractions
$\mathcal{B}(D_s^+\to K^-K^+\pi^+\pi^0)=(5.42\pm0.10\pm0.17)$\%, $\mathcal{B}(D_s^+\to \phi\rho(770)^+)=(5.59\pm0.15\pm0.30)$\% and $\mathcal{B}(D_s^+\to \bar{K}^{*}(892)^0K^{*}(892)^+)=(5.64\pm0.23\pm0.27)$\% have much better precision compared to the previous world average values.
The measured $\mathcal{B}(D_s^+\to \phi\rho(770)^+)$ is consistent with the theoretical prediction~\cite{Bedaque:1993fb} (5.70\%),
while that of $D_s^+\to \bar{K}^{*}(892)^0 K^{*}(892)^+$ is still much larger than its prediction (1.5\%).
The ratio ${\cal R}_{K_1(1270)}=\frac{{\cal
		B}(K_1(1270)^0\to K^{*}(892)\pi)}{{\cal B}(K_1(1270)^0\to K\rho)}$
is determined to be $0.99\pm0.15\pm0.18$, which is consistent with the LHCb~\cite{LHCb:2018mzv} and CLEO-c~\cite{CLEO:2012beo} measurements.

\begin{figure*}
\centering
\includegraphics[width=0.225\textwidth]{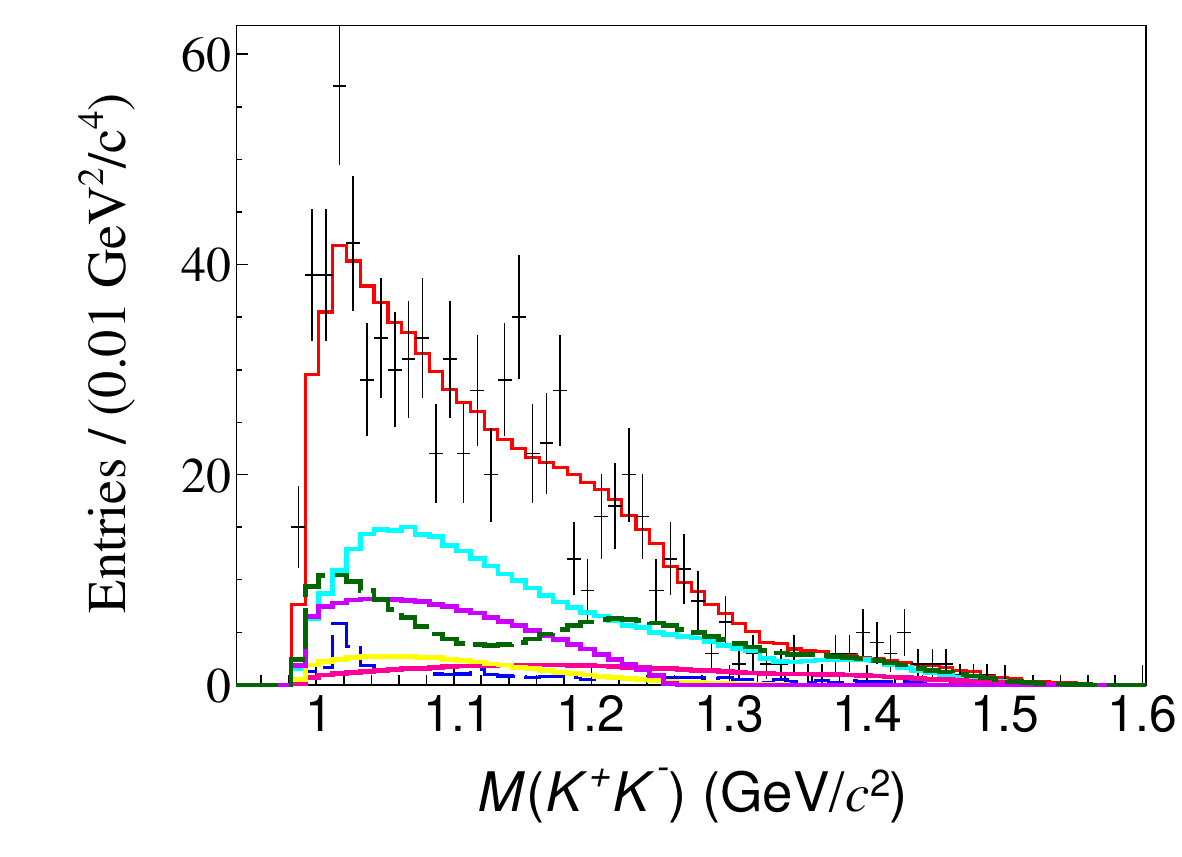}
\includegraphics[width=0.225\textwidth]{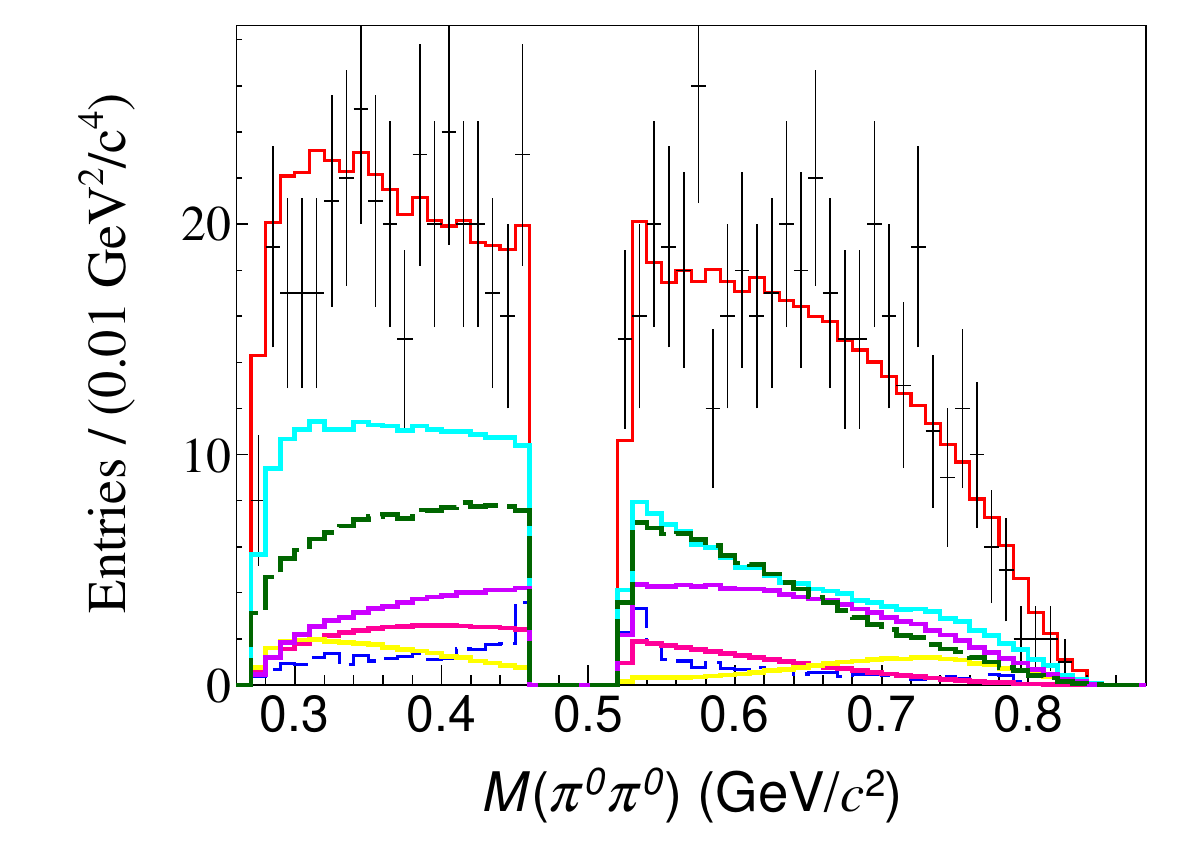}
\includegraphics[width=0.225\textwidth]{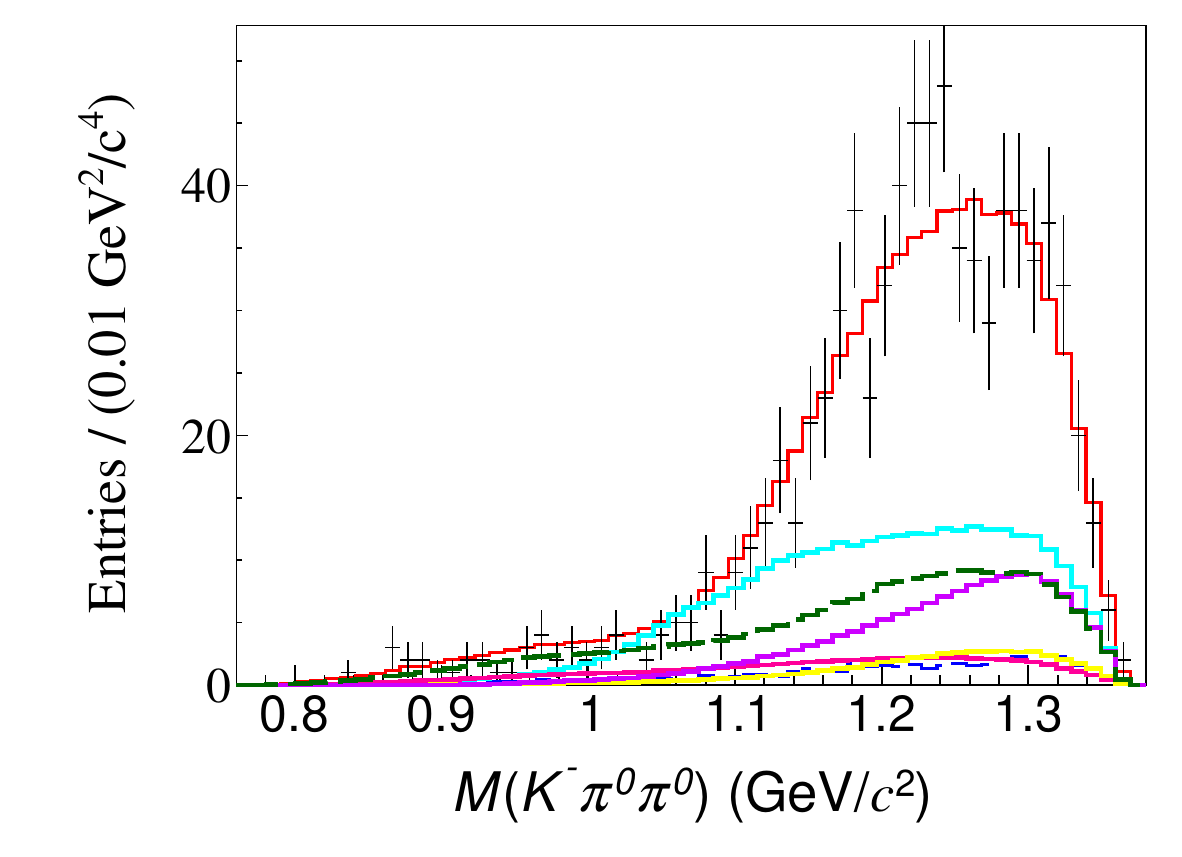}
\includegraphics[width=0.225\textwidth]{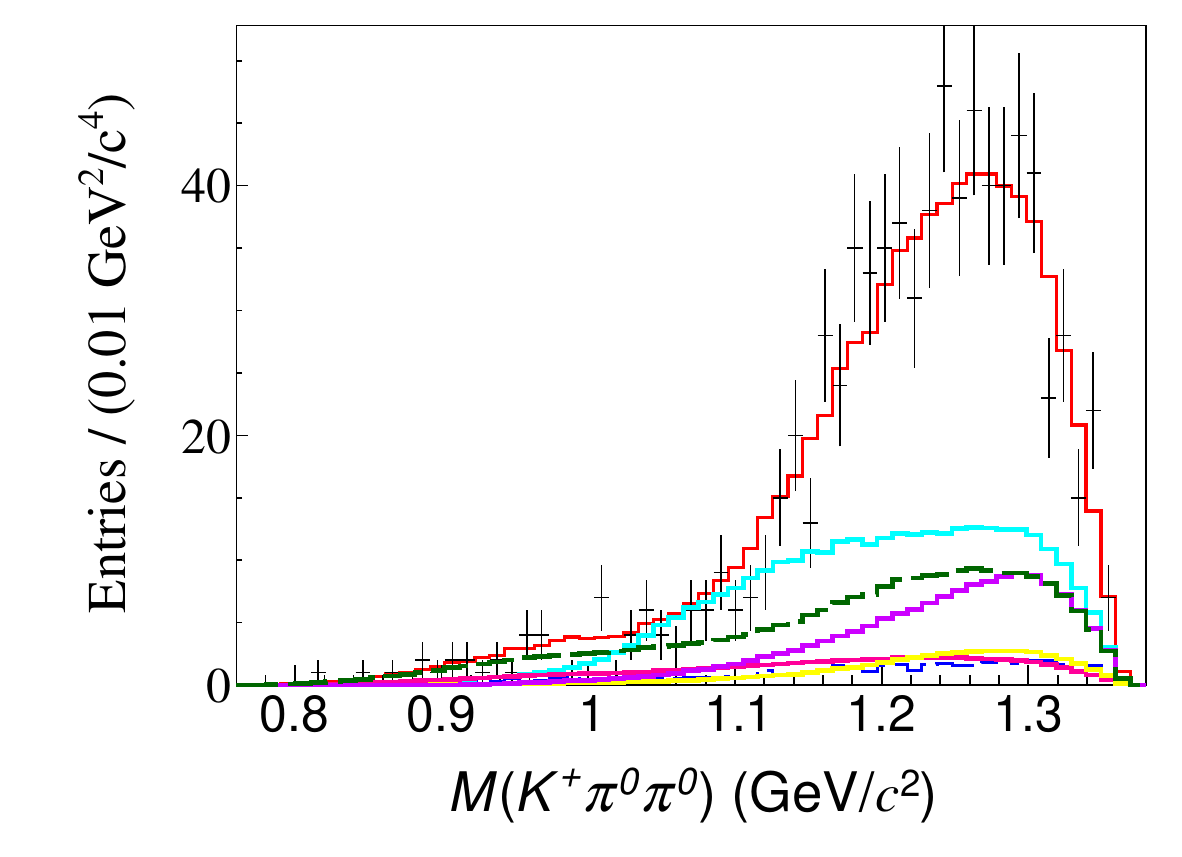}
\includegraphics[width=0.225\textwidth]{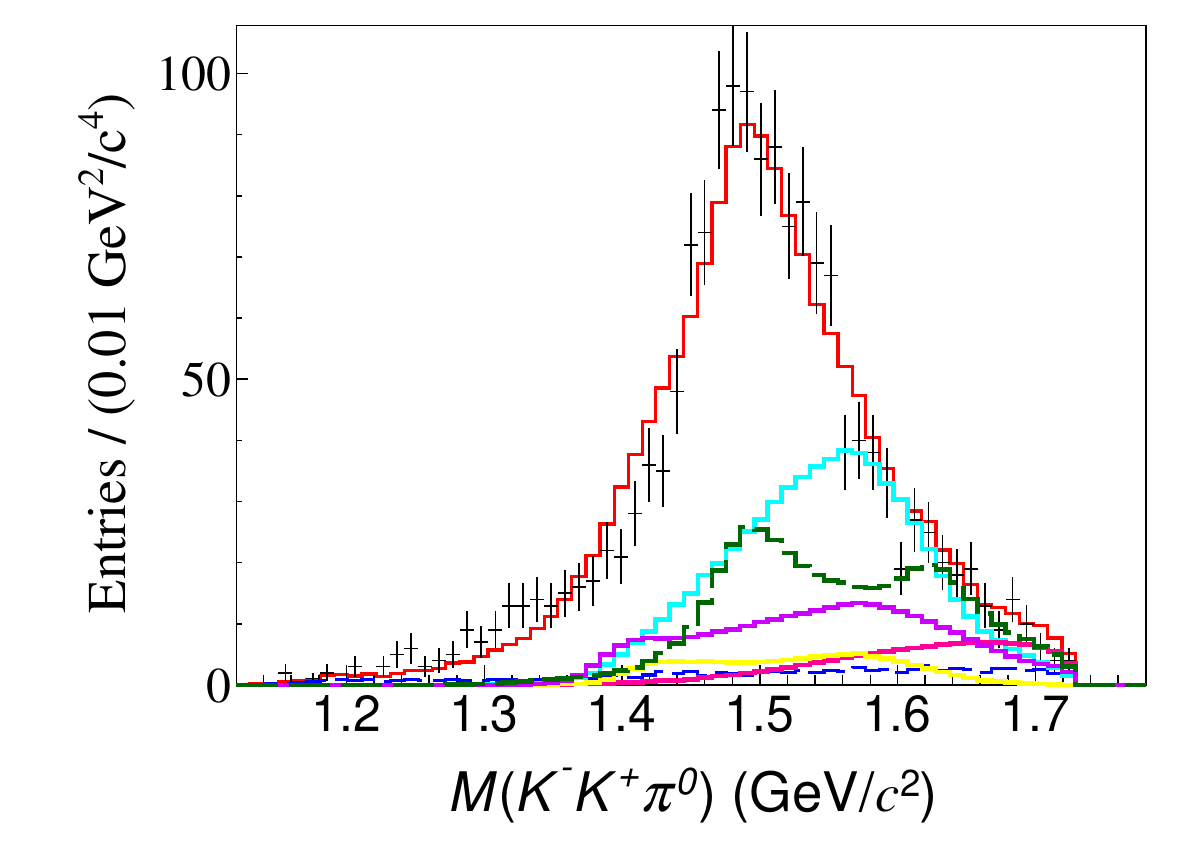}
\includegraphics[width=0.225\textwidth]{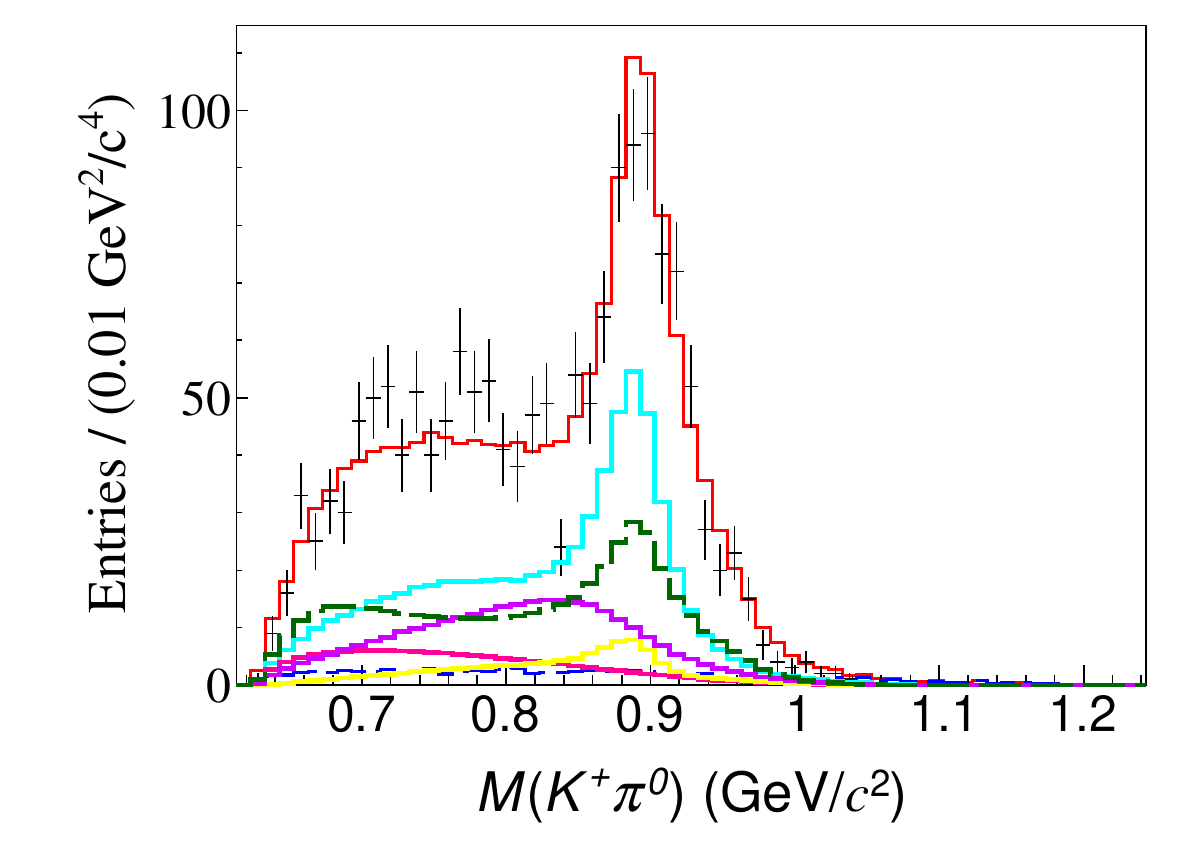}
\includegraphics[width=0.225\textwidth]{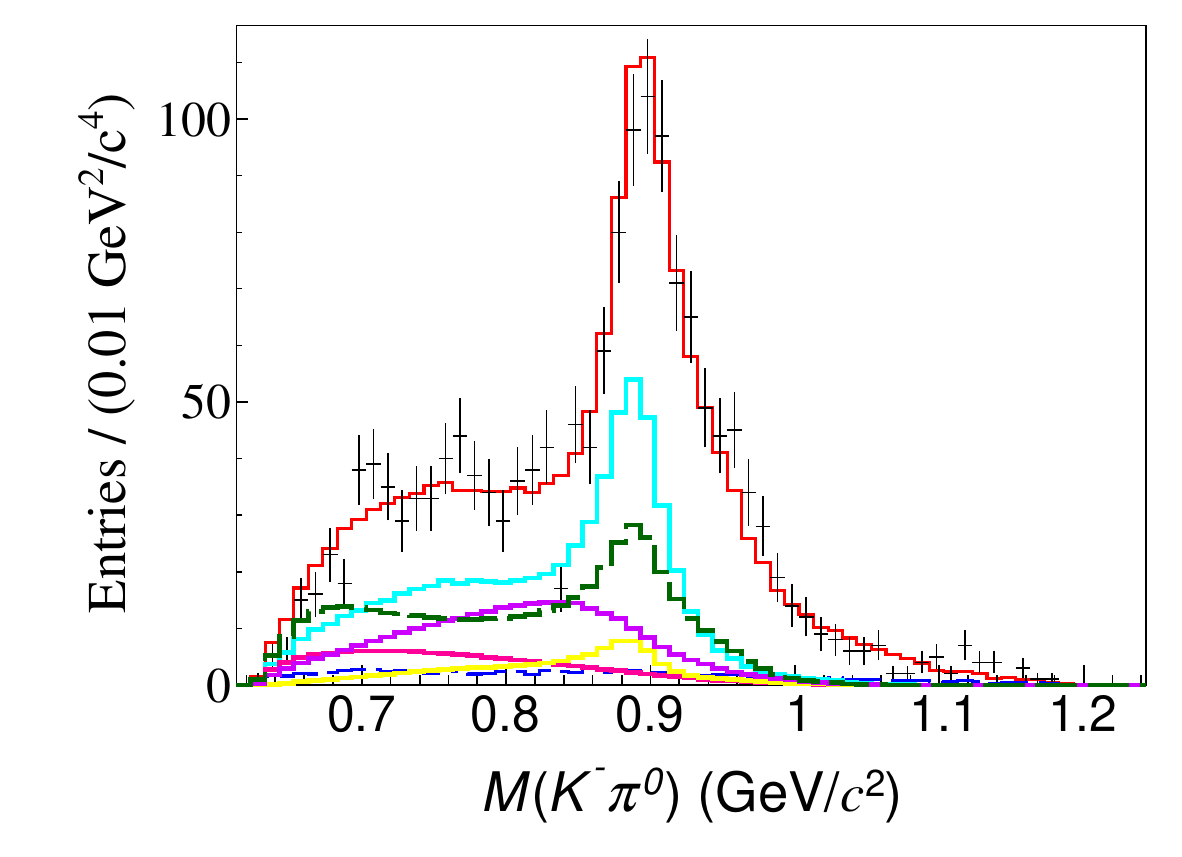}
\includegraphics[width=0.225\textwidth]{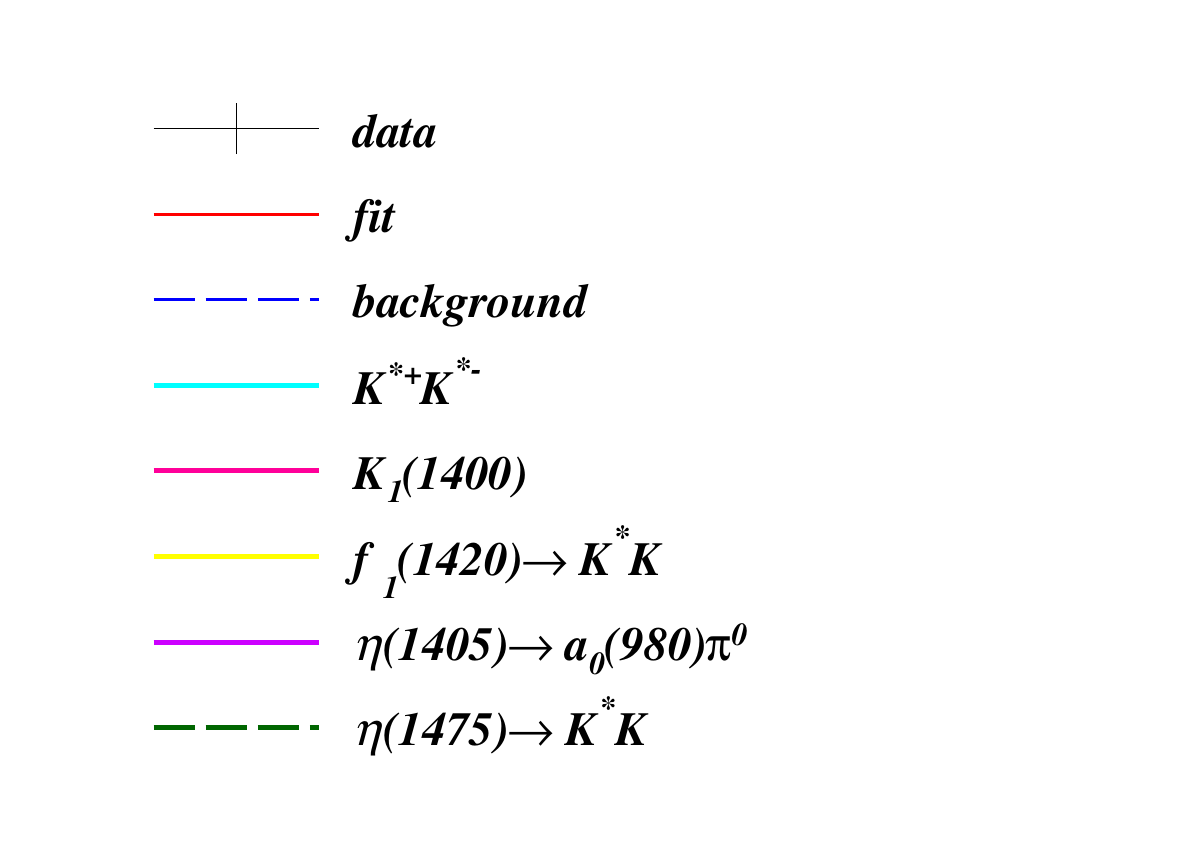}
 \caption{
The projections of the amplitude analysis fit of $D^0\to K^+K^-2\pi^0$ on
    two-body and three-body particle mass distributions~\cite{BESIII:2026dwz}.
 }
    \label{fig:D0_KKpi0pi0}
\end{figure*}

Both non-relativistic constituent quark models, with and without FSIs, predict transverse-polarization dominance in the $D^0 \to K^{*}(892)^+ K^{*}(892)^-$ process, with longitudinal polarization fractions of $0.32^{+0.02}_{-0.01}$ and 0.31, respectively~\cite{Cao:2023csx}. Amplitude analysis of the SCS decay $D^0 \to K^+K^-2\pi^0$  offers an opportunity to
determine its polarization fractions in $D^0 \to K^{*}(892)^+ K^{*}(892)^-$, thus testing theoretical predictions. Furthermore, the $D^0 \to K^+K^-2\pi^0$ decay also encompasses numerous intermediate $D \to AP$ processes, which provide unique information to study axial-vector mesons, such as $K_1(1270)$, $K_1(1400)$, $f_1(1420)$, etc.~\cite{Cheng:2013cwa,Cheng:2011pb}.
Recently, Ref.~\cite{BESIII:2026dwz} reported an amplitude analysis of $D^0 \to K^+ K^- 2\pi^0$ for the first time.
The analysis is conducted with a total of 791 events with a signal purity of 94.2\%.
The amplitude analysis fit projections on two-body and three-body particle mass distributions are shown in Fig.~\ref{fig:D0_KKpi0pi0}.
The branching fraction of $D^0 \to K^+ K^- 2\pi^0$ is measured to be $(0.73\pm0.03\pm0.01)\times 10^{-3}$.
Especially, the dominant intermediate process is $D^0 \to K^{*}(892)^+K^{*}(892)^-$, with a branching fraction of $(2.79 \pm 0.13 \pm 0.11) \times 10^{-3}$.
It is significantly lower than those from Refs.~\cite{Cao:2023csx,Cheng:2010rv}, but in agreement with the prediction based on the flavor SU(3) symmetry model~\cite{Kamal:1990ky}.
Amplitude analysis reveals that the $D^0 \to K^{*}(892)^+K^{*}(892)^-$ decay is $\cal S$-wave dominant. Its longitudinal polarization fraction is measured to be $0.468\pm0.046\pm0.011$,  which is higher than the prediction in Ref.~\cite{Cao:2023csx} by more than three standard deviations.

\subsubsection{Analyses of $D^+_{(s)}\to K\pi\pi\pi$}

\begin{figure*}[!htbp]
 \centering
    \includegraphics[width=0.225\textwidth]{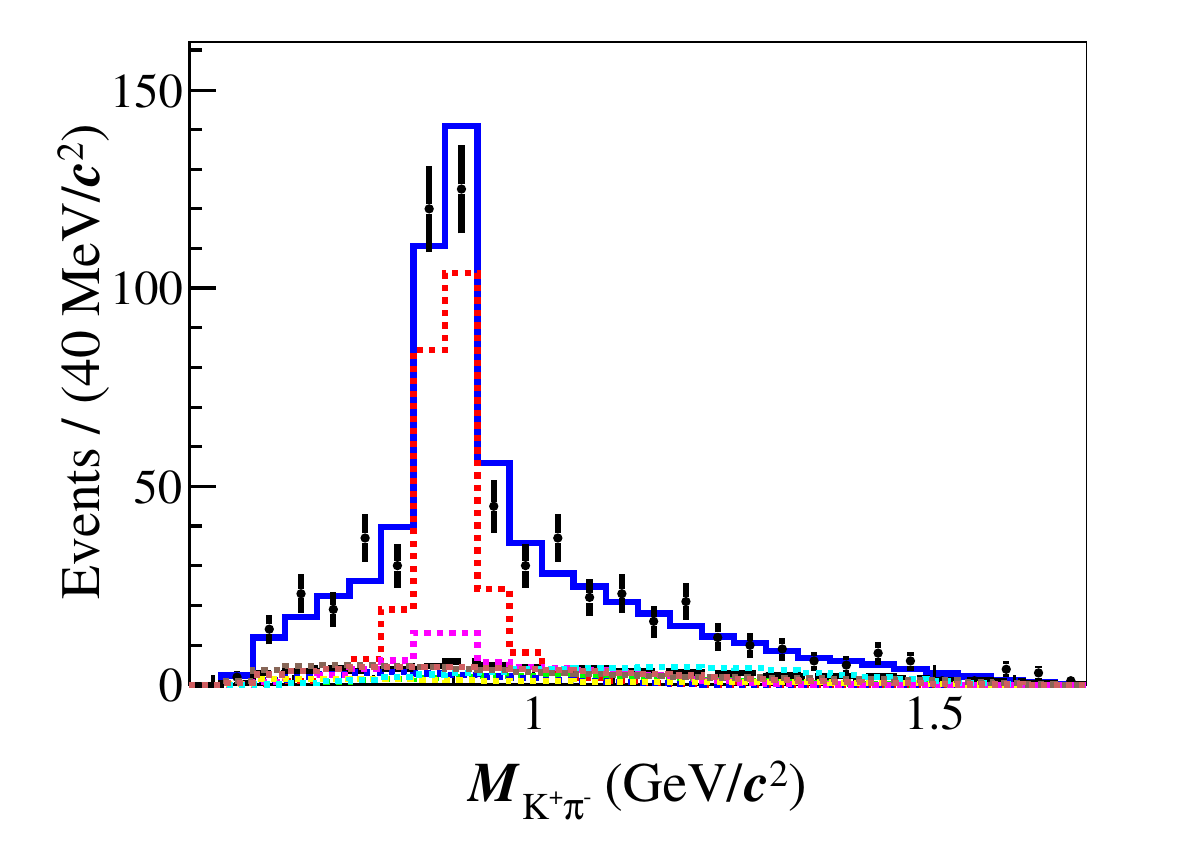}
    \includegraphics[width=0.225\textwidth]{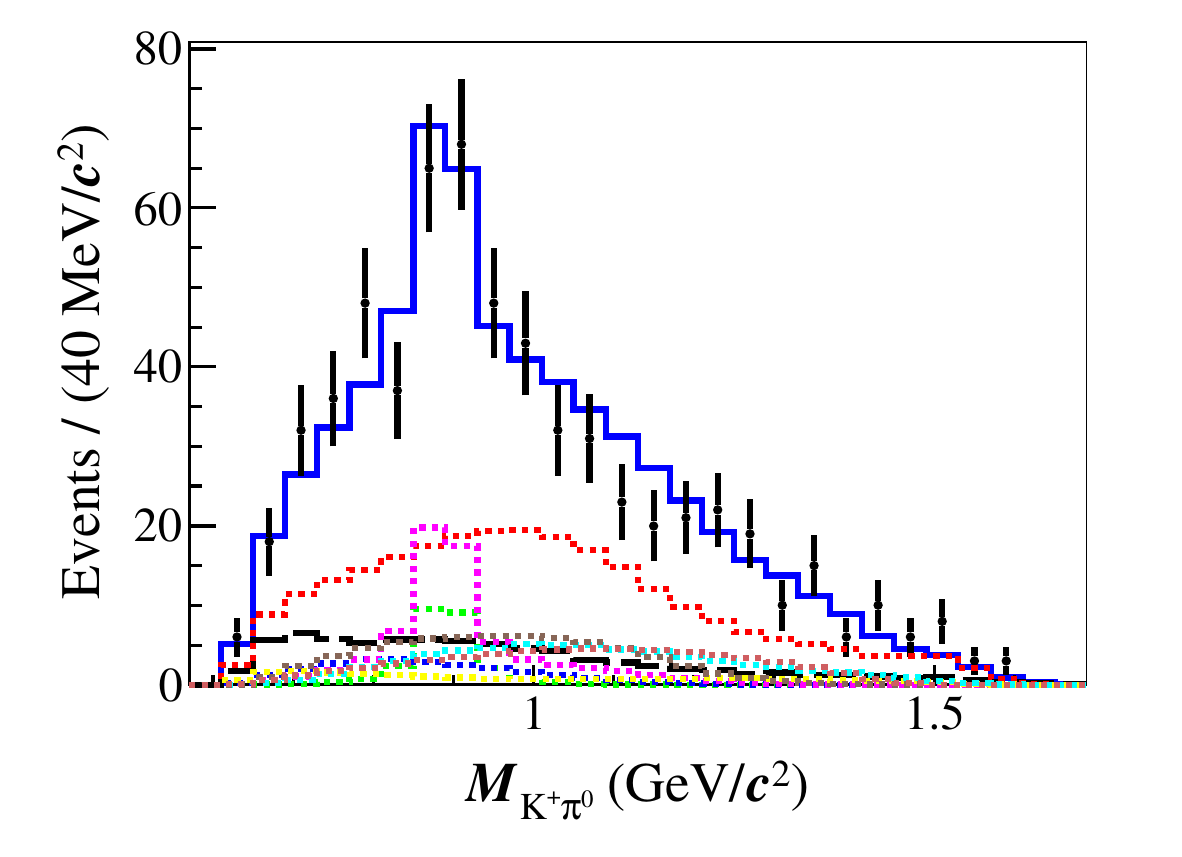}
    \includegraphics[width=0.225\textwidth]{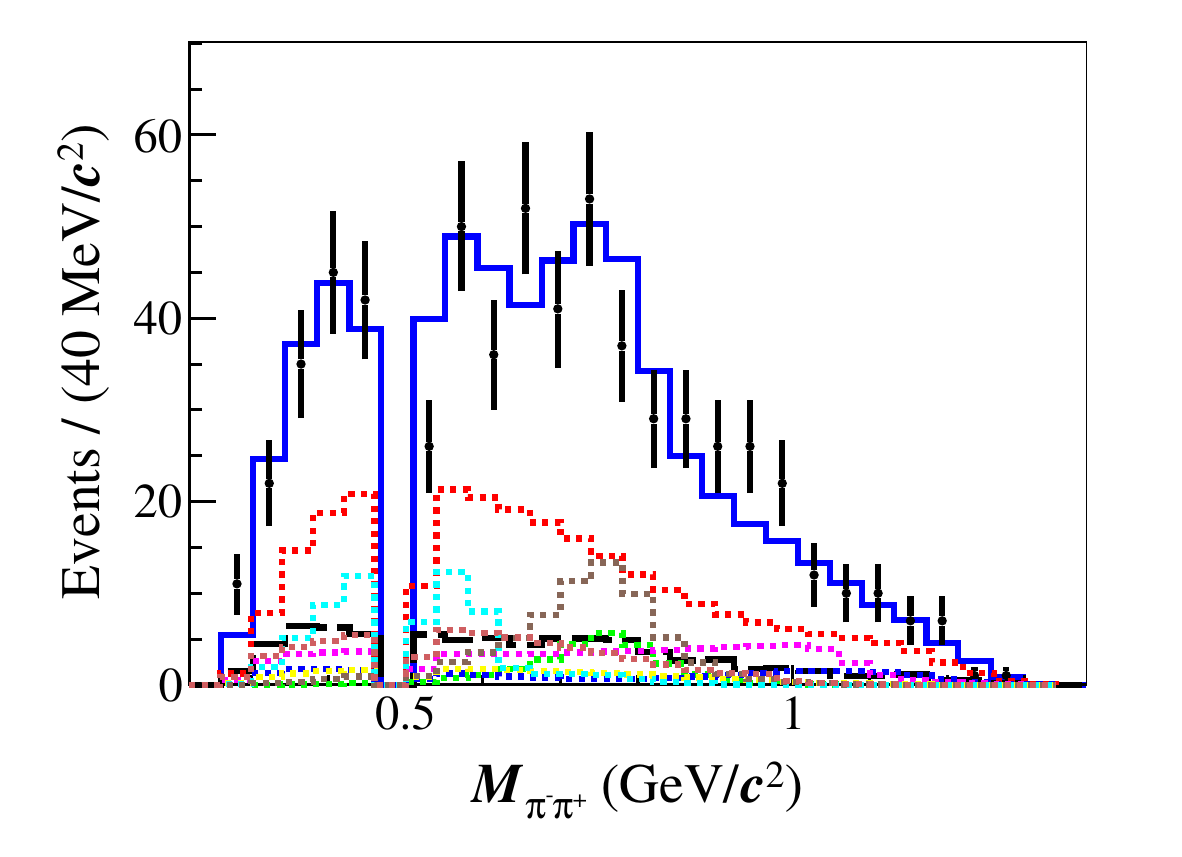}
    \includegraphics[width=0.225\textwidth]{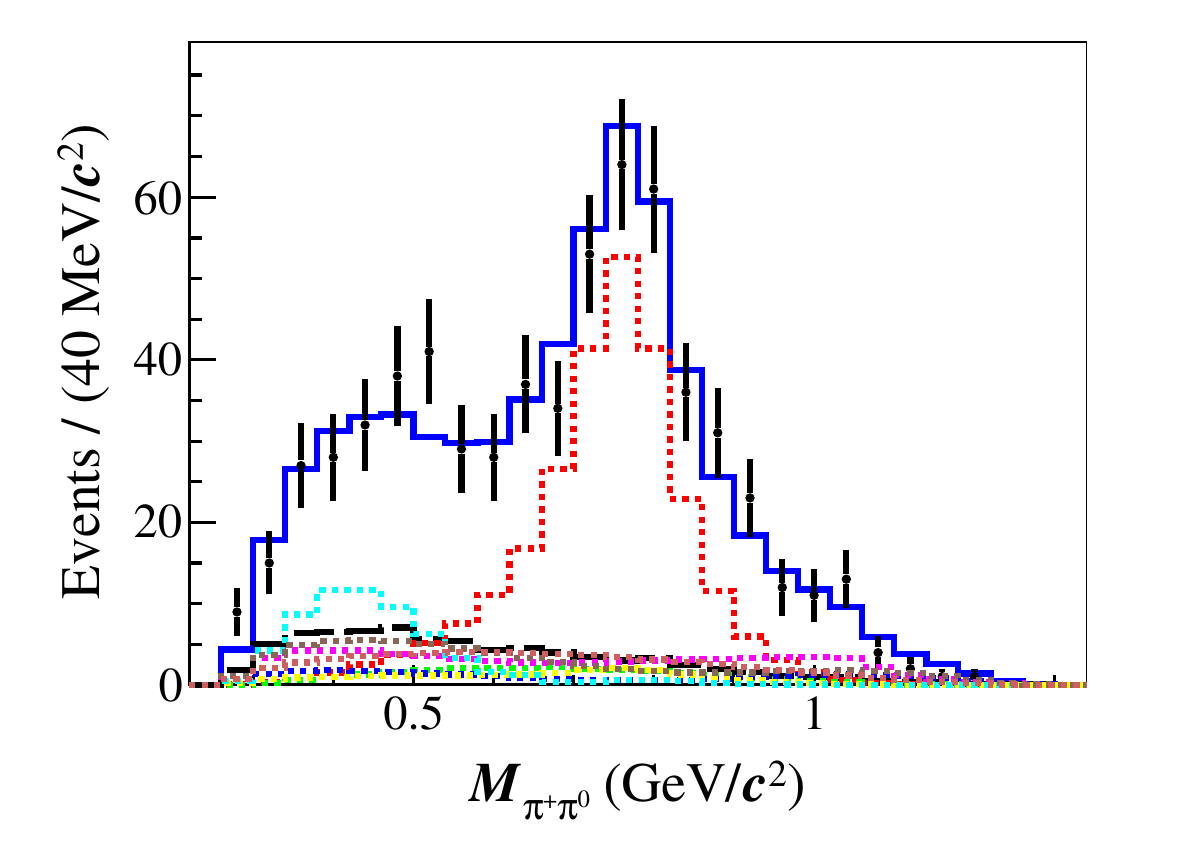}
    \includegraphics[width=0.225\textwidth]{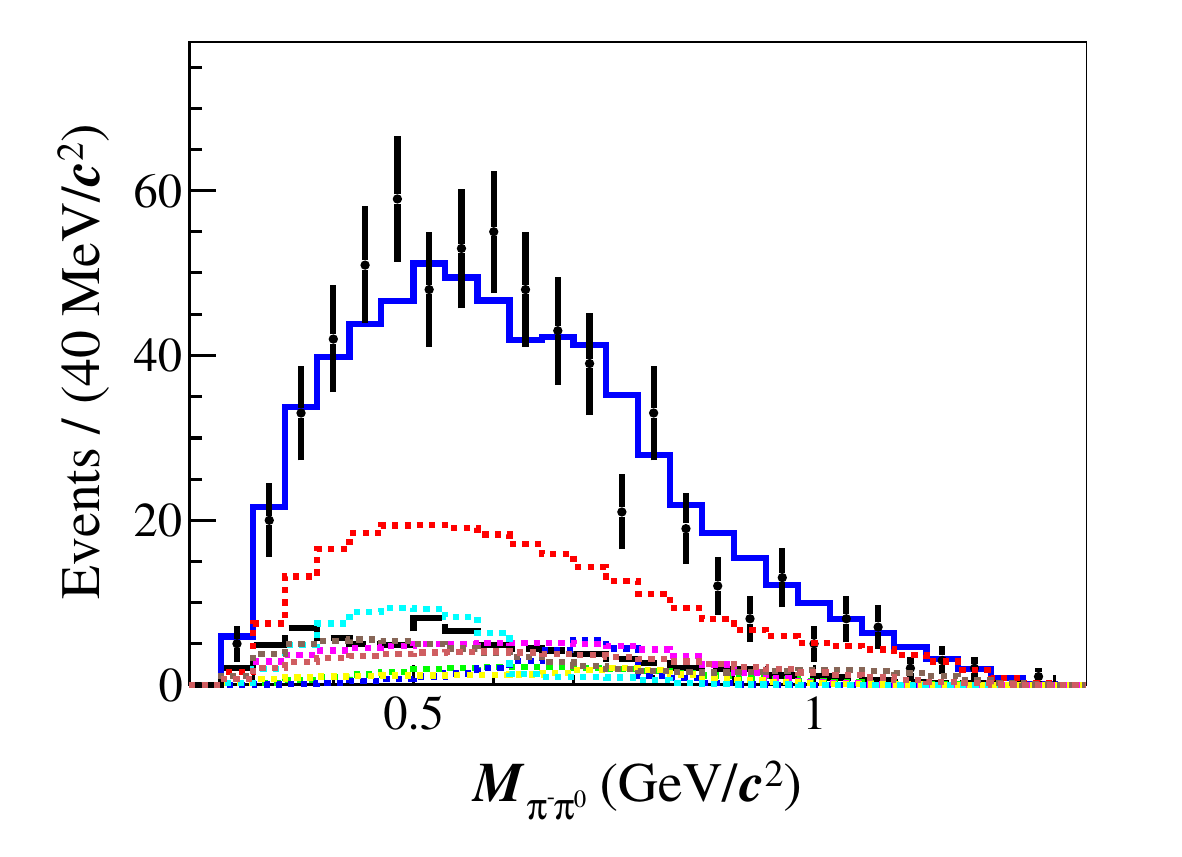}
    \includegraphics[width=0.225\textwidth]{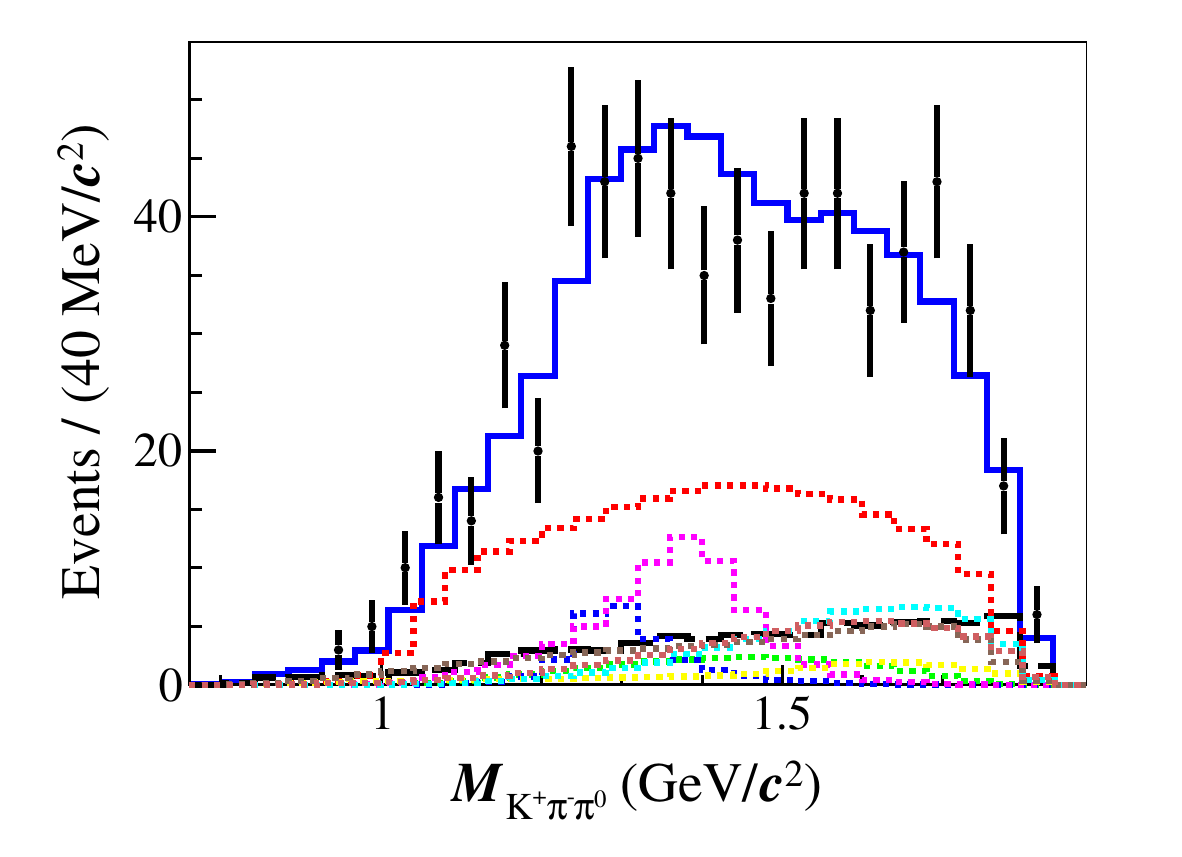}
    \includegraphics[width=0.225\textwidth]{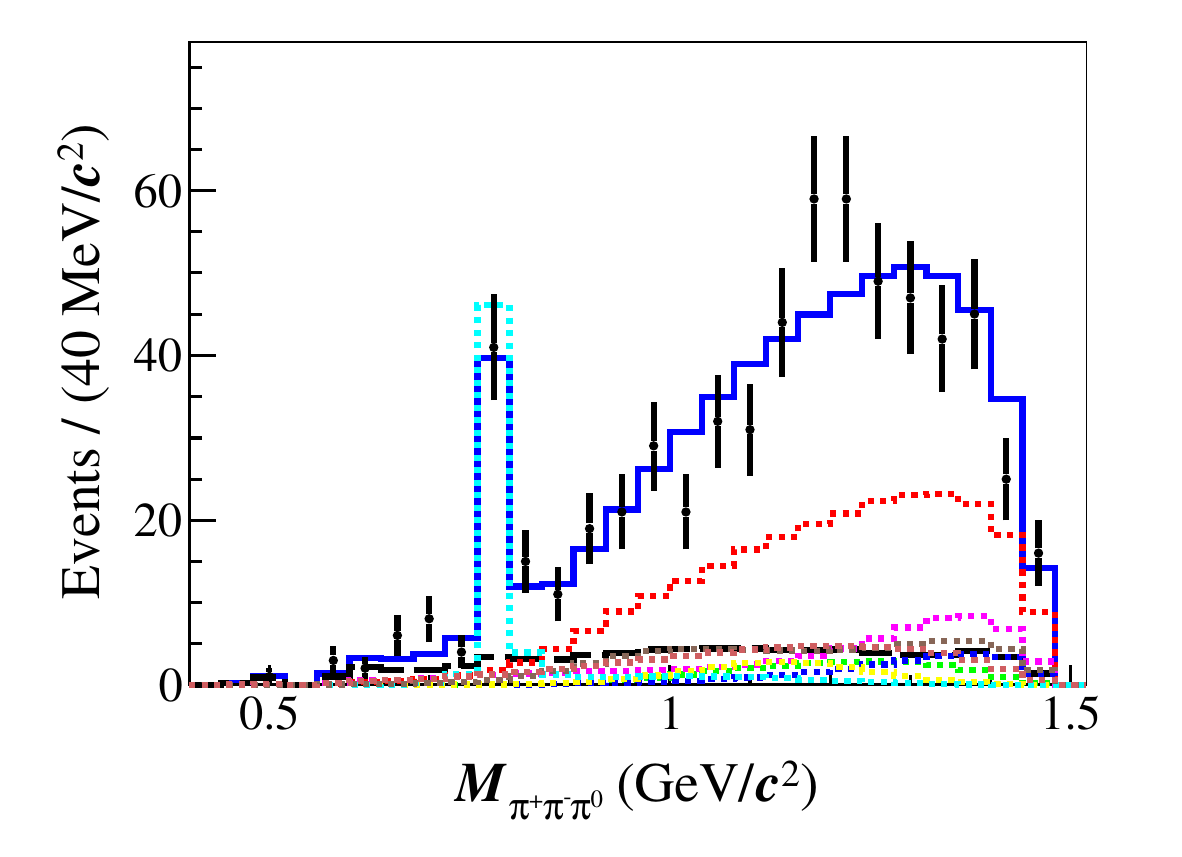}
    \includegraphics[width=0.225\textwidth]{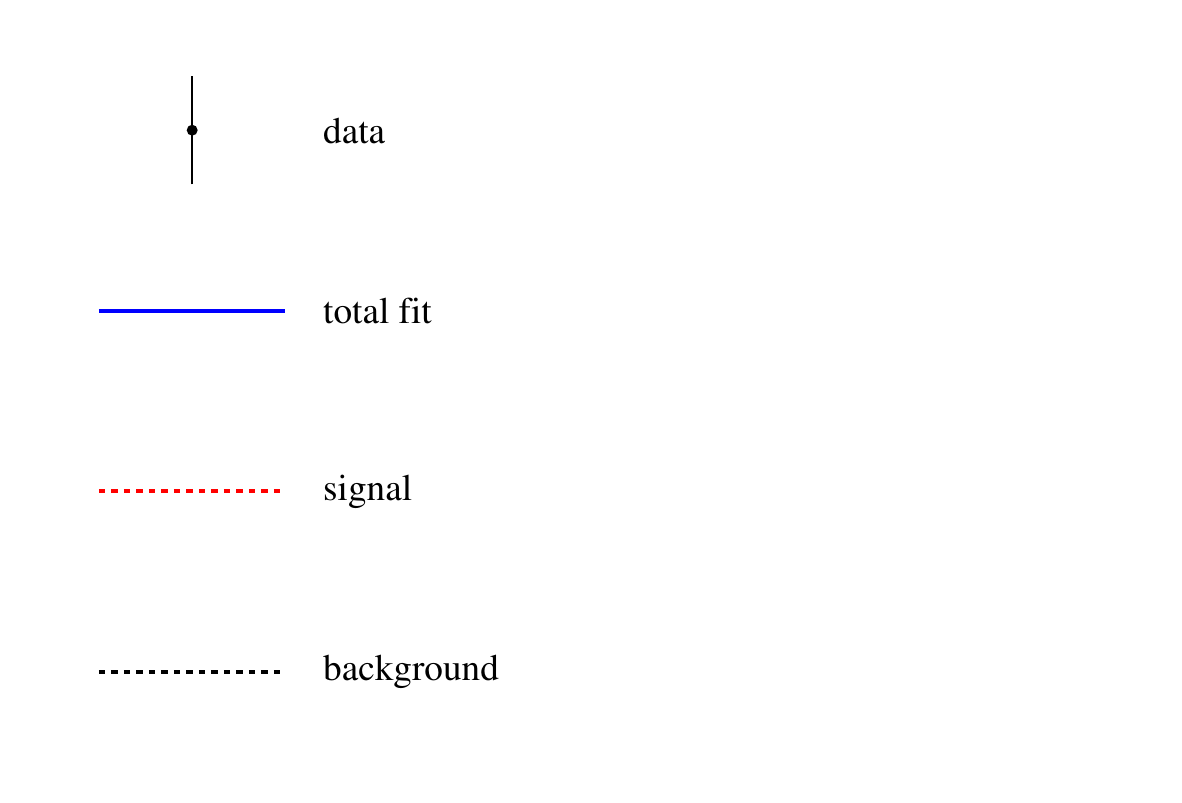}
 \caption{
 The projections of the amplitude analysis fit of $D^{+}_{s}\to K^+\pi^+\pi^-\pi^0$ on
    two-body and three-body particle mass distributions~\cite{BESIII:2022bvv}.
}
    \label{fig:Ds_K2pipi0}
\end{figure*}

Amplitude analysis of the SCS decay $D_{s}^{+} \to K^+\pi^{+}\pi^{-}\pi^{0}$ accesses two-body intermediate processes such as $D^{+}_s \to VV$, $AP$, and $VP$,
e.g., $D_s^+ \to K^*(892)^{0}\rho(770)^+$, $K_1(1270)^0\pi^+$, and $D_s^+ \to \omega K^+$,
thereby refining
theoretical predictions~\cite{Bauer:1986bm,Kamal:1990ky,Hinchliffe:1995hz,ElHassanElAaoud:1999min,Kang:2009iy}, probing $CP$ violation~\cite{Kang:2009iy,Qin:2013tje}, and understanding of $K_1$ mixing~\cite{Cheng:2003bn,Cheng:2010vk,Cheng:2010rv,Cheng:2011pb,Cheng:2013cwa,Guo:2018orw}.
The knowledge of $D_s^+ \to \omega K^+$ is important to investigate the W-annihilation contribution in $D^{+}_s \to VP$ decays and improve the understanding of SU(3)$_F$ flavor symmetry breaking effects in hadronic decays of charmed mesons~\cite{Cheng:2016ejf,Cheng:2019ggx,Cheng:2021yrn}.
Previously, BESIII reported the first evidence for $D_s^+ \to \omega K^+$ with a branching fraction of
$(0.87\pm0.25\pm0.07)\times 10^{-3}$~\cite{BESIII:2018mwk}, much smaller than the initial prediction of $2.12\times 10^{-3}$~\cite{Cheng:2019ggx}.
Reference~\cite{BESIII:2022bvv} presented the first observation of $D_{s}^{+} \to K^+\pi^{+}\pi^{-}\pi^{0}$.
With 630 candidates with a signal purity of 87\%, the first amplitude analysis on this decay was performed.
The amplitude analysis fit projections on two-body or three-body particle mass distributions  are shown in Fig.~\ref{fig:Ds_K2pipi0}.
The dominant intermediate process in this decay channel is $D_s^+ \to K^*(892)^0\rho(770)^+$, with a fraction of $(40.5\pm2.8\pm1.5)\%$.
Besides, significant contributions of $D^+_s\to K^+\omega$, $D^+_s\to K_1(1270)^0\pi^+$, $D^+_s\to K_1(1400)^0\pi^+$,
$D^+_s\to K^+a_1(1260)^0$, $D^+_s[S]\to(K^+\pi^0)_V\rho(770)^0$, and $D^+_s\to K^+\pi^0)_{{\cal S}{\rm -wave}}(\pi^+\pi^-)_{{\cal S}{\rm -wave}}$ are also observed.
The branching fraction of $D_s^+ \to K^+\pi^+\pi^-\pi^0$ is determined to be $\mathcal{B}(D_s^+ \to K^+\pi^+\pi^-\pi^0) =
(9.75\pm0.54\pm0.17)\times 10^{-3}$.
The asymmetry for the branching fractions of $D_{s}^{+} \to K^+\pi^{+}\pi^{-}\pi^{0}$ and
$D_{s}^{-} \to K^-\pi^{-}\pi^{+}\pi^{0}$ is determined to be $(6.6 \pm 5.4\pm 0.7)$\%,
and no evidence for $CP$ violation is found under the current sample size.

\begin{figure*}[hbtp]
\centering
\begin{minipage}[b]{0.225\textwidth}
\epsfig{width=1.00\textwidth,clip=true,file=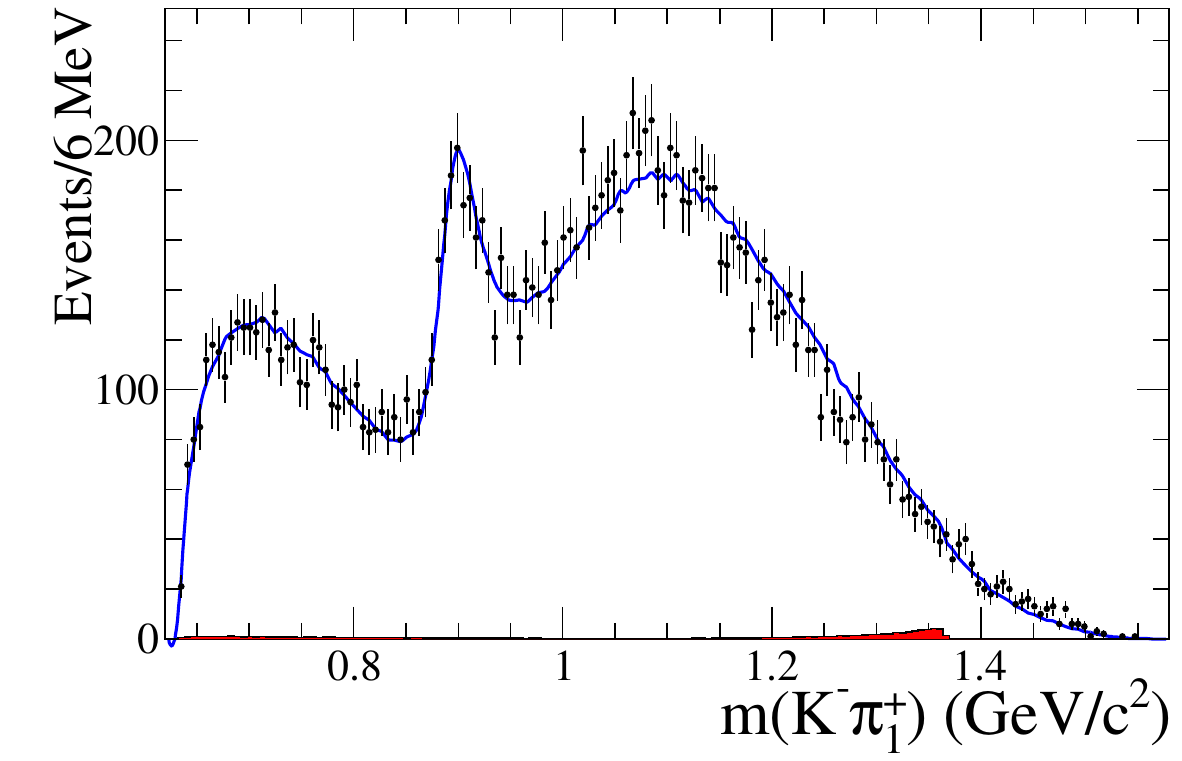}
\put(-25,60){(a)}
\end{minipage}
\begin{minipage}[b]{0.225\textwidth}
\epsfig{width=1.00\textwidth,clip=true,file=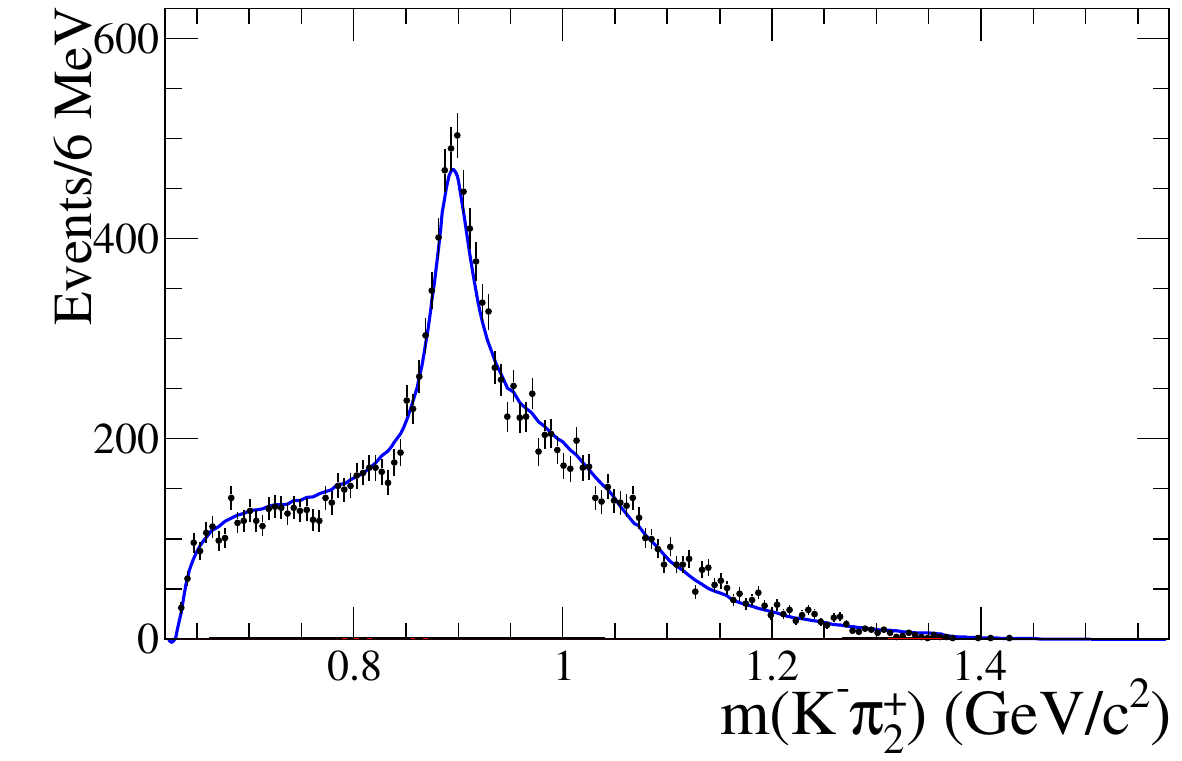}
\put(-25,60){(b)}
\end{minipage}
\begin{minipage}[b]{0.225\textwidth}
\epsfig{width=1.00\textwidth,clip=true,file=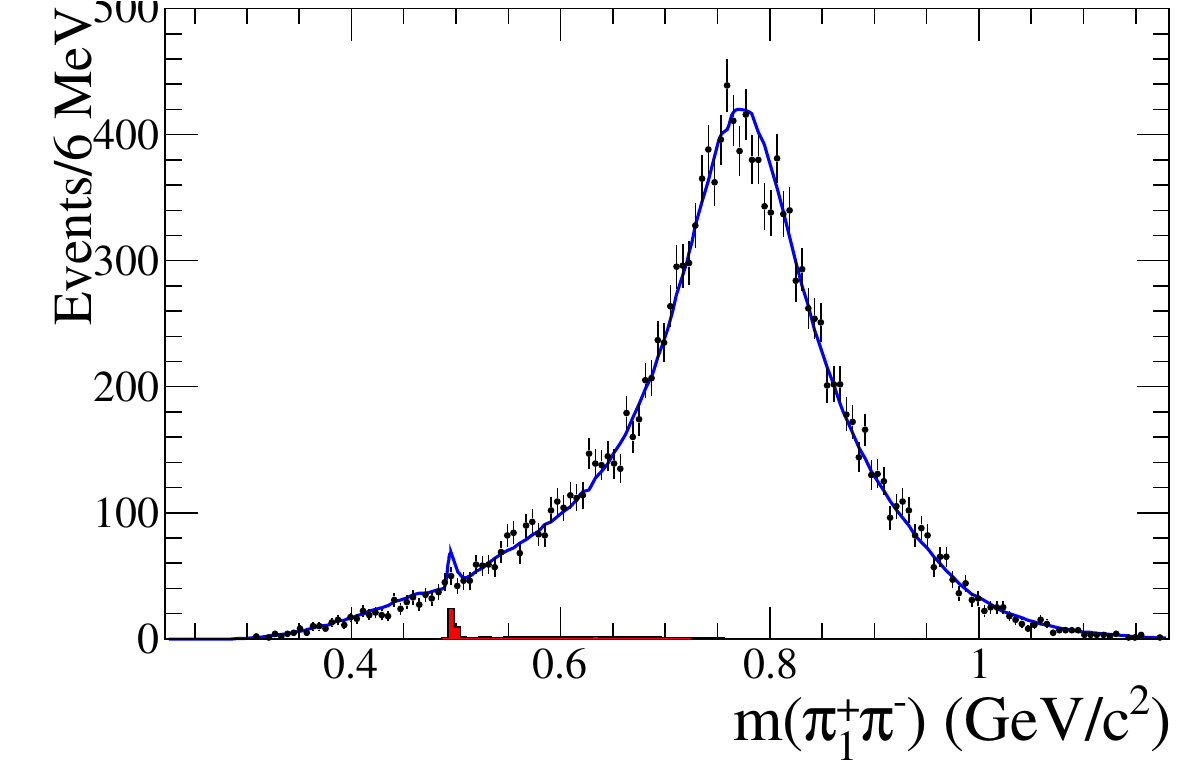}
\put(-25,60){(c)}
\end{minipage}
\begin{minipage}[b]{0.225\textwidth}
\epsfig{width=1.00\textwidth,clip=true,file=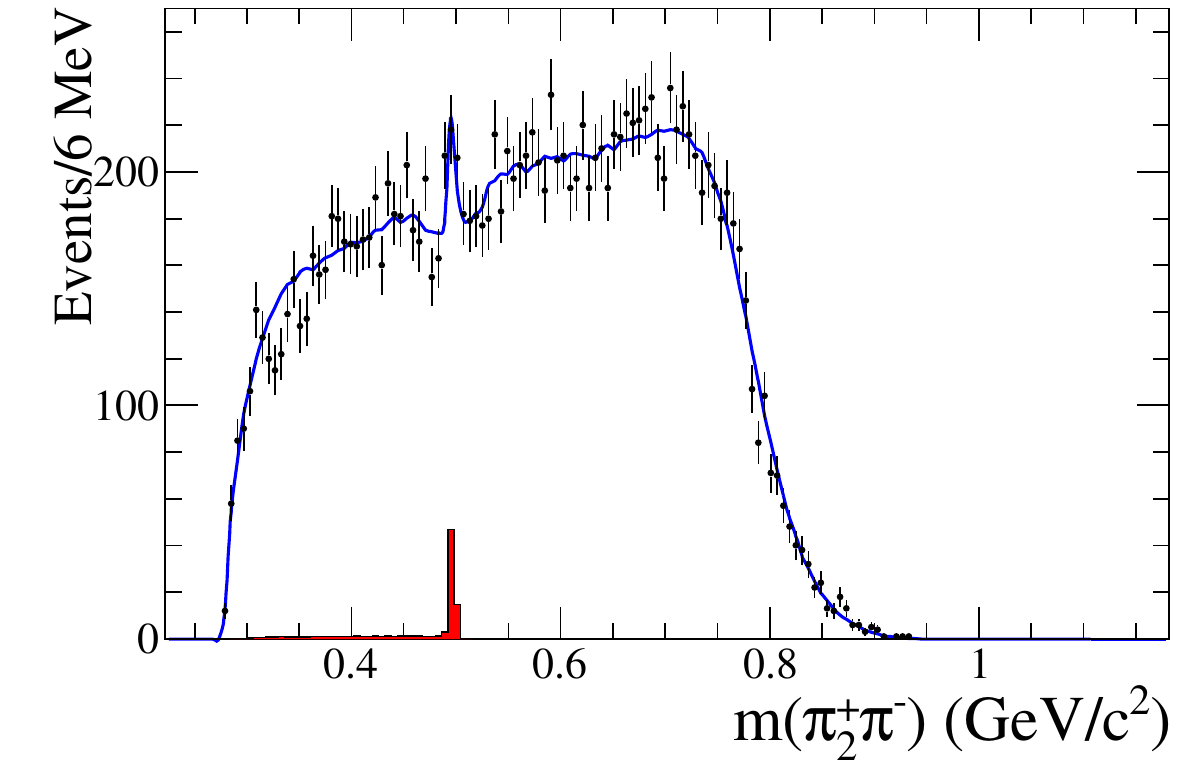}
\put(-25,60){(d)}
\end{minipage}
\begin{minipage}[b]{0.225\textwidth}
\epsfig{width=1.00\textwidth,clip=true,file=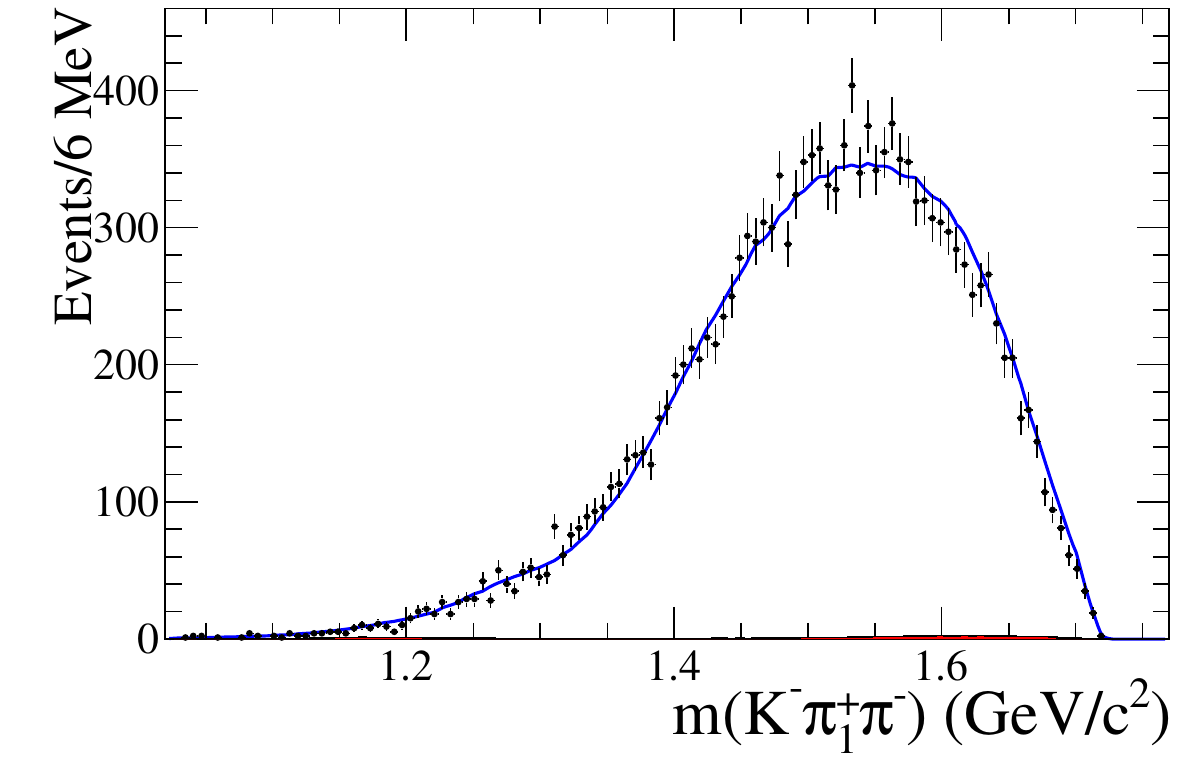}
\put(-25,60){(e)}
\end{minipage}
\begin{minipage}[b]{0.225\textwidth}
\epsfig{width=1.00\textwidth,clip=true,file=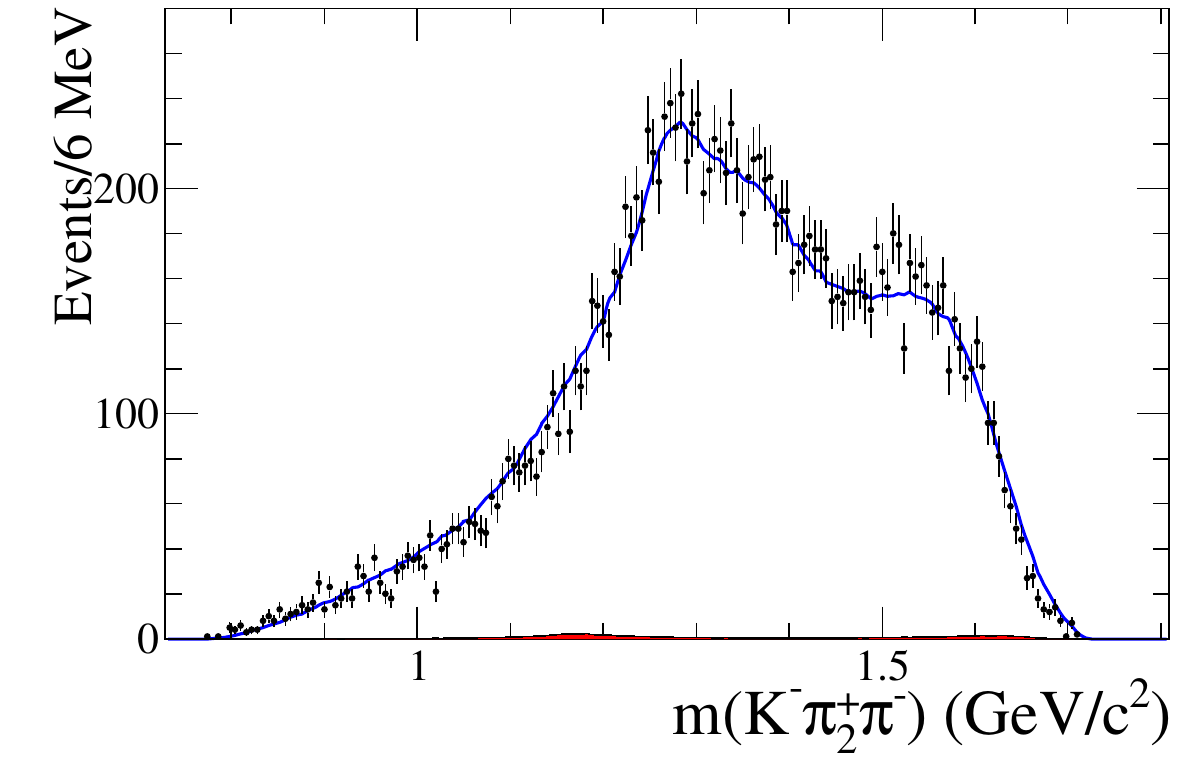}
\put(-25,60){(f)}
\end{minipage}
\begin{minipage}[b]{0.225\textwidth}
\epsfig{width=1.00\textwidth,clip=true,file=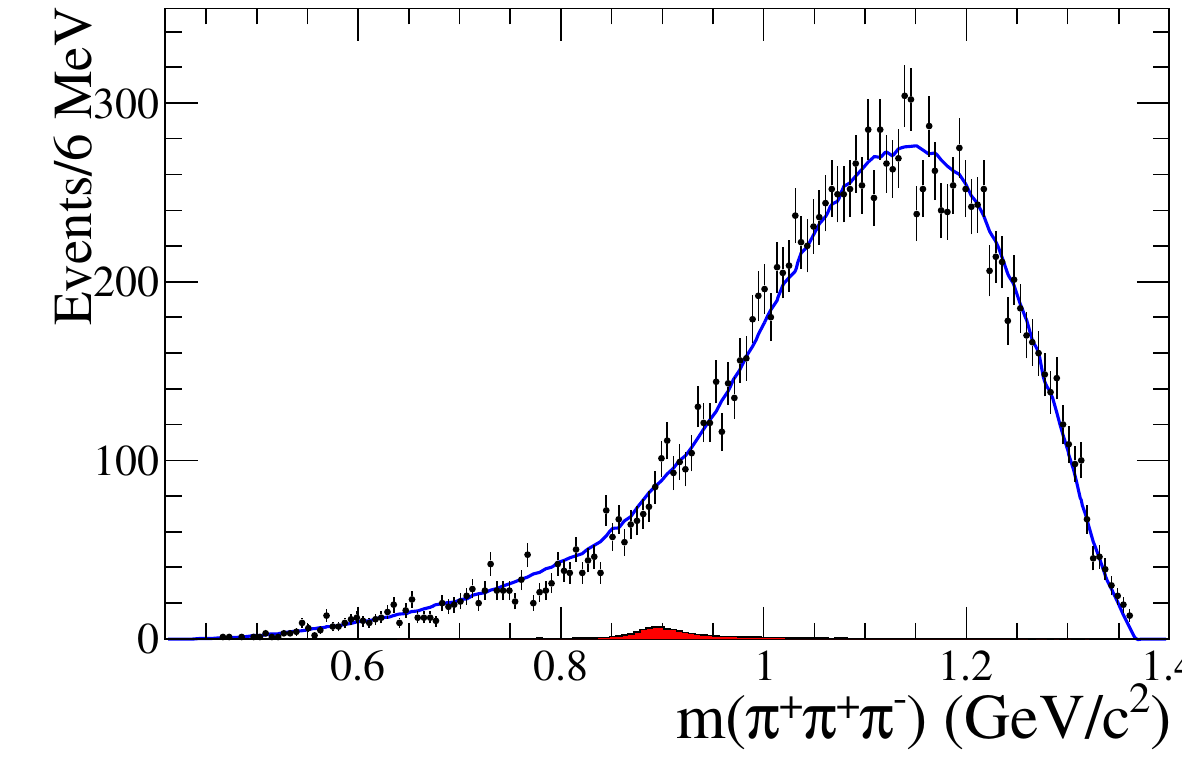}
\put(-25,60){(g)}
\end{minipage}
\begin{minipage}[b]{0.225\textwidth}
\epsfig{width=1.00\textwidth,clip=true,file=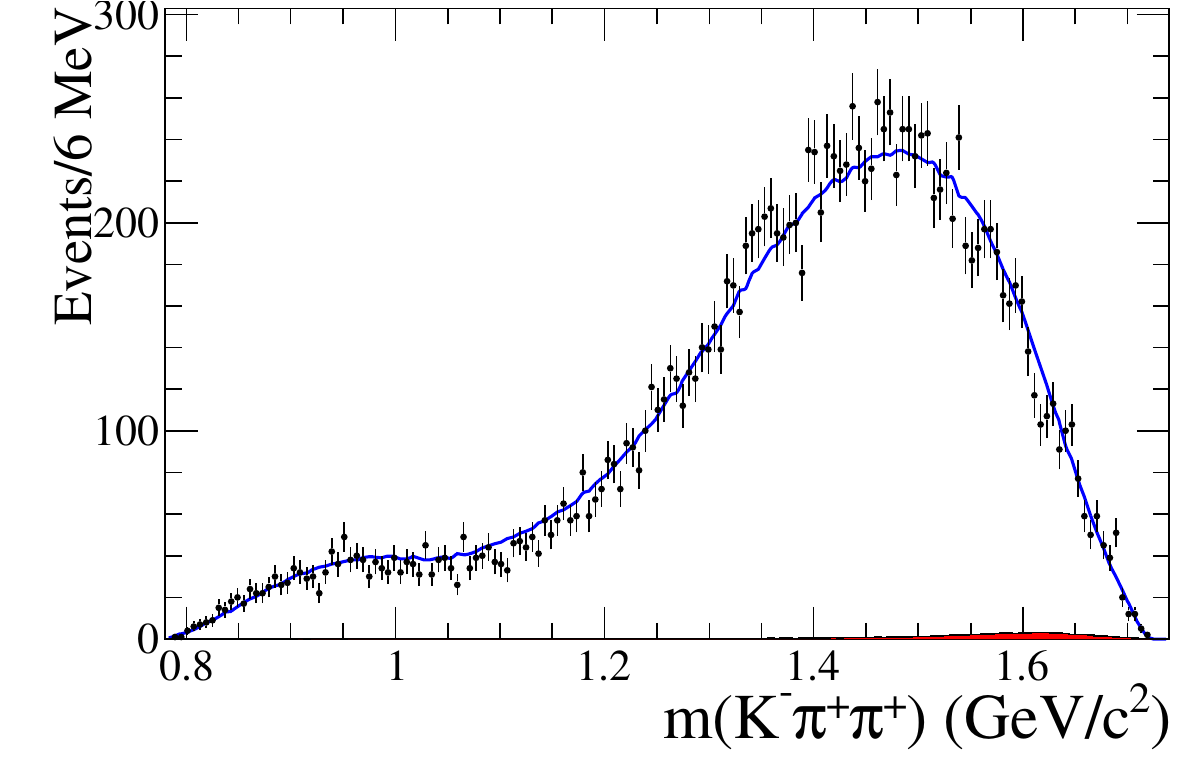}
\put(-25,60){(h)}
\end{minipage}
\caption{
The projections of the amplitude analysis fit of $D^{+}\to K^-2\pi^+\pi^-$ on
    two-body and three-body particle mass distributions~\cite{BESIII:2017jyh}.
}
\label{fig:D0_K3pi}
\end{figure*}

Insights into two-body decays $D \to \bar{K}^*(892) \rho(770)$,
$D \to \bar K a_{1}(1260)$, $D \to \bar K_1(1270) \pi$, and $D \to \bar K_1(1400) \pi$
which can be extracted from amplitude analyses of CF hadronic decays $D^{0(+)}\to \bar K\pi\pi\pi$, are important to deepen the understanding of
$D^{0} \to VV$ and $D \to AP$.
Results about the two $K_1$ states are important inputs to further investigations of the mixing of the axial-vector kaon mesons~\cite{Cheng:2003bn,Cheng:2010vk,Cheng:2010rv,Cheng:2011pb,Cheng:2013cwa,Guo:2018orw}.
Especially, knowledge of the decay substructure of $D^{0} \to K^{-} 2 \pi^{+} \pi^{-}$ in combination with a precise measurement
of strong phases can also help to improve the measurement of the CKM angle
$\gamma$ (the phase of $V_{cb}$ relative to $V_{ub}$)~\cite{Atwood:1996ci}.
In the measurement of $\gamma$, the parametrization model is an important input in a model dependent
method and also can be used to generate Monte Carlo simulated events to check the sensitivity
in a model independent method~\cite{Harnew:2014zla}.
Furthermore, the branching fractions of intermediate processes can be used to
understand the $D^{0}$-$\bar{D}^{0}$ mixing in theory~\cite{Falk:2001hx,Cheng:2010rv}.
Previous amplitude analyses of $D^{0} \to K^{-} 2\pi^{+} \pi^{-}$ were presented by
MARKIII~\cite{MARK-III:1991fvi} and E691~\cite{FNAL-691:1992exu} in the early 1990s.
Using about $1300$ signal events,
MARKIII obtained the branching fractions for $D^{0} \to K^{-} a^{+}_{1}(1260)$,
$D^{0} \to \bar{K}^*(892)^{0} \rho(770)^{0}$, $D^{0} \to K_{1}(1270)^{-}\pi^{+}$,
as well as for the three- and four-body nonresonant decays.
Based on $1745$ signal events and $800$ background events,
E691 obtained a similar result but without considering $D^{0} \to K_{1}(1270)^{-}\pi^{+}$.
All of them have large uncertainties.
In 2017, an amplitude analysis of $D^{0} \to K^{-} 2 \pi^{+} \pi^{-}$ was performed,
with a nearly background free sample of about 16k candidate events tagged by $\bar D^0\to K^+\pi^-$~\cite{BESIII:2017jyh}.
The amplitude analysis fit projections on two-body or three-body particle mass distributions  are shown in Fig.~\ref{fig:D0_K3pi}.
The amplitude model includes the two-body decays $D^{0} \to \bar K^*(892)^0\rho(770)^{0}$,
$D^{0} \to K^{-}a_{1}(1260)^+$ and $D^{0} \to K_{1}(1270)^{-}\pi^{+}$,
the three-body decays $D^{0} \to \bar K^*(892)^0\pi^{+}\pi^{-}$,
$D^{0} \to K^{-}\pi^{+}\rho(770)^{0}$,
the four-body nonresonant decay $D^{0} \to K^{-}2\pi^{+}\pi^{-}$,
with fit fractions of
$(12.3\pm0.4\pm0.5)\%$,
$(54.6\pm2.8\pm3.7)\%$,
$(0.8\pm0.2\pm0.2)\%$,
$(3.4\pm0.3\pm0.5)\%$,
$(8.4\pm1.1\pm2.5)\%$,
$(7.0\pm0.4\pm0.5)\%$,
$(21.9\pm0.6\pm0.6)\%$, respectively.
The obtained results improve upon the earlier results from MARKIII and are
consistent with them within corresponding uncertainties.
Combined with the world average of the branching fraction of
$D^0\to K^-2\pi^+\pi^-$,
the branching fractions for each component are also presented.
Later, LHCb~\cite{LHCb:2017swu} reported amplitude analyses of
$D^{0} \to K^{-} 2\pi^{+} \pi^{-}$ and
$D^{0} \to K^{+} 2\pi^{-} \pi^{+}$,
with about 0.89 million and 3k signal events with signal purities of
99.6\% and 82.4\%, respectively.
Measurements of the strong phase difference in $D\to K^\pm\pi^\mp\pi^+\pi^-$ were presented in separate papers~\cite{BESIII:2021eud,BESIII:2026oav}.
 
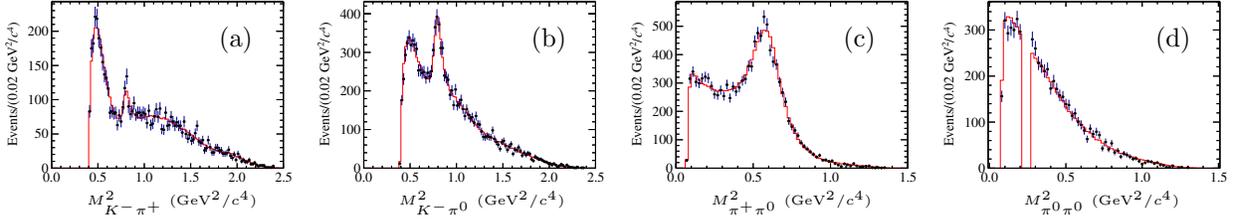
\begin{figure*}[!htp]
\centering
\begin{minipage}[b]{0.225\textwidth}
\epsfig{width=0.98\textwidth,file=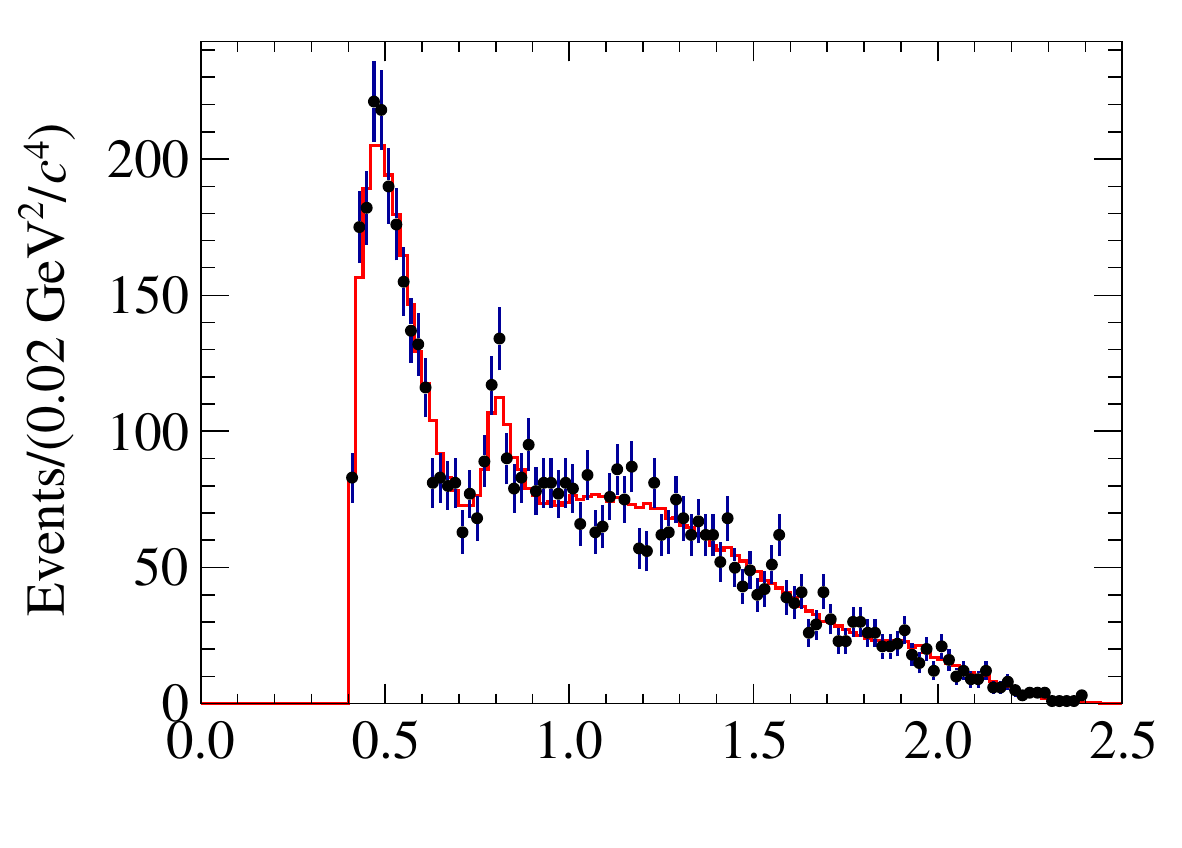}
\put(-80,0){\tiny $M^2_{K^-\pi^+}$ (GeV$^2/c^4$)}
\put(-30,60){(a)}
\end{minipage}
\begin{minipage}[b]{0.225\textwidth}
\epsfig{width=0.98\textwidth,file=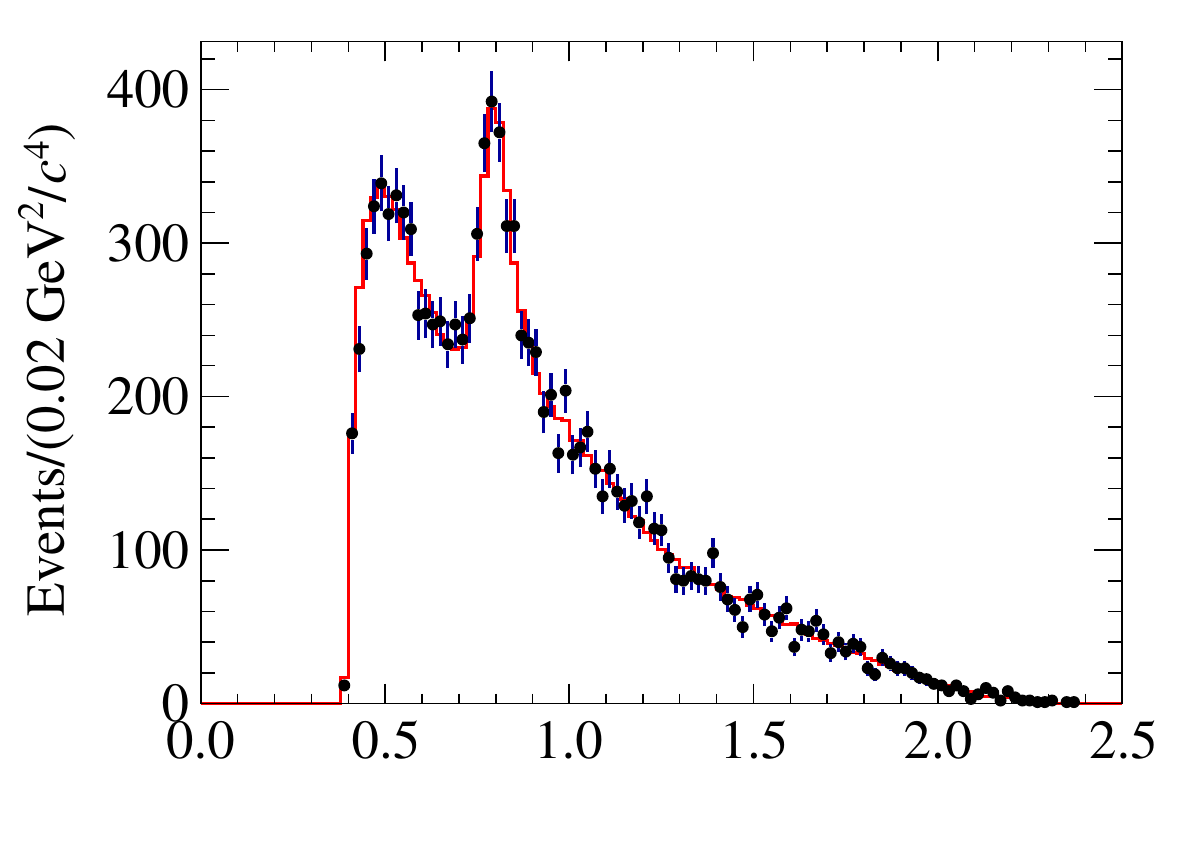}
\put(-80,0){\tiny $M^2_{K^-\pi^0}$ (GeV$^2/c^4$)}
\put(-30,60){(b)}
\end{minipage}
\begin{minipage}[b]{0.225\textwidth}
\epsfig{width=0.98\textwidth,file=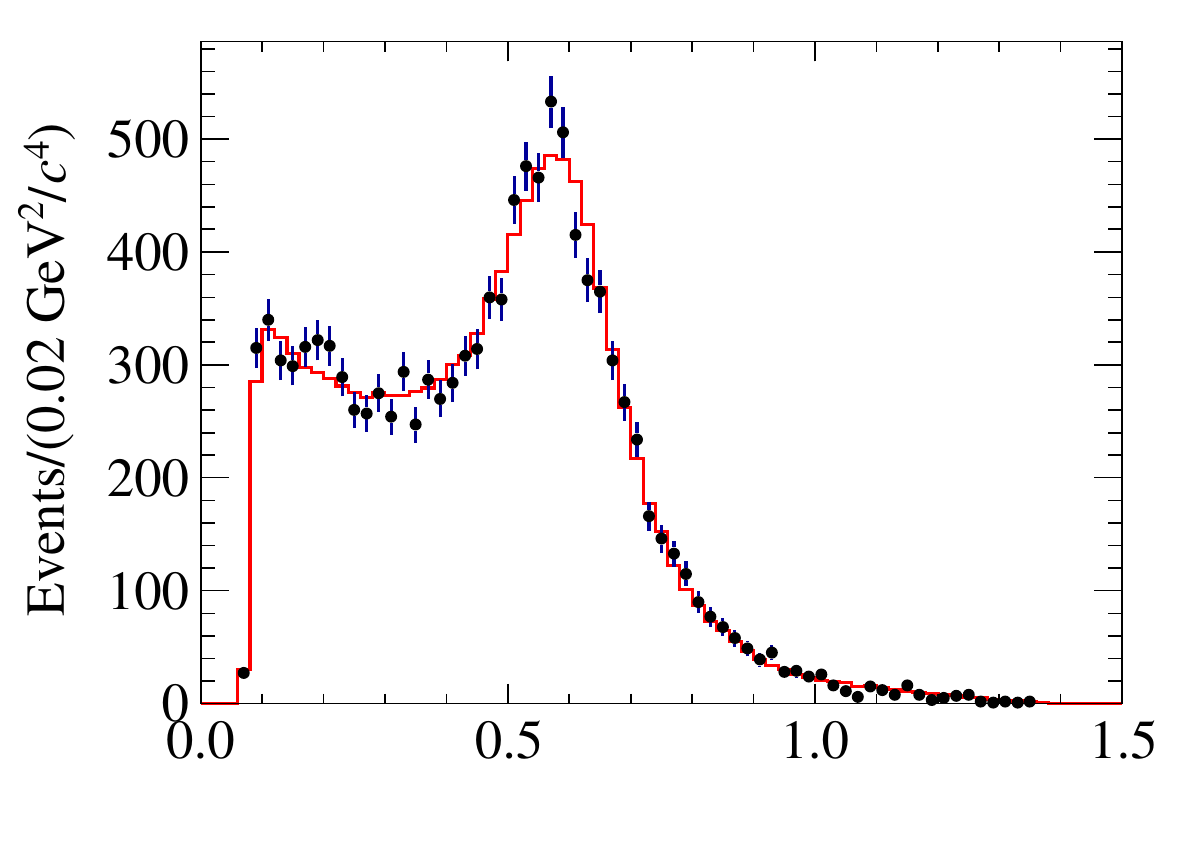}
\put(-80,0){\tiny $M^2_{\pi^+\pi^0}$ (GeV$^2/c^4$)}
\put(-30,60){(c)}
\end{minipage}
\begin{minipage}[b]{0.225\textwidth}
\epsfig{width=0.98\textwidth,file=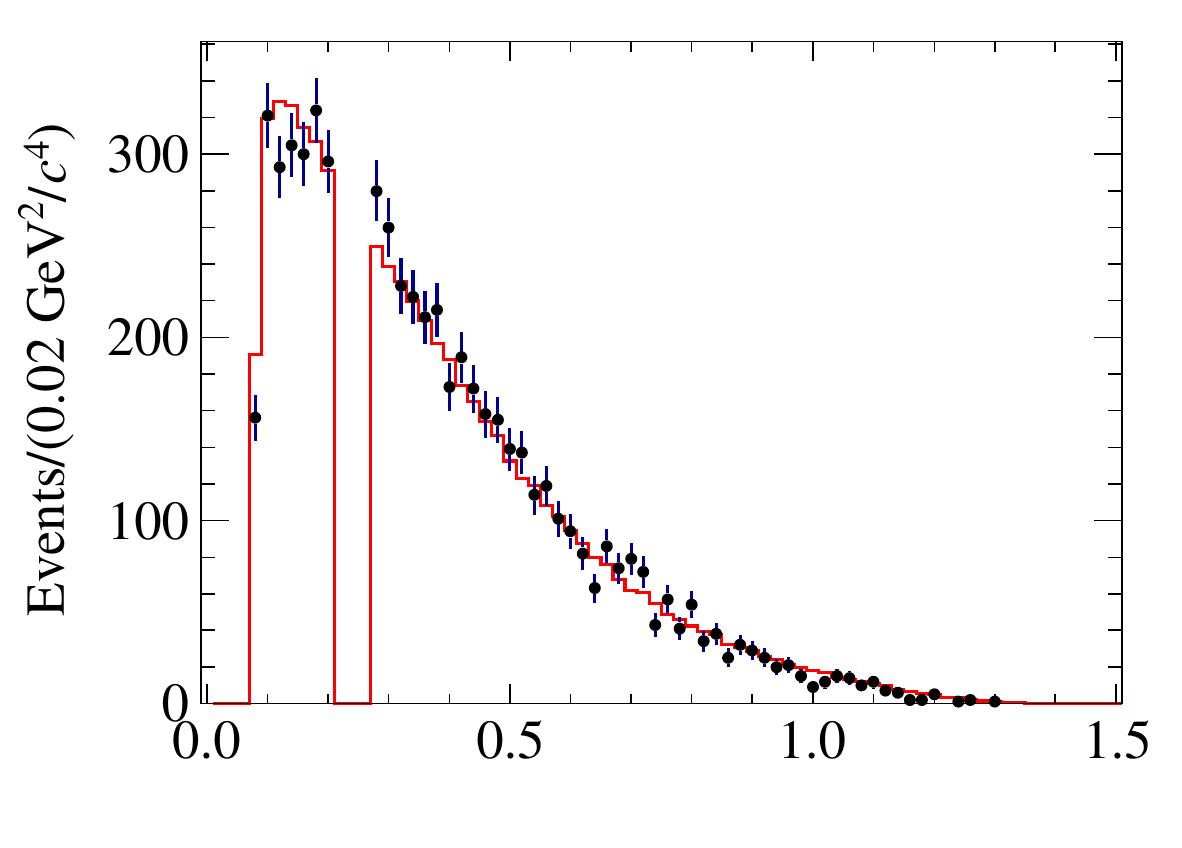}
\put(-80,0){\tiny $M^2_{\pi^0\pi^0}$ (GeV$^2/c^4$)}
\put(-30,60){(d)}
\end{minipage}
\caption{
The projections of the amplitude analysis fit of $D^0\to K^-\pi^+2\pi^0$ on
    two-body particle masses squared distributions~\cite{BESIII:2019lwn}.
}
\label{fig:D0_Kpi2pi0}
\end{figure*}

The CF decay $D^0\to K^-\pi^+2\pi^0$, which was previously used as a ``single-tag" mode in some analyses of CLEO-c and BESIII, 
is expected to have a large branching fraction, but knowledge of its subdecays is absent before.
In 2019,
Ref.~\cite{BESIII:2019lwn} reported the amplitude analysis and branching fraction measurement of $D^0\to K^-\pi^+2\pi^0$,
based on about 6.0k candidates with a signal purity of 99\%.
Figure~\ref{fig:D0_Kpi2pi0} shows the amplitude analysis fit projections on two-body particle masses squared distributions.
The $D^0\to K^-a_1(1260)^+$ decay is the dominant amplitude occupying $28\%$ of the total fit fraction (98.54\%)
and other important amplitudes are $D\to K_1(1270)^-\pi^+$, $D\to (K^-\pi^0)_{{\cal S}{\rm -wave}}\rho(770)^+$,
and $D\to K^*(892)^-\rho(770)^+$, which are similar, in general, to the amplitudes in
$D^0\to K^-2\pi^+\pi^-$~\cite{BESIII:2017jyh}.

\begin{figure*}[hbtp]
\centering
\begin{minipage}[b]{0.225\textwidth}
\epsfig{width=1.00\textwidth,clip=true,file=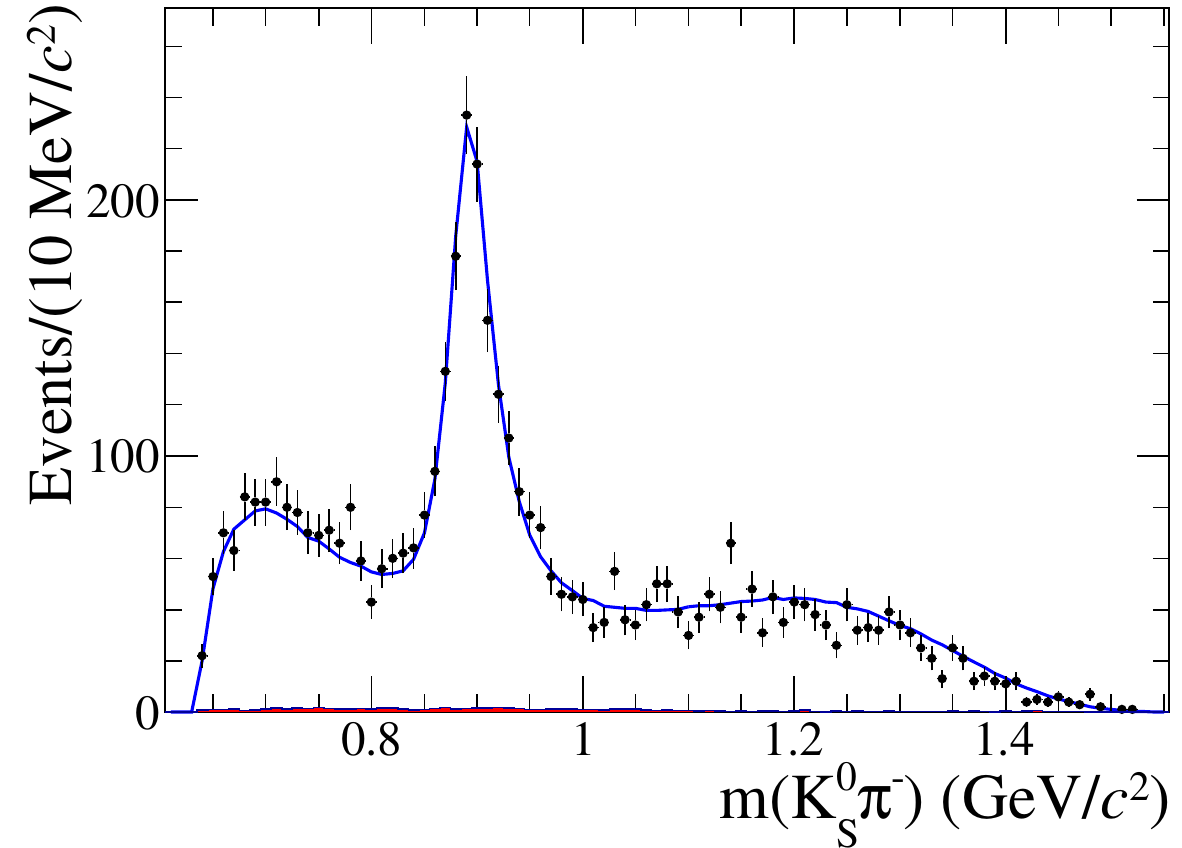}
\put(-25,70){(a)}
\end{minipage}
\begin{minipage}[b]{0.225\textwidth}
\epsfig{width=1.00\textwidth,clip=true,file=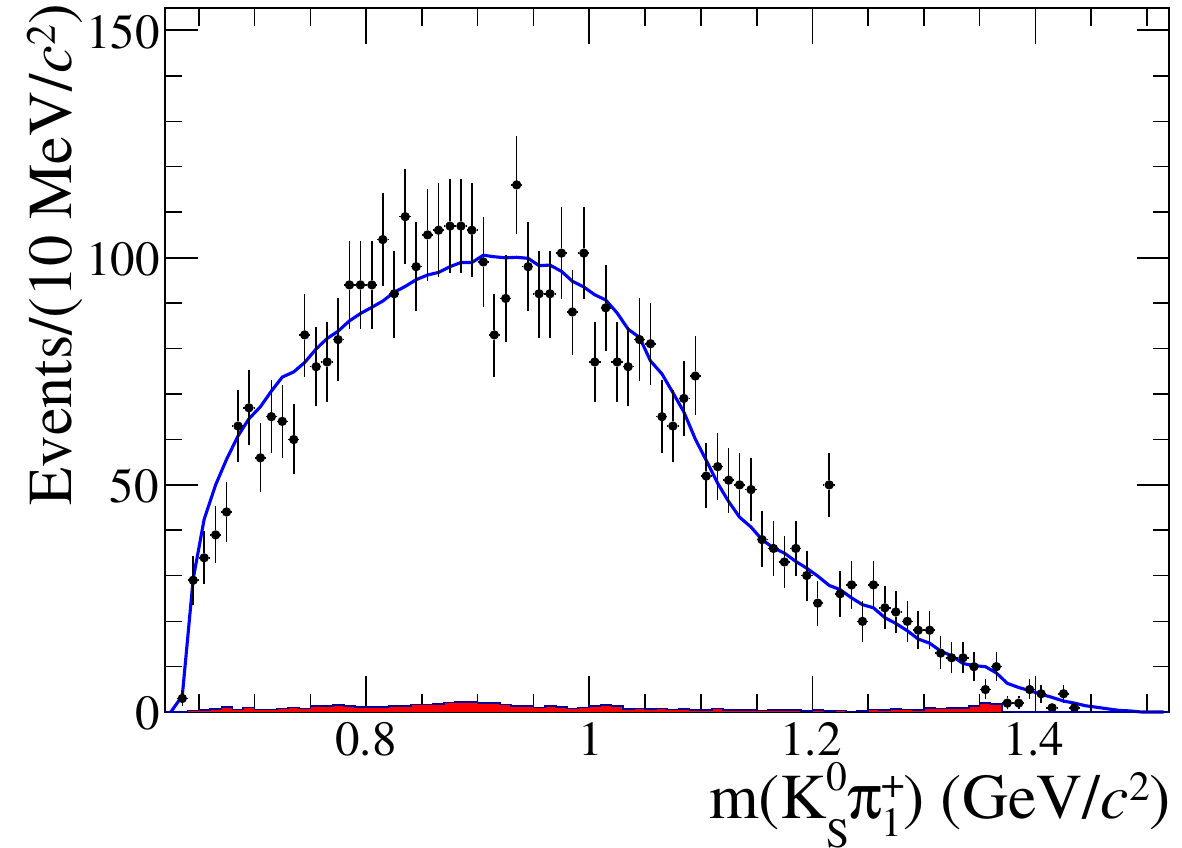}
\put(-25,70){(b)}
\end{minipage}
\begin{minipage}[b]{0.225\textwidth}
\epsfig{width=1.00\textwidth,clip=true,file=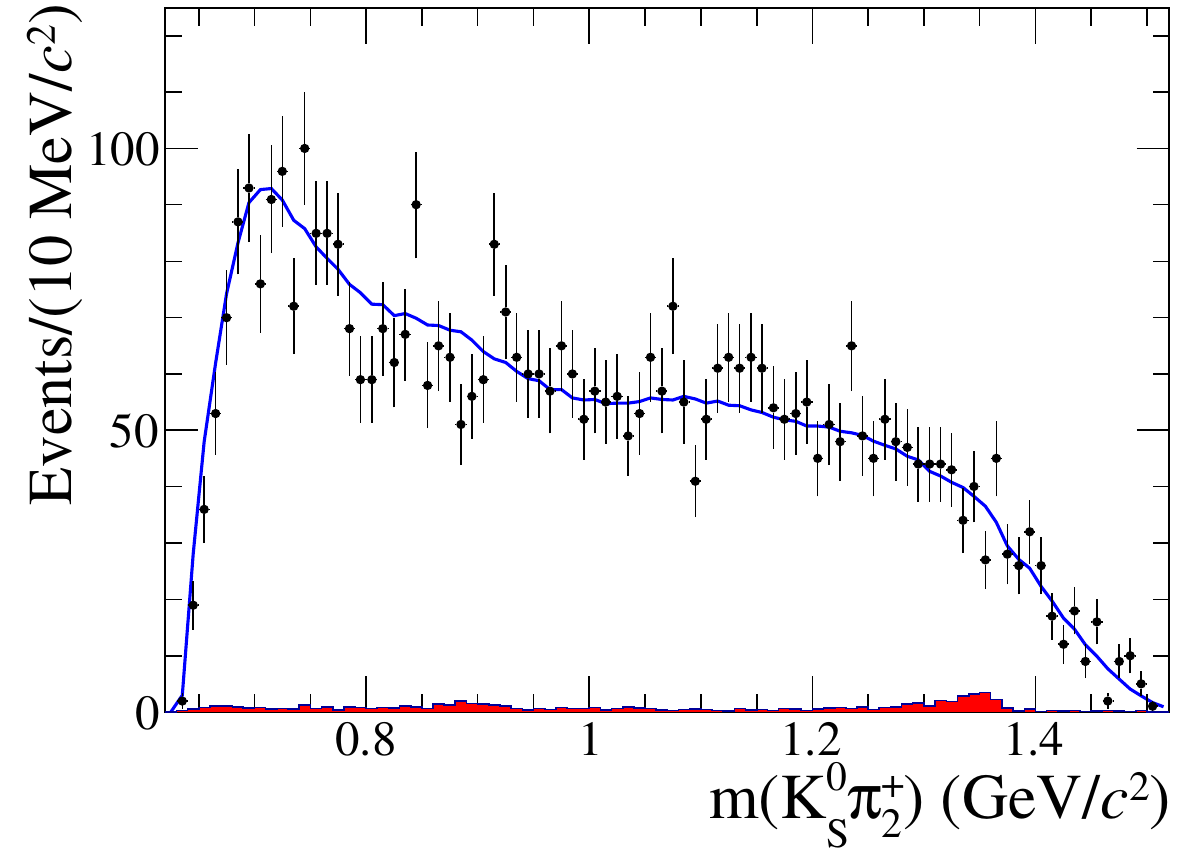}
\put(-25,70){(c)}
\end{minipage}
\begin{minipage}[b]{0.225\textwidth}
\epsfig{width=1.00\textwidth,clip=true,file=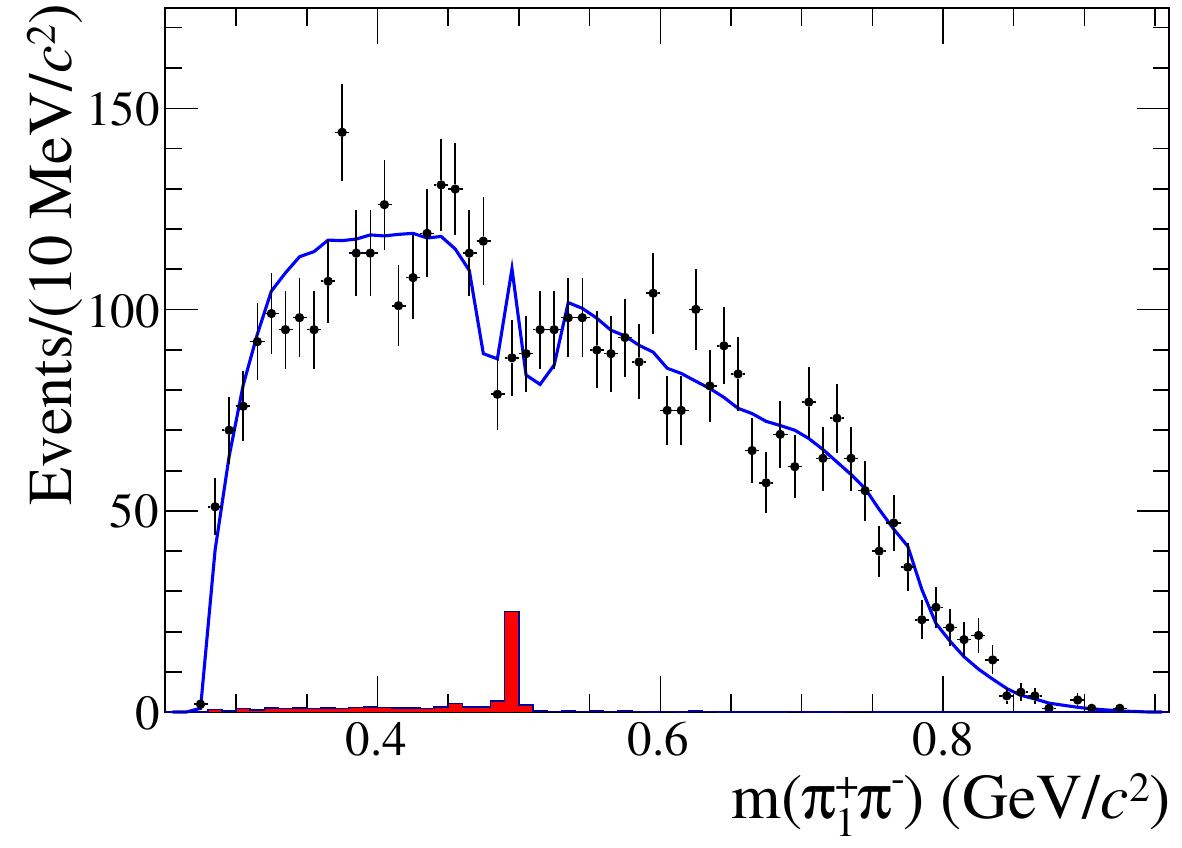}
\put(-25,70){(d)}
\end{minipage}
\begin{minipage}[b]{0.225\textwidth}
\epsfig{width=1.00\textwidth,clip=true,file=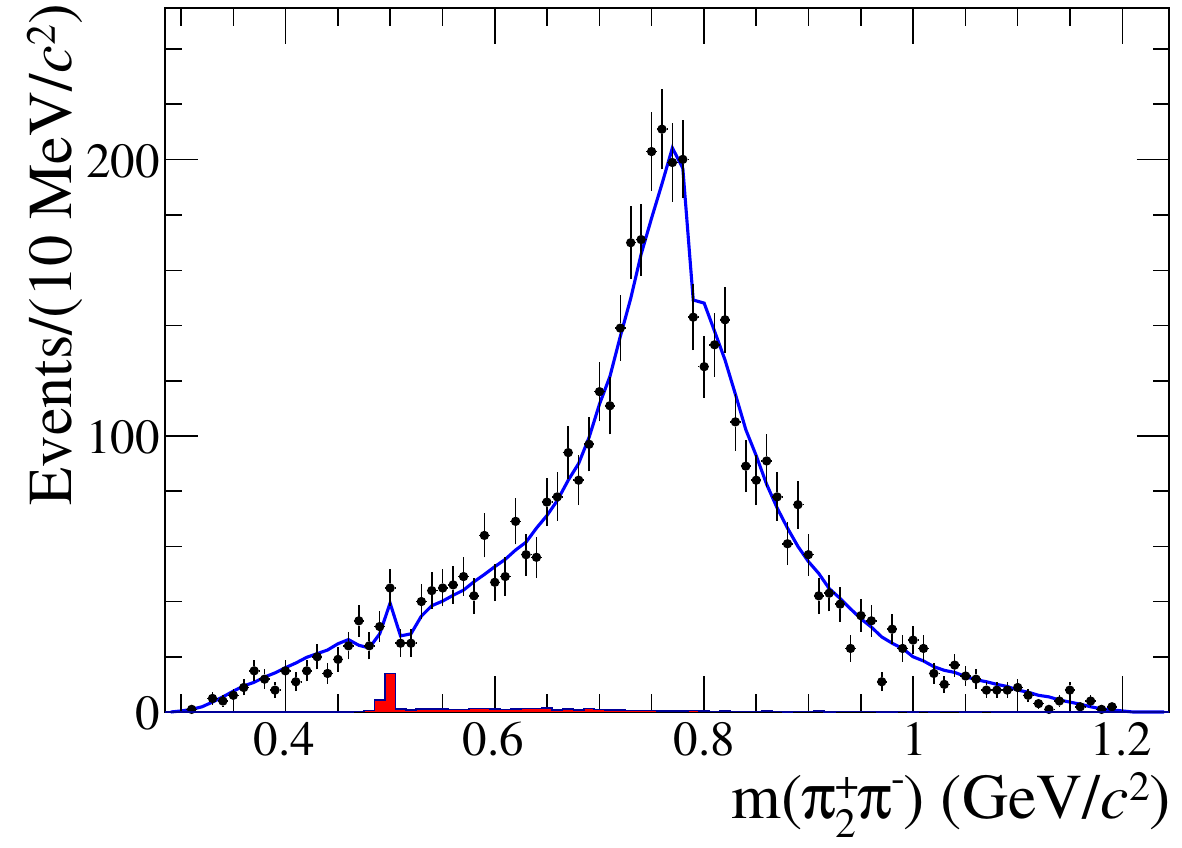}
\put(-25,70){(e)}
\end{minipage}
\begin{minipage}[b]{0.225\textwidth}
\epsfig{width=1.00\textwidth,clip=true,file=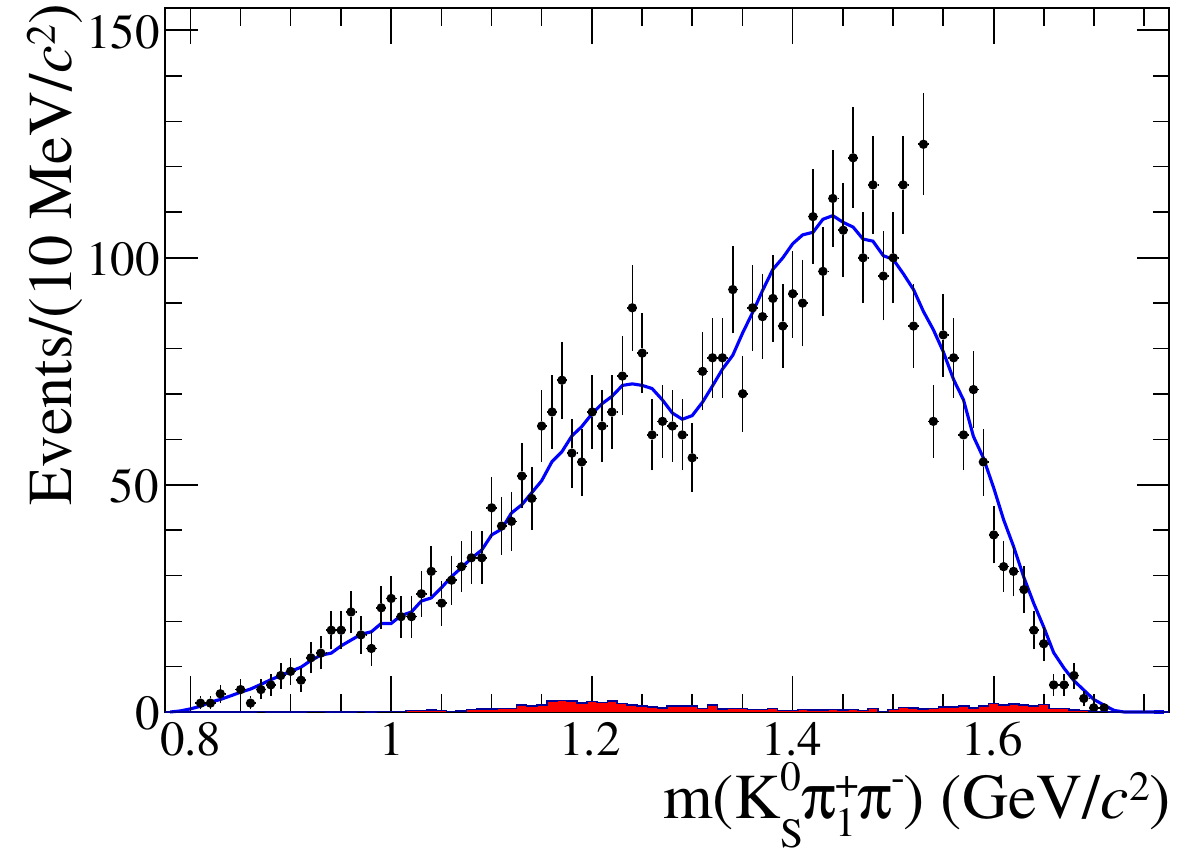}
\put(-25,70){(f)}
\end{minipage}
\begin{minipage}[b]{0.225\textwidth}
\epsfig{width=1.00\textwidth,clip=true,file=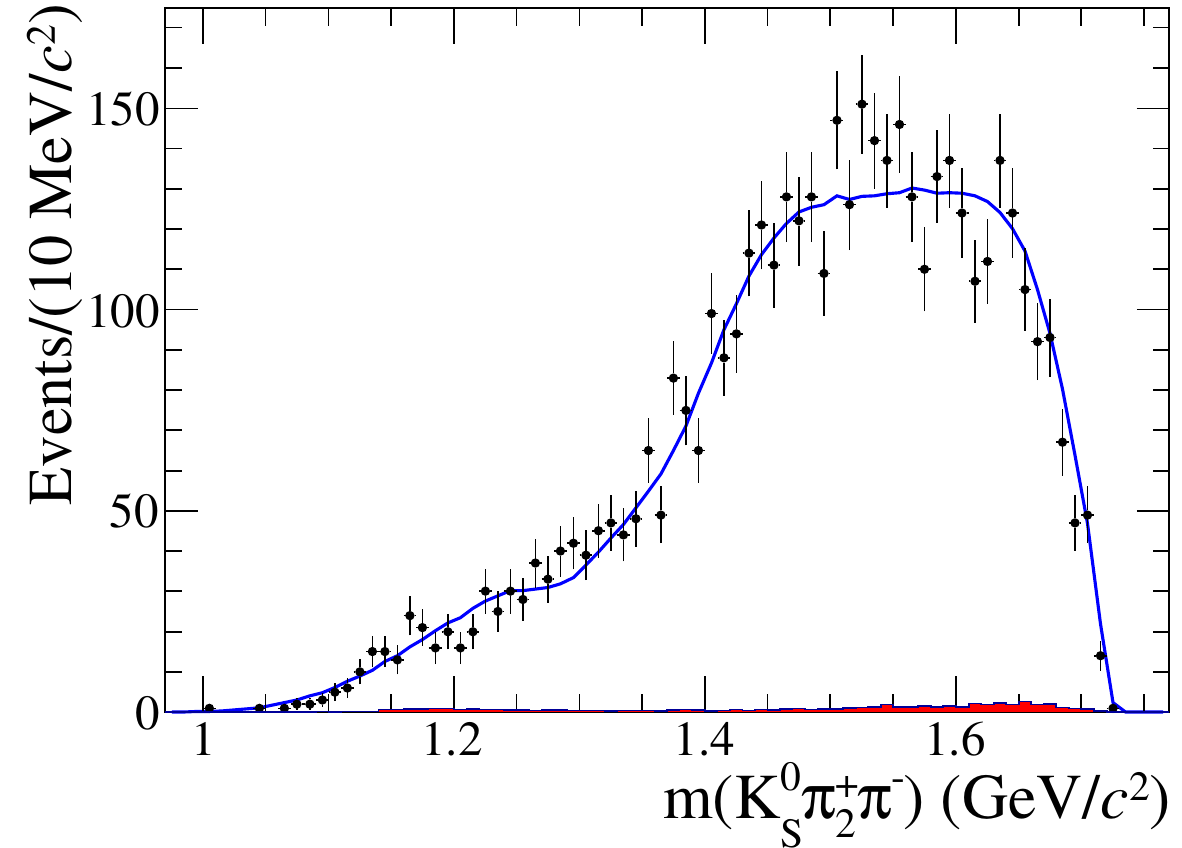}
\put(-25,70){(g)}
\end{minipage}
\begin{minipage}[b]{0.225\textwidth}
\epsfig{width=1.00\textwidth,clip=true,file=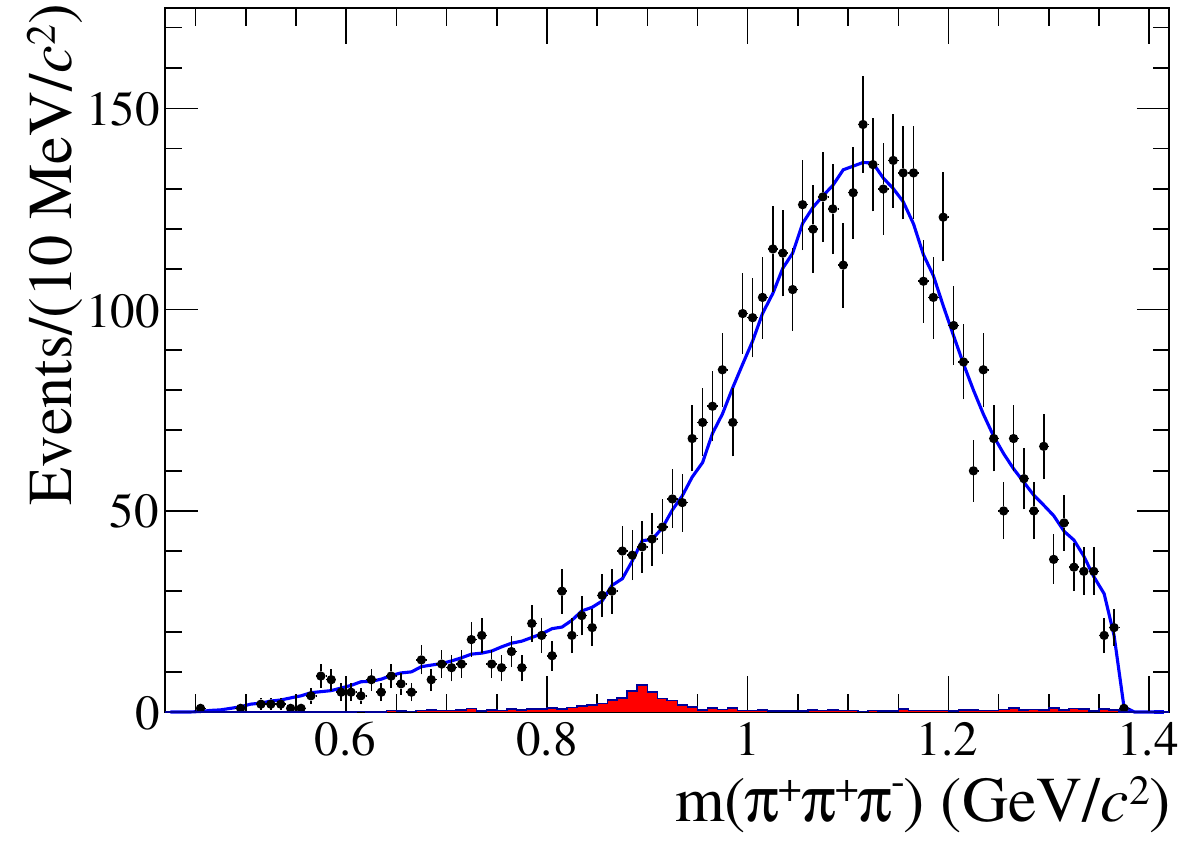}
\put(-25,70){(h)}
\end{minipage}
\caption{
The projections of the amplitude analysis fit of $D^+\to K^0_S2\pi^+\pi^-$ on
    two-body and three-body particle mass distributions~\cite{BESIII:2019ymv}.
}
\label{fig:Dp_KS3pi}
\end{figure*}

Reference~\cite{BESIII:2019ymv} reported an amplitude analysis of $D^{+} \to K_{S}^{0}2\pi^{+} \pi^{-}$,
by using about 4.6k candidates with a signal purity of about 97.5\%.
The amplitude analysis fit projections on two-body or three-body particle mass distributions  are shown in Fig.~\ref{fig:Dp_KS3pi}.
The significant amplitudes, which contribute to the model that best fits the data, are composed of
five quasi-two-body decays $ K_{S}^{0} a_{1}(1260)^{+}$,
$ \bar K_{1}(1270)^{0} \pi^{+}$,
$ \bar K_{1}(1400)^{0} \pi^{+}$,
$ \bar K_{1}(1650)^{0} \pi^{+}$,
and $ \bar K(1460)^{0} \pi^{+}$,
a three-body decay $K_{S}^{0}\pi^{+}\rho(770)^{0}$, as well as
a non-resonant component $ K_{S}^{0}2\pi^{+}\pi^{-}$.
The dominant amplitude  is $ K_{S}^{0} a_{1}(1260)^{+}$,
with a fit fraction of $(40.3\pm2.1\pm2.9)\%$.
The branching fractions for each intermediate process are also presented by using the world average of ${\cal B}(D^{+} \to K_{S}^{0}2\pi^{+} \pi^{-})$.
Compared with the results of $D^{0} \to K^{-}2\pi^{+}\pi^{-}$~\cite{BESIII:2017jyh}, the 
$D \to K a_{1}(1260)$ decay is found to be the dominant substructure in both $D^{0}$ and $D^{+}$ decays.
For the two $K_{1}$ states, the contributions from $D \to K_{1}(1270) \pi$ are at the same level for both $D^{+}$ and $D^{0}$ decays.
For $D \to K_{1}(1400) \pi$, the related branching fraction in $D^{+}$ decays is found to be greater than that in $D^{0}$ decay by one order of magnitude.
The branching fractions of $D^{+}\to K_{S}^{0}a_{1}(1260)^{+}(\to f_{0}(500)\pi^{+})$,
$D^{+}\to \bar K_{1}(1400)^{0}(\to K^*(892)^-\pi^{+})\pi^{+}$, and
$D^{+}\to \bar K_{1}(1270)^{0}(\to K_{S}^{0}\rho(770)^{0})\pi^{+}$ are determined for the first time,
while the other intermediate decays are measured with significantly improved precision
compared with the previous MARKIII measurements~\cite{MARK-III:1991fvi}.
Similar to the results of $D^{0} \to K^{-}2\pi^{+}\pi^{-}$ reported
by BESIII~\cite{BESIII:2017jyh} and LHCb~\cite{LHCb:2017swu},
the $D \to \bar K a_{1}(1260)$ decay is found to be the dominant substructure.
The contributions from $D \to K_{1}(1270) \pi$ are at the same level for both $D^{+}$ and $D^{0}$ decays,
while the branching fraction of $D^+ \to K_{1}(1400) \pi$ is greater than its counterpart in $D^{0}$ decay by one order of magnitude.

\begin{figure*}[htbp]
          \centering
          \includegraphics[width=0.225\textwidth]{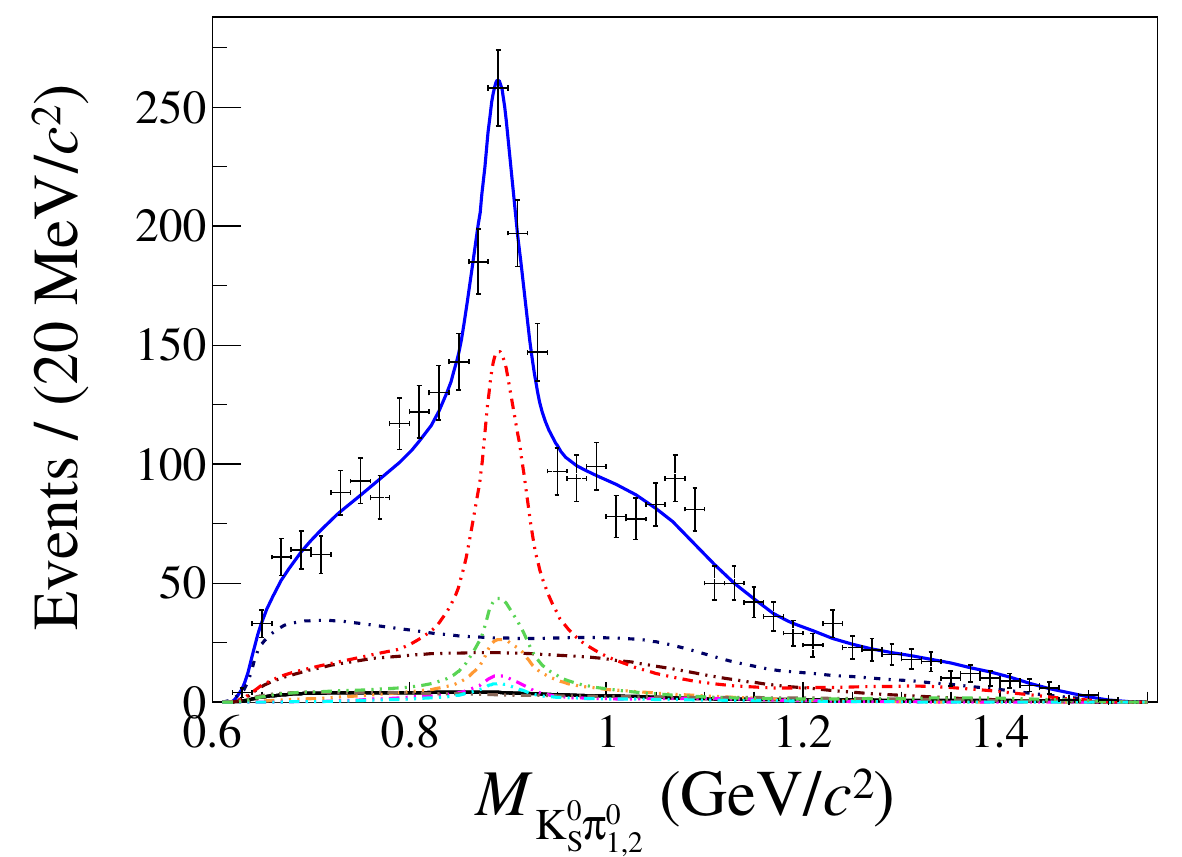}
          \includegraphics[width=0.225\textwidth]{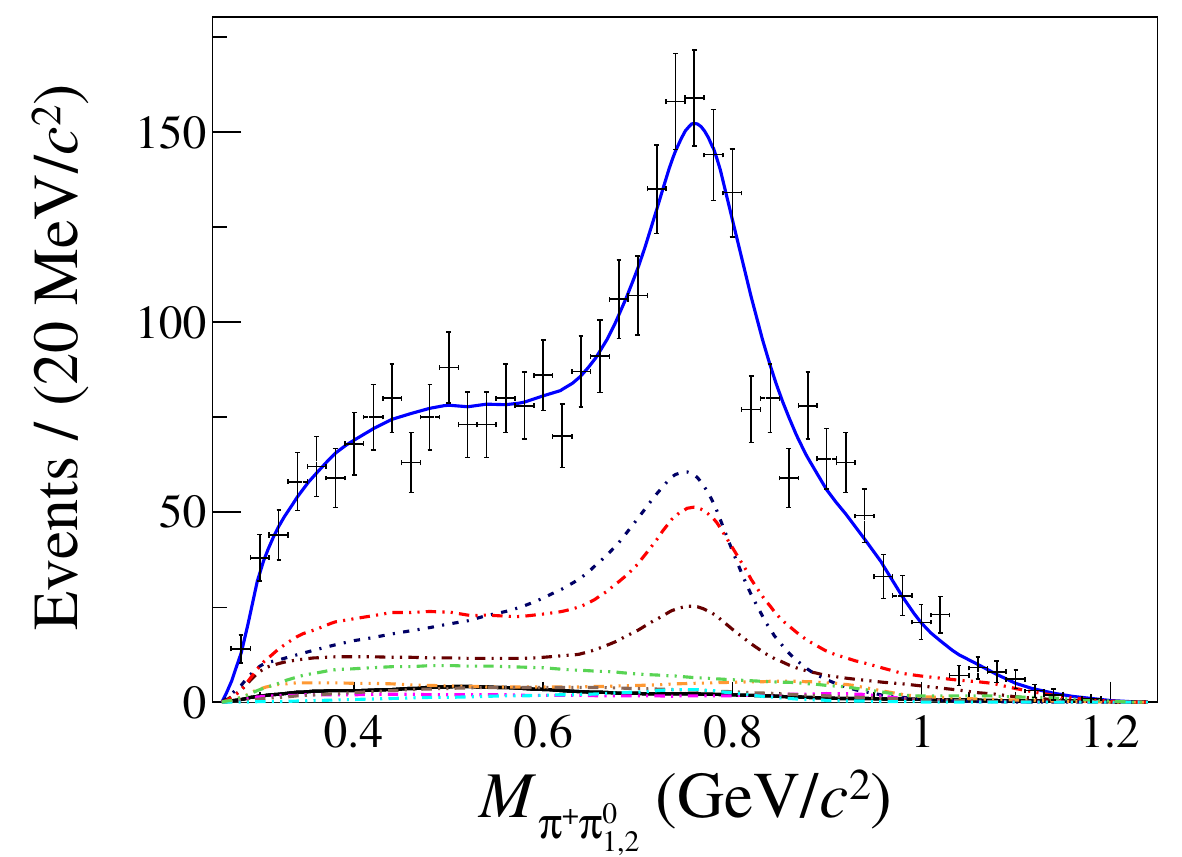}
          \includegraphics[width=0.225\textwidth]{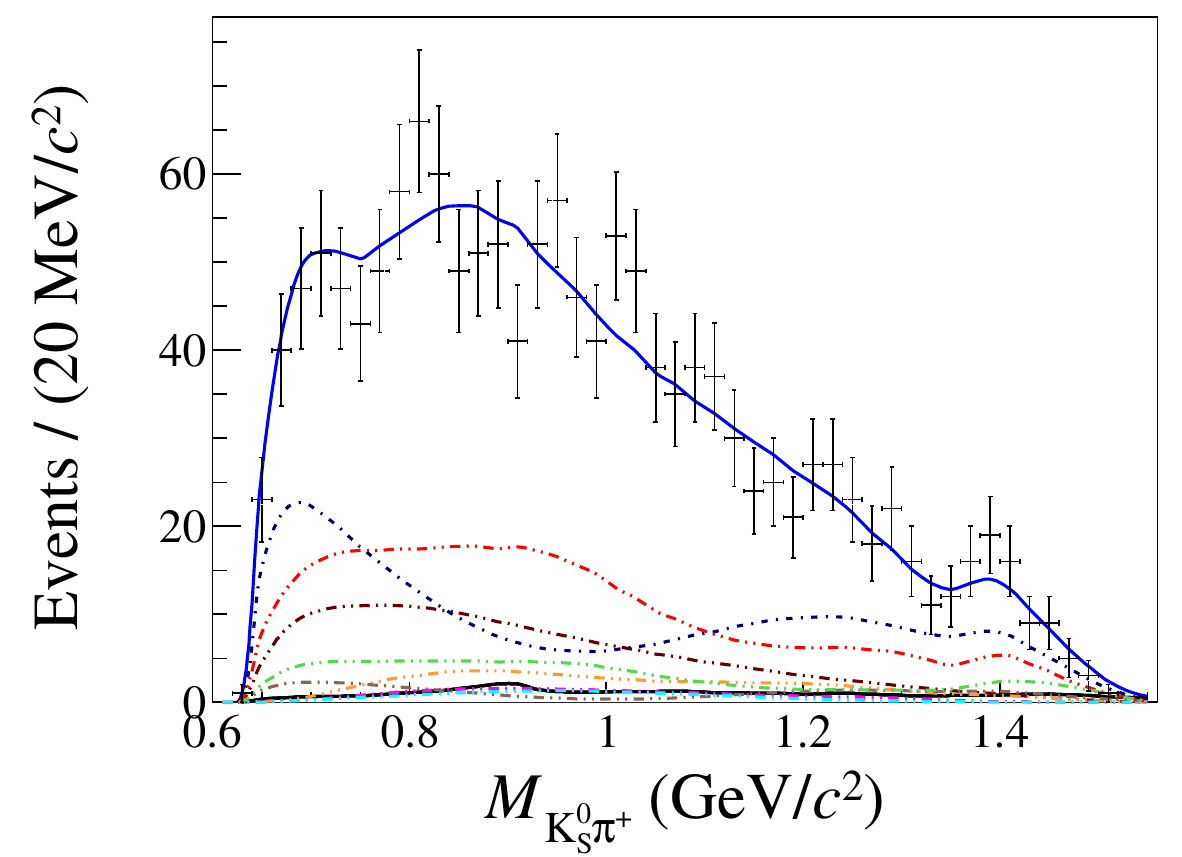}
          \includegraphics[width=0.225\textwidth]{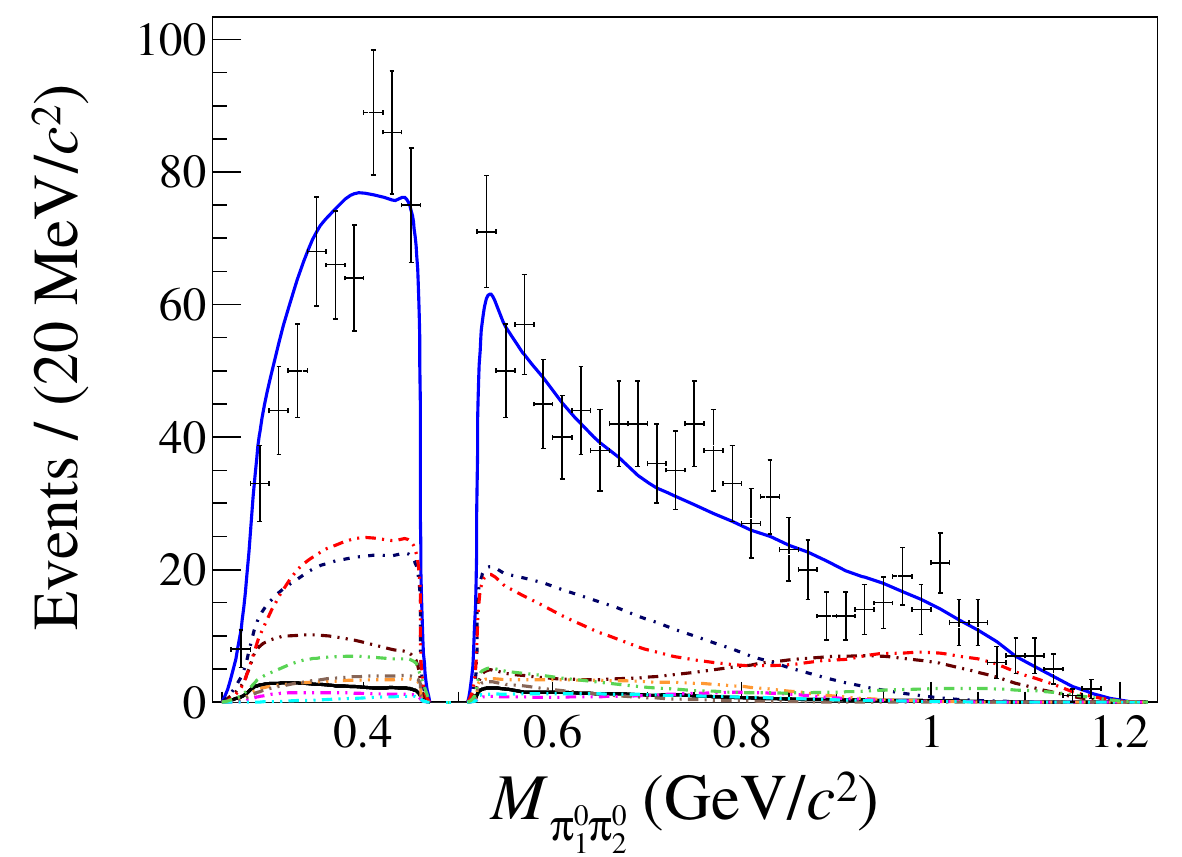}
          \includegraphics[width=0.225\textwidth]{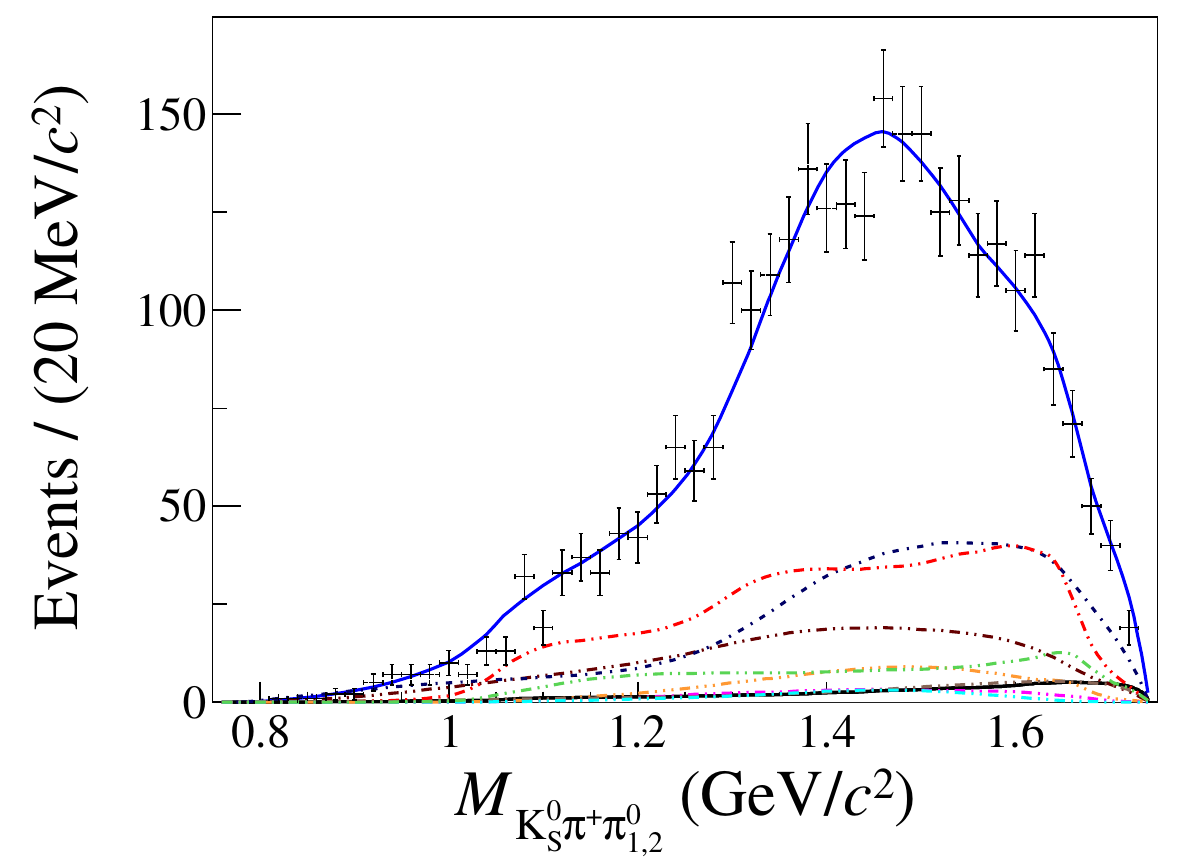}
          \includegraphics[width=0.225\textwidth]{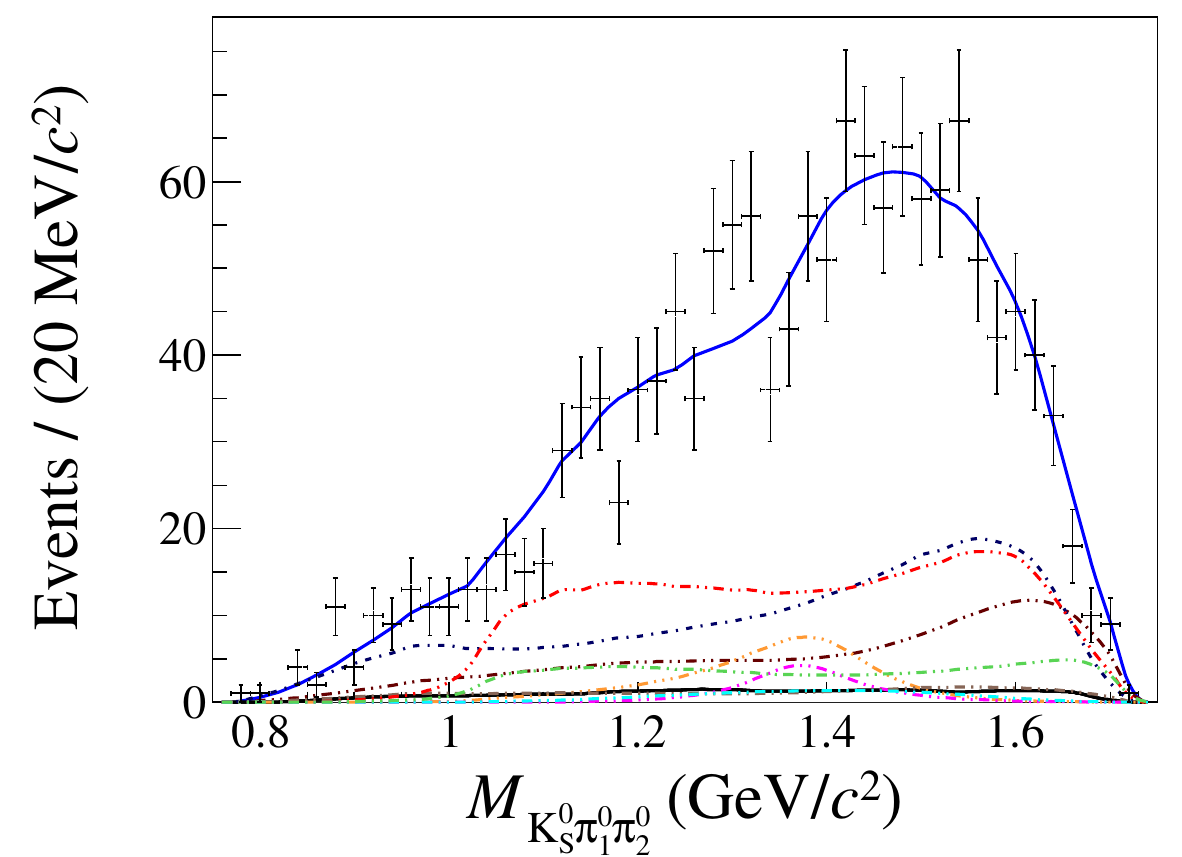}
          \includegraphics[width=0.225\textwidth]{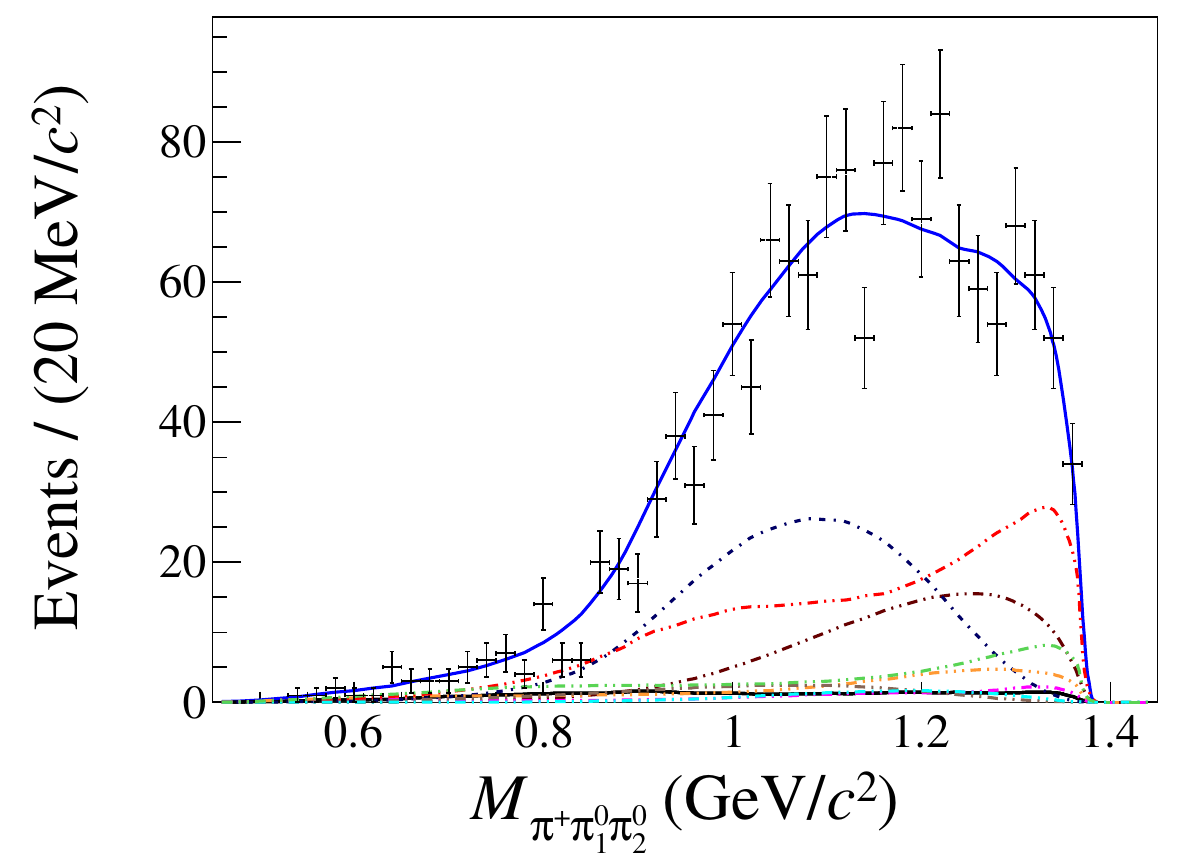}
          \includegraphics[width=0.225\textwidth]{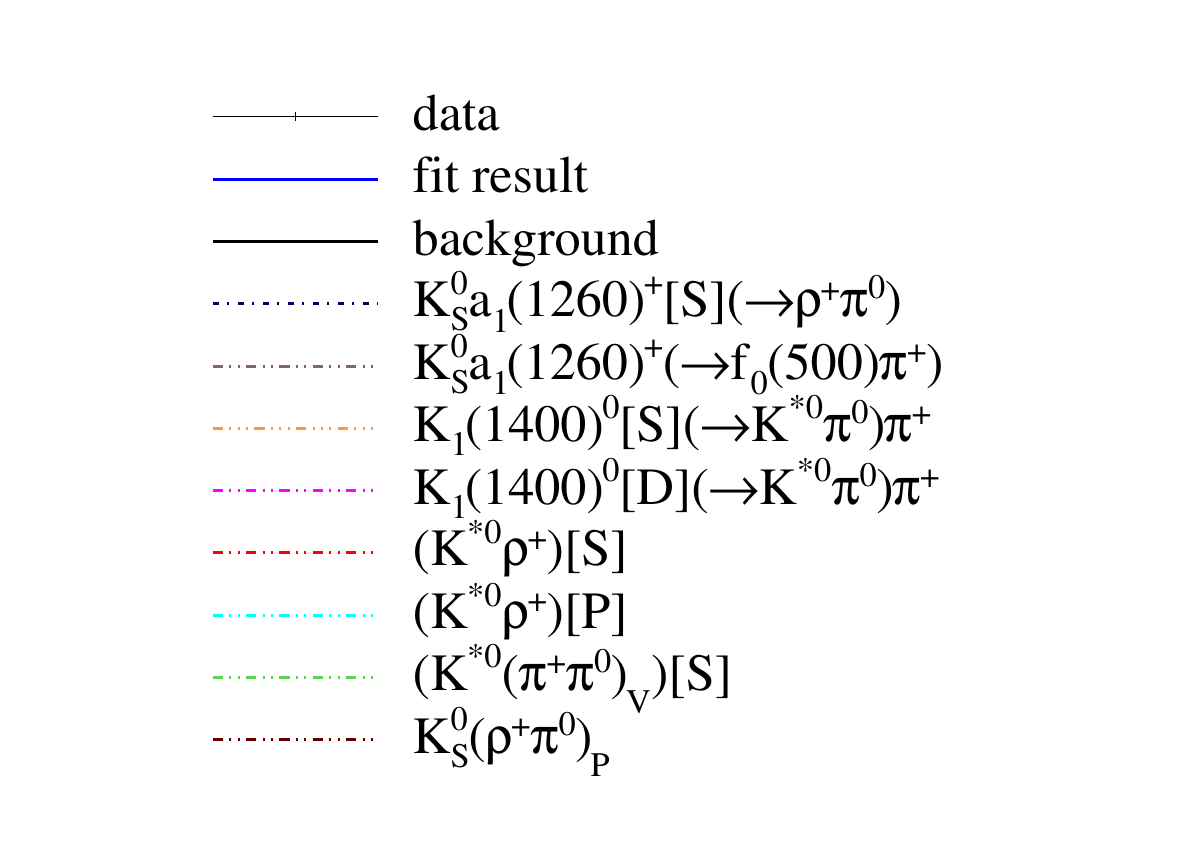}
          \caption{
The projections of the amplitude analysis fit of $D^+\to K^0_S\pi^+2\pi^0$ on
    two-body and three-body particle mass distributions~\cite{BESIII:2023qgj}.
 }
    \label{fig:Dp_KSpi2pi0}
\end{figure*}

From the analysis of about 1.5k candidates with a purity of about 97\%,
the first amplitude analysis of $D^+\to K^0_S\pi^+2\pi^0$ was presented in Ref.~\cite{BESIII:2023qgj}.
The projections of the amplitude analysis fit on two-body or three-body particle mass distributions are given in Fig.~\ref{fig:Dp_KSpi2pi0}.
Its decay amplitude can be well described by the subprocesses of
    $D^+\to K_S^0a_1(1260)^+[S](\to \rho(770)^+\pi^0)$,
    $D^+\to K_S^0a_1(1260)^+(\to f_0(500)\pi^+)$,
    $D^+\to K_1(1400)^0[S](\to \bar K^*(892)^0\pi^0)\pi^+$,
    $D^+\to K_1(1400)^0[D](\to \bar K^*(892)^0\pi^0)\pi^+$,
    $D^+\to K_1(1400)^0(\to \bar K^*(892)^0\pi^0)\pi^+$,
    $D^+[S]\to \bar K^*(892)^0\rho(770)^+$,
    $D^+[P]\to \bar K^*(892)^0\rho(770)^+$,
    $D^+\to  \bar K^*(892)^0\rho(770)^+$,
    $D^+[S]\to \bar K^*(892)^0(\pi^+\pi^0)_V$, and
    $D^+\to K_S^0(\rho(770)^+\pi^0)_P$.
The $D^+\to \bar K^*(892)^0\rho(770)^+$ and $D^+\to K_S^0a_1(1260)^+[S](\to \rho(770)^+\pi^0)$ decays dominate in $D^+\to K_S^0\pi^+2\pi^0$ with fractions
of $(33.6\pm2.7\pm1.4)\%$ and $(30.0\pm3.6\pm4.2)\%$, respectively. The branching fractions for each intermediate process are also presented.
The branching fraction of $D^+\to K^0_S\pi^+2\pi^0$  is determined to be $\mathcal{B}(D^+\to K_S^0\pi^+2\pi^0) = (2.888\pm0.058\pm0.069)\%$,
which is consistent with the previous BESIII result $(2.904\pm0.062\pm0.087)\%$~\cite{BESIII:2022mji}.
The branching fractions of each component are also presented.
The measured branching fraction of $D^+\to \bar{K}^{*}(892)^0\rho(770)^+$ is determined to be $(5.82\pm0.49\pm0.28)\%$,
which is consistent with the MARKIII result $(4.8\pm1.2\pm1.4)\%$~\cite{MARK-III:1991fvi} but with much improved precision.
The measured branching fraction of $D^+\to K_S^0a_1(1260)^+[S](\to \rho(770)^+\pi^0)$ is also consistent with the previous BESIII result~\cite{BESIII:2019ymv}.
An obvious $D^+\to \bar{K}_1(1400)^0(\to \bar{K}^{*}(892)^0\pi^0)\pi^+$ signal is observed, but no significant $D^+\to K_1(1270)^+\pi^0$ contribution is found.
This phenomenon is consistent with the theoretical prediction~\cite{Cheng:2003bn} and similar to that in $D^+\to K_S^02\pi^+\pi^-$, where the fit fraction of $D^+\to K_1(1400)\pi$ is about 10 times that of $D^+\to K_1(1270)\pi$~\cite{BESIII:2019ymv}.

\begin{figure*}
	        \centering
    \includegraphics[width=0.225\textwidth]{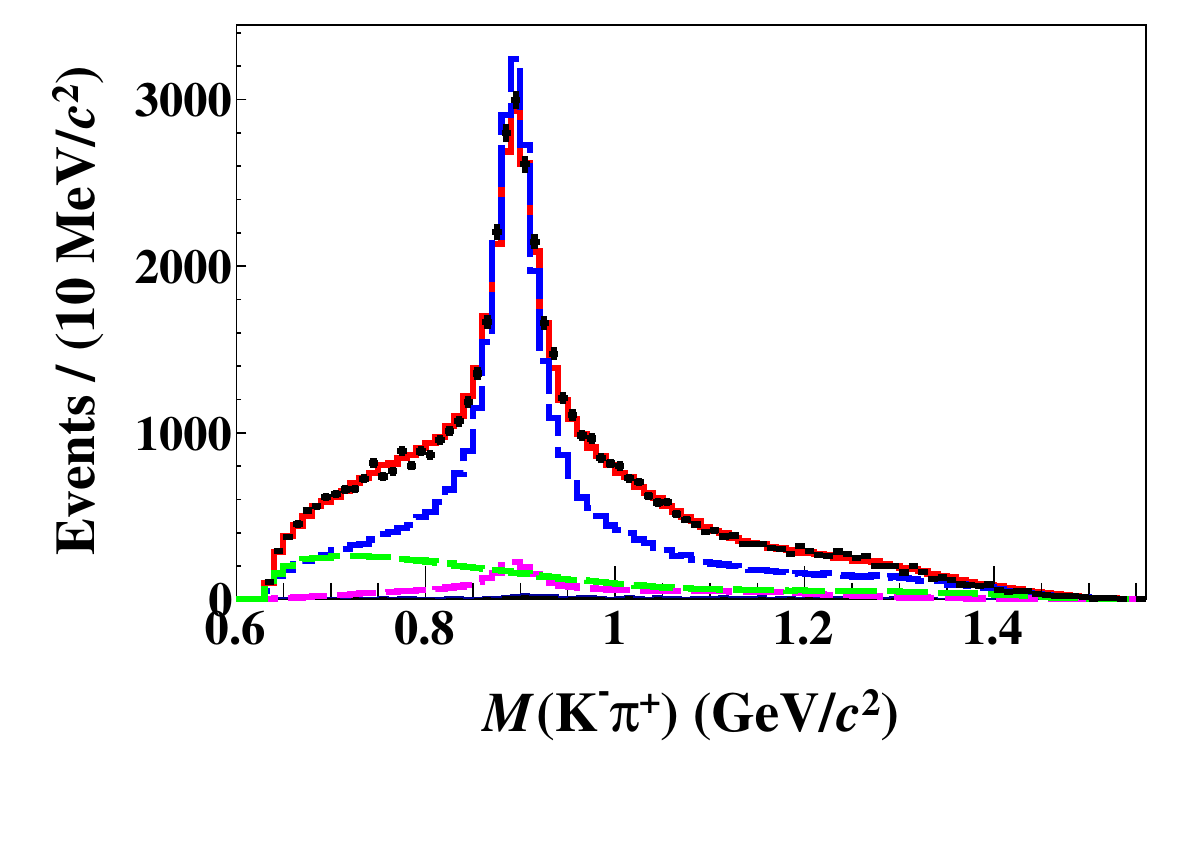}
    \includegraphics[width=0.225\textwidth]{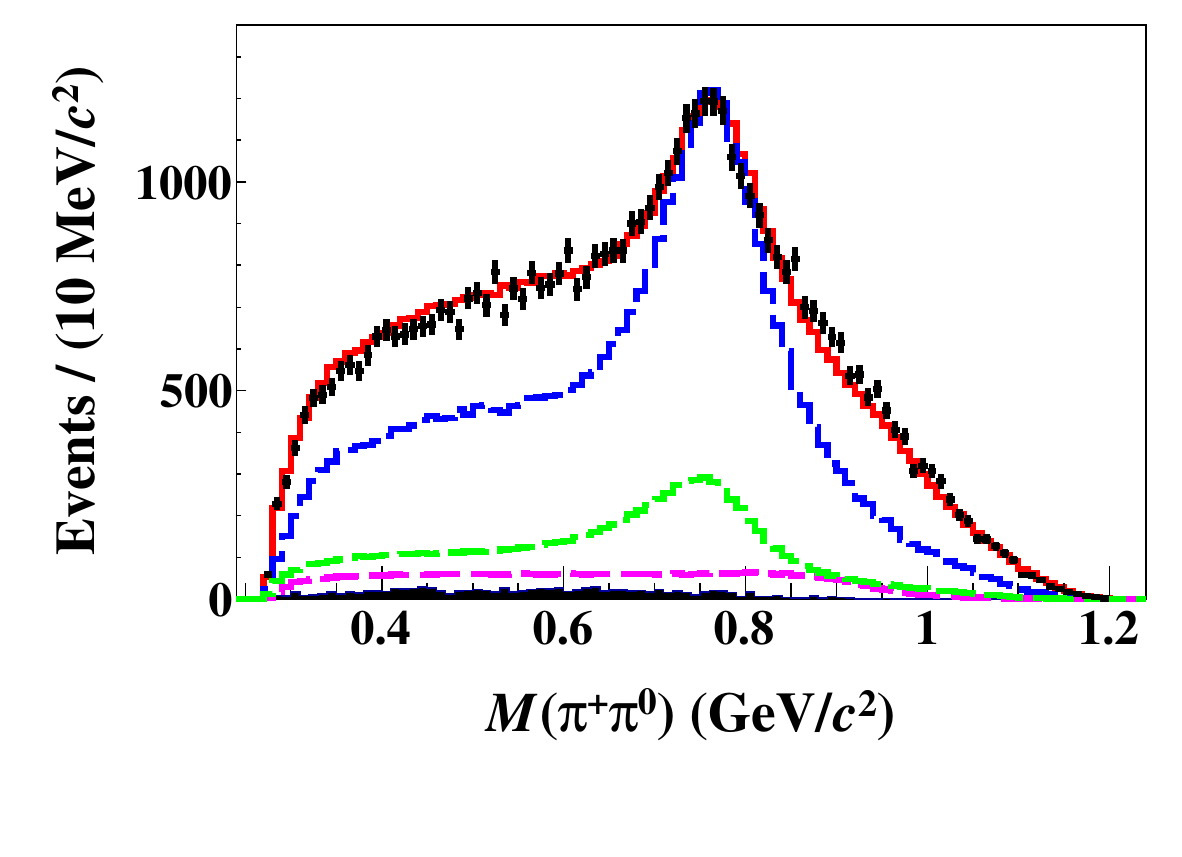}
    \includegraphics[width=0.225\textwidth]{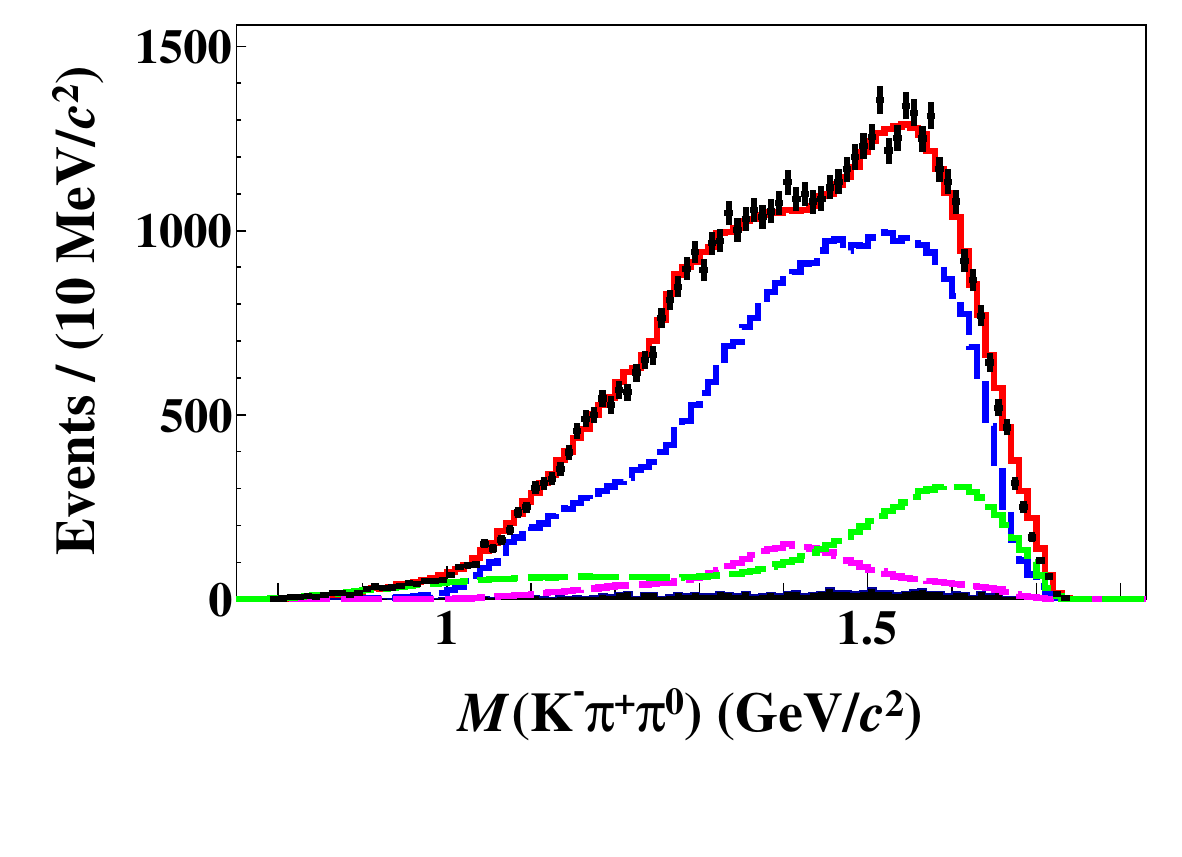}
    \includegraphics[width=0.225\textwidth]{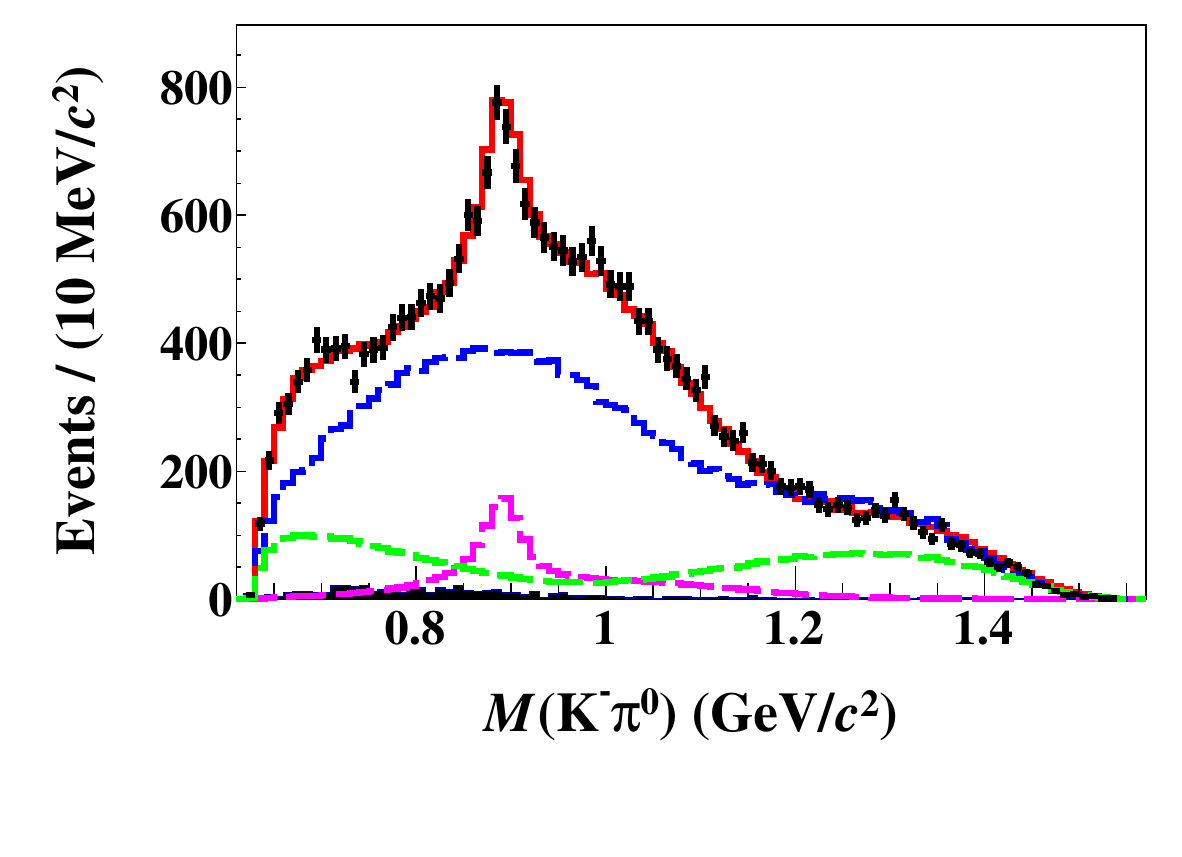}
    \includegraphics[width=0.225\textwidth]{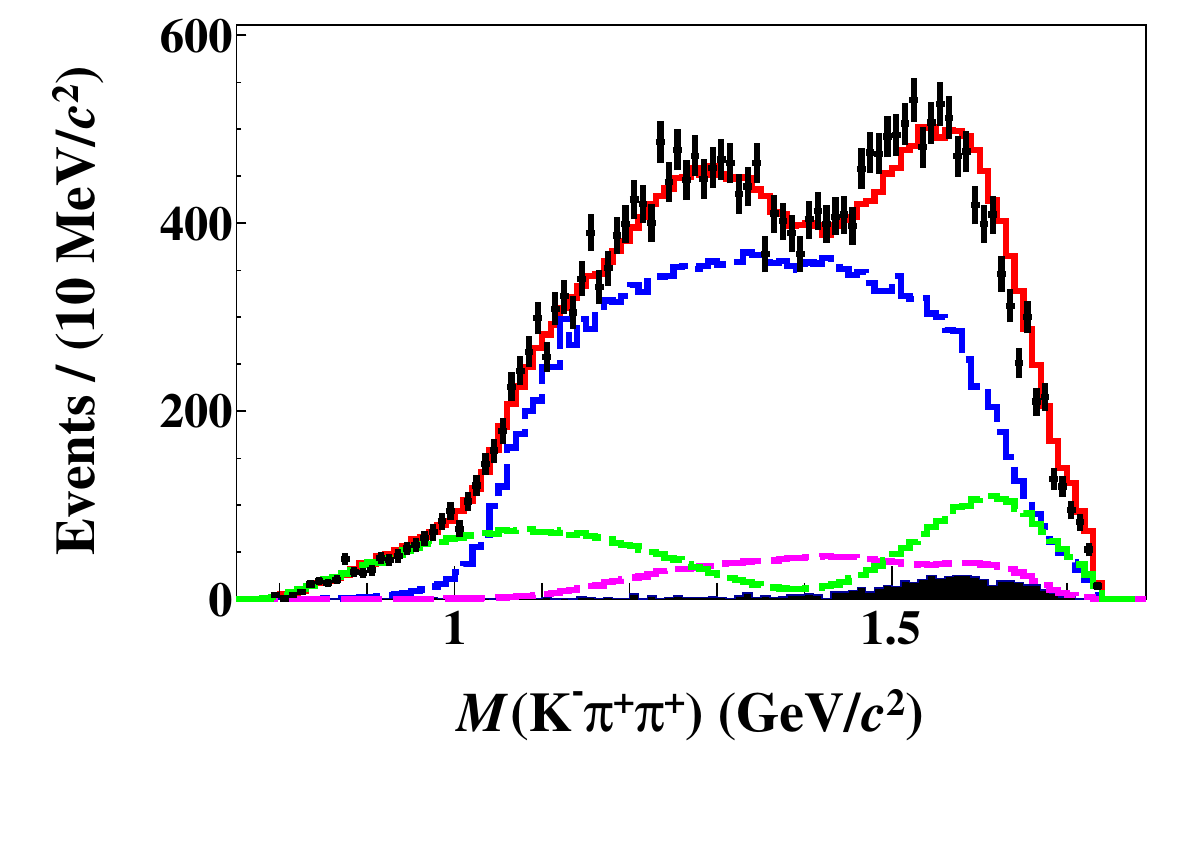}
     \includegraphics[width=0.225\textwidth]{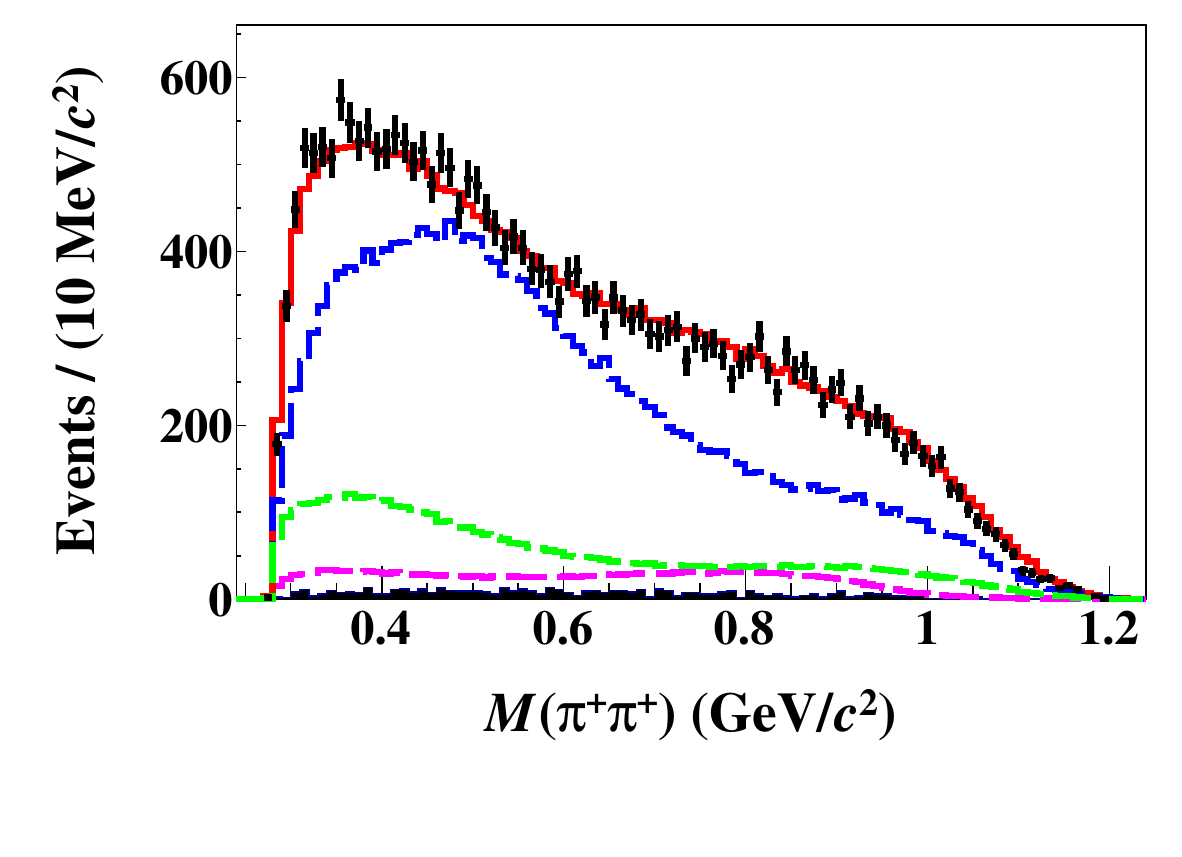}
     \includegraphics[width=0.225\textwidth]{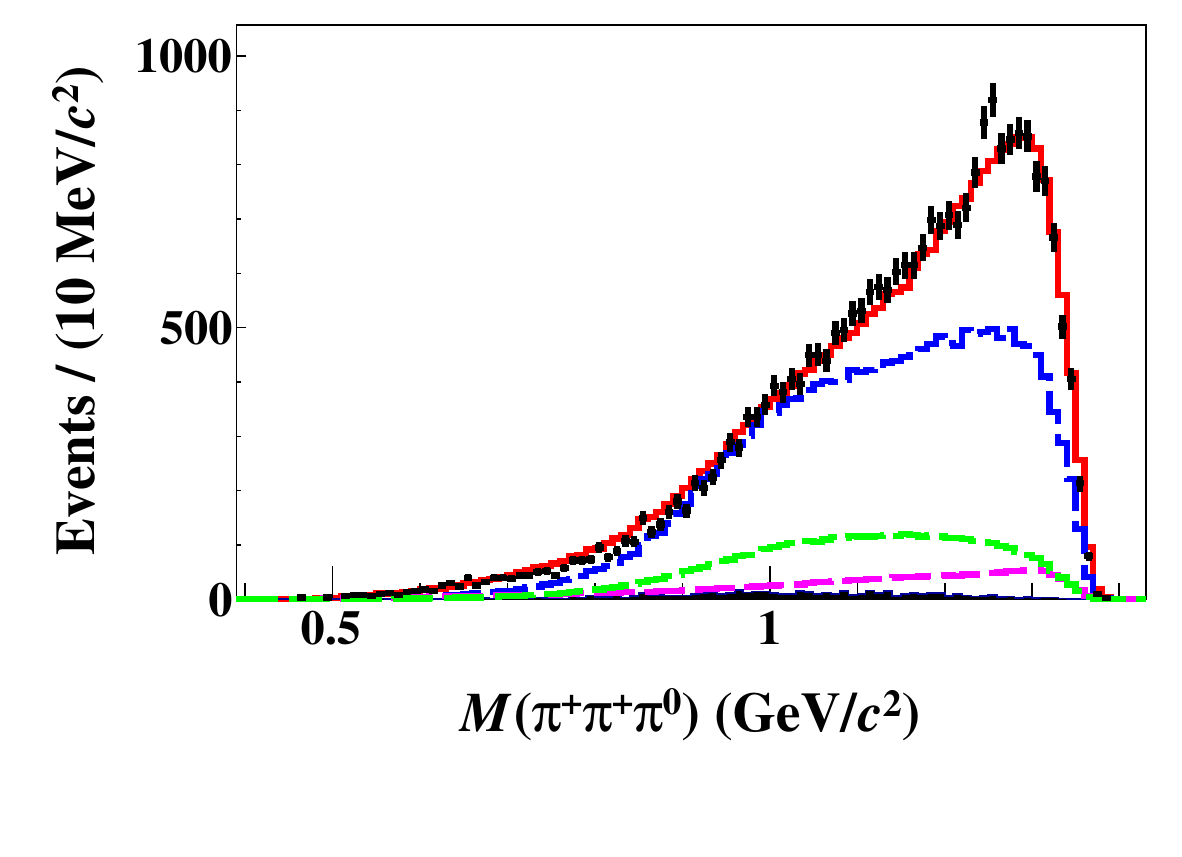}
    \includegraphics[width=0.225\textwidth]{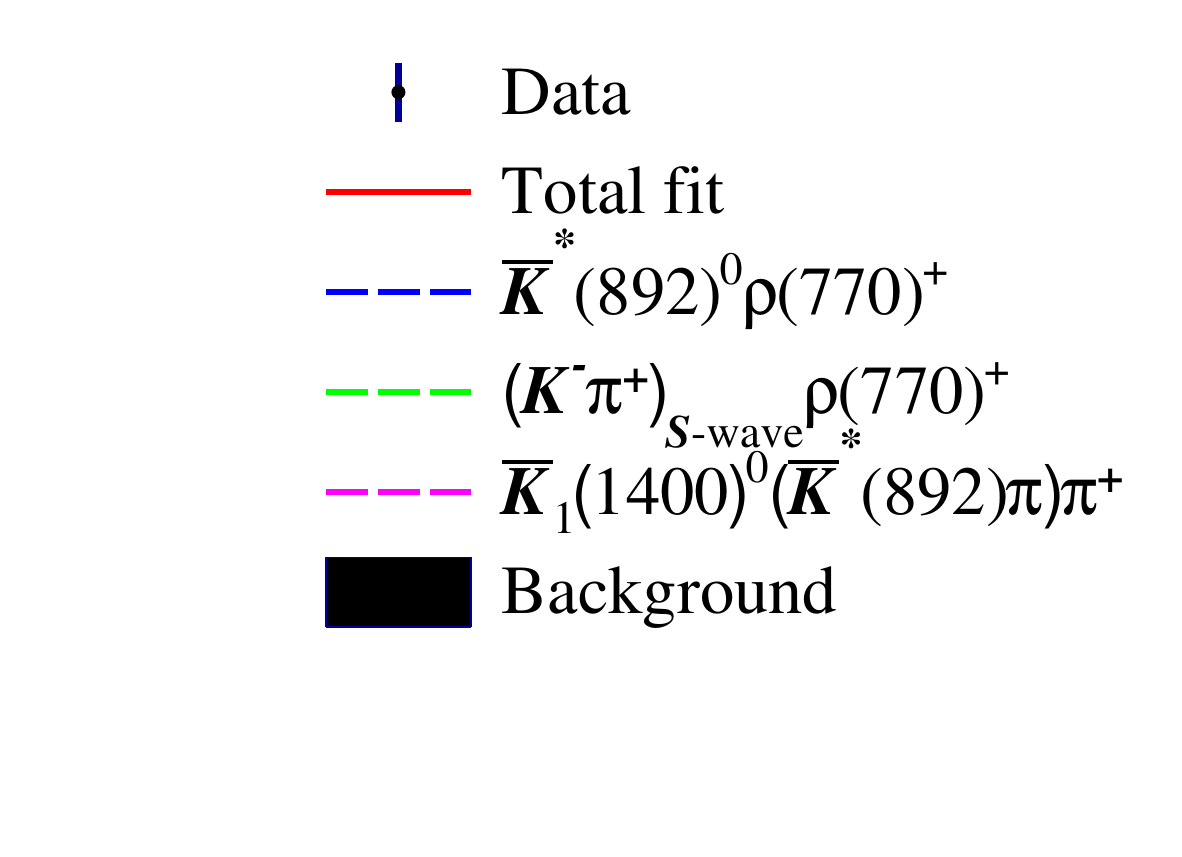}
	\caption{
The projections of the amplitude analysis fit of $D^+\to K^-2\pi^+\pi^0$ on
    two-body and three-body particle mass distributions~\cite{BESIII:2024waz}.
}
    \label{fig:Dp_K2pipi0}
\end{figure*}

In 2024, based on about 26.7k candidates with a signal purity of 98.4\%,
an amplitude analysis was conducted on the hadronic decay $D^+ \to K^-2\pi^+\pi^0$~\cite{BESIII:2024waz}.
Figure~\ref{fig:Dp_K2pipi0} exhibits the projections of the amplitude analysis fit on two-body or three-body particle mass distributions.
From the amplitude analysis, the dominant intermediate process is $D^+ \to \bar K^{*}(892)^0\rho(770)^+\to K^-2\pi^+\pi^0$,
with a fraction of $(68.4\pm1.1\pm2.6)\%$. In addition, significant contributions of
 $D^+\to \bar K_1(1270)^0\pi^+$, $D^+\to \bar K_1(1400)^0\pi^+$, $D^+\to \bar K_1(1460)^0\pi^+$,
 $D^+\to [K^-\rho(770)^+]^{L=1}\pi^+$,
  $D^+\to [\bar K^*(892)\pi]^{L=1}\pi^+$,
  $D^+\to [\bar K^*(892)^0\pi^+]^{L=1}\pi^0$,
  $D^+\to [K^-[\pi^+\pi^0]^{L=1}]^{L=0}\pi^0$, and
  $D^+[S]\to [K^-\pi^+]^{L=1}\rho(770)^+$ are also observed.
The branching fraction of $D^+\to K^-2\pi^+\pi^0$ is determined to be $(6.35\pm0.04\pm0.07)\%$,
which is consistent with the previous CLEO-c measurement, $(5.98\pm0.08\pm0.16)\%$~\cite{CLEO:2007rrw}.
The  branching fraction of each intermediate process is also presented.
The obtained ${\cal B}(D^+ \to \bar{K}^{*}(892)^0\rho(770)^+)=(6.23\pm0.11\pm0.25)\%$
is consistent with previous measurements from MARKIII~\cite{MARK-III:1991fvi} and BESIII~\cite{BESIII:2023qgj} with much improved precision.
In addition, the obtained ${\cal B}(D^+ \to \bar{K}_1(1400)^0\pi^+)$ is consistent with the previous BESIII result~\cite{BESIII:2023qgj} within 1.5$\sigma$,
with precision improved by a factor of 4.4.

\subsubsection{Analyses of $D^+_{(s)}\to \pi\pi\pi\pi$}

\begin{figure*}[htp]
  \begin{center}
    \includegraphics[width=0.45\textwidth]{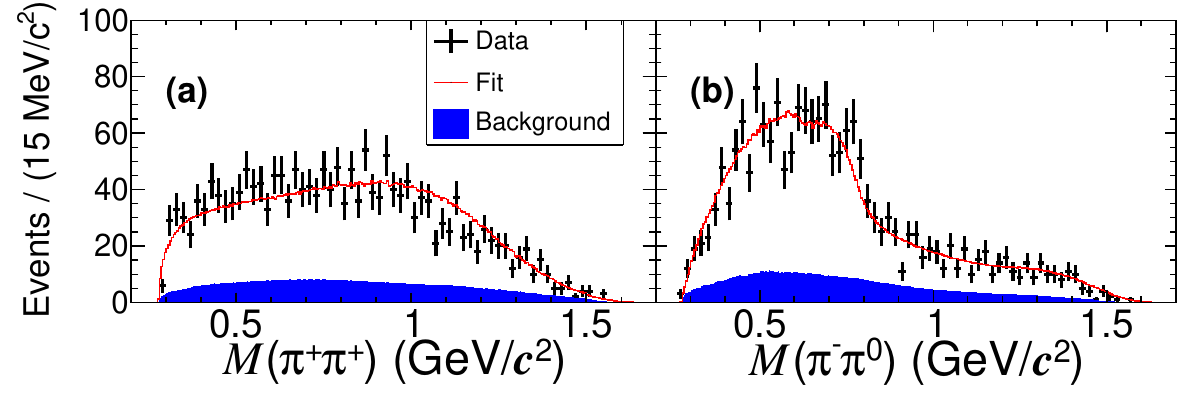}
    \includegraphics[width=0.45\textwidth]{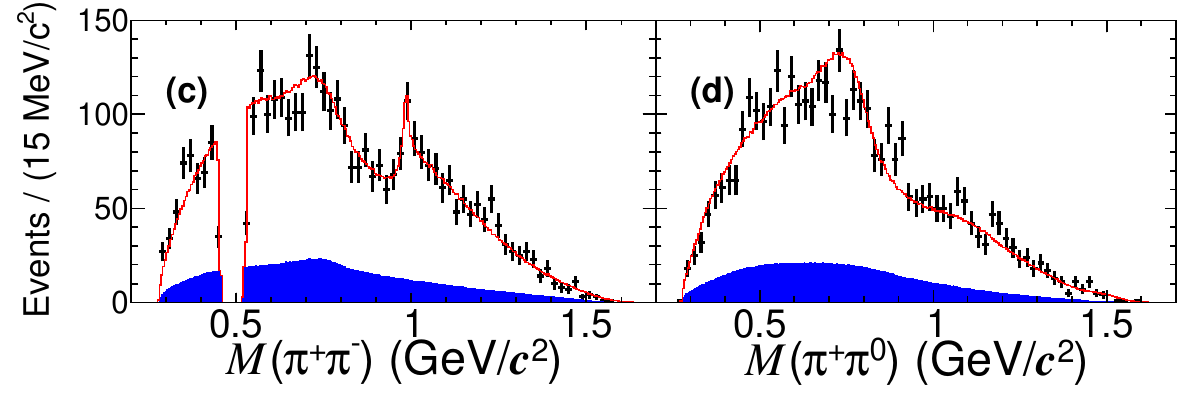}
    \includegraphics[width=0.45\textwidth]{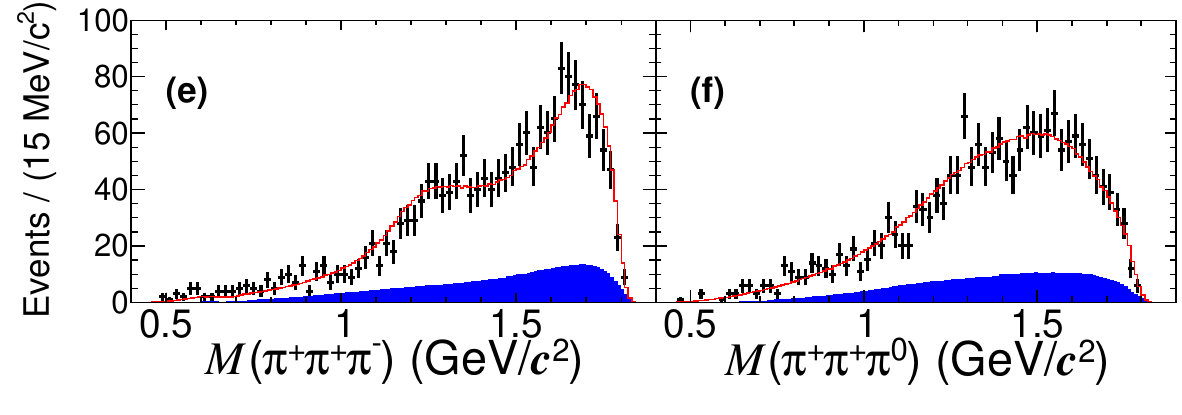}
    \includegraphics[width=0.23\textwidth]{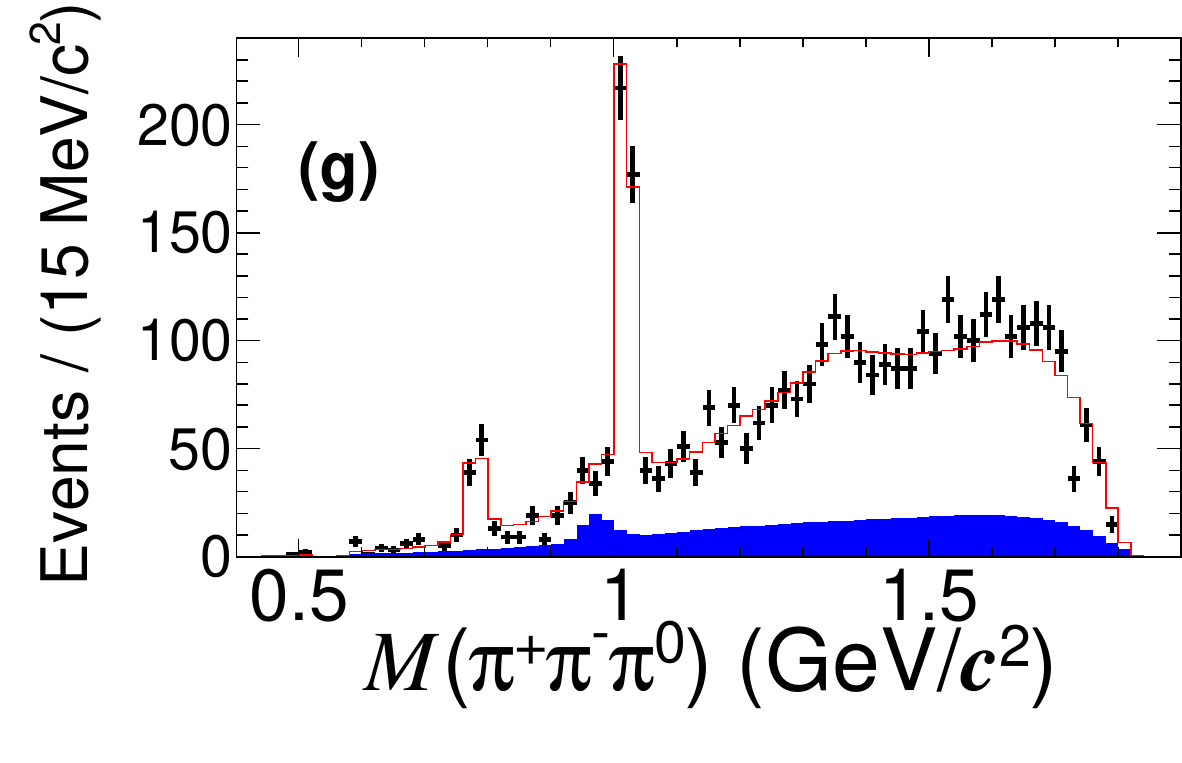}
    \caption{The projections of the amplitude analysis fit of $D^{+}_{s}\to 2\pi^+\pi^-\pi^0$ on
    two-body and three-body particle mass distributions~\cite{BESIII:2024muy}.
      }
    \label{fig:Ds_3pipi0}
  \end{center}
\end{figure*}

Compared to the relatively well-advanced progress in studies of $D_{(s)}\to VP$, $PP$, and $VV$ decays,
the $D_{(s)}\to SV$ modes remain less explored.
Among these, precise knowledge of the $D^+_s \to f_0(980)\rho(770)^+$ decay, which proceeds dominantly via a $W$-external-emission diagram, allows for precise theoretical predictions in the absence of FSIs~\cite{Hsiao:2023qtk, Hsiao:2019ait, Yu:2021euw, Zhang:2022xpf}.
Because FSI effects are essential for understanding light scalar mesons like $f_0(980)$, whose nature was debated,
an amplitude analysis of $D^+_s \to 2\pi^+\pi^-\pi^0$ is important to constrain FSI effects and shed light on the $f_0(980)$ structure.
Moreover, precision measurement of $D_{s}^{+} \to \omega\pi^+$, which mediates solely via $W$-annihilation, helps to improve theoretical predictions for $D_{(s)}\to VP$~\cite{Cheng:2012wr, Qin:2013tje, Cheng:2010ry, Cheng:2016ejf}.
In addition, the $D^+_s \to 2\pi^+\pi^-\pi^0$ decay offers a clean environment to measure the ratio
${\cal R}_\phi = \mathcal{B}(\phi\to 3\pi)/\mathcal{B}(\phi\to K^+K^-)$, which will complement previous studies in $e^+e^-$ and $K^-p$ scattering.
Precise $\phi$ decay branching fractions are important for both strong-interaction studies~\cite{Yukawa:1935xg,STAR:2022fan} and $B$-decay analyses~\cite{LHCb:2023exl,CDF:2005apk,BaBar:2003mjy,BaBar:2006ahy}.
Previously, BESIII reported ${\cal B}(D_{s}^{+} \to \omega\pi^+)=(1.77\pm0.32\pm0.13)\times 10^{-3}$ with 3.19 fb$^{-1}$ of data at 4.178 GeV~\cite{BESIII:2018mwk}.
In addition, the $D_{s}^{+} \to 2\pi^+\pi^-\pi^0$ channel also contains contributions from
$D_{s}^{+} \to \eta \pi^{+}$, $D_{s}^{+} \to f_0(500)\rho(770)^{+}$,
$D_{s}^{+} \to f_0(1370)\rho(770)^{+}$, $D_{s}^{+} \to f_2(1270)\rho(770)^{+}$,
$D_{s}^{+} \to \rho(770)^0\rho(770)^{+}$, $D_{s}^{+} \to a_1(1260)\pi$, etc,
offering a comprehensive probe of low-energy strong interactions and the $D_s^+$ weak-decay mechanism.
Using about 1.6k candidates with a signal purity of 80\%, Ref.~\cite{BESIII:2024muy} reported
an amplitude analysis on $D^{+}_{s}\to 2\pi^+\pi^-\pi^0$.
Figure~\ref{fig:Ds_3pipi0} presents the projections of the amplitude analysis fit on two-body or three-body particle mass distributions.
Its amplitude can be well described by
      $f_{0}(1370)\rho(770)^{+}$,
      $f_{0}(980)\rho(770)^{+}$,
      $f_{2}(1270)\rho(770)^{+}$,
      $(\rho(770)^{+}\rho(770)^{0})_{S}$,
      $(\rho(1450)^{+}\rho(770)^{0})_{S}$,
      $(\rho(770)^{+}\rho(1450)^{0})_{P}$,
      $\phi((\rho(770)\pi)\to\pi^+\pi^-\pi^0)\pi^{+}$,
      $\omega((\rho(770)\pi)\to\pi^+\pi^-\pi^0)\pi^{+}$,
      $a_{1}(1260)^+(\rho(770)^{0}\pi^{+})_{S}\pi^{0}$,
      $a_{1}(1260)^0((\rho(770)\pi)_{S}\to\pi^+\pi^-\pi^0)\pi^{+}$, and
      $\pi(1300)^{0}((\rho(770)\pi)_{P}\to\pi^+\pi^-\pi^0)\pi^{+}$,
      with fractions of
      $(24.9\pm3.8\pm2.1)\%$,
      $(12.6\pm2.1\pm1.0)\%$,
      $(9.5\pm1.7\pm0.6)\%$,
      $(3.5\pm1.2\pm0.6)\%$,
      $(4.6\pm1.3\pm0.8)\%$,
      $(8.6\pm1.3\pm0.4)\%$,
      $(24.9\pm1.2\pm0.4)\%$,
      $(6.9\pm0.8\pm0.3)\%$,
      $(12.5\pm1.6\pm1.0)\%$,
      $(6.3\pm1.9\pm1.2)\%$, and
      $(11.7\pm2.3\pm2.2)\%$, respectively.
The $D_{s}^{+} \to f_0(980)\rho(770)^{+}$ decay is observed with a statistical significance
greater than 10$\sigma$.
The obtained branching fractions of
$D^+_s \to \eta\pi^+$ and
$D^{+}_{s}\to2\pi^+\pi^-\pi^0|_{{\rm non}-\eta}$
as well as some main subdecays
$D^+_s \to f_{0}(1370)(\to\pi^+\pi^-)\rho(770)^+(\to\pi^+\pi^0)$,
$D^+_s \to \phi(\to\pi^+\pi^-\pi^0)\pi^+$, and
$D^+_s \to f_{0}(980)(\to\pi^+\pi^-)\rho(770)^+(\to\pi^+\pi^0)$
are
$(1.56\pm0.09\pm0.04)\%$,
$(2.04\pm0.08\pm0.05)\%$,
$(5.08\pm0.80\pm0.43) \times 10^{-3}$,
$(5.08\pm0.32\pm0.10) \times 10^{-3}$, and
$(2.57\pm0.44\pm0.20) \times 10^{-3}$,
respectively.
Taking the branching fraction of $D^+_s \to \phi(\to K^+K^-)\pi^+$ from Ref.~\cite{BESIII:2020ctr},  the relative branching fraction
 between
$\phi$ decays into $\pi^+\pi^-\pi^0$ and $K^+K^-$ is calculated to be
\mbox{${\cal R}_\phi=\frac{\mathcal{B}(\phi(1020) \to \pi^+\pi^-\pi^0)}{\mathcal{B}(\phi(1020) \to K^+K^-)}=0.230 \pm 0.014\pm 0.010$}.
It significantly deviates from the world average value \mbox{${\cal R}^{\text{PDG}}_\phi=\frac{\mathcal{B}(\phi(1020) \to \pi^+\pi^-\pi^0)}{\mathcal{B}(\phi(1020) \to K^+K^-)}=0.313\pm0.010$} by more than $4\sigma$~\cite{ParticleDataGroup:2024cfk}.
The data in the PDG on the OZI suppressed $\phi \to \rho \pi$ decays are mostly from $e^+e^-$ collisions on the $\phi$ peak~\cite{ParticleDataGroup:2024cfk}.
The possible interference effect between $e^+e^- \to \gamma^{*} \to \rho \pi$ and $e^+e^- \to \phi \to \rho \pi$ may not be well considered,
and therefore, BESIII $D_s$ decay provides an independent test of this ratio, in which no such interference arises.
This is the first measurement of ${\cal R}_\phi$ in hadronic decays of charmed mesons, and the lower than expected value motivates further studies.
The branching fraction of the $W$-annihilation decay
$D^+_s \to \omega(\to\pi^+\pi^-\pi^0)\pi^+$ is determined to be
$(1.41\pm0.17\pm0.06) \times 10^{-3}$, with precision improved by a factor of two compared with
previous measurements.

\begin{figure*}[htbp]
  \centering
  \begin{overpic}[width=0.225\textwidth]{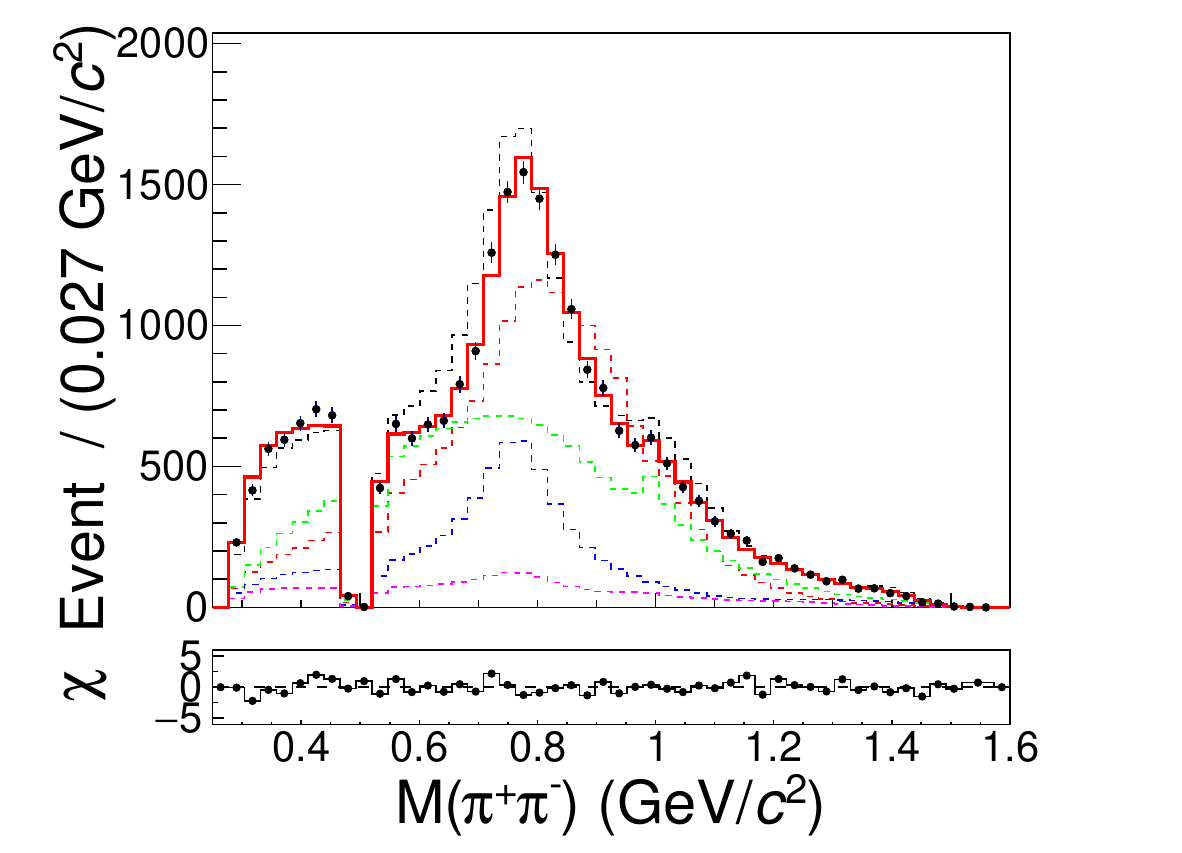}
  \end{overpic}
  \begin{overpic}[width=0.225\textwidth]{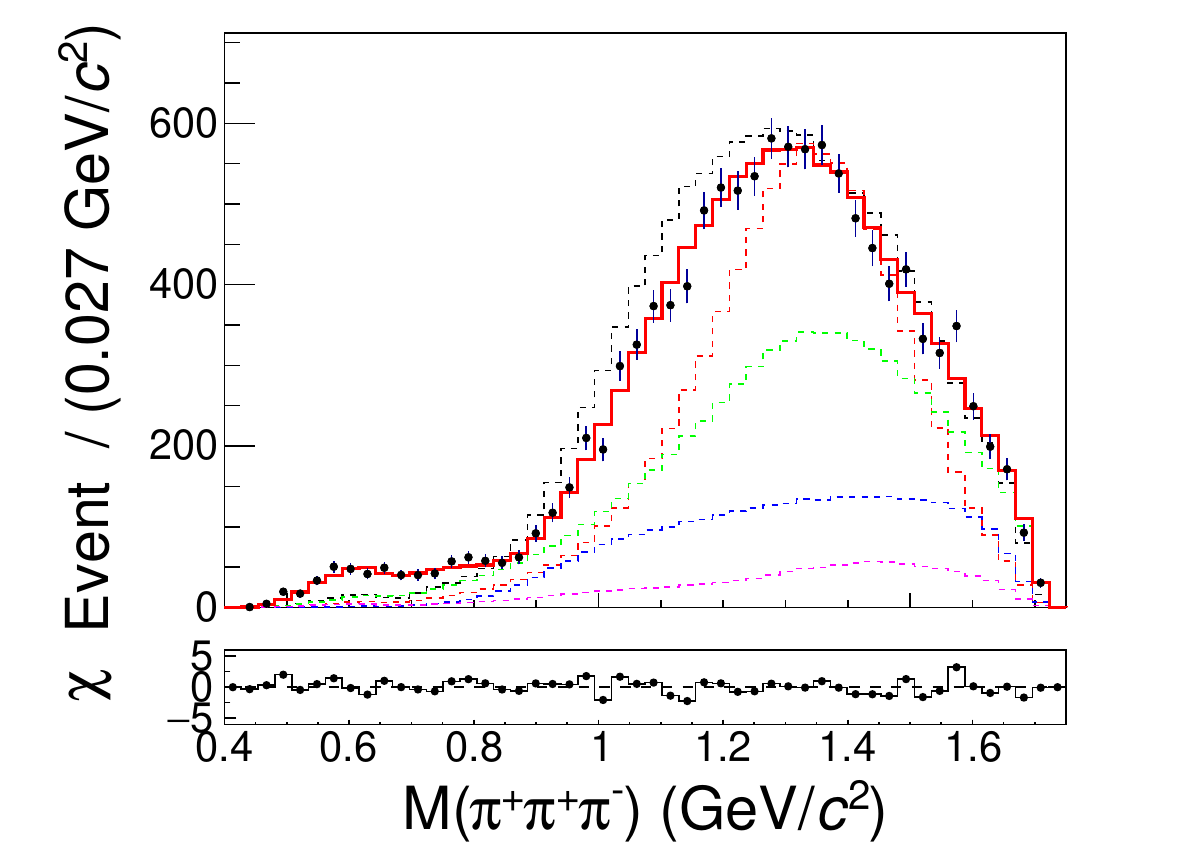}
  \end{overpic}
  \begin{overpic}[width=0.225\textwidth]{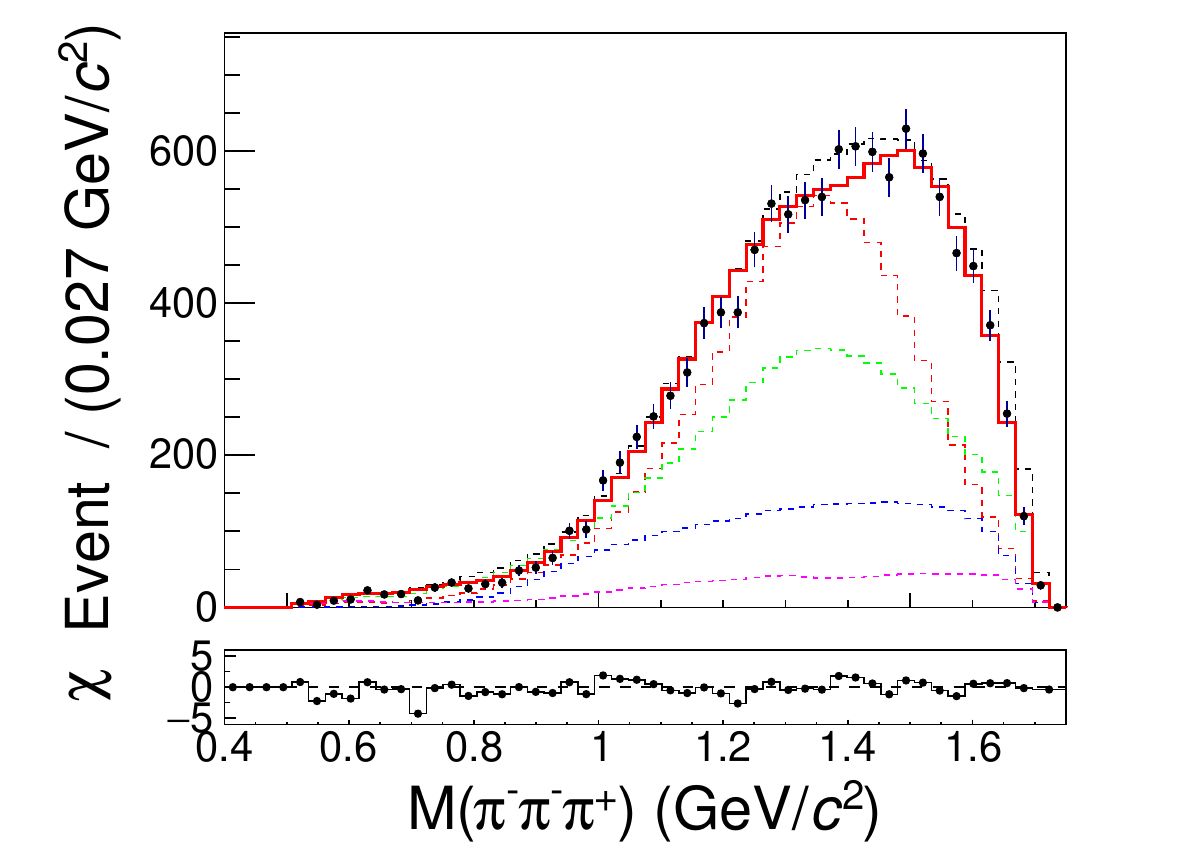}
  \end{overpic}
\begin{overpic}[width=0.225\textwidth]{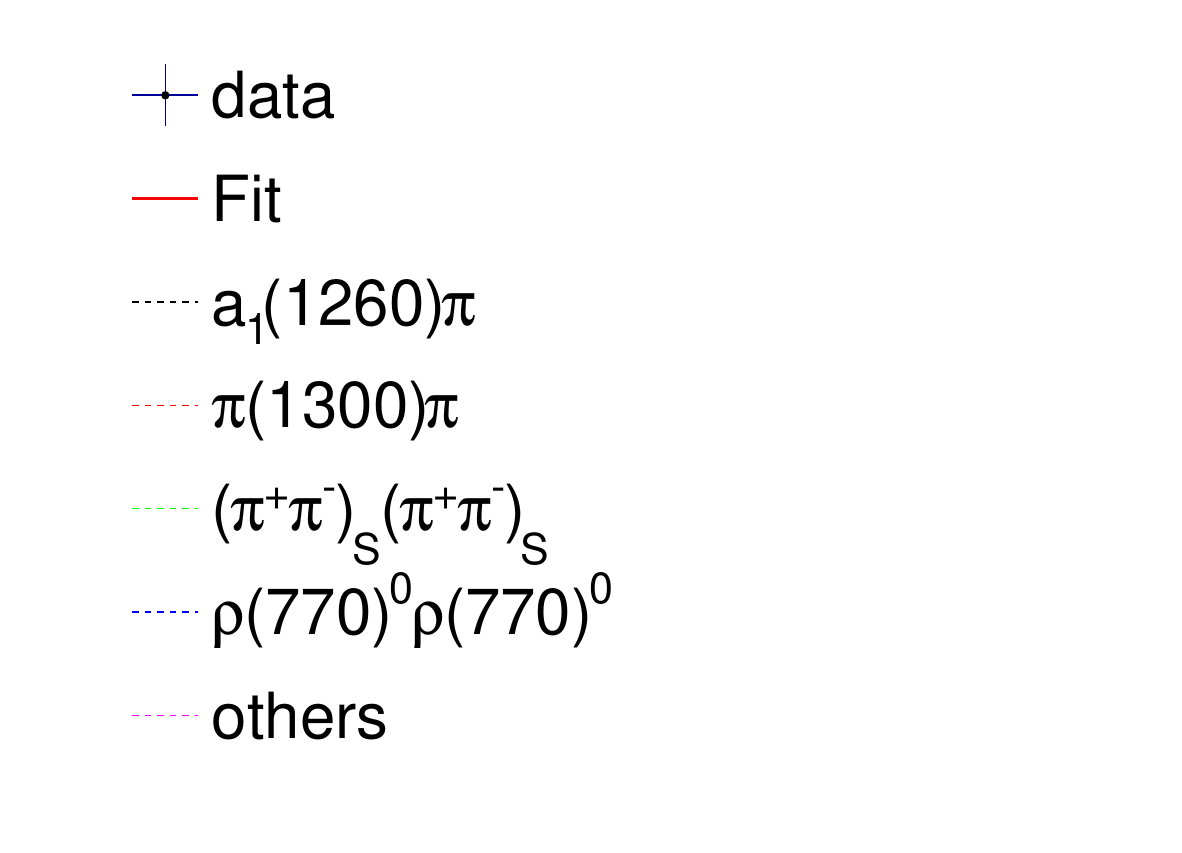}
  \end{overpic}
  \caption{
The projections of the amplitude analysis fit of $D^+\to 2\pi^+2\pi^-$ on
    two-body and three-body particle mass distributions~\cite{BESIII:2023exz}.
  }
  \label{fig:D0_4pi}
\end{figure*}

\begin{figure*}[htbp]
  \centering
 \begin{overpic}[width=0.225\textwidth]{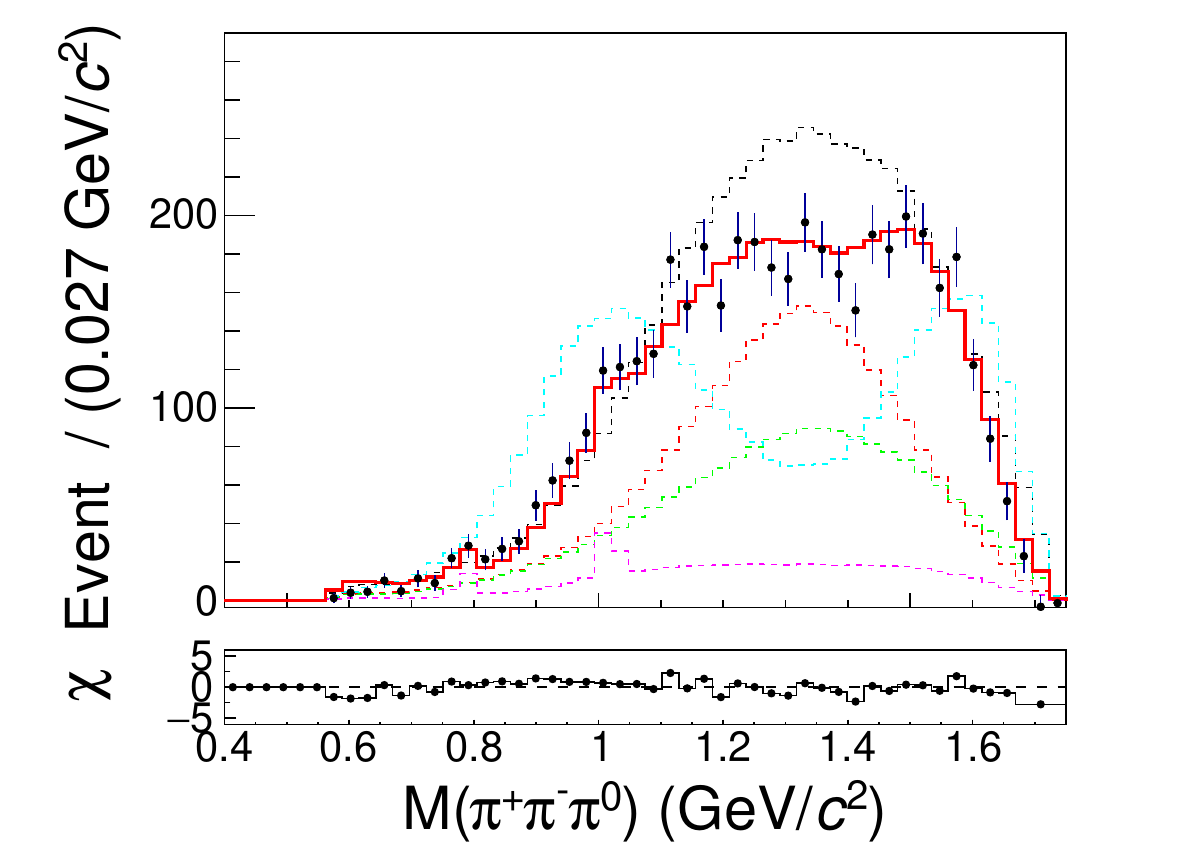}
  \end{overpic}
  \begin{overpic}[width=0.225\textwidth]{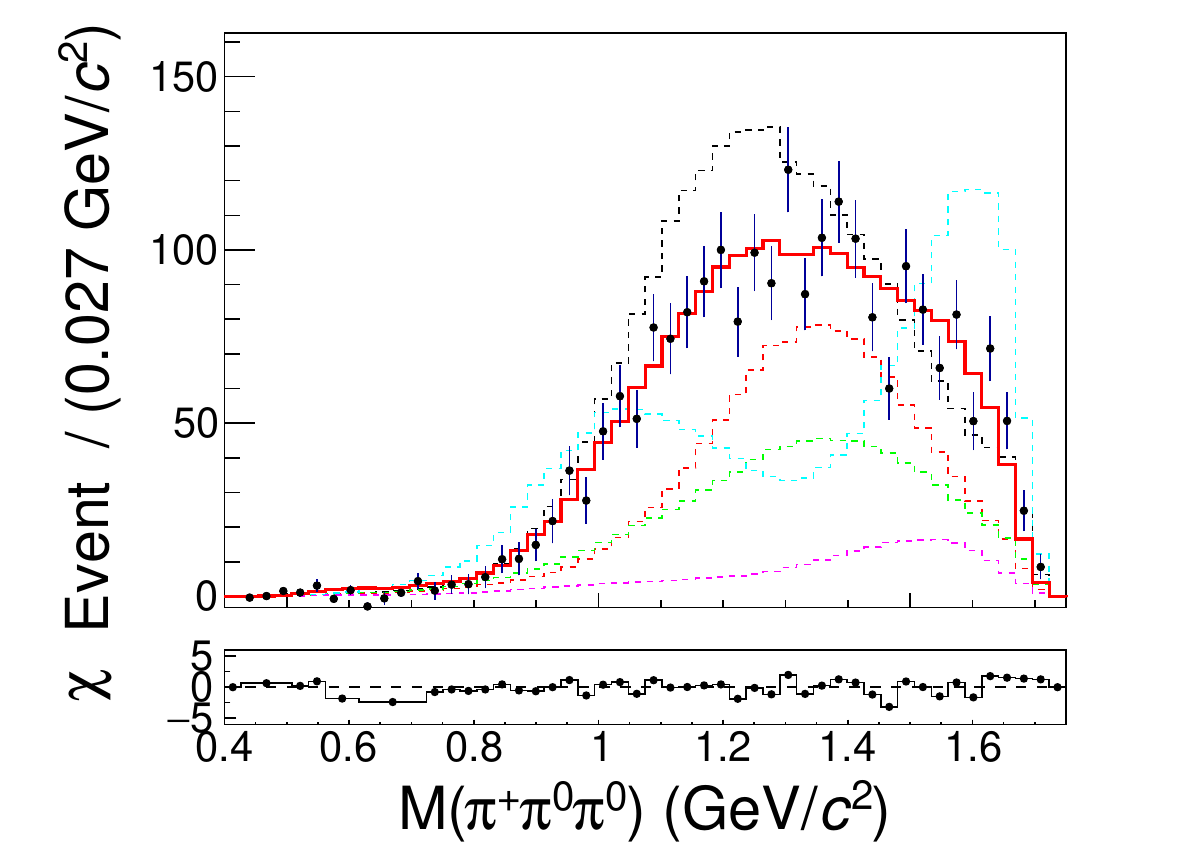}
  \end{overpic}
\begin{overpic}[width=0.225\textwidth]{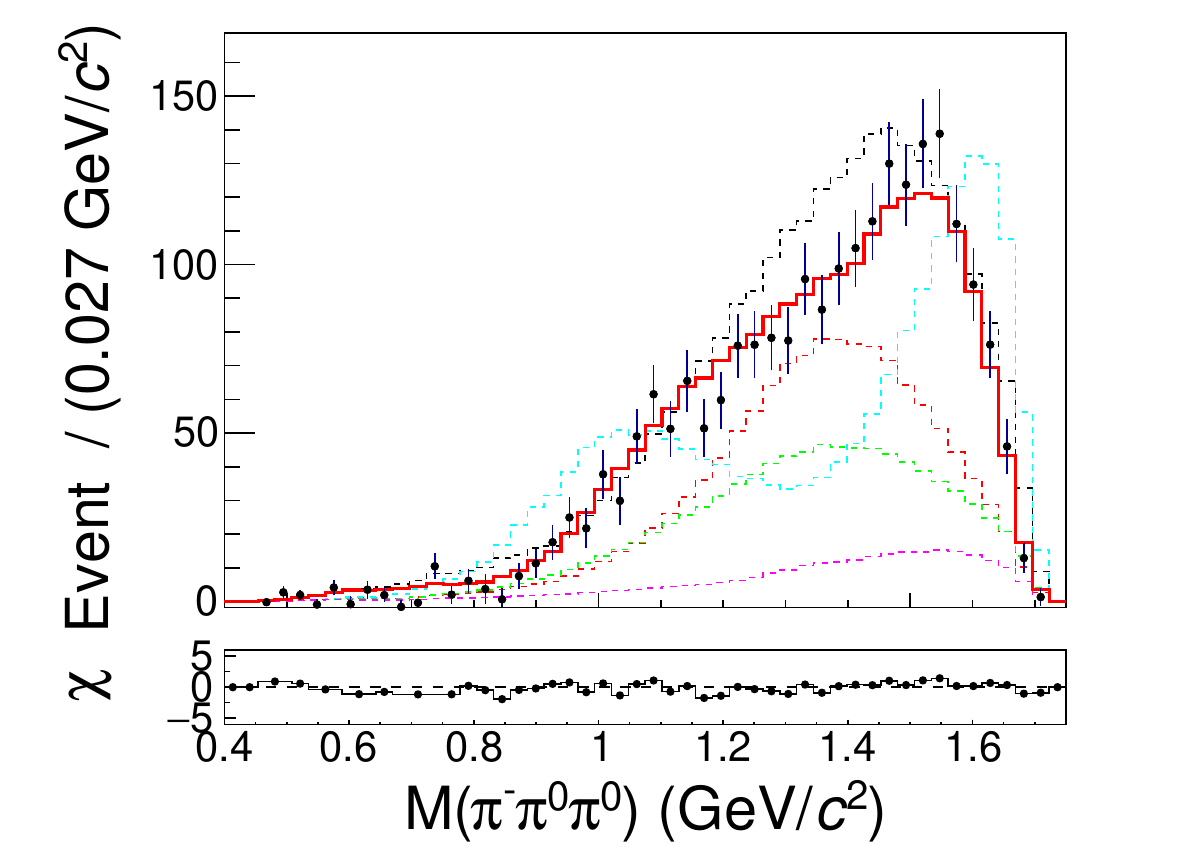}
\end{overpic}
   \begin{overpic}[width=0.225\textwidth]{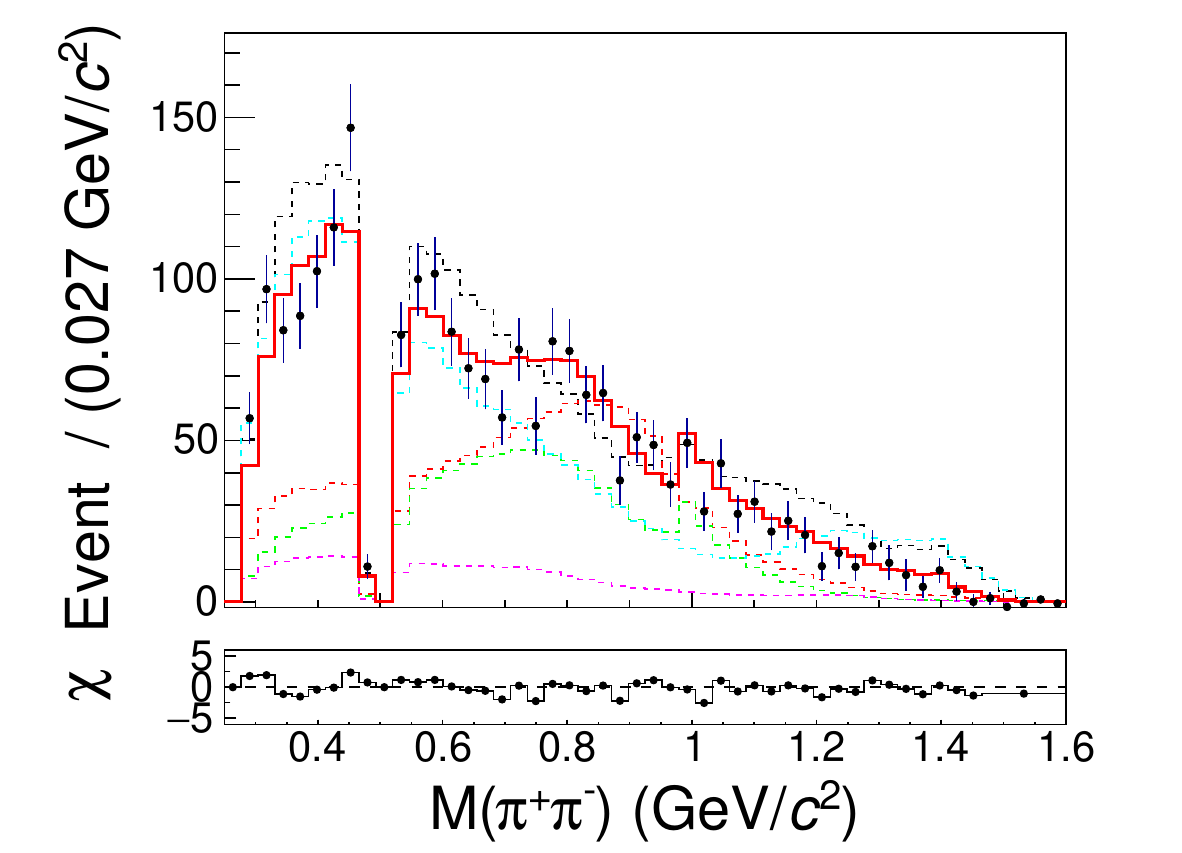}
  \end{overpic}
 \begin{overpic}[width=0.225\textwidth]{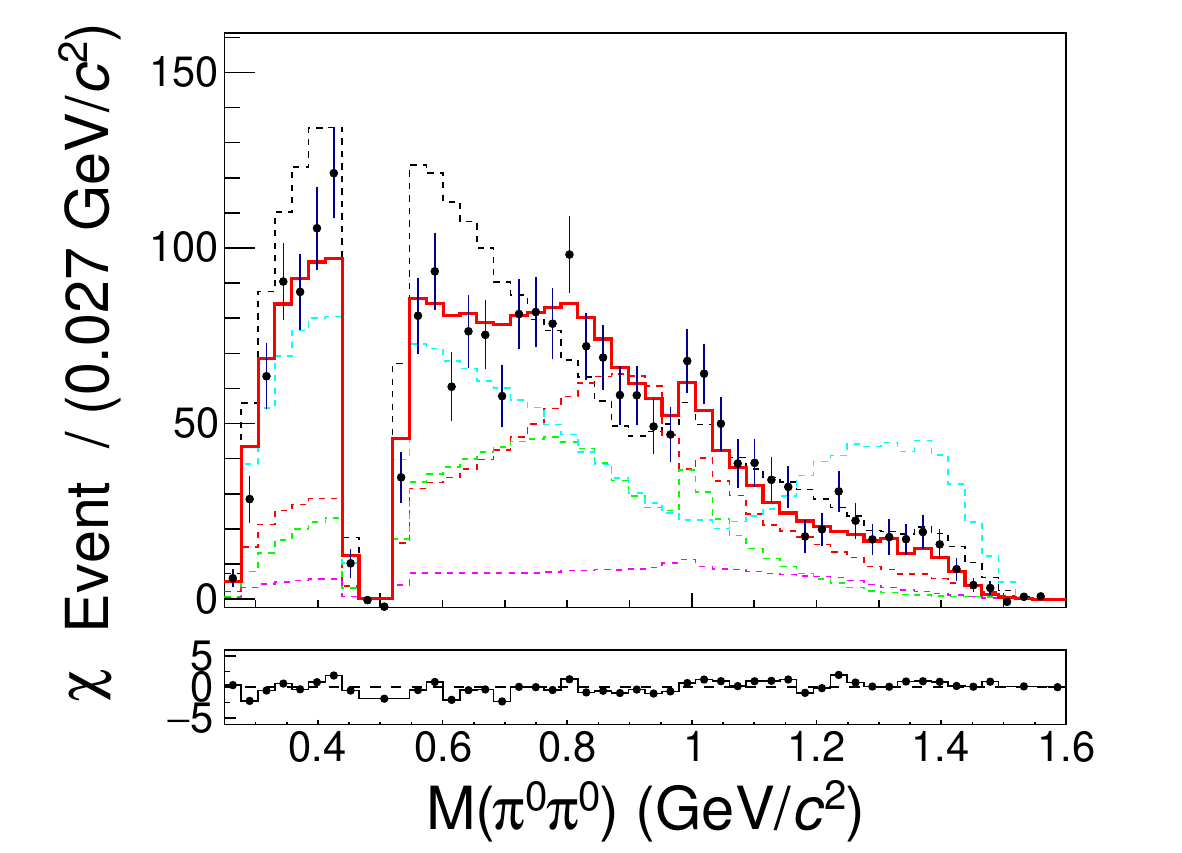}
  \end{overpic}
 \begin{overpic}[width=0.225\textwidth]{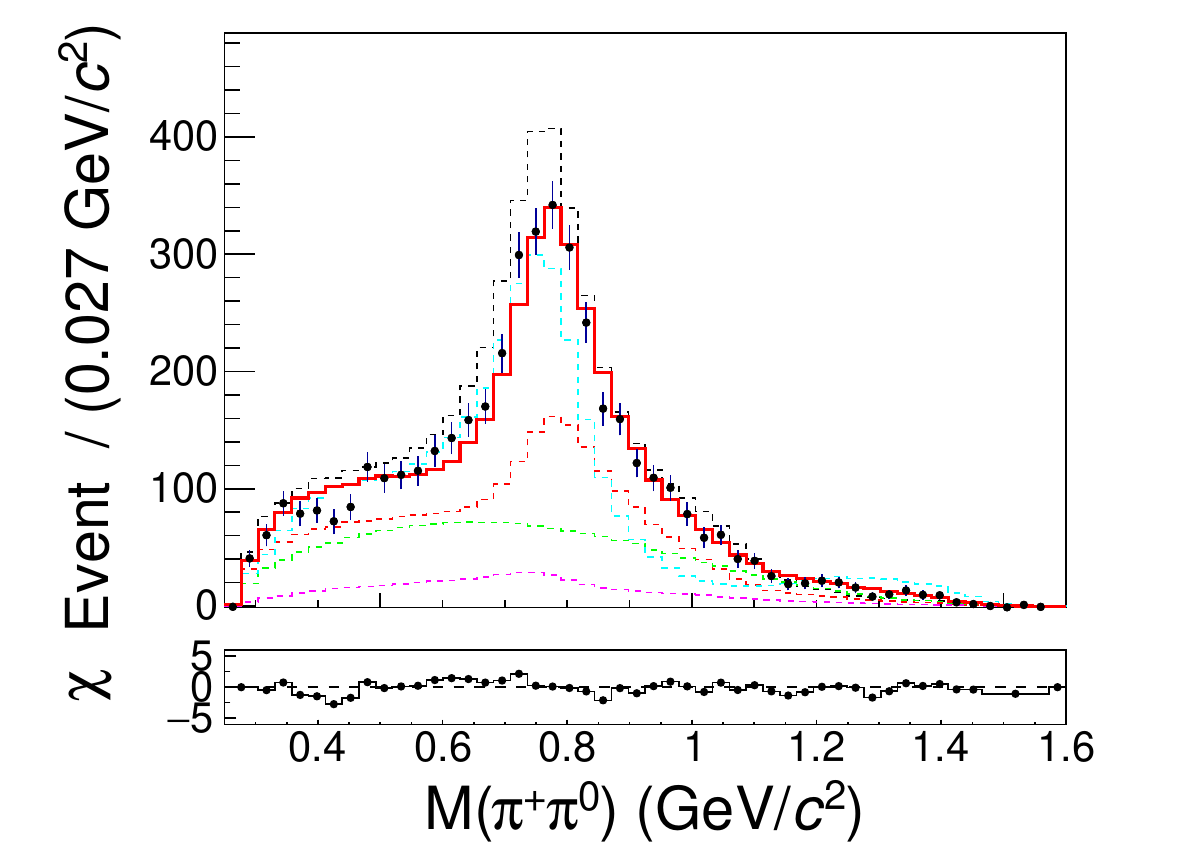}
  \end{overpic}
 \begin{overpic}[width=0.225\textwidth]{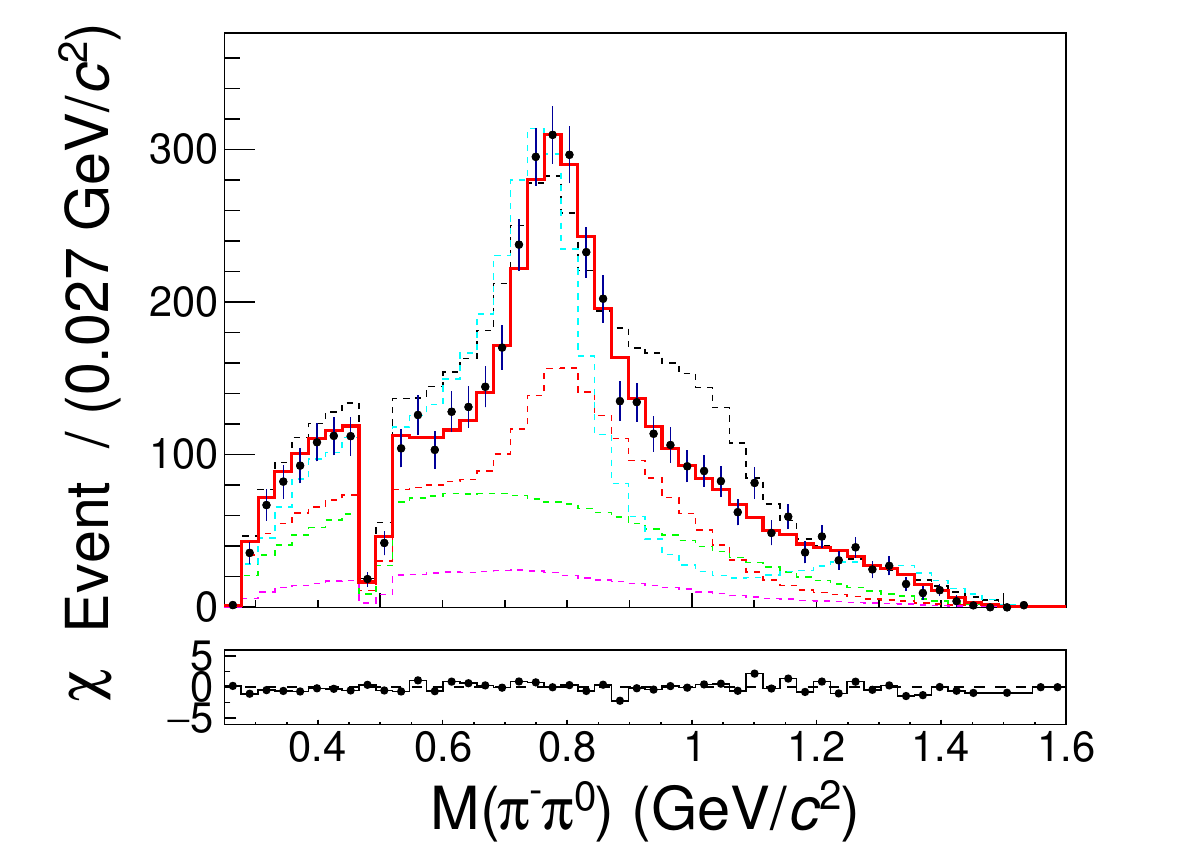}
  \end{overpic}
\begin{overpic}[width=0.225\textwidth]{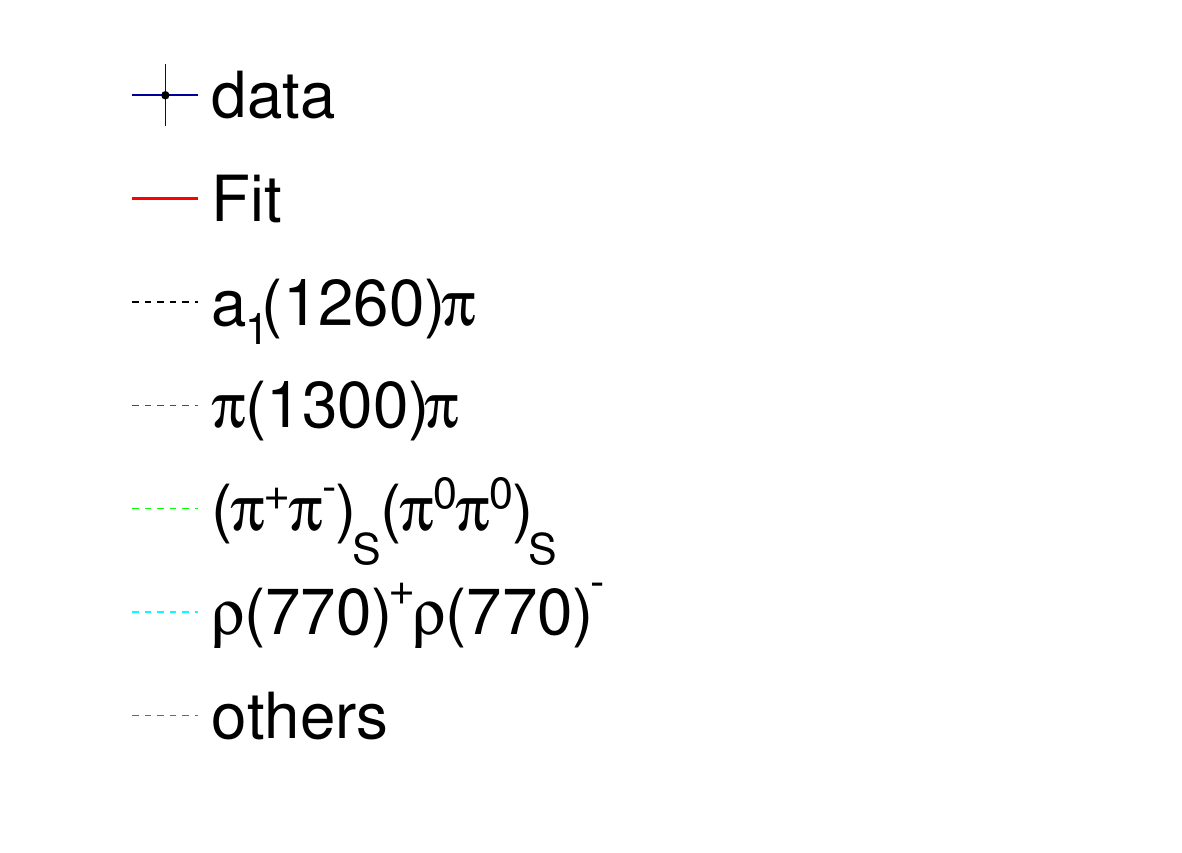}
\end{overpic}
  \caption{
The projections of the amplitude analysis fit of  $D^0\to \pi^+\pi^-2\pi^0$ on
    two-body and three-body particle mass distributions~\cite{BESIII:2023exz}.
  }
\label{fig:D0_2pi2pi0}
\end{figure*}

The decay $D^{0}\to4\pi$ is a sensitive mode~\cite{Harnew:2017tlp, LHCb:2019yan} to extract the CKM angle $\gamma$ via $B^{-}\to D K^{-}$.
Their $CP$-even fractions and relative strong phase parameters in the different phase space bins ($c_i$/$s_i$) serve as the direct inputs in the GLW~\cite{Gronau:1991dp} and BPGGSZ~\cite{Giri:2003ty,Belle:2004bbr} methods, respectively.
A reliable decay amplitude model of $D^{0}\to4\pi$ is critical for precisely extracting the $CP$-even fractions. It is also essential for obtaining the model-independent $c_i$/$s_i$ parameters~\cite{Harnew:2017tlp} and for searching for $CP$ violation in $D^0\to 4\pi$~\cite{LHCb:2016qbq}.
Reference~\cite{BESIII:2023exz} reported a joint amplitude analysis on the two hadronic decays $D^0\to 2\pi^+2\pi^-$ and $D^0\to\pi^+\pi^-2\pi^0$(non-$\eta$).
This analysis was based on about 5.8k candidates for $D^0\to 2\pi^+2\pi^-$ and 2.2k candidates for $D^0\to \pi^+\pi^-2\pi^0$.
The fit projections on two-body and three-body particle mass distributions for $D^0\to 2\pi^+2\pi^-$ and $D^0\to\pi^+\pi^-2\pi^0$(non-$\eta$)
are shown in Figs.~\ref{fig:D0_4pi} and \ref{fig:D0_2pi2pi0}, respectively.
The fit fractions of individual components are obtained, and large interferences among the dominant components of the decays $D^{0}\to a_{1}(1260)\pi$, $D^{0}\to\pi(1300)\pi$, $D^{0}\to\rho(770)\rho(770)$ and $D^{0}\to2(\pi\pi)_{S}$ are found in both channels. With the obtained amplitude model, the $C\!P$-even fractions of $D^0\to 2\pi^+2\pi^-$ and $D^0\to\pi^+\pi^-2\pi^0$(non-$\eta$) are determined to be $(75.2\pm1.1\pm1.5)\%$ and $(68.9\pm1.5\pm2.4)\%$, respectively. The branching fractions of $D^0\to 2\pi^+2\pi^-$ and $D^0\to\pi^+\pi^-2\pi^0$(non-$\eta$) are measured to be $(0.688\pm0.010\pm0.010)\%$ and $(0.951\pm0.025\pm0.021)\%$, respectively. The amplitude analysis provides an important model for binning strategy in the measurements of the strong phase parameters of $D^0 \to 4\pi$ when used to determine the CKM angle $\gamma (\phi_{3})$ via the $B^{-}\to D K^{-}$ decay.

\subsubsection{Analyses of $D^+_{(s)}\to \pi\pi\pi\eta$}

Although the light scalar mesons $f_{0}(500)$, $f_{0}(980)$, and $a_{0}(980)$ are experimentally well established, their classification as conventional $q\bar{q}$ states remains highly debated~\cite{ParticleDataGroup:2024cfk}. Persistent puzzles include the heavy $a_0(980)$ mass, the OZI-suppressed $\phi \to \gamma a_{0}(980)^0$ decay~\cite{Braghin:2022uih}, and anomalously large branching fractions in decays involving scalars~\cite{BESIII:2018sjg,BESIII:2019jjr,BESIII:2021aza,BESIII:2024tpv}. These motivate alternative interpretations such as
compact tetraquark states~\cite{Hsiao:2023qtk, Hsiao:2019ait,Yu:2021euw,Jaffe:1976ig,Brito:2004tv,Klempt:2007cp,Alexandrou:2017itd,Humanic:2022hpq}, two-meson molecule bound states~\cite{Weinstein:1982gc,Dai:2014lza,Sekihara:2014qxa,Duan:2020vye, Ikeno:2021kzf}, and mixed states~\cite{Braghin:2022uih}.
The production of these exotic structures typically relates to FSI, whose mechanisms are predominantly governed by non-perturbative QCD.
Charmed meson decays, receiving significant FSI contributions, offer a unique probe of scalar meson structure.
Studies of $D_{(s)}\to \pi\pi\pi\eta$ decays, with possible contributions of $D_{(s)}\to SS$, $D_{(s)}\to SP$, $D_{(s)}\to SV$, and $D_{(s)}\to VP$,
offer crucial insight into scalar meson structures.

\begin{figure}[htbp]
	\flushleft
	\includegraphics[width=0.45\textwidth]{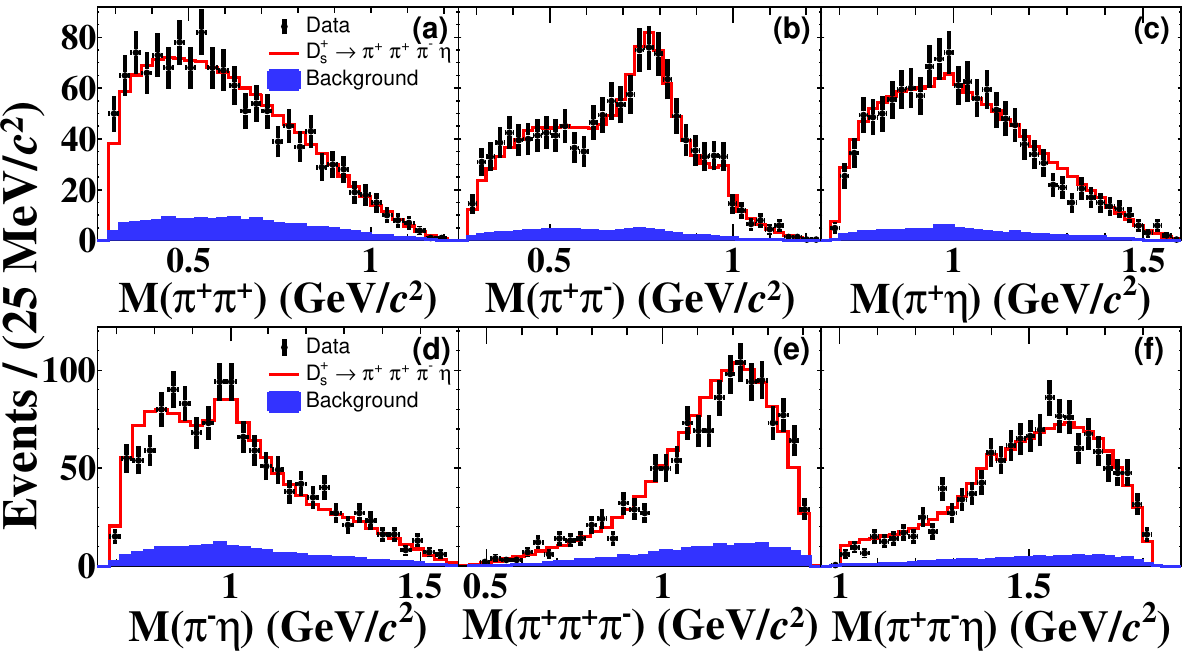}
	\caption{
The projections of the amplitude analysis fit of $D^{+}_{s}\to 2\pi^+\pi^-\eta$ on
    two-body and three-body particle mass distributions~\cite{BESIII:2021aza}.
}
		  	\label{fig:Ds_3pieta}
\end{figure}

Topological analyses highlight the significance of weak annihilation~\cite{Chau:1987tk,Rosner:1999xd}, with amplitudes comparable to tree-level contributions~\cite{Cheng:2010ry,Rosner:1999xd}.
The branching fraction of the pure WA $PP$ final state $D^+_s \to \pi^0\pi^+$ is $<0.037\%$~\cite{CLEO:2009fiz}, and for the $VP$ final state $D^+_s \to \rho(770)^0\pi^+$ is 0.019\%~\cite{BaBar:2008nlp,ParticleDataGroup:2024cfk}, whereas the $SP$ mode $D^+_s \to a_0(980)^{+(0)}\pi^{0(+)}$ reaches $\sim$1.46\%~\cite{BESIII:2019jjr}. This enhancement is puzzling since direct $a_0(980)\pi$ production violates G-parity~\cite{Achasov:2017edm}, and its origin remains debated: either via dynamically generated $a_0(980)$ from $K\bar K$ interactions~\cite{Molina:2019udw} or via $\rho\eta$ triangle rescattering~\cite{Hsiao:2019ait}. The WA process with a $VS$ final state, $D^+_s \to a_0(980)^+\rho(770)^0$, conserves G-parity and is experimentally unexplored.
Its measurement would critically test weak annihilation mechanisms and help clarify the nature of $a_0(980)^+$~\cite{ParticleDataGroup:2024cfk}.
The first amplitude analysis and branching fraction measurement of $D^{+}_{s} \to 2\pi^{+}\pi^{-}\eta$ were presented in Ref.~\cite{BESIII:2021aza}.
The amplitude analysis fit projections on two-body or three-body particle mass distributions are shown in Fig.~\ref{fig:Ds_3pieta}.
The amplitude analysis finds the amplitudes of
  $a_1(1260)^+(\to\rho(770)^0\pi^+)\eta$,
    $a_1(1260)^+(\to f_0(500)\pi^+)\eta$,
    $a_0(980)^+\rho(770)^0$,
    $\eta(1405)(\to a_0(980)^-\pi^+)\pi^+$,
    $\eta(1405)(\to a_0(980)^+\pi^-)\pi^+$,
    $f_1(1420)(\to a_0(980)^-\pi^+)\pi^+$,
    $f_1(1420)(\to a_0(980)^+\pi^-)\pi^+$,
    $[a_0(980)^-\pi^+]_S\pi^+$,
    $[a_0(980)^+\pi^-]_S\pi^+$,
    $[f_0(980)\eta]_S\pi^+$, and
    $[f_0(500)\eta]_S\pi^+$,
    with fractions of
$(55.4\pm3.9\pm2.0)\%$
$(8.1\pm1.9\pm2.1)\%$
$(6.7\pm2.5\pm1.5)\%$
$(0.7\pm0.2\pm0.1)\%$
$(0.7\pm0.2\pm0.1)\%$
$(1.9\pm0.5\pm0.3)\%$
$(1.7\pm0.5\pm0.3)\%$
$(5.1\pm1.2\pm0.9)\%$
$(3.4\pm0.8\pm0.6)\%$
$(6.2\pm1.7\pm0.9)\%$ and
$(12.7\pm2.6\pm2.0)\%$, respectively.
The branching fraction of this decay is measured to be $\mathcal {B}(D^{+}_{s} \to
2\pi^{+}\pi^{-}\eta)$ = ($3.12\pm0.13\pm0.09$)\%, and the branching fractions for the intermediate
processes are also presented.
The WA decay $D^{+}_{s} \to a_0(980)^+\rho(770)^0, ~a_0(980)^+ \to \pi^+ \eta$ is observed with a branching fraction
of ($0.21\pm0.08\pm0.05)$\%.
This branching fraction and that of $D^{+}_{s} \to a_0(980)^+\pi^0$ obtained in Ref.~\cite{BESIII:2019jjr} are both larger than those of the
pure WA processes $D^{+}_{s} \to \rho(770)^0\pi^+$ and $D^{+}_{s} \to \pi^0\pi^+$ by one order of magnitude.
These measurements indicate that long-distance weak annihilation may play an essential role, and provide
a good opportunity to study the final-state rescattering in the WA process~\cite{Cheng:2010ry, Molina:2019udw, Hsiao:2019ait}.
 
\begin{figure}
  \begin{center}
    \includegraphics[width=0.45\textwidth]{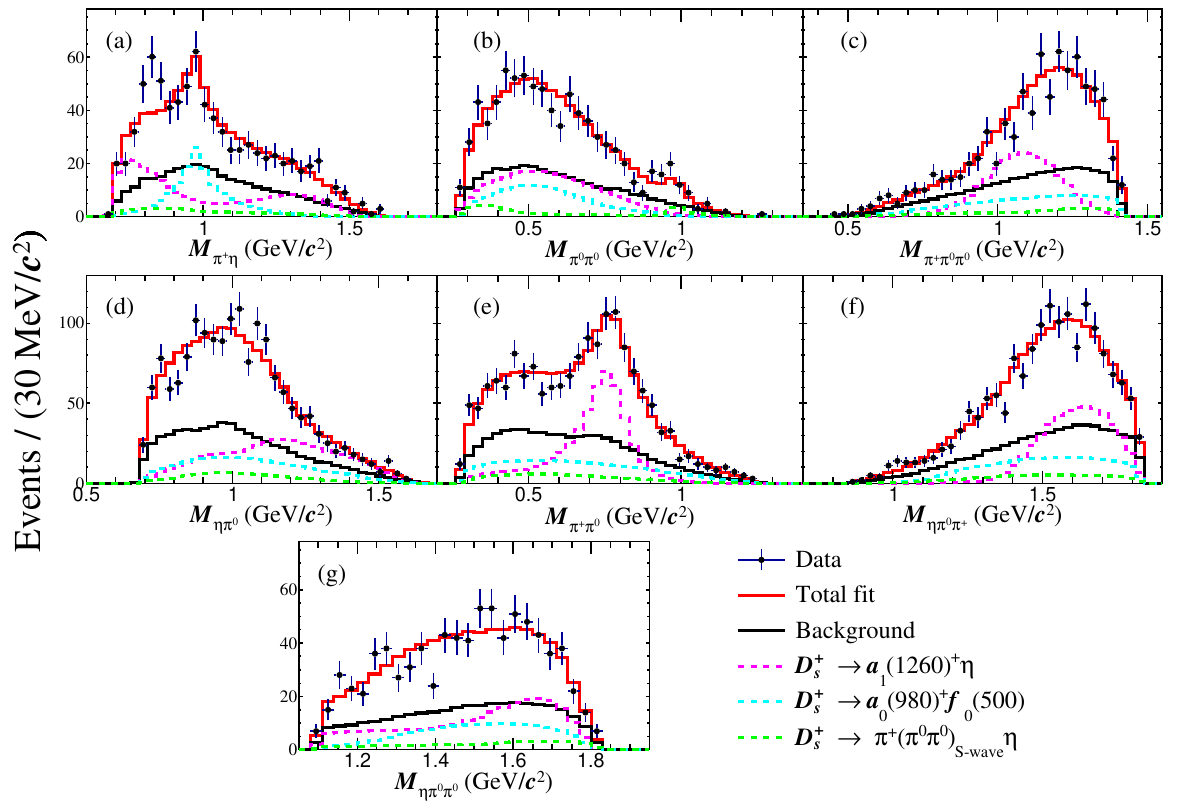}
    \caption{
The projections of the amplitude analysis fit of $D^{+}_{s}\to \pi^+2\pi^0\eta$ on
    two-body and three-body particle mass distributions~\cite{BESIII:2026mtz}.
    }
    \label{fig:Ds_pipi0pi0eta}
  \end{center}
\end{figure}

The $D_s^+ \to \pi^+2\pi^0\eta$ decay, with possible contributions of $D^+_s\to SS$ ($D_s^+ \to a_0(980)^+ f_0(500)$),
$D^+_s\to SP$ ($D_{(s)}^+ \to a_1(1260)\eta$) and $D^+_s\to VS$ ($D_s^+\to a_0(980)^{+}\rho(770)^{0}(a_0(980)^{0}\rho(770)^{+})$),
also offers a valuable opportunity to probe the internal structure of scalar mesons:
$f_0(500)$, $a_0(980)/f_0(980)$, and  $a_1(1260)$. The obtained results are important
to understand the nature of these scalar particles~\cite{Cheng:2002mk, Cheng:2002ai, Boito:2008zk, Cheng:2022vbw, Dai:2023jix}.
Previously, its isospin-symmetric partner channel $D_s^+ \to a_1(1260)^+\eta, a_1(1260)^+ \to \rho(770)^0\pi^+$,
had already been experimentally observed in $D_s^+\to 2\pi^+\pi^-\eta$~\cite{BESIII:2021aza}.
The $D_s^+\to a_0(980)^{+}\rho(770)^{0}(a_0(980)^{0}\rho(770)^{+})$ decay is predicted by Ref.~\cite{Yu:2021euw},
and could be accessed via $D_s^+ \to \pi^+2\pi^0\eta$.
Recently, Ref.~\cite{BESIII:2026mtz} reported the first observation and amplitude analysis of $D_s^+ \to \pi^+2\pi^0\eta$.
The  projections of the amplitude analysis fit on two-body or three-body particle mass distributions are given in Fig.~\ref{fig:Ds_pipi0pi0eta}.
This decay is found to be dominated by
$D_s^{+}\to a_1(1260)^+\eta, a_1(1260)^+\to \rho(770)^+\pi^0$,
$D_s^{+}\to a_0(980)^+f_0(500)$, and
$D_s^{+}\to \pi^+(\pi^0\pi^0)_{\mathcal{S}-\rm{wave}}\eta$ with fit fractions of
$(59.7\pm5.2\pm2.9)\%$, $(33.1\pm4.9\pm6.8)\%$, and $(9.4\pm1.9\pm4.1)\%$, respectively.
The  $D_s^+ \to a_0(980)^+f_0(500)$ decay is observed for the first time, and
a structure around 0.8 GeV/$c^2$ is observed in the $M_{\pi^+\eta}$ spectrum as shown in Fig.~\ref{fig:Ds_pipi0pi0eta}(a). This structure cannot be described by known resonances, which is similar to structures visible in $D_s^+ \to 2\pi^+\pi^-\eta$~\cite{BESIII:2021aza}.
We obtain the branching fractions
$\mathcal{B}(D_{s}^{+} \to\pi^{+}2\pi^{0}\eta|_{\rm non-\eta^\prime})=(2.97 \pm 0.23\pm 0.14)\%$
and $\mathcal{B}(D_{s}^{+} \to\pi^{+}\eta^\prime)=(3.66 \pm 0.36\pm 0.09\pm 0.08)\%$,
with a branching fraction of $(0.98 \pm 0.16\pm 0.22)\%$ with a significance exceeding 10$\sigma$.
The obtained branching fraction is unexpectedly large (of order $10^{-2}$), which may indicate a substantial tetraquark
component in the $a_0(980)$ or significant contributions from FSIs of triangle rescattering~\cite{Wang:2025mdn}.
The branching fraction of $D_s^+\to a_1(1260)^+\eta, a_1(1260)^+\to\rho(770)^+\pi^0$ is
determined to be $(1.77\pm0.21\pm0.12)\%$. Compared with
$\mathcal{B}(D_s^+\to a_1(1260)^+\eta, a_1(1260)^+\to\rho(770)^0\pi^+) = (1.73\pm0.14\pm0.08)\%$ measured by BESIII~\cite{BESIII:2021aza}, the isospin symmetry is
validated for the decays of $a_1(1260)^+\to \rho(770)^0\pi^+$ and $a_1(1260)^+\to \rho(770)^+\pi^0$.

\begin{figure*}[htbp]
  \centering
  \includegraphics[width=0.48\textwidth]{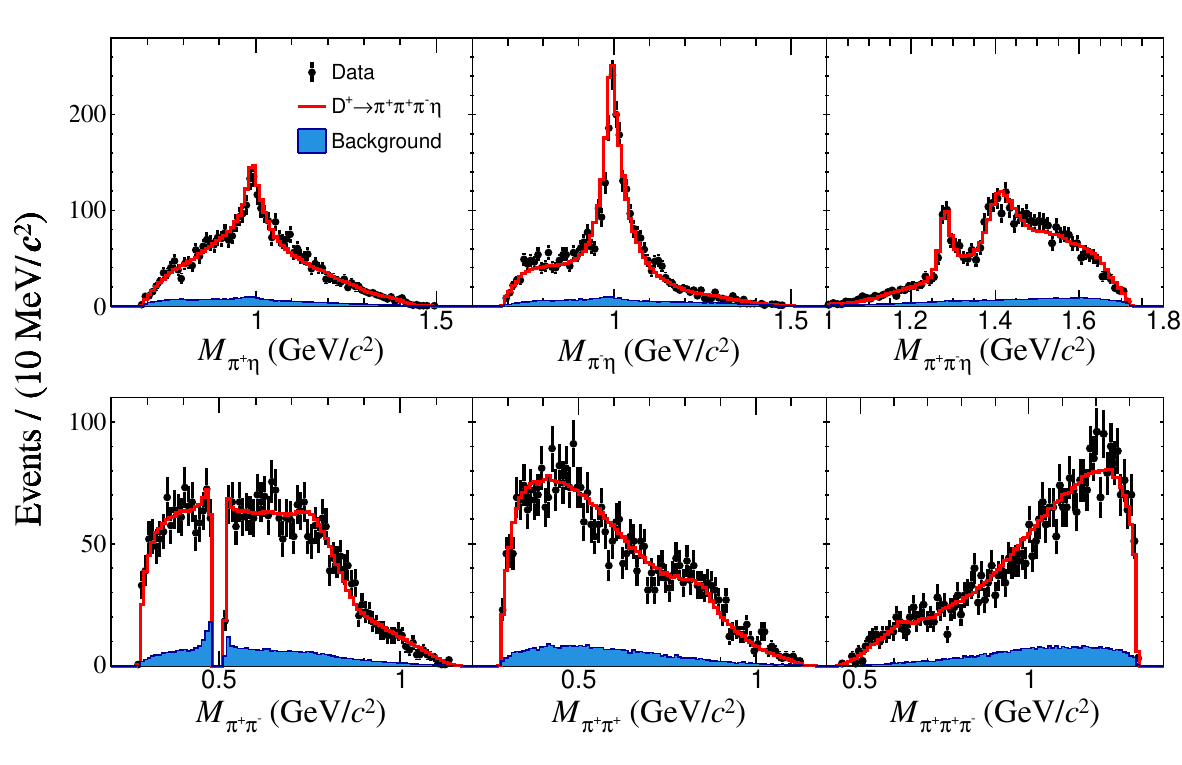}
  \includegraphics[width=0.48\textwidth]{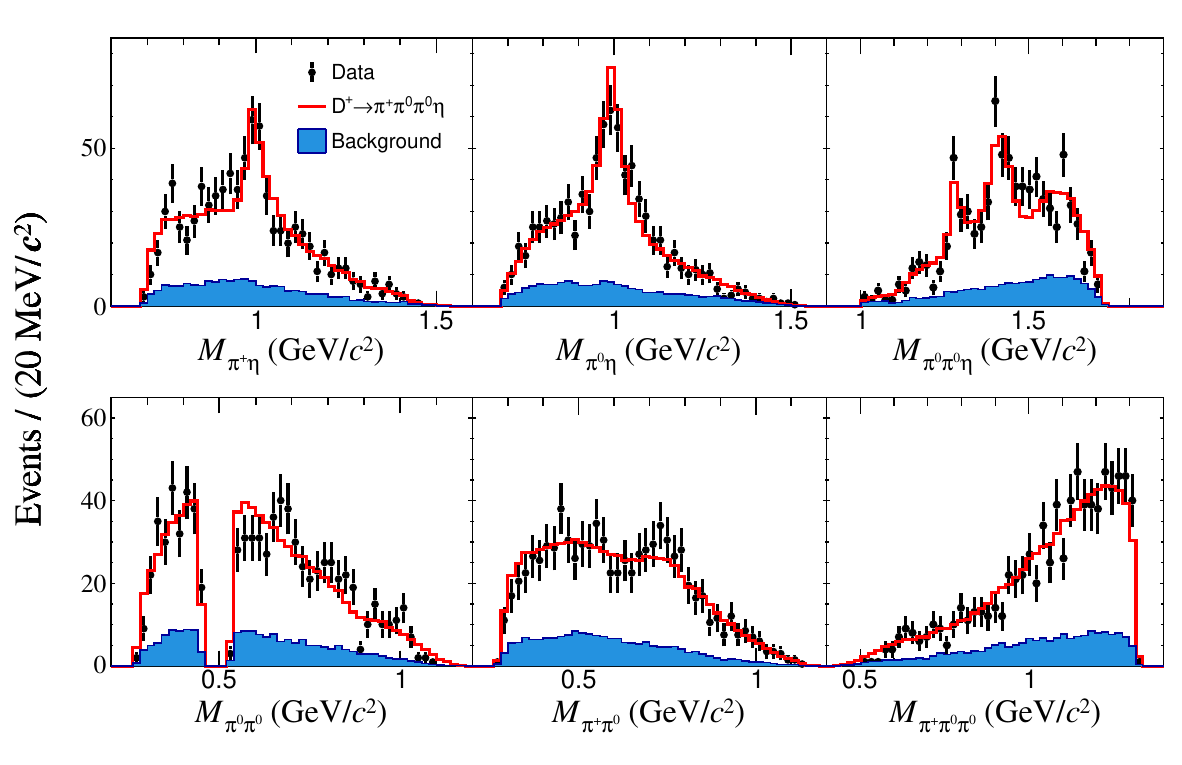}
  \caption{
The projections of the amplitude analysis fit of (left) $D^+\to 2\pi^+\pi^-\eta$ and (right) $D^+\to \pi^+2\pi^0\eta$ on
    two-body and three-body particle mass distributions~\cite{BESIII:2026mbo}.
  }
  \label{fig:Dp_3pietas}
\end{figure*}

Previously,
BESIII reported unexpectedly large branching fractions of $D_s^+\to a_0(980)^{0(+)}\pi^{+(0)}$~\cite{BESIII:2019jjr}, $D_s^+\to a_0(980)^{0(+)}\rho(770)^{+(0)}$~\cite{BESIII:2021aza}, and $D^{0(+)}\to a_0(980)^{+}\pi^{-(0)}$~\cite{BESIII:2024tpv}. Notably, $\mathcal{B}(D^{+}\to a_{0}(980)^{+}\pi^{0})/\mathcal{B}(D^{+}\to a_{0}(980)^{0}\pi^{+})=2.6\pm0.6\pm0.3$~\cite{BESIII:2024tpv}, while
the $q\bar{q}$ model predicts a value well below unity for $\mathcal{B}(D^{+}\to a_{0}(980)^{+}\pi^{0})/\mathcal{B}(D^{+}\to a_{0}(980)^{0}\pi^{+})$, due to suppressed $W^+-a_0(980)^+$ coupling in external $W$-emission. This strongly indicates substantial FSI effects in $D\to SP$ decays~\cite{Cheng:2024zul}.
Similar enhancements are expected in $D^+\to SV$ modes.
In addition, the $D\to SS$ process $D^+ \to a_0(980)^+ f_0(500)$ is expected to be suppressed in the $q\bar{q}$ picture but could be significantly enhanced via tetraquark diagrams~\cite{Cheng:2024zul}. Observation of this process would indicate a potential tetraquark component in the scalar meson.
Recently, the first amplitude analysis of $D^+ \to 2\pi^+ \pi^{-} \eta$ and $D^+ \to \pi^+ 2\pi^{0} \eta$ was reported in Ref.~\cite{BESIII:2026mbo},
based on 3.1k and 0.65k candidates with signal purities of 89.1\% and 78.0\%, respectively.
The amplitude analysis fit projections on two-body or three-body particle mass distributions  are shown in Fig.~\ref{fig:Dp_3pietas}.
The branching fractions of the $D^+ \to 2\pi^+ \pi^- \eta$ and $D^+ \to \pi^+ 2\pi^0 \eta$ decays are measured to be $(3.20\pm0.06\pm0.03)\times 10^{-3}$ and $(2.43 \pm 0.11 \pm 0.04) \times 10^{-3}$, respectively. Both achieve three times better precision than the previous BESIII measurements~\cite{BESIII:2020pxp}.
The  $D^{+}\to a_0(980)^{+}f_0(500)$ decay is observed for the first time with an unexpectedly large branching fraction at a level of $10^{-3}$,
supporting the tetraquark interpretation of light scalar particles within the topological diagrams frameworks~\cite{Cheng:2024zul,Chau:1986jb,Cheng:2010ry}.
Moreover, we observe the decays $D^+ \to a_0(980)^{+(0)} \rho(770)^{0(+)}$ and measure the ratio $r_{+/0} \equiv \frac{\mathcal{B}(D^+ \to a_0(980)^+ \rho(770)^0)}{\mathcal{B}(D^+ \to a_0(980)^0 \rho(770)^+)}$ for the first time to be $0.55\pm0.08\pm0.05$. This result
together with the earlier BESIII result in $D \to a_0 \pi$ decays~\cite{BESIII:2024tpv}, implies substantial FSI effects, and provides a new perspective on the nature of the $a_0(980)$ states.
Notably, no significant $a_1(1260)\eta$ signal is observed, which contradicts the abundant $a_1(1260)$ typically seen in other $D\to P 3\pi$ decays~\cite{BESIII:2023qgj,BESIII:2023exz}, suggesting anomalous dynamics in this final state.

\subsection{Five-body decays}

\subsubsection{Analysis of $D^+_s\to K^+K^-2\pi^+\pi^-$}

\begin{figure*}[htp]
          \centering
          \includegraphics[width=0.19\textwidth]{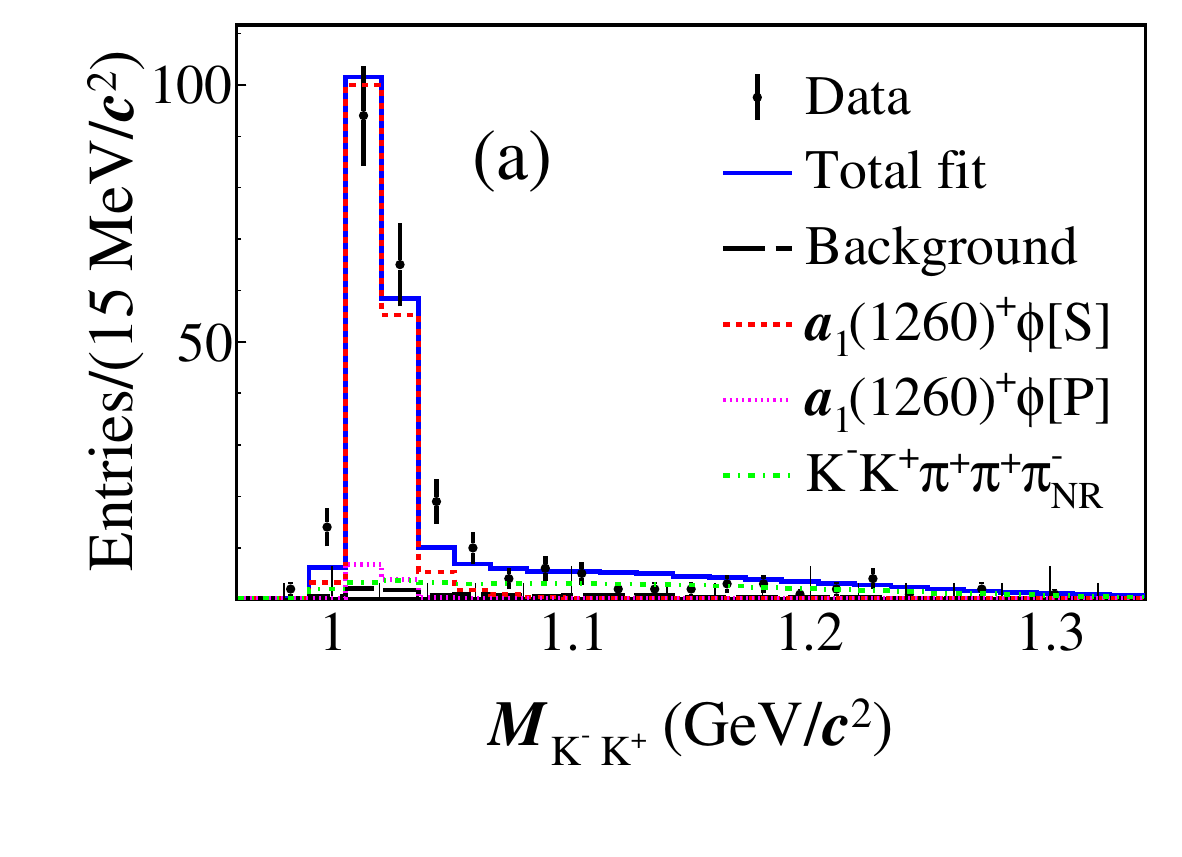}
          \includegraphics[width=0.19\textwidth]{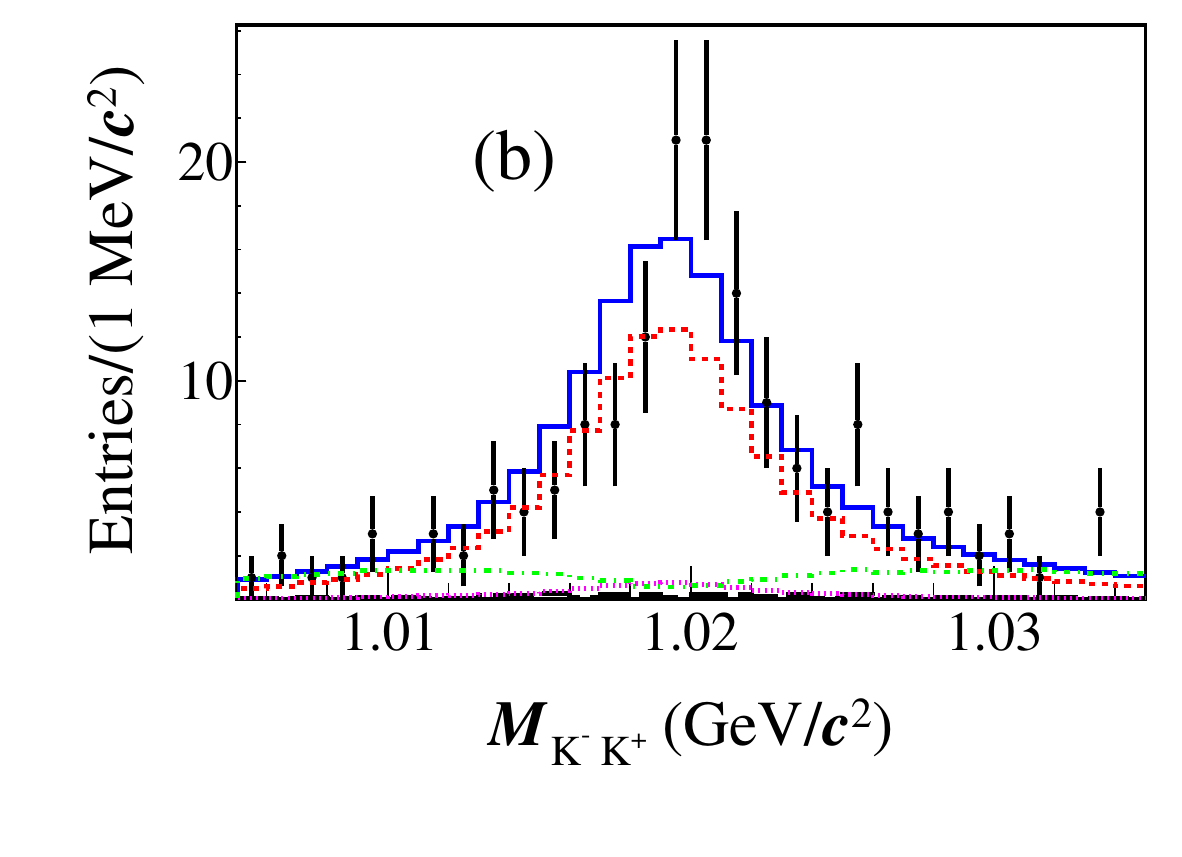}
          \includegraphics[width=0.19\textwidth]{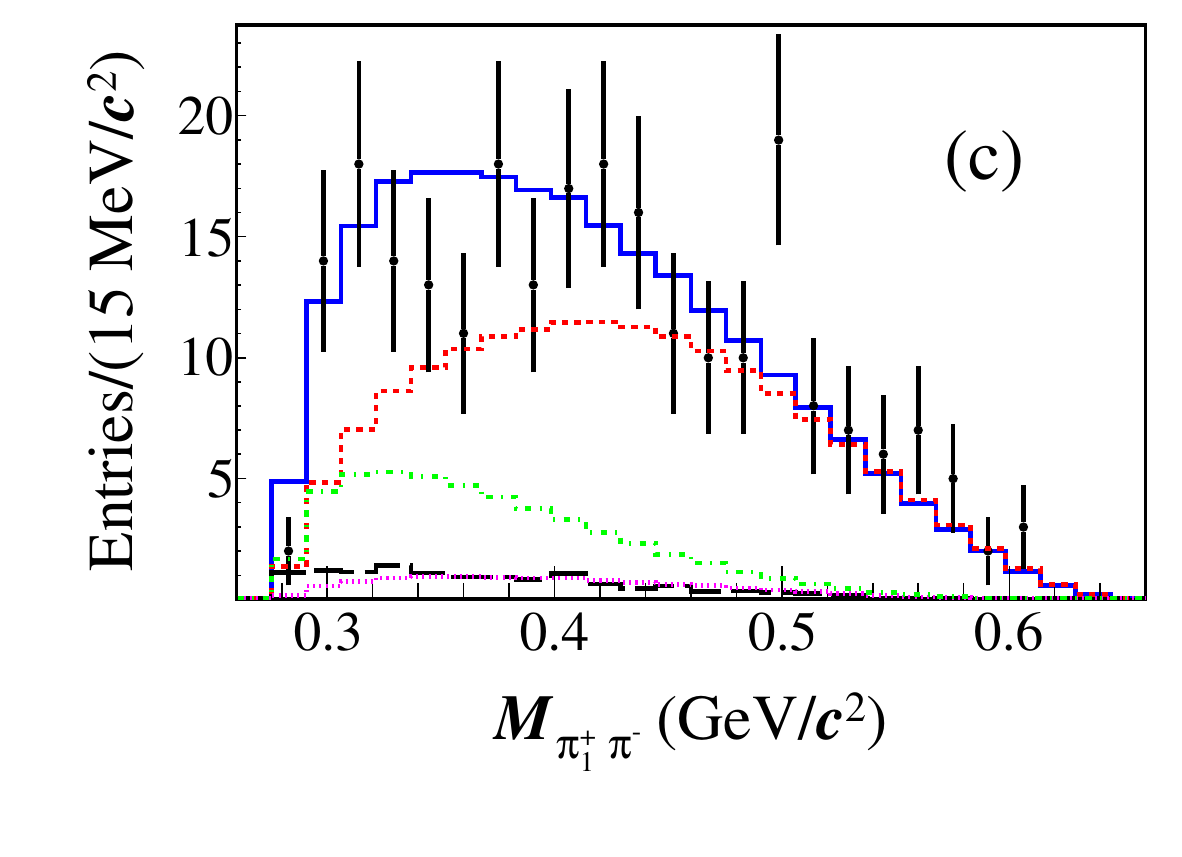}
          \includegraphics[width=0.19\textwidth]{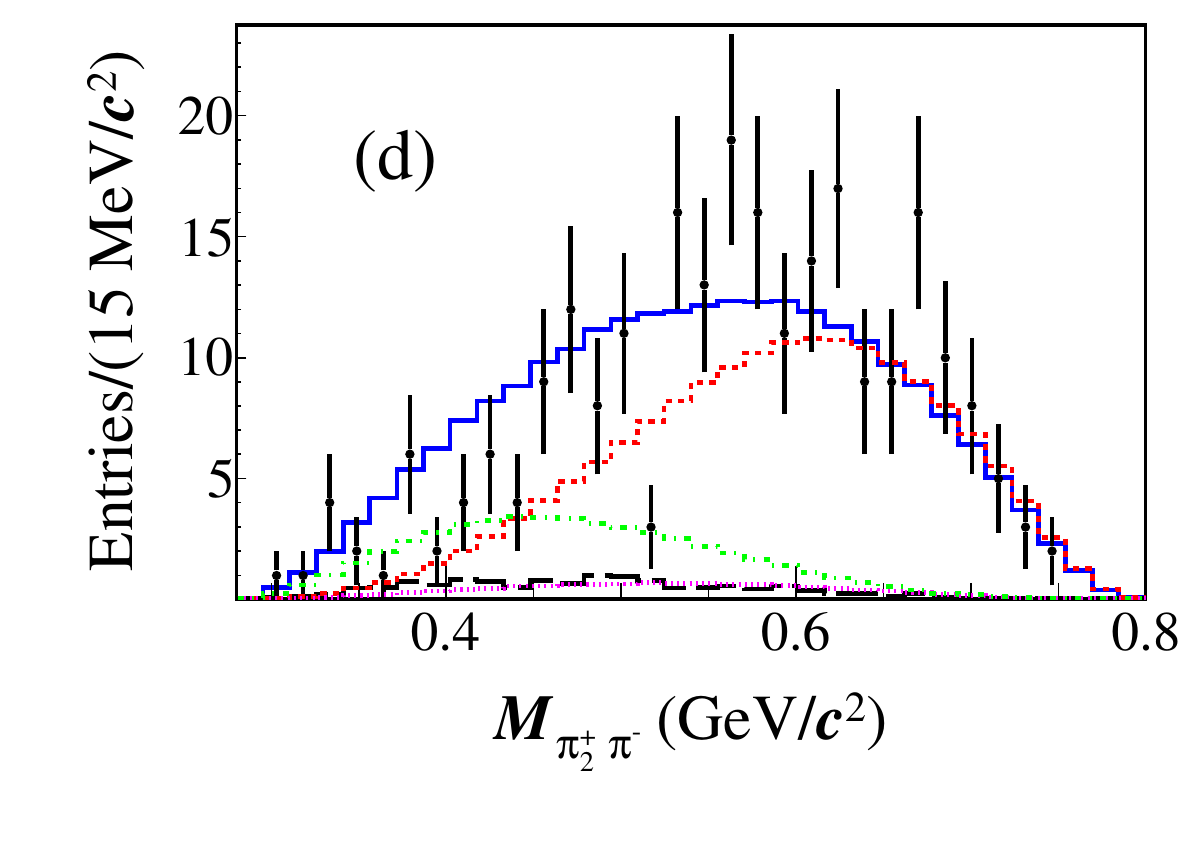}
          \includegraphics[width=0.19\textwidth]{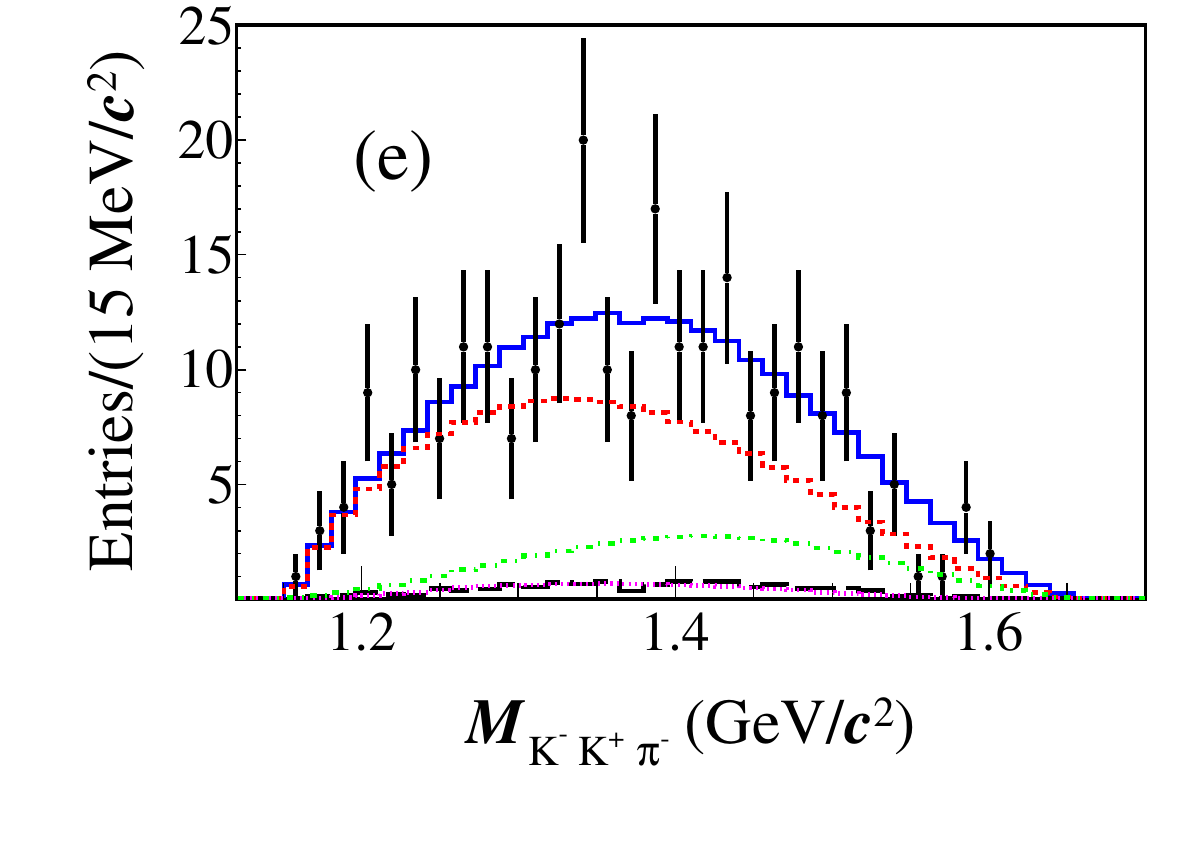}
          \includegraphics[width=0.19\textwidth]{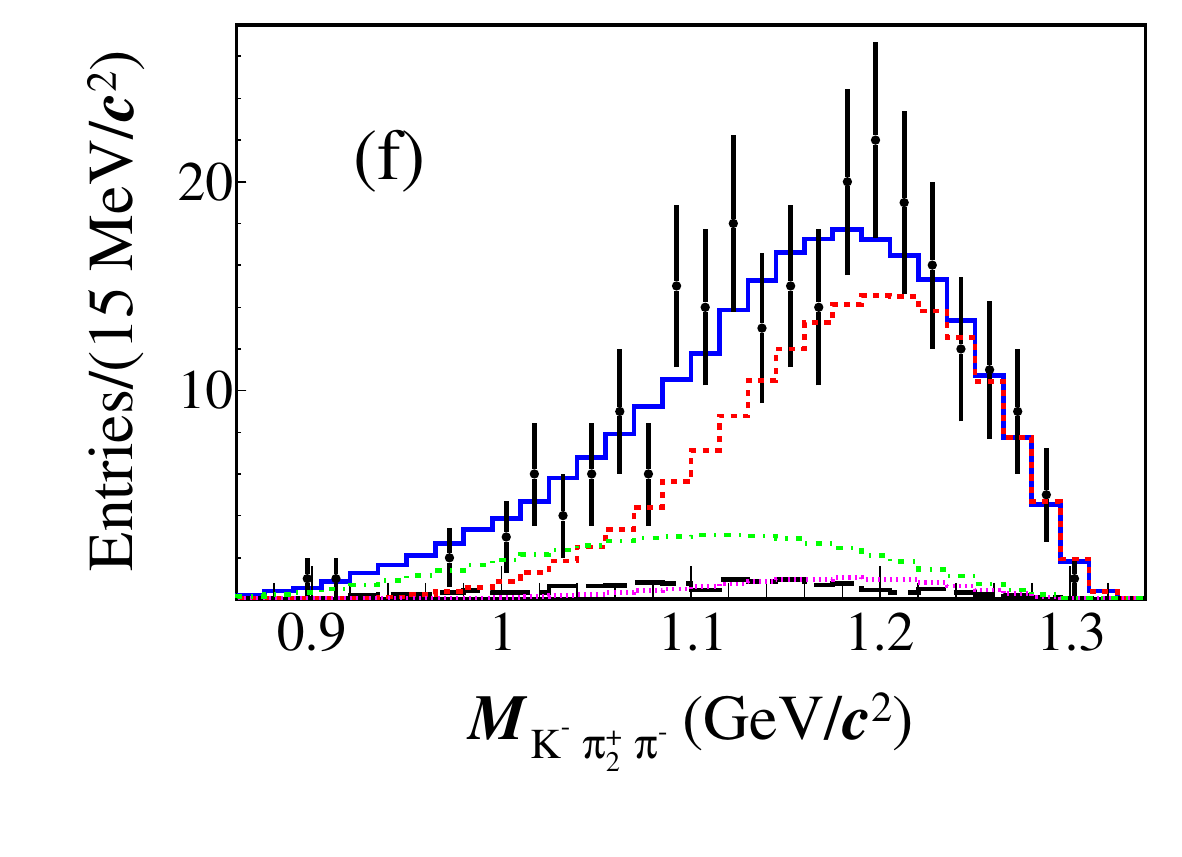}
          \includegraphics[width=0.19\textwidth]{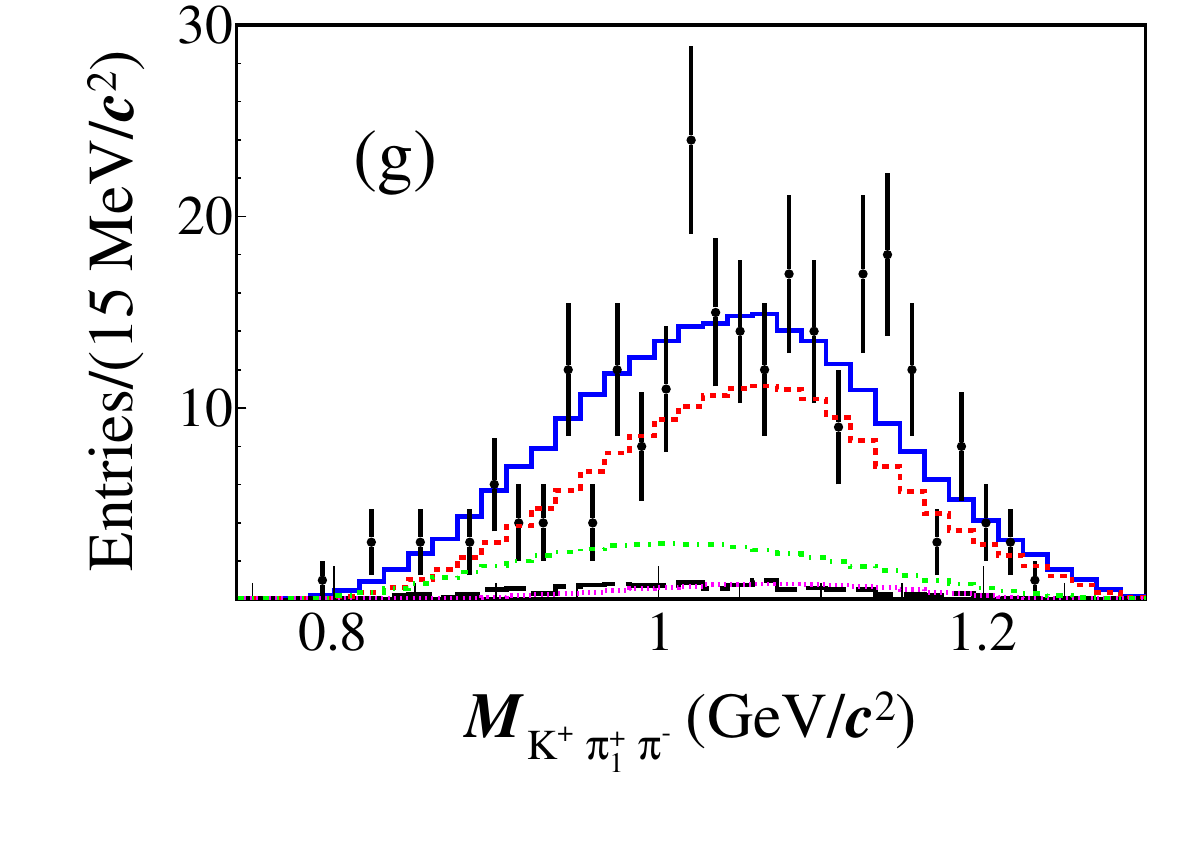}
          \includegraphics[width=0.19\textwidth]{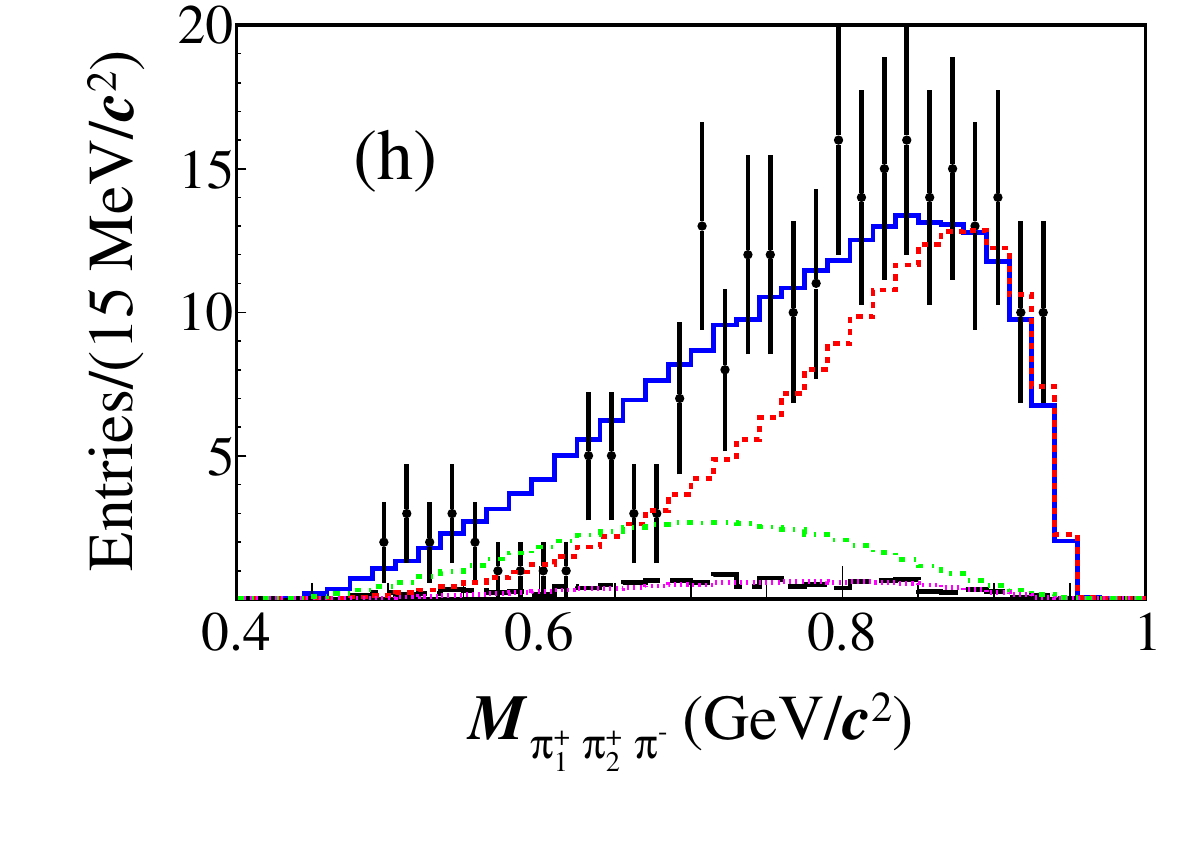}
          \includegraphics[width=0.19\textwidth]{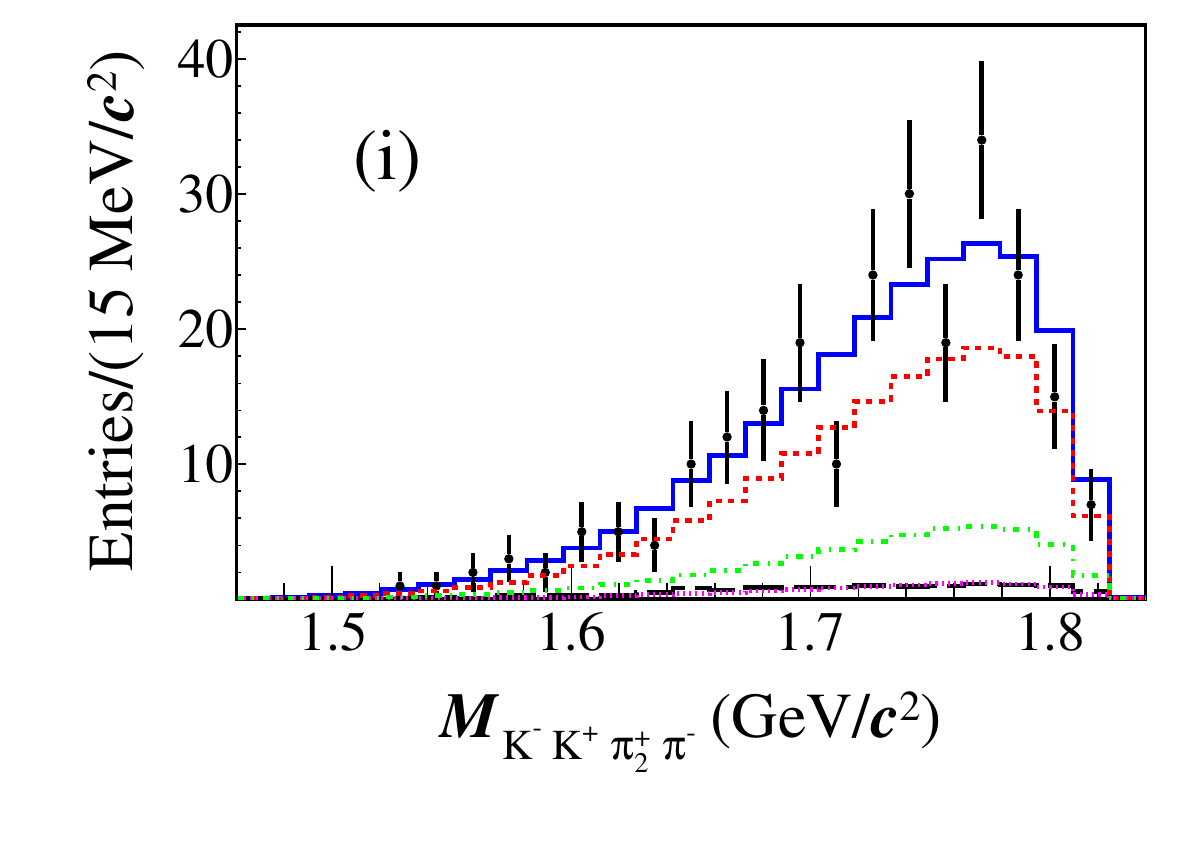}
          \includegraphics[width=0.19\textwidth]{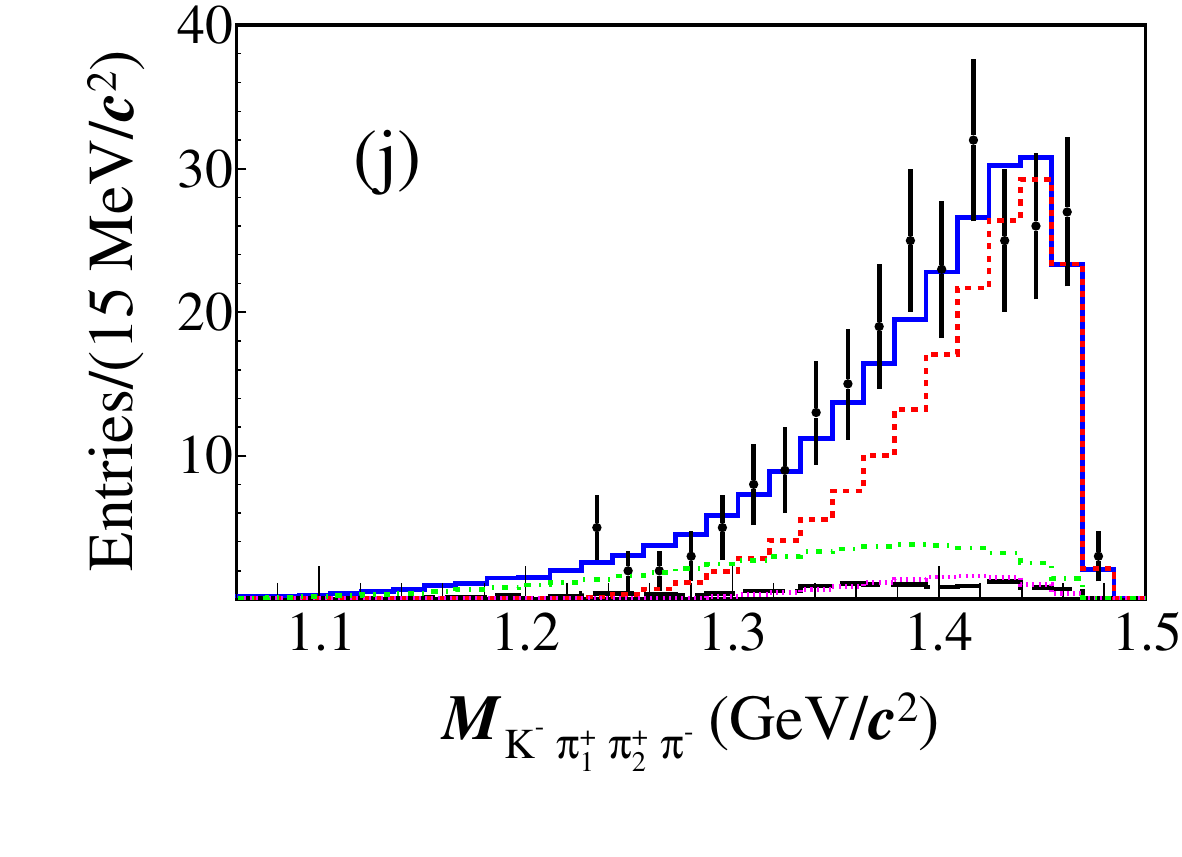}
          \caption{
The projections of the amplitude analysis fit of $D^+_s\to K^+K^-2\pi^+\pi^-$ on
    two-body, three-body, and four-body particle mass distributions~\cite{BESIII:2022jts}.
     }
    \label{fig:Ds_KK3pi}
\end{figure*}

Previously, no study relative to $D_s^+ \to AV$ decays was reported.
One $D_s^+ \to AV$ process $D^+_s \to a_1(1260)^+\phi$ is hopefully accessed via the $D_s^+ \to K^+K^-2\pi^+\pi^-$ decay.
Experimentally, only E687~\cite{E687:1997dng} and FOCUS~\cite{FOCUS:2002psb} reported the branching fractions of $D^+_s\to K^+K^-2\pi^+\pi^-$
relative to $D^+_s\to K^+K^-\pi^+$ with large uncertainties.
The amplitude analysis of $D^+_s\to K^+K^-2\pi^+\pi^-$ was performed by
analyzing double-tag candidates from 6.3 fb$^{-1}$ of data at 4.178-4.226 GeV.
Based on 243 candidates with a signal purity of 96.6\%,
the amplitude analysis of $D^+_s\to K^+K^-2\pi^+\pi^-$~\cite{BESIII:2022jts}
finds that it is dominated by $D^+_s\to a_1(1260)^+\phi$ [$(78.1\pm2.9\pm1.6)\%$]
with the nonresonant decay $D^+_s\to (K^-K^+2\pi^+\pi^-)_{\rm NR}$ [$(21.8\pm2.9\pm0.8)\%$].
This is the first information about $D_s^+ \to AV$ decays.
The amplitude analysis fit projections on two-body, three-body, and four-body particle mass distributions are shown in Fig.~\ref{fig:Ds_KK3pi}.
The branching fraction of $D^+_s\to K^-K^+2\pi^+\pi^-$ is measured to be $(6.60\pm0.47\pm0.35)\times 10^{-3}$,
and the branching fractions of the intermediate processes are also presented.

\subsubsection{Analysis of $D^+_s\to 2\pi^+\pi^-2\pi^0$}

The polarization information of heavy-flavor mesons decaying into $VV$ final states has long been studied for its unique sensitivity to new physics and hadron structure~\cite{Dunietz:1990cj,Valencia:1988it}. The discrepancy between experimental measurements and theoretical predictions for $B\to\phi K^{*}(892)$,
known as the ``polarization puzzle", motivated extensive $B\to VV$ investigations. Various models have successfully explained this phenomenon~\cite{BaBar:2003zor,Kagan:2004uw,Zou:2015iwa,Alvarez:2004ci,Yu:2024vlz}, whereas charm weak decays remain more controversial due to the charm quark mass lying between applicability of heavy quark symmetry and chiral perturbation theory~\cite{Cheng:2010ry}.

In the charm sector, transverse polarization is generally predicted to dominate $D_{(s)} \to VV$ decays in naive factorization~\cite{ElHassanElAaoud:1999min} and Lorentz-invariant symmetry models~\cite{Hiller:2013cza}. Although qualitatively supported by measurements like $D^0 \to \bar{K}^{*}(892)^0\rho(770)^0$~\cite{MARK-III:1991fvi}, quantitative discrepancies persist, e.g., the incomplete transverse polarization in $D^0 \to \omega\phi$~\cite{BESIII:2021raf,LHCb:2018mzv}. A long-distance FSI approach~\cite{Cao:2023csx} improves consistency for some modes, yet puzzles remain for longitudinally dominated decays such as $D^0 \to \rho(770)^0\rho(770)^0$~\cite{FOCUS:2007ern}.
Theorists usually discussed polarizations in terms of partial-wave amplitudes ($\mathcal{S}$, $\mathcal{P}$, $\mathcal{D}$ waves)
and concluded $\mathcal{S}$-wave dominance. However, experimental measurements reveal discrepancies with
$\mathcal{D}$-wave dominant in $D^0 \to K^{*}(892)^-\rho(770)^+$, $\bar{K}^{*}(892)^0\rho(770)^0$, $\rho(770)^+\rho(770)^-$, $\rho(770)^0\rho(770)^0$ and $\mathcal{P}$-wave dominant in $D_s^+ \to K^{*}(892)^0\rho(770)^+, K^{*}(892)^+\rho(770)^0$~\cite{BESIII:2019lwn, MARK-III:1991fvi, FOCUS:2007ern,BESIII:2022bvv}.
The pure WA decay, $D_s^+ \to \omega\rho(770)^+$,
provides an ideal counterpart to the $W$-emission decay $D_s^+ \to \phi\rho(770)^+$ for investigating the polarization puzzle~\cite{Song:2025tog}.

In addition, WA calculations suffer large uncertainties from non-factorizable effects, making experimental WA amplitude determinations crucial for approaches like the diagrammatic method~\cite{Cheng:2024hdo,Cheng:2022vbw}.
Small branching fractions of $D_s^+ \to \rho(770)^0\pi^+$ and $D_s^+ \to \omega\pi^+$ suggest a suppressed WA diagram for $D \to VP$ decays, whereas large branching fractions of $D_s^+ \to a_0(980)^+\pi^0$ and
$a_0(980)^+\rho(770)^0$~\cite{BESIII:2019jjr, Hsiao:2019ait, Yu:2021euw} imply significant WA contributions for $D \to SP/SV$ decays.
The first measurement of $D_s^+ \to \omega\rho(770)^+$ benefits the understanding of the WA contribution in $D_s^+ \to VV$ decays.

Furthermore, recent measurements of $\mathcal{B}(D_s^+ \to \phi\pi^+)$ via $\phi\to K^+K^-$~\cite{BESIII:2020ctr} and $\phi\to\pi^+\pi^-\pi^0$~\cite{BESIII:2024muy} show tension with the existing measurements of $\phi$ decay branching fractions. A precise measurement of $D_s^+ \to \phi(\to\pi^+\pi^-\pi^0)\rho(770)^+$ would thus serve as an independent check on $\phi$ decay branching fractions.

\begin{figure}[htp]
  \begin{center}
    \includegraphics[width=0.48\textwidth]{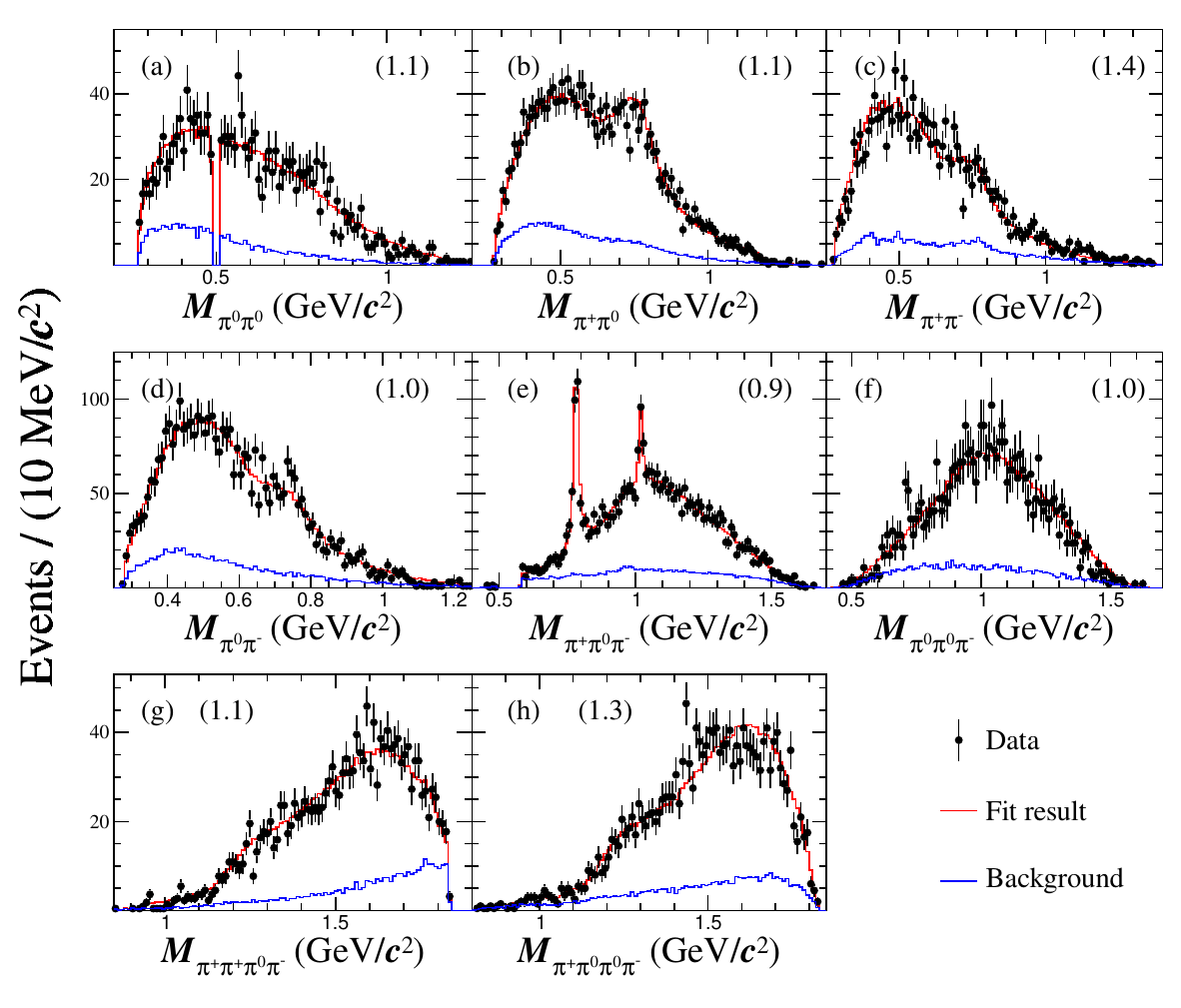}
    \caption{
The projections of the amplitude analysis fit of $D^+_s\to 2\pi^+\pi^-2\pi^0$ on the two-body, three-body, and four-body particle mass distributions~\cite{BESIII:2022jts}.
}
    \label{fig:Ds_5pi}
  \end{center}
\end{figure}

Previously, only CLEO-c reported $\mathcal{B}(D_s^+ \to \omega\pi^+\pi^0) = (2.78\pm0.65\pm0.25)\%$ with a relative $D_s^+ \to \omega\rho(770)^+$ fraction of $0.52\pm0.30$~\cite{CLEO:2009nsf}.
By making use of 1.9k candidates with a signal purity of 79.3\%, Ref.~\cite{BESIII:2022jts} reported
the first amplitude analysis and branching fraction measurement of this decay channel.
Double-tag candidates were used in the amplitude analysis of $D^+_s\to 2\pi^+\pi^-2\pi^0$,
from 7.33 fb$^{-1}$ of data at 4.128-4.226 GeV.
Figure~\ref{fig:Ds_5pi} provides the projections of the amplitude analysis fit on two-body, three-body, and four-body particle mass distributions.
The fit fractions of each component in the amplitude analysis are
$D_s^{+}\to \omega\rho(770)^+$ [$(20.0\pm1.4^{+0.9}_{-1.2})\%$],
$D_s^{+}\to \phi\rho(770)^+$ [$(13.9\pm1.0^{+0.5}_{-0.4})\%$],
$D_s^{+}\to \rho(770)(1450)^+(\to \omega\pi^+)\pi^0$ [$(7.8\pm0.8^{+0.5}_{-0.6})\%$],
$D_s^{+} \to a_1(1260)^0(\to \rho(770)^+\pi^-)\rho(770)^+$ [$(11.4\pm0.7\pm0.4)\%$],
$D_s^{+} \to a_1(1260)^0( \to \rho(770)^-\pi^+)\rho(770)^+$ [$(7.4\pm0.4\pm0.3)\%$],
$D_s^{+} \to a_1(1260)^+(\to \rho(770)^+\pi^0)\rho(770)^0$ [$(16.5\pm1.4\pm1.5)\%$],
$D_s^{+}\to b_1(1235)^+(\to \omega\pi^+)\pi^0$ [$(10.8\pm1.0\pm0.7)\%$], and
$D_s^{+}\to b_1(1235)^0(\to \omega\pi^0)\pi^+$  [$(14.6\pm1.2\pm0.5)\%$].

Especially, the pure $W$-annihilation decay $D_s^+ \to \omega\rho(770)^+$ is observed for the first time,
with a branching fraction of $(0.99\pm0.08^{+0.05}_{-0.07})\%$.
It is of the same order of magnitude as $D_s^+ \to a_0(980)^{+(0)}\pi^{0(+)}$ and far larger than other WA processes.
In comparison to the low significance of the $\mathcal{D}$ wave in the $D_s^+ \to \phi\rho(770)^+$ decay,
the dominance of the $\mathcal{D}$ wave over the $\mathcal{S}$ and $\mathcal{P}$ waves, with a fraction of $(51.9\pm7.3{\ ^{+4.8}_{-7.9}})\%$ is observed in the decay $D_s^+ \to \omega\rho(770)^+$. This fraction deviates from the expectation of the naive factorization model~\cite{Cheng:2022vbw},
offering important insights for unraveling the ``polarization puzzle".
The branching fraction of $D^+_s\to 2\pi^+\pi^-2\pi^0$ is measured to be ($4.41\pm0.15\pm0.13$)\%,
and the branching fractions of each subdecay are also presented.
In addition, the obtained ${\cal B}(D_s^+\to \omega\pi^+\pi^0)=(2.31\pm0.13{\ ^{+0.10}_{-0.11}})\%$ is consistent with the CLEO-c measurement~\cite{CLEO:2009nsf} within uncertainties.
The branching fraction of $D_s^+ \to \phi\rho(770)^+$ is measured to be $(3.98\pm0.33^{+0.21}_{-0.19})\%$, and the ${\cal R}_{\phi}= {\mathcal{B}(\phi\to\pi^+\pi^-\pi^0)}/{\mathcal{B}(\phi\to K^+K^-)}$ is determined to be $0.222\pm0.019{\ ^{+0.016}_{-0.016}}$, which is consistent with the previous measurement based on charmed meson decays~\cite{BESIII:2024muy}, but deviates from the results from $e^+e^-$ annihilation and $K$-$N$ scattering experiments by more than 3$\sigma$.

\section{Rare decays}
\label{sec:raredecays}

In the SM, the flavor-changing-neutral-current (FCNC) processes are strongly  suppressed by the GIM mechanism~\cite{Glashow:1970gm}  and can only occur at loop level. Such processes have been observed in $K$ and $B$ meson decays, {\it e.g.}, the decays $K^{\pm} \to
\pi^{\pm}\mu^{+}\mu^{-}$~\cite{NA482:2010zrc}, and $B^{0} \to K^*(892)^0\mu^{+}\mu^{-}$~\cite{LHCb:2013zuf}, where heavy virtual quarks, particularly the top quark, contribute significantly in the loops.
By contrast, in the charm sector there is no heavy particle like the top quark in the loop diagram $c \to u\ell^+\ell^-$ ($\ell$ = $e$ or $\mu$), $c \to u \gamma$ or  $c \to u\nu\bar \nu$. Consequently, the GIM suppression in charm FCNC decays is much stronger than that in the corresponding processes of $B$ and $K$ mesons, leading to a theoretically very small branching fraction that would not exceed the level of $10^{-9}$~\cite{Fajfer:2001sa, Paul:2011ar,Cappiello:2012vg}.
On the other side, possible new physics can significantly increase the decay rates of these short-distance  processes. Hence, they can serve as clean channels in experiments to search for new physics~\cite{Fajfer:2001sa, Paul:2011ar}.
However, these decay rates of $D$ mesons also receive substantial contributions from long distance effects through vector meson decays, like $D \to h V,V \to \ell^+\ell^-$, where $h$ is a hadron.
These long-distance effects can mimic the signal and typically yield branching fractions of order
$10^{-6}$~\cite{Paul:2011ar,Cappiello:2012vg}, thereby overshadowing the short-distance component.
To disentangle the short-distance effects, a measurement of the angular dependence or $C\!P$ asymmetry~\cite{Cappiello:2012vg} is essential. Moreover, comparing the electron and muon channels in these decays provides a clean test of lepton flavor universality, another key prediction of the SM~\cite{Cappiello:2012vg,Gisbert:2020vjx}.
References~\cite{Burdman:1995te,Fajfer:2015zea} expect that long-distance processes or potential contributions from new physics could enhance the branching fractions  of these decays by two to three orders of magnitude relative to short-range interactions.

Before BESIII, searches for $D^0\to \gamma\gamma$ came from
CLEOII~\cite{CLEO:2002yyt}, BaBar~\cite{BaBar:2011bxy}, and Belle~\cite{Belle:2015pzk}.
Searches for $D^0\to p \ell^-$ and $D^0\to \bar p \ell^+$ came from CLEO-c~\cite{CLEO:2009apb} and Belle~\cite{Belle:2023mao}.
Searches for $D^0\to \ell^+\ell^-$ came from
MARKII~\cite{Riles:1986jg},
MARKIII~\cite{MARK-III:1987wjm,MARK-III:1987jdg},
ACCMOR~\cite{ACCMOR:1987dby},
EMC~\cite{EuropeanMuon:1985mxw},
SPEC~\cite{Biino:1985qc},
CLEO~\cite{CLEO:1987exm},
ARGUS~\cite{ARGUS:1988yhb},
E789~\cite{E789:1994bzi},
E653~\cite{E653:1995rpz},
BEATRICE~\cite{BEATRICE:1995siq,BEATRICE:1997duj},
CLEOII~\cite{CLEO:1996jxx},
E771~\cite{E771:1994ytb},
E791~\cite{E791:1999vct,E791:1999vct},
E789~\cite{E789:1999lnm},
FOCUS~\cite{E791:1999vct},
CDF~\cite{CDF:2003guv},
BaBar~\cite{BaBar:2004uxd,BaBar:2012ujp},
HERAB~\cite{HERA-B:2004dun},
Belle~\cite{Belle:2010ouj},
CDF~\cite{CDF:2010vpg}, and
LHCb~\cite{LHCb:2013jyo,LHCb:2015pce,LHCb:2022jaa}.
Searches for $D^0\to h \ell \ell$, $D^0\to hh\ell \ell$, $D^+\to h \ell\ell$, and $D^+_s\to h\ell \ell$ came from
E653~\cite{E653:1995rpz},
CLEO~\cite{CLEO:1987exm},
MARKII~\cite{Weir:1989sq},
MARKIII~\cite{MARK-III:1988gbl},
CLEOII~\cite{CLEO:1996jxx},
E687~\cite{E687:1997alg},
E791~\cite{E791:1995por,E791:1999vct},
FOCUS~\cite{FOCUS:2003gvx},
CLEO-c~\cite{CLEO:2005hbm,CLEO:2010ksb},
BaBar~\cite{BaBar:2011ouc,BaBar:2018bwm,BaBar:2020faa},
E791~\cite{E791:2000jkj}, and LHCb~\cite{LHCb:2013hxr,LHCb:2017uns,LHCb:2020car}.
In addition,
BaBar~\cite{BaBar:2008kjd} determined the branching fraction
of $D^0\to \gamma \phi$ and $D^0\to \gamma \bar K^*(892)^0$ relative to $D^0\to K^-\pi^+$;
Belle~\cite{Belle:2016mtj} determined the branching fraction
of $D^0\to \gamma \phi$ relative to $D^0\to K^+K^-$,
that of $D^0\to \gamma \rho(770)^0$ relative to $D^0\to \pi^+\pi^-$  and
that of $D^0\to \gamma \bar K^*(892)^0$ relative to $D^0\to K^-\pi^+$.

\begin{figure*}[htbp]
\includegraphics[width=0.8\linewidth, trim=0 10pt 0 0, clip]{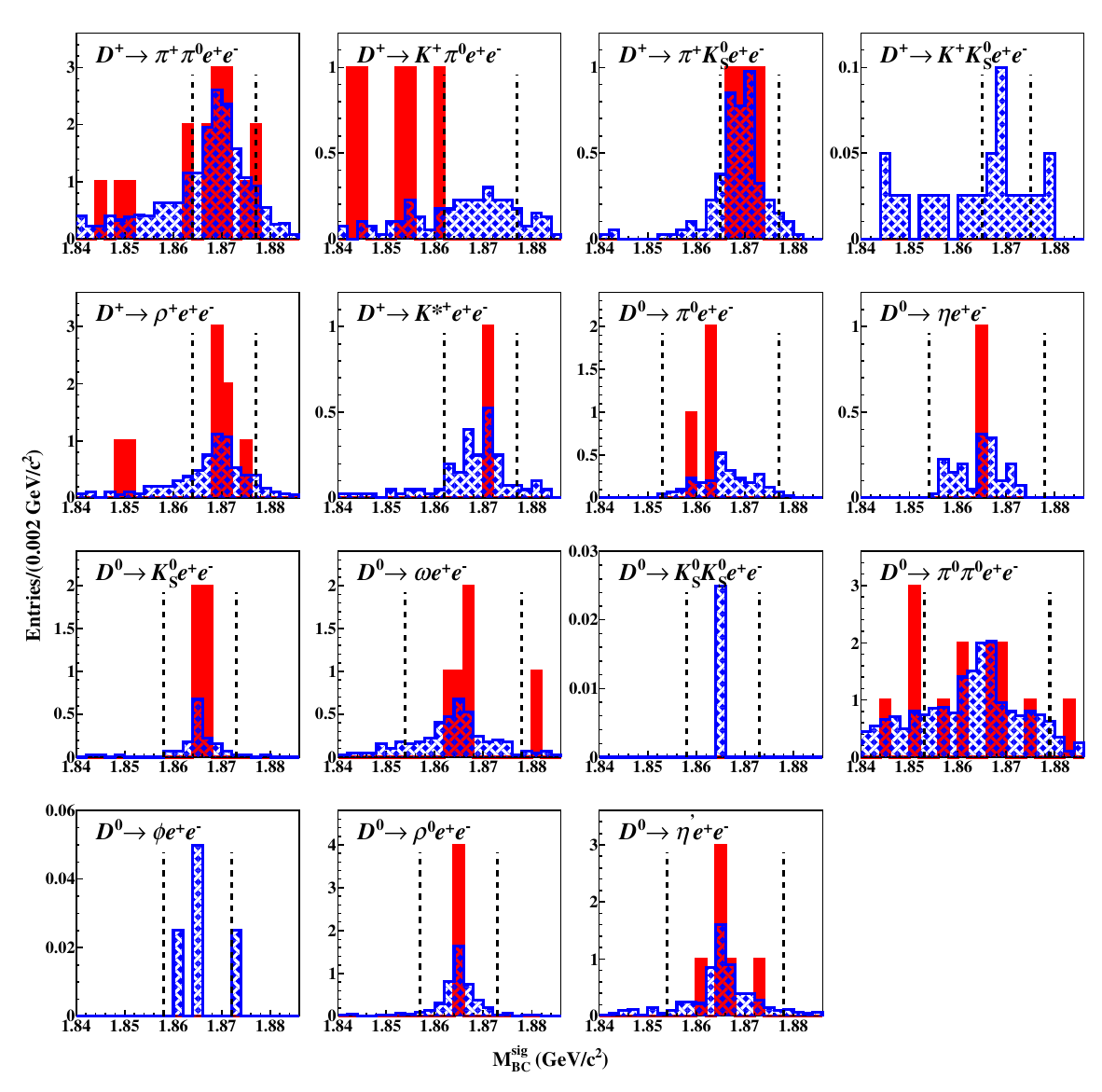}
\caption{The distributions of $M_{\rm BC}^{\rm sig}$ of accepted candidates for $D^{0(+)}\to he^+e^-$ and $D^{0(+)}\to hh^\prime e^+e^-$~\cite{BESIII:2026ujm}.
The solid histograms are data, the hatched ones are the simulated background events, and the black dashed lines denote the signal region.
 }
\label{fig:D_hhee}
\end{figure*}

At BESIII, the FCNC decays
$D^0\to\gamma\gamma$~\cite{BESIII:2015kvk},
$D^0\to \nu\nu$~\cite{BESIII:2021slf},
$D^+\to \omega(\gamma)\gamma^\prime$~\cite{BESIII:2024rkp}, and
$D\to h(h^\prime)e^+e^-$~\cite{BESIII:2018hqu,BESIII:2026ujm},
are searched with the double-tag method,
by analyzing 2.93, 2.93, 7.9 and 20.3 fb$^{-1}$ of data at 3.773 GeV, respectively.
No significant signals of these decays are observed, and the upper limits on the branching fractions of these decays at the 90$\%$ confidence level
are set, as shown in Table~\ref{tab:NP_related}.
The results on $D^0\to \nu\nu$, $D^+\to \omega\gamma^\prime$, $D^+\to\gamma\gamma^\prime$,
$D^+ \to \pi^+\pi^0 e^+ e^- $, $D^+ \to K^+\pi^0 e^+ e^- $,
$D^+ \to K^0_S\pi^+ e^+ e^- $, $D^+ \to K^0_SK^+ e^+ e^- $, $D^+\to \rho(770)^{+}  e^+ e^-$, $D^+\to K^*(892)^+  e^+ e^-$, $D^0\to K_S^0 K_S^0 e^+ e^-$, $D^0\to 2\pi^0 e^+ e^-$, and
$D^0\to \eta^{\prime}  e^+ e^-$ offer the first constraints;
while the upper limits on
$D^0\to \pi^0 e^+ e^-$, $D^0\to \eta e^+ e^-$, $D^0\to \omega e^+ e^-$, $D^0\to K_S^0 e^+ e^-$, $D^+\to \pi^+ \pi^0 e^+ e^-$, $D^+\to K^+ \pi^0 e^+ e^-$, $D^+\to
\pi^+ K_S^0 e^+ e^-$, and $D^+\to K^+ K_S^0 e^+ e^-$
are significantly improved compared with previous measurements.
Also, the results reported in Ref.~\cite{BESIII:2024rkp} provide the most stringent constraint on the new physics energy scale associated with $cu\gamma'$ coupling in the world, with the new physics energy scale related parameter $|\mathbb{C}|^2+|\mathbb{C}_5|^2<8.2\times10^{-17}{\rm GeV}^{-2}$ at the 90\% confidence level.
As examples, Fig.~\ref{fig:D_hhee} shows the distributions of $M_{\rm BC}^{\rm sig}$ for the signal decays $D^{0(+)}\to he^+e^-$ and $D^{0(+)}\to hh^\prime e^+e^-$ in data.

\begin{figure*}[htp]
    \centering
    \includegraphics[width=0.3\textwidth]{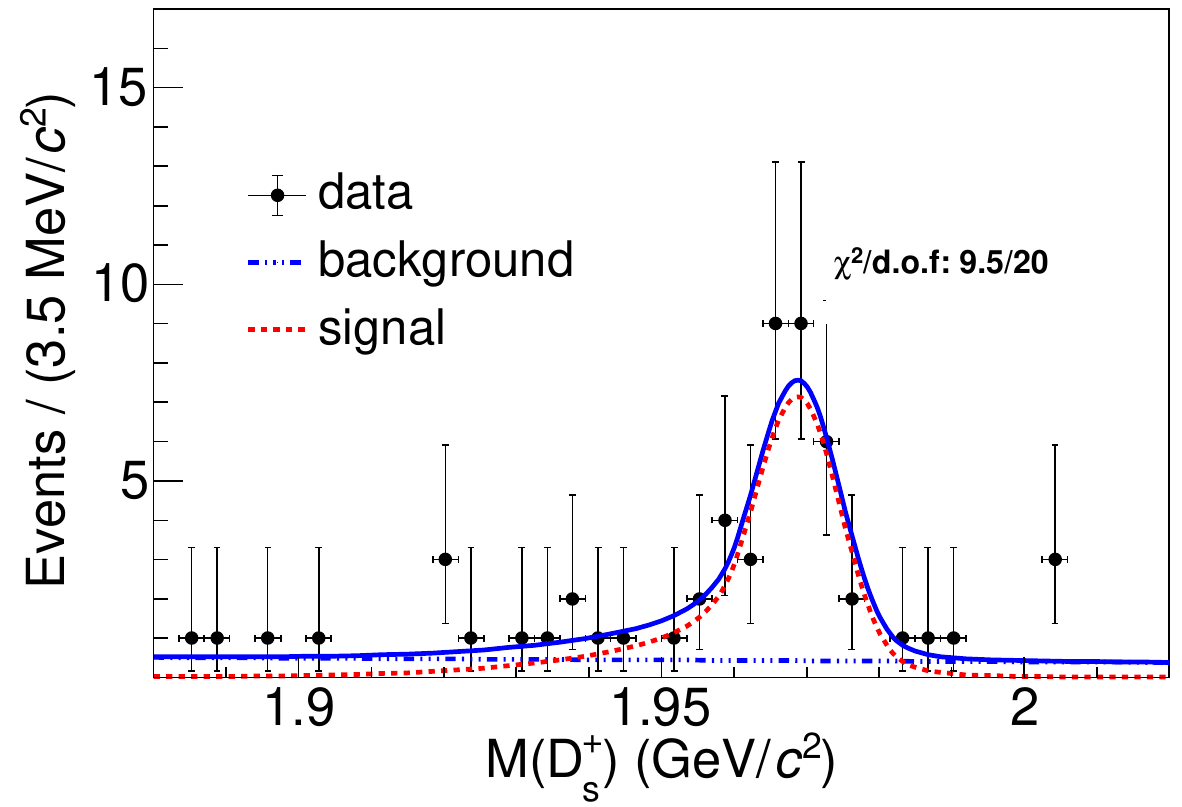}
    \put(-105,90){\color{black}{\scriptsize $D_s^+\to\pi^+\phi,\phi\to e^+e^-$}}
    \includegraphics[width=0.3\textwidth]{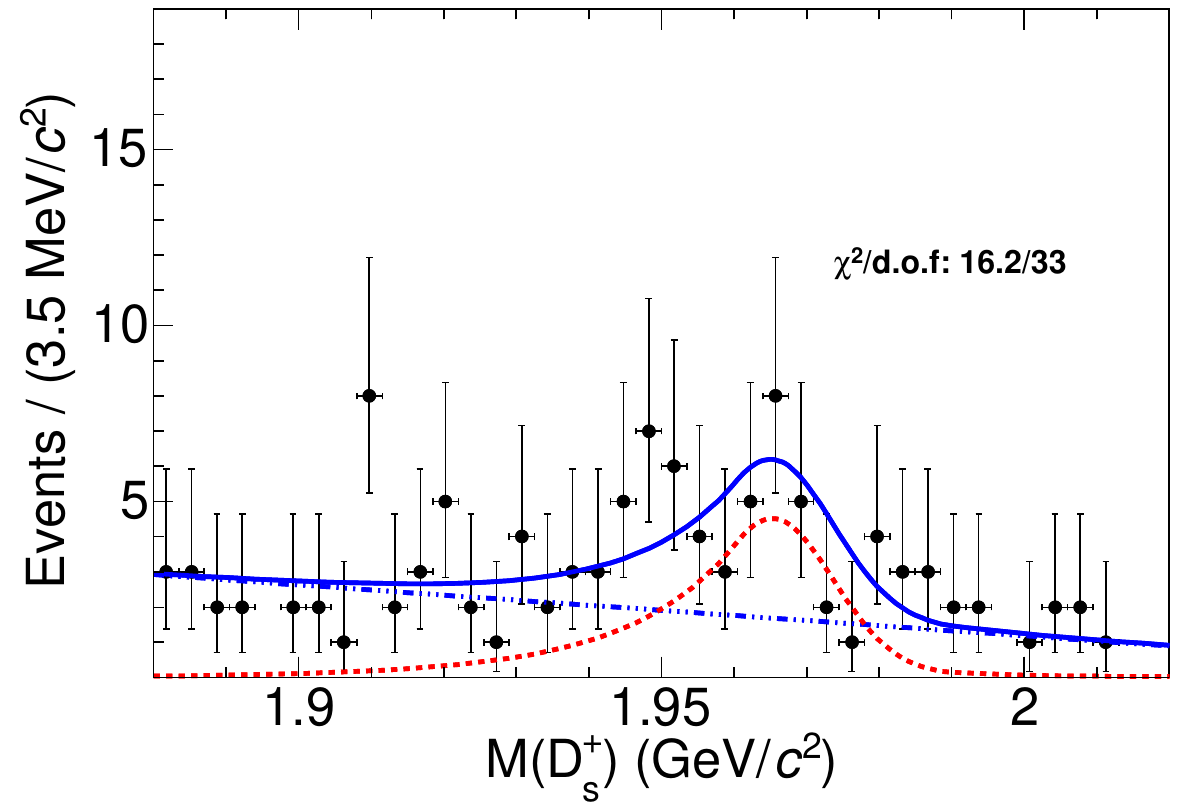}
    \put(-130,90){\color{black}{\scriptsize $D_s^+\to\rho(770)^+\phi$}}
    \put(-105,80){\color{black}{\scriptsize $\rho(770)^+\to\pi^+\pi^0,\phi\to e^+e^-$}}
\\[0.1cm]
    \includegraphics[width=0.3\textwidth]{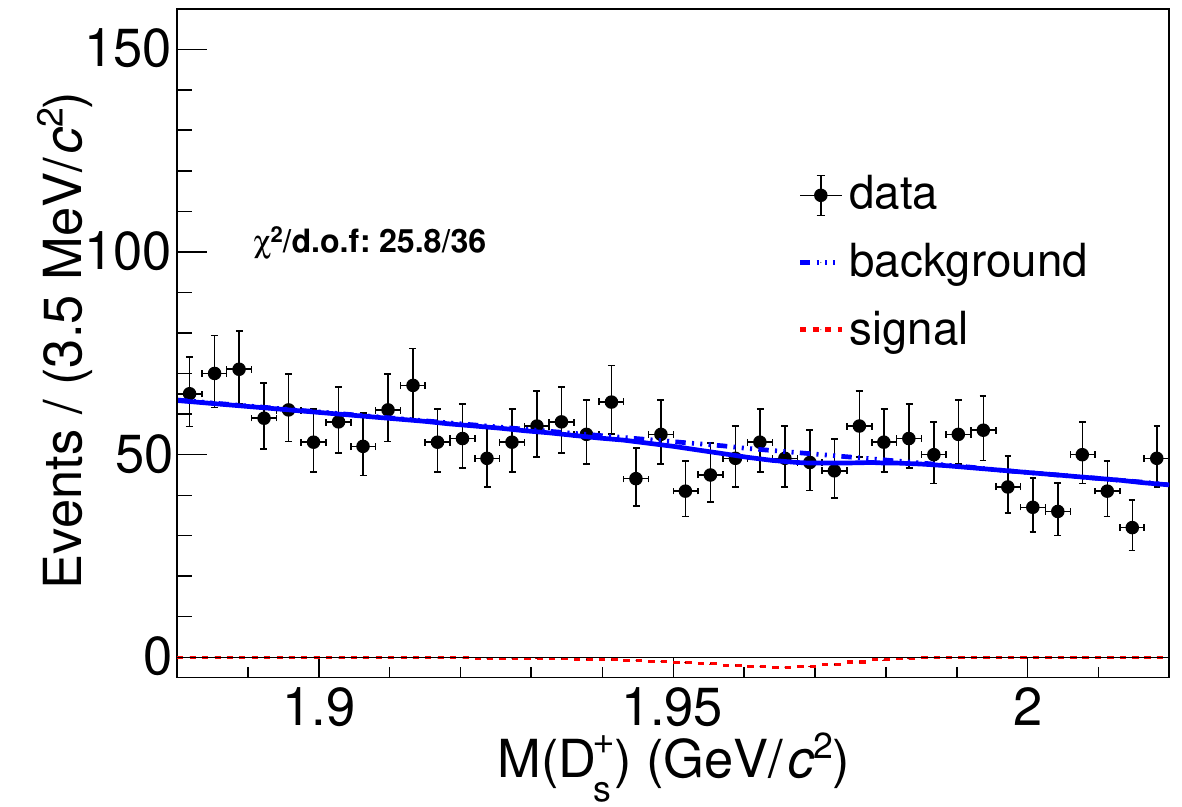}
    \put(-100,90){\color{black}{\scriptsize $D_s^+\to\pi^+\pi^0e^+e^-$}}
    \includegraphics[width=0.3\textwidth]{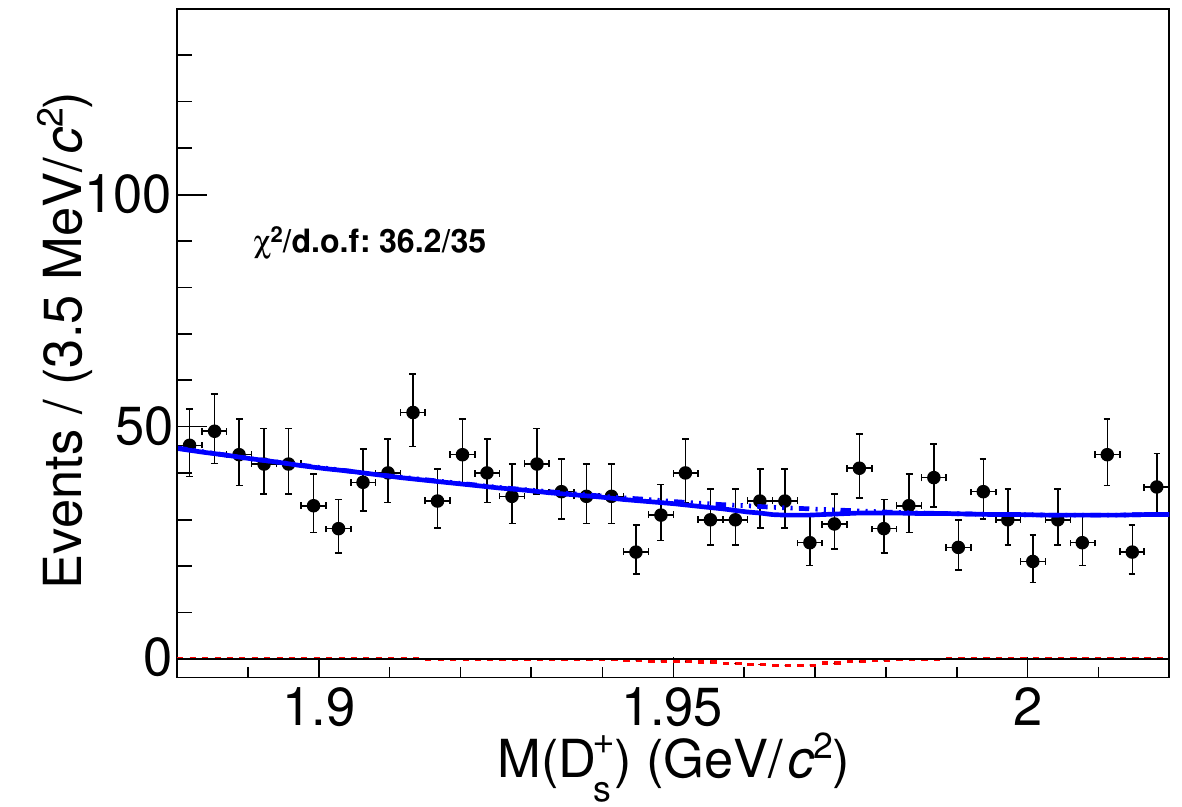}
    \put(-100,90){\color{black}{\scriptsize $D_s^+\to K^+\pi^0e^+e^-$}}
    \includegraphics[width=0.3\textwidth]{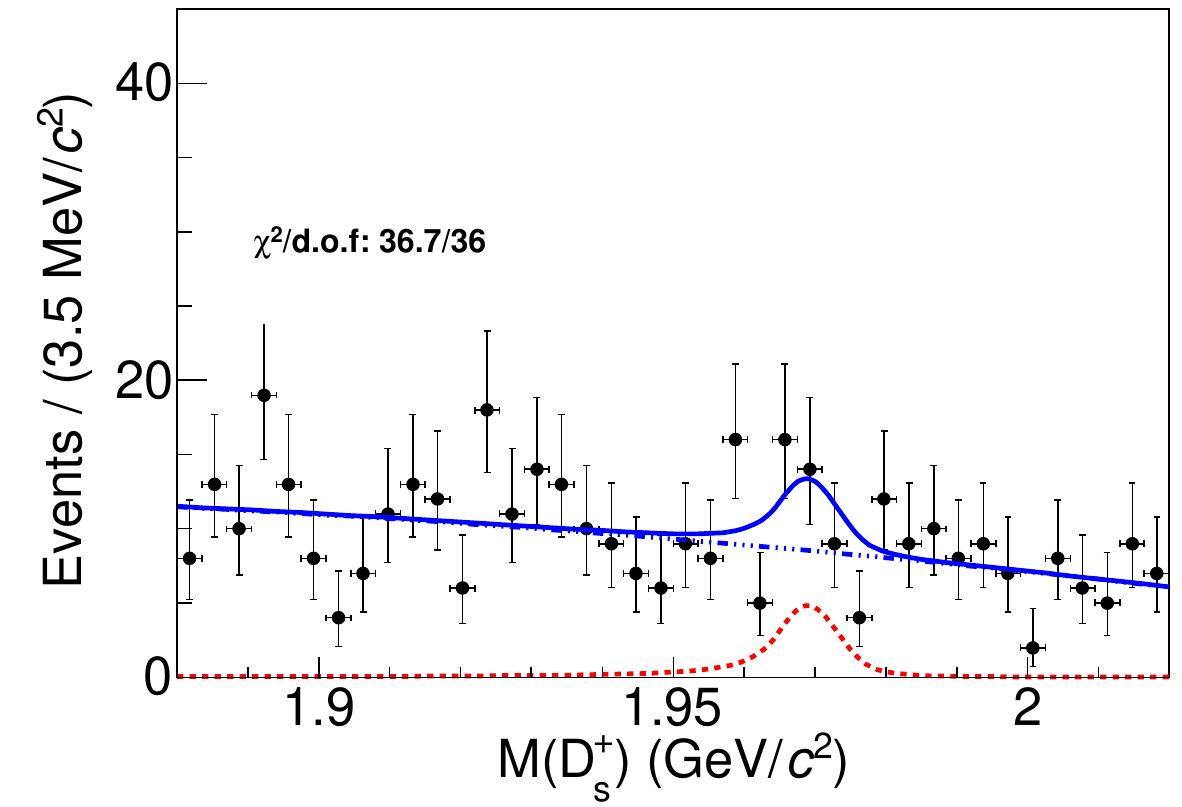}
   \put(-100,90){\color{black}{\scriptsize $D_s^+\to K_S^0\pi^+e^+e^-$}}
    \caption{
    Fits to the $M(D_s^+)$ distributions of accepted candidates for $D^+_s\to he^+e^-$ and $D^+_s\to hh^\prime e^+e^-$~\cite{BESIII:2024nrw}.
    }
    \label{fit:pwa-center}
    \end{figure*}

Using 7.33 fb$^{-1}$ of data at 4.128-4.226 GeV, the rare decays $D_s^+\to h^+(h^0)e^+e^-$ are searched~\cite{BESIII:2024nrw}.
The $D_s^+\to\pi^+\phi,\phi\to e^+e^-$ decay is observed with a statistical significance of 7.8$\sigma$
and evidence for the $D_s^+\to \rho(770)^+\phi,\phi\to e^+e^-$ decay is found for the first time with a statistical significance of 4.4$\sigma$.
Their branching fractions are measured to be
$\mathcal{B}(D_s^+\to\pi^+\phi,\phi\to e^+e^-)=(1.17^{+0.23}_{-0.21}\pm0.03)\times 10^{-5}$ and
$\mathcal{B}(D_s^+\to\rho(770)^+\phi, \phi\to e^+e^-)=(2.44^{+0.67}_{-0.62}\pm0.16)\times 10^{-5}$.
Both branching fractions are consistent with the products of the PDG values, $\mathcal{B}(D_s^+\to\pi^+\phi)\cdot\mathcal{B}(\phi\to e^+e^-)$ and $\mathcal{B}(D_s^+\to\rho(770)^+\phi)\cdot\mathcal{B}(\phi\to e^+e^-)$ within uncertainties.
No significant signal of the four-body rare decays is observed, and the upper limits on the branching fractions of these decays at the 90\% confidence level
are set to be $ \mathcal{B}(D_s^+\to\pi^+\pi^0e^+e^- )<7.0\times 10^{-5}$, $\mathcal{B}(D_s^+\to K^+\pi^0 e^+e^- )<7.1\times 10^{-5}$, and $ \mathcal{B}(D_s^+\to K_S^0\pi^+e^+e^-)<8.1\times 10^{-5}$, which are the first upper limits on these decays. The fits to the $M(D_s^+)$ distributions of accepted candidates for $D^+_s\to he^+e^-$ and $D^+_s\to hh^\prime e^+e^-$ in data are shown in Fig.~\ref{fit:pwa-center}.

The FCNC radiative decays $D^{+} \to \gamma \rho(770)^+$ and $D^{+} \to \gamma K^*(892)^+$~\cite{BESIII:2024zfv} as well as
$D^0\to \gamma\bar K_1(1270)^0$ and $D^+\to \gamma K_1(1270)^+$~\cite{BESIII:2026xmx}
are searched with 20.3~fb$^{-1}$ of data at 3.773 GeV, while
the $D^+_s\to \gamma \rho(770)^+$ decay is searched with 7.33 fb$^{-1}$ of data at 4.128-4.226 GeV~\cite{BESIII:2024thr}.
No significant signals are observed and their branching fraction upper limits at the 90\% confidence level are set to be
$1.3\times10^{-5}$, $1.8\times10^{-5}$, and $6.1\times 10^{-5}$, respectively. These results are beneficial for further understanding the long-range effects in rare weak radiative decays and for testing theoretical calculations of non-perturbative QCD~\cite{Burdman:1995te,Fajfer:2015zea}.

Based on 2.93 fb$^{-1}$ of data at 3.773 GeV,
the BNV and LNV decays
$D^+\to \Lambda e^+$,
$D^+\to \bar \Lambda e^+$,
$D^+\to \Sigma^0 e^+$, and
$D^+\to \bar \Sigma^0 e^+$ are searched with the single-tag method~\cite{BESIII:2019udi};
the BNV and LNV decays
$D^0\to pe^-$ and $D^0\to \bar pe^+$~\cite{BESIII:2021krj}
as well as
$D^+\to ne^+$ and $D^+\to \bar ne^+$\cite{BESIII:2022svy} are searched;
the LNV $\Delta L =2$ decays of $D\to \bar K \pi e^{+}e^{+}$~\cite{BESIII:2019oef}
are searched with the double-tag method.
No obvious signals are found, and the upper limits on the branching fractions of these
decays at the 90\% confidence level are set, as shown in Table~\ref{tab:NP_related}.
All of them are the first constraints.
Reference~\cite{BESIII:2019oef} also searched for a Majorana neutrino with different mass hypotheses.
Upper limits, which are of the level $10^{-7}\sim10^{-6}$, on the branching fractions are set at the 90\% confidence level for the decays $D^0\to K^- e^+ \nu_m (\pi^- e^+)$ and $D^+ \to K_S^0 e^+ \nu_m (\pi^- e^+)$ with different $m_{\nu_m}$ hypotheses within the range 0.25 to 1.0 GeV/$c^2$. The constraints on the mixing element $|V_{e\nu_m}|^{2}$ depending on $m_{\nu_m}$ are also evaluated based on the related variables and the measured branching fractions. The results provide the supplementary information in the study of mixing between the heavy Majorana neutrino and the SM neutrino $\nu_e$ in $D$ meson decays.

\begin{table*}[htbp]
	\caption{Results on searches for rare decays of charmed mesons. The superscript $*$ denotes the first search and the $i$ denotes the searching sensitivity
is improved.}
	\centering
		\centering
		\begin{tabular}{ccccc}
			\hline
			\hline
			Decay  & Data (fb$^{-1}$ @ GeV) & ${\cal B}^{\rm UL}$ & Ref. &  Case \\ \hline
			$D^0 \to \gamma\gamma$ & 2.93 @3.773 & $3.8\times10^{-6}$ & \cite{BESIII:2015kvk} & FCNC  \\ \hline
			$D^{0} \to \pi^{0} \nu \bar{\nu}$ & 2.93 @3.773 & $2.1\times10^{-4}$$^*$ & \cite{BESIII:2021slf} & FCNC \\ \hline
			$D^0\to \omega\gamma^\prime$  & 7.9 @3.773 & $1.1\times 10^{-5}$$^*$ & \cite{BESIII:2024rkp} & FCNC  \\
			$D^0\to \gamma\gamma^\prime$ & 7.9 @3.773 & $2.0\times 10^{-6}$$^*$ & \cite{BESIII:2024rkp} & FCNC \\ \hline

			$D^0 \to K^-K^+ e^+ e^- $ & 2.93 @3.773 & $1.1\times10^{-5}$$^i$ & \cite{BESIII:2018hqu}  & FCNC \\
			$D^0 \to \pi^+\pi^- e^+ e^- $ & 2.93 @3.773 & $0.7\times10^{-5}$$^i$ & \cite{BESIII:2018hqu}  & FCNC \\
			$D^0 \to K^-\pi^+ e^+ e^- $ & 2.93 @3.773 & $4.1\times10^{-5}$$^i$ & \cite{BESIII:2018hqu}  & FCNC \\
			$D^+ \to \pi^+\pi^0 e^+ e^- $ & 20.3 @3.773 & $2.4\times10^{-6}$$^*$ & \cite{BESIII:2026ujm}  & FCNC \\
			$D^+ \to K^+\pi^0 e^+ e^- $ & 20.3 @3.773 & $1.1\times10^{-6}$$^*$ & \cite{BESIII:2026ujm}  & FCNC \\
			$D^+ \to K^0_S\pi^+ e^+ e^- $ & 20.3 @3.773 & $4.8\times10^{-6}$$^*$ & \cite{BESIII:2026ujm}  & FCNC \\
			$D^+ \to K^0_SK^+ e^+ e^- $ & 20.3 @3.773 & $3.7\times10^{-6}$$^*$ & \cite{BESIII:2026ujm}  & FCNC \\

			$D^+ \to \rho(770)^+ e^+ e^- $ & 20.3 @3.773 & $2.1\times10^{-6}$$^*$ & \cite{BESIII:2026ujm}  & FCNC \\
			$D^+ \to K^*(892)^+ e^+ e^- $ & 20.3 @3.773 & $2.5\times10^{-6}$$^*$ & \cite{BESIII:2026ujm}  & FCNC \\

			$D^0 \to \pi^0 e^+ e^- $ & 20.3 @3.773 & $0.7\times10^{-6}$$^i$ & \cite{BESIII:2026ujm}  & FCNC \\
			$D^0 \to \eta e^+ e^- $ & 20.3 @3.773 & $1.4\times10^{-6}$$^i$ & \cite{BESIII:2026ujm}  & FCNC \\
			$D^0 \to \omega e^+ e^- $ & 20.3 @3.773 & $1.6\times10^{-6}$$^i$ & \cite{BESIII:2026ujm}  & FCNC \\
			$D^0 \to K^0_S e^+ e^- $ & 20.3 @3.773 & $2.6\times10^{-6}$$^i$ & \cite{BESIII:2026ujm}  & FCNC \\

			$D^0 \to 2K^0_S e^+ e^- $ & 20.3 @3.773 & $7.8\times10^{-6}$$^*$ & \cite{BESIII:2026ujm}  & FCNC \\
			$D^0 \to 2\pi^0 e^+ e^- $ & 20.3 @3.773 & $1.9\times10^{-6}$$^*$ & \cite{BESIII:2026ujm}  & FCNC \\

			$D^0 \to \phi e^+ e^- $ & 20.3 @3.773 & $4.6\times10^{-6}$$^i$ & \cite{BESIII:2026ujm}  & FCNC \\
			$D^0 \to \rho(770)^0 e^+ e^- $ & 20.3 @3.773 & $0.7\times10^{-6}$$^i$ & \cite{BESIII:2026ujm}  & FCNC \\
			$D^0 \to \eta^\prime e^+ e^- $ & 20.3 @3.773 & $2.5\times10^{-6}$$^*$ & \cite{BESIII:2026ujm}  & FCNC \\ \hline
			$D_s^+\to\pi^+\phi(\to e^+e^-)$ & 7.33 @(4.128-4.226) & $(1.17^{+0.23}_{-0.21}+0.03)\times10^{-5}$ & \cite{BESIII:2024nrw} & FCNC  \\
			$D_s^+\to\rho(770)^+\phi(\to e^+e^-)$ & 7.33 @(4.128-4.226) & $(2.24^{+0.67}_{-0.62}+0.16)\times10^{-5}$$^*$ & \cite{BESIII:2024nrw} & FCNC  \\
			$D_s^+\to\pi^+\pi^0e^+e^-$ & 7.33 @(4.128-4.226) & $7.0\times10^{-5}$$^*$ & \cite{BESIII:2024nrw} & FCNC  \\
			$D_s^+\to K^+\pi^0e^+e^-$ & 7.33 @(4.128-4.226) & $7.1\times10^{-5}$$^*$ & \cite{BESIII:2024nrw} & FCNC  \\
			$D_s^+\to K^0_S\pi^+e^+e^-$ & 7.33 @(4.128-4.226) & $8.1\times10^{-5}$$^*$ & \cite{BESIII:2024nrw} & FCNC  \\ \hline
			$D^+\to\gamma\rho(770)^+$ & 20.3 @ 3.773 & $1.3\times 10^{-5}$$^*$ & \cite{BESIII:2024zfv} & FCNC,Radiative \\
			$D^+\to\gamma K^*(892)^+$ & 20.3 @ 3.773 & $1.8\times 10^{-5}$$^*$ & \cite{BESIII:2024zfv} & FCNC,Radiative \\
			$D^0\to\gamma \bar K_1(1270)^0$ & 20.3 @ 3.773 & $7.7\times 10^{-4}$$^*$ & \cite{BESIII:2026xmx} & FCNC,Radiative \\
			$D^+\to\gamma K_1(1270)^+$ & 20.3 @ 3.773 & $3.9\times 10^{-5}$$^*$ & \cite{BESIII:2026xmx} & FCNC,Radiative \\
			\hline
			$D^+_s \to \gamma \rho(770)^+$ & 7.33 @(4.128-4.226) & $6.1\times 10^{-5}$$^*$ & \cite{BESIII:2024thr} & FCNC,Radiative  \\  \hline
			 $D^+\to \Lambda e^+$ & 2.93 @3.773 & $1.1\times 10^{-6}$$^*$ & \cite{BESIII:2019udi}  & BNV,LNV \\
			 $D^+\to \bar\Lambda e^+$ & 2.93 @3.773 & $6.5\times 10^{-7}$$^*$ & \cite{BESIII:2019udi} & BNV,LNV  \\
			 $D^+\to \Sigma^0e^+$ & 2.93 @3.773 & $1.7\times 10^{-6}$$^*$ & \cite{BESIII:2019udi} & BNV,LNV  \\
			 $D^+\to \bar\Sigma^0e^+$ & 2.93 @3.773 & $1.3\times 10^{-6}$$^*$ & \cite{BESIII:2019udi} & BNV,LNV  \\  \hline
			$D^0 \to\bar p e^-$ & 2.93 @3.773 & $1.2\times 10^{-6}$$^i$ & \cite{BESIII:2021krj} & BNV,LNV \\
			$D^0 \to p e^+$ & 2.93 @3.773 & $2.2\times 10^{-6}$$^i$ & \cite{BESIII:2021krj} & BNV,LNV \\  \hline			
			$D^{+} \to \bar n e^+$ & 2.93 @3.773 & $1.43\times 10^{-5}$$^*$ & \cite{BESIII:2022svy} & BNV,LNV  \\
			$D^{+} \to n e^+$ & 2.93 @3.773 & $2.91\times 10^{-5}$$^*$ & \cite{BESIII:2022svy} & BNV,LNV  \\  \hline
			$D^0\to K^- \pi^- 2e^+$ & 2.93 @3.773 & $2.8\times 10^{-6}$ & \cite{BESIII:2019oef} & LNV \\
			$D^+\to K^0_S \pi^- 2e^+$ & 2.93 @3.773 & $3.3\times 10^{-6}$ & \cite{BESIII:2019oef} & LNV \\
			$D^+\to K^- \pi^0 2e^+$ & 2.93 @3.773 & $8.5\times 10^{-6}$ & \cite{BESIII:2019oef} & LNV \\
 \hline
			\hline
		\end{tabular}
\label{tab:NP_related}
\end{table*}

\section{Other measurements of $D^{*0}$,
$D^{*+}$ and $D_{s}^{*+}$}
\label{sec:others}

Before BESIII, MARKIII~\cite{Adler:1988fz}, CLEOII~\cite{CLEO:1992xqa}, and ARGUS~\cite{ARGUS:1994bem} reported the branching fraction of $D^{*0}\to D^0\gamma$; while LHCb~\cite{LHCb:2021jfy} and BaBar~\cite{BaBar:2005wmf} reported the branching fraction of $D^{*0}\to D^0\pi^0 $ relative to $D^{*0}\to D^0\gamma $.
In addition, CLEOII~\cite{CLEO:1992xqa,CLEO:1997rew} and ARGUS~\cite{ARGUS:1994bem} reported the branching fraction of $D^{*+}\to D^0\pi^+$; and CLEOII also reported the branching fractions of $D^{*+}\to D^+\pi^0$ and $D^{*+}\to D^+\gamma$; while CLEOII~\cite{CLEO:1995ewe} and BaBar~\cite{BaBar:2005wmf} reported the branching fraction of $D^{*+}_s\to \pi^0D^+_s$ relative to $D^{*+}_s\to \gamma D^+_s$.

Using 0.48~fb$^{-1}$ of data at $4.009$~GeV, BESIII measured
the branching fractions of the decays of $D^{*0}$ into
$D^0\pi^0$ and $D^0\gamma$ to be ${\cal B}(D^{*0}
\to D^0\pi^0)=(65.5\pm 0.8\pm 0.5)\%$ and ${\cal B}(D^{*0} \to
D^0\gamma)=(34.5\pm 0.8\pm 0.5)\%$, respectively\cite{BESIII:2014rqs}, by assuming that the $D^{*0}$ decays only into these two modes.
The ratio of the two branching fractions is determined to be ${\cal B}(D^{*0} \to D^0\pi^0)/{\cal B}(D^{*0} \to D^0\gamma) =1.90\pm 0.07\pm 0.05$, which is independent of the aforementioned
assumption. This measurement is consistent with the previous ones~\cite{ParticleDataGroup:2014cgo} within uncertainties but with much better precision. These much improved results are helpful for updating the parameters in the effective models mentioned above, such as the mass of the charm quark~\cite{Eichten:1979ms,Aliev:1994nq}, the effective coupling constant~\cite{Cheng:1993gc}, and the magnetic moment of the charm quark~\cite{Miller:1988tz}. 

Based on 7.33~fb$^{-1}$ of data at $4.128-4.226$~GeV, BESIII measured
the branching fraction of $D^{*+}_s\to D^+_s\pi^0$ relative to that of $D^{*+}_s\to D^+_s\gamma$ to be
$(6.16\pm 0.43\pm 0.19)\%$.
Using the world average value of the branching fraction of $D^{*+}_s\to D^+_se^+e^-$
leads to the branching fractions of $D^{*+}_s\to D^+_s\gamma$ and $D^{*+}_s\to D^+_s\pi^0$
to be $\mathcal B(D^{*+}_s\to D^+_s\gamma)=(93.57\pm0.44\pm0.19)\%$ and $\mathcal B(D^{*+}_s\to D^+_s\pi^0)=(5.76\pm0.44\pm0.19)\%$~\cite{BESIII:2022kbd}, which are consistent with previous measurements by CLEO-c~\cite{CLEO:2011mla} and BaBar~\cite{BaBar:2005wmf}, and those theoretical calculations based on the covariant model~\cite{Cheung:2015rya}.

The spin and parity of the charmed mesons $D_{s}^{*+}$, $D^{*0}$ and
$D^{*+}$~\cite{Gell-Mann:1964ewy,DeRujula:1975qlm,Gaillard:1974mw,Glashow:1970gm} are determined for the first time to be $J^P=1^{-}$ with
significances greater than 10$\sigma$ over other hypotheses of $2^{+}$
and $3^{-}$~\cite{BESIII:2023abz}, using
3.19~fb$^{-1}$ of data at $4.178$~GeV. This is the first experimental determination of the
spin and parity of the $D_{(s)}^*$ mesons, which are the cornerstone for the exploration of the properties of heavier
charm and beauty mesons~\cite{Asner:2008nq}.

Based on 3.19~fb$^{-1}$ of data at $4.178$~GeV,
the electromagnetic Dalitz decay $D^{*0}\to D^0e^+e^-$ is observed for the first time with a statistical significance of $13.2\sigma$.
The ratio of the branching fraction of $D^{*0}\to D^0e^+e^-$ to that of $D^{*0}\to D^0\gamma$ is measured to be
$(11.08\pm0.76\pm0.49)\times 10^{-3}$. Using the world average value~\cite{ParticleDataGroup:2022pth} of the branching fraction of $D^{*0}\to D^0\gamma$,
the branching fraction of $D^{*0}\to D^0e^+e^-$ leads to $(3.91\pm0.27\pm0.17\pm0.10)\times 10^{-3}$~\cite{BESIII:2021vyq}.

\section{Prospect and summary}
\label{sec:summary}

Currently, the BESIII experiment has achieved a series of world's leading and important breakthroughs
in several key areas, including (semi)leptonic $D$ decays,
the strong-phase difference in hadronic decays of neutral charmed mesons, amplitude analyses of multi-body
hadronic $D$ decays, absolute branching fraction measurements of hadronic $D$ decays, and searches for rare
$D$ decays. Some elementary physical quantities have been determined with remarkable precision: for instance,
the decay constants of charmed mesons have been measured to about 1\%, the best measurement of hadronic transition form
factors in semileptonic charmed meson decays to better than 0.3\%, and the CKM matrix elements $|V_{cs}|$ and
$|V_{cd}|$ to 0.5\% and 1\%, respectively. By reweighting measurements of (semi)leptonic $D$ decays from all experiments, we have also provided the most precise averages for the CKM matrix elements  $|V_{cs}|=0.9686\pm0.0009\pm0.0038$ and $|V_{cd}|=0.2259\pm0.0014\pm0.0013$, the decay constants $f_{D^+}=(213.1\pm2.0\pm1.5)$ MeV and $f_{D^+_s}=(253.2\pm1.2\pm1.6)$ MeV, as well as the hadronic transition form factors  $f^{D\to K}_+(0)=0.7376\pm0.0007\pm0.0013$, $f^{D\to \pi}_+(0)=0.6337\pm0.0053\pm0.0037$, $f^{D\to \eta}_+(0)=0.351\pm0.009\pm0.005$, $f^{D\to \eta^\prime}_+(0)=0.263\pm0.025\pm0.006$, $f^{D_s\to \eta}_+(0)=0.4653\pm0.0058\pm0.0069$, $f^{D_s\to \eta^\prime}_+(0)=0.535\pm0.020\pm0.011$, and $f^{D_s\to K^0}_+(0)=0.627\pm0.036\pm0.009$. A series of strong-phase difference parameters for neutral hadronic $D$ mesons have been extracted with the
world's highest precision, and amplitude analyses of dozens of multi-body hadronic decays have been performed.
Moreover, dozens of new decay modes of charmed mesons have been observed;
tests of lepton flavor universality with (semi)leptonic $D$ decays have been reported;
DCS $D$ decays and $W$-annihilation $D$ decays with anomalously large branching fractions are observed,
along with an anomalous behavior in the branching fraction of the $\phi$ meson and the manifestation of vector meson
polarization in $D\to VV$ decays.

However, compared with CF $D$ decays, the progress in studying SCS and DCS $D$ decays has been relatively slower.
In the future, the BESIII experiment will continue experimental investigations of various $D$ decays~\cite{Asner:2008nq,BESIII:2020nme},
with particular emphasis on: precise measurements of SCS semileptonic $D$ decays; accurate determinations
of hadronic transition form factors; precision measurements of hadronic decays with final states containing
$\gamma$, $\pi^0$, $\eta$, and $K^0_L$; studies of asymmetries in exclusive $D\to K^0_SX$ and $D\to K^0_LX$ decays;
and the exploration of intermediate structures in multi-body DCS $D$ decays. In addition, searches for rare $D$ decays
will be carried out with higher sensitivity. Especially,
the application of machine learning techniques in particle physics experiments opens up
new and promising avenues for discovering rare processes~\cite{Li:2026krn}.

In 2025, the BEPCII was upgraded and the inner chamber of BESIII was upgraded to CGEM~\cite{BESIII:2020nme,Amoroso:2021glb,Melendi:2026gji,Gramigna:2025lci,Gramigna:2024gyu,Balossino:2022ywn}.
The data-taking capability of the BESIII experiment above 4 GeV is expected to increase by a factor of 2 to 4.
The operating energy of the BEPCII collider will be extended to around 5.6 GeV, allowing for the continued accumulation
of high-quality charmed meson samples. These enhancements will further expand the potential for charmed meson
physics studies at BESIII, opening up new opportunities for precision measurements and rare decay searches in the charm sector~\cite{BESIII:2020nme}.

\section{Acknowledgement}

This work is supported in part by National Natural Science Foundation of China (NSFC) under Contracts Nos. 12192264 and 12305089.

\bibliography{reference}

\end{document}